%% file: main.tex
\documentclass[12pt]{report}

\usepackage[nottoc,numbib]{tocbibind}
\usepackage{uhthesis2023}
\usepackage{mathptmx}
\usepackage[T1]{fontenc}
\usepackage[pdftex]{graphicx}
\usepackage{amsmath}
\usepackage{amssymb}
\usepackage{siunitx}
\usepackage{booktabs}
\usepackage{subcaption}
\usepackage{multirow}
\usepackage{longtable}
\usepackage{xspace}
\usepackage{physics}
\graphicspath{{./}}

\usepackage{hyperref}

\usepackage[%
	backend=biber,
	style=trad-abbrv,%
	citestyle=numeric-comp,%
	sorting=none,%
	minnames=5,%
	maxnames=5,%
	doi=false,%
	url=false%
]{biblatex}

\AtEveryBibitem{
    \clearfield{day}
    \clearfield{month}
    \clearfield{urlyear}
    \clearfield{urlmonth}
}

\usepackage[toc,acronym]{glossaries}
\makenoidxglossaries{}

\newacronym{2x2}{2\ensuremath{\times}2\xspace}{DUNE Near Detector 2-by-2 Prototype}
\newacronym{2p2h}{2p2h}{two-particle, two-hole}
\newacronym{argoneut}{ArgoNeuT}{Argon Neutrino Teststand}
\newacronym{bert}{BERT}{Bertini}
\newacronym{bnb}{BNB}{Booster Neutrino Beam}
\newacronym{caf}{CAF}{Common Analysis File}
\newacronym{cc}{CC}{charged-current}
\newacronym{cp}{CP}{Charge-Parity}
\newacronym{crt}{CRT}{Cosmic Ray Tagger}
\newacronym{dis}{DIS}{deep inelastic scattering}
\newacronym{dune}{DUNE}{Deep Underground Neutrino Experiment}
\newacronym{fd}{FD}{far detector}
\newacronym{fhc}{FHC}{Forward Horn Current}
\newacronym{fnal}{FNAL}{Fermi National Accelerator Laboratory}
\newacronym{fsi}{FSI}{final-state interactions}
\newacronym{ftfp}{FTFP}{Fritiof with Precompound}
\newacronym{geant}{GEANT4}{\emph{GE}ometry \emph{AN}d \emph{T}racking}
\newacronym{hp}{HP}{hadron production}
\newacronym{icarus}{ICARUS}{Imaging Cosmic And Rare Underground Signals}
\newacronym{lartpc}{LArTPC}{liquid argon time-projection chamber}
\newacronym{mc}{MC}{Monte Carlo}
\newacronym{uboone}{MicroBooNE}{Micro Booster Neutrino Experiment}
\newacronym{mec}{MEC}{meson exchange current}
\newacronym{minos}{MINOS}{Main Injector Neutrino Oscillation Search}
\newacronym{minerva}{MINER\kern-0.05em$\nu$\kern-0.09em A}{Main Injector Neutrino Experiment $\nu$-A}
\newacronym{nova}{NO\kern-0.05em$\nu$\kern-0.09em A}{NuMI Off-Axis $\nu_e$ Appearance}
\newacronym{nc}{NC}{neutral-current}
\newacronym{nd}{ND}{Near Detector}
\newacronym{numi}{NuMI}{Neutrinos at the Main Injector}
\newacronym{pmns}{PMNS}{Pontecorvo-Maki-Nakagawa-Sakata}
\newacronym{pot}{POT}{protons-on-target}
\newacronym{ppfx}{PPFX}{Package to Predict the Flux}
\newacronym{qe}{QE}{quasi-elastic}
\newacronym{rhc}{RHC}{Reverse Horn Current}
\newacronym{sbn}{SBN}{Short-Baseline Neutrino}
\newacronym{sbnd}{SBND}{Short-Baseline Near Detector}
\newacronym{sm}{SM}{Standard Model}
\newacronym{surf}{SURF}{Sanford Underground Research Facility}
\newacronym{tki}{TKI}{Transverse Kinematic Imbalance}

\newcommand{\numi}{\acrshort{numi}\xspace}
\newcommand{\numu}{\ensuremath{\nu_\mu}\xspace}
\newcommand{\numub}{\ensuremath{\bar{\nu}_\mu}\xspace}
\newcommand{\nue}{\ensuremath{\nu_e}\xspace}
\newcommand{\nueb}{\ensuremath{\bar{\nu}_e}\xspace}

\newcommand{\Enu}{\ensuremath{\textup{E}_\nu}\xspace}
\newcommand{\fhc}{\acrshort{fhc}\xspace}
\newcommand{\fnal}{\acrshort{fnal}\xspace}
\newcommand{\geant}{\acrshort{geant}\xspace}
\newcommand{\hp}{\acrshort{hp}\xspace}
\newcommand{\icarus}{\acrshort{icarus}\xspace}
\newcommand{\icarusangle}{\SI{100.1}{\milli\radian}\xspace}

\newcommand{\MINERvA}{\acrshort{minerva}\xspace}
\newcommand{\NOvA}{\acrshort{nova}\xspace}
\newcommand{\pot}{\acrshort{pot}\xspace}
\newcommand{\ppfx}{\acrshort{ppfx}\xspace}

\newcommand{\rhc}{\acrshort{rhc}\xspace}

\begin{document}
\pagenumbering{roman}

\include{frontmatter}

\pagenumbering{arabic}

\include{introduction}

\clearpage
\include{numi_ppfx}
\clearpage
\include{analysis}

\clearpage
\include{2x2}
\clearpage
\include{trigger_emulation}
\clearpage
\include{conclusion}
\clearpage
\include{bibliography}
\clearpage
\include{appendices}

\end{document}

%% file: frontmatter.tex
\title{The NuMI Neutrino Flux Prediction at ICARUS}
\author{Anthony P. Wood}
\degree{Doctor of Philosophy}{dissertation}
\department{Physics}
\college{College of Natural Sciences and Mathematics}
\chair{Daniel D. Cherdack}
\cochairfalse
\firstreader{Lisa W. Koerner}
\secondreader{Claudia Ratti}
\thirdreader{Anthony R. Timmins}
\fourthreader{Ricardo Vilalta}
\threereaderstrue
\fourreaderstrue
\fivereadersfalse
\submitdate{December 2024}
\tablelisttrue
\figurelisttrue
\titlep
\copyrightpage

\begin{dedication}
	\centering
	I dedicate this thesis in honor of my parents, my brother, Kristena, and my friends for their love and support throughout this journey.
\end{dedication}

\begin{acknowledgements}
	I would like to express my sincerest gratitude to my advisor, Daniel Cherdack, for his guidance and support.
	I would also like to thank my research group, my committee, and the members of the ICARUS collaboration for all of their valuable feedback toward shaping this work.\\

	\noindent This material is based upon work supported by the U.S. Department of Energy, Office of Science, Office of Basic Energy Sciences Energy Frontier Research Centers program under Award Number DE-SC0021407.\\

	\noindent This material is based upon work supported by the U.S. Department of Energy, Office of Science, Office of Workforce Development for Teachers and Scientists, Office of Science Graduate Student Research (SCGSR) program.
	The SCGSR program is administered by the Oak Ridge Institute for Science and Education for the DOE under contract number DE-SC0014664.\\

	\noindent This material is based upon work that is supported by the Visiting Scholars Award Program of the Universities Research Association under Award Number 22-F-17.
	Any opinions, findings, and conclusions or recommendations expressed in this material are those of the author and do not necessarily reflect the views of the Universities Research Association, Inc.
\end{acknowledgements}

\begin{abstract}
	\input{abstract}
\end{abstract}

\begin{singlespace}
\newpage
\tableofcontents
\newpage
\listoftables
\newpage
\listoffigures
\end{singlespace}

\newpage
\begin{singlespace}
	\printnoidxglossary[type=\acronymtype,title=\centering{\large{ABBREVIATIONS}}, toctitle=\large{ABBREVIATIONS},sort=def]
\end{singlespace}

%% file: abstract.tex
The \acrfull{dune} is a next-generation long-baseline neutrino oscillation experiment seeking to probe fundamental symmetries within the structure of the \acrfull{pmns} mixing matrix, and perform precision measurements its parameters including the neutrino mass ordering via the sign of $\Delta m^2_{31}$, and the charge-parity violating phase, $\delta_{CP}$.
To make these measurements with high precision, \acrshort{dune} will require external $\nu$-Ar scattering cross section data as a crucial input to the oscillation fit.
\acrfull{icarus} is a 476~t liquid argon neutrino detector located at \acrfull{fnal} where it is serving as the far detector for the \acrfull{sbn} program along the \acrfull{bnb} axis.
\acrshort{icarus} additionally lies 795~m downstream and 100.1~mrad off-axis of the \acrfull{numi} neutrino beam.
From this position, \acrshort{icarus} is exposed to a large flux of \acrshort{numi} (anti-)electron and (anti-)muon neutrinos, and poses a unique opportunity to provide high-statistics measurements of quasi-elastic and single pion-production cross sections for four neutrino flavors ($\nu_\mu$, $\nu_e$, $\bar{\nu}_\mu$, $\bar{\nu}_e$).
This dissertation is centered around accurately characterizing the models and estimating their precision for use in making these measurements.
This includes identifying major sources of uncertainty in the models such that they can be properly propagated to the cross section measurements.
Specifically, this work focused on the model of the \acrshort{numi} beamline and its impact on the neutrino fluxes, but also delved into the detector response model and its impact on reconstructed observables in the detector.
Significant efforts were made to improve the characterization to enhance precision and thus reduce the level of propagated uncertainty.
In particular, the \acrshort{numi} flux was determined to be composed of 57\% $\nu_\mu$, 38\% $\bar{\nu}_\mu$, 3\% $\nu_e$, and 2\% $\bar{\nu}_e$ while the horns are operating in the positive-particle focusing configuration.
The total uncertainty on the $\nu_\mu + \bar{\nu}_\mu$ ($\nu_e + \bar{\nu}_e$) flux while operating in the forward horn operating mode was determined to be 10.84\% (9.04\%).
Compared to the on-axis flux, mesons that eventually decay to neutrinos more frequently reinteract within the NuMI structure, resulting in elevated uncertainty as these processes are not well-constrained by existing hadron interaction cross section measurements.
Covariance matrices were calculated to propagate the flux uncertainty characterization to NuMI analyses.

%% file: introduction.tex
\chapter{Introduction}

The \acrfull{dune}~\cite{DUNE_tdr,abi_deep_2020,abi_long-baseline_2020} is a next-generation long-baseline neutrino oscillation experiment seeking to perform precision measurements of the \acrfull{pmns} mixing matrix parameters including the \acrfull{cp}-violating phase, $\delta_{CP}$, and to determine the neutrino mass ordering via the sign of the atmospheric mass-splitting, $\Delta m_{32}^2$.
\acrshort{dune} will employ a \SI{40}{\kilo\tonne} \acrfull{lartpc} \acrfull{fd} located at the Sanford Underground Research Facility (SURF), which will lie \SI{1300}{\kilo\meter} downstream of a \SI{1.2}{\mega\watt} neutrino beam at \acrfull{fnal}.
To make these measurements with high precision, \acrshort{dune} will require external measurements of $\nu$-Ar scattering cross sections as a critical prior in the oscillation fit.

\begin{figure}[htbp]
    \centering
    \includegraphics[width=0.7\textwidth]{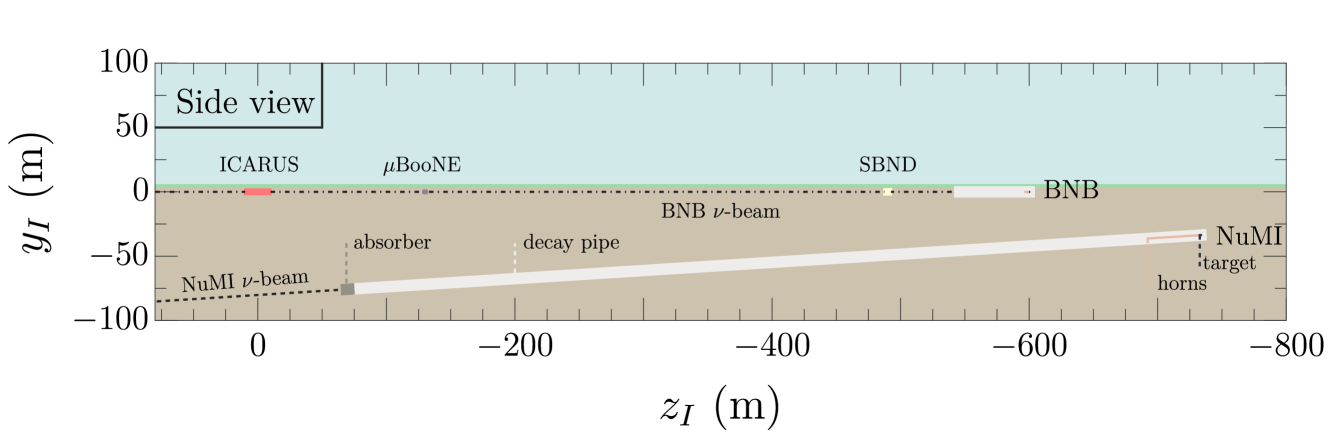}
    \includegraphics[width=0.4\textwidth]{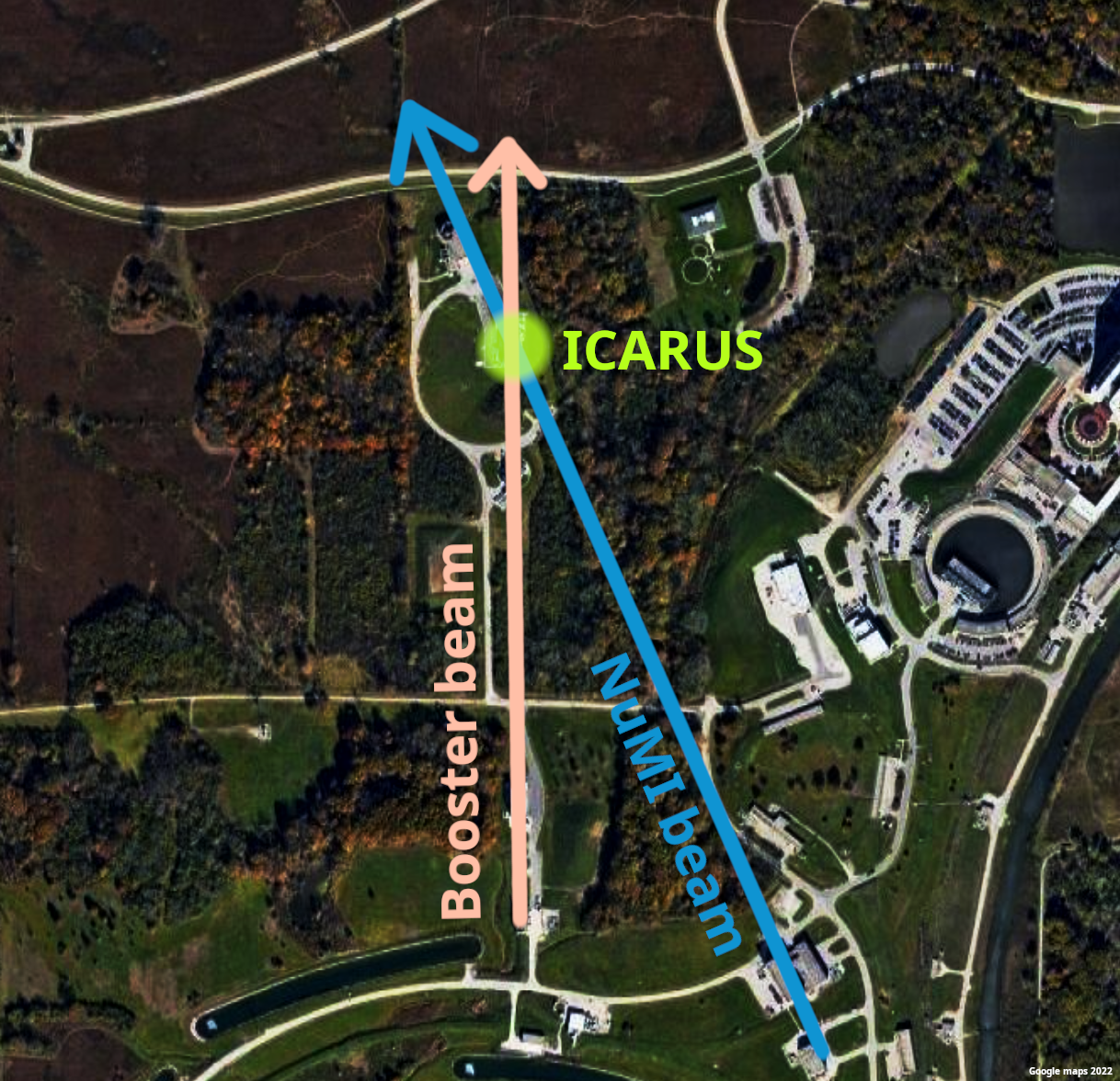}
    \caption[ICARUS at FNAL Site]{Side (top) and bird's-eye (bottom) views of \icarus{}'s position relative to both \acrshort{bnb} and \numi beams on the \fnal site. Wilson Hall can be seen on the right-hand side of the bottom image. Side view reproduced from~\citeauthor{chatterjee2024heavyneutralleptonsearches} (\citeyear{chatterjee2024heavyneutralleptonsearches}).}%
    \label{fig:beams_birdseye}
\end{figure}

\begin{figure}[htbp]
    \centering
    \includegraphics[width=0.5\textwidth]{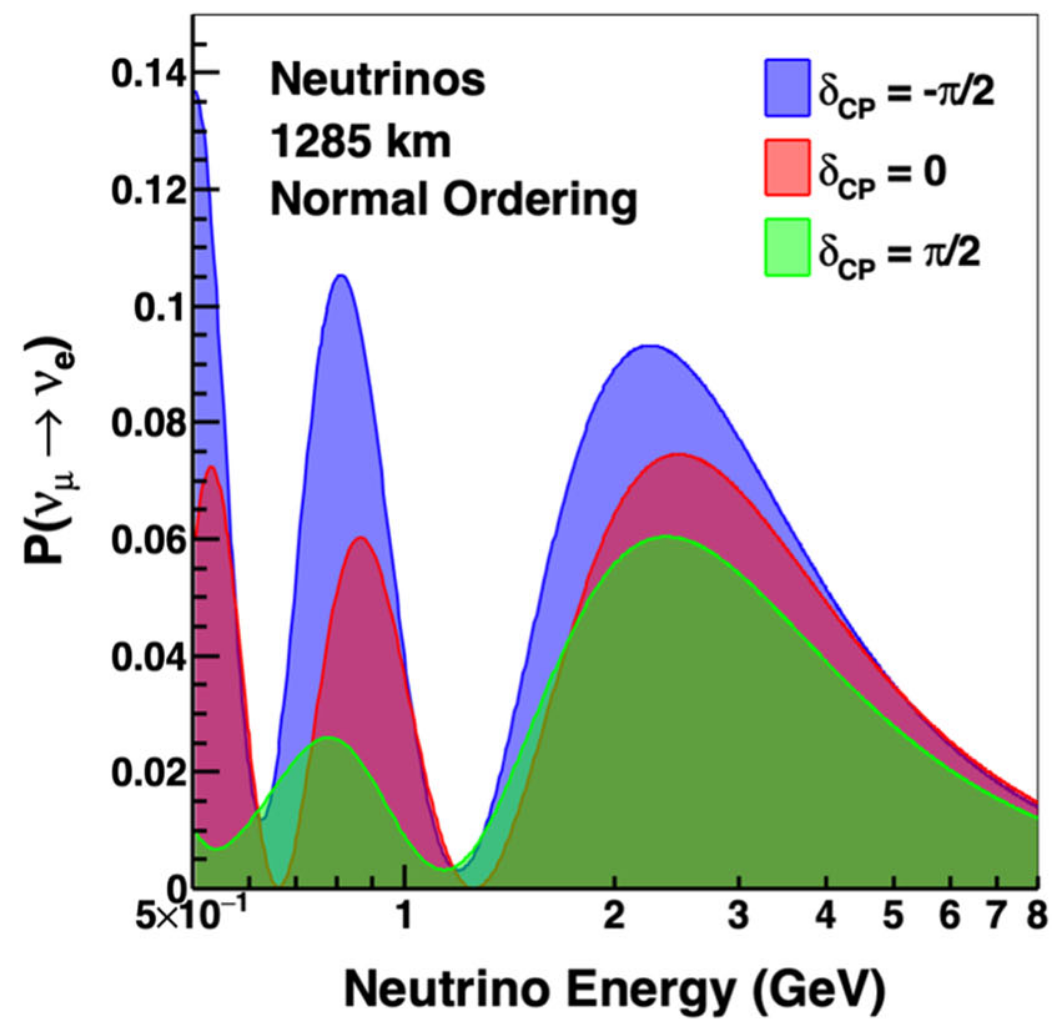}
    \caption[DUNE Oscillation Probabilities]{\acrshort{dune} $\numu \to \nue$ oscillation probabilities for possible values of $\delta_{CP}$, assuming a boff-axis ofaseline of $L = \SI{1285}{\kilo\meter}$ and normal mass-ordering. The probability peaks below \SI{1}{\GeV} and again between 1--\SI{4}{\GeV}. Reproduced from \citeauthor{abi_long-baseline_2020} (\citeyear{abi_long-baseline_2020}).}%
    \label{fig:dune_osc_prob}
\end{figure}

\begin{figure}[htbp]
    \centering
    \includegraphics[width=0.6\textwidth]{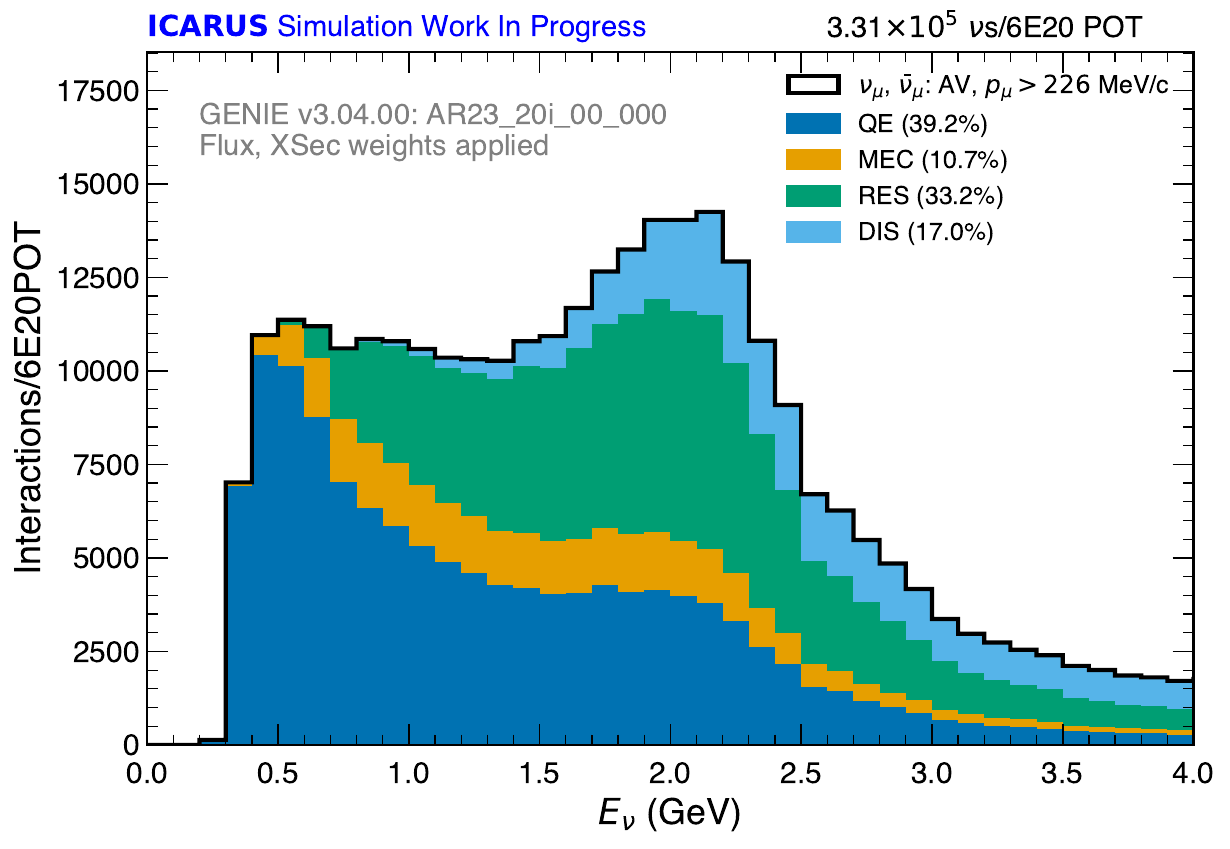}
    \caption[NuMI-ICARUS \numu\ CC Event Rate]{Expected $\numu+\numub$ CC event rate at \icarus\ in the combined Run 1 and Run 2 data.%
     The NuMI-ICARUS interaction peak at $\approx \SI{2}{\GeV}$ is within the dune oscillation maximal region, and will provide a valuable sample of cross section data. Reproduced from \citeauthor{howard_numi_offaxis_2024} (\citeyear{howard_numi_offaxis_2024})}%
    \label{fig:numi_event_rate}
\end{figure}

The \acrfull{icarus}~\cite{antonello_icarus_2014,abratenko_icarus_2023} experiment is a \SI{476}{\tonne} (active) \acrshort{lartpc} neutrino detector currently located at \fnal\ where it is serving as the \acrshort{fd} for the \acrfull{sbn}~\cite{machado_short-baseline_2019,abratenko_icarus_2023} program. The detector is 700 m down the central axis of the \acrfull{bnb}.
\icarus\ also lies \SI{795}{\meter} downstream and \icarusangle\ (\SI{5.75}{\degree}) off-axis of the \acrfull{numi} neutrino beam, as shown in Figure~\ref{fig:beams_birdseye}.
At this position, \icarus\ is expected to see a large flux of \numi\ \mbox{(anti-)}\hspace{0pt}electron and \mbox{(anti-)}\hspace{0pt}muon neutrinos in a kinematic region that is relevant for \acrshort{dune}.
The full \icarus\ physics run ($\sim 1.8 \times 10^{21}$ \acrfull{pot} over a 3-year period) will provide a high-statistics sample of quasi-elastic and single pion-production interactions for four neutrino flavors (\numu, \nue, \numub, \nueb).
Figure~\ref{fig:dune_osc_prob} shows the $\numu \to \nue$ oscillation probabilities at the \acrshort{dune} \acrshort{fd} as a function of neutrino energy~\cite{abi_long-baseline_2020}.
The probability peaks below \SI{1}{\GeV} and between 1--\SI{4}{\GeV}, the latter of which overlaps with the
predicted event rate of \numu\ and \numub\ \acrfull{cc} interactions in the combined Run 1 and Run 2 data ($6\times10^{20}$ total POT) shown in Figure~\ref{fig:numi_event_rate}~\cite{howard_numi_offaxis_2024}.

The core of this doctoral thesis project is centered around identifying major sources of uncertainty for these cross section measurements, especially those related to the neutrino flux, evaluating them, and reducing them such that the inputs for \acrshort{dune} analyses are as accurate and precise as possible.
A crucial ingredient to making neutrino cross section measurements at \icarus\ is a thorough understanding of the \numi\ flux simulation and related uncertainties, as will be discussed in Section~\ref{sec:xsec_uncert}.
The subsequent chapters include a comprehensive characterization of the \numi\ off-axis flux by studying the effects of beamline geometry and run conditions, along with hadron production modeling in the \numi\ beamline.
In the \acrshort{dune} oscillation maximum region that lies between $\sim$1--\SI{4}{\GeV}, where the \numi\ neutrino-mode flux at \icarus\ is predicted to be composed of $57\%$ \numu, $38\%$ \numub, $3\%$ \nue, and $2\%$ \nueb~while the horns are operating in the positive current (positive-particle focusing) configuration.

The large wrong-sign (antineutrino) component of the neutrino-mode flux is a consequence of \icarus{}'s high off-axis angle relative to the \numi\ beam.
At this angle, neutrinos arriving at \icarus\ are the result of broad spectrum kaon decays, and unfocused pion decays, mostly occurring in the region around the \numi\ target and magnetic focusing horns rather than the decay pipe.
On average, these mesons reinteract more often than those that produce on-axis neutrinos.
Thus, meson interaction uncertainties ($\sim 10\%$) are a leading systematic in the \numi\ flux uncertainty for \icarus\ in the 1--4 GeV region as will be shown in Chapter~\ref{sec:analysis}.
This region of hadron interaction phase space is not yet constrained by data in the underlying hadron production modeling, so there is also a strong potential for improvement of the flux uncertainties should differential cross section measurements in the relevant kinematic region become available.

Chapter~\ref{sec:2x2} contains an extension of the analysis presented in Chapter~\ref{sec:analysis} to the \acrfull{2x2} prototype, which will be exposed to the on-axis \acrshort{numi} beam as a means to test and validate the \acrshort{dune} \acrfull{nd} design. Finally, Chapter~\ref{sec:triggeremu} reports the outcome of a parallel project to improve the emulation of the \icarus\ trigger system in the \icarus\ \acrfull{mc} simulation.

\section{Neutrinos in the Standard Model}

The \acrfull{sm}~\cite{Workman:2022ynf,novaes2000standardmodelintroduction} of particle physics describes the elementary constituents of matter and the mechanisms by which they interact with one another via fundamental forces--electromagnetic, weak, and strong--mediated by exchange particles.
Fundamental forces are mediated by integer spin particles which obey Bose-Einstein statistics, or `bosons'.
The mediators include the photon, $\gamma$, for electromagnetic interactions, the $W^\pm$ and $Z^0$ bosons for the weak interactions, and the gluon, $g$, for the strong interaction.
The final boson, the Higgs boson, $H$, is responsible for the generation of mass in the \acrshort{sm}.
Matter is composed of half-integer spin particles which obey Fermi-Dirac statistics, or `fermions', and are divided into two families: quarks and leptons.
There are six types of quarks that can be grouped into three generations: up ($u$) and down ($d$), charm ($c$) and strange ($s$), and top ($t$) and bottom ($b$).
Quarks are bound together by the strong force to form hadrons, which may be classified further into baryons, a bound-state of an odd-numbered quark configuration, such as protons ($uud$) and neutrons ($udd$), or mesons, a bound state of even-numbered quark configuration, such as pions (e.g., $u\bar{d}$ for $\pi^+$) and kaons (e.g., $u\bar{s}$ for $K^+$).
Leptons are similarly divided into three generations of electromagnetically charged and neutral doublets: the electron ($e^-$) and electron neutrino ($\nu_e$), the muon ($\mu^-$) and muon neutrino ($\nu_\mu$), and the tau ($\tau^-$) and tau neutrino ($\nu_\tau$).
However, in contrast to hadronic matter, leptons do not participate in the strong interaction and are only subject to the electromagnetic and weak forces.

\begin{figure}[htbp]
    \centering
    \begin{subfigure}{0.49\textwidth}
        \includegraphics[width=\textwidth]{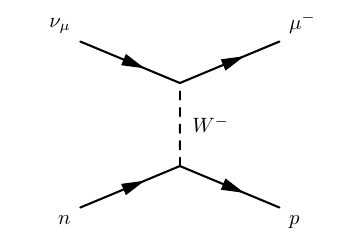}
        \caption{Charged Current}%
        \label{fig:feyn_cc}
    \end{subfigure}
    \begin{subfigure}{0.49\textwidth}
        \includegraphics[width=\textwidth]{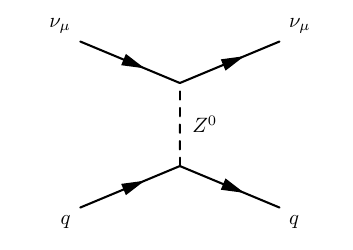}
        \caption{Neutral Current}%
        \label{fig:feyn_nc}
    \end{subfigure}
    \caption[$\numu CC 1\mu1\pi$ and $\numu NC$ Feynman Diagrams]{Example of a \numu \acrshort{cc} interaction producing a $\mu^-$ and a proton in the final state (left), and a \numu \acrshort{nc} interaction with an arbitrary fermion, $q$ (right).}%
\end{figure}

In the \acrshort{sm}, neutrinos interact via the weak force, which may occur through either \acrfull{cc} processes, via the exchange of $W^\pm$, or \acrfull{nc}, via the exchange of $Z^0$.
As an example, consider the most common process expected at \icarus\ as shown previously in Figure~\ref{fig:numi_event_rate}: a \numu\ which interacts with a neutrino quasi-elastically (QE), i.e., same number of initial and final state particles, but enough momentum transfer to produce the outgoing $\mu^-$.
The Feynman diagram for this process is available in Figure~\ref{fig:feyn_cc}.
Here, an incident \numu\ exchanges a $W^-$ boson converting the \numu\ into a $\mu^-$, and a neutron into a proton via conversion of a $d$ quark to a $u$ quark.
As for the \acrshort{nc} process, neutrinos may scatter elastically off any fermion, $q$, in the detector via exchange of the $Z^0$, as shown in Figure~\ref{fig:feyn_nc}.

\subsection{Neutrino Mass Mixing and Oscillations}

Over the past several decades, evidence~\cite{PhysRevLett.81.1562,PhysRevLett.90.021802} has been accruing that neutrinos undergo mixing and oscillate between flavor states as they propagate.
A prerequisite condition for oscillation to occur is that neutrino are massive particles, contrary to the predictions of the \acrshort{sm}.
The weak flavor eigenstates of neutrinos, $\ket{\nu_\alpha}$, can be described as a superposition of mass eigenstates, $\ket{\nu_i}$, related by the \acrshort{pmns} mixing matrix, $U$~\cite{Giunti_2005,kayser2013neutrinooscillationphysics}:
\begin{align}
    \ket{\nu_\alpha} &= \sum_{i} U_{\alpha i}^* \ket{\nu_i}{,} \\
    \text{and inversely,} \quad \ket{\nu_i} &= \sum_{\alpha} U_{\alpha i} \ket{\nu_\alpha}{.}
\end{align}
where $\alpha$ corresponds to a definite flavor state $\alpha \in \{e, \mu, \tau\}$ and $i$ represents the definite mass states $i \in \{1, 2, 3\}$.
In the plane wave description, the definiteness of the mass and energy of the mass eigenstates implies that they experience different phases as they evolve in time:
\begin{equation}
    \ket{\nu_i(t)} = \sum_{i} e^{-iE_i t} \ket{\nu_i}{.}
\end{equation}
Similarly, applying time evolution to the flavor states and re-expressing in the flavor basis yields
\begin{equation}
    \ket{\nu_\alpha(t)} = \sum_{i} U_{\alpha i}^* e^{-iE_i t}\ket{\nu_i} = \sum_{\beta}\sum_{i}U_{\alpha i}^* e^{-i E_i t} U_{\beta i} \ket{\nu_\beta}{,}
\end{equation}
and thus the neutrino of definite flavor $\alpha$ at time $t$ is a superposition, or mixing, of all possible flavor states.

The \acrshort{pmns} matrix, $U$, describes the degree of mixing, and can be parameterized by three mixing angles, $\theta_{12}$, $\theta_{13}$, and $\theta_{23}$, and a \acrshort{cp}-violating phase, $\delta_{CP}$:

\begin{equation}
\centering
U = \begin{pmatrix}
1 & 0 & 0 \\
0 & c_{23} & s_{23} \\
0 & -s_{23} & c_{23}
\end{pmatrix}
\begin{pmatrix}
c_{12}c_{13} & 0 & s_{13}e^{-i\delta_{CP}} \\
0 & 1 & 0 \\
-s_{12}c_{13}e^{i\delta_{CP}} & 0 & c_{13}
\end{pmatrix}
\begin{pmatrix}
c_{12} & s_{12} & 0 \\
-s_{12} & c_{12} & 0 \\
0 & 0 & 1
\end{pmatrix}.
\end{equation}

The probability for a neutrino of flavor $\nu_\alpha$ to oscillate into a neutrino of flavor $\nu_\beta$ is given by the squared amplitude
\begin{equation}
    P_{\nu_\alpha \to \nu_\beta}(t,E) = \left|\braket{\nu_\beta}{\nu_\alpha(t)}\right|^2 = \sum_{i} \sum_{j} U_{\alpha i}^* U_{\beta i} U_{\alpha j} U_{\beta j}^* \exp(-i\frac{\Delta m_{ij}^2 t}{2 E}){,}
\end{equation}
where the mass squared difference term, $\Delta m_{ij}^2 / 2E$, arises in the ultrarelativistic limit ($p \gg m_i$) and the mass states are assumed to propagate with the same momentum.

For a given baseline, $L$, in this same limit, the time becomes $t \approx L$ such that
\begin{equation}
    P_{\nu_\alpha \to \nu_\beta}(L, E) = \sum_{i} \sum_{j} U_{\alpha i}^* U_{\beta i} U_{\alpha j} U_{\beta j}^* \exp(-i\frac{\Delta m_{ij}^2 L}{2 E}){.}
\end{equation}
Under the assumed unitarity of $U$, the usual form of the three-flavor oscillation probability is obtained:
\begin{multline}\label{eq:osc_prob}
    P_{\nu_\alpha \to \nu_\beta}(L, E) = \delta_{\alpha\beta} - 4\sum_{i>j} \Re(U_{\alpha i}^* U_{\beta i} U_{\alpha j} U_{\beta j}^*)\sin^2\left(\Delta m_{ij}^2 \frac{L}{4E}\right)\\
    + 2\sum_{i>j} \Im(U_{\alpha i}^* U_{\beta i} U_{\alpha j} U_{\beta j}^*)\sin(\Delta m_{ij}^2 \frac{L}{2E}){,}
\end{multline}
where the probability oscillates as a function of $L/E$.

Altogether, there are a total of seven parameters describing neutrino oscillations: three mixing angles, three mass-squared differences, and one \acrshort{cp}-violating phase.
Their current status is summarized in Table~\ref{tab:osc_params}\cite{esteban_fate_2020,nufit2024}.

\begin{table}[htbp]
\centering
    \caption[Neutrino Oscillation Parameters (Best-fit)]{Best-fit values of the neutrino oscillation parameters. Table from NuFit 5.3 (2024). }
    \includegraphics[width=0.9\textwidth]{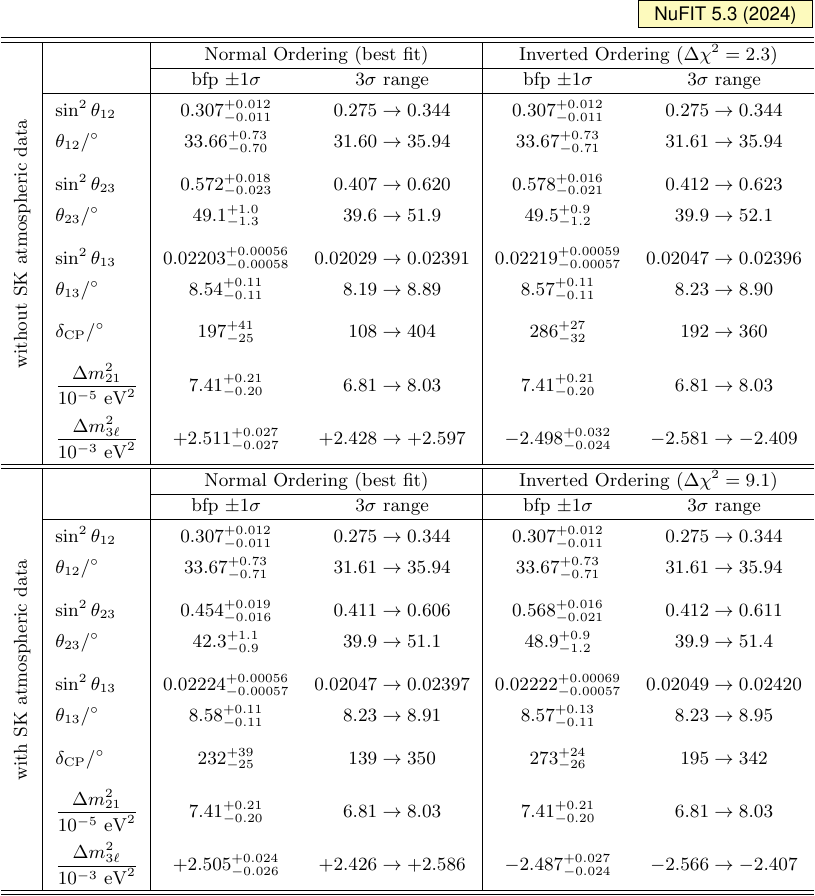}%
    \label{tab:osc_params}
\end{table}

\clearpage
\section{Neutrino Cross Section Measurements}\label{sec:xsec_uncert}
\input{propagation_formulae}

%% file: propagation_formulae.tex
Long-baseline neutrino experiments rely on the ability to count the number of neutrino interactions occurring within both a near and far detector to measure the oscillation probability.
The number of predicted interactions in the far detector as a function of the reconstructed neutrino energy, $E_\nu^{reco}$, is given by:
\begin{equation}
	N_{pred}\qty(E_\nu^{reco}) = \Phi\qty(E_\nu^{true})\sigma\qty(E_\nu^{true})P_{\nu_\alpha \to \nu_\beta}\qty(E_\nu^{true})\varepsilon\qty(E_\nu^{true})S\qty(E_\nu^{true},E_\nu^{reco}){,}
\end{equation}
where $\Phi$ is the neutrino flux, $\sigma$ is the interaction cross section, $P_{\nu_\alpha \to \nu_\beta}$ is the oscillation probability from Equation~\ref{eq:osc_prob} for a given baseline, $L$, $\varepsilon$ is the selection efficiency, and $S$ is the smearing matrix accounting for the energy resolution of the detector.
The number of observed interactions in the detector is interpretable as an oscillation probability, provided that the flux and cross-section are well-understood.
Employing a near detector assists in constraining uncertainties related to the flux and cross-section, as the common sources of uncertainty will cancel in the ratio between near and far-detector measurements.
However, dependence on the flux does not fully cancel in the ratio as the far detector is placed at a baseline corresponding to an oscillation maximum where \nue\ appearance is expected, resulting in shifts in the flux composition and neutrino energy distribution relative to the near detector.
Similarly, the cross section does not fully cancel as the two detectors are not necessarily identical in design, response, or acceptances.
Furthermore, intranuclear effects, such as \acrfull{fsi}--discussed in more detail in later chapters--are not well-modeled and distort outgoing particle kinematics, introducing biases in the reconstructed energy spectrum.
The aforementioned quantities are functions of the true neutrino energy, $E_\nu^{true}$, which is not directly observable and consequently translates to a large model-dependence of the oscillation fit parameterization.
Providing external measurements where possible can help to constrain and reduce model-dependence of the fit.
The basis for this thesis is primarily concerned with the potential for \icarus\ to provide precision measurements of the neutrino-argon cross section at \acrshort{dune}-relevant energy scales.

To ascertain the importance of the \numi\ flux prediction on \icarus\ neutrino cross section measurements, the relevant formulae are derived.
The number of observed interactions, or \emph{signal} events, $N^{sig}_{obs}$, of a given neutrino species $\nu_x$ is given by the following integral over the neutrino energy, $E$:
\begin{equation}
	N^{sig}_{obs} = \int_{E_{min}}^{E_{max}} \sigma(E) \phi_x(E) \varepsilon_x(E) M dE\ {,}
\end{equation}
where $M$ is the total number of nuclear targets.
Extracting the cross section and efficiency averaged over the energy range, the following is obtained:
\begin{equation}
	N^{sig}_{obs} = \langle \sigma_x \rangle \langle \varepsilon_x \rangle M \int_{E_{min}}^{E_{max}} \phi_x(E) dE = \langle \sigma_x \rangle \langle \varepsilon_x \rangle M \Phi_x\ {.}
\end{equation}
From here, the general form for the energy-averaged interaction cross-section is given by the following equation
\begin{equation}
	\langle \sigma_x \rangle =
	\frac{N^x_\textup{int} - B^x}{\langle \varepsilon^x \rangle M \Phi_x}{,}
\end{equation}
where $N^x_\textup{int}$ is the measured number of interactions for neutrino $\nu_x$, $B^x$ is the predicted number of background interactions.
From here, it can be seen that similar to the oscillation measurements, measuring the cross section depends on the neutrino flux, $\Phi_x$, which is estimated based on a model of the source beam.
As such, a high-quality flux prediction is a crucial foundation upon which to perform a cross section measurement.

The uncertainty on the flux prediction is propagated to the cross section by applying the general propagation formula for a multivariate function $f$ of variables $x_i$ is given by:
\begin{equation}\label{eq:propagation}
	\sigma^2_f =f^2 \sum_{i,j} \left[ \left(\frac{\partial f}{\partial x_i}\right)^2 \sigma^2_{i} + \left(\frac{\partial f}{\partial x_j}\right)^2 \sigma^2_{j} + 2\frac{\partial f}{\partial x_i}\frac{\partial f}{\partial x_j} \sigma_{ij} \right]\ {.}
\end{equation}
Here, $\sigma_{i}$, not to be confused with the cross section, represents the uncertainty in variable $x_i$, $\sigma_{ij}$ is the covariance between variables $x_i$ and $x_j$.
The integrated flux uncertainty for a single neutrino, $\nu_x$, is therefore
\begin{equation}\label{eq:flux_uncert}
	\sigma_x^2 = \Phi_x^2 \sum_{i,j} (\sigma_{i}^x)^2 + (\sigma_{j}^x)^2 + 2\sigma_{ij}^{x}\ {.}
\end{equation}

Of particular interest for the \numi\ cross section program is the potential for performing ratio measurements.
The ratio of cross sections of two neutrinos $\nu_x$ and $\nu_y$ is:
\begin{equation}
	R_{xy} = \frac{\sigma_x}{\sigma_y}
	= \frac{N^x_\textup{int}}{N^y_\textup{int}} \cdot  \frac{\varepsilon^y}{\varepsilon^x} \cdot \frac{\Phi_y}{\Phi_x}\ {,}
\end{equation}
where the ratio $\varepsilon^y/\varepsilon^x$ may not equal 1 if measurements of $\nu_x$ and $\nu_y$ come from different data sets (e.g., forward and reversed horn current runs).

Computing partial derivatives of $R$ yields:
\begin{equation}
	\frac{\partial R}{\partial \phi^x_k} = -\frac{R}{\Phi_x}\ ;\quad
	\frac{\partial R}{\partial \phi^y_k} = \frac{R}{\Phi_y}\ {.}
\end{equation}

Applying the result to the uncertainty propagation formula, Eq.~\ref{eq:propagation}
\begin{equation}
	\begin{split}
		\sigma_R^2 &
		= \sum_{i,j}\left[
			\frac{\partial R}{\partial \phi^x_i}\frac{\partial R}{\partial \phi^x_j}\sigma_{ij}^x 
			+ \frac{\partial R}{\partial \phi^y_i}\frac{\partial R}{\partial \phi^y_j}\sigma_{ij}^y 
			+2\frac{\partial R}{\partial \phi^x_i}\frac{\partial R}{\partial \phi^y_j}\sigma_{ij}^{xy} 
		\right]                  \\
		           & = R^2 \cdot
		\sum_{i,j}\left[
			\frac{1}{\Phi_y^2} \cdot \sigma_{ij}^x
			+ \frac{1}{\Phi_x^2} \cdot \sigma_{ij}^y
			-\frac{2}{\Phi_x\Phi_y} \sigma_{ij}^{xy} 
			\right]\ ,
	\end{split}
\end{equation}
where $\sigma_{ij}^{xy}$ is the covariance between fluxes $\phi^x$ and $\phi^y$ in energy bins $i$ and $j$.
Refer to Eq.~\eqref{eq:cov} for more information.
As the last term in the formula has a negative sign, positive covariance between flux bins decreases the total uncertainty in the ratio measurement.
This is advantageous of the absolute flux, as shown in Eq.~\eqref{eq:flux_uncert}, where positive covariance increases the total uncertainty.
In later chapters, it will be demonstrated that the \numi{}-\icarus\ flux in the neutrino flavor-energy space is expected to be highly correlated, which motivates the use of ratio measurements to reduce the total uncertainty in the cross section measurements.

%% file: numi_ppfx.tex
\chapter{The NuMI Beam and PPFX}\label{sec:numi_ppfx}
The \acrfull{numi} beam~\cite{adamson_numi_2016} installed at \fnal\ was originally built for the \acrfull{minos} experiment~\cite{minos_adamson_first_2006}.
In the years since, the beam has been utilized by many other experiments, including \acrfull{argoneut}~\cite{soderberg2009argoneutliquidargontime}, \acrfull{nova}~\cite{habig_nova_2012}, \acrfull{minerva}~\cite{aliaga_soplin_neutrino_2016}.
As such, the on- and near-on-axis neutrino flux is well-studied and validated.
However, the flux in the highly off-axis region at which \icarus\ sits relative to \numi\ has not been as well characterized and is the principal focus of this thesis.
This chapter will provide an overview of the structure and operating principles of the \numi\ beamline, followed by a discussion of the \numi\ simulation and corrections applied therein.

\begin{figure}[htbp]
	\centering
	\includegraphics[width=0.8\textwidth]{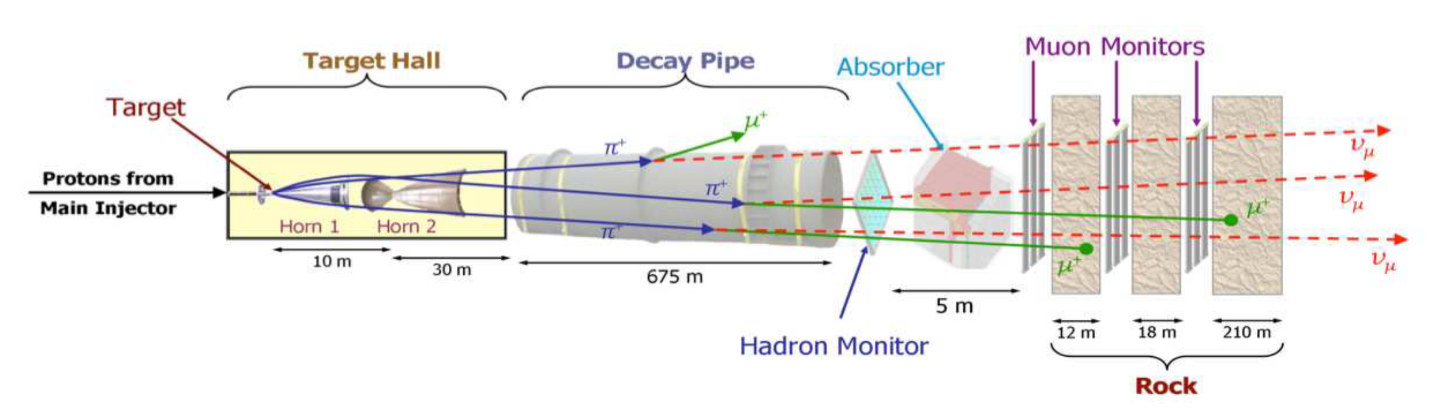}
	\caption[The \numi\ Beamline at \fnal{}]{The \numi\ beamline at \fnal{}. Reproduced from Adamson \emph{et al.} (2016).}
	\label{fig:numi_beam}
\end{figure}

To produce the beam, $\mathrm{H}^-$ are accelerated to \SI{400}{\MeV} in a linear accelerator, and then injected into the Booster, where they are stripped of their electrons, converting them to protons, and accelerated to \SI{8}{\GeV} in \SI{1.6}{\micro\second} bunches.
A fraction of these bunches are redirected to a beryllium target, producing the \acrshort{bnb} beam.
The rest are further accelerated to \SI{120}{\GeV} in the Main Injector, and impinged upon a graphite target in the \numi\ target hall shown in Figure~\ref{fig:numi_beam}.
The target is chased by two magnetic horns, which focus the positively charged hadrons produced in the target into a \SI{675}{\meter} decay volume, while deflecting the negatively charged particles away from the beamline.
In the \acrfull{fhc} operating mode, neutrinos are predominantly produced through the decays of $\pi^+ \to \mu^+\numu$ and $K^+ \to \mu^+ \numu$.
Finally, absorbers are placed at the downstream end of the decay volume to attenuate undecayed hadrons and muons, yielding a high-purity \numu\ beam.

In its original design, \numi\ operated at \SI{400}{\kilo\watt} with a peak intensity of $4\times10^{13}$ protons per spill and spill width of \SI{10}{\micro\second}.
The beam has undergone several upgrades since its inception, with the most recent beginning in 2019~\cite{yonehara_megawatt_2022}, which increased the beam power from \SI{700}{\kilo\watt} to \SI{1}{\mega\watt} achieving a peak intensity of $6.3\times10^{13}$ protons per spill.
To support the higher intensity beam and mitigate the increased thermal load, a new \SI{120}{\centi\meter} target was installed with wider graphite fins, and the beam spot was increased from \SI{1.3}{\centi\meter} to \SI{1.5}{\centi\meter}.
Impacts on the neutrino flux in the context of \icarus\ off-axis measurements due to the \SI{1}{\mega\watt} upgrade are explored in more detail in Section~\ref{sec:megawatt}.

\section{The NuMI Simulation}\label{sec:beamsim}
The \numi\ simulation is performed using software based upon the \acrfull{geant} toolkit~\cite{agostinelli_geant4_2003}, which uses a combination of the \acrfull{ftfp} nuclear string model and the \acrfull{bert} hadronic cascade model to transport and interact simulated hadrons within the \numi\ geometry.
The output of the simulation, and the input to the analysis presented in this thesis, is stored in a ROOT TTree following the Dk2Nu specification~\cite{dk2nu_proposal}.
A single Dk2Nu entry contains a detailed record of each interaction in the neutrino's ancestry, beginning with the initial \SI{120}{\GeV} beam proton and ending with the neutrino itself.
As an example, consider the following decay chain:

\begin{equation*}
	p + \mathrm{C} \to \pi^+ \to \nu_\mu + \mu^+ \to \nue + \numub + e^+.
\end{equation*}
Here, the proton $p$ interacts with a carbon nucleus $\mathrm{C}$ in the \numi\ target to produce a $\pi^+$ meson, which subsequently decays into a $\nu_\mu$ and $\mu^+$.
The $\mu^+$ then decays into a $\nue$, $\numub$, and $e^+$.
The chain leading up to each produced neutrino, $\nu_\mu$, $\nue$, and $\numub$, is stored in separate entries of the form:
\begin{enumerate}
	\item $ \overset{0}{p} \to \overset{1}{\pi^+} \to \overset{2}{\numu} $,
	\item $ \overset{0}{p} \to \overset{1}{\pi^+} \to \overset{2}{\mu^+} \to \overset{3}{\nue} $,
	\item $ \overset{0}{p} \to \overset{1}{\pi^+} \to \overset{2}{\mu^+} \to \overset{3}{\numub} $,
\end{enumerate}
where an index of $0$ corresponds to the beam proton, and the index of the neutrino is the highest in the chain.

The neutrino's characteristics are reported in the decay frame of its parent hadron, rather than for a specific detector location.
A probabilistic approach for calculating the neutrino flux at a particular location using the primary hadron's kinematic information is applied.
The benefit of this approach is that it allows the simulation to be reused for multiple detectors and locations, as the flux can be calculated for any location by applying the appropriate transformations outlined in the next section.

A summary of the \numi\ simulation samples used in this analysis is presented in Table~\ref{tab:flux_samples}.
\begin{table}[htbp]
	\centering
	\caption[Summary of Flux Productions]{Summary of the \numi\ simulation samples used in this analysis.}%
	\begin{tabular}{l l p{0.5\textwidth}}
		\toprule\toprule\\
		GEANT Version & POT & Note \\
		\midrule\\
		4.9.2 & $5\times10^8$ & Generated for the \acrshort{nova} 3rd analysis.\\
		4.9.6 & $7\times10^8$ & Generated by N.~Bostan incorporating the \SI{1}{\mega\watt} target geometry. RHC sample is only 50M POT.\\
		4.9.6 & $2.5\times10^8$ & Generated by \acrshort{uboone} to study geometry and model changes. See Section~\ref{sec:missing_geom}.\\
		4.10.4 & $1\times10^9$ & Generated by the author for improved statistics.\\
		\bottomrule\bottomrule
	\end{tabular}%
	\label{tab:flux_samples}
\end{table}

\subsection{Flux Weights}
To calculate the number of neutrinos that reach the detector and their energy, $E_\nu^{\textup{LAB}}$, primary hadron decay kinematics are used to transform the neutrino's energy into the lab frame.
In the case of (pseudo-) scalar hadron parents ($\pi^\pm$, $K^\pm$, $K^0_L$), neutrinos are emitted isotropically from the decay vertex.
The flux weight--i.e., the number of neutrinos per square meter that reach the detector--is determined by calculating the fractional solid angle of the unit sphere enclosing the decay vertex, out of $4\pi$, formed by the neutrino and parent momentum vectors, under the hypothesis that the neutrino is directed toward the detector location.
A correction is applied in the case of muon parents to account for their polarization.
Finally, an importance weight, $w_\textup{imp}$, is applied to mitigate over-representation of certain decay processes.

\subsubsection{Isotropic Decays}
A Lorentz boost factor, $M$, is calculated from the angle between the parent's decay momentum and the decay vertex position with respect to the detector position, $\cos\theta_{\textup{parent}-\nu}$, as follows:
\begin{equation}
	M = \frac{1}{\gamma\left( 1 - \beta \cos\theta_{\textup{parent}-\nu}\right)},
\end{equation}
where $\beta$ is the parent's velocity in units of the speed of light, and $\gamma = \left(1 - \beta^2 \right)^{-1/2}$.

The neutrino's energy in the lab frame is therefore given by:
\begin{equation}
	E_\nu^{\textup{LAB}} = M E_{\nu}^{\textup{CM}},
\end{equation}

where $E_{\nu}^{\textup{CM}}$ is the neutrino's energy in the decay frame.

Similarly, the total flux weight, $w_{\textup{flux}}$ is calculated via:

\begin{equation}
	w_{\textup{flux}} = w_{\textup{imp}} \frac{4\pi M^2}{\left|\vec{r}_{\textup{detector} - \nu}\right|^2}\quad \left[\si{\meter^{-2}}\right].
\end{equation}

The total flux and its inherent uncertainty due to random sampling is then calculated by the sum of the flux weights for each neutrino event, $i$:
\begin{align}
	\Phi                   & = \sum_{i} w_{\textup{flux},i}           \\
	\sigma_{\textup{stat}} & = \sqrt{\sum_{i} w_{\textup{flux},i}^2}.\label{eq:statistical_uncertainty}
\end{align}

\clearpage
\subsubsection{Anisotropic Decays}
In the case of a muon parent, the flux weight must be modified to account for the polarization of the muon.
First, the neutrino is boosted to the muon decay frame:

\begin{equation}
	\vec{p}^{\,CM}_\nu = \vec{p}_\nu - \gamma\vec{\beta}\left(E_{\nu}^{\textup{LAB}} - \frac{\gamma\vec{\beta}\cdot\vec{p}_\nu}{\gamma + 1 }\right)
\end{equation}

where

\begin{equation*}
	\vec{p}_\nu = \frac{\vec{r}_{\textup{detector} - \nu}}{\left|\vec{r}_{\textup{detector}-\nu}\right|}E_{\nu}^{\textup{LAB}}
\end{equation*}
Next, the parent particle of the muon is boosted to the muon's production frame
\begin{equation*}
	\vec{p}^{CM}_{\mu-\textup{par}} = \vec{p}_{\mu-\textup{par}} - \gamma_{\textup{prod}}\vec{\beta}_{\textup{prod}}\left(E_{\mu-\textup{par}} - \frac{\gamma_{\textup{prod}}\vec{\beta}_{\textup{prod}}\cdot\vec{p}_{\mu-\textup{par}}}{\gamma_{\textup{prod}} + 1 }\right),
\end{equation*}
where $\vec{\beta}_\textup{prod}$ is the muon's velocity at its production point.

Finally, the new decay angle is calculated with respect to the (anti-) spin direction and the flux weight is modified according to neutrino flavor
\begin{equation*}
	\cos\theta = \frac{\vec{p}_{\nu}^{\,CM} \cdot \vec{p}_{\mu-\textup{par}}^{\,CM}}{\left|\vec{p}_{\nu}^{\,CM}\right|^{2} \left|\vec{p}_{\mu-\textup{par}}^{\,CM}\right|^{2}}
\end{equation*}
For $\nu_e$ or $\bar{\nu}_e$, the flux weight becomes
\begin{equation*}
	w_{\textup{flux}}' = w_{\textup{flux}}\left(1 - \cos\theta \right).
\end{equation*}
For $\nu_\mu$ and $\bar{\nu}_\mu$
\begin{equation}
	w_{\textup{flux}}' = w_{\textup{flux}}\frac{ \left(3 - 2x_{\nu}\right) - \left( 1- 2 x_\nu\right)\cos{\theta}}{3 - 2x_\nu},
\end{equation}
where
\begin{equation}
	x_{\nu}= \frac{2 E_\nu^{CM}}{m_\mu}.
\end{equation}

\section{The NuMI Off-Axis Flux}\label{sec:offaxis_flux}
In this section, the base simulation of the \numi\ off-axis flux for the \icarus\ detector location is presented.
The \icarus\ detector location in the \numi\ coordinate system and variation of the flux within the angular acceptance region spanning $\SI{98.4}{\milli\radian}-\SI{102.3}{\milli\radian}$ was studied in detail in~\cite{antoni_detector_location_note,cherdack_prediction_2023,a_menegolli_m_torti_icarus_nodate}.
It was determined that the differential flux across the solid angle of the detector is stable relative to statistical limitations of the simulated sample, and the flux characterization using the single ray connecting the \numi\ origin to the \icarus\ geometric center\footnote{$x=\SI{4.5}{\meter},y=\SI{79.92}{\meter},z=\SI{795.13}{\meter}$} is sufficient for this analyis.

\begin{figure}[htbp]
	\centering
	\includegraphics[width=0.7\textwidth]{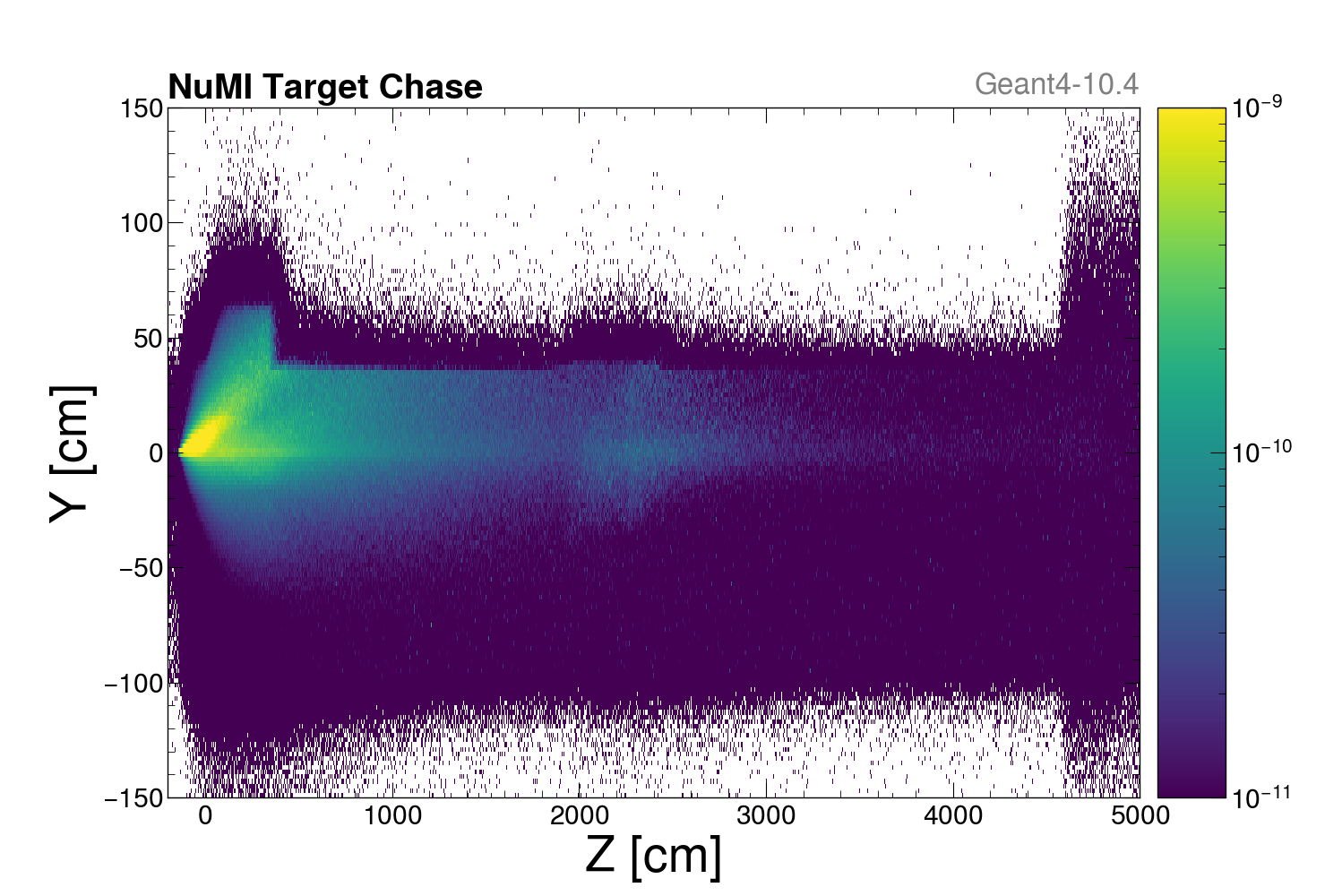}
	\includegraphics[width=0.7\textwidth]{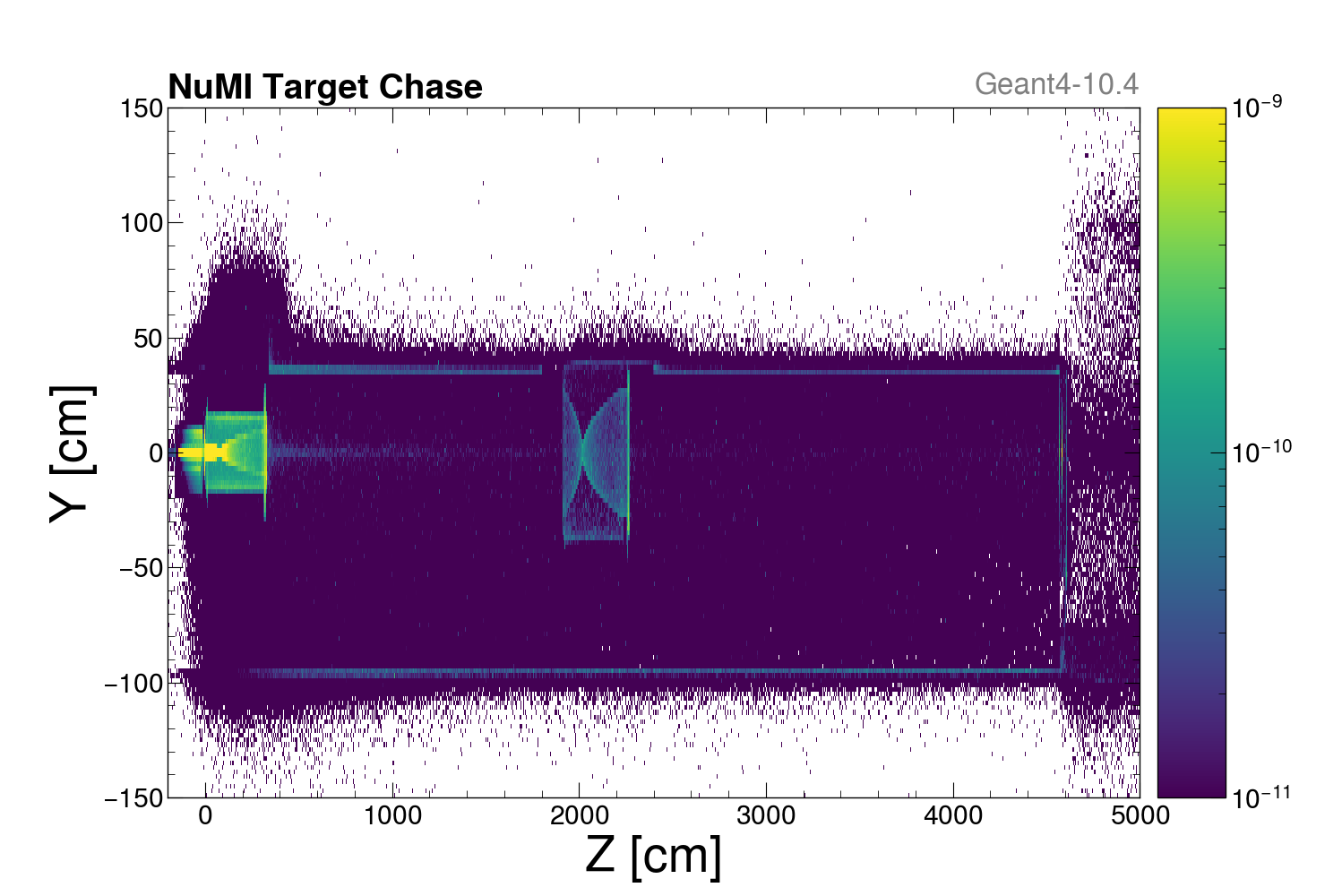}
	\caption[$\nu$ and Parent Hadron Production Vertices]{Production vertices in the \numi\ target chase volume for neutrinos that arrive at \icarus\ (top) and their parent hadrons (bottom). Most primary hadrons are unfocused by either horn, as such the majority of \icarus neutrinos are produced in the region before or within the first horn, beginning a few centimeters to the right of $z=\SI{0}{\centi\meter}$.}%
	\label{fig:decay_vertices}
\end{figure}

\begin{figure}[htbp]
	\centering
	\begin{subfigure}{\textwidth}
		\includegraphics[width=0.49\textwidth]{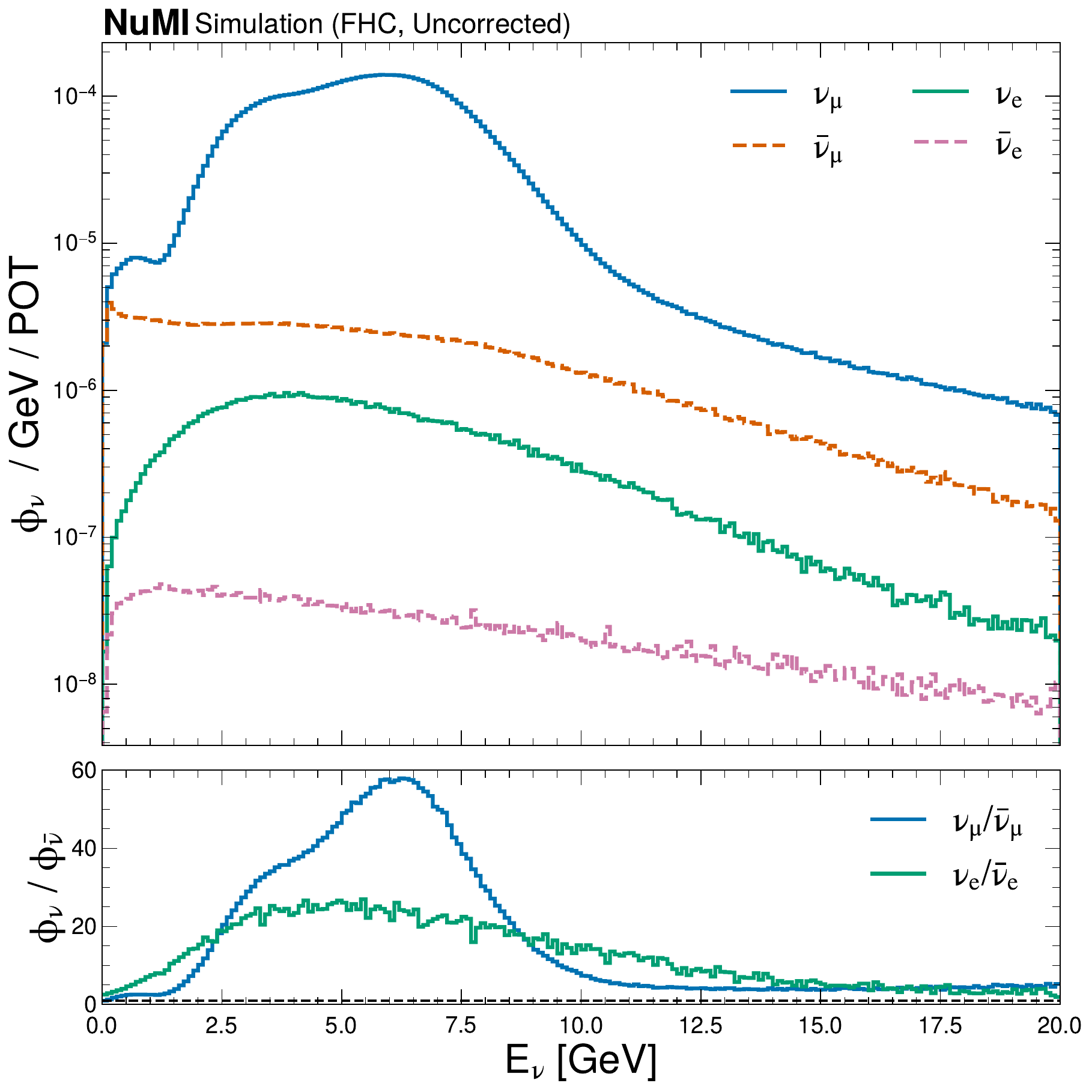}
		\includegraphics[width=0.49\textwidth]{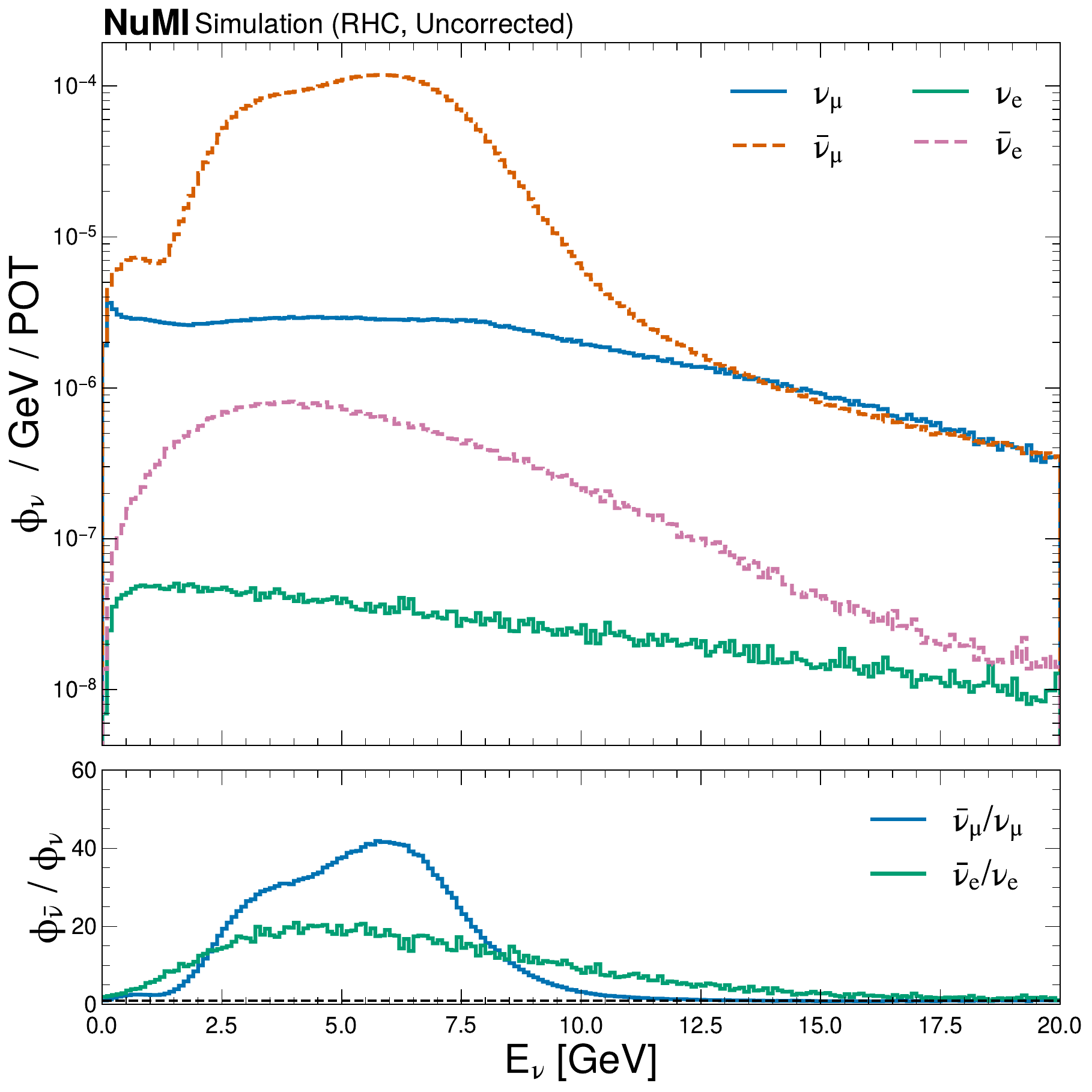}
		\caption{On-axis}%
		\label{fig:uncorrected_flux_onaxis}
	\end{subfigure}
	\begin{subfigure}{\textwidth}
		\includegraphics[width=0.49\textwidth]{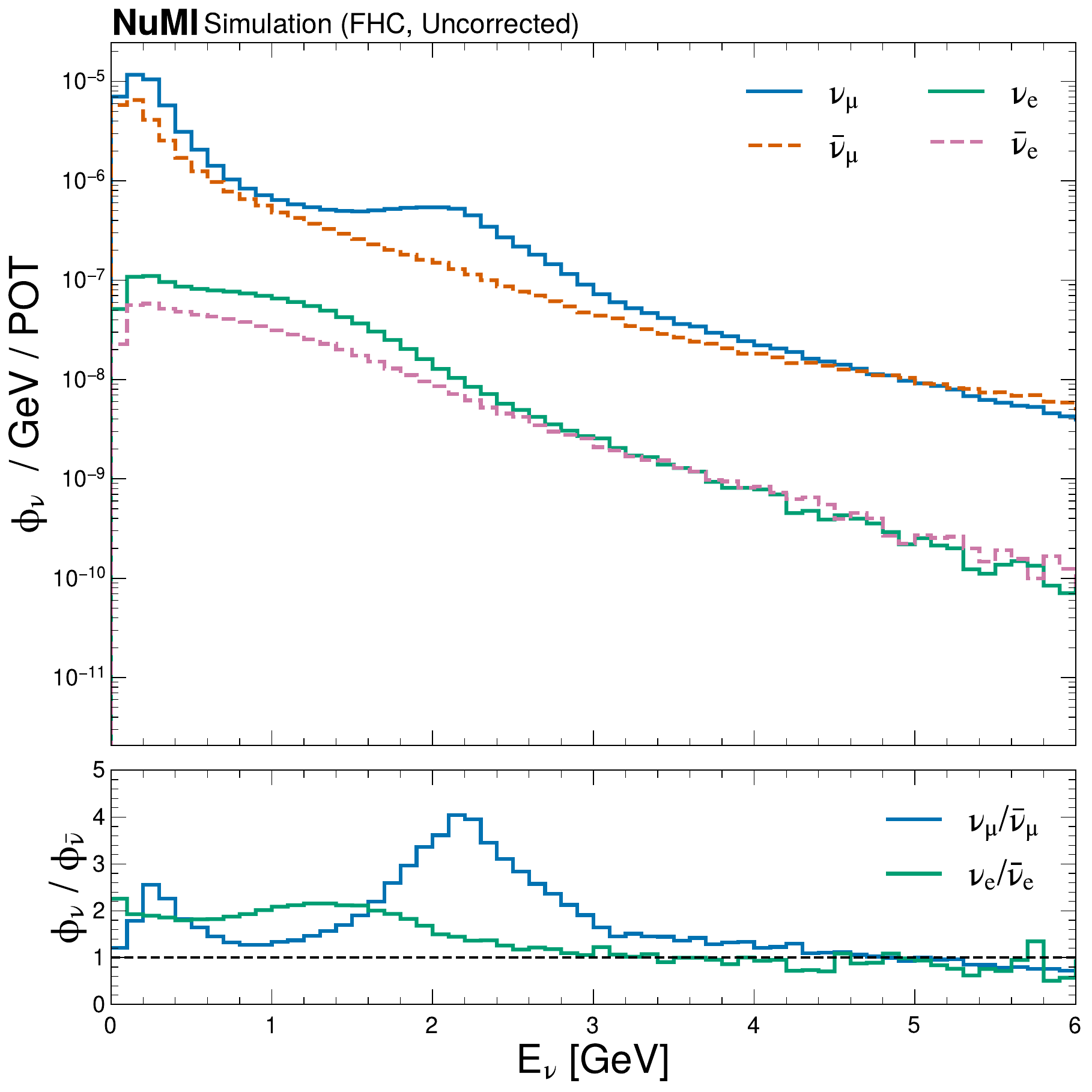}
		\includegraphics[width=0.49\textwidth]{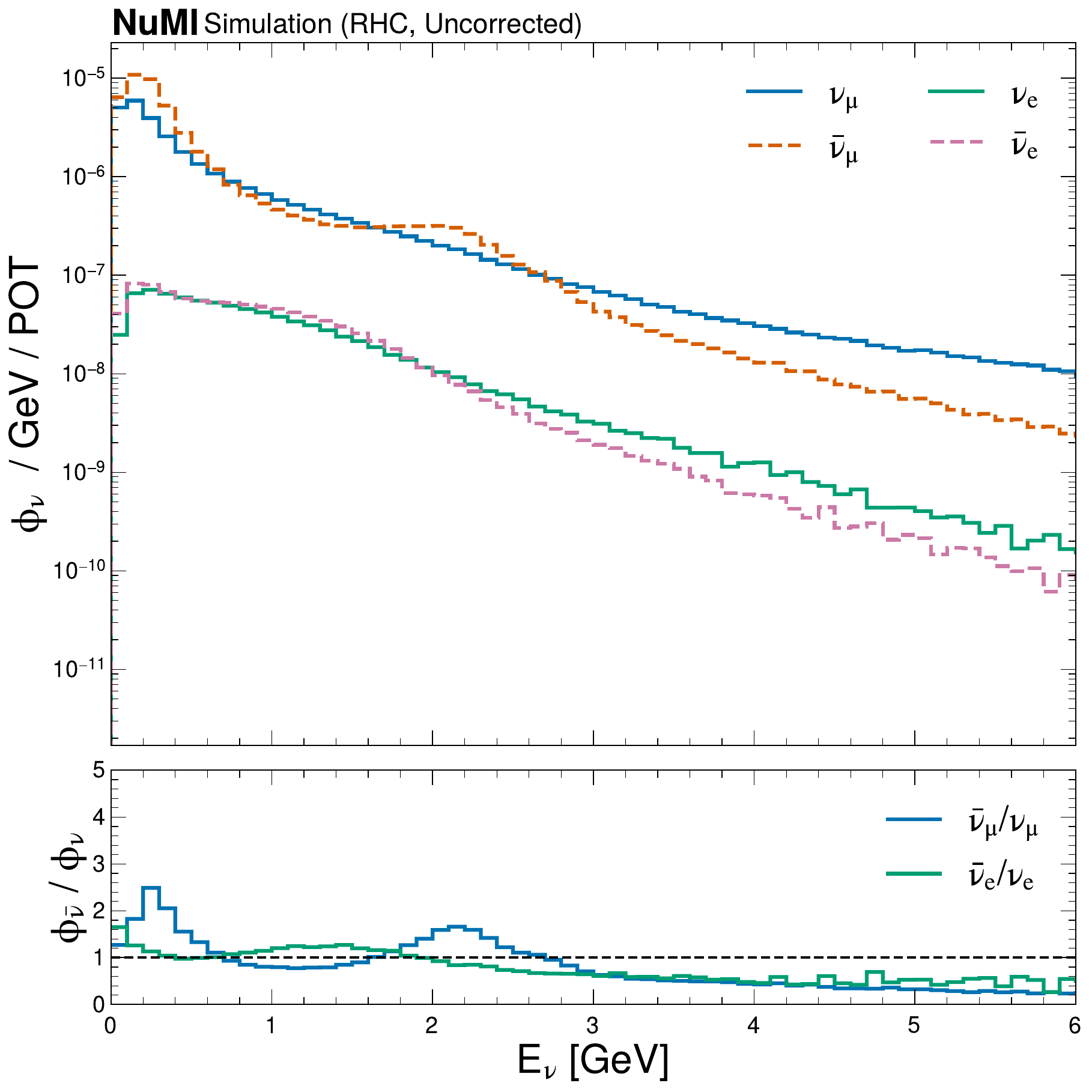}
		\caption{Off-axis at \icarusangle}%
		\label{fig:uncorrected_flux_offaxis}
	\end{subfigure}
	\caption[The NuMI On- and Off-Axis Flux Simulation]{The \numi\ on-axis (top) and off-axis (bottom) flux in \fhc (left) and \rhc (right).}%
\end{figure}

At \icarusangle\ off-axis, much of the \numi\ beam geometry does not impact the predicted flux.
As shown in Figure~\ref{fig:decay_vertices}, the majority of neutrinos are produced upstream of, or within, the first horn.
For this reason, the flux is produced from the decays of unfocused hadrons, which creates lower purity beam composed of a larger fraction of \nue\ and \numub\ relative to the on-axis flux.
While there is a small component of hadrons that are focused by the horns, these particles usually undergo secondary interactions with helium in the decay pipe or with the enclosing material (steel, concrete, etc.) before they produce \icarus{}-bound neutrinos.
Figures~\ref{fig:uncorrected_flux_onaxis}~and~\ref{fig:uncorrected_flux_offaxis} show the raw simulated flux for the \numi\ \fhc\ and \rhc\ operating modes for both on-axis and at the \icarus\ off-axis location, respectively.
Compared to the on-axis flux, the wrong-sign contamination is significantly larger, ranging from approximately 20\% to 80\% of the right-sign component in key regions of interest.

\section{The Package to Predict the Flux}\label{sec:ppfx}

The \acrfull{ppfx} is a software package developed by L.~Aliaga and the \MINERvA~Collaboration~\cite{aliaga_soplin_neutrino_2016,PPFX_github} to apply corrections to the simulated neutrino flux and propagate systematic uncertainties according to experimental \hp cross section data.
To extract the correction, \ppfx calculates a ratio between the measured cross section and a data-like template generated from the Geant4 models mentioned in Section~\ref{sec:beamsim}:
\begin{equation}
	w_{\textup{ppfx}} = \frac{\sigma_{\textup{data}}}{\sigma_{\textup{MC}}}.
\end{equation}
This ratio is calculated for each interaction in the neutrino's ancestry chain and applied as a product of weights to each neutrino event in the simulation.
For interactions that are not covered by data, a weight of $1$ is applied.

To propagate uncertainties, \ppfx\ employs a multiverse approach to produce a set of weights for each neutrino event.
\ppfx\ generates random throws, or \emph{universes}, of the weights modified within the limits of the data uncertainties.
For interactions unsupported by data, \ppfx\ assumes a conservative uncertainty estimate and generates universes about the central value of $1$.
This distribution of universes is then used to calculate the uncertainty on the flux correction, which is propagated to downstream analyses.
Analysis of the distribution flux universes as it pertains to \icarus\ will be discussed in more detail in Chapter~\ref{sec:analysis}.

\subsection{Hadron Interaction Channels}\label{sec:ppfx_channels}
\ppfx\ maps each interaction in the neutrino's ancestry to a corresponding hadron interaction channel.
These channels are outlined below along with the applied experimental data sets.

\begin{itemize}
	\item \textbf{p + C $\boldsymbol{\to{} \pi^{\pm}}$ + X:}
	      inclusive charged pion production in p+C interactions,
          \begin{itemize}
	         \item NA61 31 GeV/$c$~\cite{na61_pCpi31}, Barton \emph{et al.} 100 GeV/$c$~\cite{Barton} MIPP 120 GeV/$c$~\cite{MIPP}, NA49 158 GeV/$c$~\cite{NA49_pCpi160}
          \end{itemize}

	\item \textbf{p + C $\boldsymbol{\to{} \textup{K}}$ + X:}
	      inclusive kaon production in p+C interactions,
          \begin{itemize}
	        \item MIPP 120 GeV/$c$~\cite{MIPP}, NA49 158 GeV/$c$~\cite{NA49_pCK160}
          \end{itemize}

	\item \textbf{n + C $\boldsymbol{\to{} \pi^{\pm}}$ + X:}
	      inclusive charged pion production in n+C interactions,
	      \emph{spectra estimated based on p+C data and isospin symmetry,}

	\item \textbf{p + C $\boldsymbol{\to{} \textup{N}}$ + X:}
	      inclusive nucleon production in p+C interactions,
          \begin{itemize}
	        \item NA49 158 GeV/$c$~\cite{NA49_pCp160}
          \end{itemize}

	\item \textbf{($\boldsymbol{\pi^\pm}$, K) + A ($\boldsymbol{\to \pi^\pm}$, K, N) + X:}
	      meson interactions,
	      \textit{no experimental data,}

	\item \textbf{N + A $\boldsymbol{\to}$ X:}
	      interactions of nucleons on nuclei, not included in other channels; this contribution includes p+C (50\%) and n+C (20\%) interactions  not covered by experimental data,
	      \textit{no experimental data\footnote{Unless the interaction qualifies for A-scaling. See following paragraph.},}

	\item \textbf{others:}
	      interactions not included in the other channels; primarily interactions of $\Lambda$, $\overline{\textup{p}}$ and $\overline{\textup{n}}$,
	      \textit{no experimental data,}

	\item \textbf{attenuation:}
	      correction for the probability of a particle interacting in a given volume, or passing without interacting, based on data on total inelastic cross section measurements of p+C, $\pi^\pm$+C, $\pi^\pm$+Al, K$^\pm$+C and K$^\pm$+Al interactions.
\end{itemize}

\ppfx\ determines whether a particular dataset is applicable to an interaction by first checking the identities of the incoming and outgoing particles.
Scaling of the NA49 data collected at \SI{158}{\GeV/c} to energies relevant to \numi\ (\SI{12}{\GeV/c}--\SI{120}{\GeV/c}) is performed; however, the scaling can be unreliable below \SI{12}{\GeV/c}.
Therefore, for the N+C data sets, incident nucleons must have momenta greater than \SI{12}{\GeV/c}.
The interaction must be within the kinematic phase space of the datasets, which are binned in terms of the transverse momentum of the outgoing particle, $p_T$, and the Feynman-X, $x_F$.
Finally, rather than checking the identity of the nuclear target of the incident particle, \ppfx\ assumes a carbon target for all interactions that occur within the \numi\ target volume, and non-carbon nuclei outside this volume.
For interactions that otherwise meet the previous criteria but occur outside the target volume, \ppfx\ will attempt to apply a scale factor to the data to account for the difference in nuclear composition.
If none of the criteria are met, the interaction will be assigned to either the \textit{meson}, \textit{N+A}, or \textit{other} channels where appropriate.

Each of these requirements discretizes the interaction phase space into bins of incident particle, outgoing particle, target nucleus, and interaction kinematics.
As an example, the incident meson interaction channel is defined by bins of 5 incident mesons ($\pi^\pm$, $K^\pm$, $K^0_L$), 9 possible outgoing particles ($p$, $n$, $\mu^\pm$, $\pi^\pm$, $K^\pm$, $K^0_L$), and 8, $0.25$-wide, $x_F$ bins ranging $-1 \leq x_F \leq 1$ for a total of 360 bins.

\subsection{PPFX Output}
The \ppfx\ output that is utilized in the next chapter is a ROOT file containing the following information:
\begin{itemize}
    \item Flux vs. \Enu\ histograms without corrections applied for each neutrino mode, weighted for the \icarus\ detector location.
    \item Flux vs. \Enu\ histograms for the central value about which the universes are sampled.
    \item Flux vs. \Enu\ histograms for each universe.
    \item The number of \pot\ simulated, for normalizing the results.
\end{itemize}

%% file: analysis.tex
\chapter{NuMI Neutrino Flux Prediction and Uncertainties for ICARUS}\label{sec:analysis}

\begin{figure}[htbp]
	\centering
	\includegraphics[width=0.49\textwidth]{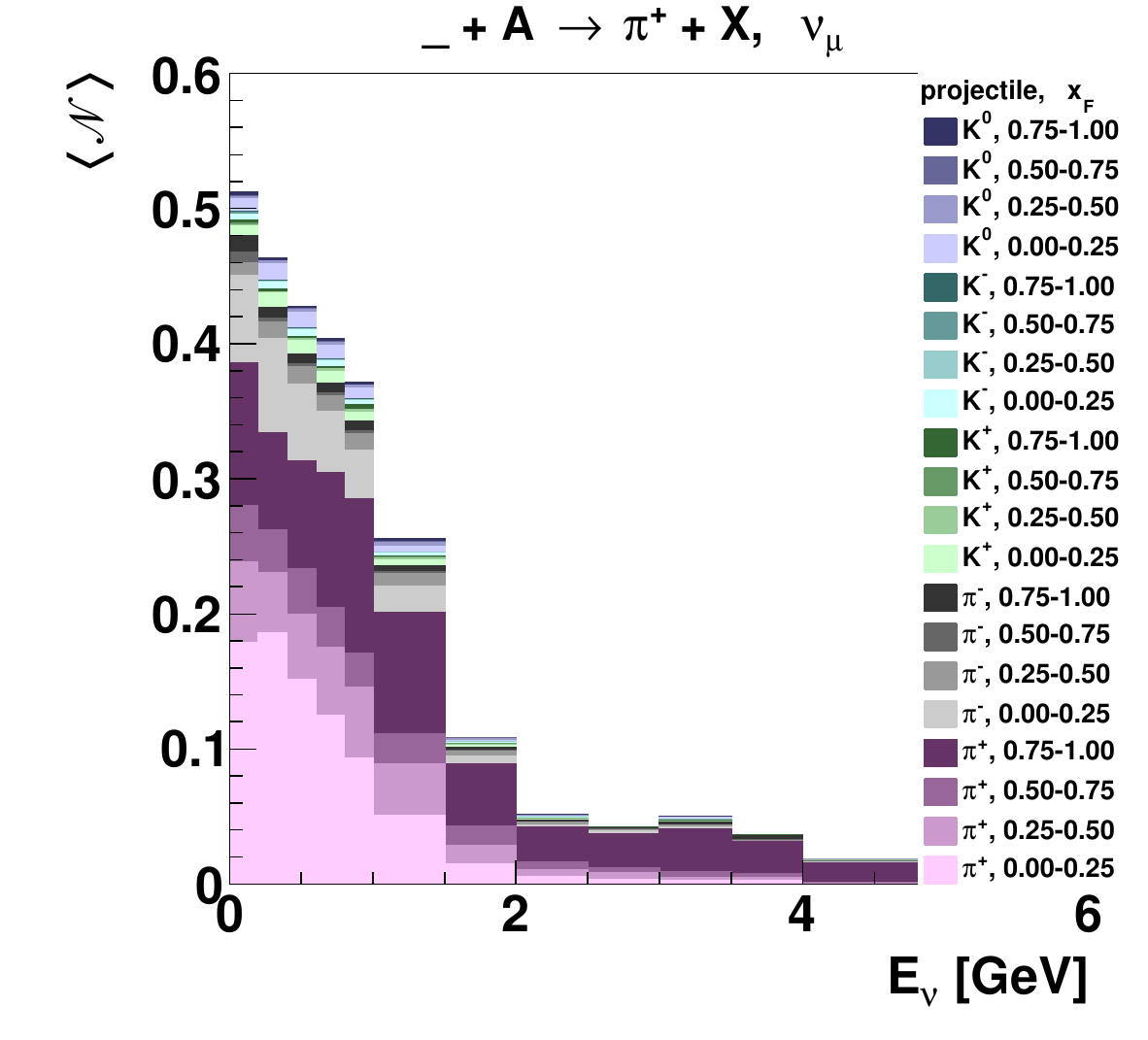}
	\includegraphics[width=0.49\textwidth]{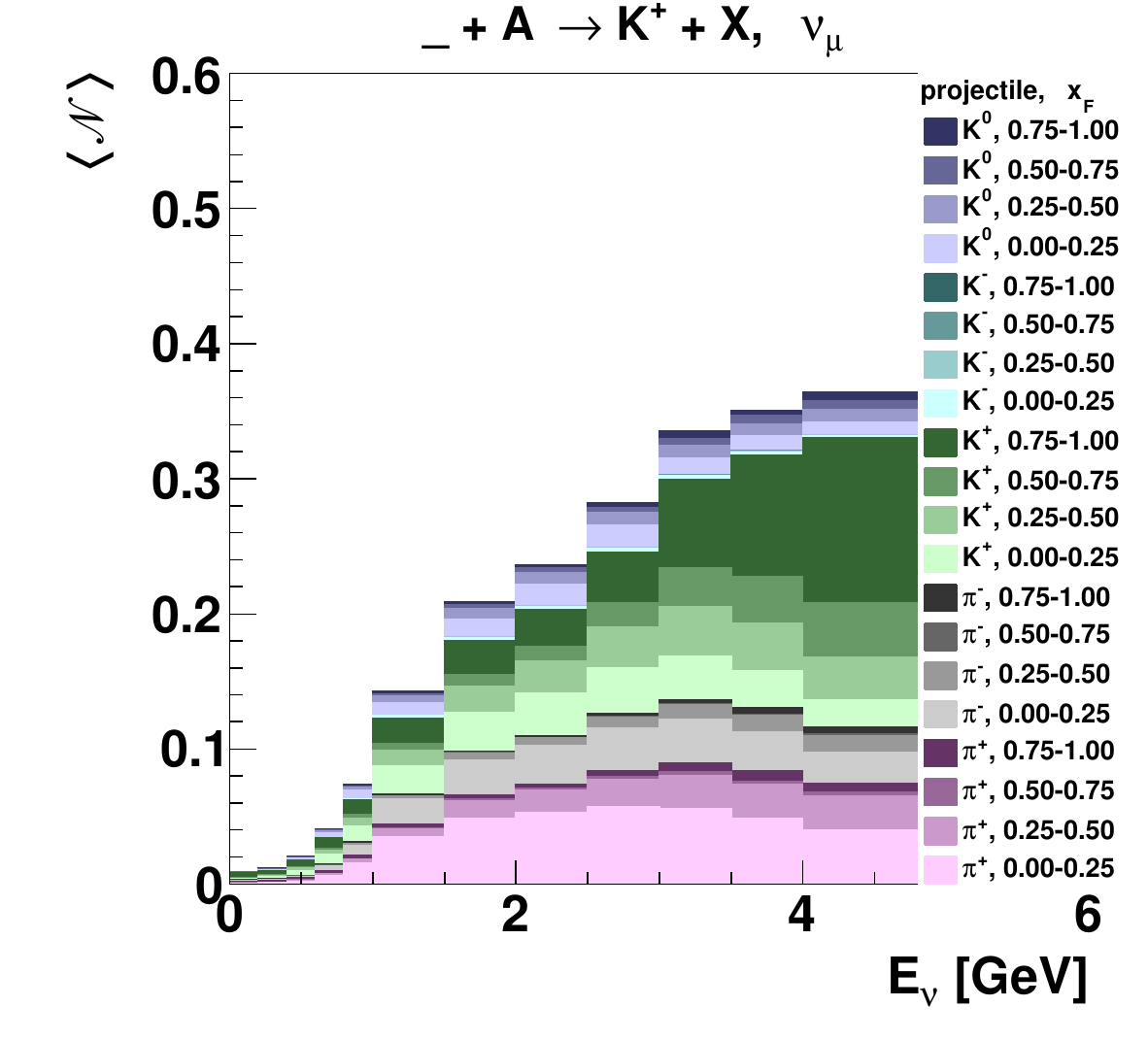}
	\caption[Number of Incident Meson Interactions ($x_F$ Bins)]{The number of interactions of incident mesons in bins of $x_F$, producing $\pi^+$ (left) and $K^+$ (right), in the decay chain leading to the production of \numu{}. \%
	 Some bins in the $x_F$ space distribute broadly across when mapped to neutrino energy, e.g., $\pi^+ \to \pi^+$ for $0.75 \leq x_F < 1.00$ and $\pi^+ \to K^+$ for $0 \leq x_F < 0.25$.
	}%
	\label{fig:meson_interactions}
\end{figure}

This section contains a discussion of the \numi\ flux prediction, as well as its related uncertainties, calculated from simulated data prepared as described in Chapter~\ref{sec:numi_ppfx}.
The characterization of the flux is carried out in terms of the neutrino energy and eight neutrino modes, $(\mathrm{FHC}, \mathrm{RHC})\otimes(\numu, \numub, \nue, \nueb)$, as these are the two relevant variables used by the GENIE neutrino event generator~\cite{Andreopoulos:2009rq} to simulate neutrino interactions in the \icarus\ detector.
As discussed at the end of Section~\ref{sec:ppfx}, the hadron interaction phase space is binned according incoming/outgoing particle type, interaction kinematics, etc.
This analysis is performed with the initial expectation that the hadron interaction phase space is highly degenerate when mapped to the neutrino energy-flavor space, and can potentially produce large correlations between bins of neutrino energy and flavor.
As an example, Figure~\ref{fig:meson_interactions} demonstrates the manner in which some of the hadron interaction phase space distributes broadly across across the \numu\ energy spectrum.
Resulting correlations between bins of neutrino energy and flavor can be exploited to make higher precision measurements using the arguments described in Section~\ref{sec:xsec_uncert}.
The results of this analysis have been made available to the \icarus\ collaboration for application to \numi\ analyses~\cite{sbndata}, and the code to reproduce it is publicly available~\cite{FluxAnalysis}.
The inputs used in this section are as follows.
\begin{enumerate}
	\item The nominal flux prepared as described in Section~\ref{sec:beamsim}
	\item The PPFX central value flux produced by the PPFX reweighting procedure discussed in Section~\ref{sec:ppfx}
	\item The set of $N = 100$ PPFX universes created using the procedure described in Section~\ref{sec:ppfx}.
	\item A set of flux simulations created using alternate geometry configurations\footnote{These alternate geometries cannot be estimated through reweighting. Instead, a set of statistically independent files must be generated for each. The variations are informed based on the design tolerances and operating conditions of the beam.}.
\end{enumerate}

\section{Methods for Estimating and Combining Uncertainties}
\subsection{Methods to Extract the Hadron Production Corrections and Uncertainties from PPFX}
As discussed in Section~\ref{sec:ppfx}, PPFX was used to perform hadronic interaction reweighting on the input Monte Carlo simulation based on experimental hadron production data, which produced a set of flux universes varied according to the underlying hadron interaction cross section model.
The distribution of flux universes was studied and used to calculate the corrected \numi\ flux prediction.
Systematic uncertainty based on data, or model if no data is available, was propagated to the flux prediction.
After verifying that the universes were distributed normally (see Section~\ref{sec:had-uncerts})--which is a prerequisite to estimate the flux prediction based on the central moments of the flux universe distribution described below--the predicted flux in each bin of neutrino flavor and energy was calculated as the mean flux across $N$ PPFX universes:

\begin{equation}
	\overline{\phi}_i = \frac{1}{N}\sum_{k = 1}^N \phi_i^k,
\end{equation}
where \(\overline{\phi}_i\) is the mean flux in bin \(i\),
and \(\phi_i^k\) is the neutrino flux in bin \(i\) and universe \(k\).
Similarly, the systematic uncertainty due to hadron production modelling in a particular bin, \(\sigma_i\),
was calculated as the width of the flux universe distribution:

\begin{equation}
	\sigma_i = \sqrt{\frac{\sum_{k = 1}^N{\left({\phi_i^k - \overline{\phi}_i}\right)}^2}{N - 1}}.
\end{equation}

To quantify the impact of the hadron interaction reweighting across the entire neutrino energy-flavor space,
a covariance matrix, $\mathbf{V}_\textup{hp}$, whose entries are the covariance between energy-flavor bins $i$ and $j$:
\begin{equation}\label{eq:cov}
	\sigma_{ij} = \frac{1}{N-1}\sum_{k = 1}^N \frac{\left(\phi_i^k - \overline{\phi}_i\right)\left(\phi_j^k - \overline{\phi}_j\right)}{\overline{\phi}_i \overline{\phi}_j},
\end{equation}
where, as a reminder, $\sigma_{ij}$ is the matrix element corresponding to the covariance between energy-flavor bins, \(i\) and \(j\).

A principal component analysis of the total hadron production uncertainty covariance matrix, $\mathbf{V}_{hp}$, was performed to extract its eigenvalues, \(\lambda\), and normalized eigenvectors, \(\vb{\hat{u}}\).
\begin{equation}
	\mathbf{V}_{hp} = \begin{bmatrix}\vb{\hat{u}}_1 & \cdots & \vb{\hat{u}}_n\end{bmatrix}
	\begin{bmatrix}\lambda_1 & \cdots & 0 & 0 \\ \vdots & \ddots & 0 & 0 \\ 0 & \cdots & \lambda_{n-1} & 0\\ 0 & \cdots & 0 & \lambda_n\end{bmatrix}
	\begin{bmatrix}\vb{\hat{u}}_1 & \cdots & \vb{\hat{u}}_n\end{bmatrix}^{-1}
\end{equation}
The eigenvalue and eigenvector pairs were sorted in order of decreasing eigenvalue.
The eigenvectors, $\vb{\hat{u}}$, are linear combinations of energy-flavor bins forming an orthogonal basis along each axis of which the variance of the matrix is maximized.
Each eigenvalue, $\lambda$, encodes the total variance described by the corresponding eigenvector.
It is possible to use these quantities to identify and reduce the effects of degenerate parameters and statistical noise by selecting a threshold for the fractional eigenvalue, \(\alpha\), below which the eigenvalue-vector pairs are removed, and the covariance matrix is reconstructed.
\begin{equation}
	\left(\frac{\lambda_k}{g_n} < \alpha \to \lambda_k = 0\right) \; \forall \; k \leq n,
\end{equation}
where $g_n$ is the sum of all positive, real eigenvalues:
\begin{equation}
	g_n = \sum_{k=1}^{n} \lambda_{k}.
\end{equation}

For this analysis, all positive, real eigenvalues were kept, while negative eigenvalues corresponding to statistical noise were discarded.
A modified covariance matrix, $\mathbf{V}_{PC}'$, is constructed using the remaining eigenvalues.

\begin{equation}
	\mathbf{V}_{hp}' = \begin{bmatrix}\vb{\hat{u}}_1 & \cdots & \vb{\hat{u}}_n\end{bmatrix}
	\begin{bmatrix}\lambda_1 & \cdots & 0 & 0 \\ \vdots & \ddots & 0 & 0 \\ 0 & \cdots & \color{red}0 & 0\\ 0 & \cdots & 0 & \color{red}0\end{bmatrix}
	\begin{bmatrix}\vb{\hat{u}}_{1} & \cdots & \vb{\hat{u}}_n\end{bmatrix}^{-1}
\end{equation}

\subsection{Methods to Estimate the Uncertainties from Beamline Mismodeling and Variations in Operating Conditions}\label{sec:beam_uncert_methods}
To estimate the uncertainty due to focusing of the NuMI beam, statistically independent samples of the neutrino flux were produced according to the variations specified in the \NOvA\ third analysis~\cite{NOvA_MC_3rd}.
Appendix~\ref{sec:appendix_beam_samples} contains a complete description of each sample.
Each variation is taken to be a $1 \sigma$ fluctuation from the nominal, and treated as separate systematic on the flux.
Additional datasets to study the impact of Earth's magnetic field were also considered.
The effect was found to be consistent with the statistical uncertainties; thus it was not to include it in the flux systematic uncertainties.

The difference in flux between each Run \(8-32\) and the nominal sample (Run 15) was calculated, and a covariance matrix was constructed according to Eq.~\eqref{eq:cov}, where \(n\) is taken to be \(n \in \{\phi_i^+, \phi_i^\textup{nom}, \phi_i^- \}\) and \(\overline{\phi}_i \to \phi_i^\textup{nom}\).
\(\phi_i^\textup{nom}\) represents the nominal flux in \(i^\textup{th}\) bin  of neutrino energy and flavor, and \(\phi_i^\pm\) is the flux for the \(\pm 1\sigma\) variant.
It was discovered that large statistical fluctuations were present in the samples, hindering interpretation of systematic effects.
To compensate, smoothing of the statistical fluctuations was accomplished by applying a median filter, implemented in the ROOT function \verb+TH1::Smooth()+~\cite{SmoothAlg}, to the fractional flux differences,
\begin{equation}
	\frac{\Delta \phi^i_k}{\phi_{nom}^i} = \frac{\left(\phi_{k}^i - \phi_{nom}^i\right)}{\phi_{nom}^i}{.}
\end{equation}

A smoothed version of the absolute flux differences was calculated by multiplying through by the nominal flux.
Smoothed and non-smoothed absolute flux differences were compared against the statistical uncertainty of the nominal sample to discriminate
true systematic effects.
Additionally, Run 30, i.e., a constantly applied magnetic field in the decay medium, is used as a benchmarking sample.

As discussed in Section~\ref{sec:offaxis_flux}, decays of neutrino parent particles occur before the decay pipe, therefore observed effects in Run 30 are purely statistical in origin.
Beamline samples or portions thereof that did not meet the following criteria were removed from the final covariance calculation:
\begin{enumerate}
	\item $\Delta \phi_k \geq \sigma_{stat}$, see Eq.~\eqref{eq:statistical_uncertainty}
	\item $\Delta \phi_k$ exhibits similar trends between smoothed and non-smoothed spectra
	\item Dissimilar from Run 30
\end{enumerate}

\begin{figure}[htbp]
	\centering
	\begin{subfigure}[b]{0.99\textwidth}
		\centering
		\includegraphics[width=\textwidth]{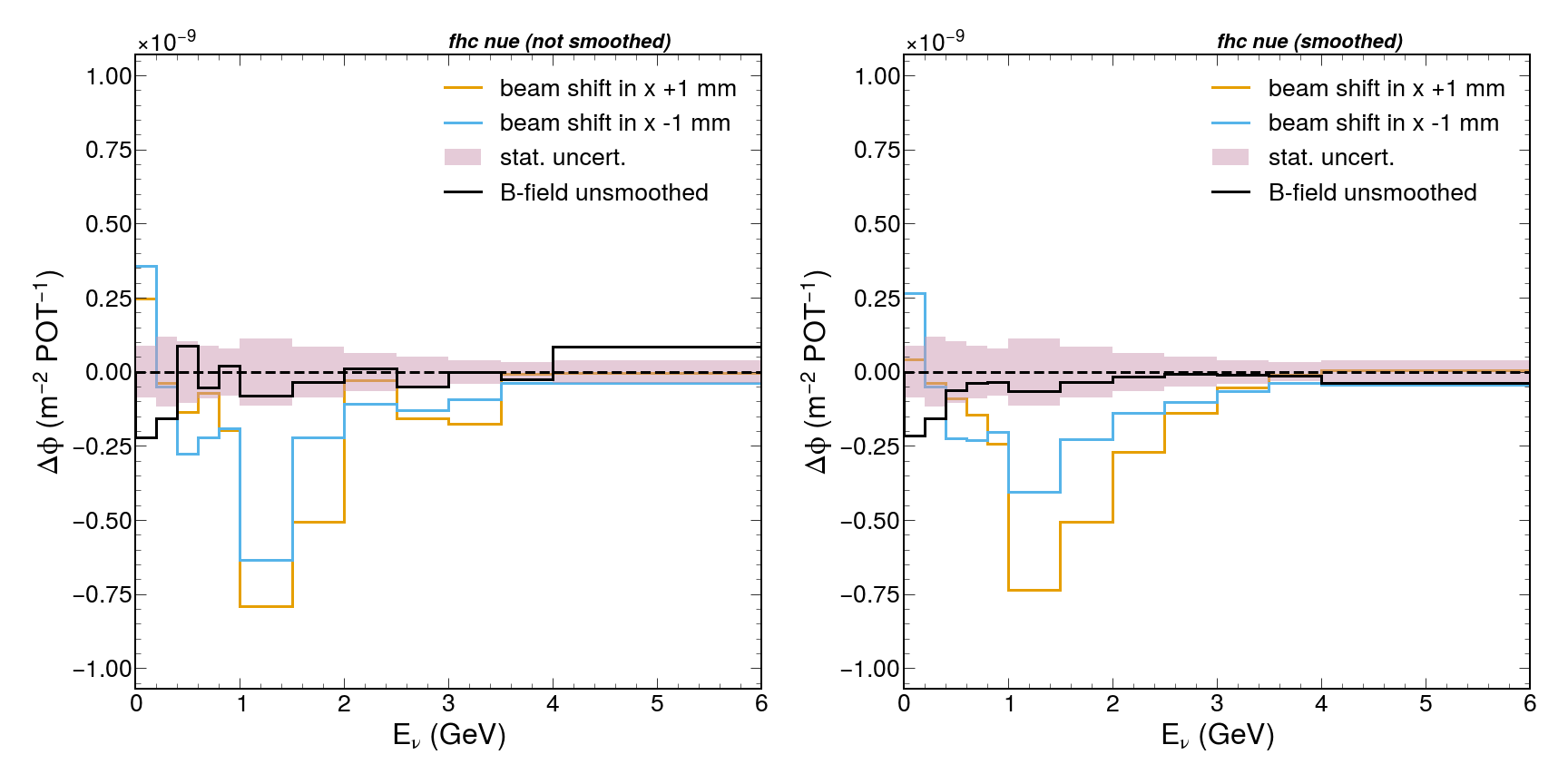}
		\caption[Beam Shift $x \pm \SI{1}{\milli\meter}$]{The effect of shifting the x-position of the NuMI beam is significant between \(0.6 \leq E_{\nu} \leq 3.6 \si{\GeV}\) and is present between both smoothed and non-smoothed spectra. %
		Therefore, this sample was included in the analysis.}%
		\label{fig:did-meet-criteria}
	\end{subfigure}
	\begin{subfigure}[b]{0.99\textwidth}
		\centering
		\includegraphics[width=\textwidth]{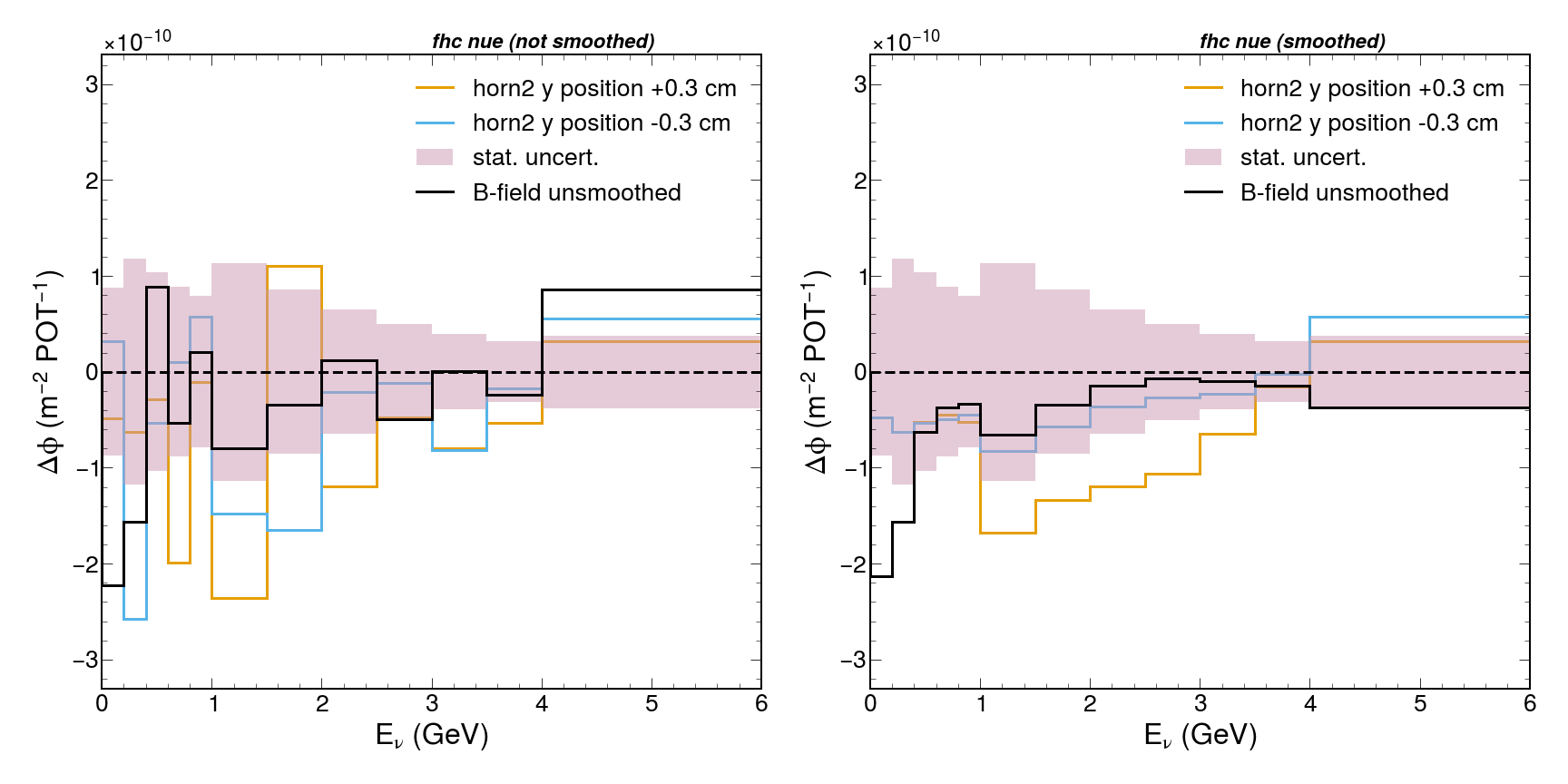}
		\caption[Horn2 $y$-Position $\pm \SI{0.3}{\centi}{\meter}$]{Shift in the $y$-positioning of NuMI Horn2 was excluded due to the presence of large statistical fluctuations in the pre-smoothed spectrum, especially for \(E_\nu\)
			up to \SI{2}{\GeV}. %
			Averaging over the bins between $0.6-\SI{2.5}{\GeV}$ of the $+\SI{0.3}{\centi\meter}$ shift (orange), yields a spectrum closer to the 0-point, e.g., the average of the two largest bins between $1-\SI{2}{\GeV}$ is $\approx -0.6$. %
			In contrast, the smoothing indicates a large shift in the same region, which could be the result of the smoothing algorithm artificially enhancing the effect.}%
		\label{fig:did-not-meet-criteria}
	\end{subfigure}
	\caption[Beam Focusing Systematics Smoothing Procedure]{Smoothed and non-smoothed flux comparison for forward horn current \nue{}, demonstrating examples of samples selected for inclusion
		in (\ref{fig:did-meet-criteria}) and rejection from (\ref{fig:did-not-meet-criteria}) this analysis.}
\end{figure}

\noindent Examples of samples which did and did not meet the criteria are shown in Figures~\ref{fig:did-meet-criteria} and~\ref{fig:did-not-meet-criteria}, respectively.
The remaining spectra have been included in Appendix~\ref{sec:appendix-beam_frac_shifts}.
Finally, covariance matrices were calculated once again using the samples that passed the selection process. The total systematic effect due to beam focusing is calculated from the addition of each individual covariance matrix, \(\mathbf{V}_x\), to yield a total:
\begin{equation}
	\mathbf{V}_\textup{beam} = \sum \mathbf{V}_x.
\end{equation}

\clearpage
\subsection{Combining Uncertainties From All Sources}
Finally, the total uncertainty on the flux prediction is fully represented by adding together the hadron covariance matrix, \(\mathbf{V}_\textup{hp}\), the beam focusing matrix, \(\mathbf{V}_\textup{beam}\), a covariance matrix characterizing the differences between the older and updated beam geometry, \(\mathbf{V}_\textup{geom}\)\footnote{Refer to Section~\ref{sec:megawatt} for more details.},
and a diagonal matrix containing the statistical uncertainty on each neutrino energy-flavor bin, \(\mathbf{V_\textup{stat}}\).

\begin{equation}
	\mathbf{V}_\textup{tot} = \mathbf{V}_\textup{hp} + \mathbf{V}_\textup{beam} + \mathbf{V}_\textup{geom} + \mathbf{V}_\textup{stat}.
\end{equation}

\noindent Since the variances are being added, this is equivalent to a quadrature sum of the correlated $1 \sigma$ uncertainties.
The full covariance matrix is used to calculate the total uncertainty on the flux for various combinations and ratios of the four neutrino modes according to the formula described in Section~\ref{sec:xsec_uncert}.
Both the total uncertainties and each contribution are included in the analysis data products so that analyzers can use and manipulate the components as needed, or use the total for simpler use cases.

\subsection{Flux Binning Scheme}
By default, PPFX outputs flux histograms with \SI{100}{\MeV}-wide bins spanning 0--\SI{20}{\GeV} of neutrino energy.
A lower-resolution, variable bin scheme was adopted to suppress the large statistical fluctuations present in the nominal flux (Run 15).
The fluctuations are especially noticeable where the total flux is relatively small, e.g., the \nue\ and the \numu\ flux in the high-energy tail.
Bin widths were chosen to be monotonically increasing, doubling approximately where the bin-to-bin flux dropped off by a factor of 2 or more.
Care was taken to avoid creating large bins in regions where the flux rapidly increases or decreases.
The binning scheme is outlined in Table~\ref{tab:binning}, and will be used for plots binned in \Enu\ for the remainder of the Note. Additionally, Table~\ref{tab:mat_binning} details how these bins are organized within covariance matrices presented in this note.

\begin{table}[htbp]
	\centering
	\caption[The Neutrino Energy Binning Scheme]{The neutrino energy binning scheme applied to this analysis with the PPFX-corrected NuMI neutrino flux in the corresponding bin. This binning was selected using a heuristic approach in order to enhance energy resolution while simultaneously minimizing the effects of statistical fluctuations present in the flux simulation. \textsuperscript{*}Note that the 12.0--\SI{20.0}{\GeV} bin is removed in the case of both \nue\ and \nueb\ spectra, as this bin was consistently found to be empty across both horn operating modes.}
	\begin{tabular}{l l   c c c c   c c c c}
		\toprule\toprule                                                                                                                                                                                                                                                             \\
		\multicolumn{2}{c}{}  & \multicolumn{8}{c}{Flux [$\si{\meter}^{-2} \text{ } \si{\GeV}^{-1} \text{ } \si{POT}^{-1}$]}                                                                                                                                                         \\
		\cline{3-10}                                                                                                                                                                                                                                                                 \\
		\multicolumn{2}{c}{}  & \multicolumn{4}{c}{FHC}                                                                      & \multicolumn{4}{c}{RHC}                                                                                                                               \\
		Bin                   & Interval                                                                                     & \nue{}                  & \nueb{}         & \numu{}         & \numub{}        & \nue{}          & \nueb{}         & \numu{}         & \numub{}        \\
		                      & [GeV]                                                                                        & $\times 10^{8}$         & $\times 10^{8}$ & $\times 10^{6}$ & $\times 10^{6}$ & $\times 10^{8}$ & $\times 10^{8}$ & $\times 10^{6}$ & $\times 10^{6}$ \\
		\midrule                                                                                                                                                                                                                                                                     \\
		1                     & $\left[0.0, 0.2\right)$                                                                      & 1.417                   & 0.595           & 1.975           & 1.304           & 0.667           & 1.140           & 1.179           & 1.842           \\
		2                     & $\left[0.2, 0.4\right)$                                                                      & 1.861                   & 0.877           & 1.949           & 0.779           & 1.049           & 1.412           & 0.752           & 1.848           \\
		3                     & $\left[0.4, 0.6\right)$                                                                      & 1.566                   & 0.871           & 0.769           & 0.364           & 1.022           & 1.153           & 0.379           & 0.714           \\
		4                     & $\left[0.6, 0.8\right)$                                                                      & 1.462                   & 0.820           & 0.322           & 0.218           & 0.937           & 1.078           & 0.238           & 0.285           \\
		5                     & $\left[0.8, 1.0\right)$                                                                      & 1.374                   & 0.727           & 0.192           & 0.151           & 0.833           & 1.018           & 0.173           & 0.157           \\
		6                     & $\left[1.0, 1.5\right)$                                                                      & 2.693                   & 1.383           & 0.304           & 0.248           & 1.590           & 2.000           & 0.292           & 0.230           \\
		7                     & $\left[1.5, 2.0\right)$                                                                      & 1.417                   & 0.833           & 0.255           & 0.149           & 0.969           & 1.089           & 0.181           & 0.171           \\
		8                     & $\left[2.0, 2.5\right)$                                                                      & 0.603                   & 0.483           & 0.205           & 0.089           & 0.578           & 0.496           & 0.112           & 0.134           \\
		9                     & $\left[2.5, 3.0\right)$                                                                      & 0.296                   & 0.283           & 0.086           & 0.049           & 0.347           & 0.248           & 0.065           & 0.057           \\
		10                    & $\left[3.0, 3.5\right)$                                                                      & 0.165                   & 0.169           & 0.039           & 0.030           & 0.224           & 0.136           & 0.042           & 0.027           \\
		11                    & $\left[3.5, 4.0\right)$                                                                      & 0.092                   & 0.114           & 0.023           & 0.021           & 0.147           & 0.080           & 0.030           & 0.017           \\
		12                    & $\left[4.0, 6.0\right)$                                                                      & 0.114                   & 0.155           & 0.038           & 0.042           & 0.238           & 0.103           & 0.068           & 0.026           \\
		13                    & $\left[6.0, 8.0\right)$                                                                      & 0.014                   & 0.023           & 0.008           & 0.015           & 0.037           & 0.016           & 0.026           & 0.006           \\
		14                    & $\left[8.0, 12.0\right)$                                                                     & 0.003                   & 0.005           & 0.002           & 0.005           & 0.007           & 0.003           & 0.008           & 0.002           \\
		15\textsuperscript{*} & $\left[12.0, 20.0\right]$                                                                    & 0                       & 0               & 9.9e-5          & 3.36e-4         & 0               & 0               & 4.89e-4         & 7.7e-5          \\
		\bottomrule\bottomrule
	\end{tabular}%
	\label{tab:binning}
\end{table}

\begin{table}[htbp]
	\centering
	\caption[Horn-Flavor-Energy Row/Column Ordering]{Horn-flavor-energy element ordering for covariance matrices, read from left to right (columnwise), or bottom to top (row-wise). Refer to Table~\ref{tab:binning} for the \Enu\ bin stops of each neutrino species.}
	\begin{tabular}{r c c r}
		\toprule\toprule
		Row/Column No. & Horn Current & Neutrino Flavor & \Enu\ Range [GeV] \\
		\midrule
		1--14          & FHC          & \nue{}          & [0, 12]           \\
		15--28         & FHC          & \nueb{}         & [0, 12]           \\
		29--43         & FHC          & \numu{}         & [0, 20]           \\
		44--58         & FHC          & \numub{}        & [0, 20]           \\
		59--72         & RHC          & \nue{}          & [0, 12]           \\
		73--86         & RHC          & \nueb{}         & [0, 12]           \\
		87--101        & RHC          & \numu{}         & [0, 20]           \\
		102--116       & RHC          & \numub{}        & [0, 20]           \\
		\bottomrule\bottomrule
	\end{tabular}%
	\label{tab:mat_binning}
\end{table}

\clearpage
\subsection{Validating the Suitability of the Chosen Number of PPFX Universes}\label{sec:had-uncerts}
\begin{figure}
	\centering
	\includegraphics[width=\textwidth]{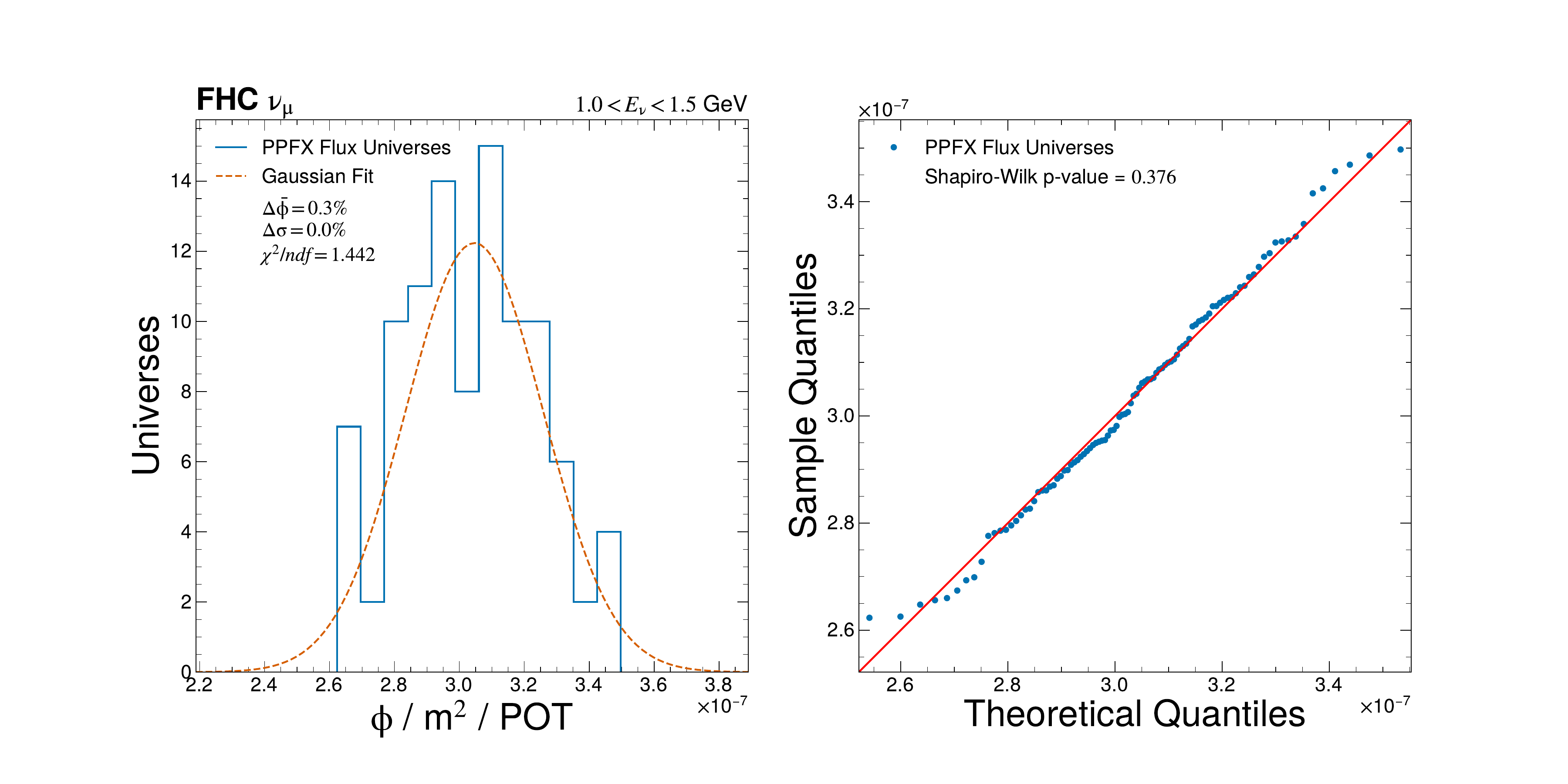}
	\caption[PPFX Universe Gaussianity Tests]{An example distribution of muon neutrino PPFX flux universes (blue) in the 1.0--\SI{1.5}{\GeV}, demonstrating normal behavior. From the Gaussian fit (left), the mean and width of the distribution are consistent with the fit at the sub-percent level. On the right, the Shapiro-Wilk p-value is well beyond a 0.05 significance, and the quantile comparison tracks linearly. Taken collectively, this evidence supports that the flux universes in this bin are approximately normally distributed.}%
	\label{fig:gaussian-fit}
\end{figure}

For this analysis, PPFX was configured to generate 100 flux universes.
To assess the normality of the distribution of flux universes, a fit to a Gaussian function was applied to a histogram of the universes, for each bin of neutrino energy and flavor.
In addition, the Shapiro-Wilk (SW) test and a quantile comparison of the universe distribution with a Gaussian was performed for each bin.
An example of one such fit is shown in Figure~\ref{fig:gaussian-fit}.
An majority ($91\%$) of bins were found to have an SW p-value $> 0.05$, consistent with  Gaussianity.
Fitting Gaussian functions to the distributions revealed a difference in the estimated mean of less than 1\%, on average, across all neutrino modes, an average 7\% difference in the distribution width, a mean $\chi^2 / ndf$ of 0.98, and also followed a linear trend in the quantile comparisons.
See Appendix~\ref{sec:gaussian_fits_to_universes} for the full suite of results in each bin.
These findings indicate that the PPFX universes are normally distributed, 100 universes were sufficient to accurately describe the means and widths, and thus the hadron interaction correction to the NuMI flux extracted from the universes represents the universes' mean and the distsribution widths.

\section{Flux Uncertainties}
\subsection{PPFX Corrections and Uncertainties}\label{sec:PPFX_corr_and_uncert}

\begin{figure}
	\centering
	\includegraphics[width=0.48\textwidth]{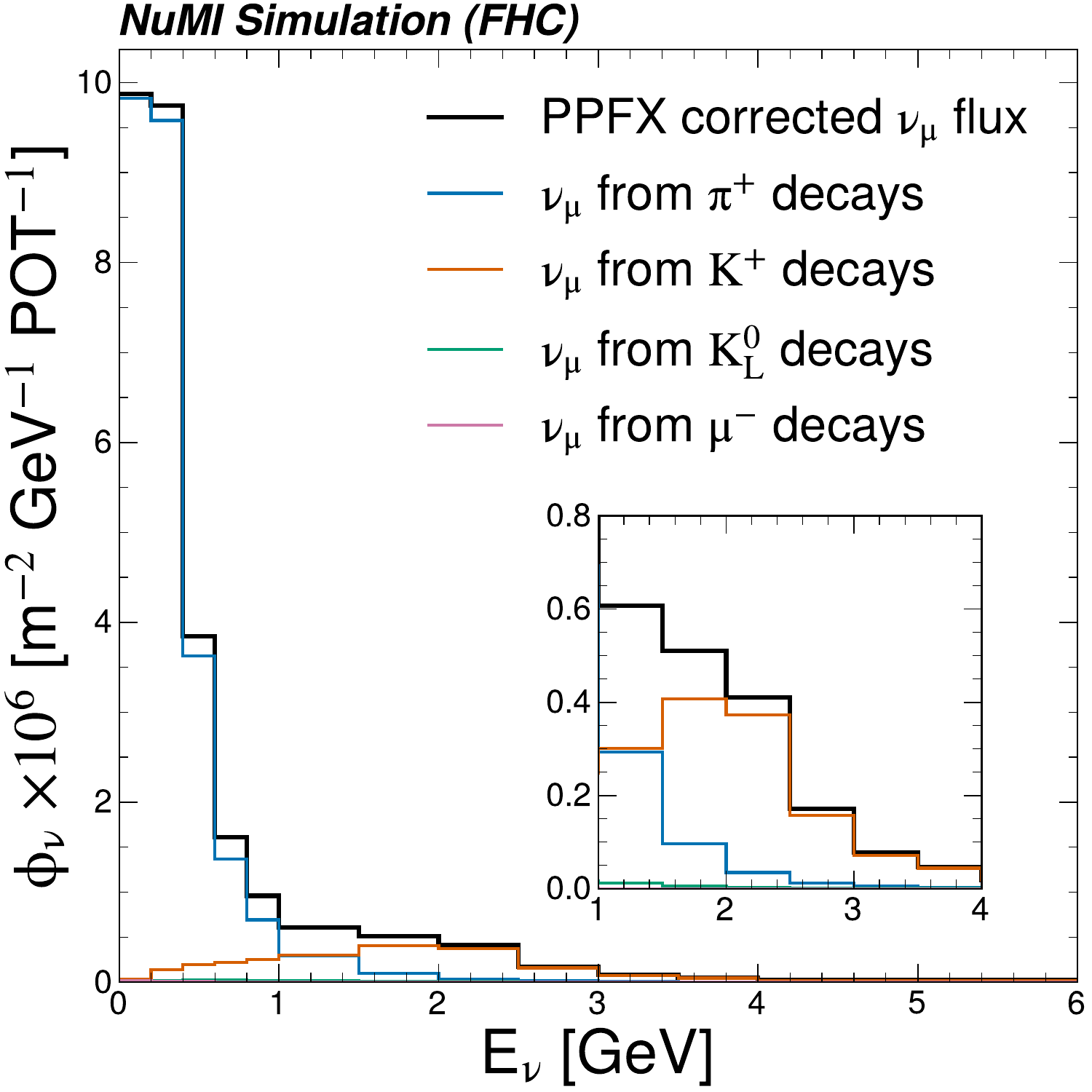}
	\includegraphics[width=0.48\textwidth]{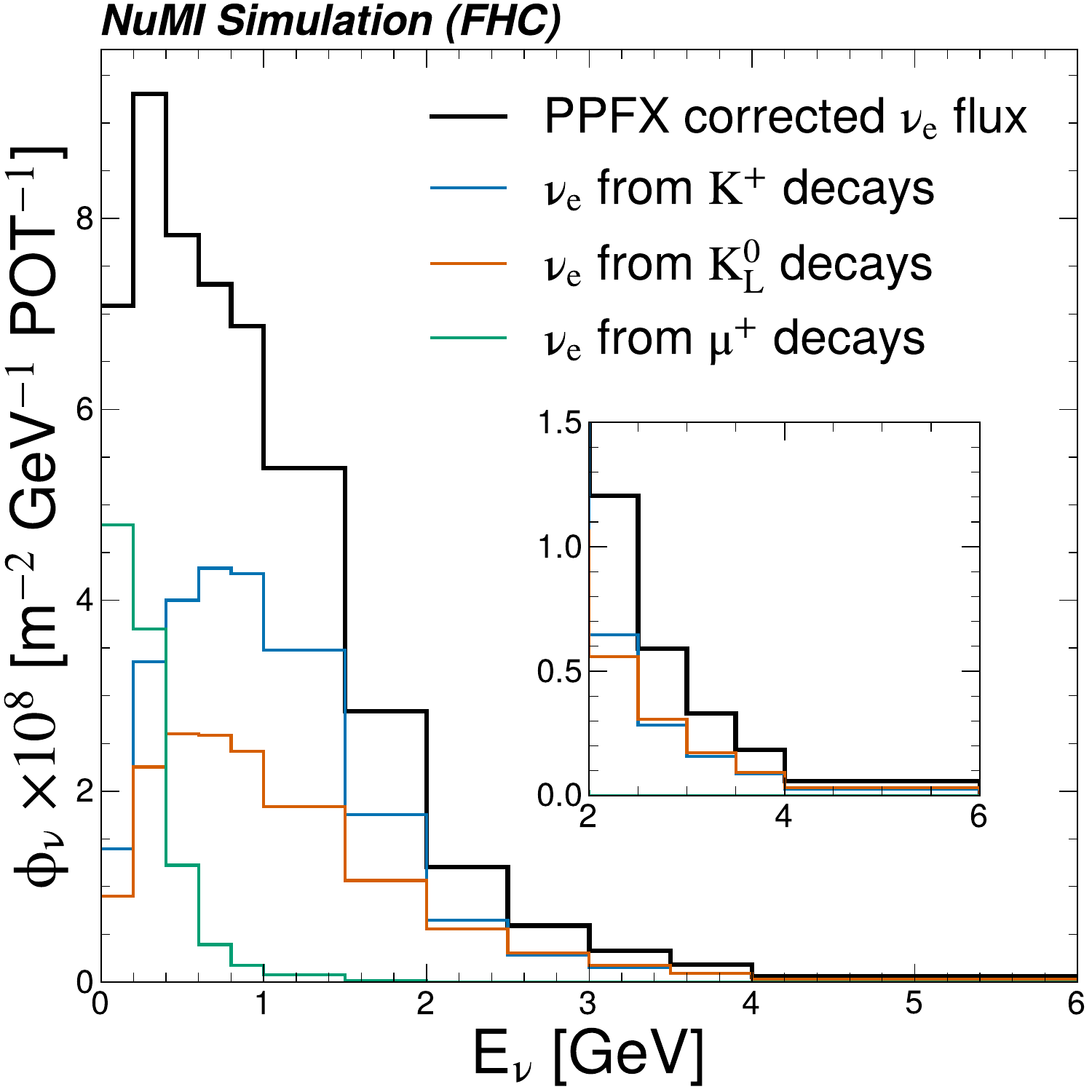}
	\caption[Corrected Flux Spectra for FHC \numu and \nue]{Corrected flux spectra for FHC muon (left) and electron (right) neutrinos with the contributions from the decays of their parent particle.}%
	\label{fig:parent_composition}
\end{figure}

The total corrected flux in each \Enu\ bin is determined from the mean of the 100 flux universes. Figure~\ref{fig:parent_composition} shows the flux after PPFX corrections were applied. The breakdown of the flux by neutrino parent particle is also shown to help decipher the impact of the various hadronic interaction channels on the systematic uncertainty.

\begin{figure}
	\centering
	\includegraphics[width=0.70\textwidth]{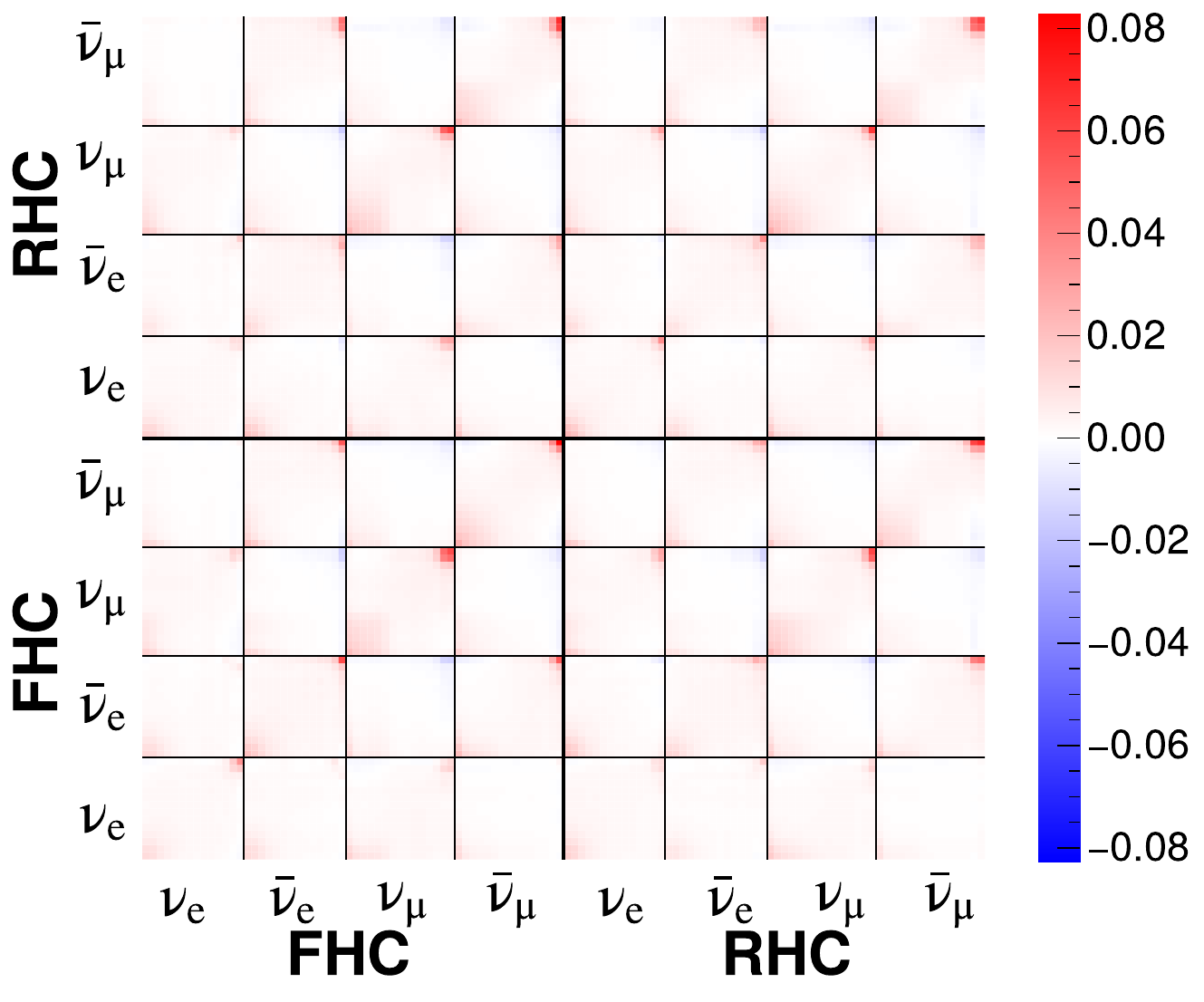}
	\includegraphics[width=0.70\textwidth]{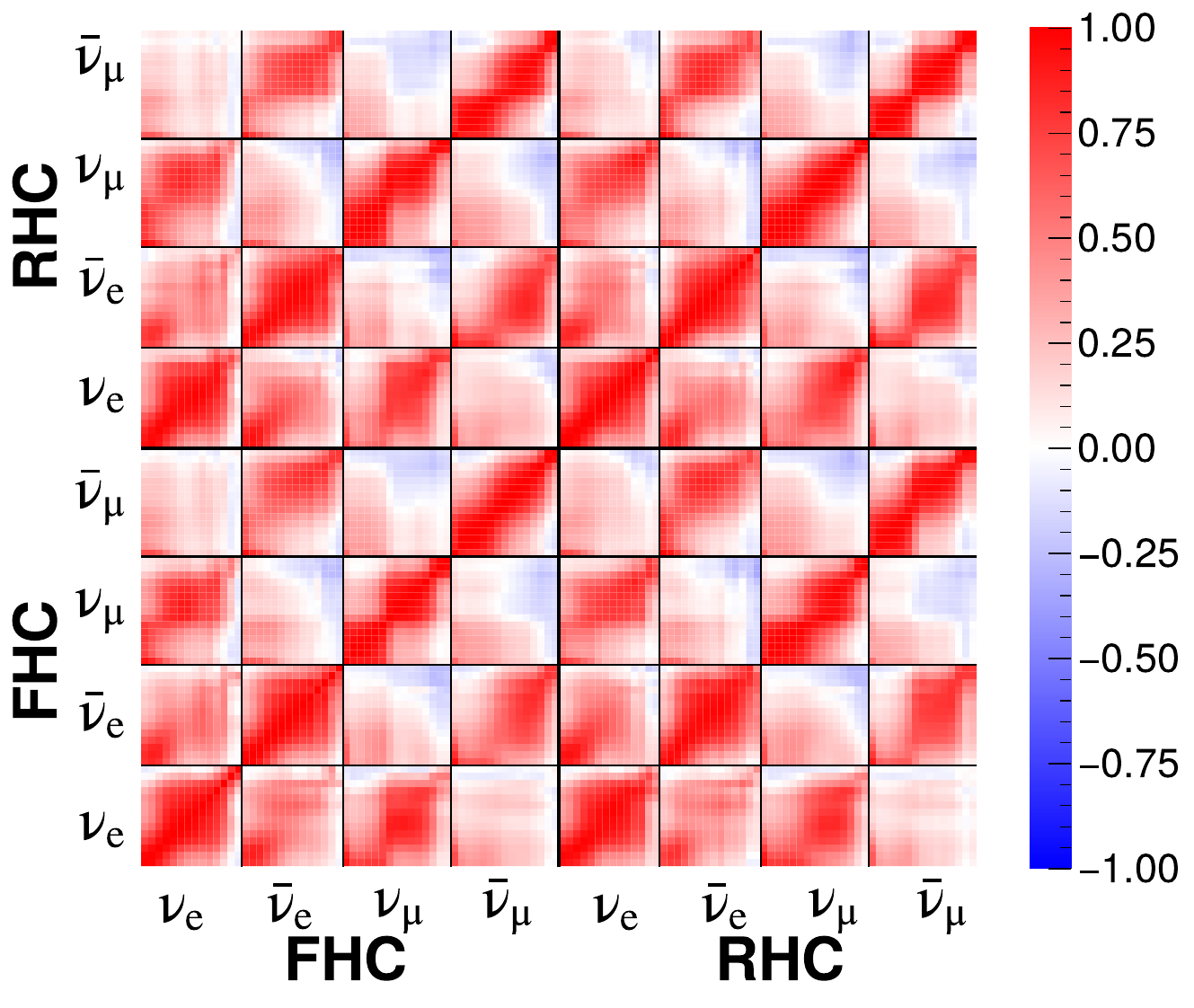}
	\caption[Hadron Interaction Correlation Matrix]{Hadron interaction covariance (upper) and correlation (lower) matrix.
		Each bin in the 2D histogram represents the correlation between corresponding energy bins of fluxes, for a given pair of neutrino modes.
		Positive correlations are present between the majority of bin pairs, while anti-correlations can be found in the 6--20~\si{\GeV} region for certain neutrino-antineutrino bins. Refer to
		Tables~\ref{tab:binning}~and~\ref{tab:mat_binning} for the identities of each matrix row/column.
	}%
	\label{fig:hp-cov}
\end{figure}

For each interaction channel described in Section~\ref{sec:ppfx_channels}, the universes were used to calculate a covariance matrix according to Eq.~\eqref{eq:cov}.
Figure~\ref{fig:hp-cov} shows the total covariance matrix.
Boxes along the diagonal represent the neutrino energy spectrum covariance for a particular neutrino flavor in a specific running mode.
Off-diagonal boxes in the upper-right (RHC) and lower-left (FHC) quadrants give the covariance between energy spectra of different flavors in the same running mode.
The upper-left and lower-right quadrants give the covariance between flavors across the two different running modes.
For analysis of \numu\ interactions with data taken in \fhc, only the corresponding sector located at row 3, column 3 (taking (0, 0) to be the bottom left of the matrix) is required to calculate the total uncertainty.
Similarly, for combined \numu\ and \numub\ analyses, the set of three sectors located at (3, 3), (3, 4), and (4, 4) are required.
Only three are required, because the covariance matrix is symmetric about the diagonal, therefore sector (4, 3) is redundant.
Finally, an analysis involving all horn modes and flavors requires the use of the entire matrix.

As can be seen in the small boxes, strong positive bin-to-bin correlations were found for neighboring and near-neighboring bins across all flavors in both running modes. The strength of the correlations tends to diminish as the distance between bins increases, and there is almost no correlation between the highest energy bins and the lower energy bins. Regions of stronger correlation tend to be defined by the transitions between where different parent particles dominate the neutrino production.

This trend is also seen for neutrinos of the same flavor produced in different running modes, and to a lesser degree for off-diagonal boxes that give correlations for either neutrino-neutrino pairs, or antineutrino-antineutrino pairs. This is likely due to the fact that neutrinos are mostly produced by positive particles, while antineutrinos are mostly produced by negatively charged particles. Off-diagonal boxes that give correlations between neutrino-antineutrino pairs have the weakest correlations, and even exhibit negative correlations in high-energy bins.
Individual covariance matrices for each interaction channel have been included in Appendix~\ref{sec:appendix-had-covs}.

\begin{figure}[htbp]
	\centering
	\includegraphics[width=0.63\textwidth]{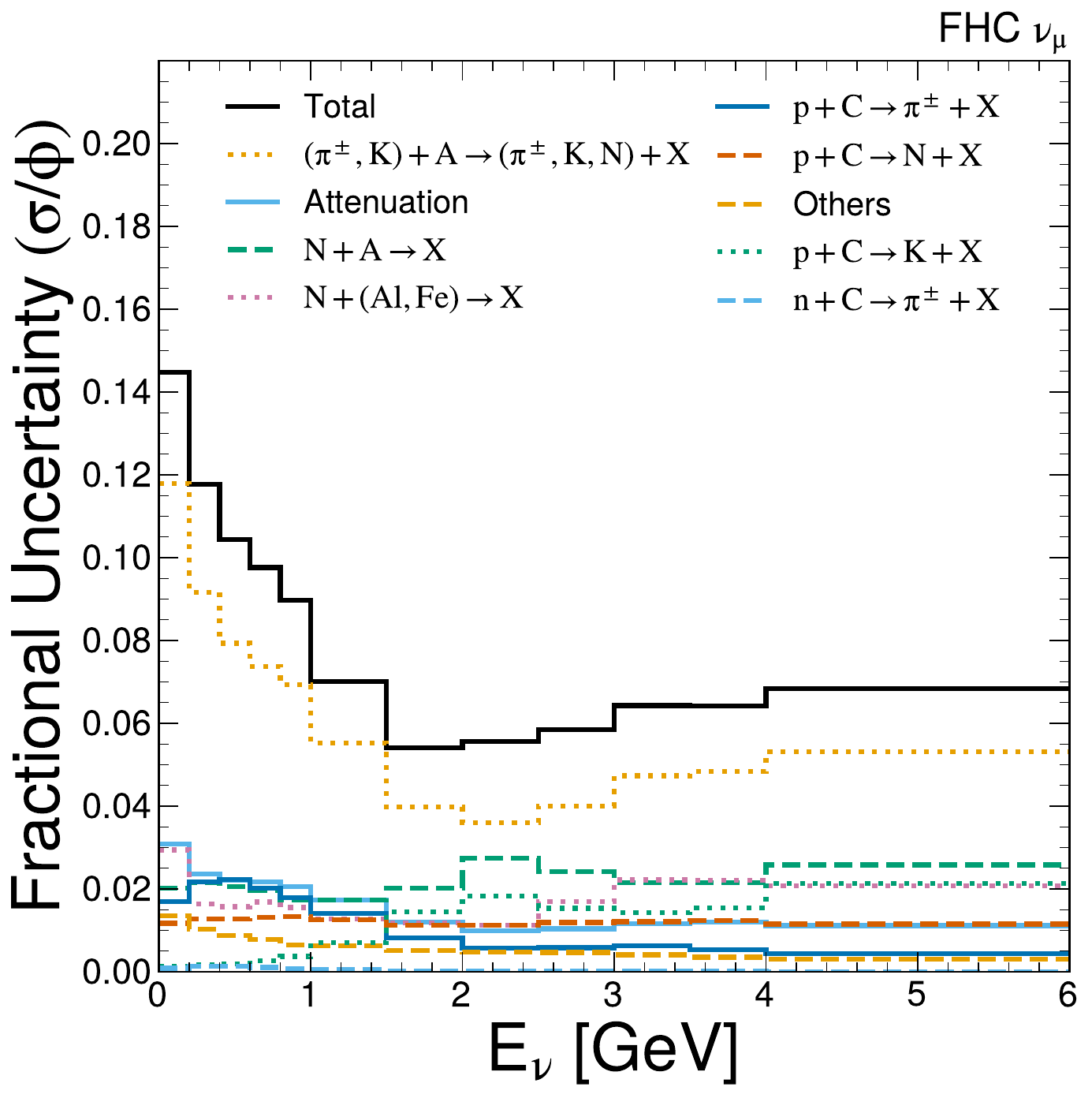}
	\includegraphics[width=0.63\textwidth]{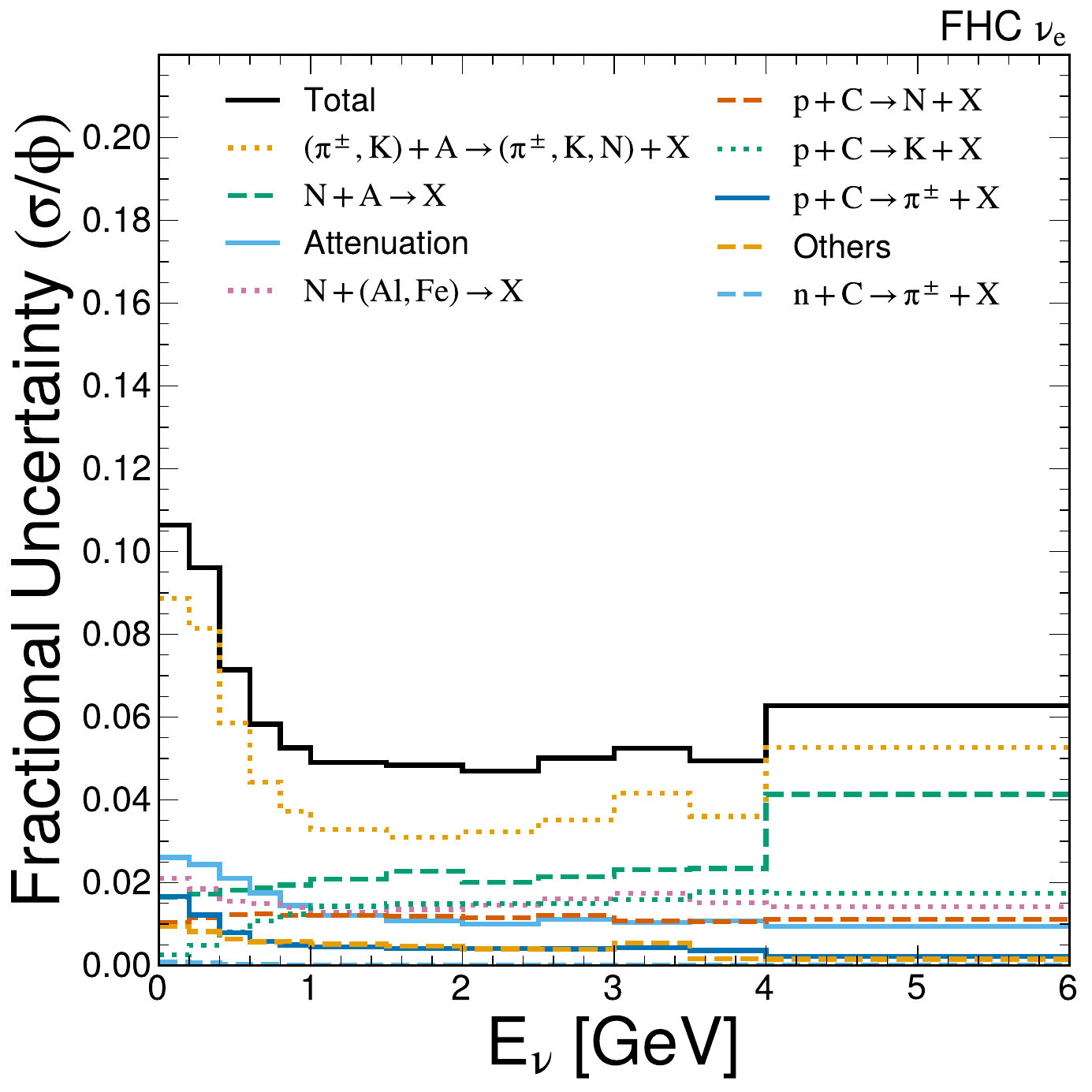}
	\caption[Hadron Interaction Fractional Uncertainties for \numu and \nue]{Hadron interaction fractional uncertainties for muon (top) and electron (bottom) neutrinos. Reinteracting mesons constitute the largest source of uncertainty. Note that the legend is ordered according to the contribution to the total uncertainty within the displayed range of neutrino energy.}%
	\label{fig:frac_uncert}
\end{figure}

The variance of the flux as a function of E$_{\nu}$ is extracted from the diagonals of each component covariance matrices to compute the contributions of each systematic effect to the total. Figure~\ref{fig:frac_uncert} shows the fractional contribution of each interaction channel to the systematic uncertainty for both FHC electron and muon neutrinos.
Appendix~\ref{sec:appendix-had-uncerts} contains fractional uncertainty contributions to the remaining neutrino modes. The total \numu systematic uncertainty is largest in the lower-energy regions of the spectra, where decays from high angle pions dominate. Low energy kaons, many of which undergo secondary interactions on materials other than carbon (see Figure~\verb|\ref{fig:ancestor_interactions}|) also contribute. For \nue the lowest energy bins are dominated by neutrinos from muon decays, but above 500 MeV the kaon induced contribution begins to dominate. The uncertainty peaks at 15\% (11\%) for muon (electron) neutrinos in the 0--\SI{200}{\MeV} bin, and is lowest at the 5\% level in the 1.5--\SI{2.0}{\GeV} bin.
From this study, it was found that the dominant contribution to the uncertainty is the incident meson channel, which characterizes uncertainties on the cross section of hadron reinteractions.
The remaining uncertainty comes from the quadrature sum of multiple channels at 1–3\% level.

\begin{figure}[htbp]
	\centering
	\begin{subfigure}{\textwidth}
		\centering
		\includegraphics[width=0.48\textwidth]{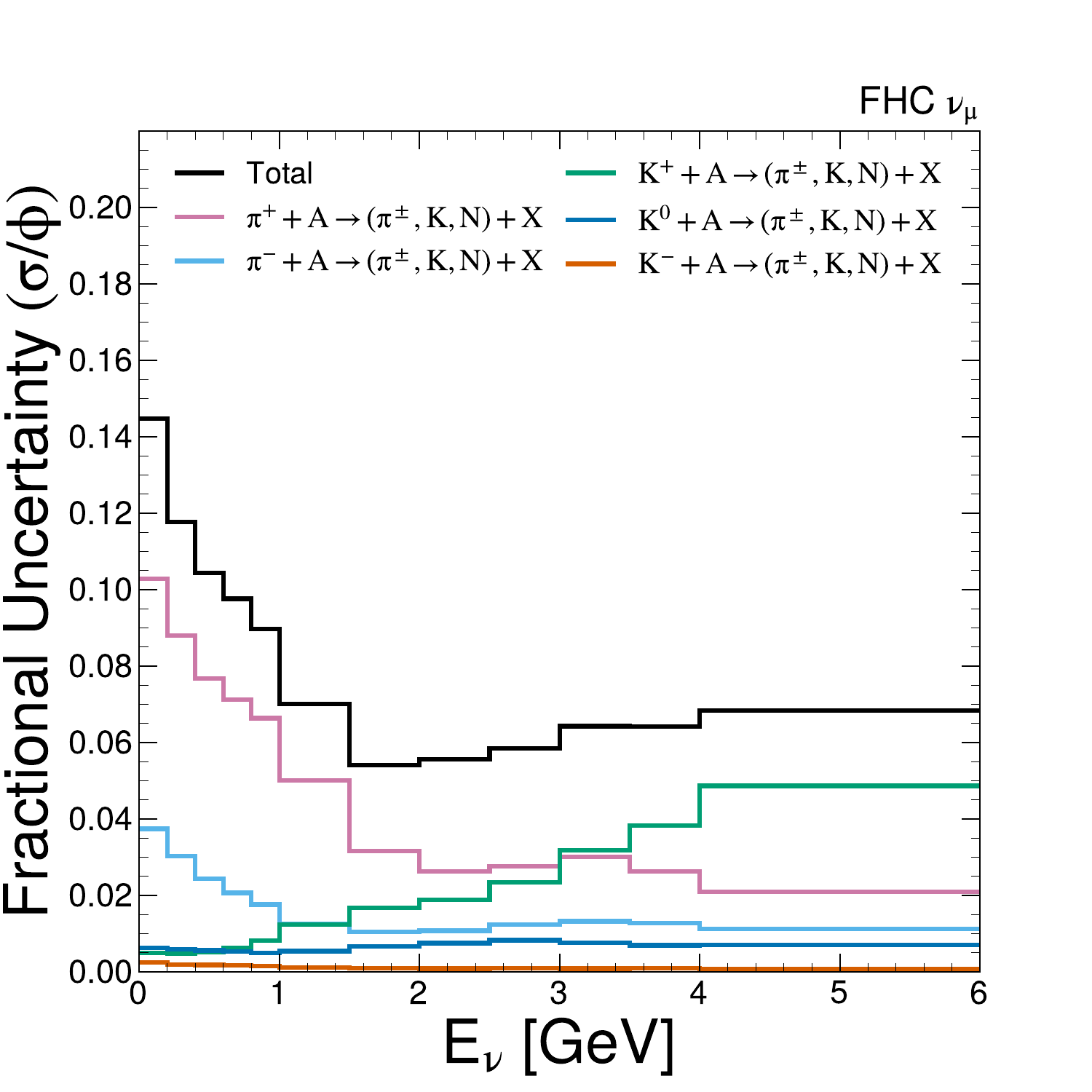}
		\includegraphics[width=0.48\textwidth]{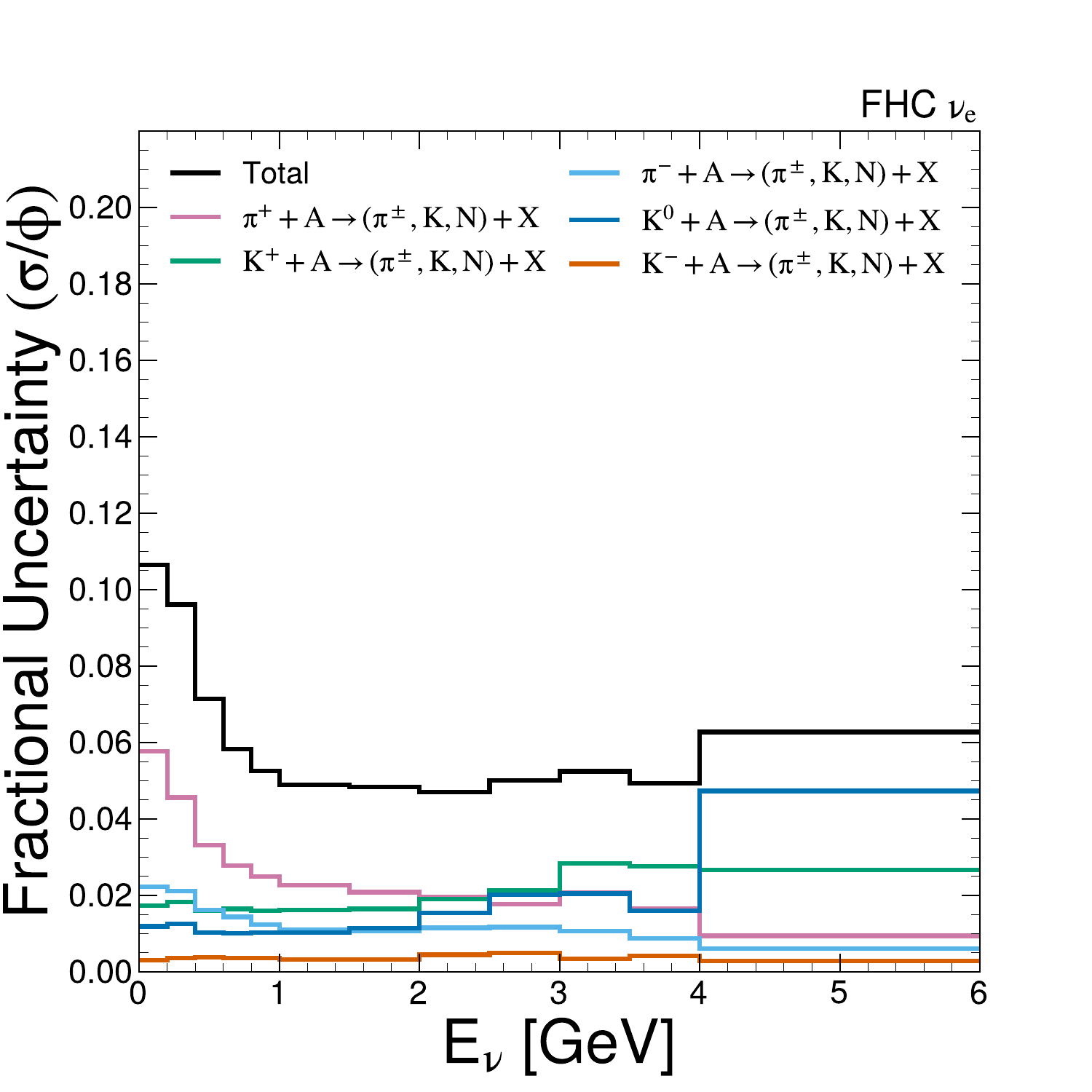}
		\caption{Uncertainty contribution by incoming meson.}%
		\label{fig:meson_breakdown_a}
	\end{subfigure}
	\begin{subfigure}{\textwidth}
		\centering
		\includegraphics[width=0.48\textwidth]{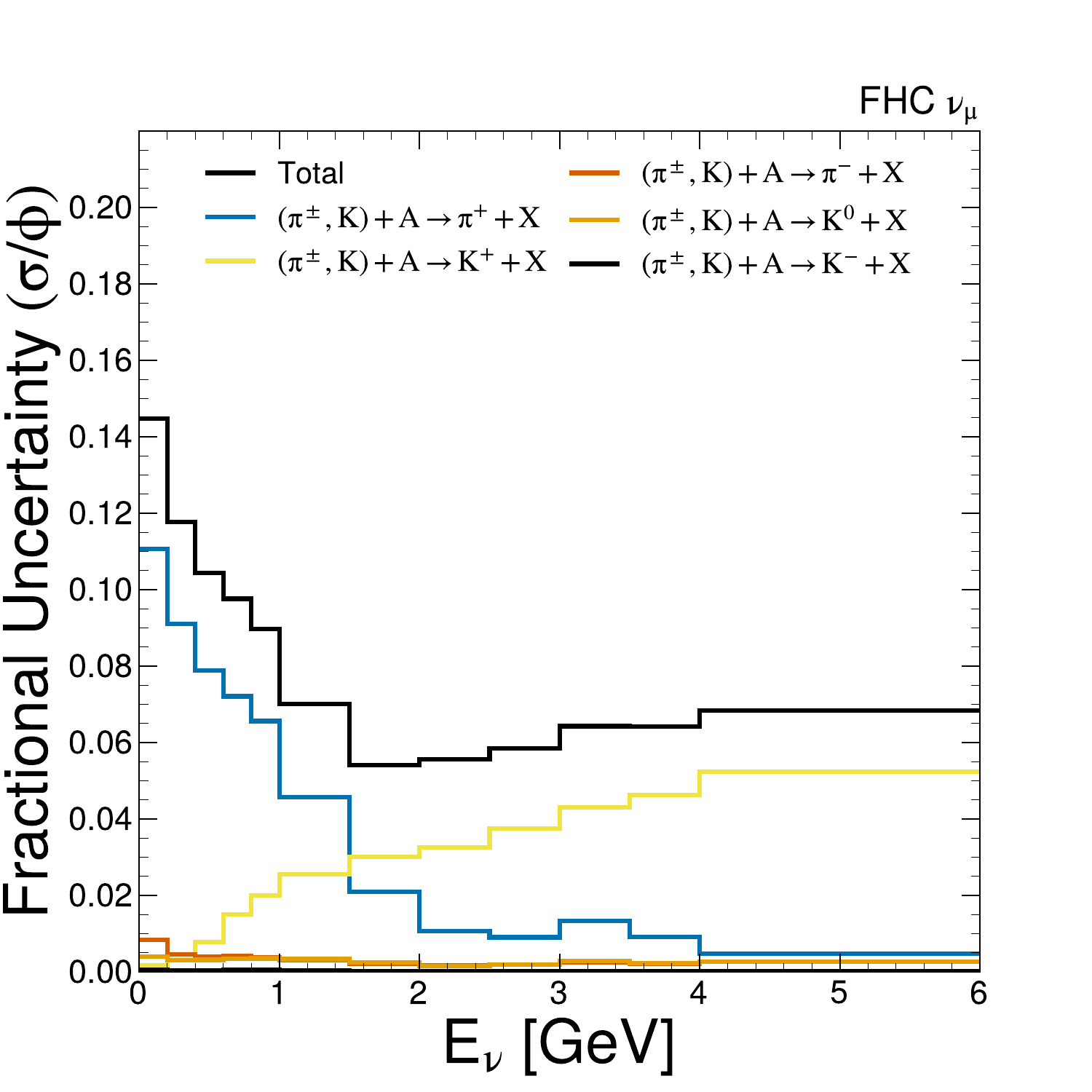}
		\includegraphics[width=0.48\textwidth]{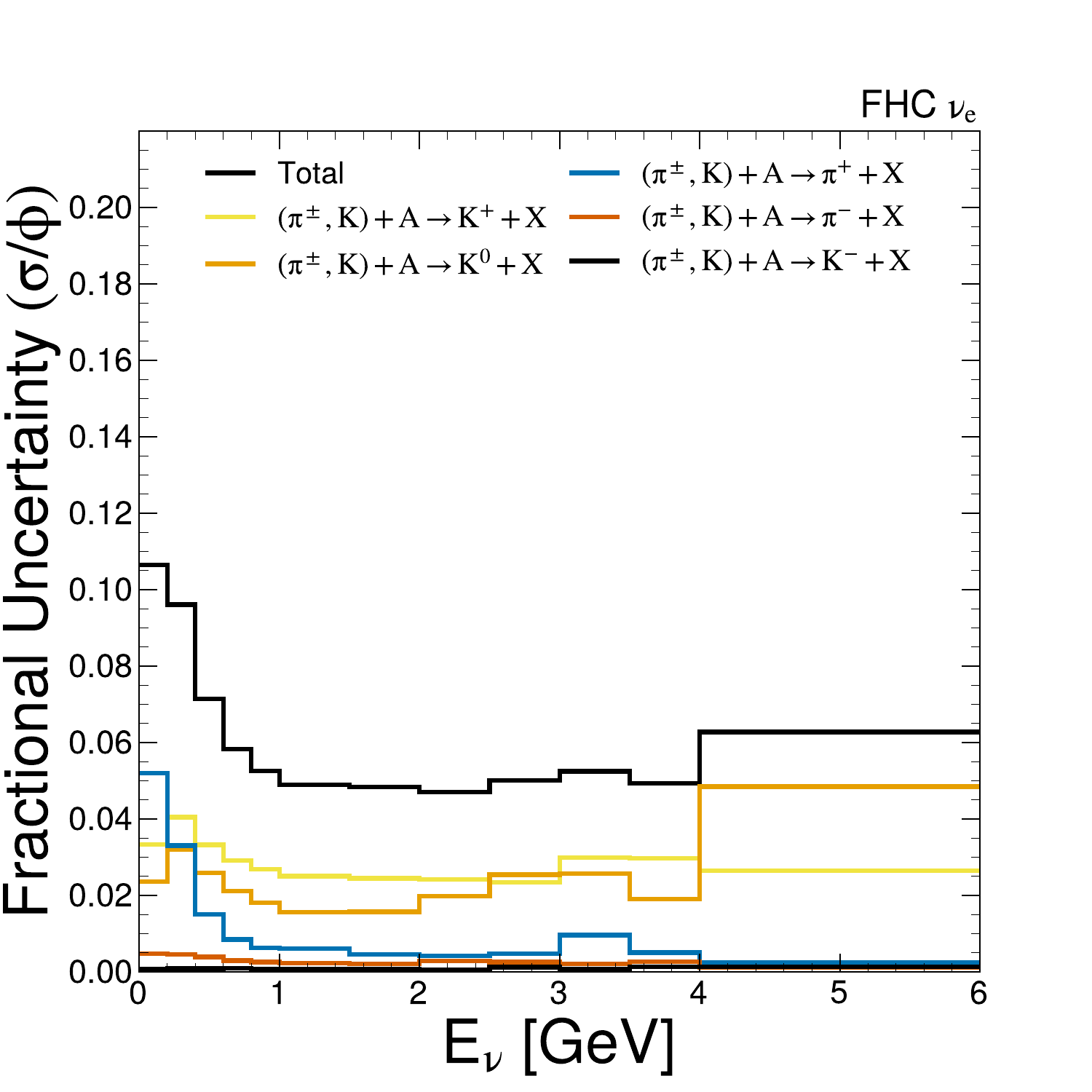}
		\caption{Uncertainty contribution by outgoing meson.}%
		\label{fig:meson_breakdown_b}
	\end{subfigure}
	\caption[Uncertainty Contributions to the Incident Meson Channel]{Uncertainty contributions to the meson-incident channel, grouped by the incoming meson flavor (a) and the outgoing meson flavor (b). %
	 For elastic collisions the categorization is the same, however inelastic collision can change the flavor of the incoming particle, or produce multiple outgoing particles. %
	 Black is the Total HP uncertainty accounting for all effects as shown in Figure~\ref{fig:frac_uncert}. %
	 In the lower set of plots, outgoing charged pions (blue) and kaons (yellow) track according to their contribution to the flux in Figure~\ref{fig:parent_composition}.}%
	\label{fig:meson_breakdown}
\end{figure}

As described in Section~\ref{sec:ppfx}, hadrons produced in the NuMI beam may interact multiple times before decaying to a neutrino that arrives at \icarus.
The cross section on light meson interaction are not currently supported by experimental data in \ppfx{}, therefore these multiply interacting hadrons are subject to the conservatively assigned 40\% uncertainty on the cross section for each reinteraction in the hadron production chain culminating in $\nu$ production.
Figure~\ref{fig:meson_breakdown} further decomposes the inclusive meson channel according to the incoming projectile (Figure~\ref{fig:meson_breakdown_a}) and outgoing particle (Figure~\ref{fig:meson_breakdown_b}).
The contributions from individual mesons track with the expected flux composition, i.e., for \numu, charged pions contribute more significantly at lower values of \Enu, while kaons contribute at higher \Enu.
The crossover between uncertainty contributions is consistent with the crossover point between hadron parents.

\clearpage
\subsection{Principal Component Analysis of the Total Hadron Production Covariance Matrix}
\begin{figure}
	\centering
	\centering
	\includegraphics[width=0.94\textwidth]{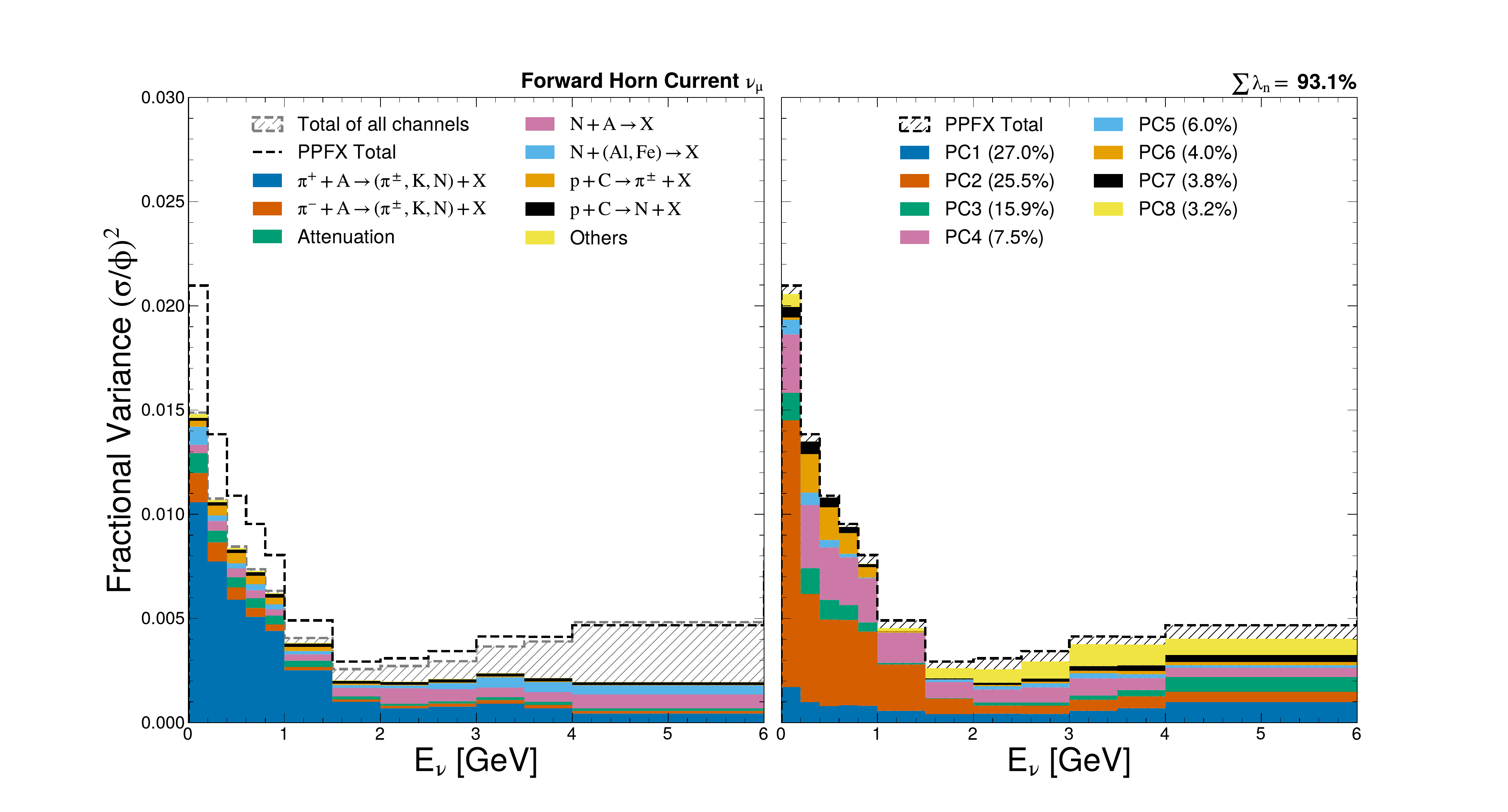}
	\includegraphics[width=0.94\textwidth]{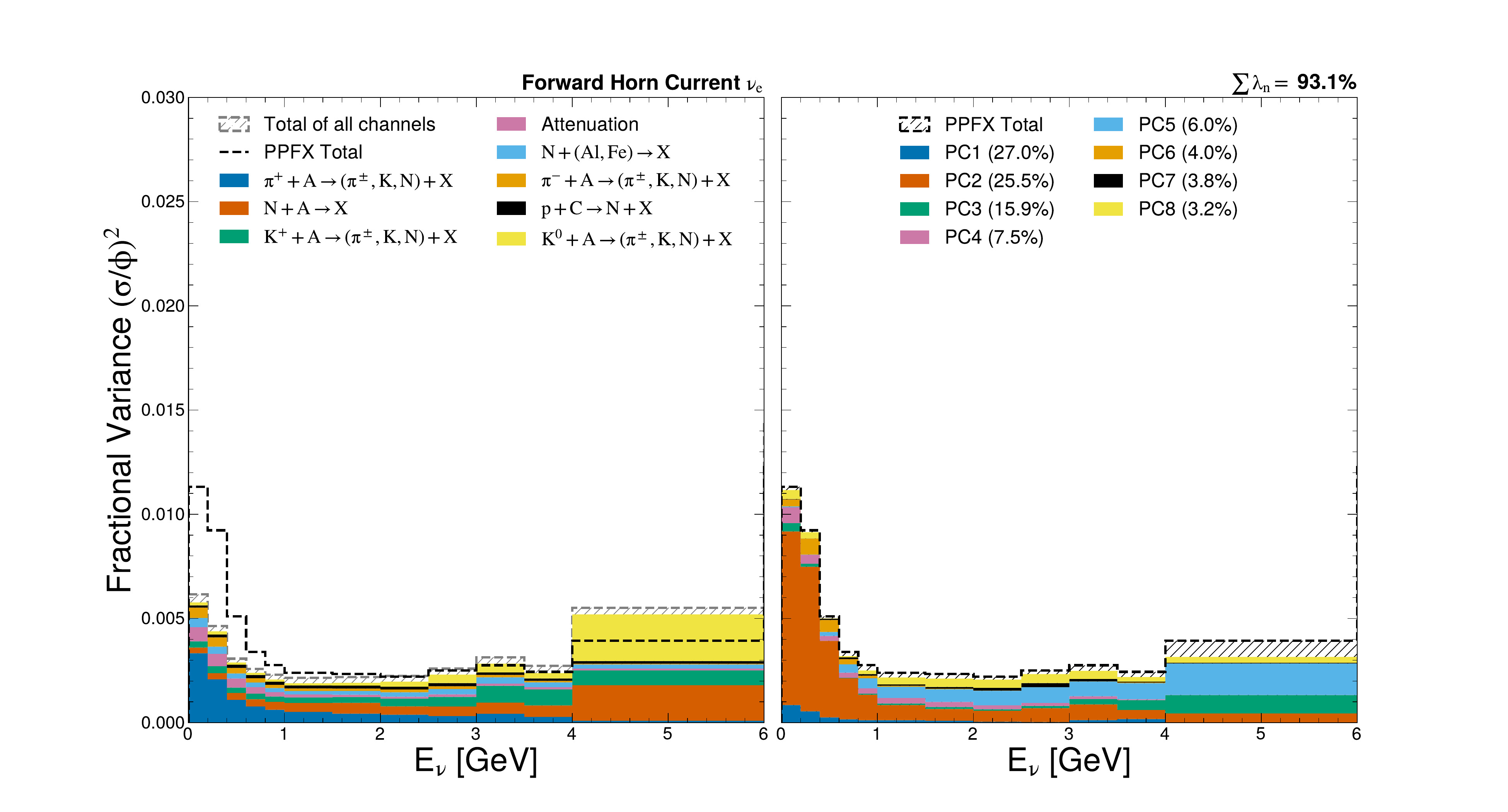}
	\caption[Variance Contributions from Hadron Production vs. Principal Components]{Variance contributions from the top 8 Hadron Production Channels (HPCs, left), and top 8 principal components (PCs, right) for FHC \numu\ (top) and \nue\ (bottom). The ``PPFX Total'' represents the variance across universes, which correctly accounts for correlations across the HPCs, and thus matches between the HPCs and PCs. The ``Total of all channels'' is the sum of the variances of each individual HPCs, which does not properly account for correlations between HPCs. In contrast, the right-hand plots, demonstrate that, because of their linear independence, the PCs fully describe the total variance in the flux from hadron model uncertainties. The discrepancy between these two quantities highlights how channel-to-channel correlations affect bin-to-bin correlations in \Enu. One can also compare the HPC effects with the PCs to estimate which physical effects are captured by each of the PCs and where strong correlations between HPCs allow them to be grouped into a single PC.}%
	\label{fig:pca_variances}
\end{figure}

The PCA description of the hadron production uncertainties has distinct advantages in analyses.
Since the PCs are not correlated with each other, see Figure~\ref{fig:pca_variances}, there is no need to invert a matrix to compute a penalty term for a test statistic.
The small cost is that each parameter now adjusts the weights of events across the full \Enu\ spectrum, and the correct weights in each bin need to be stored and applied correctly.

\begin{figure}
	\centering
	\includegraphics[width=\textwidth]{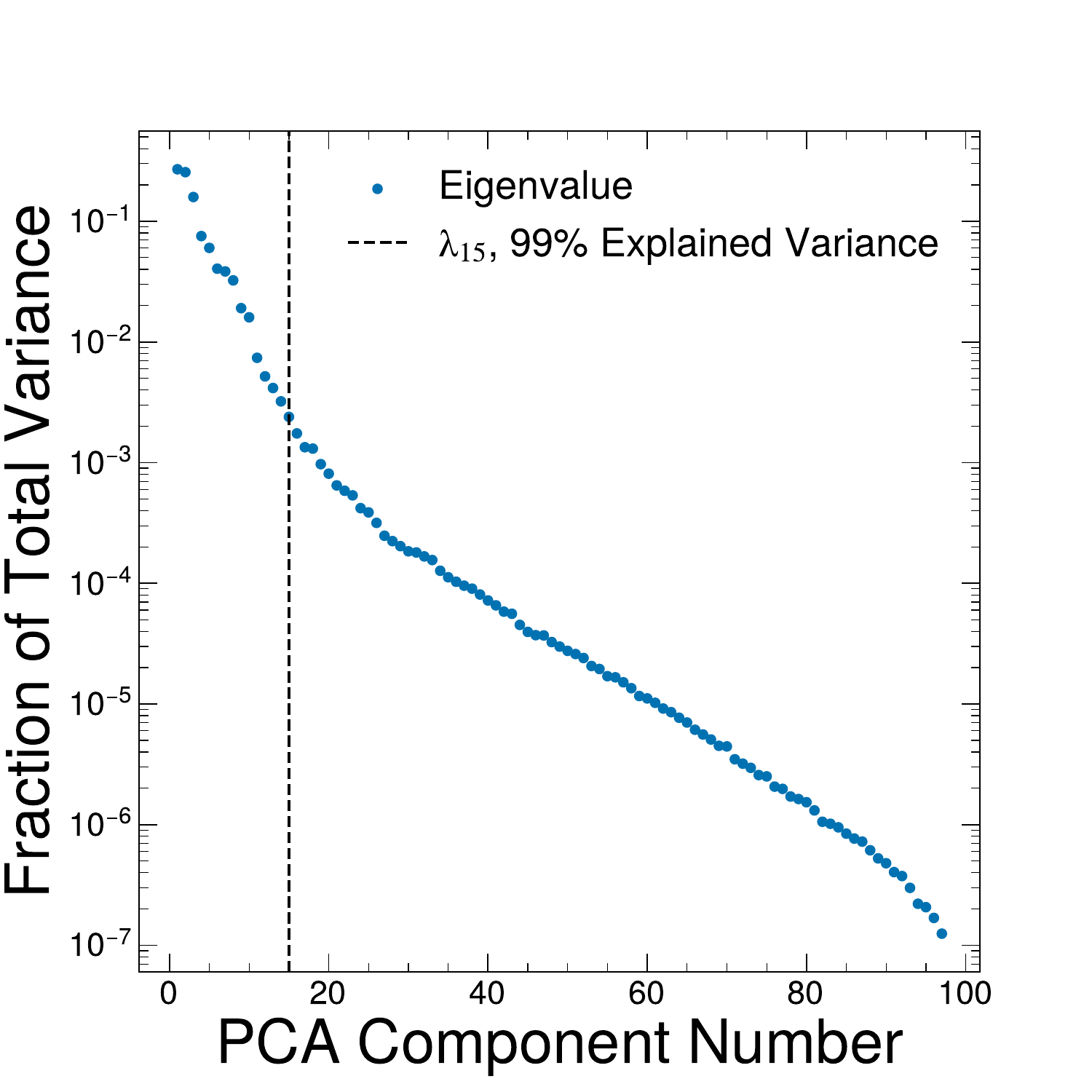}
	\caption[PCA Scree Plot]{Scree plot from the PCA of the hadron covariance matrix. As an example, a dashed line is drawn to demonstrate a threshold eigenvalue, $\lambda_{15}$, to the left of which 99\% of the variance is described, cumulatively.}%
	\label{fig:eigenvalues}
\end{figure}

\begin{figure}
	\centering
	\includegraphics[width=\textwidth]{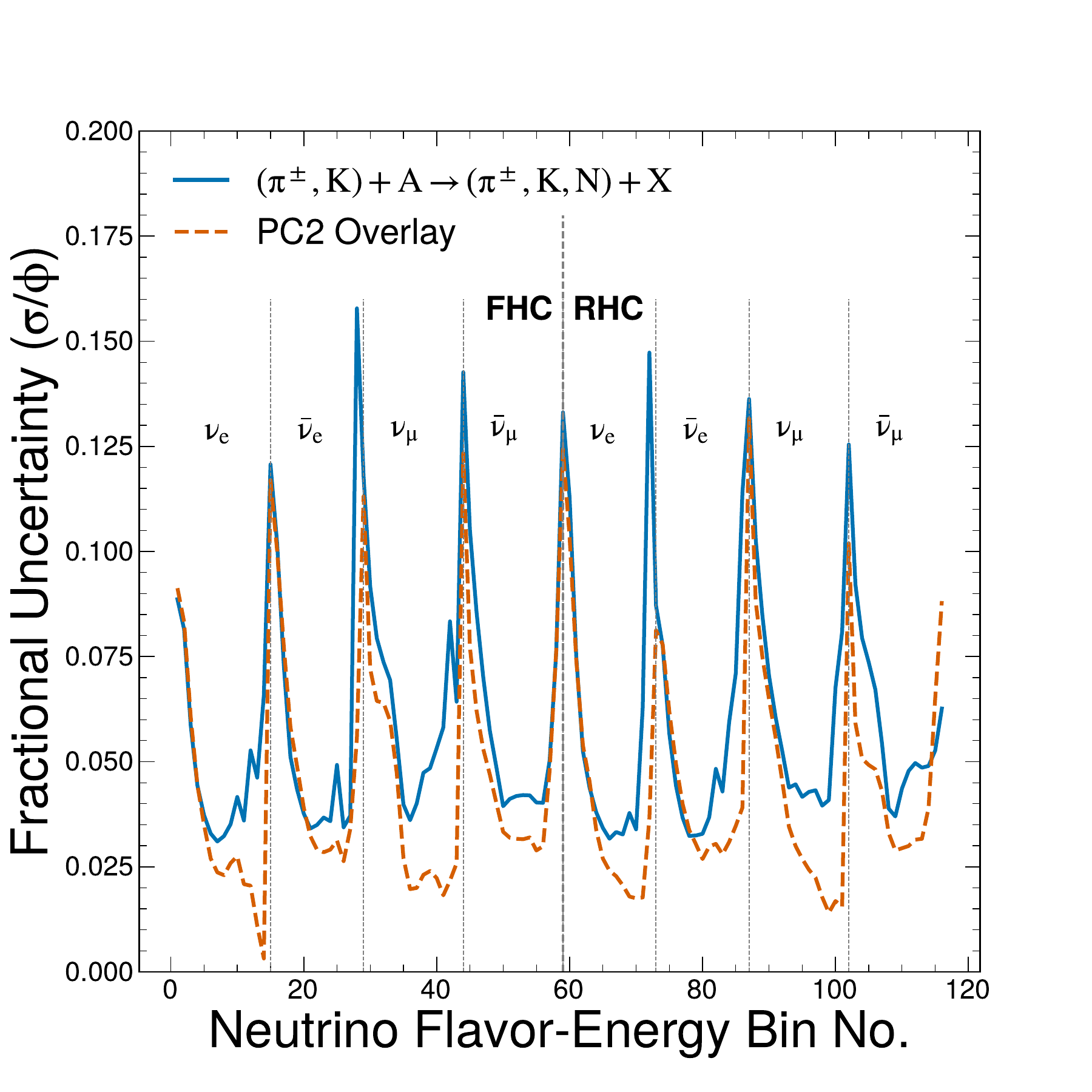}
	\caption[Comparison of the Incident Meson and the Second Principal Component]{Uncertainty due to meson interactions (blue) transposed from bins of neutrino energy and flavor to component numbers. An overlay of the second component (orange) indicates this component captures much of the low-energy variance related to meson interactions. Refer to Tables~\ref{tab:binning}~and~\ref{tab:mat_binning} for more details about bin numbering scheme.}%
	\label{fig:mesinc_overlay}
\end{figure}

The PCA of the hadron covariance matrix yielded 116 components, in accordance with the dimensions of the matrix:

\begin{equation}
	\begin{aligned}
		N_\textup{PC} & = N_\textup{Modes} \times \left( N_\textup{\nue bins} + N_\textup{\nueb bins} + N_\textup{\numu bins} + N_\textup{\numub bins} \right) \\
		              & = 2 \times (14 + 14 + 15 + 15)                                                                                                         \\
		              & = 116,
	\end{aligned}
\end{equation}

\noindent where each PC is represented by a linear combination of all 116 horn-flavor-energy bins. Eigenvalues associated with components 108--116 were found to be negative, where the largest of these 8 eigenvalues (\(\approx -2.90\times10^{-17}\)) was 16 orders of magnitude smaller than the largest positive eigenvalue (\(\approx 0.33\)).
Negative eigenvalues of this scale are likely artifacts of floating-point arithmetic, and, for this reason, these 8 eigenvalues were considered to be consistent with 0 and discarded.
The remaining eigenvalues and their corresponding explained variances, in the fractional scale, are given in Figure~\ref{fig:eigenvalues}.
It was found that 99\% of the variance in the matrix can be described by the first 15 components, while the remaining 85 components encode the final 1\%.
Regardless, all PCs are retained in an output file, which gives analyzers freedom to adjust the number of PCs used.
To demonstrate the features identified by the PCA, Figure~\ref{fig:mesinc_overlay} shows how some components can follow the trends of the individual channels shown in Figure~\ref{fig:frac_uncert}.
Appendix~\ref{sec:appendix_pca} contains the complete PCA results.

\subsection{Uncertainty Due to Focusing of the NuMI Beam}\label{sec:beam-uncerts}

\begin{figure}
	\centering
	\includegraphics[width=0.48\textwidth]{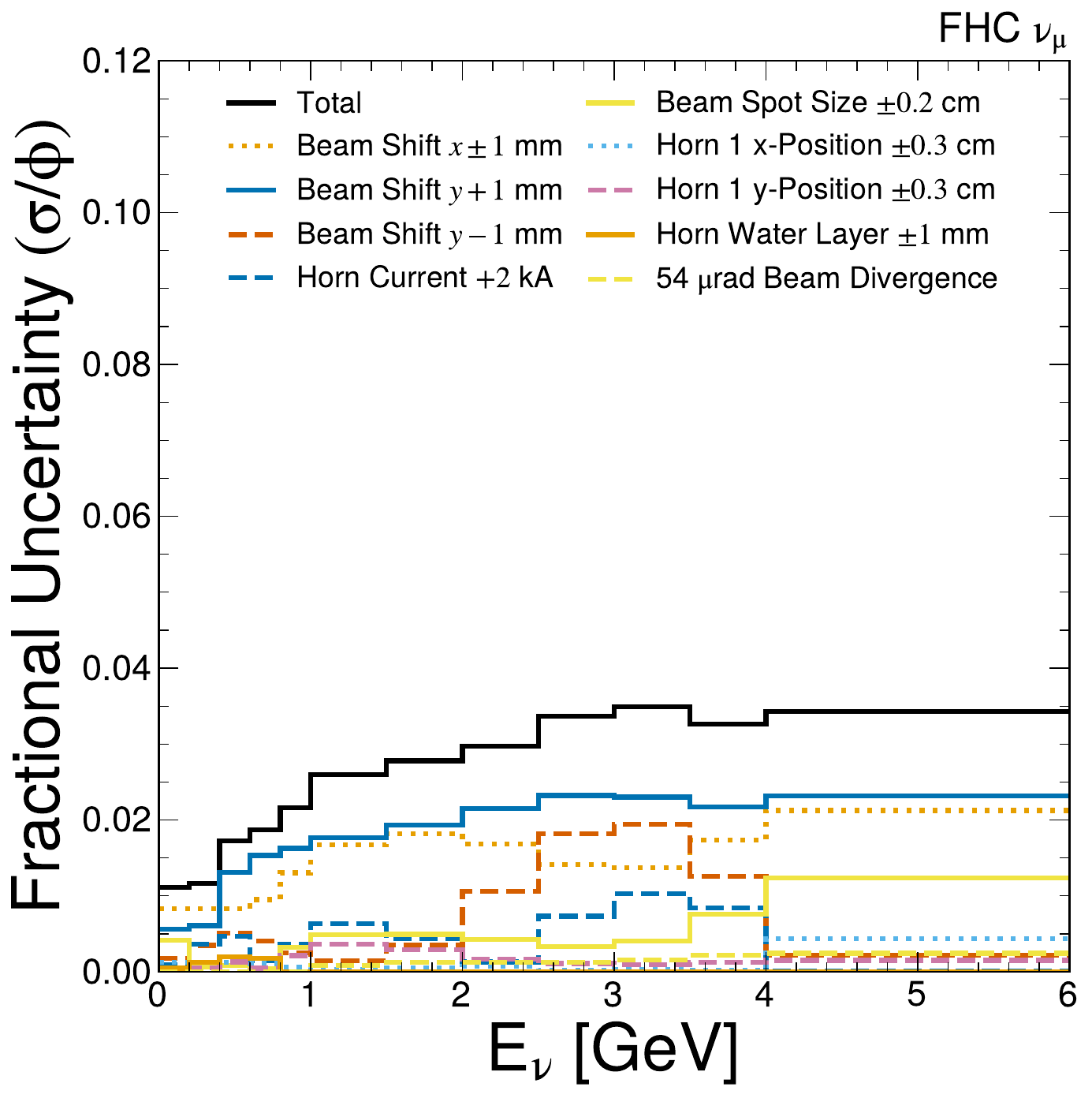}
	\includegraphics[width=0.48\textwidth]{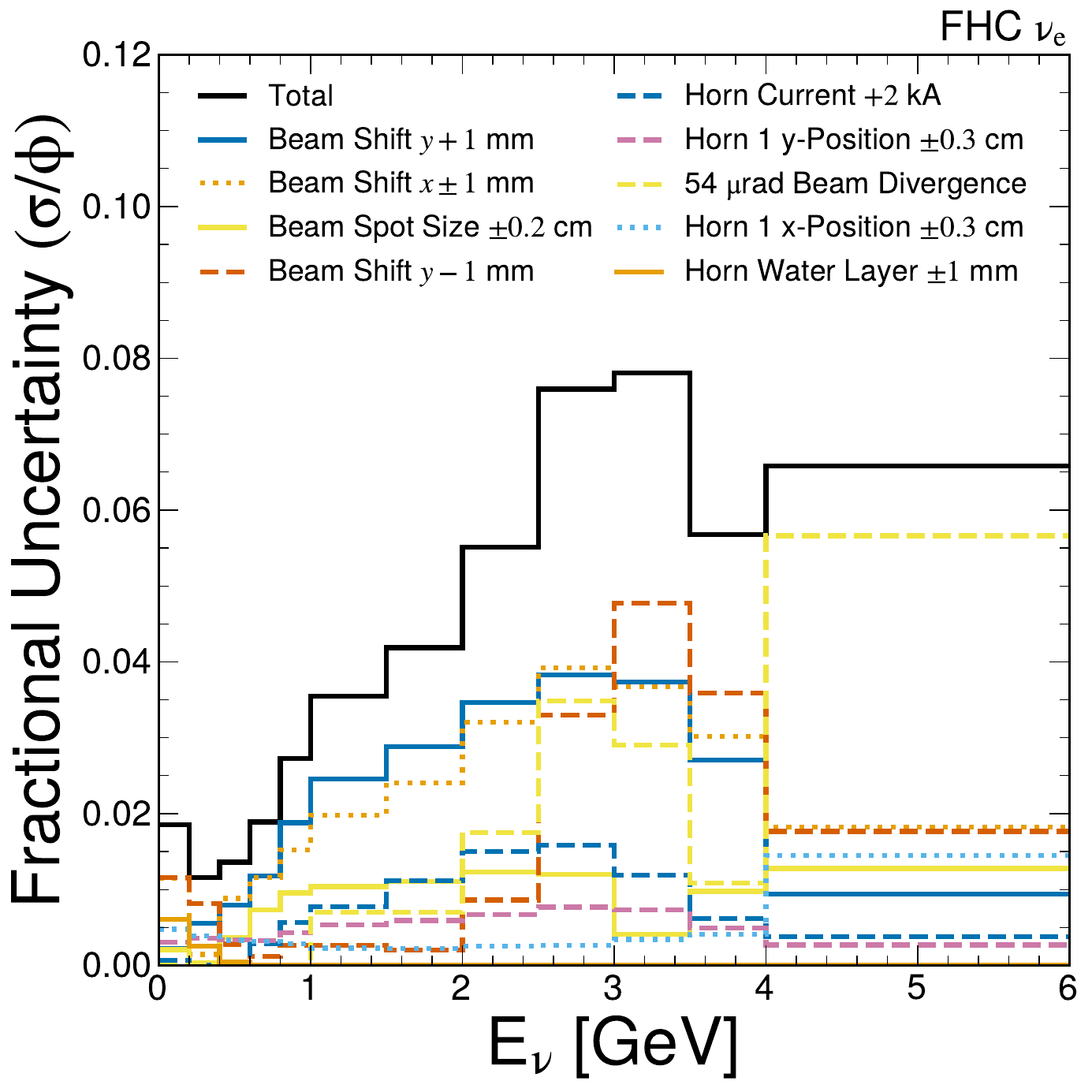}
	\caption[Beamline Focusing Fractional Uncertainties for \numu and \nue (FHC)]{Beamline focusing fractional uncertainties for both FHC electron and muon neutrino modes.
		In both cases, shifting of the proton beam spot on the target in the $x$/$y$ direction constitutes the largest contribution to the total beamline uncertainty.}%
	\label{fig:beam-frac-uncert}
\end{figure}

\begin{figure}
	\centering
	\includegraphics[width=\textwidth]{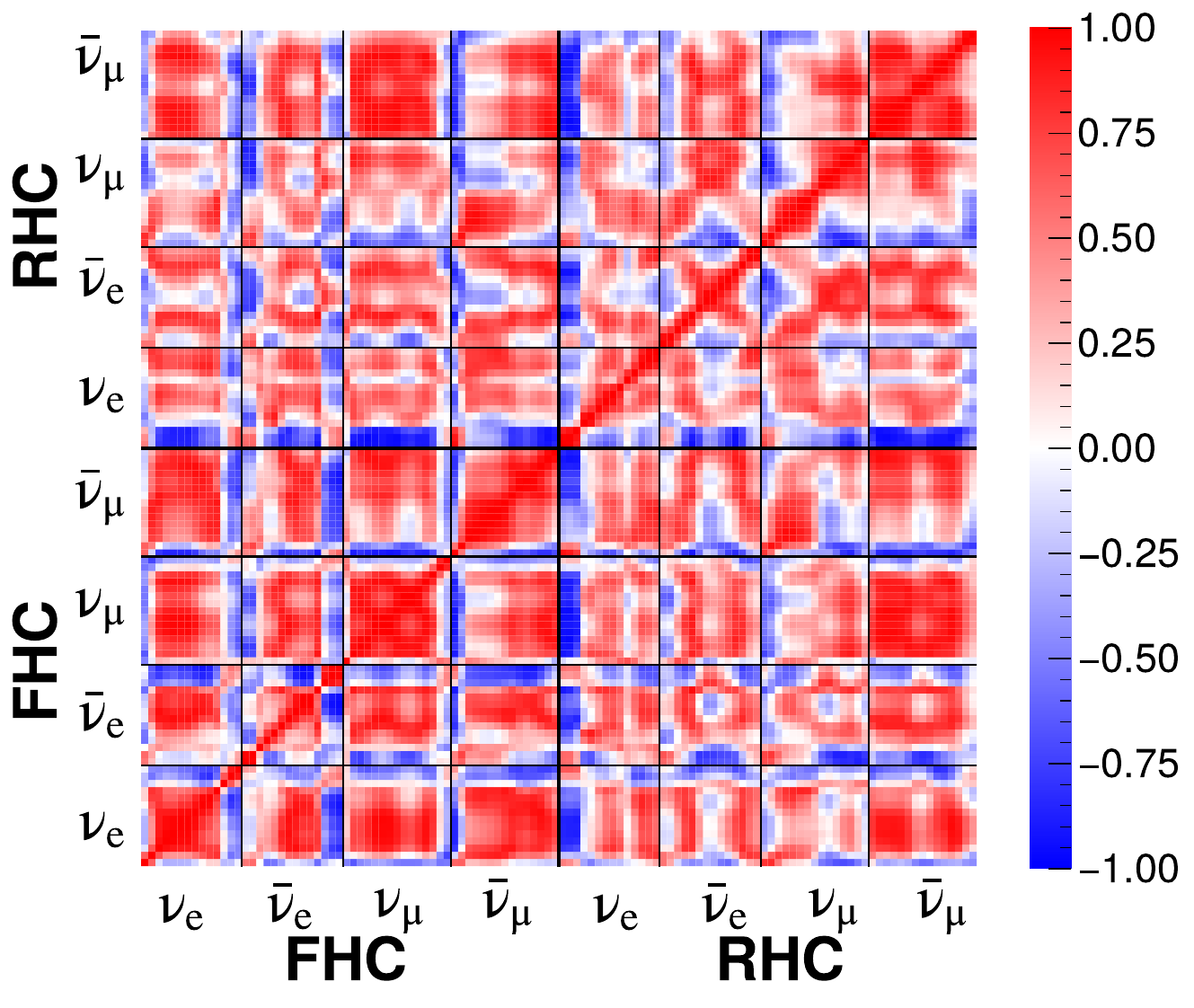}
	\caption[Beamline Focusing Correlation Matrix]{Beamline focusing correlation matrix, incorporating all systematic effects. Strong positive correlations are found between most neighboring bins. Strong positive correlations are most prevalent in the lowest energy ($\sim$0--0.5 \si{\GeV}), medium energy ($\sim$0.5-- 3.5 \si{\GeV}) and high energy ($\sim$5.5--20 \si{\GeV}) regions. Correlations between these regions tend to be small or negative. Significant negative correlations are also found between the low energy RHC \nue\ bins and nearly all other neutrino flavor-energy bins. The RHC \nueb\ flux exhibits complex behaviors in the correlations with other flux components.}%
	\label{fig:beam_total_corr_mat}
\end{figure}

Figure~\ref{fig:beam-frac-uncert} shows the fractional uncertainties due to each ``beam focusing'' systematic as extracted from the diagonals of the corresponding covariance matrices, while Figure~\ref{fig:beam_total_corr_mat} shows the correlation matrix accounting for all systematic effects.
The remaining covariance and correlation matrices have been included in Appendix~\ref{sec:appendix_beam_matrices}.
For muon neutrino modes, the error lies at the 1--2\% level for \Enu\ below \SI{1}{\GeV}, and 3--4\% for $1 < \Enu < \SI{3}{\GeV}$.
For electron neutrinos, the uncertainty starts at the 2--3\%, but quickly rises to 8\% for $\Enu \geq \SI{3}{\GeV}$.
The total beamline systematic uncertainty in the electron neutrino samples is, in general, higher than that of the muon neutrino samples, especially at larger values of \Enu{}.
Estimation of the beamline systematic uncertainties is limited by the relatively low statistics, especially for \nue\ and at higher energies where the flux is low.
Given that the samples are statistically independent, each included systematic carries some random fluctuations at the level of the sample statistical uncertainties, inflating the estimates in the low statistics regions.

\begin{figure}
	\centering
	\includegraphics[width=0.48\textwidth]{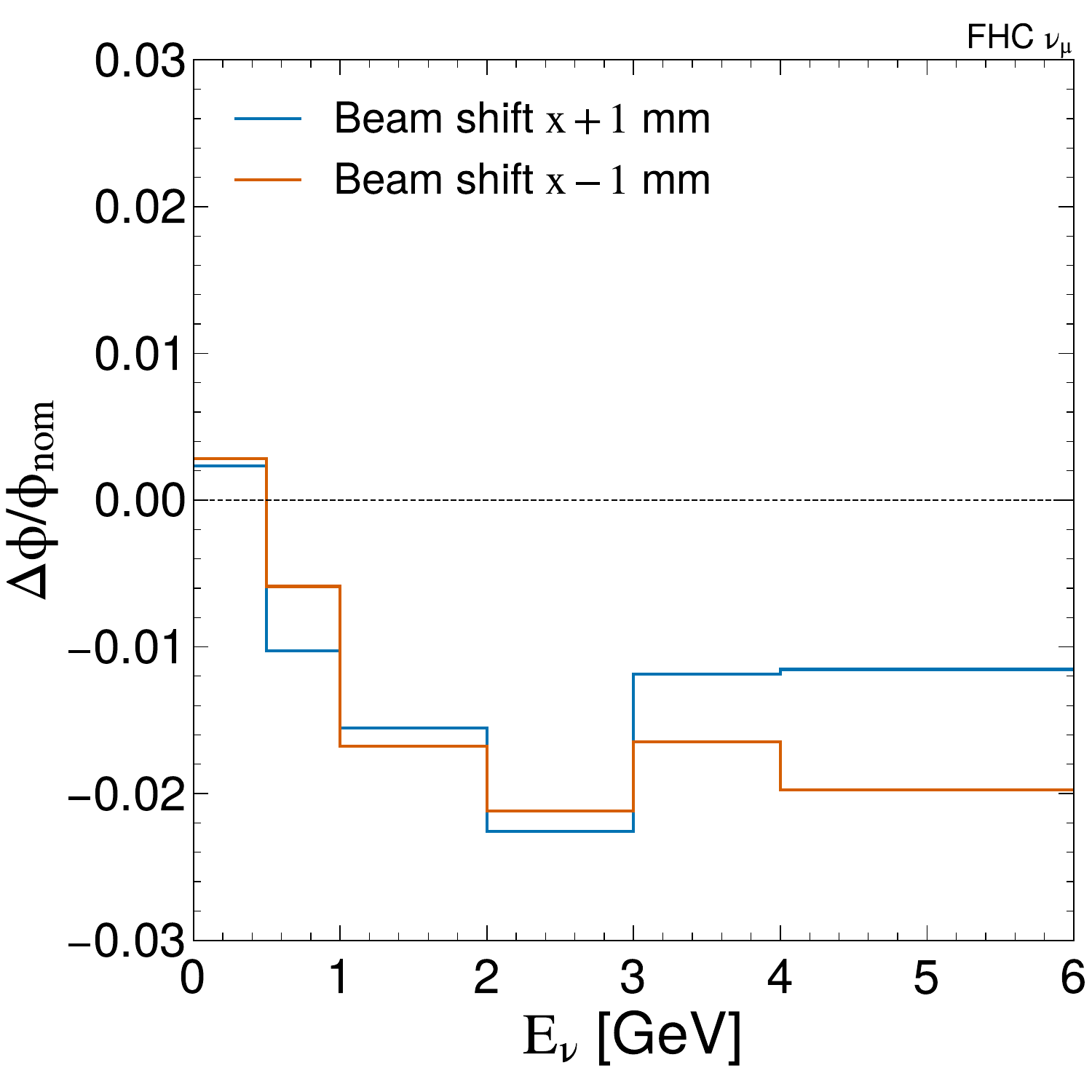}
	\includegraphics[width=0.48\textwidth]{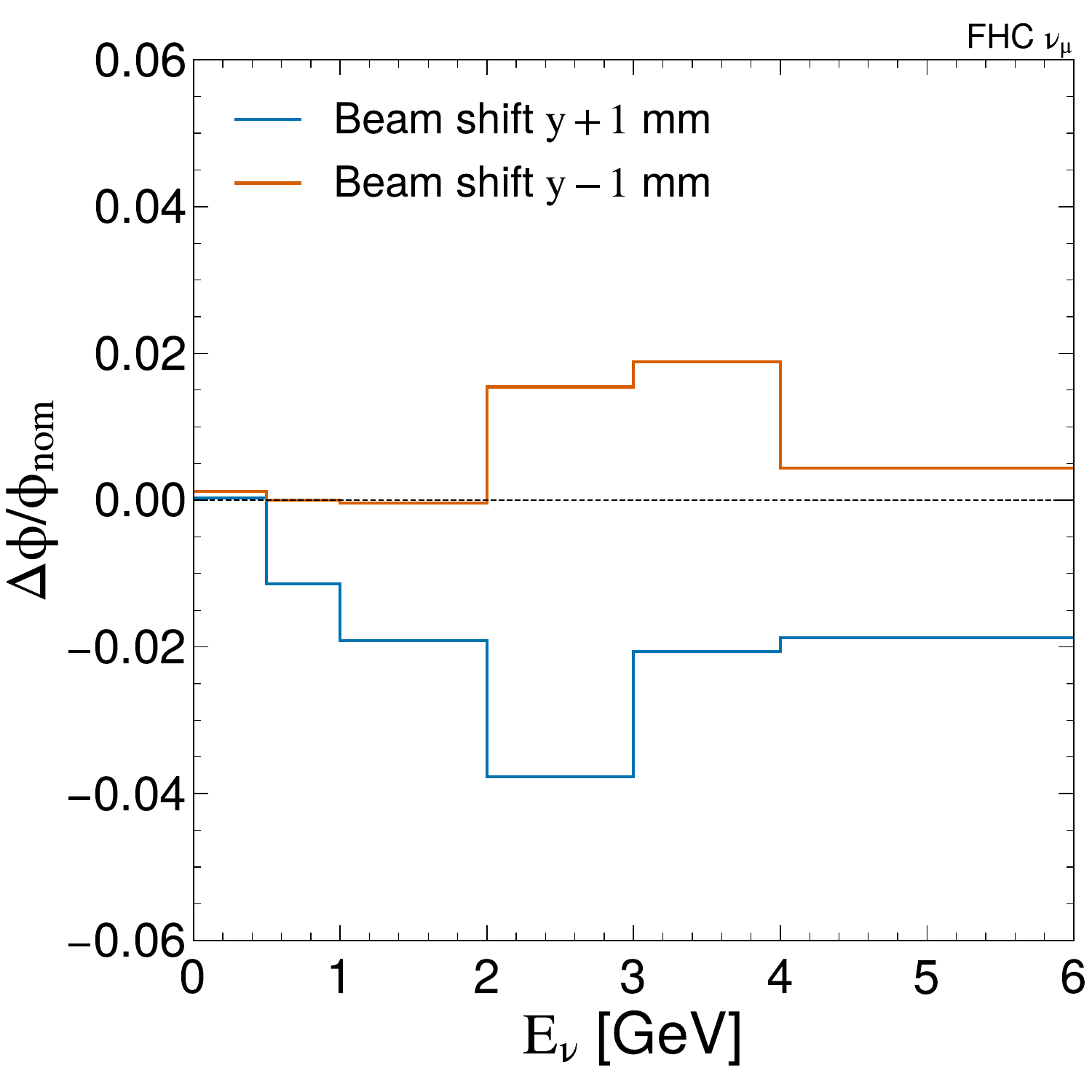}
	\caption[Beam Positioning in $x$/$y$ $\pm \SI{1}{\mm}$ (1B POT)]{Shifts in the flux due to the positioning of the NuMI beam spot on the target in the $x$/$y$ direction for FHC \numu\ simulated with 1B POT. %
	Shifting the spot by $\pm \SI{1}{\mm}$ in $x$ produces $\approx 2\%$ attenuation of the flux in either direction. %
	Shifting the spot upward by \SI{1}{\mm} results in a 2--4\% attenuation within the 1--\SI{3}{\GeV} region likely due to a portion of the beam spot being moved off-target. %
	A downward \SI{1}{\mm} shift produces a negligible affect across the majority of the neutrino energy spectrum, except in the 2--\SI{4}{\GeV} region where the flux is increased by 2\%.}%
	\label{fig:new_beam_shifts}
\end{figure}

Due to these statistical limitations of the dataset used to characterize the beamline systematics, a larger 1B POT set of altered geometry samples was produced.
These samples do not include the \SI{1}{\mega\watt} beam geometry, described in the next section.
The 1B POT samples allow for a more precise estimate of the systematic effects that does not require application of the smoothing procedure described in Section~\ref{sec:beam_uncert_methods}.
Shifts in positioning of the NuMI beam spot on the target in the $x$/$y$ direction from the nominal position are shown in Figure~\ref{fig:new_beam_shifts} for FHC \numu{}.
Fractional uncertainty contributions of the top 5 largest focusing effects contributing to the 1--\SI{3}{\GeV} region for FHC \numu{} and \nue{} are shown in Figure~\ref{fig:new_beam_uncerts}.
The remaining shifts and uncertainties have been included in Appendices~\ref{sec:appendix-beam_frac_shifts}~and~\ref{sec:appendix-beam_frac_uncerts}, respectively.
While the magnitude of focusing effects between the 1B POT and the 500M POT samples have varied at the percent-level across the neutrino energy spectrum,
in either case the largest impacts to the flux are due to the positioning and size of the spot on the \numi\ target.

\begin{figure}
	\centering
	\includegraphics[width=0.48\textwidth]{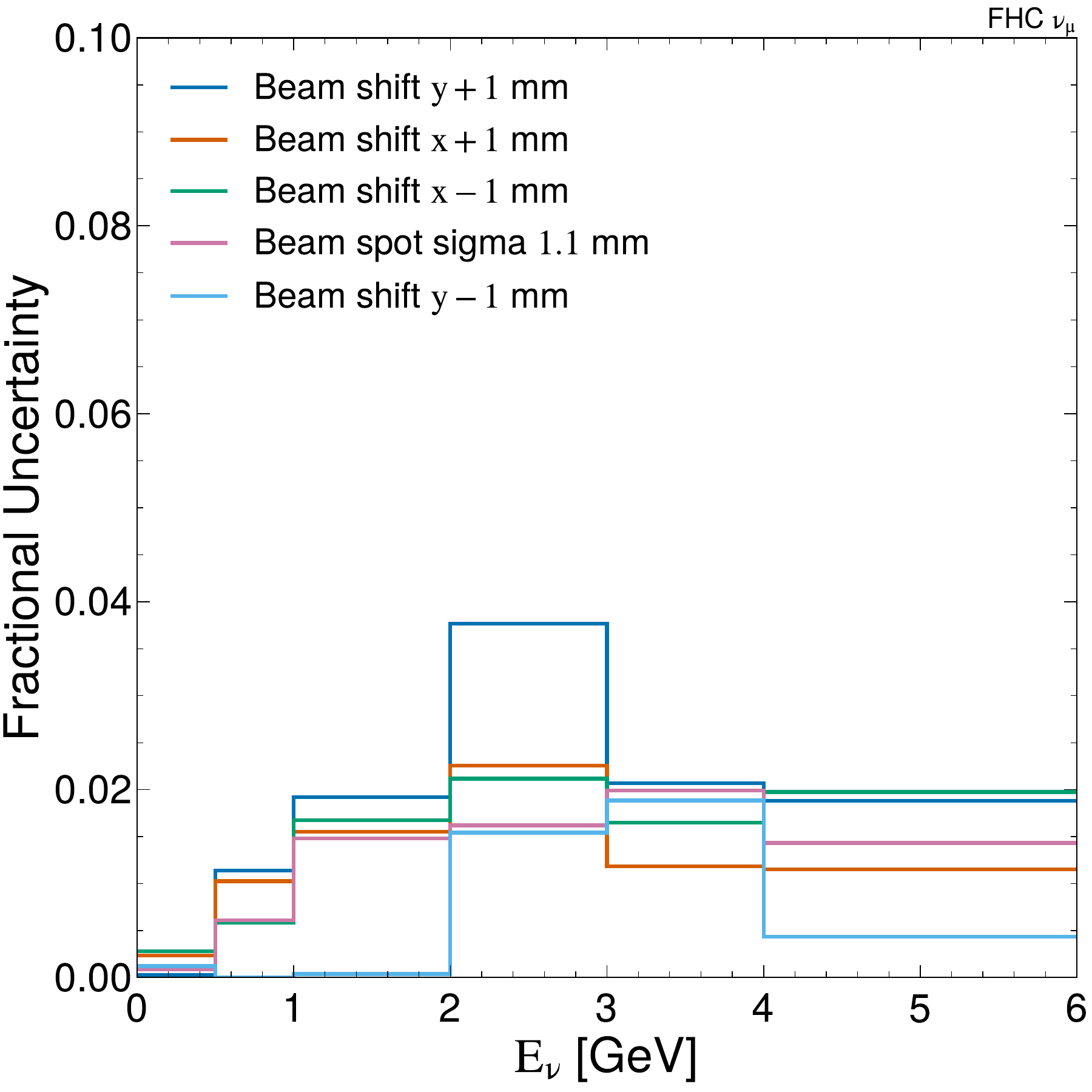}
	\includegraphics[width=0.48\textwidth]{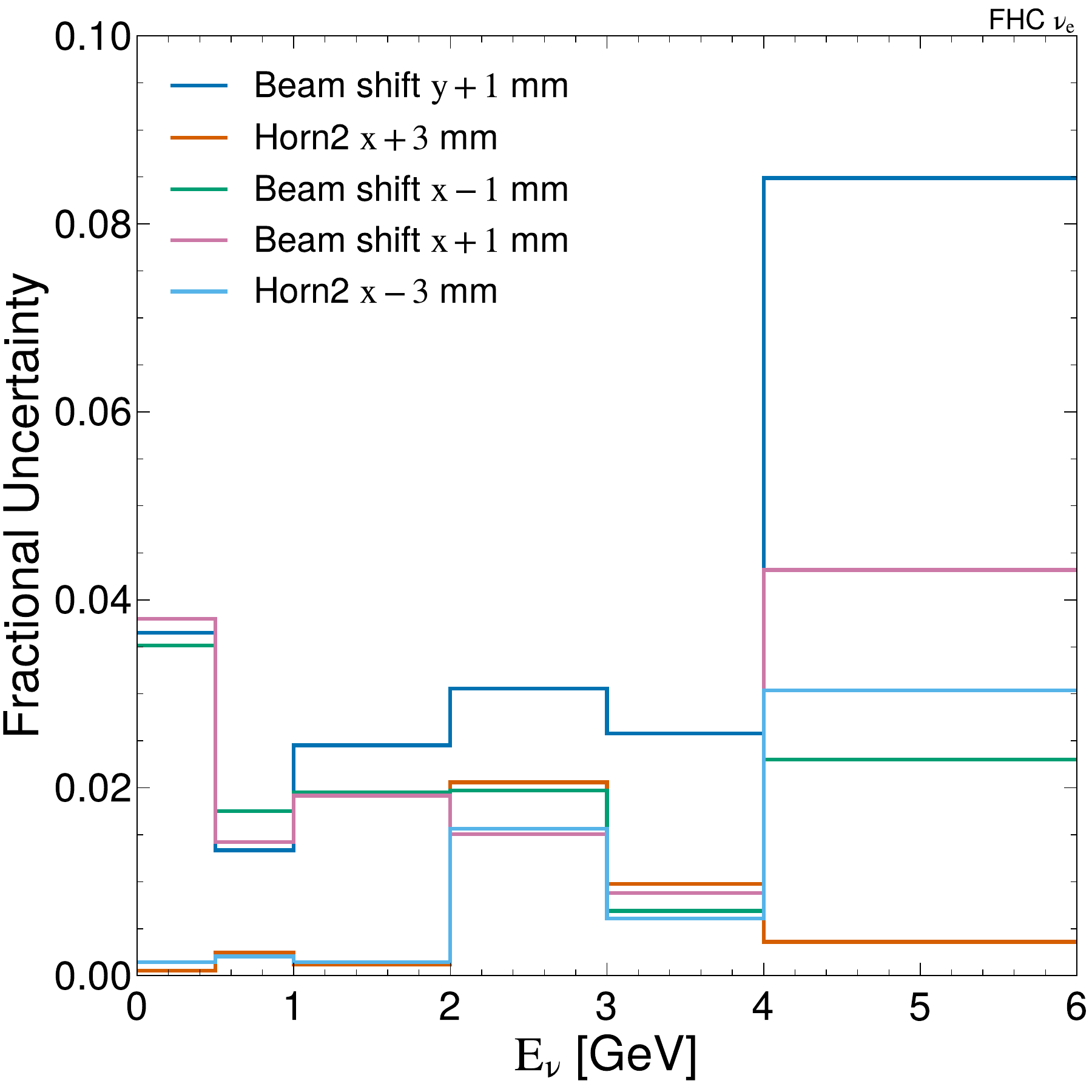}
	\caption[Beam Focusing Fractional Uncertainties (1B POT)]{Fractional uncertainties of the top 5 largest focusing effects contributing to the 1--\SI{3}{\GeV} region for FHC \numu\ (left) and \nue\ (right). %
	As was found in the lower-statistics samples, the most significant effects come from those related to the positioning or size of the beam spot. %
	For \nue, specifically, there is a 2\% contribution from the horizontal positioning of Horn2.}%
	\label{fig:new_beam_uncerts}
\end{figure}

\subsection{Difference Between the \SI{700}{kW} and \SI{1}{MW} Beam Geometries}\label{sec:megawatt}
A study was performed to estimate effects on the flux as a result of the \SI{1}{\mega\watt} upgrade to the \numi\ beam, as described in Section~\ref{sec:megawatt}.
The differences for FHC \numu\ and \nue\ are shown in Figure~\ref{fig:beam_power}, while the remaining neutrino modes have been included in Appendix~\ref{sec:appendix_megawatt_upgrade}.
It was found that the differences are small relative to the statistical uncertainty on the samples used to estimate the differences, especially in the RHC operating mode, where there is a factor 10 fewer statistics available for the \SI{1}{MW} beam geometry. However, there are regions of neutrino energy where the differences at the 2--3\% level are statistically significant. So, while it is clear that the upgraded beamline infrastructure changes the flux, the uncertainties on those changes, as extracted by this analysis, are large compared to the size of the effects.
Therefore, the flux differences between the two beam configurations is treated as an additional source of systematic uncertainty. Similar to the set of focusing uncertainties, the flux difference can be applied using via \Enu-flavor bins with a covariance matrix, or with a single parameter that adjusts the weights for each bin \Enu-flavor bin according to a vector encoding the flux ratios.

The current analysis should suffice for \numi\ analyses based on \icarus\ Run 1 and Run 2 data-taking, while \numi\ was operating in the \fhc\ mode.
However, future \numi{}-\icarus\ \acrshort{mc} productions should be based flux files generated using the \SI{1}{\mega\watt} configuration, including all relevant focusing systematics geometry alterations, which are currently being generated.

\begin{figure}
	\centering
	\includegraphics[width=0.49\textwidth]{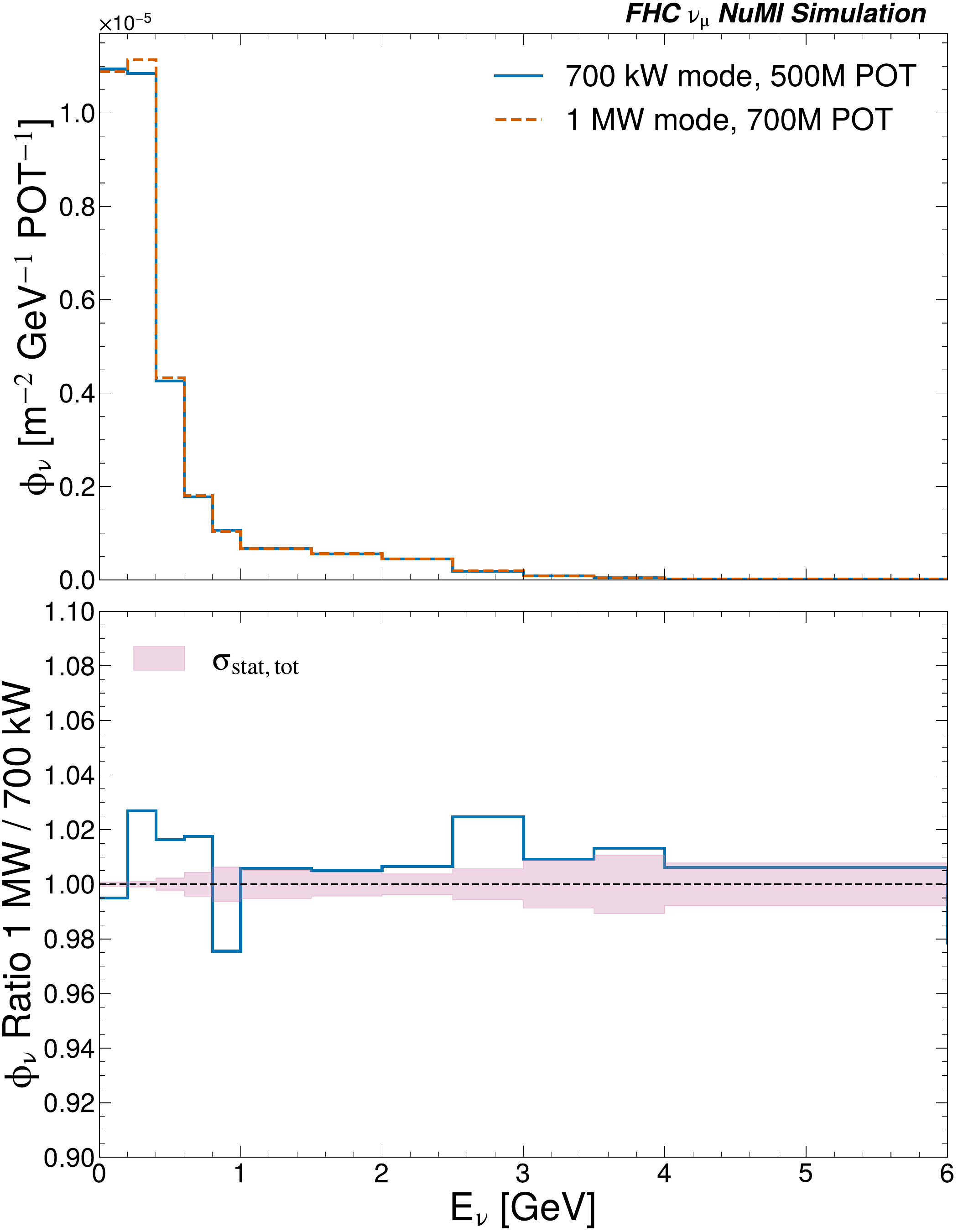}
	\includegraphics[width=0.49\textwidth]{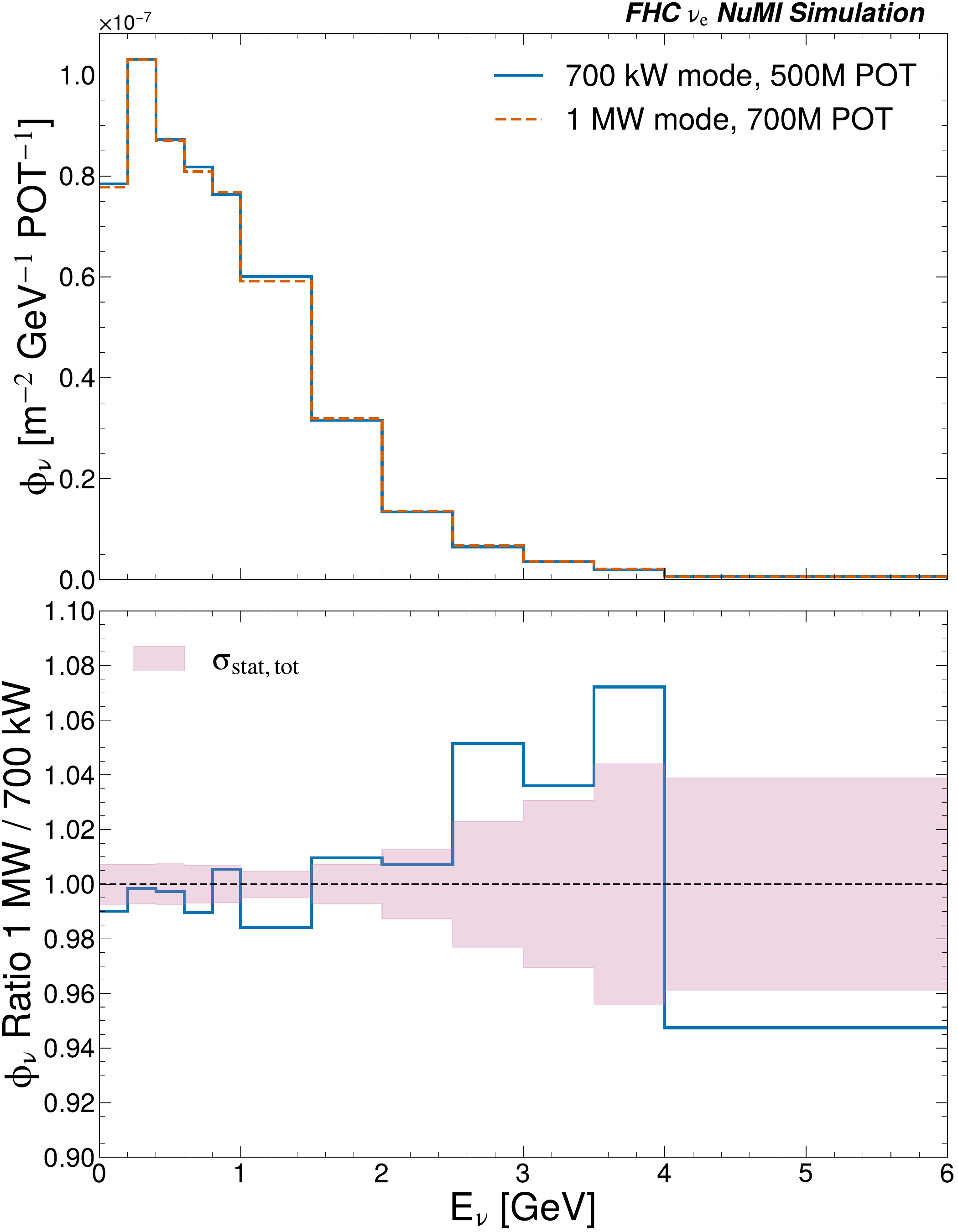}
	\caption[Comparison of the Neutrino Flux between the \SI{700}{\kilo\watt} and \SI{1}{\mega\watt} Geometries]{Comparison of neutrino flux with \SI{700}{kW} beamline geometry NuMI simulation (used in this note), and newer \SI{1}{MW} beamline geometry simulation for \numu\ (\emph{left}) and \nue\ (\emph{right}).}%
	\label{fig:beam_power}
\end{figure}

\clearpage
\section{Flux Prediction and Integrated Flux Uncertainty}

The complete flux prediction for \numu\ and \nue\ in the forward horn operating mode is given in Figure~\ref{fig:flux-pred}.
The remaining spectra can be found in Appendix~\ref{sec:appendix-flux-prediction}, and a table of the predicted neutrino flux at the geometrical center, top, and bottom of the ICARUS active volume can be found in Appendix~\ref{sec:appendix-total-uncertainties}.
The total uncertainty on the NuMI neutrino flux was calculated for various neutrino modes, as well as the flux ratio.
Table~\ref{tab:integ-uncerts} contains a selection of the total uncertainties integrated over the full $0 < \Enu < \SI{20}{\GeV}$ range, while a more complete tabulation can be found in Appendix~\ref{sec:appendix-total-uncertainties}.
These results demonstrate that hadron production systematic uncertainties are greater than or equal to the NuMI focusing uncertainties across the regions of interest.
Large flux correlations between flavors (e.g.,\ \nue{}/\numu\ or $\bar{\nu}/\nu$) would favor analyses that measured the ratios between interaction processes of the correlated flavors. The correlations would lead to cancellations in the total flux error used in the cross section extraction. Given the low levels of flavor to flavor correlations
the uncertainty on the flux ratio is not found to be particularly advantageous, (7.1 and 7.0\% for FHC and RHC, respectively), compared to the other flux modes.
In fact, anti-correlations found in the covariance matrix will tend to drive the ratio uncertainty upward.
Integration of new hadronic interaction data into PPFX from NA61\slash{}SHINE and EMPHATIC has the potential to strengthen correlations and reduce uncertainties on ratio measurements.

\begin{table}[htbp]
	\centering
	\caption[Uncertainties on the Integrated Flux (0--20 \si{\GeV})]{Uncertainties on the integrated flux in the 0--20 \si{\GeV} range. Hadron production uncertainties dominate, with all other effects contributing to sub-percent increases to the quadrature sum. FHC and RHC fractional uncertainties are similar, as are right-sign and wrong-sign errors within the same beam mode. The flux uncertainties on the sum of right and wrong-sign components are less than the uncertainties on either contribution due to mild (strong) off-diagonal flavor-to-flavor regions of the hadron production (beamline focusing) correlation matrices. The uncertainty on the $\nue{}+\nueb{}$ to $\numu{}+\numub{}$ flux ratio offers little advantage over the standalone $\nue{}+\nueb{}$ flux due to the lack of strong flavor-to-flavor correlations between the $\nue{}+\nueb{}$ and $\numu{}+\numub{}$ components.}
	\begin{tabular}{l r r r r r r r}
		\toprule\toprule
		          & \multicolumn{7}{c}{Uncertainty (\%)}                                                                                                                                                             \\
		Horn Mode & $\nu_e$                              & $\bar{\nu}_e$ & $\nu_e + \bar{\nu}_e$ & $\nu_\mu$ & $\bar{\nu}_\mu$ & $\nu_\mu + \bar{\nu}_\mu$ & $\frac{\nu_e + \bar{\nu}_e}{\nu_{\mu} + \bar{\nu}_\mu}$ \\
		\midrule                                                                                                                                                                                                     \\
		\multicolumn{8}{l}{Hadron}                                                                                                                                                                                   \\
		\cline{1-1}                                                                                                                                                                                                  \\
		FHC       & 6.63                                 & 5.84          & 5.76                  & 11.32     & 10.19           & 9.08                      & 6.83                                                    \\
		RHC       & 5.86                                 & 6.76          & 5.77                  & 10.74     & 11.27           & 9.45                      & 6.92                                                    \\
		\multicolumn{8}{l}{Beamline}                                                                                                                                                                                 \\
		\cline{1-1}                                                                                                                                                                                                  \\
		FHC       & 1.23                                 & 1.49          & 0.82                  & 1.12      & 2.37            & 1.42                      & 1.47                                                    \\
		RHC       & 2.60                                 & 1.48          & 2.15                  & 1.35      & 1.40            & 0.88                      & 1.40                                                    \\
		\multicolumn{8}{l}{Beam Power Upgrade}                                                                                                                                                                       \\
		\cline{1-1}                                                                                                                                                                                                  \\
		FHC       & 2.42                                 & 1.13          & 0.52                  & 1.37      & 1.45            & 0.35                      & 0.87                                                    \\
		RHC       & 0.67                                 & 2.12          & 1.19                  & 1.56      & 2.28            & 0.18                      & 1.37                                                    \\
		\multicolumn{8}{l}{Statistical}                                                                                                                                                                              \\
		\cline{1-1}                                                                                                                                                                                                  \\
		FHC       & 0.26                                 & 0.24          & 0.18                  & 0.09      & 0.06            & 0.05                      & 0.18                                                    \\
		RHC       & 0.21                                 & 0.29          & 0.17                  & 0.06      & 0.09            & 0.05                      & 0.18                                                    \\
		\multicolumn{8}{l}{Total}                                                                                                                                                                                    \\
		\cline{1-1}                                                                                                                                                                                                  \\
		FHC       & 7.17                                 & 6.13          & 5.85                  & 11.45     & 10.56           & 9.20                      & 7.04                                                    \\
		RHC       & 6.45                                 & 7.24          & 6.27                  & 10.94     & 11.58           & 9.50                      & 7.20                                                    \\
		\bottomrule\bottomrule
	\end{tabular}%
	\label{tab:integ-uncerts}
\end{table}

\begin{figure}[htbp]
	\centering
	\begin{subfigure}{0.49\textwidth}
		\includegraphics[width=\textwidth]{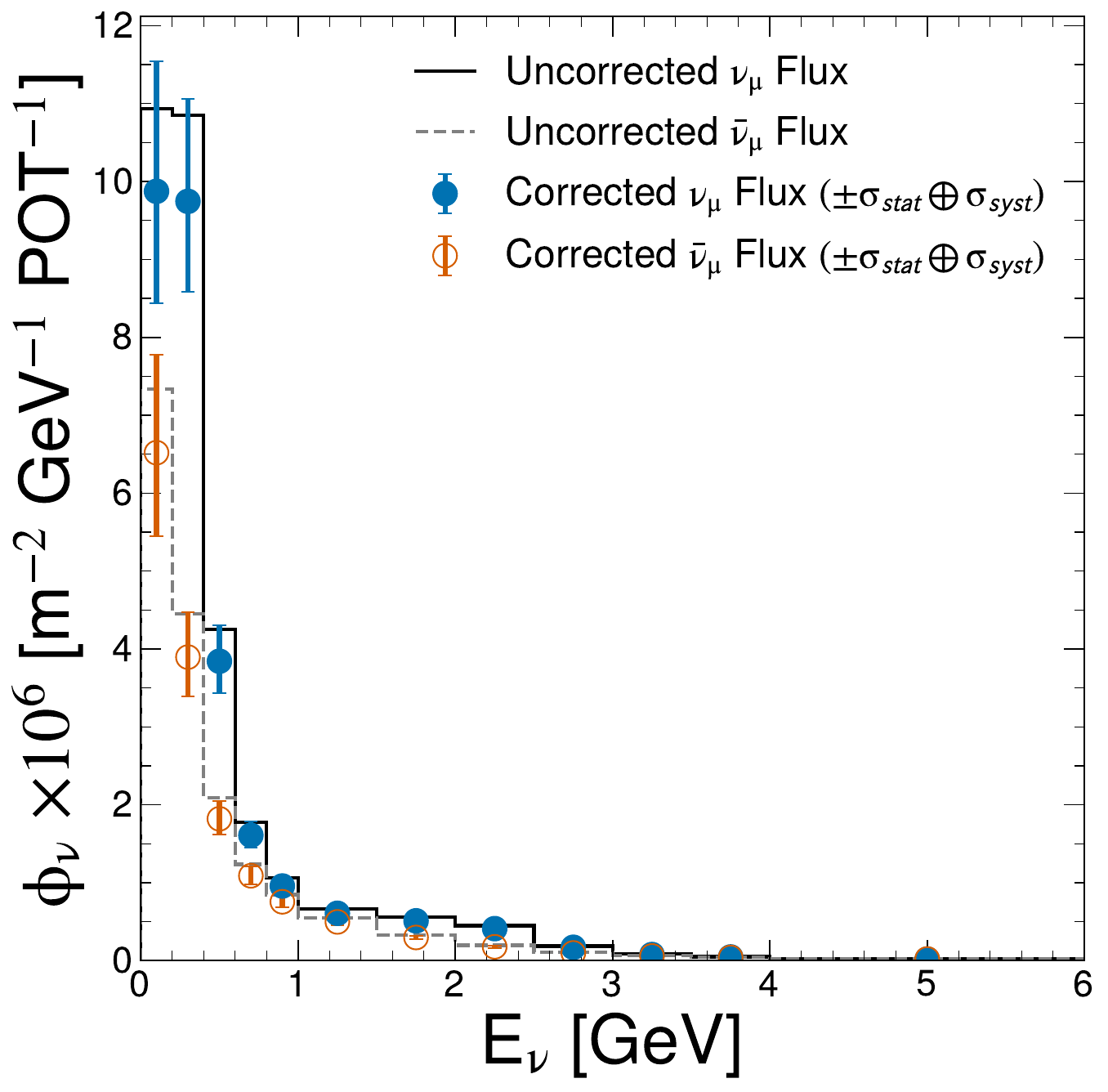}
		\caption{\numu\ and \numub}
	\end{subfigure}
	\begin{subfigure}{0.49\textwidth}
		\includegraphics[width=\textwidth]{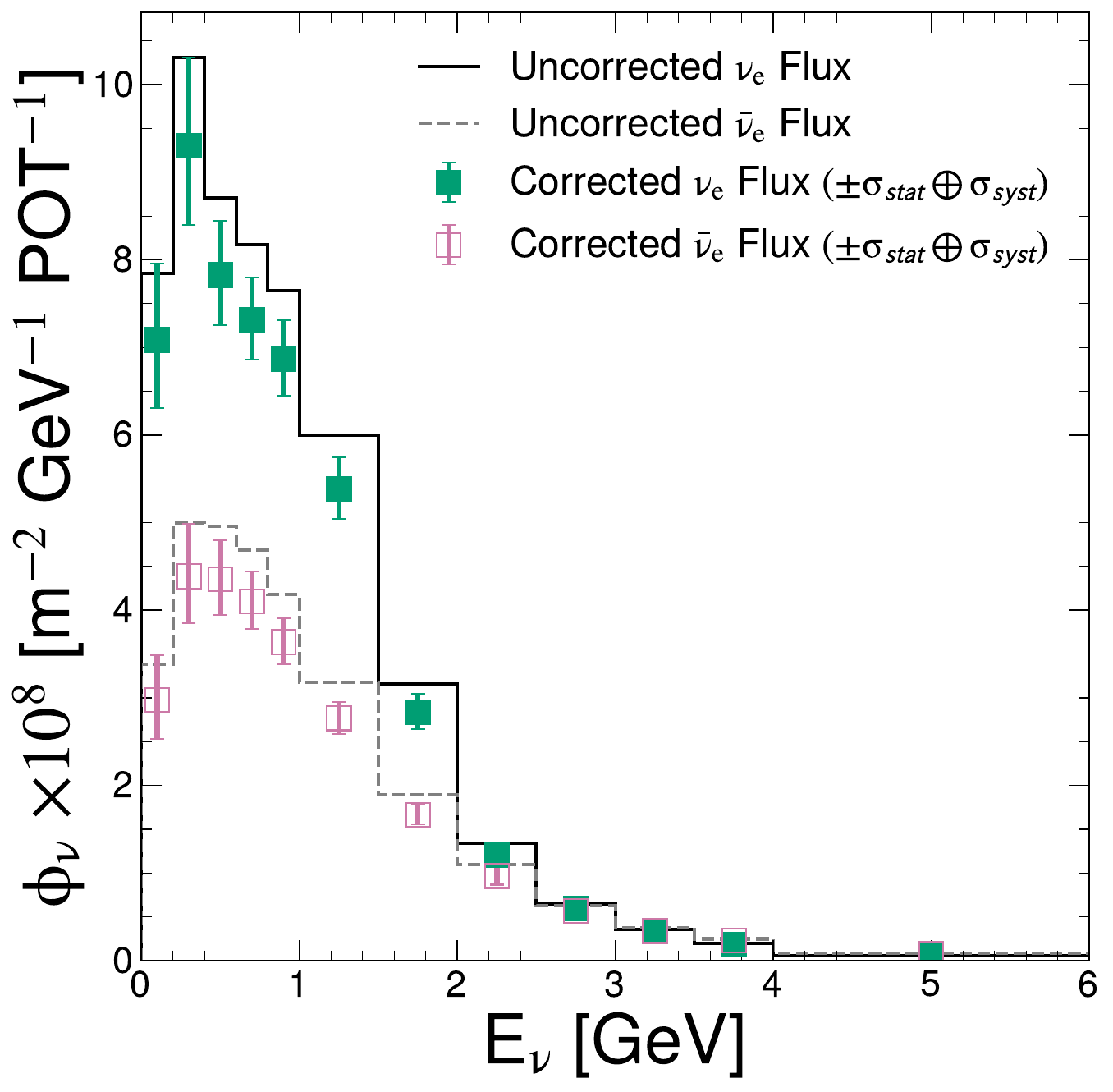}
		\caption{\nue\ and \nueb}
	\end{subfigure}
	\caption[The NuMI-ICARUS Flux Prediction with Uncertainties]{The NuMI flux spectra in the forward horn operating mode expected at ICARUS with full uncertainties accounting for hadron production, beam focusing, and statistical effects. In panel (a), the unweighted \numu\ (\numub{}) flux is shown in black (gray), while the PPFX corrected flux with total uncertainties is shown in blue (orange). In panel (b), the unweighted \nue\ (\nueb{}) flux is shown in black (gray), while the PPFX corrected flux with total uncertainties is shown in green (pink).}%
	\label{fig:flux-pred}
\end{figure}

\clearpage
\section{Investigations into Missing Geometry Components}\label{sec:missing_geom}
\input{geometry_changes.tex}

\clearpage
\section{Impacts of the Updated Hadronic Model}\label{sec:newg4}
\input{geant4_updates.tex}
\clearpage
\section{Fixes to Nucleon Interactions in the PPFX ReweightDriver}\label{sec:na_bugfix}
\input{NA_bugfix.tex}

\clearpage
\section{Updated Flux Prediction}
\input{final_flux_prediction.tex}

%% file: geometry_changes.tex
As part of an effort to identify discrepancies between their \geant and FLUKA--an alternate toolkit for modeling interactions--flux simulations, the MicroBooNE collaboration identified several components missing from the \numi\ target hall geometry description that is input to the simulation.
A boolean flag (\texttt{g3Chase}) in the simulation controlling the presence of shielding blocks installed across the top of the target chase was set to `\texttt{false}',
With the blocks included, there is more material with which hadrons may interact before decaying to neutrinos, thus simulated parent hadrons were afforded more space in which to decay relative to the \numi design specifications, resulting in an overestimation of the NuMI neutrino flux.

\begin{figure}[!ht]
	\centering
	\includegraphics[width=0.49\textwidth]{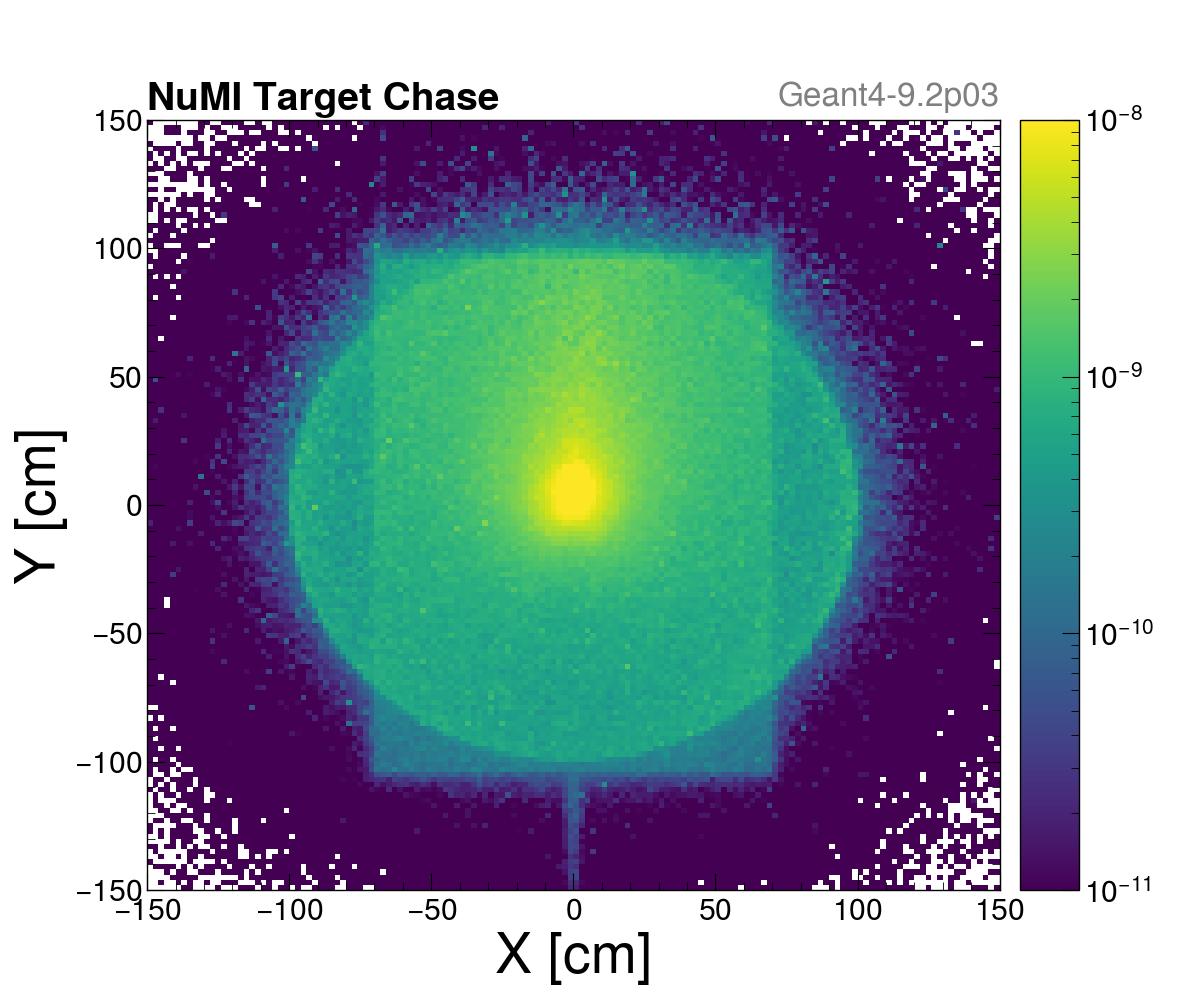}
	\includegraphics[width=0.49\textwidth]{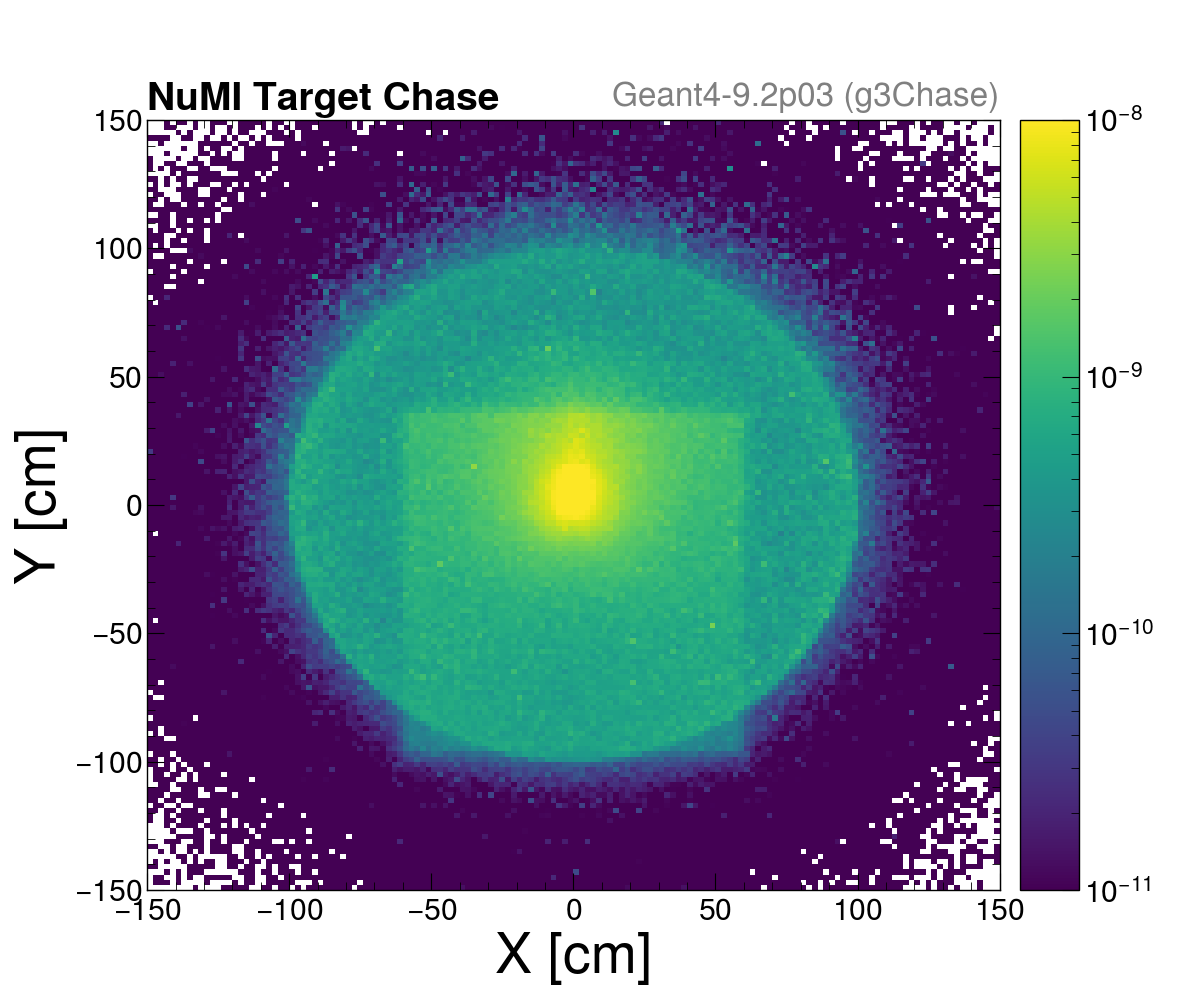}
	\includegraphics[width=0.49\textwidth]{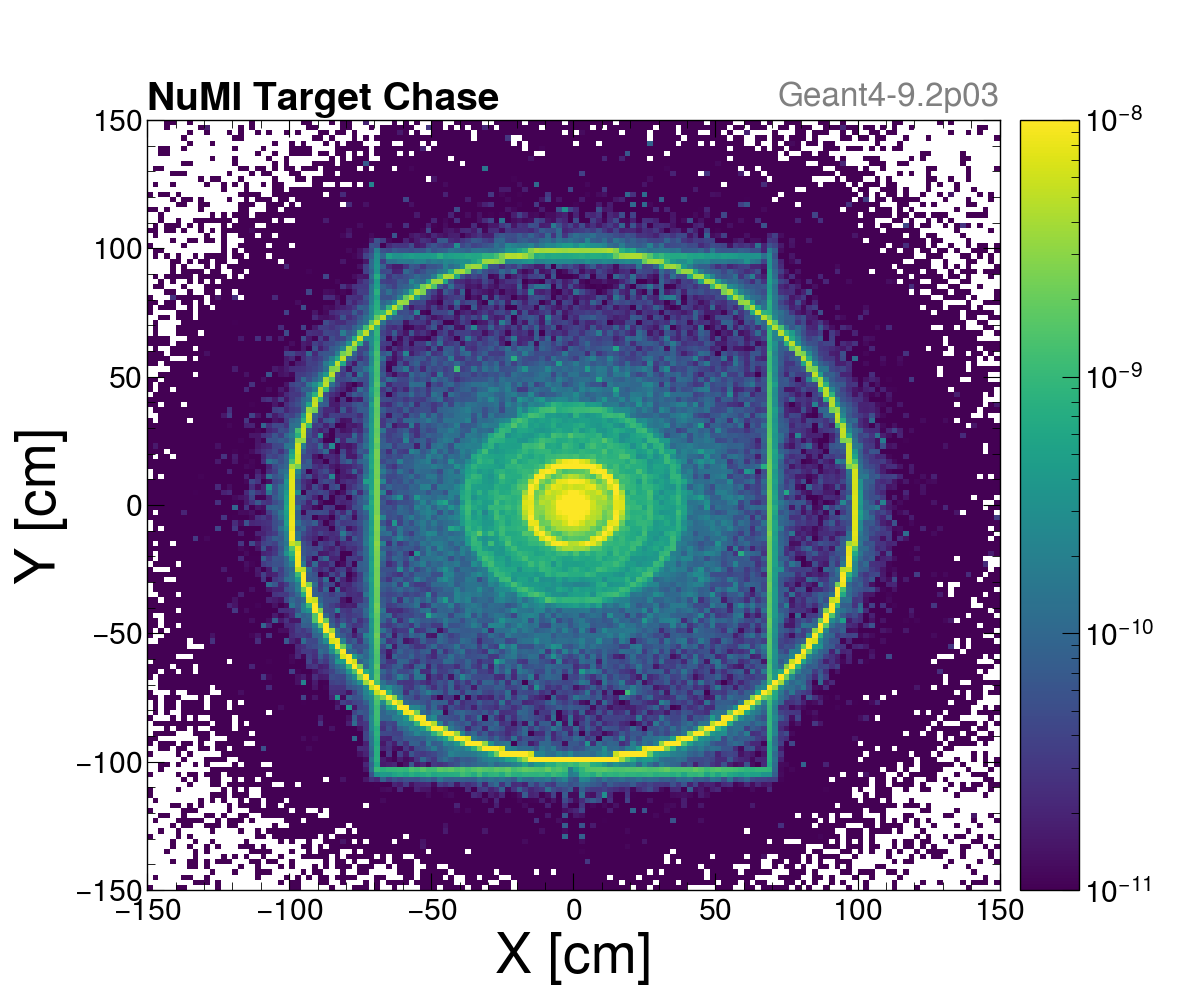}
	\includegraphics[width=0.49\textwidth]{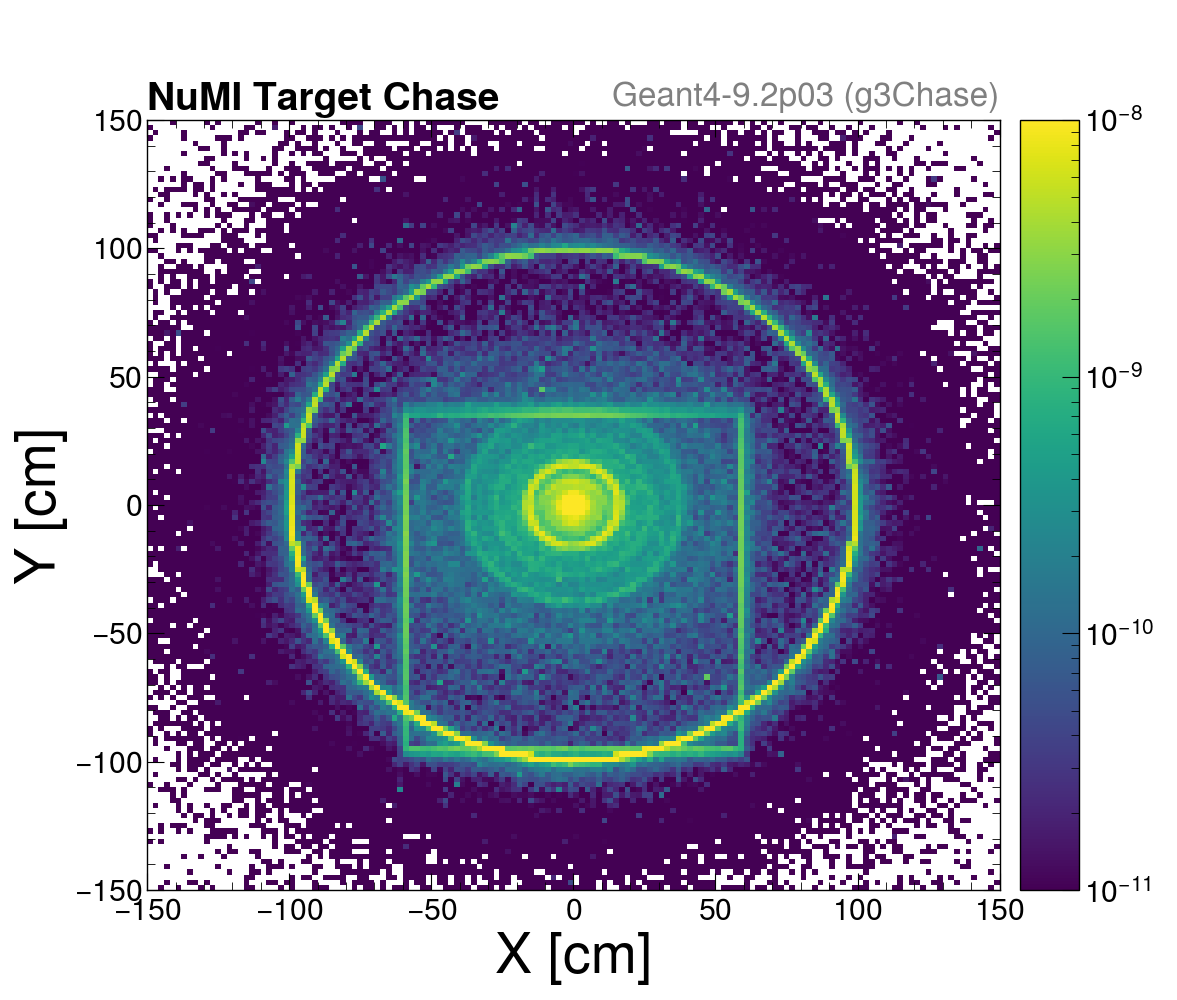}
	\caption[$\nu$ and Primary Hadron Production Vertices ($x-y$ Plane)]{ICARUS neutrino (top) and parent (bottom) production vertices in the NuMI target hall, as simulated by G4NuMI\@. The left panel shows the $x$-$y$ plane with the \texttt{g3Chase} flag set to `\texttt{false}', while the right panel shows the same plane with the flag set to `\texttt{true}'.
		Returning the shielding blocks to the geometry results in a \ensuremath{\sim}\SI{60}{\centi\meter} reduction in the vertical extent of the target chase.}%
	\label{fig:numi_target_hall_ppvxy}
\end{figure}

\begin{figure}[!ht]
	\centering
	\includegraphics[width=0.49\textwidth]{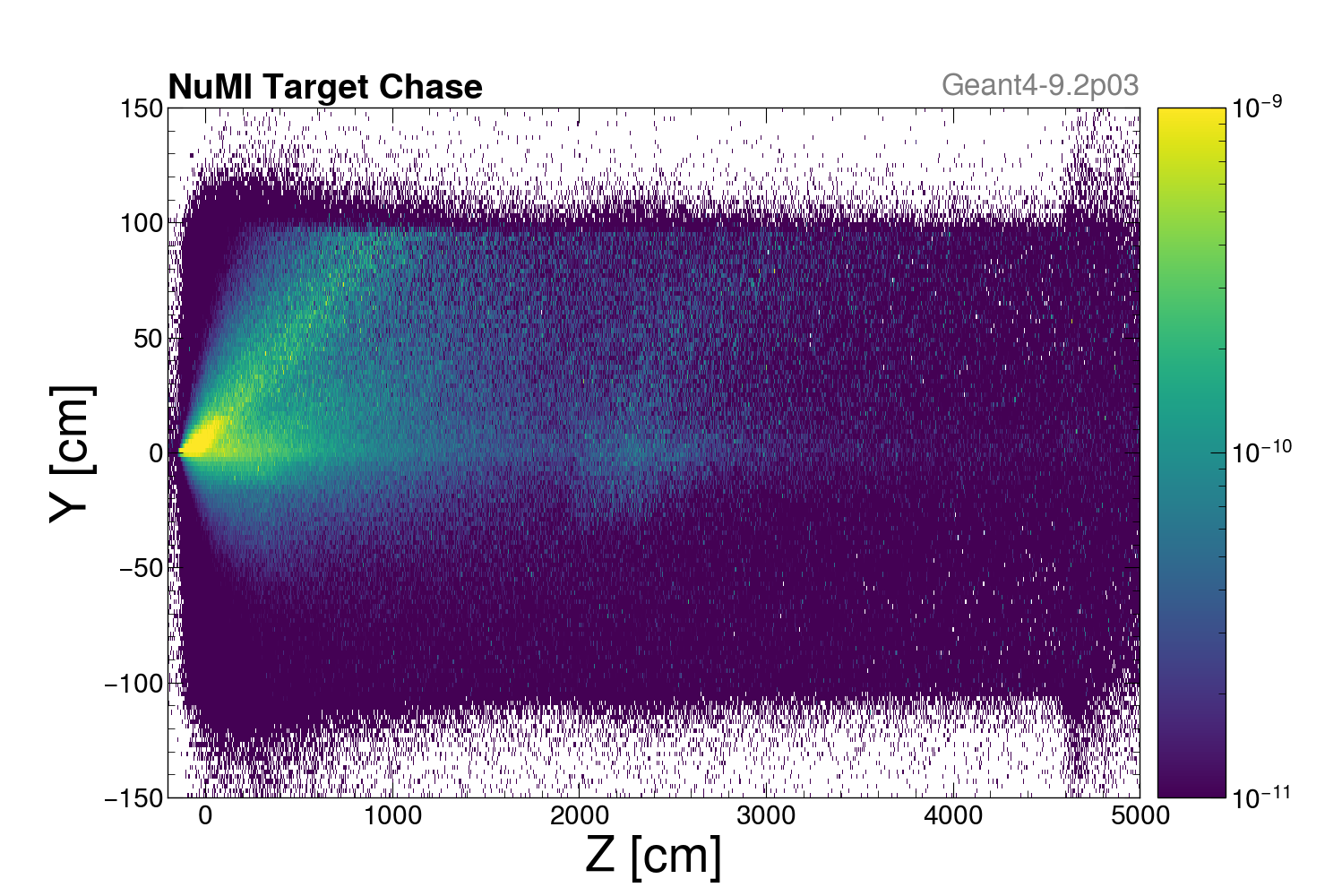}
	\includegraphics[width=0.49\textwidth]{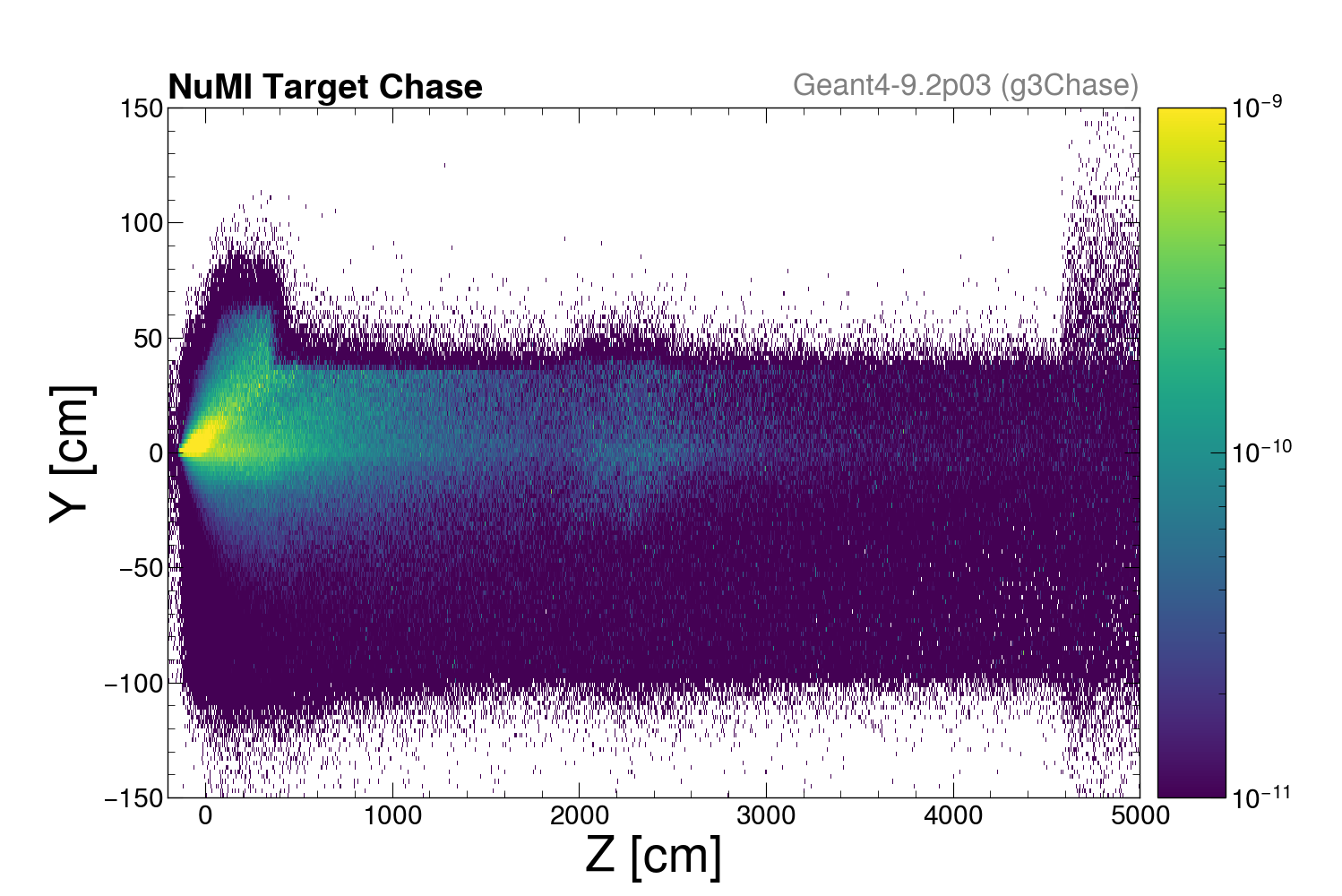}
	\includegraphics[width=0.49\textwidth]{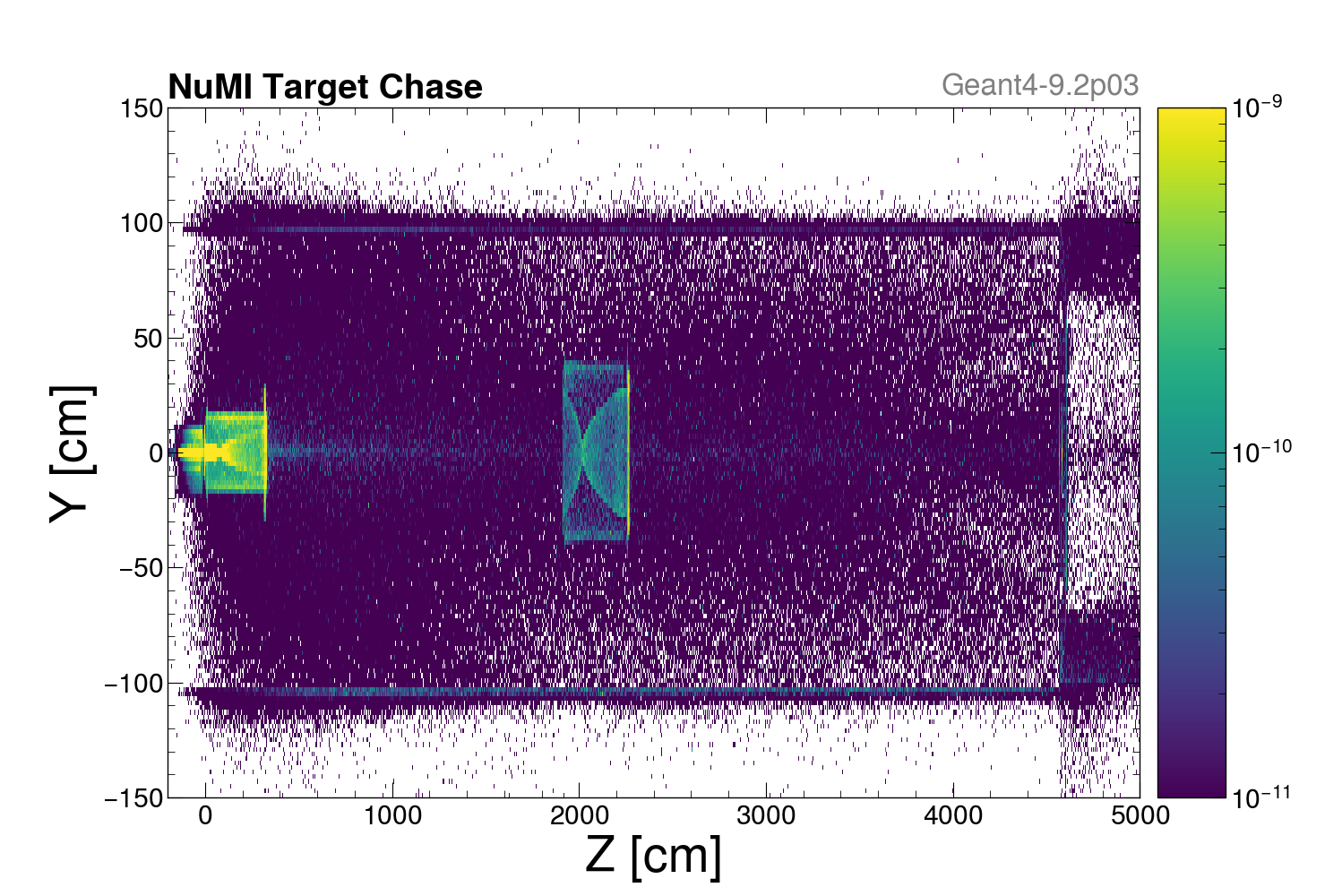}
	\includegraphics[width=0.49\textwidth]{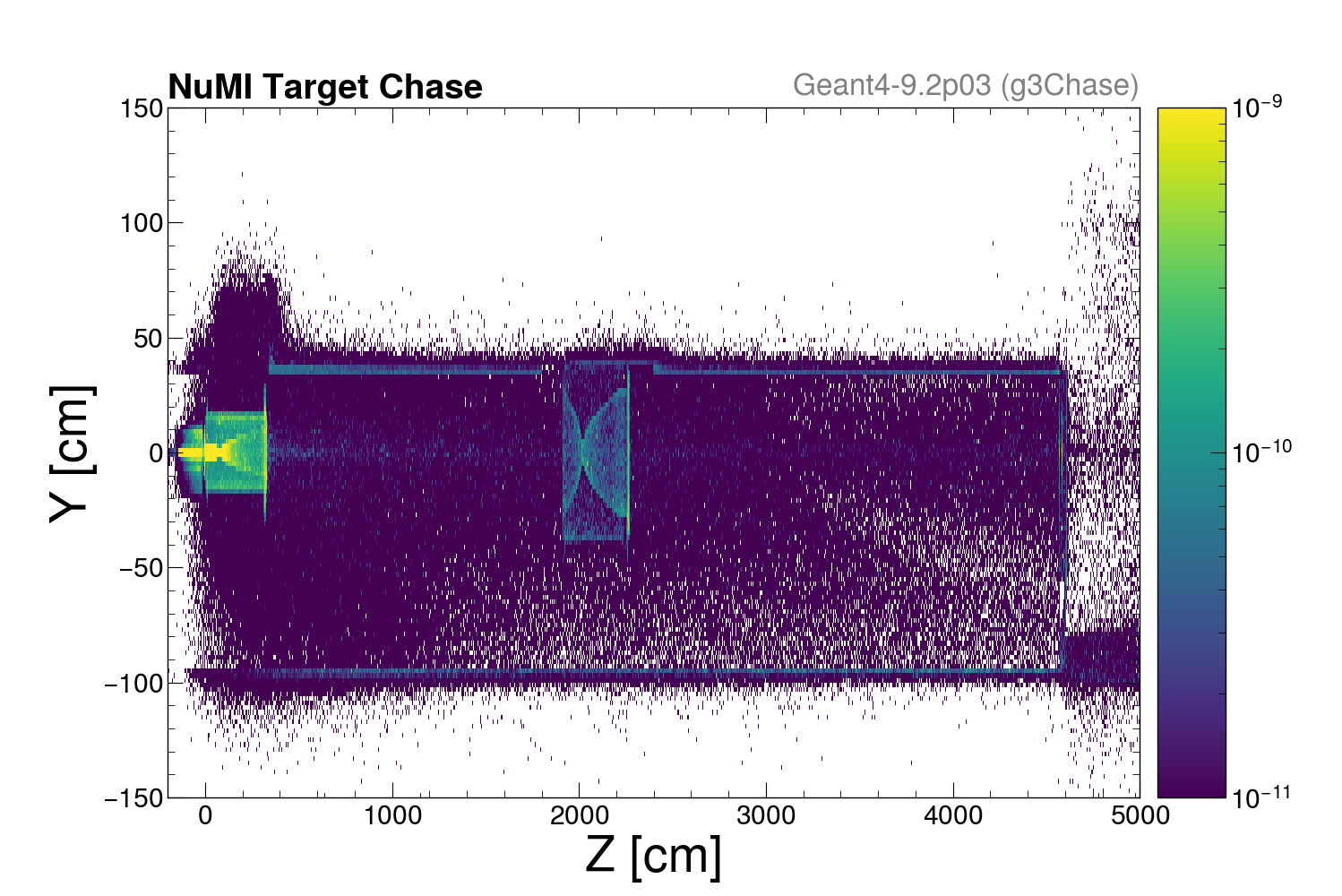}
	\caption[$\nu$ and Primary Hadron Production Vertices ($y-z$ Plane)]{ICARUS neutrino (top) and parent (bottom) production vertices in the NuMI target hall, as simulated by G4NuMI\@.
		The left panel shows the $z$-$y$ plane with the \texttt{g3Chase} flag set to `\texttt{false}', while the right panel shows the same plane with the flag set to `\texttt{true}'.
		Returning the shielding blocks to the geometry results in a \ensuremath{\sim}\SI{60}{\centi\meter} reduction in the vertical extent of the target chase. %
		Fewer neutrino decays can be seen in the region now obfuscated by the shielding (top right), and an increase in the number of hadron interactions at the base of the ceiling (bottom right).}%
	\label{fig:numi_target_hall_ppvzy}
\end{figure}

To study the impact of the missing components, two independent samples, 250M POT each, were produced with the flag toggled on and off.
Figures~\ref{fig:numi_target_hall_ppvxy} and~\ref{fig:numi_target_hall_ppvzy} show ICARUS neutrino and parent production vertices in the NuMI target hall before and after correcting the beam geometry.

\begin{figure}[ht]
	\centering

	\includegraphics[width=0.49\textwidth]{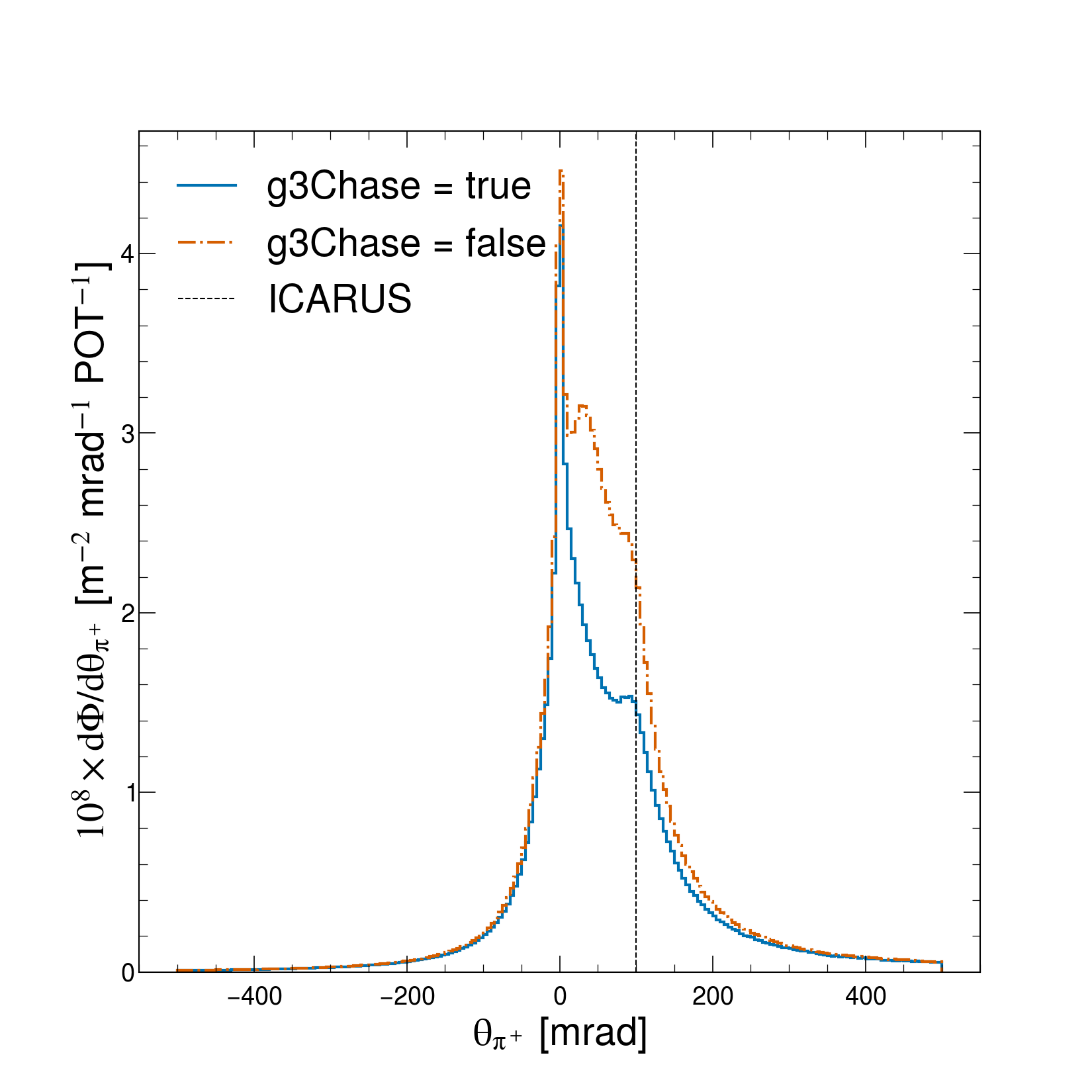}
	\includegraphics[width=0.49\textwidth]{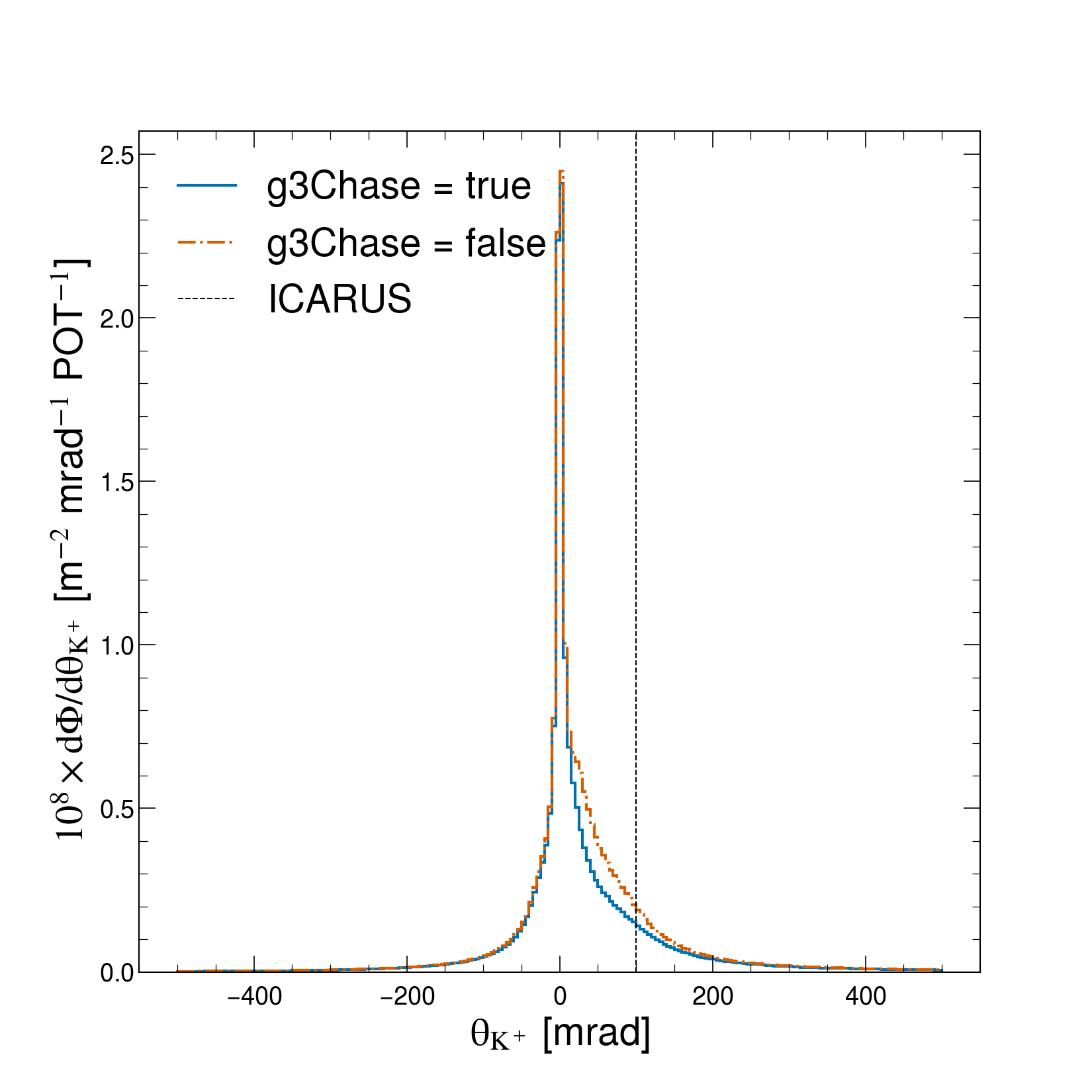}

	\caption[FHC \numu $\dd\Phi/\dd\theta_{\textup{p}}$ for Parent $\pi^+$ and $K^+$]{The angular distributions of $\pi^+$ (left) and $K^+$ (right) whose decay contribute to the muon neutrino flux, in FHC. %
	 In moving from the simulation with the shielding blocks excluded (orange) to the corrected geometry (blue), there is a significant reduction in the number of parent hadrons, especially $pi^+$, decaying to \numu, for mesons whose momenta are directed toward ICARUS (black, dashed).}%
	\label{fig:parent_angle}
\end{figure}

\begin{figure}[ht]
	\centering

	\includegraphics[width=0.49\textwidth]{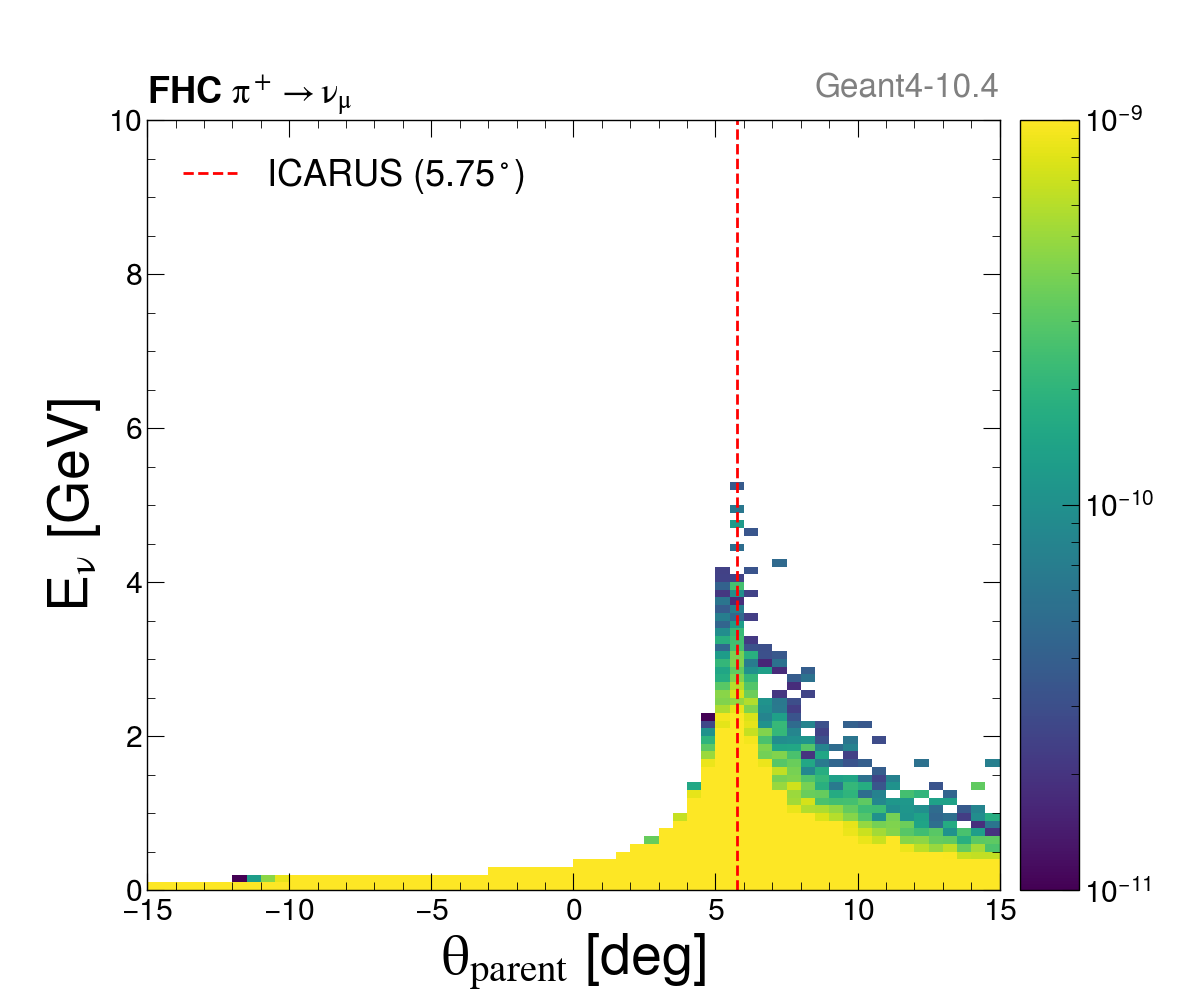}
	\includegraphics[width=0.49\textwidth]{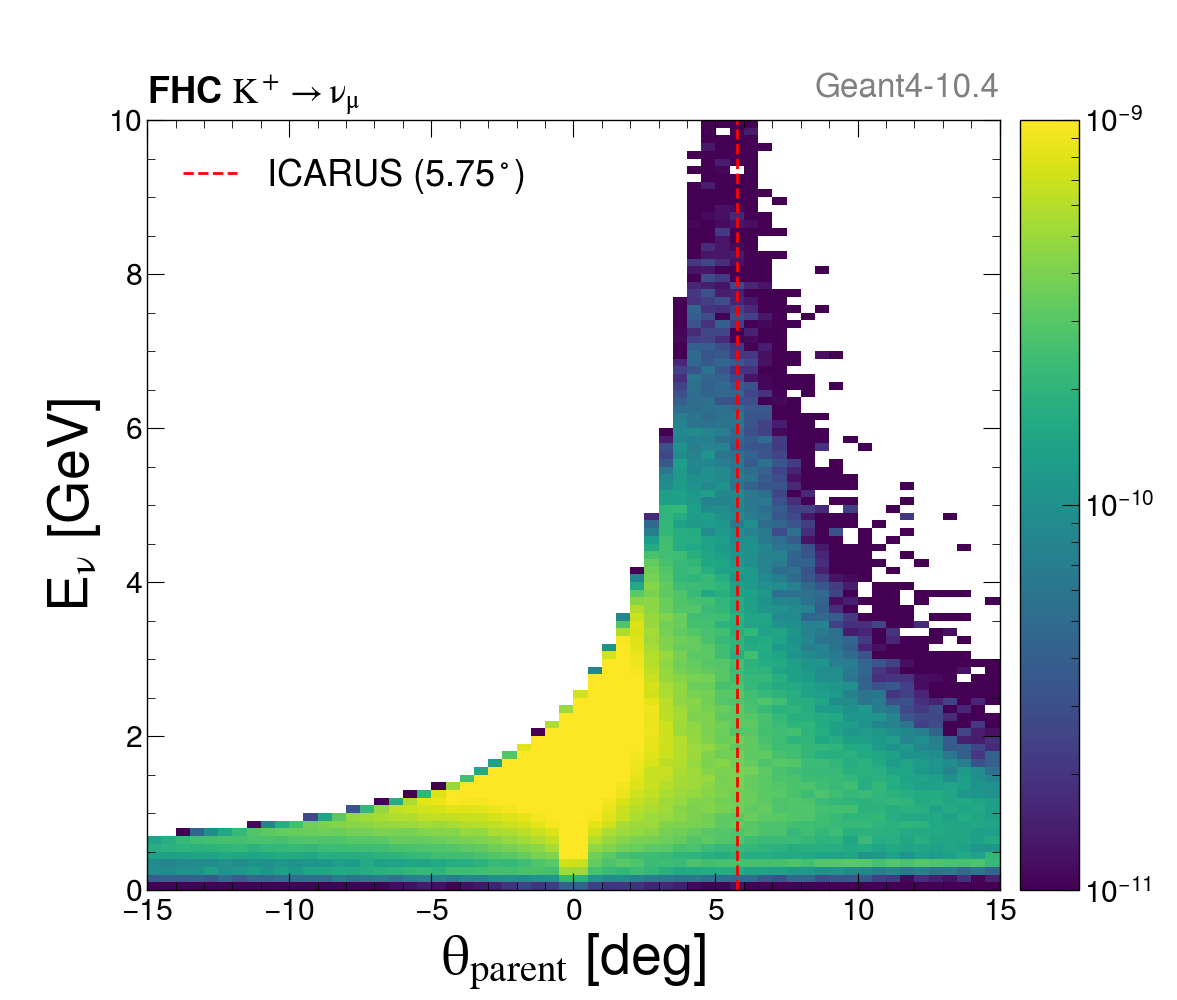}

	\caption[FHC \numu\ $\dd\Phi/\dd\Enu\dd\theta_\textup{p}$ for Parent $\pi^+$ and $K^+$]{Number of neutrinos vs. \numu\ energy and parent decay-momentum angle. %
	Primary pions contribute broadly to the energy spectrum; on-axis pions produce lower energy neutrinos ($<\SI{500}{\MeV}$). %
	The majority of primary kaons contributing to the peak region of $\sim \SI{2}{\GeV}$ are directed on-axis, while the off-axis kaons, impacted by the geometry change, contribute to the $E_\nu \gtrsim \SI{4}{\GeV}$.}%
	\label{fig:g4Update_numu_kplus_angle}
\end{figure}

Re-introducing the shielding blocks to the simulation lowers the height of the target chase ceiling by \ensuremath{\sim}\SI{60}{\centi\meter}.
In turn, this reduces the amount of available space in which upward-directed hadrons can decay.
As the ICARUS LArTPC is located \SI{100.3}{\milli\radian} ($5.75^\circ$) above the NuMI beam axis, the correction to the geometry has a large impact on the predicted neutrino flux through the detector.
Figure~\ref{fig:parent_angle} shows the angular distributions of parent pions and kaons which decay to \numu\ as a function of their momentum direction, while the remaining neutrino modes have been included in Appendices~\ref{app:parent_decay_angles} and~\ref{app:energy_vs_angles}.
The contribution from pion decays to the neutrino flux is particularly affected by the extra material, as a larger fraction of these parent mesons have their momenta directed toward ICARUS\@.
Primary kaons are predominantly directed along the beam axis and decay with wide kinematics producing ICARUS-bound neutrinos, and therefore experience less significant attenuation by the shielding blocks.
Kaons with off-axis momenta, however, contribute to the high-energy tail of the neutrino spectrum--see Figure~\ref{fig:g4Update_numu_kplus_angle}--and are more significantly impacted by the geometry correction.
These kaons are produced predominantly outside the target region, and are more likely to be directed off-axis.
The ratio of the flux with the geometry corrected to the nominal case approaches a maximum of 40\% reduction in the flux for pions and kaons with momenta directed toward \icarus\@.
This ratio was used to compute weights for the flux versus energy spectrum of each neutrino mode, decomposed by parent hadron, which may be applied to existing samples to correct for the geometry discrepancy.

%% file: geant4_updates.tex
The flux simulation and \ppfx have recently undergone development to support a newer version of \geant (4.10.4).
In the more recent version, the underlying nuclear model was fit to experimental HP data.
It is important to note is that the model parameter fit is not equivalent to the reweighting process that PPFX performs, where the flux simulation is adjusted to match the data measurements.
Fitting does not fix the model to the data, but rather constrains the model parameters to maximize consistency with the data within the bounds of experimental uncertainties and model constraints.
While the fit procedure may improve the overall data-model agreement in kinematic regions where HP data is available, it may also affect regions not currently covered by existing data.
In this section, the impact of the pre-\ppfx\ \geant update on the ICARUS flux prediction is discussed.
For a complete list of changes, refer to~\cite{geant_evol}.

\begin{figure}[htbp]
    \centering
    \begin{subfigure}{0.30\textwidth}
        \includegraphics[width=\textwidth]{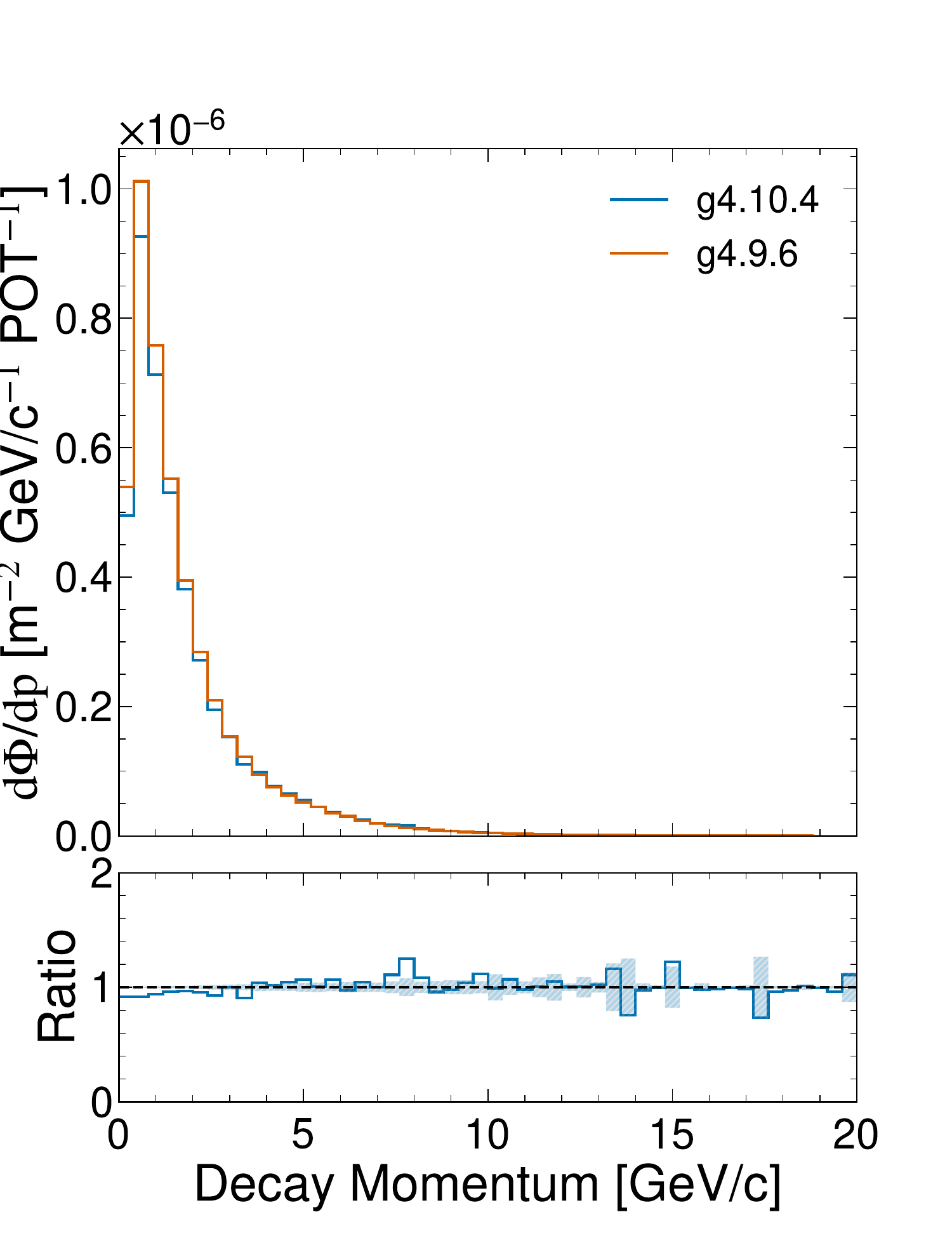}
        \caption{$\pi^+ \to \numu$}
    \end{subfigure}
    \begin{subfigure}{0.30\textwidth}
        \includegraphics[width=\textwidth]{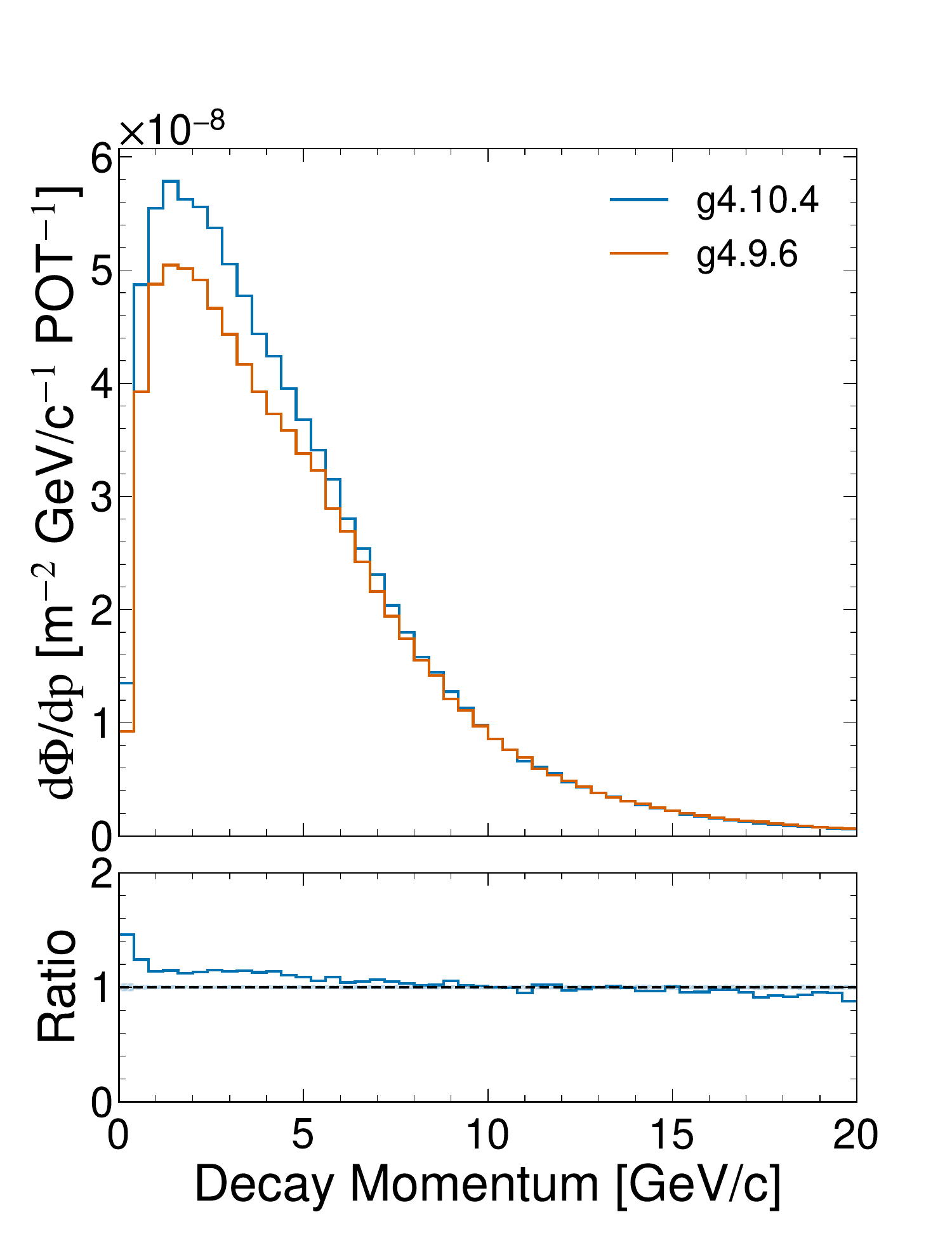}
        \caption{$K^+ \to \numu$}
    \end{subfigure}
    \begin{subfigure}{0.30\textwidth}
        \includegraphics[width=\textwidth]{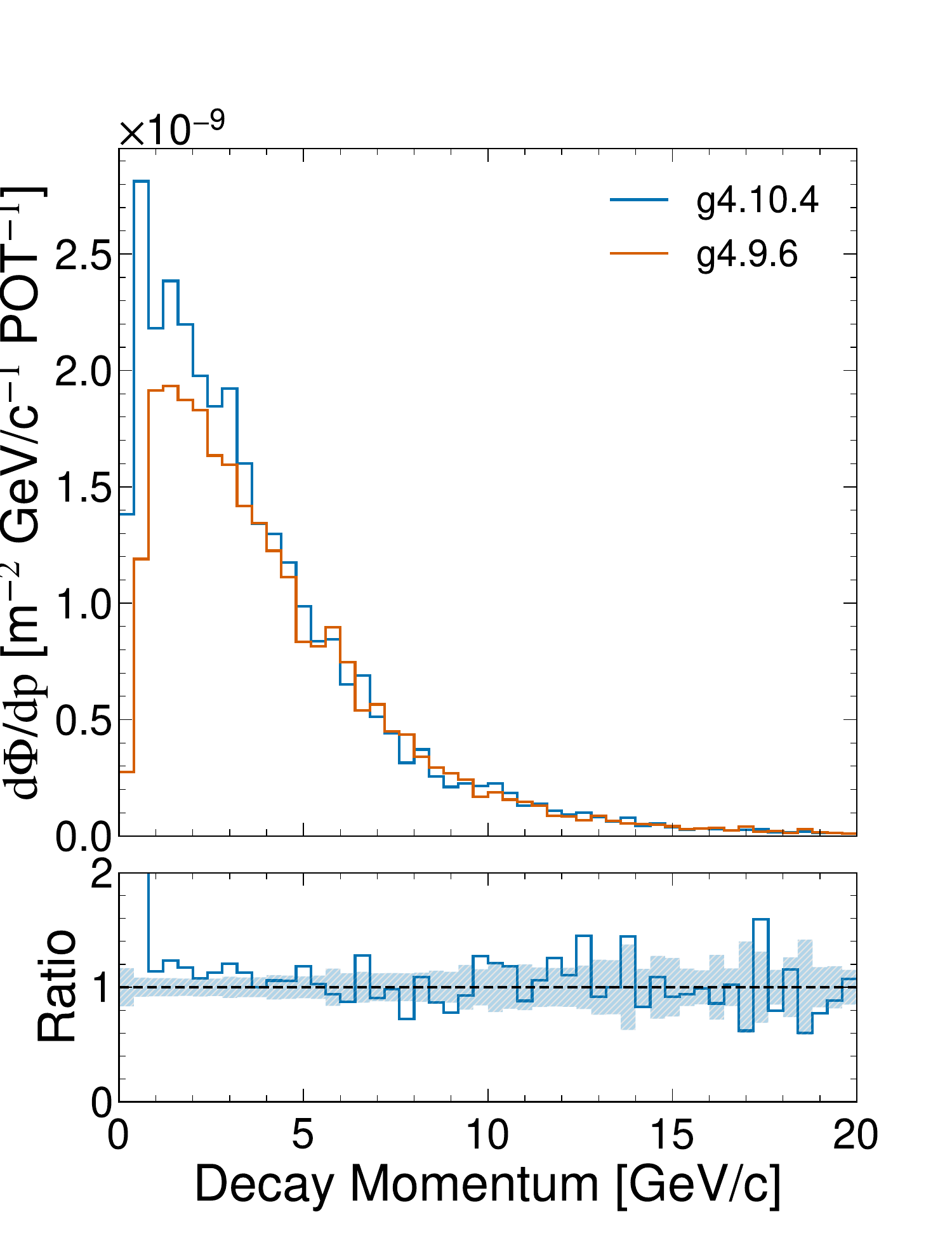}
        \caption{$K^0_L \to \numu$}
    \end{subfigure}
    \begin{subfigure}{0.30\textwidth}
        \includegraphics[width=\textwidth]{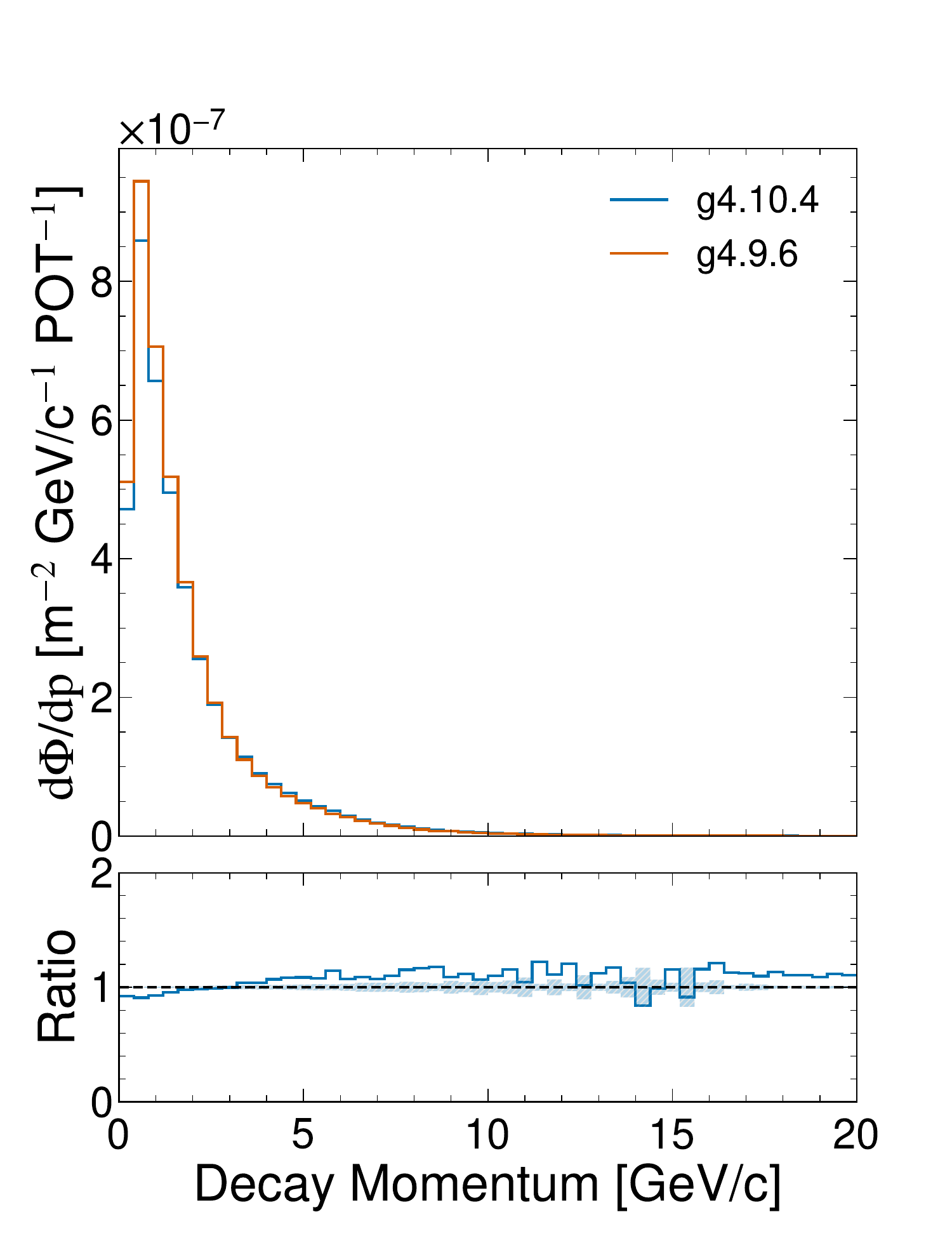}
        \caption{$\pi^- \to \numub$}
    \end{subfigure}
    \begin{subfigure}{0.30\textwidth}
        \includegraphics[width=\textwidth]{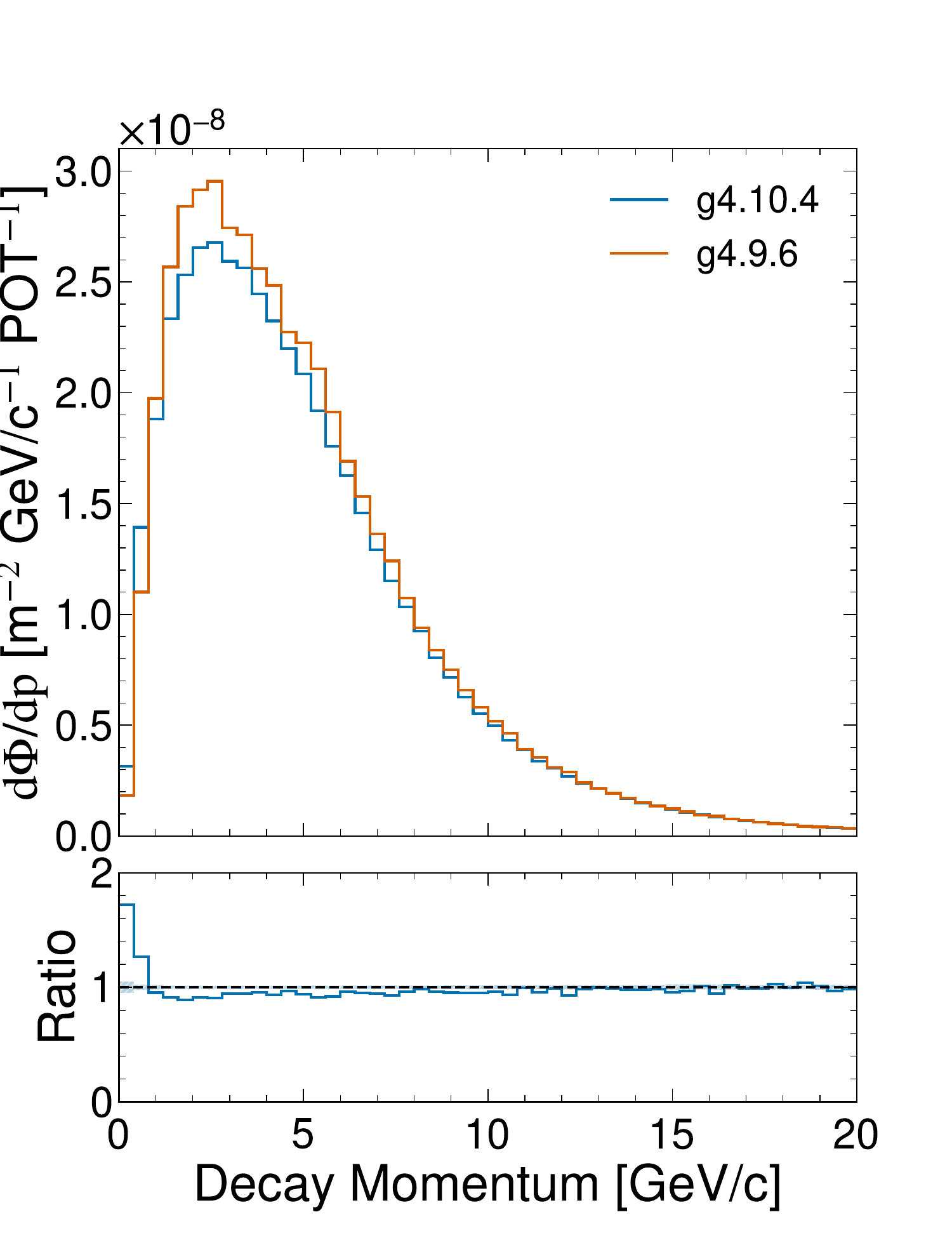}
        \caption{$K^- \to \numub$}
    \end{subfigure}
    \begin{subfigure}{0.30\textwidth}
        \includegraphics[width=\textwidth]{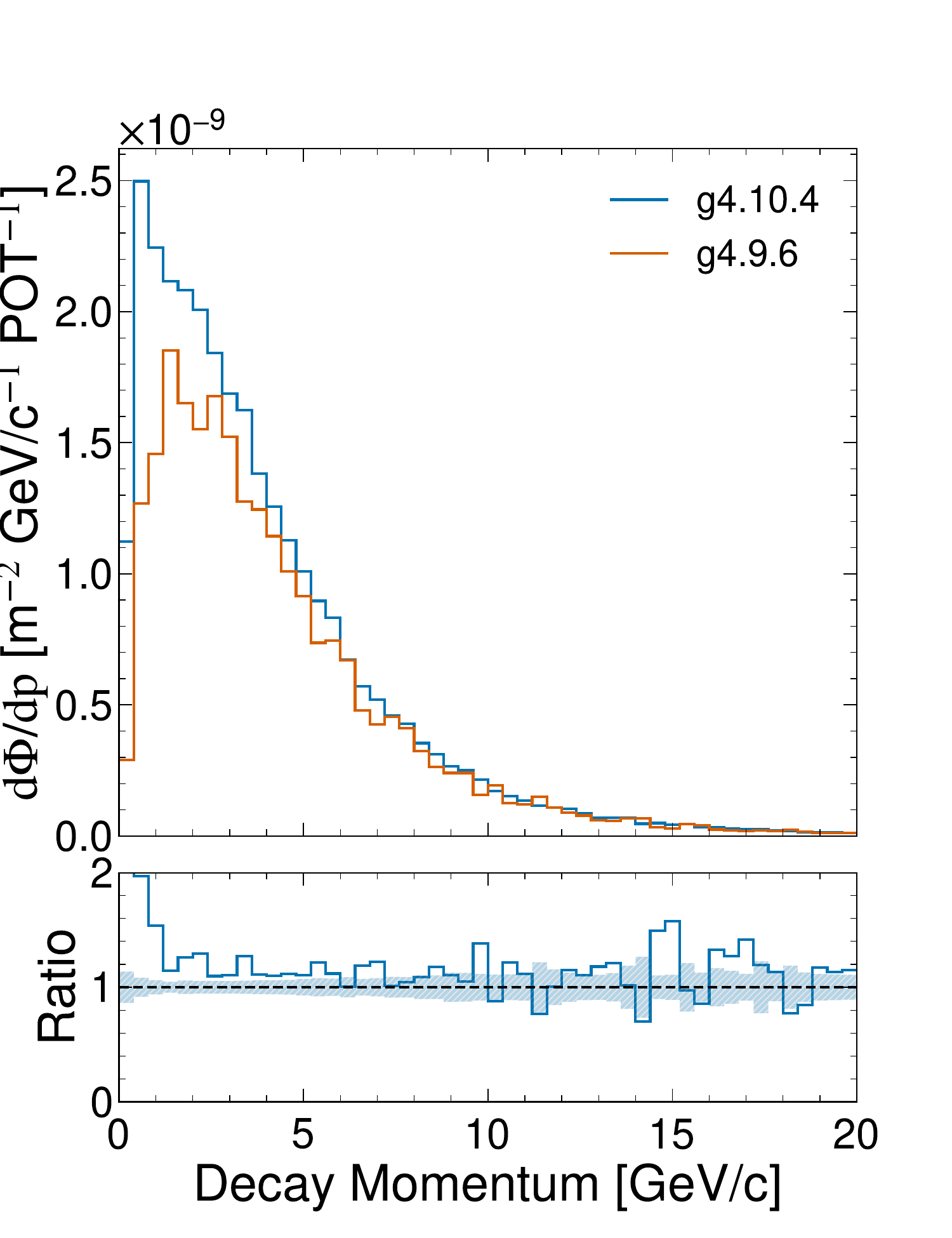}
        \caption{$K^0_L \to \numub$}
    \end{subfigure}
    \caption[Primary Hadron Decay Momenta in Geant4.10.4]{Decay momentum of parent hadrons that produce neutrinos in ICARUS, for FHC muon neutrinos (top) and RHC muon antineutrinos (bottom). Note the asymmetry between the $K^+ \to \numu$ and $K^- \to \numub$ distributions.}%
    \label{fig:parent_momentum_new_geant}
\end{figure}

Figure~\ref{fig:parent_momentum_new_geant} shows the decay momentum of the primary $\pi^\pm$, $K^\pm$, and $K^0_L$ that produce \numu\ and \numub\ arriving at ICARUS in the FHC and RHC beam operating modes, respectively.
The remaining momentum distributionsn, normalized to POT, can be found in Appendix~\ref{app:parent_decay_momenta}.
The momentum of pion parents appears to shift toward higher momenta in the updated Geant version, while the number of lower momentum kaon parents increases for \numu.
In the case of \numub\ production, in \fhc or \rhc, the momentum of the parent kaons appears to shift toward lower momenta and also exhibits attenuation of the number of higher-momentum kaons.
As the models of concern for this update are hadronic models, the production cross section should be independent of the electromagnetic charge of the parent hadron.
For this reason, the asymmetry present in the flux with respect $K^+ \to \numu$ and $K^- \to \numub$ production rates is contrary to initial expectations and requires further investigation.
\newpage

The updated \geant model introduced shifts within the interaction kinematic space, which moves some interactions from data-covered regions into ones where the hadron production data is not available.
This is exacerbated by a recently discovered bug in the \ppfx\ code impacting appropriate application of hadron uncertainties to certain processes, which will be discussed in the next section.

%% file: NA_bugfix.tex
\begin{figure}[!ht]
    \centering
    \includegraphics[width=\textwidth]{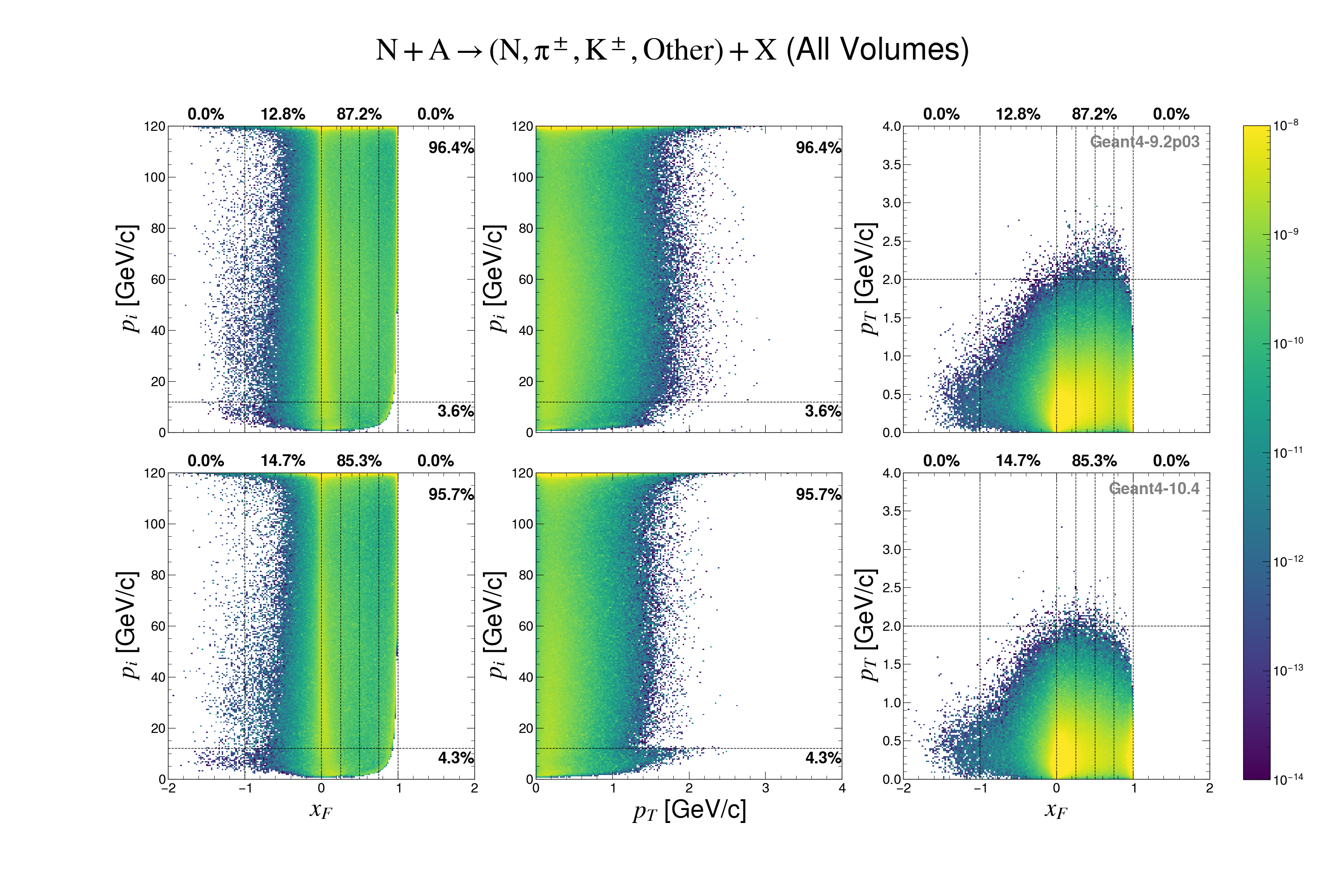}
    \caption[GEANT4.10.4 Nucleon Interaction Phase Space: $x_F-p_T-p_i$]{Phase space of the incoming momentum, $p_i$, fraction of the outgoing momentum in the longitudinal direction, in the center of mass frame, $x_F$, and the outgoing transverse momentum, $p_T$,
    for nucleon interactions leading to the production of neutrinos that arrive at ICARUS in terms of: $p_i - x_F$ (left), $p_i - p_T$ (middle), and $p_T - x_F$ (right).
    The top row was generated using GEANT4-9.2, while the bottom row was generated using GEANT4.10.4.
    Included are the primary beam proton interaction in the target as well as any secondary interactions that may occur in any NuMI volume.
    Dashed lines denote regions where PPFX performs reweighting based on available hadron production data: $p_i \geq 12 \; \si{\GeV{}/c}$, $p_T < 2 \; \si{\GeV{}/c}$, and $0 \leq x_F < 1$ in increments of 0.25.
    Other regions incur a 40\% uncertainty propagated to the neutrino flux spectra.
    Percentages indicate the fraction of interactions that occur in each region marked by the dashed lines.
    With the updated GEANT version, $ 2\%$ of interactions migrated to the $x_F < 0$ region, greater than $90\%$ of which have small $|x_F|$, where PPFX does not currently implement hadron production data.}%
    \label{fig:phase_space_NA_all_vols}
\end{figure}

A bug in PPFX affecting the treatment of out-of-phase-space nucleon interactions has been identified and corrected.
In its current implementation, interactions with $x_F < 0$ are not covered by HP data and therefore should receive a weight of $1$ and propagate a 40\% uncertainty on the differential cross section.
However, for these interactions, the 40\% uncertainty was not applied.
Figure~\ref{fig:phase_space_NA_all_vols} demonstrates percent level changes in the $x_F-p_T$ phase space of nucleon interactions leading to the production of neutrinos that arrive at ICARUS.
There is a migration of $\approx 2\%$ of interactions to the $x_F < 0$ region where PPFX does not currently implement hadron production data, which can result in an elevated estimate of the HP systematic uncertainty.

\begin{figure}[htbp]
    \centering
    \begin{subfigure}{0.49\textwidth}
        \includegraphics[width=\textwidth]{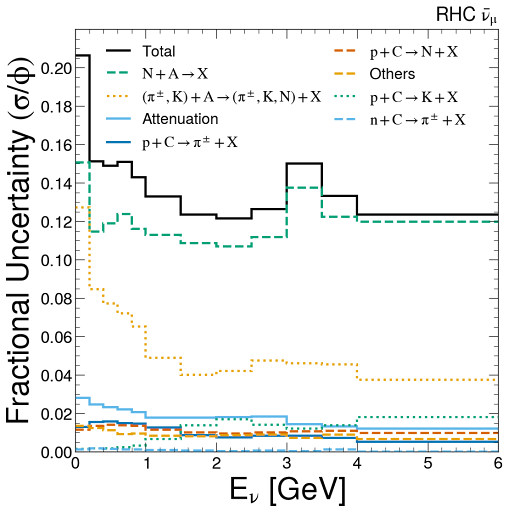}
        \caption{p+C QEL (included)}%
        \label{fig:rhc_numubar_nua_bugfix_hp_uncerts_a}
    \end{subfigure}
    \begin{subfigure}{0.49\textwidth}
        \includegraphics[width=\textwidth]{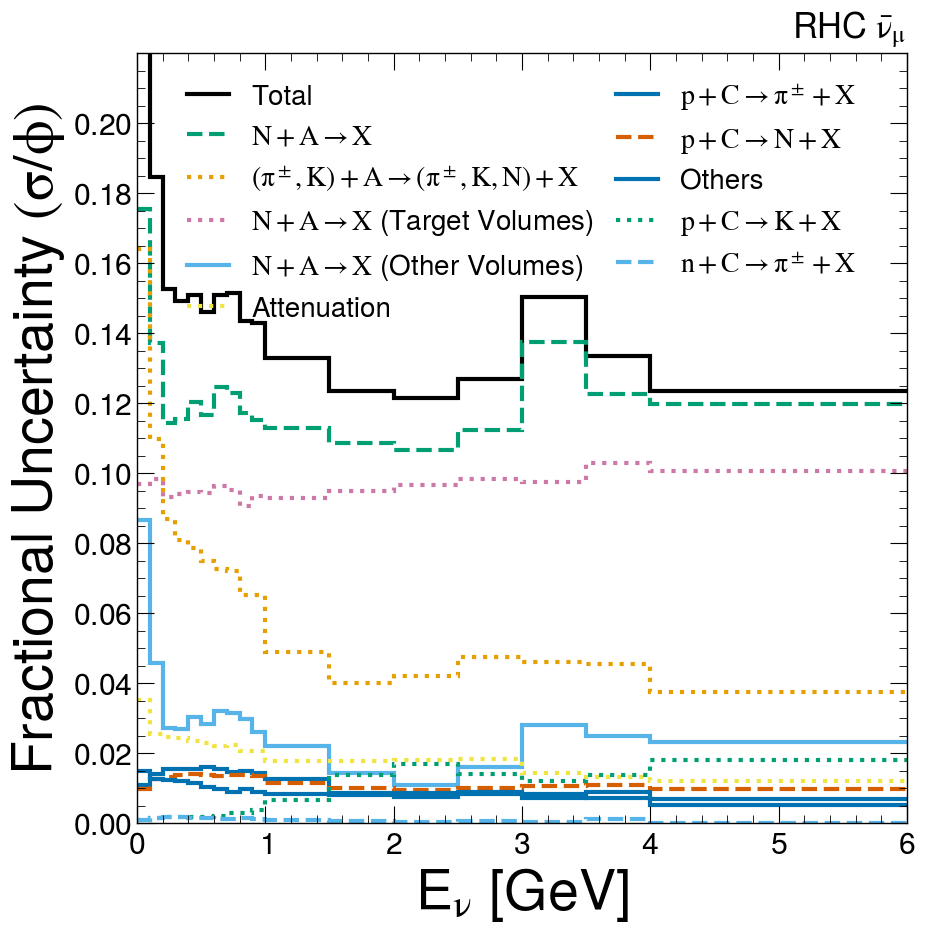}
        \caption{p+C QEL (excluded)}%
        \label{fig:rhc_numubar_nua_bugfix_hp_uncerts_b}
    \end{subfigure}
    \caption[PPFX N+A Interactions Bug Demonstration]{Fractional hadron production uncertainties for FHC \numu{}, which is dominated by nucleon interactions not covered by existing HP data (green dashed). Note that the difference in binning is due to lower (250M POT) available statistics at the outset of this study.}%
\end{figure}

Figure~\ref{fig:rhc_numubar_nua_bugfix_hp_uncerts_a} demonstrates that the $\mathrm{N + A \to{} X}$ process became the leading source of uncertainty after the bug was corrected.
Incorporating the fix re-introduced the numerous forward-going, proton (quasi-)~elastic scatters in the NuMI target as a source of uncertainty.
This class of interaction constitutes $\approx 45\%$ of the total $\mathrm{N + A \to{} X}$ interactions, which is increased from the observed rate in the previous GEANT version (36\%).
Furthermore, these elastic scatters are not covered by available data and propagate a 40\% uncertainty.
As described in~\cite{cherdack_prediction_2023}, multiple elastic scatters with low momentum transfer do not substantially influence primary hadron production, while, at the same time, result in an overly conservative inflation of the uncertainty.
For these reasons, it was decided to exclude these interactions from the uncertainty estimate, and the result is shown in Figure~\ref{fig:rhc_numubar_nua_bugfix_hp_uncerts_b}.

\begin{figure}[htbp]
    \centering
    \begin{subfigure}{0.51\textwidth}
        \includegraphics[width=\textwidth]{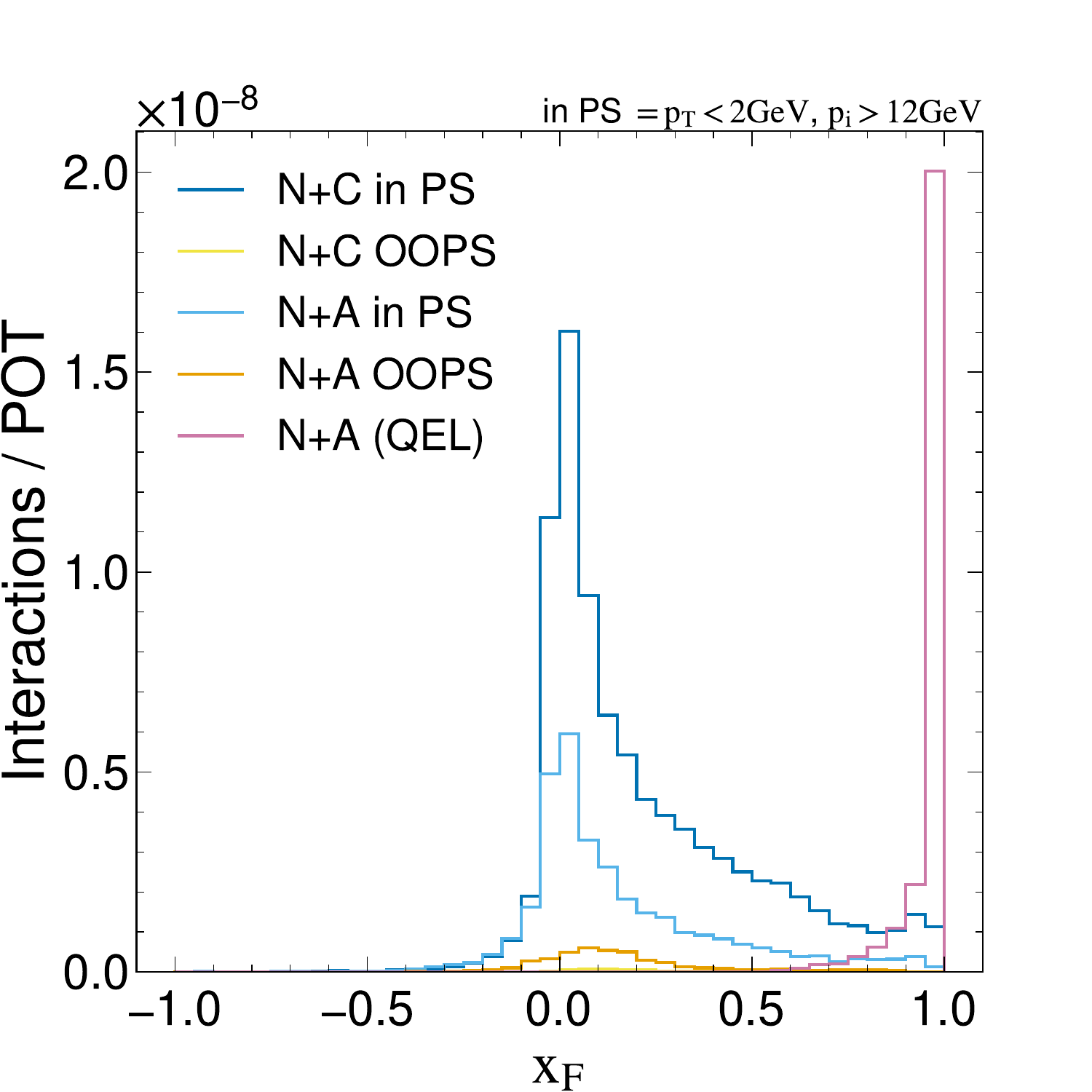}
        \caption{}%
        \label{fig:NA_phase_space_breakout}
    \end{subfigure}
    \begin{subfigure}{0.47\textwidth}
        \includegraphics[width=\textwidth]{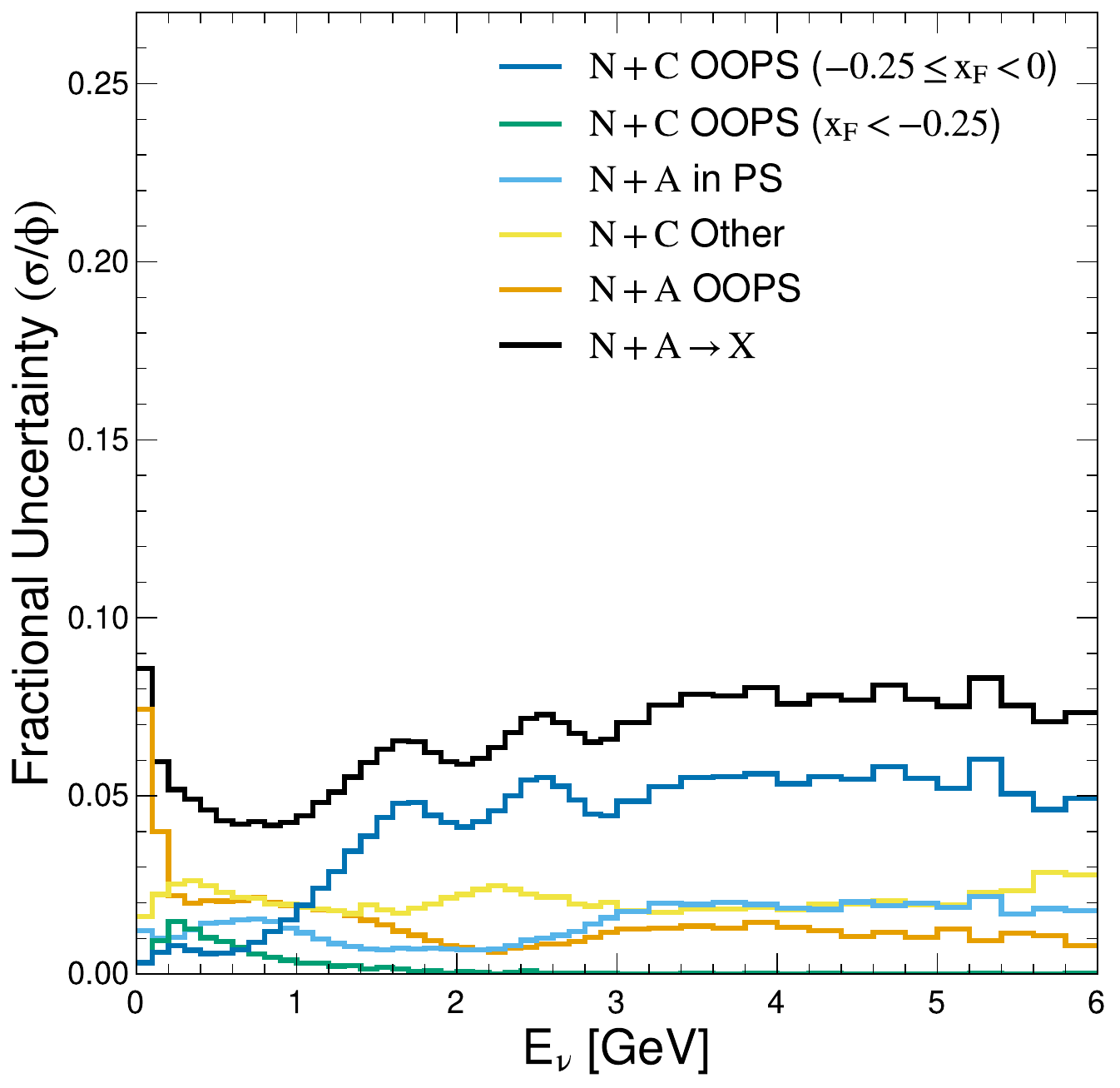}
        \caption{}%
        \label{fig:NA_uncert_breakout}
    \end{subfigure}
    \caption[PPFX N+A Interactions ($x_F$)]{Feynman-X (left) and uncertainty (right) distribution for various sub-channels of nucleon interactions that eventually yield \numub\ in their decay chain in \acrshort{rhc} operation. %
     Approximately 45\% of these interactions are (quasi-)elastic proton scatters (purple). %
     Of particular interest are in-phase-space N+C (dark blue) interactions with $x_F < 0$, 95\% of which are in the range $-0.08 \leq x_F < 0$, which contribute the majority of the uncertainty and is a potential candidate for application of new, experimental HP data. %
     Out-of-phase-space (OOPS) shown in yellow and orange, refers to interactions with either $p_T > \SI{2}{\GeV}$ or $p_i < \SI{12}{\GeV}$. %
     Interactions that are in-phase-space, but occur outside of the carbon (N+A) (light-blue) may have A-scaling applied to the cross section provided the interaction falls into data-covered $x_F$ bin.}%
\end{figure}

Despite this exclusion, the $\mathrm{N + A \to{} X}$ process remains the leading source of uncertainty at neutrino energies exceeding \SI{1.2}{\GeV}.
To elucidate the drivers of this process, Figure~\ref{fig:NA_phase_space_breakout} shows the $x_F$ distribution of several sub-channels, while Figure~\ref{fig:NA_uncert_breakout} shows the corresponding uncertainties of those populations.
Interactions that may be covered by data are nucleon-carbon (N+C) interactions with incident nucleon momenta $p_i > 12$ \si{\GeV/c}, outgoing $0 < x_F < 0.95$, $p_T < 2.0$ \si{\GeV/c}.
Additionally, PPFX applies A-scaling to interactions that occur externally to the NuMI carbon target (N+A) but otherwise meet these kinematic criteria.
Of these, proton on carbon (p+C) with $0.25 \leq x_F < 0$ is the primary source of uncertainty, 95\% of which are within $-0.08 \leq x_F < 0$.
These interactions fall into a single bin in the $x_F$ dimension, but distribute broadly across the $E_\nu$ dimension.
This manifests as an inflated, correlated uncertainty where $E_\nu > \SI{1.2}{\GeV}$, and when extended to the full horn-flavor-energy space, the correlations between bins will likely increase where the uncertainty is dominated by the N+A process.

\begin{figure}[htbp]
    \centering
    \includegraphics[width=0.49\textwidth]{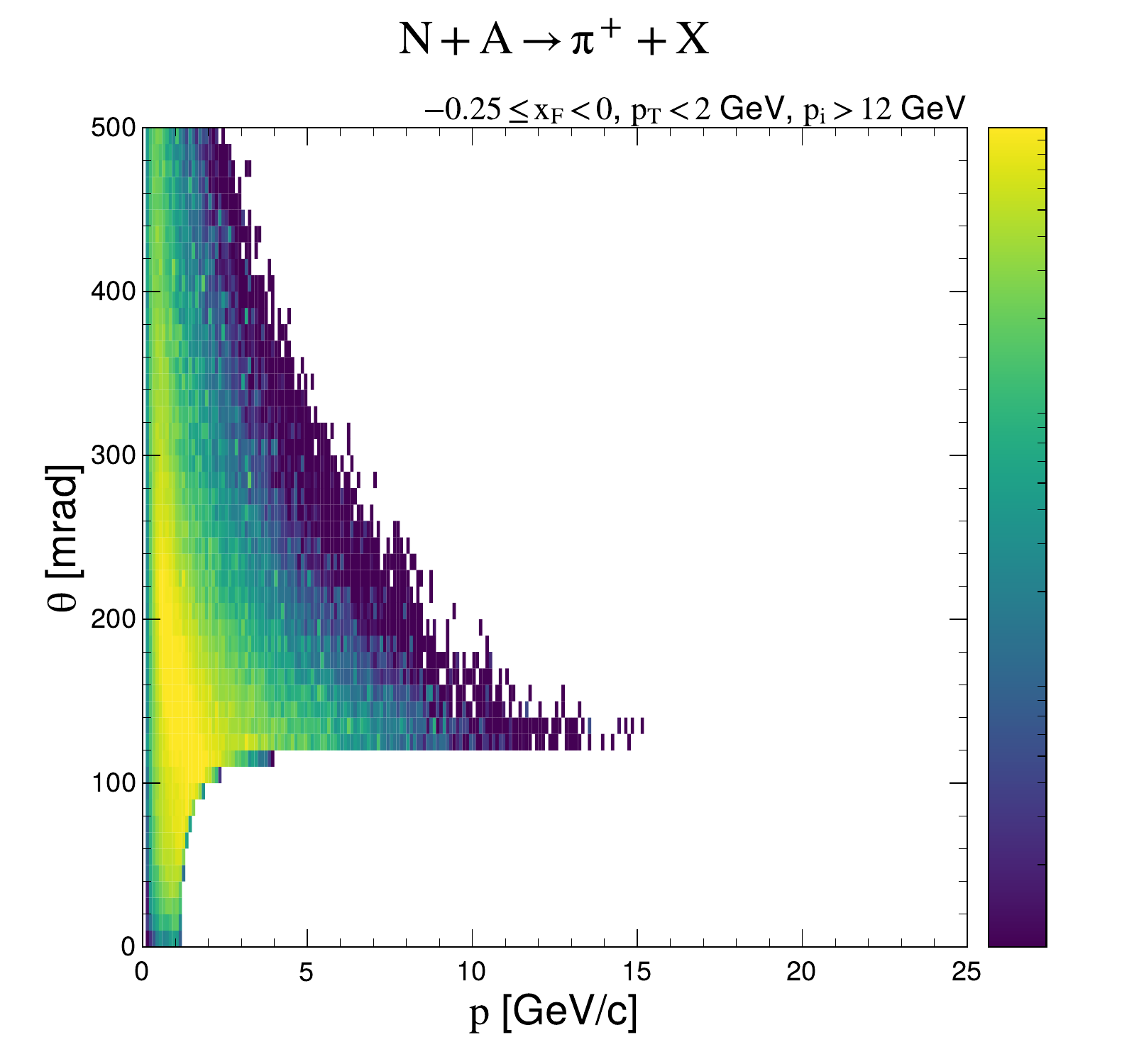}
    \includegraphics[width=0.49\textwidth]{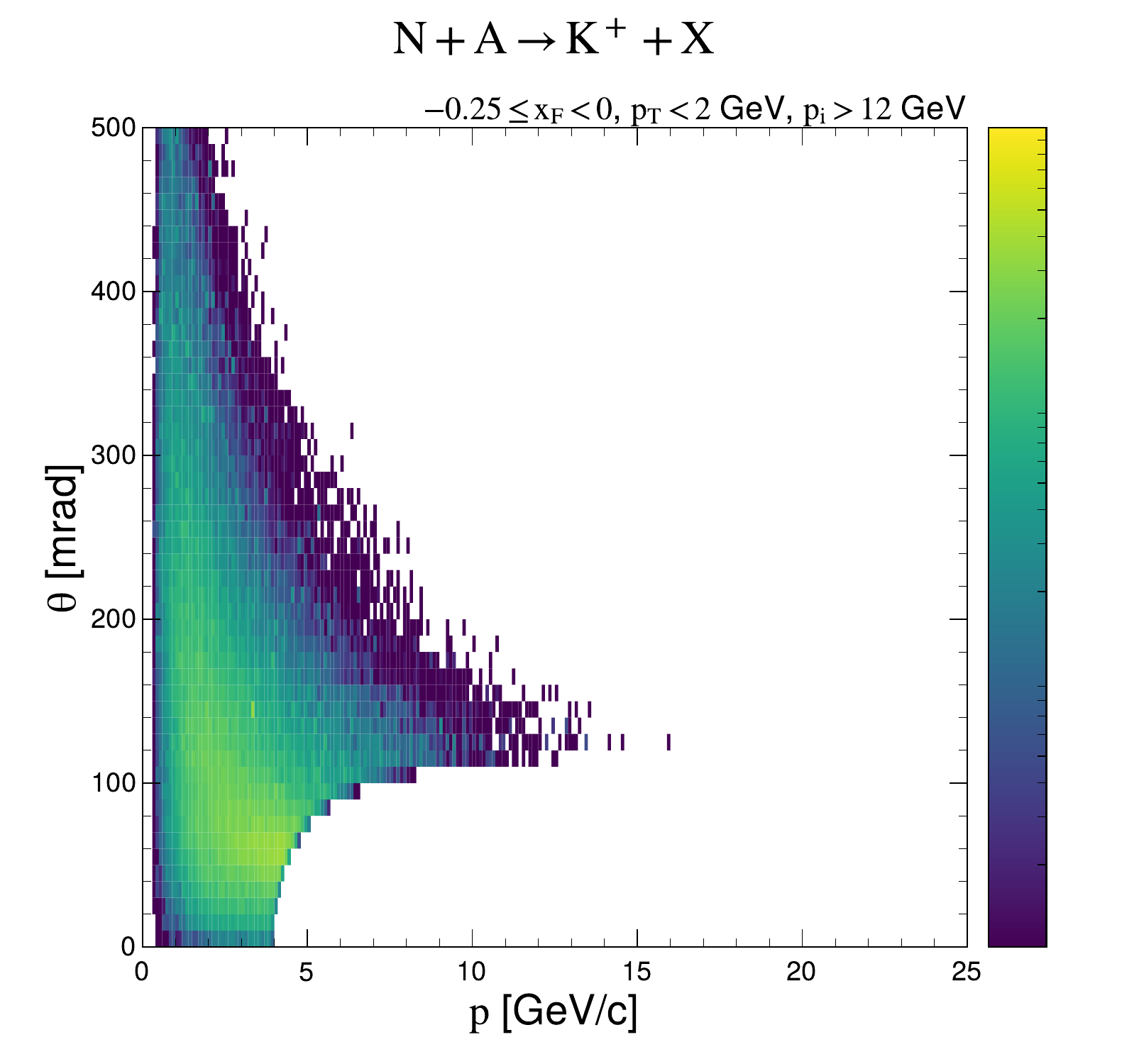}
    \includegraphics[width=0.49\textwidth]{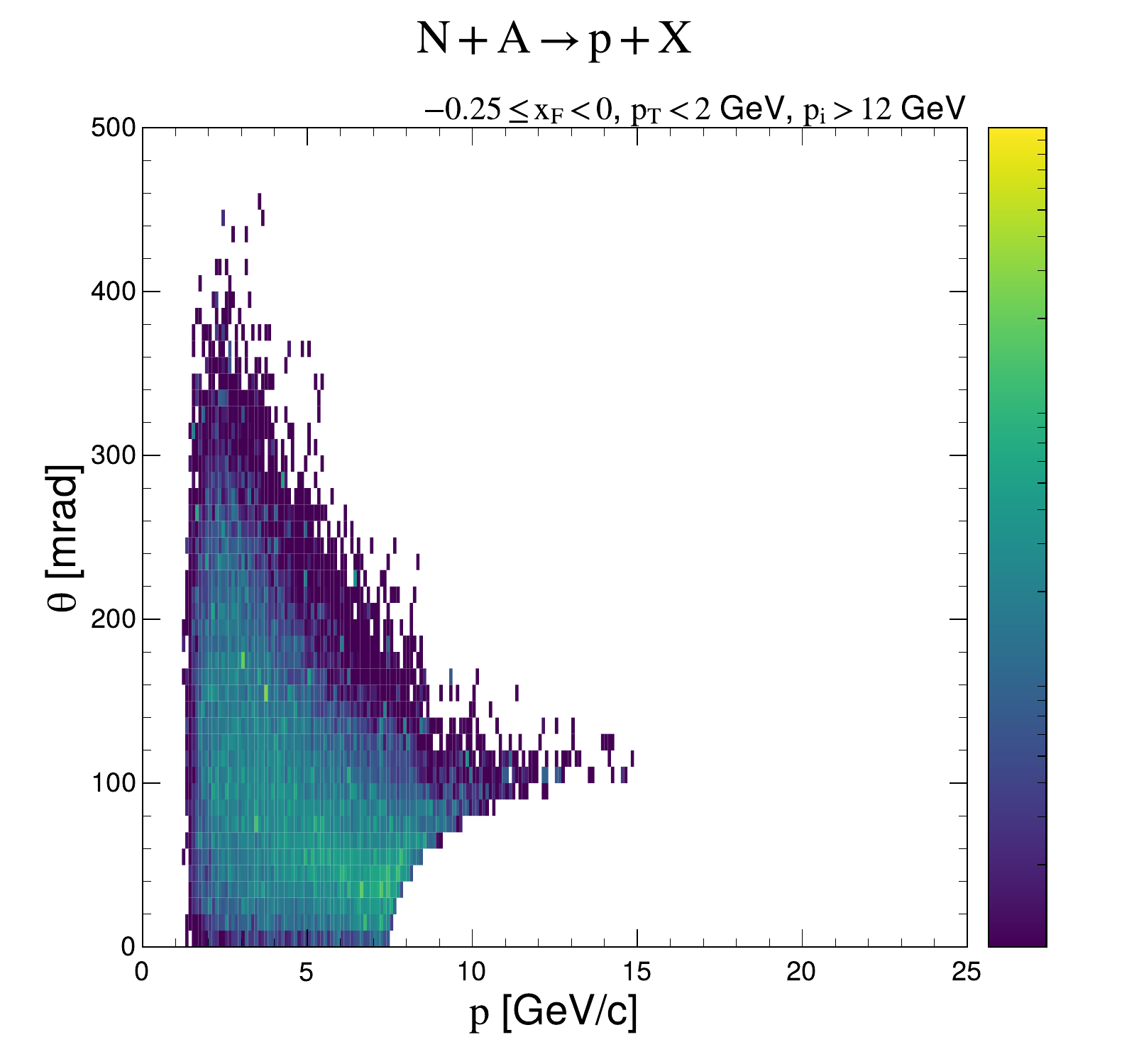}
    \caption[NA61 Phase Space Coverage]{Count of nucleon interactions, in log scale, in the NuMI target resulting in the production of $\pi^+$ (top left), $K^+$ (top right), and $p$ (bottom) within the negative $x_F$ region of interest.}%
    \label{fig:NA61_overlap_phase_space}
\end{figure}

Additional data collected by NA61/SHINE~\cite{na61shine_collaboration_measurements_2023} has been identified that could be used to constrain the component of $\mathrm{p + C \to{} \left(p, \pi^\pm, K^\pm \right)}$ for \SI{120}{\GeV/c} incident protons in the $0.25 \leq x_F < 0$ region.
The phase space of interest is shown in Figure~\ref{fig:NA61_overlap_phase_space}, and was used to make the aforementioned comparison.

At the time of this study, PPFX only has the capability to apply hadron production data collected by the MIPP and NA49 experiments, both of which report differential cross sections in terms of the outgoing hadron's $x_F$ and $p_T$.
A convenience of cross sections reported in terms of $x_F$ and $p_T$ is that these define a Lorentz invariant kinematic phase space, and therefore can be applied to any incident nucleon momentum which PPFX performs via a FLUKA-based energy scaling.
However, the NA61/SHINE data is published in terms of the outgoing hadron's $p$ and $\theta$, to which energy scaling cannot be readily applied.
To incorporate this data, PPFX will need to be updated such that either the data is transformed into $x_F$ and $p_T$ or the histograms containing the differential cross sections generated from the model should be extended to include the $p$ and $\theta$ space.
The limitation of the latter approach is that energy scaling would not be possible, and application of the data would be limited to the incident nucleon momenta of \SI{120}{\GeV/c}.
Fortunately, a significant fraction of $\mathrm{N+A}$ interactions occur at beam energies, and therefore the latter approach would be efficacious.

%% file: final_flux_prediction.tex
\begin{figure}[htbp]
    \centering
    \includegraphics[width=0.7\textwidth]{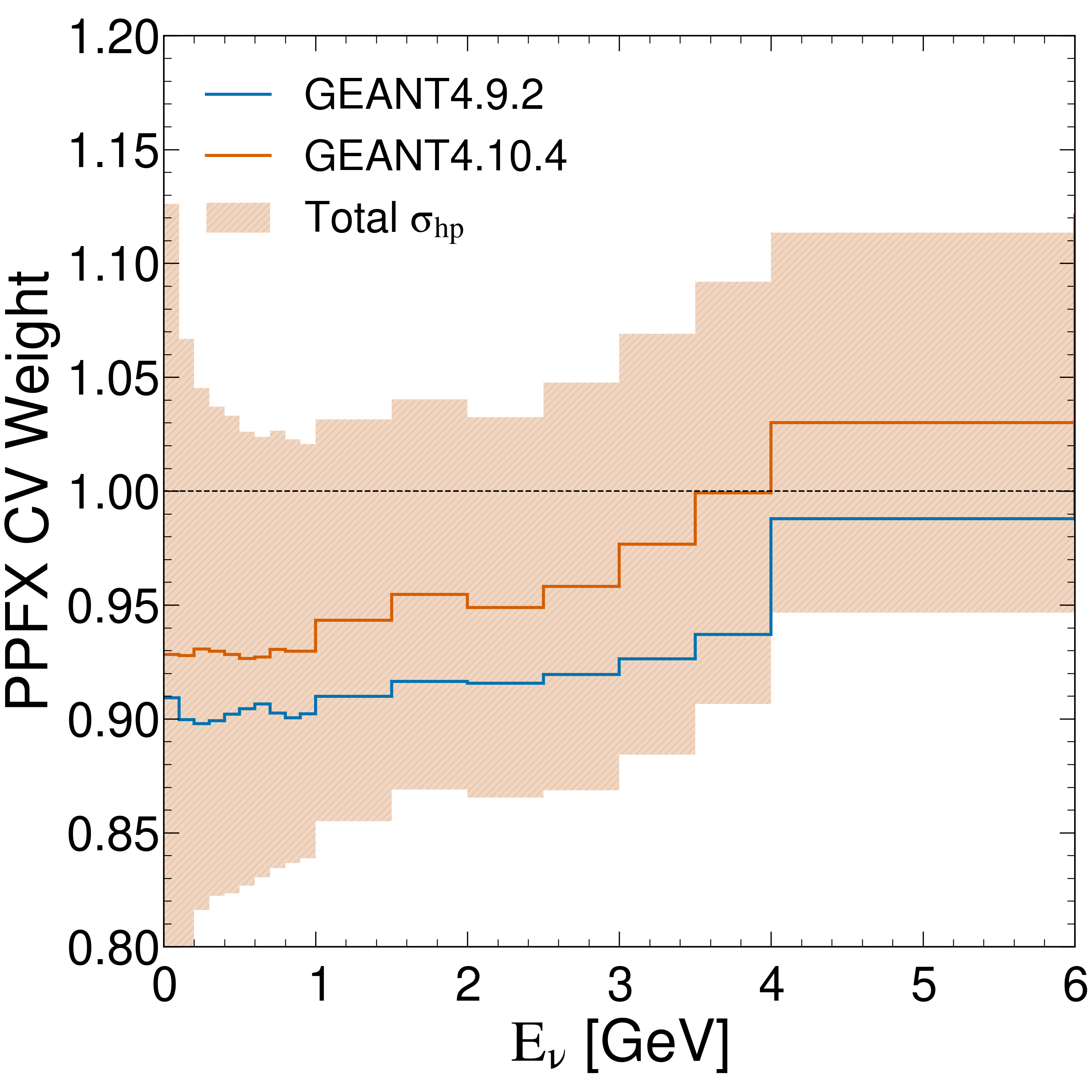}
    \caption[Old vs. Updated PPFX CV Weights]{Comparison of \ppfx central value weights with 4.9.2 model (blue) and 4.10.4 (orange). The updated weights shifted toward unity, indicating better data-model agreement, but are consistent with the old central value weights within systematic uncertainties.}%
    \label{fig:new_ppfx_weights}
\end{figure}
\begin{figure}[htbp]
    \centering
    \includegraphics[width=0.49\textwidth]{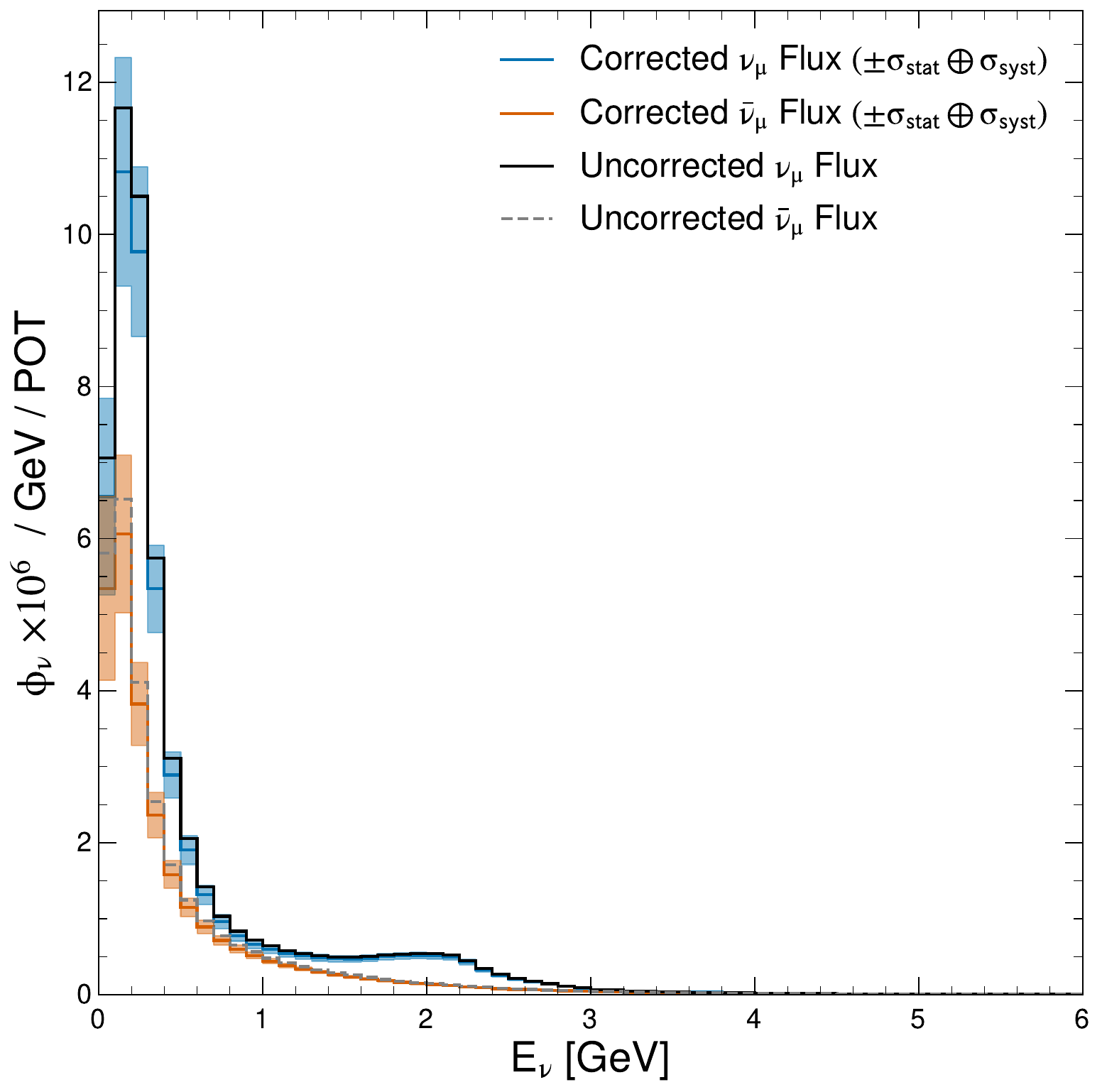}
    \includegraphics[width=0.49\textwidth]{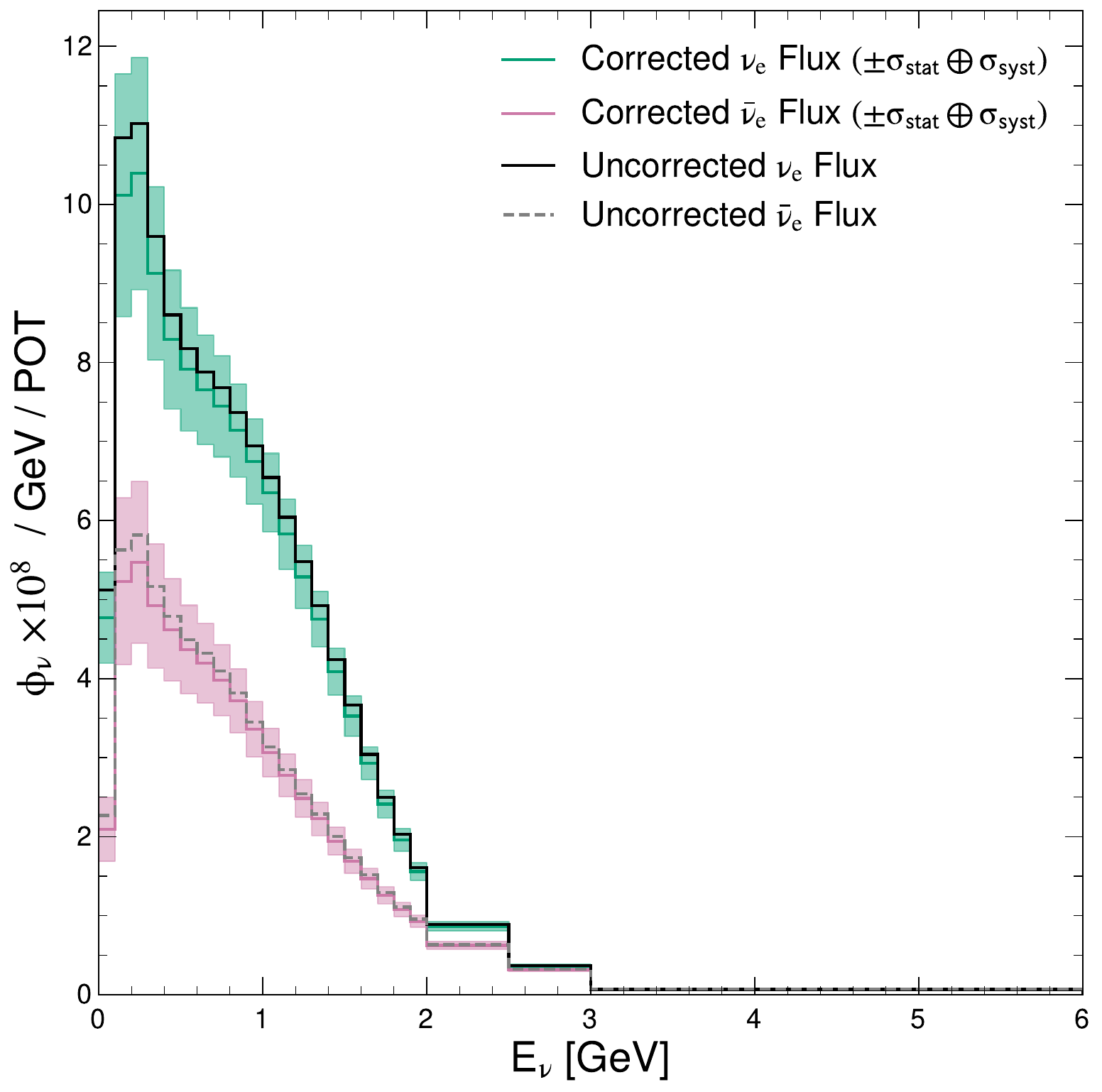}
    \caption[Geant4.10.4 Updated ICARUS Flux Prediction]{Updated \numi{}-\icarus\ flux prediction accounting for all geometry and model changes for \fhc \numu (left) and \nue (right). Note that due to higher available statistics, a finer binning is used in the interest of improved feature resolution.}%
    \label{fig:flux_pred}
\end{figure}
\begin{figure}[htbp]
    \centering
    \includegraphics[width=0.49\textwidth]{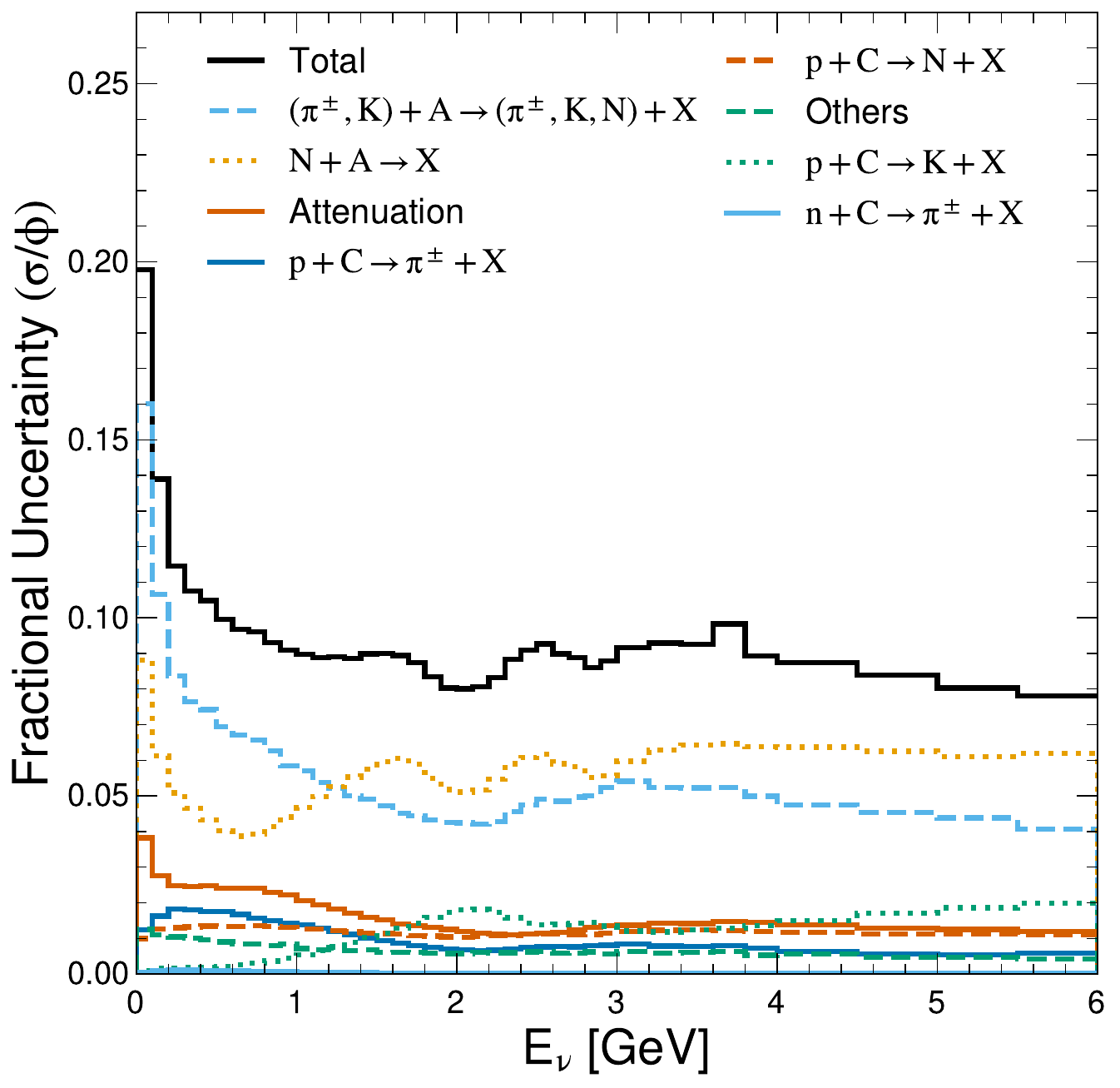}
    \includegraphics[width=0.49\textwidth]{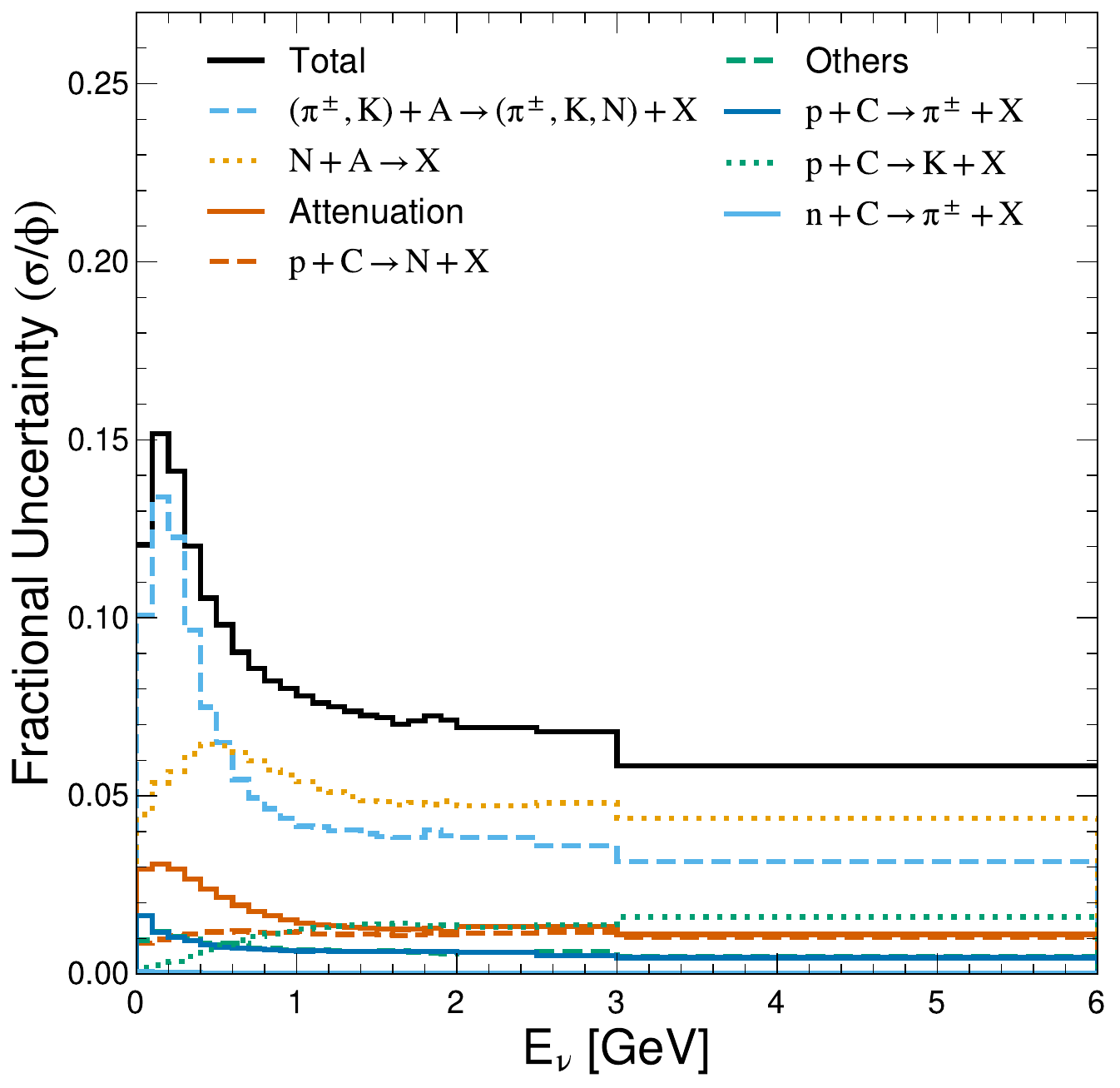}
    \caption[Geant4.10.4 Updated ICARUS Hadron Production Uncertainties]{Updated hadron production uncertainties for \fhc \numu (left) and \nue (right). %
     The incident meson channel remains the dominant source of uncertainty in the $<\SI{1}{\GeV}$ region due to a higher frequency of pion re-interactions, while the nucleon interaction channel without data coverage (N+A) drives the uncertainty in the $>\SI{1}{\GeV}$ region.}%
    \label{fig:hadron_uncertainties}
\end{figure}

In this section, the flux prediction and corresponding systematic uncertainties are reported, encapsulating the changes to the NuMI target hall geometry, the \geant\ nuclear model updates, and corrections to the PPFX reweight driver discussed across the previous sections.
The impacts to the \ppfx flux central value weights are shown in Figure~\ref{fig:new_ppfx_weights} for \fhc \numu.
The size of the weights shifted toward unity in the simulation produced with the updated model, indicating the benchmarking procedure was successful in improving data-model agreement, but the model persists in overestimating the flux across the energy range.
Between the two versions of the simulation, the central value of the flux differ at the percent-level, well-within systematic uncertainties, and is confirmed in Figure~\ref{fig:flux_pred} for \fhc \numu and \nue.
Refer to Figure~\ref{fig:flux-pred} for the previous prediction based on the 4.9.2 model.
The corresponding hadron production uncertainties can be found in Figure~\ref{fig:hadron_uncertainties}; compare to Figure~\ref{fig:frac_uncert}.
The incident meson channel remains the dominant source of uncertainty in the $<\SI{1}{\GeV}$ region due to a higher frequency of pion re-interactions, while the nucleon interaction channel without data coverage (N+A) drives the uncertainty in the $>\SI{1}{\GeV}$ region.

\begin{figure}[htbp]
\centering
\includegraphics[width=0.7\textwidth]{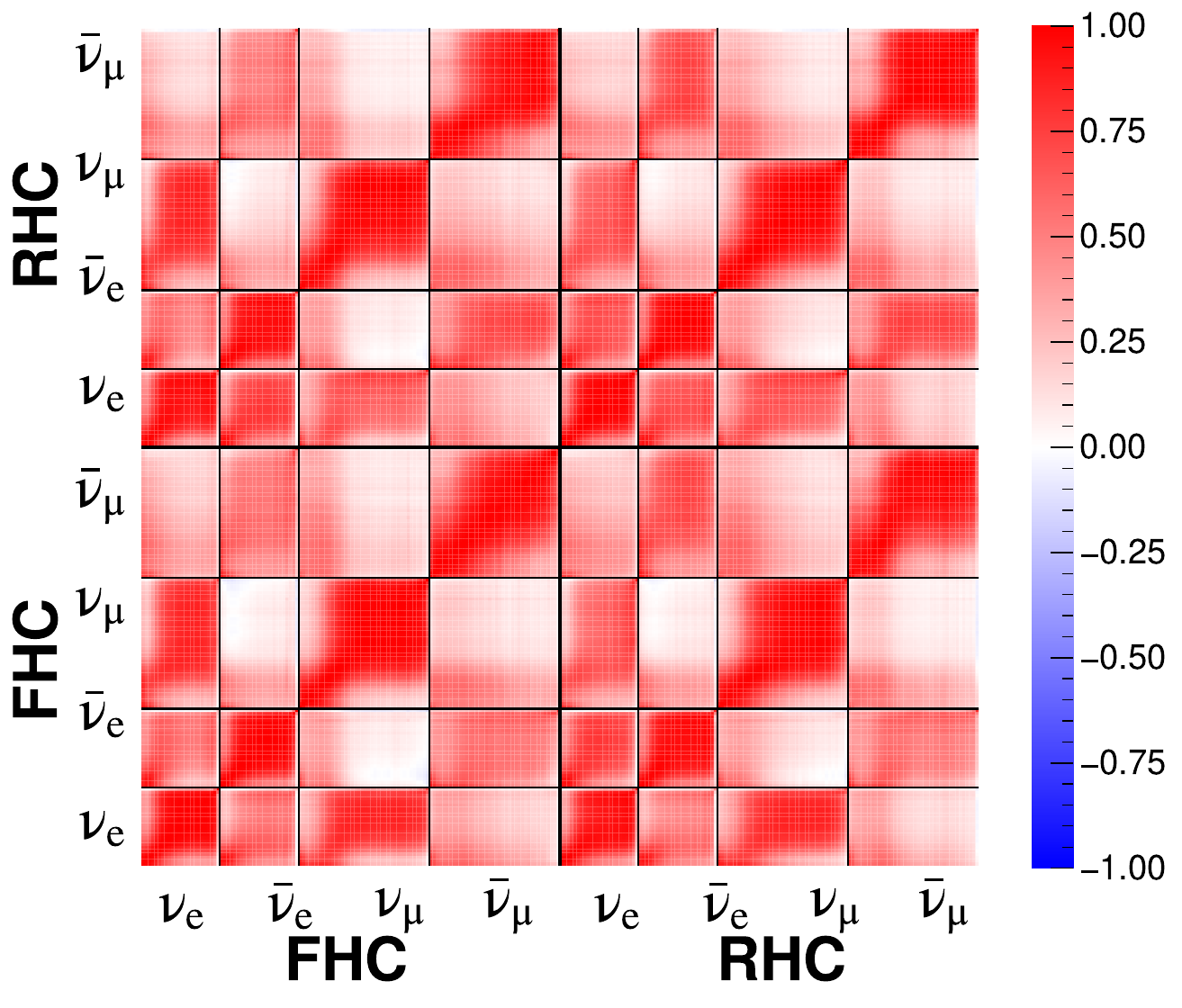}
\caption[Geant4.10.4 Total HP Correlation Matrix]{Updated Hadron Production correlation matrix for both \fhc and \rhc, and all 4 neutrino modes.}%
\label{fig:total_correlation_matrix}
\end{figure}

Figure~\ref{fig:total_correlation_matrix} shows the total correlation matrix for the updated hadron production uncertainties, and Table~\ref{tab:hadron_uncertainties} lists the integrated uncertainties for the $0 \leq E_\nu \leq \SI{20}{\GeV}$ region.
Compared to Figure~\ref{fig:hp-cov}, the correlation matrix does not exhibit significant negative correlations in the high-energy regions of the matrix, and instead has shifted toward zero due to an increased contribution of a correlated N+A uncertainty mentioned in the previous section.
A downstream effect of this is that while integrated uncertainties have increased by a few percent compared to the values presented in Table~\ref{tab:integ-uncerts}, ratio uncertainties remain relatively supressed maintaining viability for future ratio cross section measurements.

\begin{table}[htbp]
    \centering
    \caption[Updated Hadron Production Uncertainties (Integrated)]{Integrated hadron production uncertainties for $0 \leq E_\nu \leq 20$ GeV.}
    \begin{tabular}{llrrrrrrr}
        \toprule
         &  & $\nu_e$ & $\bar{\nu}_e$ & $\nu_e + \bar{\nu}_e$ & $\nu_\mu$ & $\bar{\nu}_\mu$ & $\nu_\mu + \bar{\nu}_\mu$ & $\frac{\nu_e + \bar{\nu}_e}{\nu_\mu + \bar{\nu}_\mu}$ \\
        \midrule
        \multirow[t]{2}{*}{Hadron} & FHC & 8.61 & 11.25 & 8.90 & 11.00 & 13.73 & 10.74 & 7.72 \\
         & RHC & 10.46 & 9.26 & 9.36 & 12.71 & 11.86 & 11.06 & 7.94 \\
        \cline{1-9}
        \bottomrule
    \end{tabular}%
    \label{tab:hadron_uncertainties}
\end{table}

%% file: 2x2.tex
\chapter{On-Axis Flux Prediction for the DUNE ND Prototype}\label{sec:2x2}

The \acrfull{2x2}~\cite{Russell:2024az}, is a 2-by-2 grid of ``optically segmented'', 60\% scale \acrshort{lartpc} modules, with a total active volume of \SI{2.4}{\tonne} of LAr.
Its purpose is to provide critical insights into the performance of and inform the final design of the full-scale \acrshort{dune} \acrshort{nd}.
The \acrshort{2x2} prototype will be exposed to the \rhc\ \acrshort{numi} \SI{1}{\mega\watt} beam, where it will collect approximately 300k charged-current antineutrino interactions per year in the several-GeV range.
\acrshort{ppfx}, in tandem with the tools developed for this thesis, facilitate configurable and reproducible flux predictions, which may be applied to any user-defined point relative to the \numi\ origin.
As the primary motivation for this thesis is to contribute to high-statistics $\nu-\mathrm{Ar}$ cross section measurements for \acrshort{dune},
a natural extension of the work presented in Chapter~\ref{sec:analysis} is toward upcoming \acrshort{2x2} analyses.
This chapter will focus on the application of these tools toward an on-axis flux prediction and hadron production systematics study for this purpose.

\begin{figure}[htbp]
    \centering
    \begin{subfigure}{0.49\textwidth}
        \includegraphics[width=\textwidth]{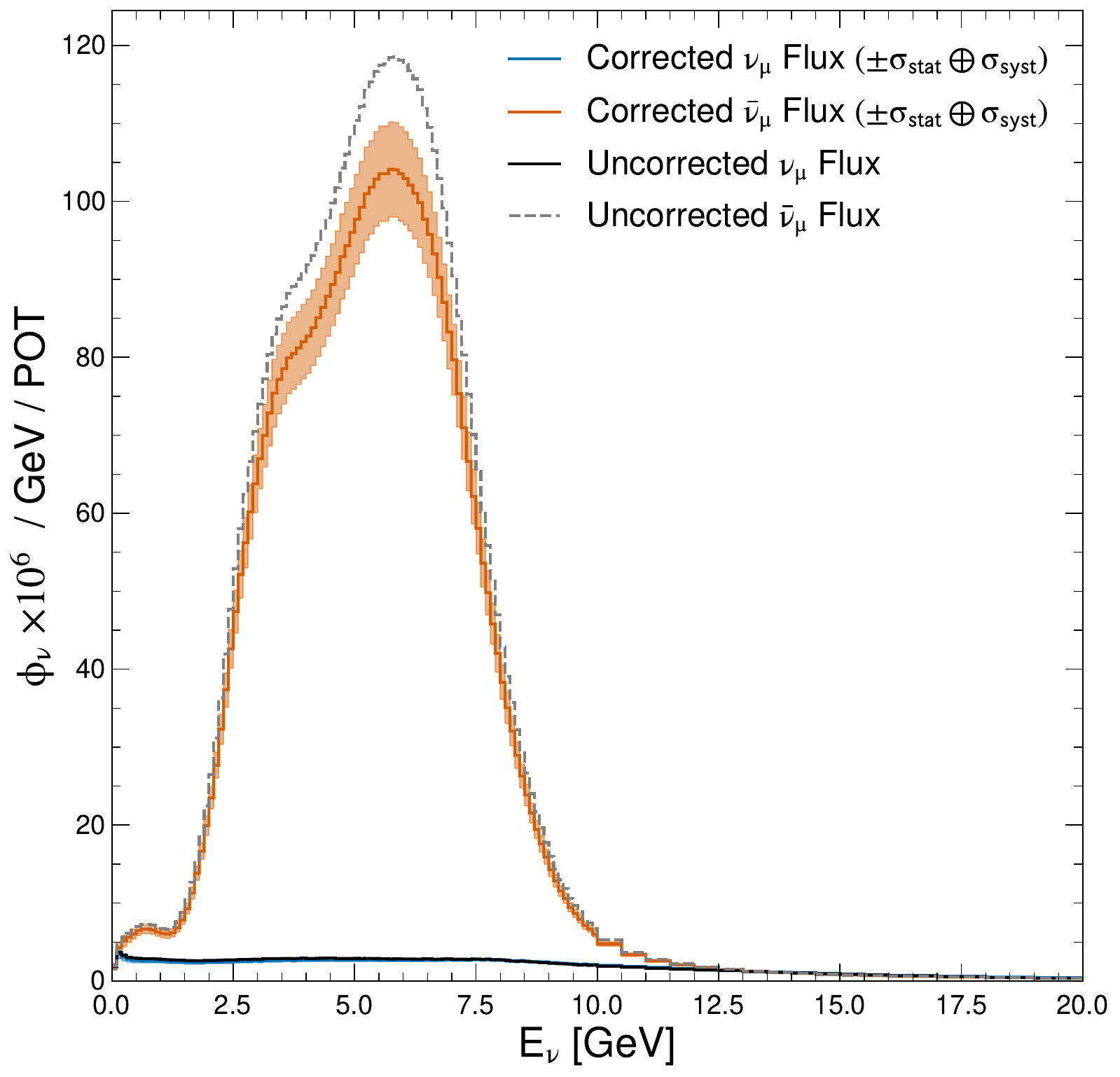}
        \caption{\numub\ and \numu}
    \end{subfigure}
    \begin{subfigure}{0.49\textwidth}
        \includegraphics[width=\textwidth]{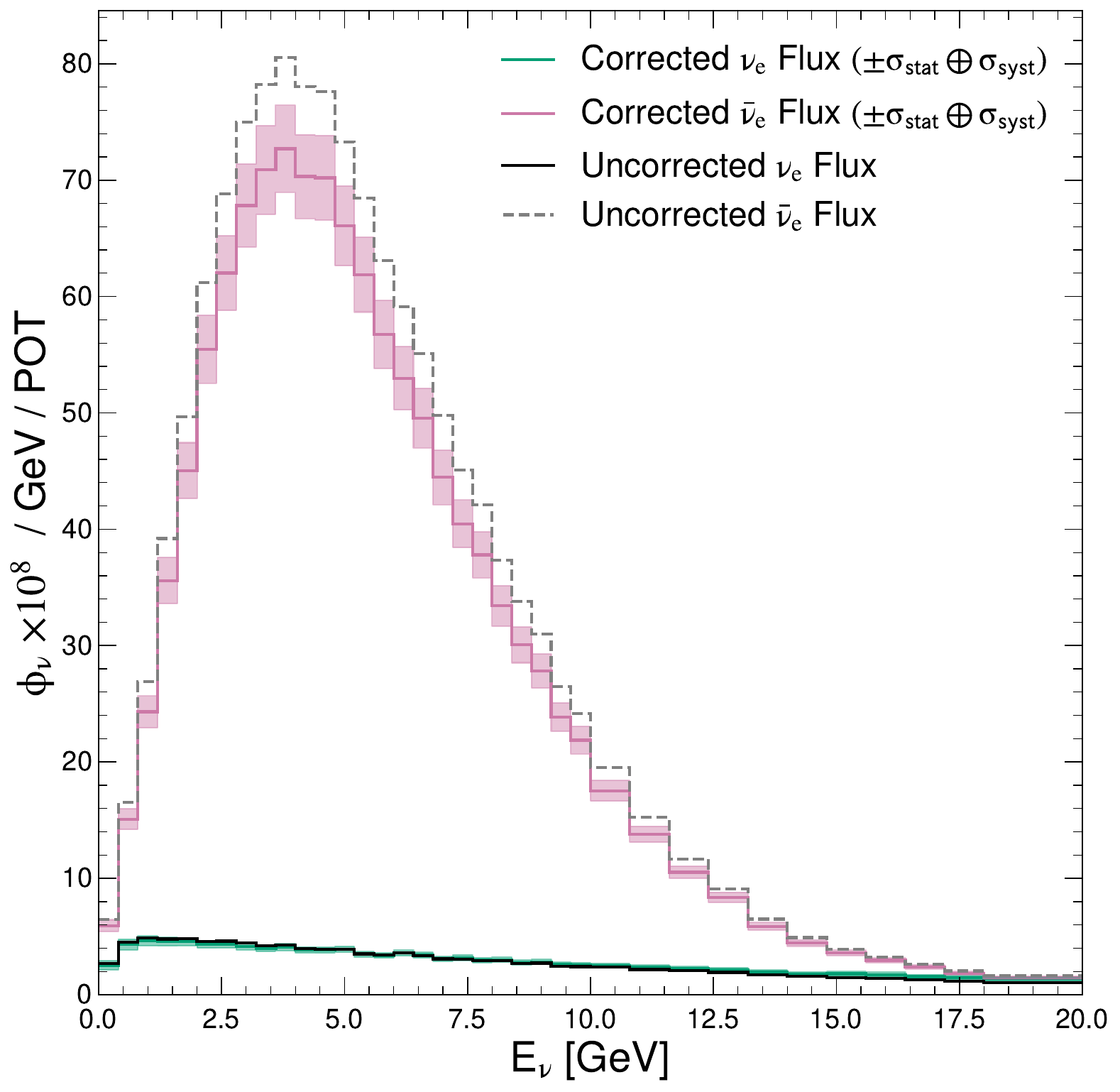}
        \caption{\nueb\ and \nue}
    \end{subfigure}
    \caption[On-Axis Flux vs. \Enu\ Spectra]{The \acrshort{ppfx}-corrected on-axis flux predictions with associated statistical and hadron production uncertainties. The \numub\ flux peaks at $\approx \SI{6}{\GeV}$, while \nueb\ peaks between $3.5-\SI{4}{\GeV}$.}%
    \label{fig:2x2_flux_prediction}
\end{figure}

Following the methodology outlined in Chapter~\ref{sec:analysis} applied to a 1-billion \pot\ \numi\ flux sample prepared using the Geant4.10.4-based simulation as described in Section~\ref{sec:newg4}.
The \acrshort{ppfx}-corrected on-axis flux prediction for the \acrshort{2x2} prototype is shown in Figure~\ref{fig:2x2_flux_prediction}.
A low-contamination flux, compared to the off-axis case shown previously, is expected with peaks at $\approx \SI{6}{\GeV}$ for \numub\ and $3.5-\SI{4}{\GeV}$ for \nueb{}.

\begin{figure}[htbp]
    \centering
    \begin{subfigure}{0.49\textwidth}
        \includegraphics[width=\textwidth]{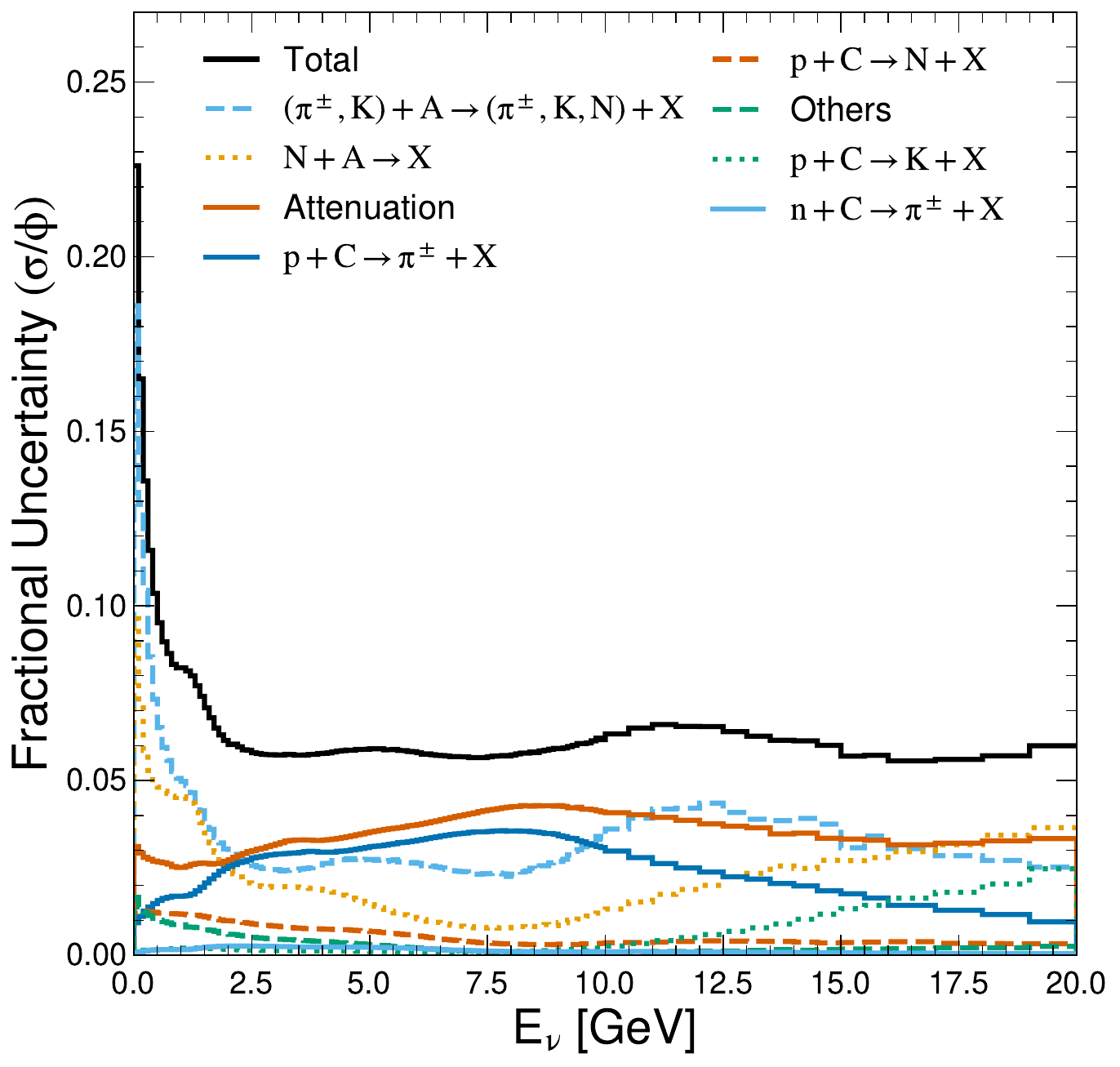}
        \caption{\numub}
    \end{subfigure}
    \begin{subfigure}{0.49\textwidth}
        \includegraphics[width=\textwidth]{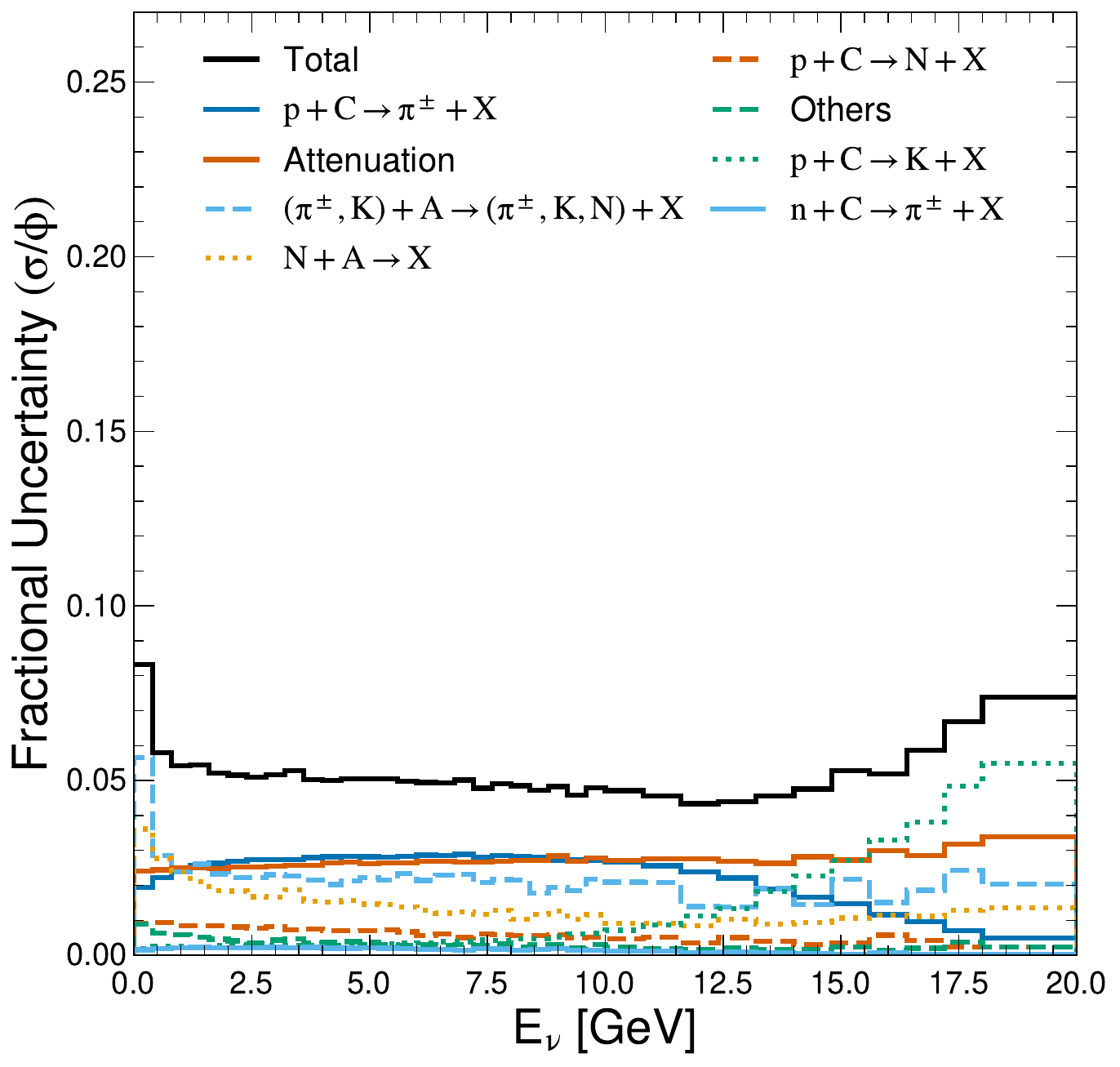}
        \caption{\nueb}
    \end{subfigure}
    \caption[On-Axis Hadron Production Uncertainties]{The fractional hadron production uncertainties for \numub\ and \nueb\ as a function of neutrino energy. The largest sources of uncertainty in the peak regions are from attenuation of the absorption cross section of the neutrino's parent hadrons based, following by pion production in the target.}%
    \label{fig:2x2_hadron_uncertainties}
\end{figure}

Figure~\ref{fig:2x2_hadron_uncertainties} shows the fractional hadron production uncertainties for \numub\ and \nueb\ as a function of neutrino energy.
In the peak regions, the hadron production systematic uncertainty is at the 6--7\% level for \numub\ and 5--6\% for \nueb{}.
Across either species, the largest sources of uncertainty in the peak regions are from attenuation of the absorption cross section of the neutrino's parent hadrons, followed by pion production in the target.
The former describes a correction factor applied to model predictions accounting for attenuation of the proton inelastic cross section as it traverses the target region.
It also includes corrections on the hadron absorption cross sections for (grand-) parent mesons in various materials.
For non-carbon media, the uncertainty on the correction can be large, see~\cite{aliaga_neutrino_2016} for more details.

\begin{figure}[htbp]
    \centering
    \includegraphics[width=0.8\textwidth]{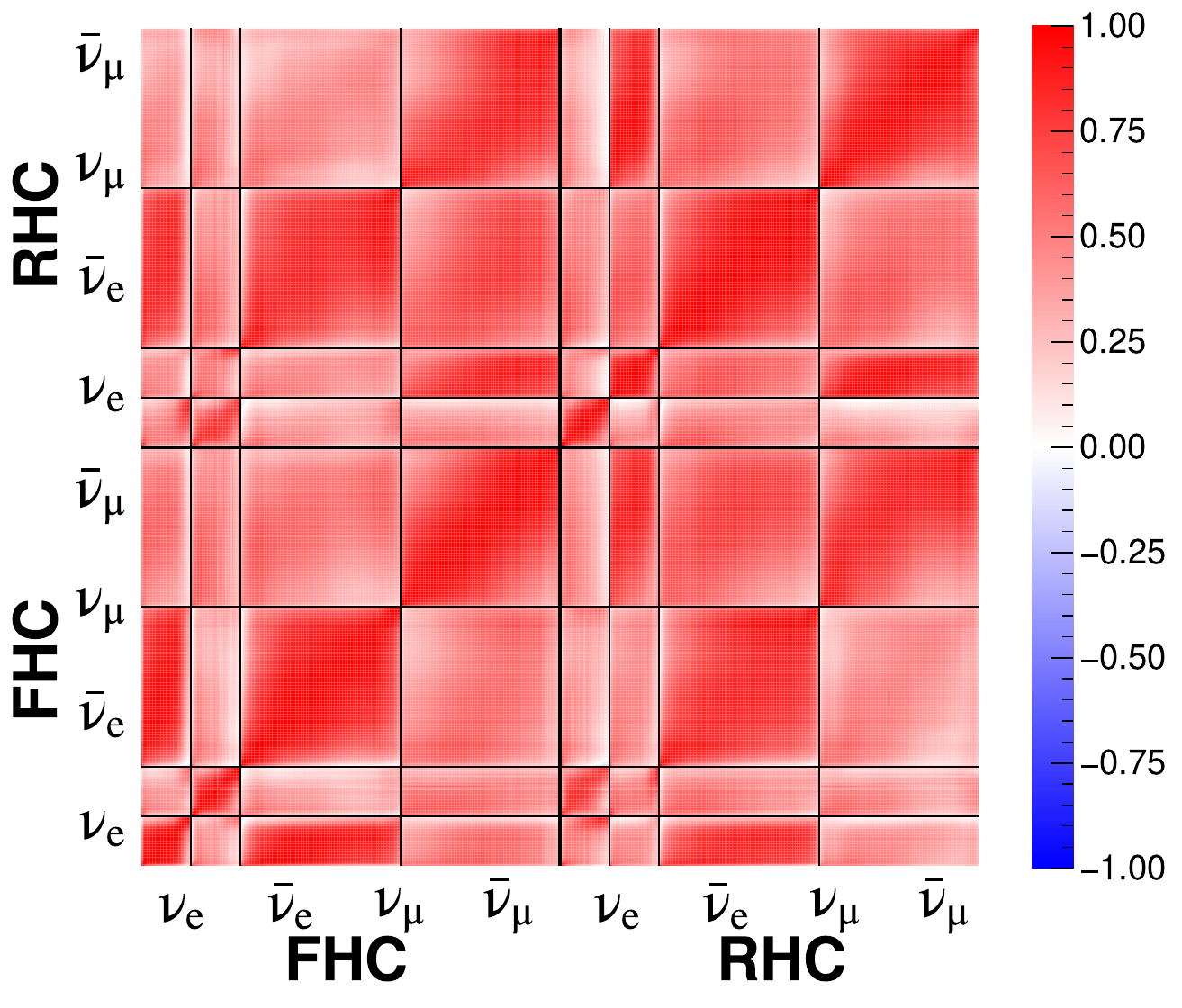}
    \caption[On-axis Correlation Matrix]{The total correlation matrix for the hadron production uncertainties. The neutrino flavor-energy space is positively correlated everywhere, with the largest correlations existing between \nue\ and \numu\, independent of horn operation }%
    \label{fig:2x2_hadron_correlation}
\end{figure}

Figure~\ref{fig:2x2_hadron_correlation} shows the total correlation matrix for the hadron production uncertainties.
The neutrino flavor-energy space is positively correlated everywhere, with the largest correlations existing between \nue\ and \numu, independent of horn operation.
Finally, Table~\ref{tab:2x2_integrated_uncertainties} shows the hadron production systematic uncertainties integrated between 0--\SI{20}{\GeV}.
Of particular revelance to \acrshort{2x2}, are the \rhc\ \numub\ and \nueb\ uncertainties, which total 5.62\% and 4.86\%, respectively.

\begin{table}[htbp]
    \centering
    \caption[On-Axis Integrated Uncertainties]{Hadron production systematic uncertainties integrated between 0--\SI{20}{\GeV}.%
     Though \fhc are calculated, here, there are no current plans to collect data in this operating mode at \acrshort{2x2}.}
    \begin{tabular}{lrrrrrrr}
    \toprule
    & $\nu_e$ & $\bar{\nu}_e$ & $\nu_e + \bar{\nu}_e$ & $\nu_\mu$ & $\bar{\nu}_\mu$ & $\nu_\mu + \bar{\nu}_\mu$ & $\frac{\nu_e + \bar{\nu}_e}{\nu_\mu + \bar{\nu}_\mu}$ \\
    \midrule
    FHC & 4.80 & 5.95 & 4.69 & 5.82 & 6.61 & 5.75 & 1.75 \\
    RHC & 6.35 & 4.86 & 4.67 & 6.37 & 5.62 & 5.53 & 1.52 \\
    \bottomrule
    \end{tabular}%
    \label{tab:2x2_integrated_uncertainties}
\end{table}

%% file: trigger_emulation.tex
\chapter{Improving the ICARUS Trigger Emulation}\label{sec:triggeremu}
During \icarus\ data-taking, the hardware trigger is responsible for initiating a read-out of the state of all detector components in response to photonic activity within the detector active volume if the activity is both concurrent with a beam spill and above some preconfigured threshold.
Waveforms are digitized on an ``electronics'' timescale relative to the firing of the hardware trigger, where $t_0$ corresponds to the firing time of the trigger in microseconds.
In simulation, generated products, are placed within a time window that matches the width of the beam gate, i.e., \SI{2.2}{\micro\second} for \acrshort{bnb} and \SI{9.5}{\micro\second} for \numi. In contrast to data, $t_0$ in simulation corresponds to the beginning of this window.
An issue arises, in simulation, as the digitization and reconstruction stages expect waveforms and downstream products to be on the electronics timescale, rather than the beam-window timescale.
This leads to a progressively worsening reconstruction as interactions occur farther out into the window.
For \numi\ \acrshort{mc} productions, in particular, 10\% of all muon tracks are reconstructed with underestimated momenta as portions of these tracks appear out-of-time with respect to the beam spill due to the timing discrepancy.

An emulated trigger was developed for \icarus{}, but had so far not been fully integrated into the simulation as of January 2024.
This work devised a remediation within the \icarus\ simulation procedure, which uses information generated by the emulated trigger to apply a shift to the timestamps of generated products from the “beam-window” timescale to the “electronics” one.
Directly modifying simulation products had to be done with careful consideration to the downstream processing, and experts on the various subsystems of the detector were consulted at various points during development to ensure the integrity of the reconstruction was maintained.
Furthermore, due to the scope and sensitivity of these changes, extensive validation of the end-to-end simulation was performed.

To shift simulated products, a new ART module was developed in the detector simulation stage after the initial run of the emulated trigger, which is required to initialize the trigger and beam gate data products.
The module reads both the emulated trigger and the beam gate data products, calculates the shift required to align the beam gate with the electronics timescale, and applies this shift to the timestamps of the following products:
\begin{enumerate}
\item simulated energy deposits,
\item simulated photons,
\item the beam gate data product itself,
\item waveforms from the optical detector simulation,
\item and simulated depositions into the \acrfull{crt} modules.
\end{enumerate}
Afterward, the trigger emulation is re-run to set the new timescale, and then proceeds to the digitization and reconstruction stages.
To ensure backward compatibility with existing analyses that depend on the original timescale, the initial trigger information is preserved and propagated to the \acrfull{caf} in addition to the new trigger information.

\begin{figure}[htbp]
	\centering
	\includegraphics[width=0.8\textwidth]{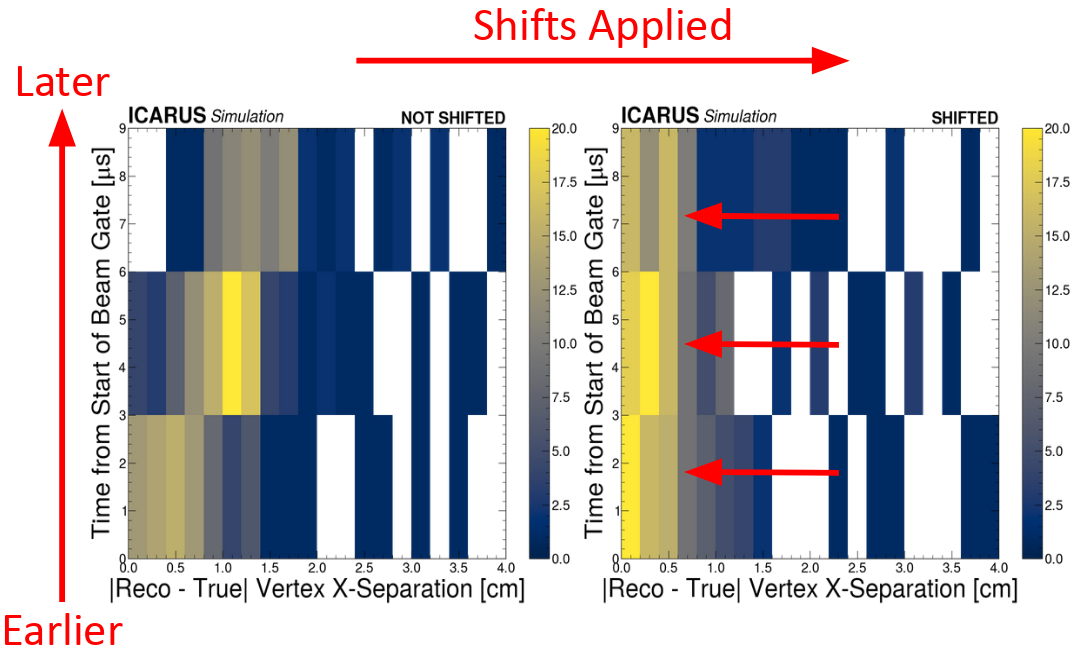}
	\caption[Vertex Reconstruction Comparison with and without Trigger Shifts]{Comparison of the difference between true and reconstructed vertex placements within the \numi\ beam window, divided into \SI{3}{\micro\second} bins, before and after applying shifts based on the emulated trigger. After shifts are applied, the vertex placement is within \SI{0.5}{\cm} of the true vertex position across the beam window.}%
	\label{fig:trigger_shifts}
\end{figure}

To evaluate the performance of the module, two \acrshort{mc} file sets of 1,500 \numu{}--\acrshort{cc} events, each, were produced, with and without the new module enabled.
The difference between the true and reconstructed $x$-coordinate of the vertex placements within the \numi\ beam window was calculated and divided into \SI{3}{\micro\second} bins.
In \icarus{}, the $x$ spatial dimension is parallel to the applied electric field and transverse to the \acrshort{bnb} axis.
The reconstruction calculates the $x$-coordinate of objects in the detector using the drift time of ionization with respect to the trigger time, as the drift velocity is a known quantity ($v_D \approx 1.6 \si{\mm/\micro\second}$ with $E_D = 500 \si{V/\cm}$).
As such, the integrity of the reconstruction is particularly sensitive to inaccuracies in the timing of the trigger.
Figure~\ref{fig:trigger_shifts} shows this difference before and after applying shifts.
Beforehand, the vertex placement is time-dependent, increasing linearly with the time from the start of the beam gate, as the events appear more out-of-time and the discrepancy widens.
After shifts are applied, the vertex placement is within \SI{0.5}{\cm} of the true vertex position across the beam window, effectively eliminating the time-dependence of vertex placement efficiency.

\begin{figure}[htbp]
	\centering
	\includegraphics[width=0.7\textwidth]{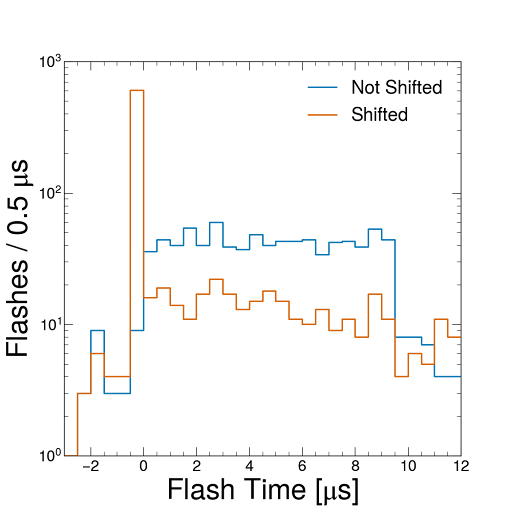}
	\caption[Reconstructed PMT Flash Times]{Reconstructed flash times in the beam (blue) and electronic (orange) timescales. The shifted flash times do not retain information about their timing with respect to the beam spill, interfering with event selections made using this quantity in existing analyses.}%
	\label{fig:flash_shifts}
\end{figure}

Some analyses require that the flash times from the optical reconstruction be on the beam-window timescale.
In Figure~\ref{fig:flash_shifts}, the difference between the true and reconstructed flash times before and after applying shifts is shown.
The shifted flash times do not retain information about their timing with respect to the beam spill.
In contrast, the unshifted distribution shows a clear distinction between in-time (between 0--\SI{9.5}{\micro\second}) and out-of-time optical activity by an order of magnitude.
To accommodate this, the timestamps of the flash times were reverted to the previous timescale during the \acrshort{caf}-making stage.

%% file: conclusion.tex
\chapter{Conclusions}
Neutrinos from the \numi\ beam provide \icarus\ with a unique means for performing high-statistics $\nu-\mathrm{Ar}$ cross section measurements in a kinematic region that is beneficial for future \acrshort{dune} long-baseline neutrino oscillation analyses.
While the \numi\ flux in the on-axis and near-on-axis regions had been well-characterized and validated, this far-off-axis region had not been previously studied in-depth.
This thesis presented a comprehensive study of the \numi\ flux for \icarus{}'s highly off-axis position at \SI{100.1}{\milli\radian}, which is a necessary first step toward these measurements.
As described in Chapter~\ref{sec:numi_ppfx}, Monte Carlo simulation of the NuMI flux was performed using the \geant toolkit, and \ppfx was used to apply a correction to the simulation and propagate uncertainties due to the hadron interaction modeling.
It was discovered that the \numi{}-\icarus\ flux is sensitive to effects specific to the off-axis location that are either subdominant or not present in the on-axis region, such as elevated contamination of the flux from the antineutrino modes,
increased frequency of hadron re-interaction in the \numi\ beam structure as well as a larger contribution to the flux from interactions in $x_F-p_T$ phase space not currently covered by existing hadron production data.

\section{The NuMI Simulated Flux Productions}
The original 500M POT sample from \NOvA was sufficient to study the impact of hadron model uncertainties initially, but sample statistics limited the capability to study the beamline uncertainties, especially above 3 GeV.
A larger statistics sample was generated to study geometry and model changes described in the final sections of Chapter~\ref{sec:analysis}, and the hadron production studies were updated accordingly.
However, at the time of writing, the beamline uncertainties have not been fully re-evaluated with the larger flux sample.

The simulations were performed for the \SI{700}{\kilo\watt} \numi configuration used between 2014--2018, while the NuMI beam operates in a 1 MW configuration since 2019.
The study presented in Section~\ref{sec:megawatt} did not reveal substantial differences between the two beam configurations, relative to available statistics.
It was recently discovered that some geometry components of the \SI{1}{\mega\watt} target design may not have been successfully included, and therefore a new set of \SI{1}{\mega\watt} files, including FHC and RHC files along with the full suite of beamline/focusing systematically altered geometries should be generated for the future studies once this issue is resolved.
Experimental uncertainties on the POT counting (Main Injector beam intensity measurements) are not covered by the simulation and must be included separately in the analysis based on POT counting uncertainties from beamline operations.

\section{Characteristics of High Off-axis Angle Neutrino Beam}
It was determined that more than 70\% of neutrinos above 0.5 GeV travelling to \icarus\ originate from decays very close to the NuMI target, see Sec.~\ref{sec:offaxis_flux}.
Hadrons undergo significantly more interactions before decaying to neutrinos travelling to \icarus\, in comparison to on-axis experiments, making the flux more sensitive to hadron model uncertainties.
Significant contribution of the wrong sign neutrino flux is present in \icarus\ for both forward and reverse horn current configurations.

\section{Hadron Model Uncertainties}
The NuMI neutrino flux systematic uncertainties are currently driven by hadron interaction uncertainties, as discussed in section~\ref{sec:had-uncerts}.
The dominant contribution to the hadron model uncertainties were found to be the interactions of kaons and charged pions at 5–\SI{40}{GeV/c} both in and outside of the target region.
Improvements to the underlying hadron interaction modeling are expected as a result of measurements being made by NA61\slash{}SHINE and EMPHATIC.
As described in Section~\ref{sec:na_bugfix}, newly published data from NA61 for proton-carbon interactions at \SI{120}{\GeV/c} could be implemented in \ppfx\ to constrain uncertainties especially for interactions with $-0.25 \leq x_F < 0$.

The flux of \nue\, \nueb\, \numu\ and \numub\ in ICARUS from the NuMI beam line in forward and reverse horn current polarities was estimated.
An analysis of the uncertainties related to modeling of hadron interactions and the beamline, as well as correlations between the flux of the various neutrino modes, was performed. The total uncertainty on the flux, accounting for statistical and systematic effects, was found to be $8.90\%$ and $11.00\%$ for focused \numu and \nue, respectively.
In the context of initial analyses at \icarus\ with the \numi\ beam, the combined $\numu+\numub$ flux uncertainty is of particular interest and was found to be $10.74\%$ in \fhc operation (relevant for Run 1 and Run 2 data sets) and $11.06\%$ in \rhc operation (relevant for Run 3 data).
The uncertainty on the flux ratio was also studied and calculated as 7.72\% in the forward horn operation, and 7.94\% in the reverse horn operation.
As such, performing ratio measurements of the neutrino interaction cross section on argon nuclei does not grant a significant advantage over other methods.
However, the integrated uncertainty due to hadron interaction modeling was discovered to be lower than initially expected.

The on-axis flux was re-evaluated for \acrshort{2x2} at high-statistics with the new hadronic model and geometry updates, and the flux was found to be consistent with the previous results within uncertainties~\cite{Aliaga:2016oaz}.
In general, the integrated flux uncertainties were found to be at the level of $\sim 5-6\%$ for all neutrino species and horn polarities.

See Ref.~\cite{FluxAnalysis, sbndata} to access the analysis code and a file containing these results.

\section{First NuMI-ICARUS Analyses}
There are several cross section analyses~\cite{howard_numi_offaxis_2024} that are currently ongoing at \icarus\ using the NuMI beam as well as Beyond Standard Model searches, one of which is pending publication~\cite{icarus_dimuon_2024}, into which the results of this thesis are a direct input.

\begin{figure}[htbp]
    \centering
    \includegraphics[width=0.49\textwidth]{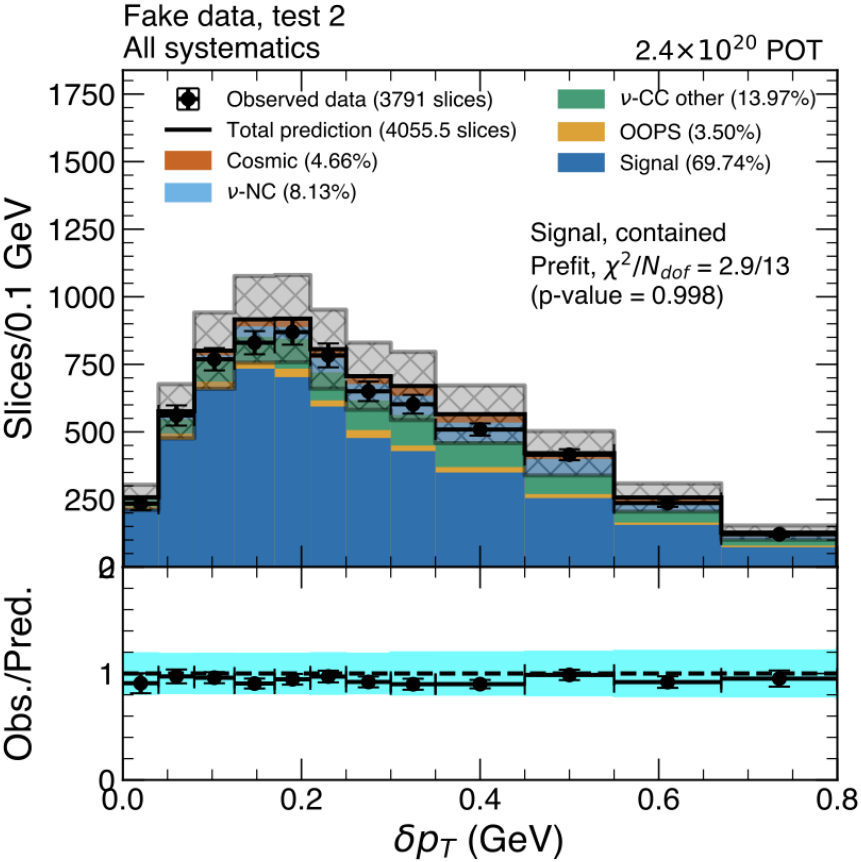}
    \includegraphics[width=0.49\textwidth]{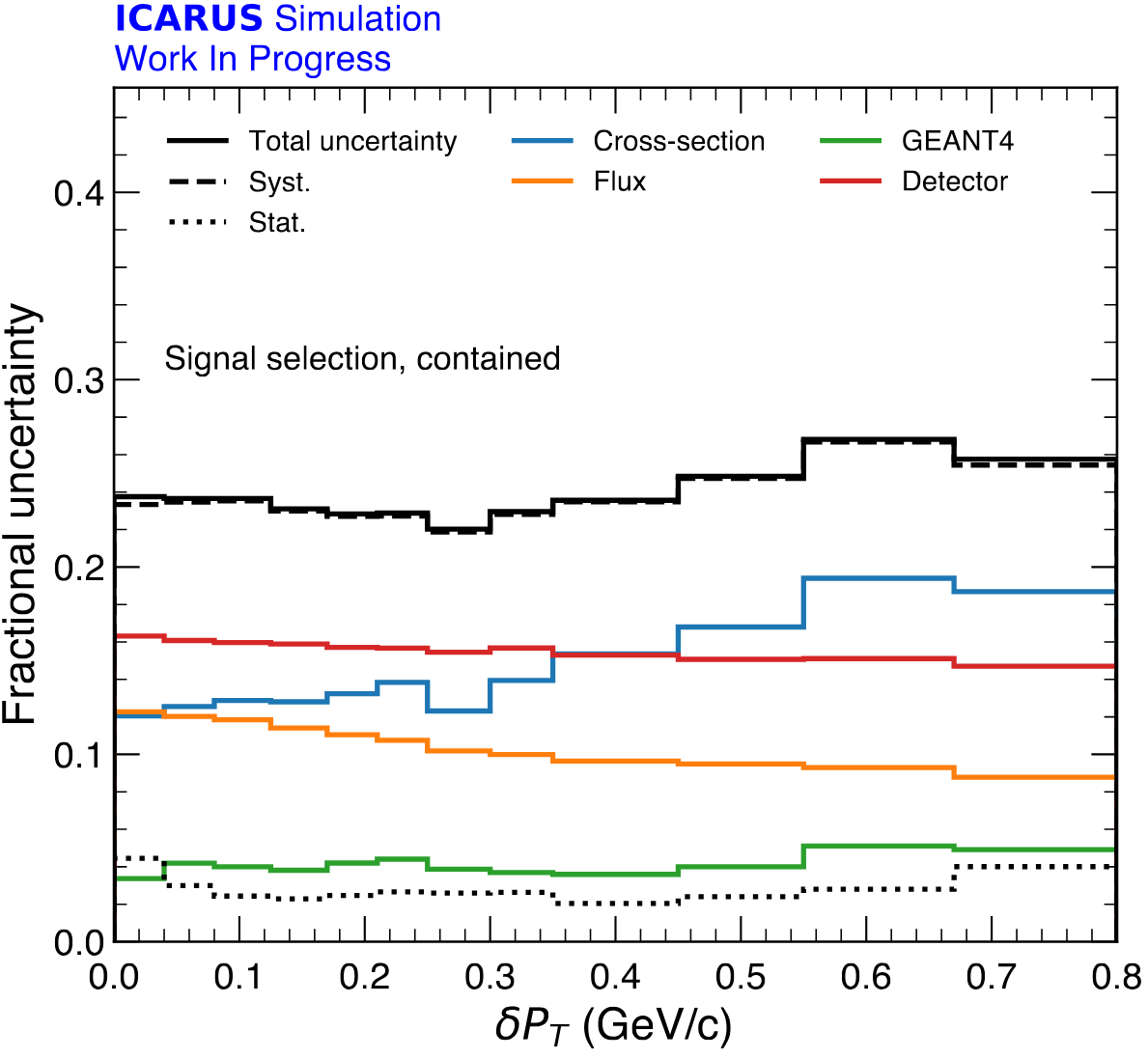}
    \caption[$\nu_\mu \mathrm{CC} 1\mu \mathrm{N}p0\pi$ Fake Data Study]{$\nu_\mu$ \acrshort{cc}$1\mu\mathrm{N}p0\pi$ fake data study uncertainties (left) and uncertainty composition (right) in terms of $\delta p_T$. \ppfx\ weights are included. Flux uncertainties calculated in this thesis are shown on the right in orange. Figures adapted from \citeauthor{howard_numi_offaxis_2024} (\citeyear{howard_numi_offaxis_2024}).}%
    \label{fig:tki_selection}
\end{figure}

The active cross section analyses are focused on the $\nu_\mu$ and $\bar{\nu}_\mu$ charged current quasi-elastic (CCQE) interactions on argon with one or more protons in the final state.
These analyses are largely being performed in terms of the \acrfull{tki}~\cite{PhysRevC.94.015503}, parameterized by the transverse momentum, $\delta p_T$, and transverse angle, $\delta \alpha_T$, imbalance between the outgoing lepton and hadrons.
This transverse kinematic system provides a useful probe of nuclear effects in neutrino interactions in a way that is minimally dependent on the neutrino energy and nuclear momentum, both of which are not directly measurable quantities and typically inferred from the reconstructed final state particles.
As final states including hadrons, in this case protons, are produced in within a nucleus, the final state particles are subject to \acrfull{fsi}, prior to exiting the nucleus, that can distort the kinematics of the interaction.
The transverse momentum imbalance, $\delta p_T$, quantifies the amount of transverse momentum carried by the initial hadronic system, while the transverse angle, $\delta \alpha_T$, yields insight into the amount of that momentum was lost to \acrshort{fsi}.
\acrshort{fsi} modeling is a significant source of uncertainty, and the \acrshort{tki} variables provide a model-independent framework in which to study these effects.
Additionally, \acrshort{tki} analyses carry the added benefit of being generally independent of the precise neutrino event rate, i.e., they are robust against normalization effects in the flux uncertainties.
Figure~\ref{fig:tki_selection} shows the selection of $\nu_\mu$ \acrshort{cc}$1\mu\mathrm{N}p0\pi$ events with respect to $\delta p_T$.
This selection incorporates flux weights and \numi\ flux uncertainties calculated in this thesis, the latter of which is shown to be at the $\sim 10\%$ level, subdominant to both cross section and detector modeling systematic uncertainties.

\begin{figure}[htbp]
    \centering
    \includegraphics[width=0.6\textwidth]{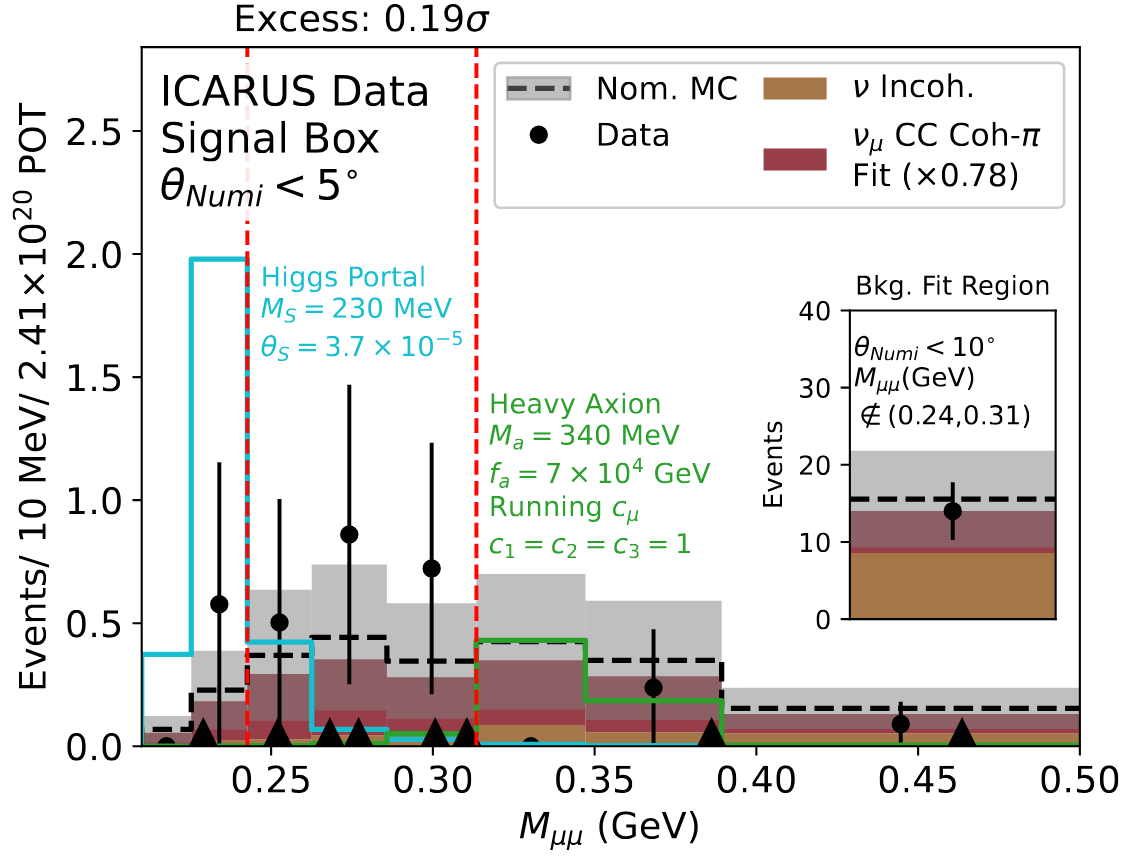}
    \caption[LLP Search Di-muon Invariant Mass]{Signal box result of the LLP to di-muon search. Spectra include flux weights and uncertainties. Figure adapted from \citeauthor{icarus_dimuon_2024} (\citeyear{icarus_dimuon_2024}).}%
    \label{fig:bump_hunt}
\end{figure}

Results of the long-lived particle (LLP) search are shown in Figure~\ref{fig:bump_hunt}, where a bump hunt was performed in the invariant mass of the di-muon final state produced from two possible beyond standard model (BSM) processes.
The two processes describe the decay of $K^\pm$ or $K^0_L$ to either a Higgs portal scalar or an axion-like particle, which subsequently decays to two muons.
The former is a model of dark matter, while the latter is proposed as a solution to the strong CP problem.
This analysis is highly dependent on an accurate and precise prediction of kaon production in the \numi\ beam as well as the flux to constrain the coherent pion background.
To that end, a set of weights were calculated specifically for this analysis to account for the geometry hadron production model changes described in Sections~\ref{sec:missing_geom}~and~\ref{sec:newg4} in the context of the kaon contribution to the flux, in particular.
Similar to the \acrshort{tki} analysis, flux weights and uncertainties are included, here.

These analyses represent the first of many that will be performed at \icarus\ using the \numi\ beam, as pion final state and \nue\ analyses begin to take shape, and in each case, the flux prediction presented in this thesis are a crucial input.

%% file: bibliography.tex
\begin{singlespace}
	\printbibliography[heading=bibintoc,title=Bibliography]
\end{singlespace}

%% file: appendices.tex
\appendix
\chapter{PPFX Universe Normality Study}\label{sec:gaussian_fits_to_universes}
\input{gaussian_fits_to_universes}

\chapter{NuMI ICARUS Flux Prediction}\label{sec:appendix-flux-prediction}
\input{flux_prediction}

\chapter{Hadron Fractional Uncertainties}\label{sec:appendix-had-uncerts}
\input{hp_uncertainties}

\chapter{Hadron Production Matrices}\label{sec:appendix-had-covs}
\input{hp_covariance_matrices}

\chapter{Principal Component Analysis}\label{sec:appendix_pca}
\input{pca_variance_plots}

\chapter{NuMI Beamline Monte Carlo Samples}\label{sec:appendix_beam_samples}
\input{beam-samples-descriptions}

\chapter{Beam Focusing Systematic Variations}\label{sec:appendix-beam_frac_shifts}
\input{beam_run_shifts}

\chapter{Beam Focusing Fractional Uncertainties}\label{sec:appendix-beam_frac_uncerts}
\input{beam_frac_uncertainties}

\chapter{Beamline Focusing Systematic Covariance Matrices}\label{sec:appendix_beam_matrices}
\input{beam_correlation_matrices}

\chapter{Differences Between the 700 kW and 1 MW NuMI Beamline Geometries}\label{sec:appendix_megawatt_upgrade}
\input{1megawatt_upgrade_differences}

\chapter{Total Flux Uncertainties}\label{sec:appendix-total-uncertainties}
\input{total_uncertainty}

\chapter{Parent Decay Momenta}\label{app:parent_decay_momenta}
\input{parent_decay_momenta}
\clearpage

\chapter{Parent Decay Momentum Angles}\label{app:parent_decay_angles}
\input{parent_decay_angles}
\clearpage

\chapter{Neutrino Energy vs. Parent Decay Momentum Angle}\label{app:energy_vs_angles}
\input{energy_vs_angles}
\clearpage

\chapter{Flux File README}
\input{flux_file_readme}

%% file: gaussian_fits_to_universes.tex
\clearpage
\section{Forward Horn Current}
\begin{figure}[!ht]
    \centering
    \includegraphics[width=0.3\textwidth]{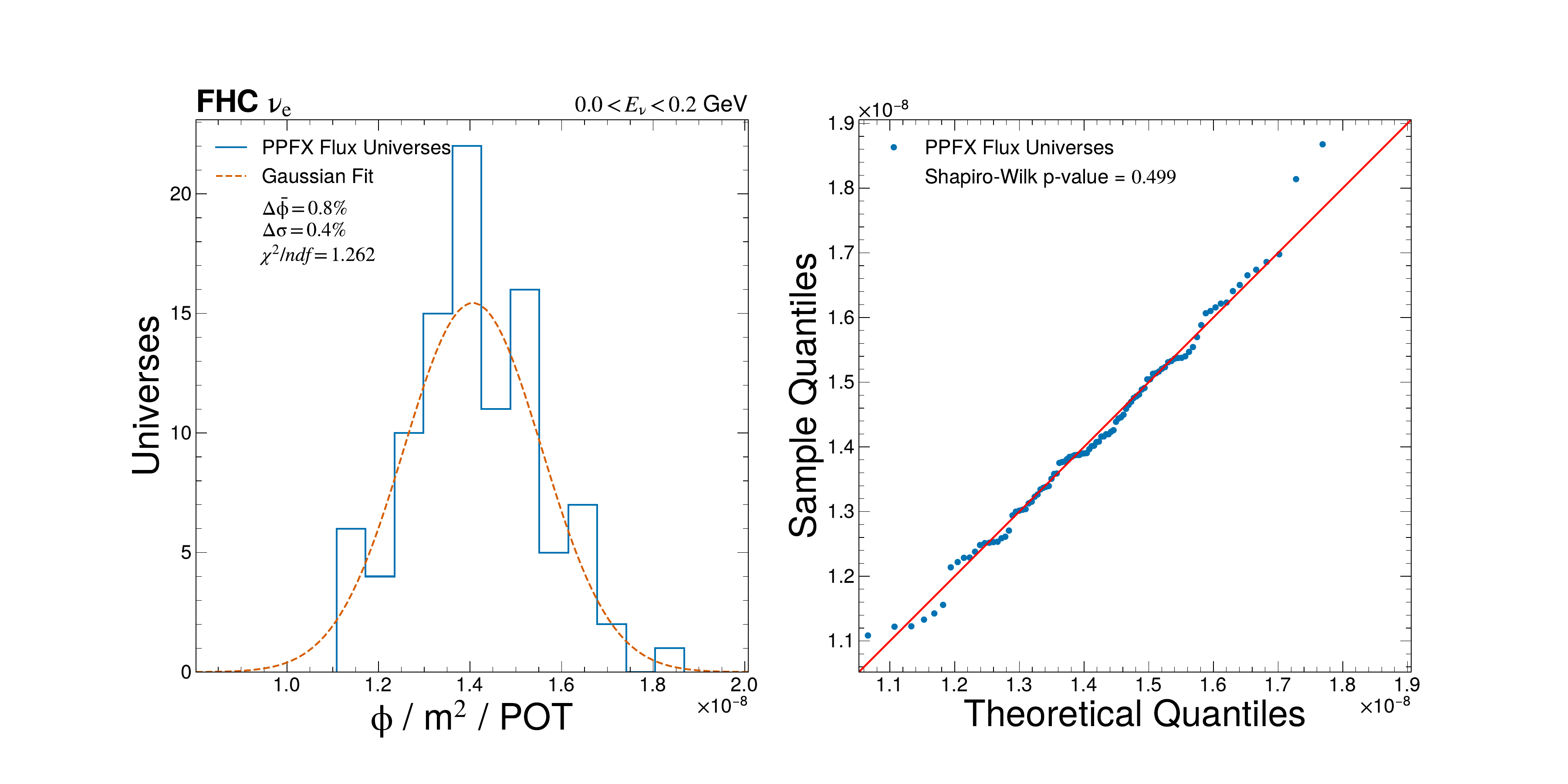}
    \includegraphics[width=0.3\textwidth]{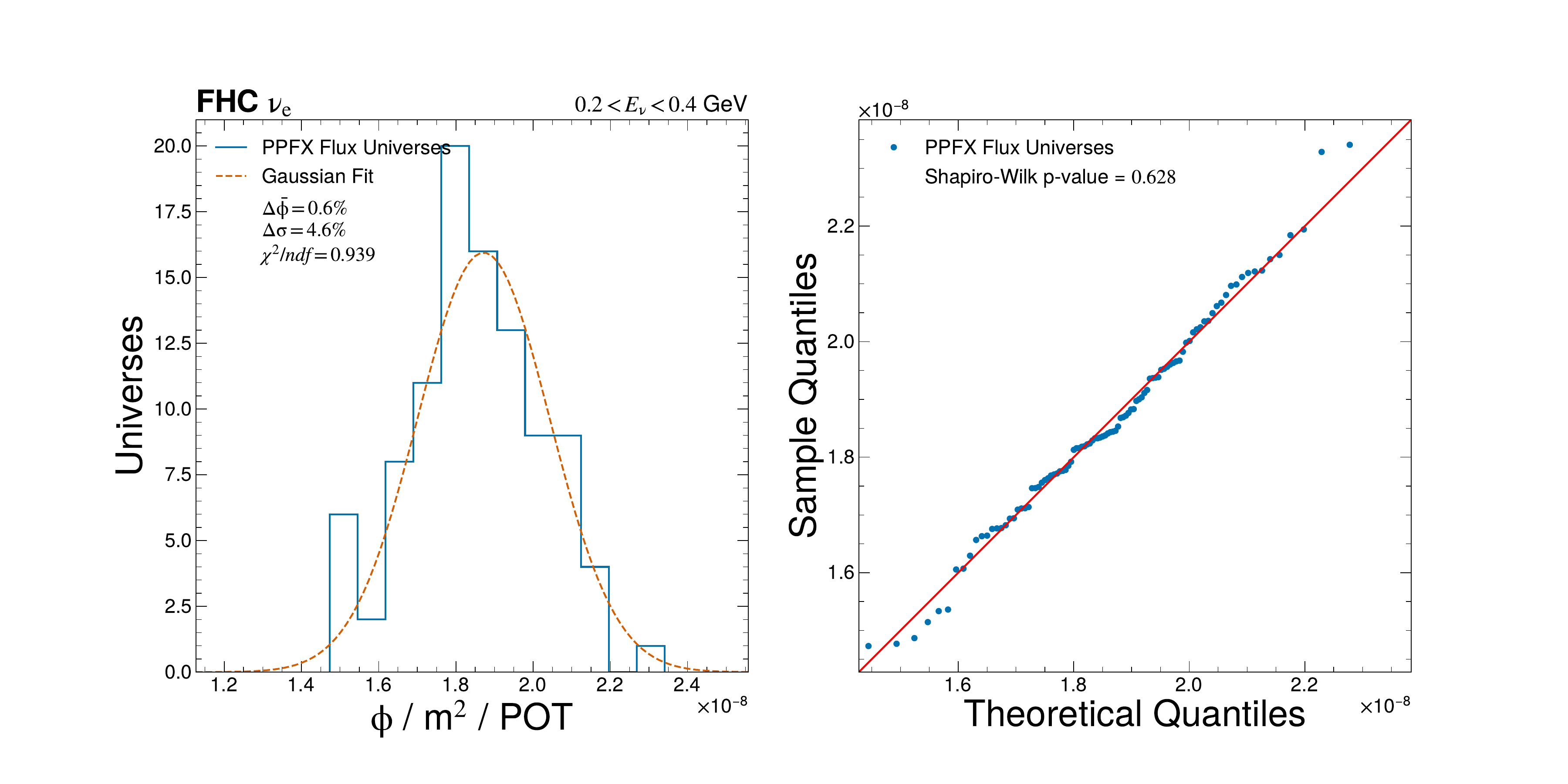}
    \includegraphics[width=0.3\textwidth]{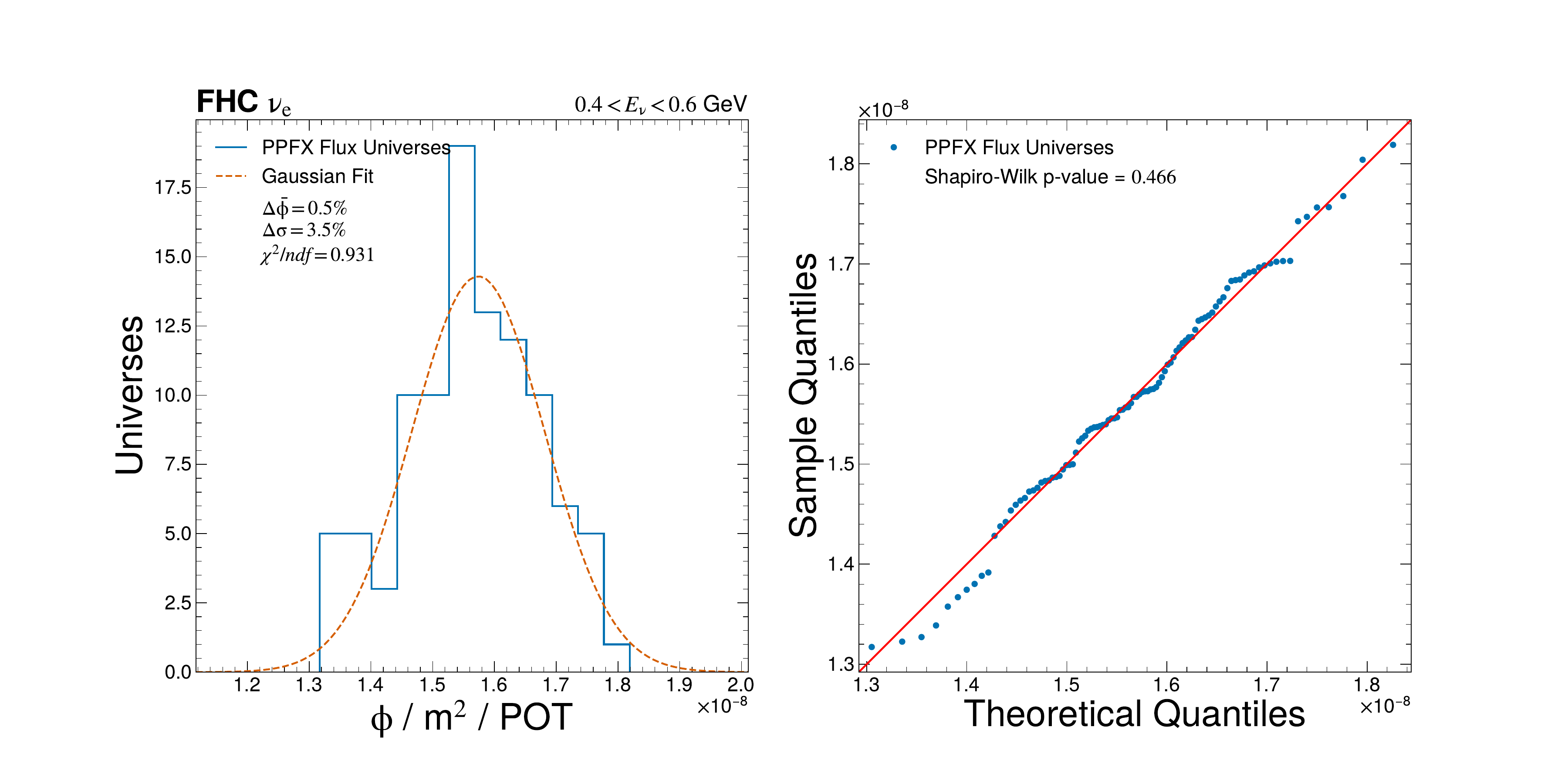}
    \includegraphics[width=0.3\textwidth]{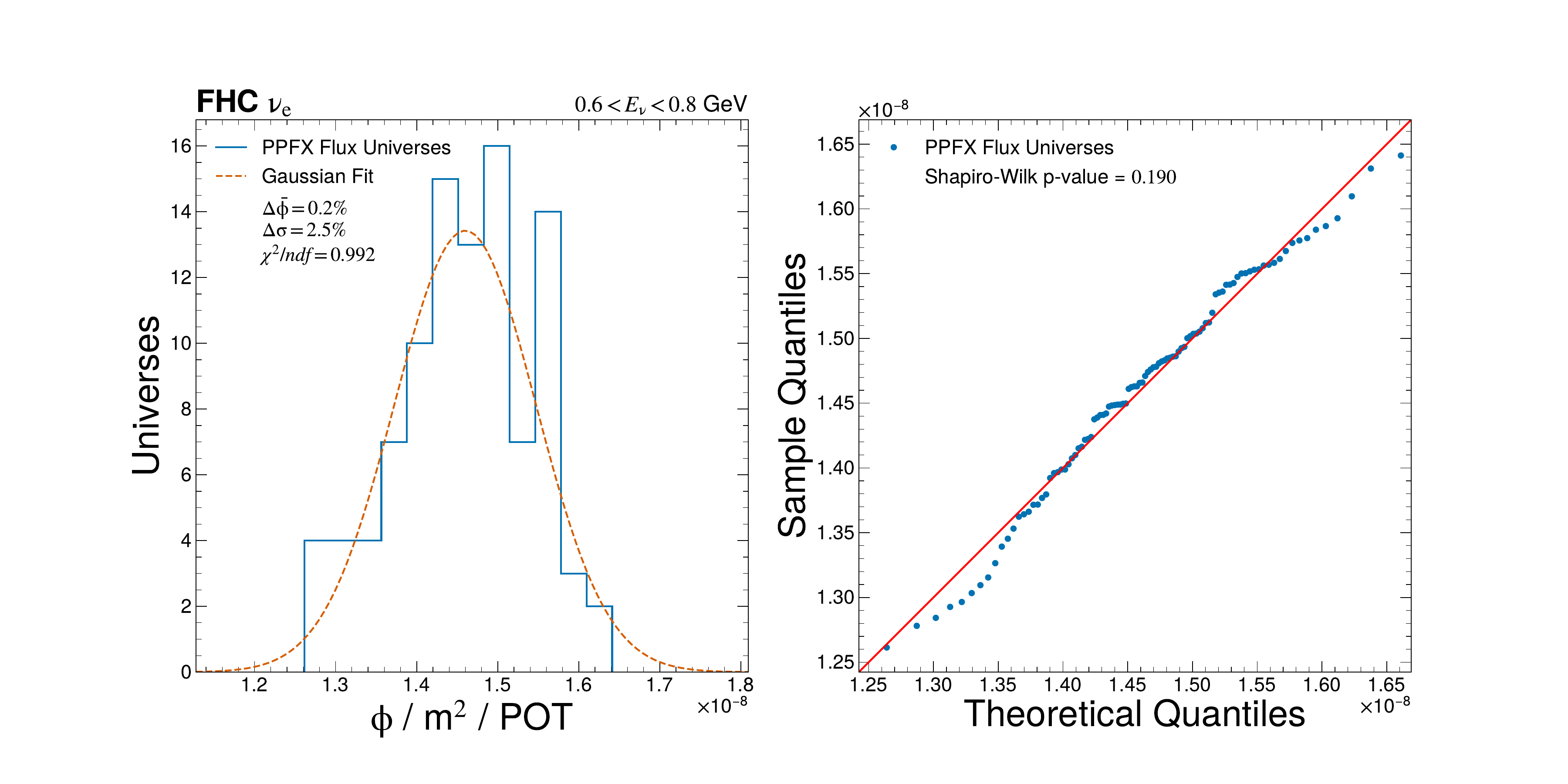}
    \includegraphics[width=0.3\textwidth]{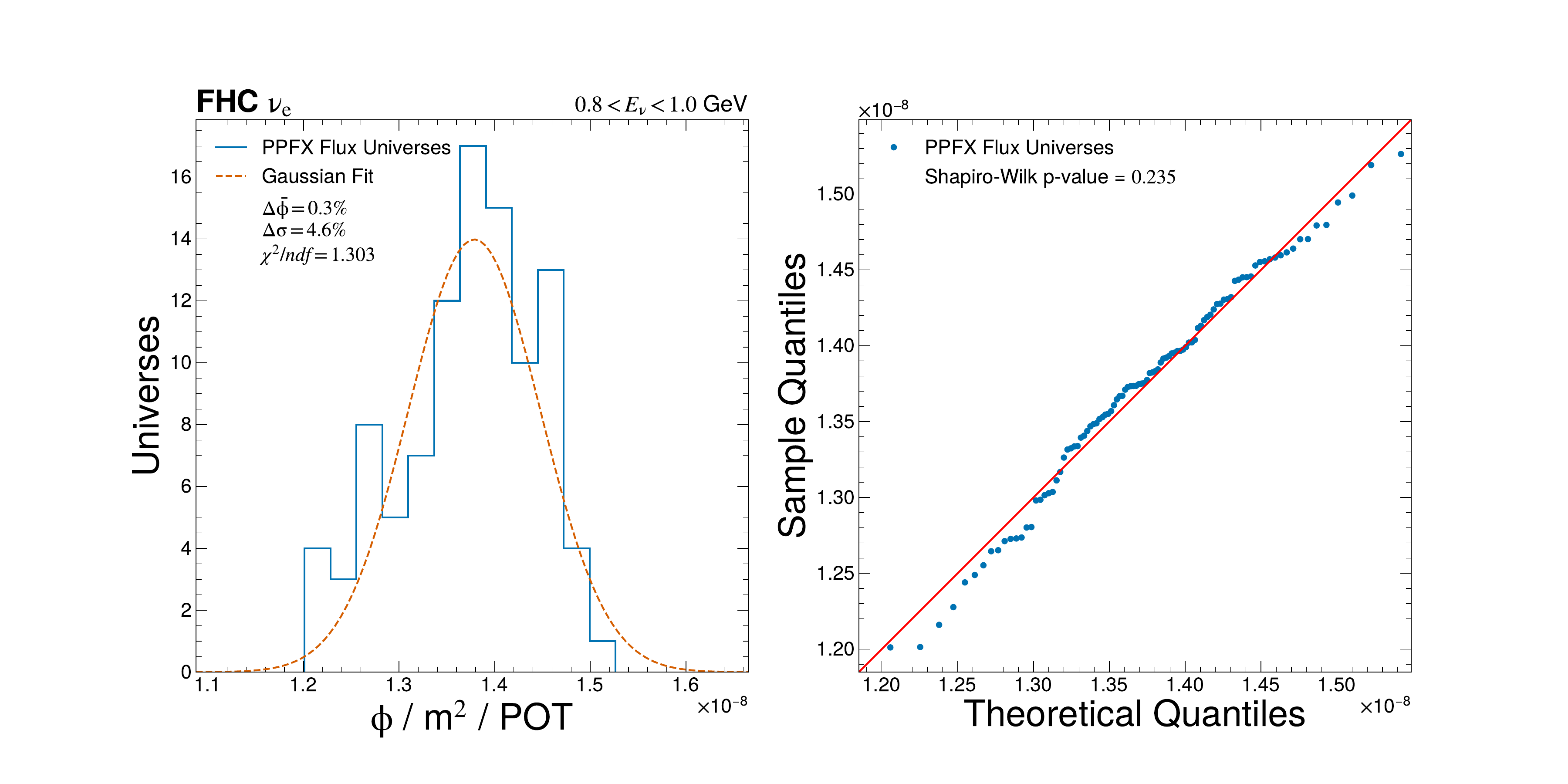}
    \includegraphics[width=0.3\textwidth]{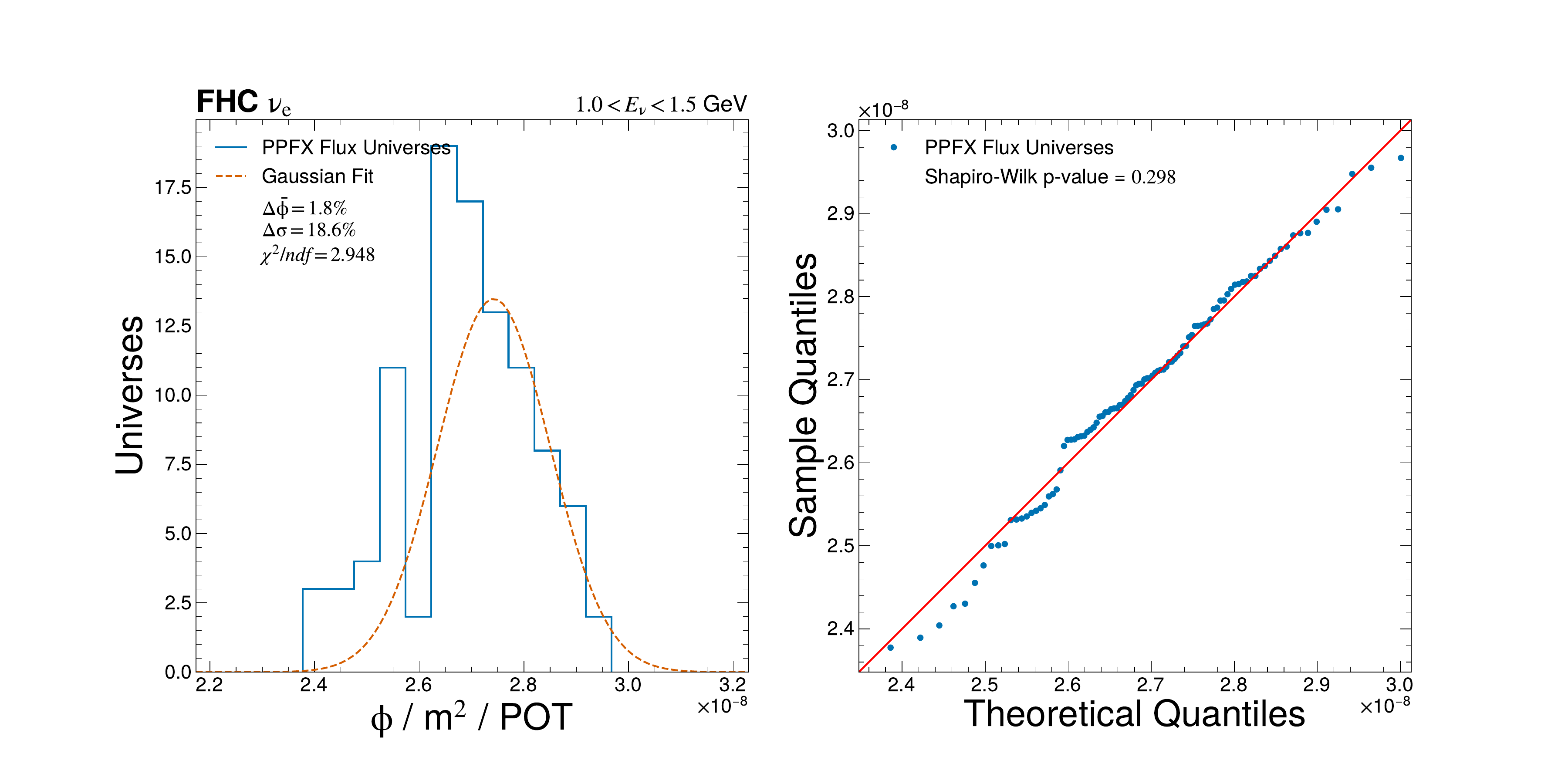}
    \includegraphics[width=0.3\textwidth]{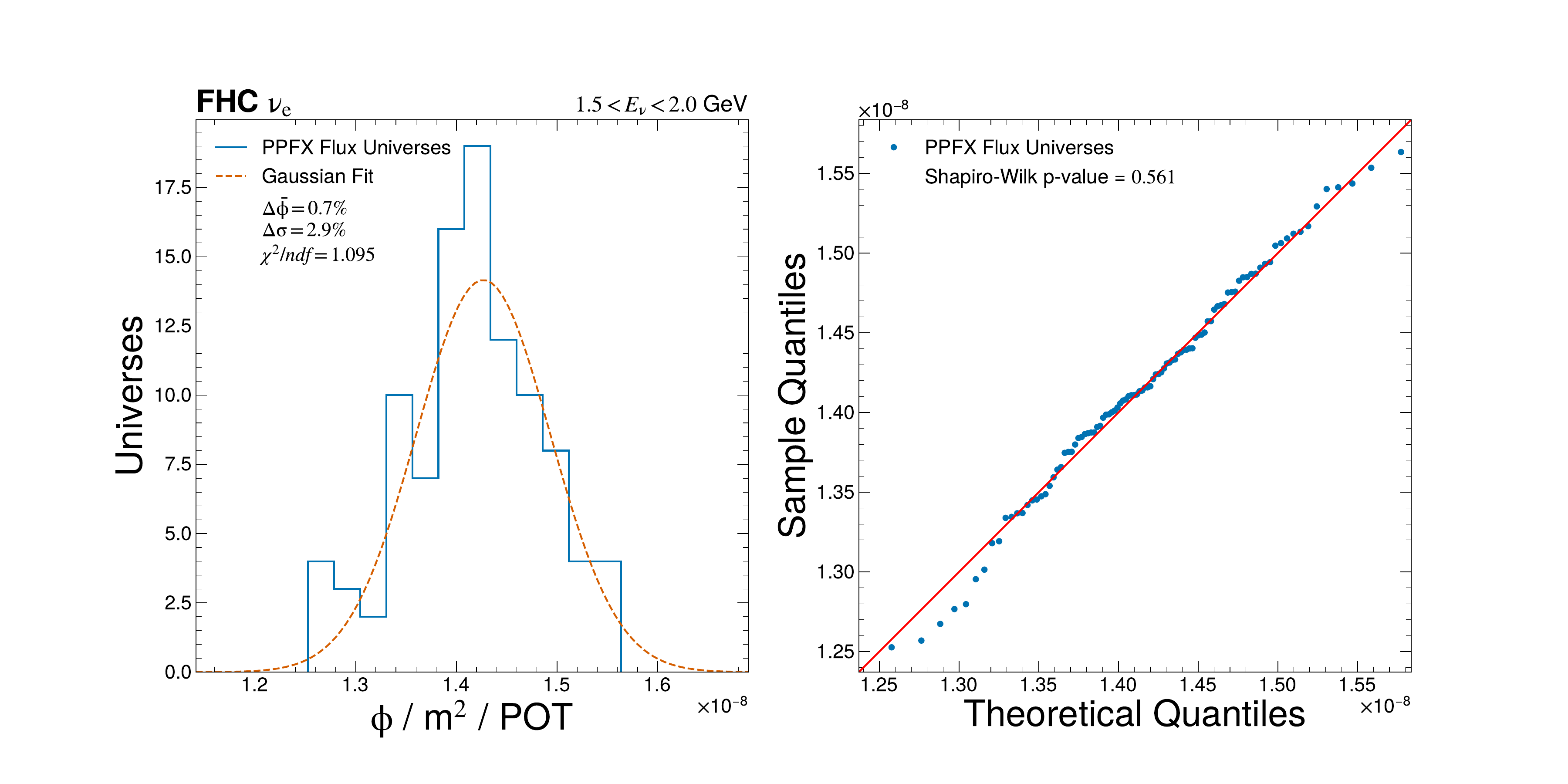}
    \includegraphics[width=0.3\textwidth]{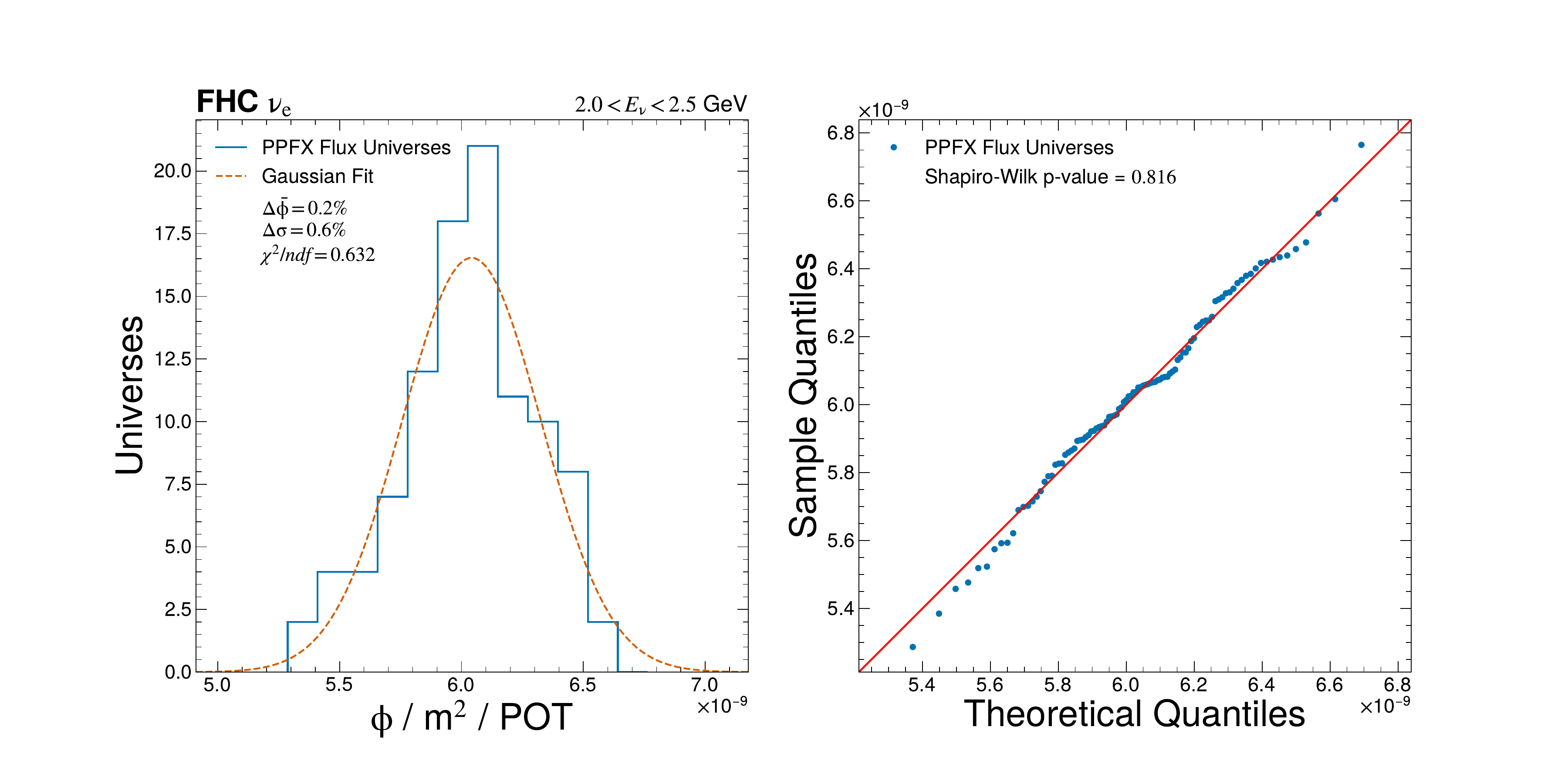}
    \includegraphics[width=0.3\textwidth]{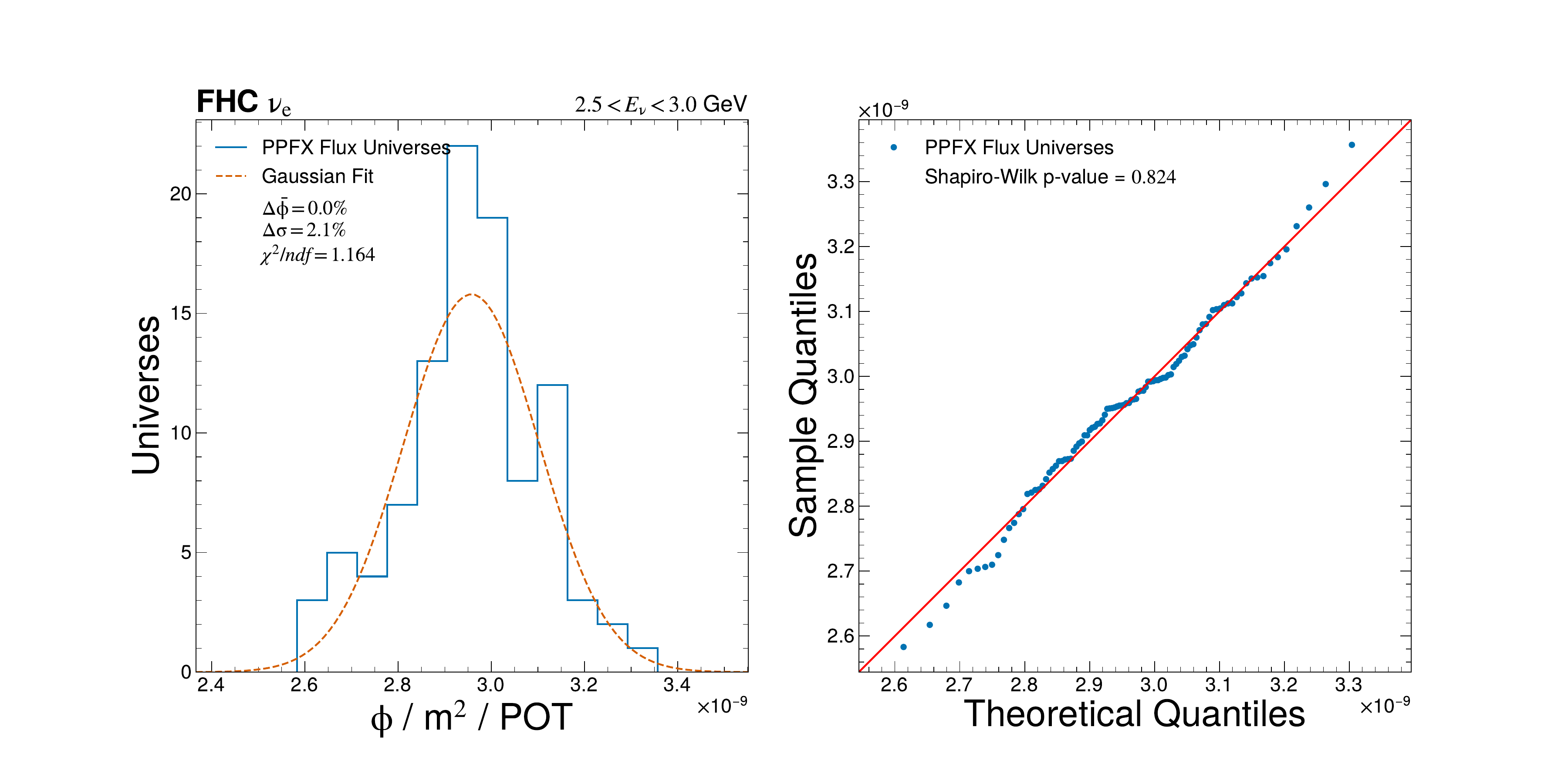}
    \includegraphics[width=0.3\textwidth]{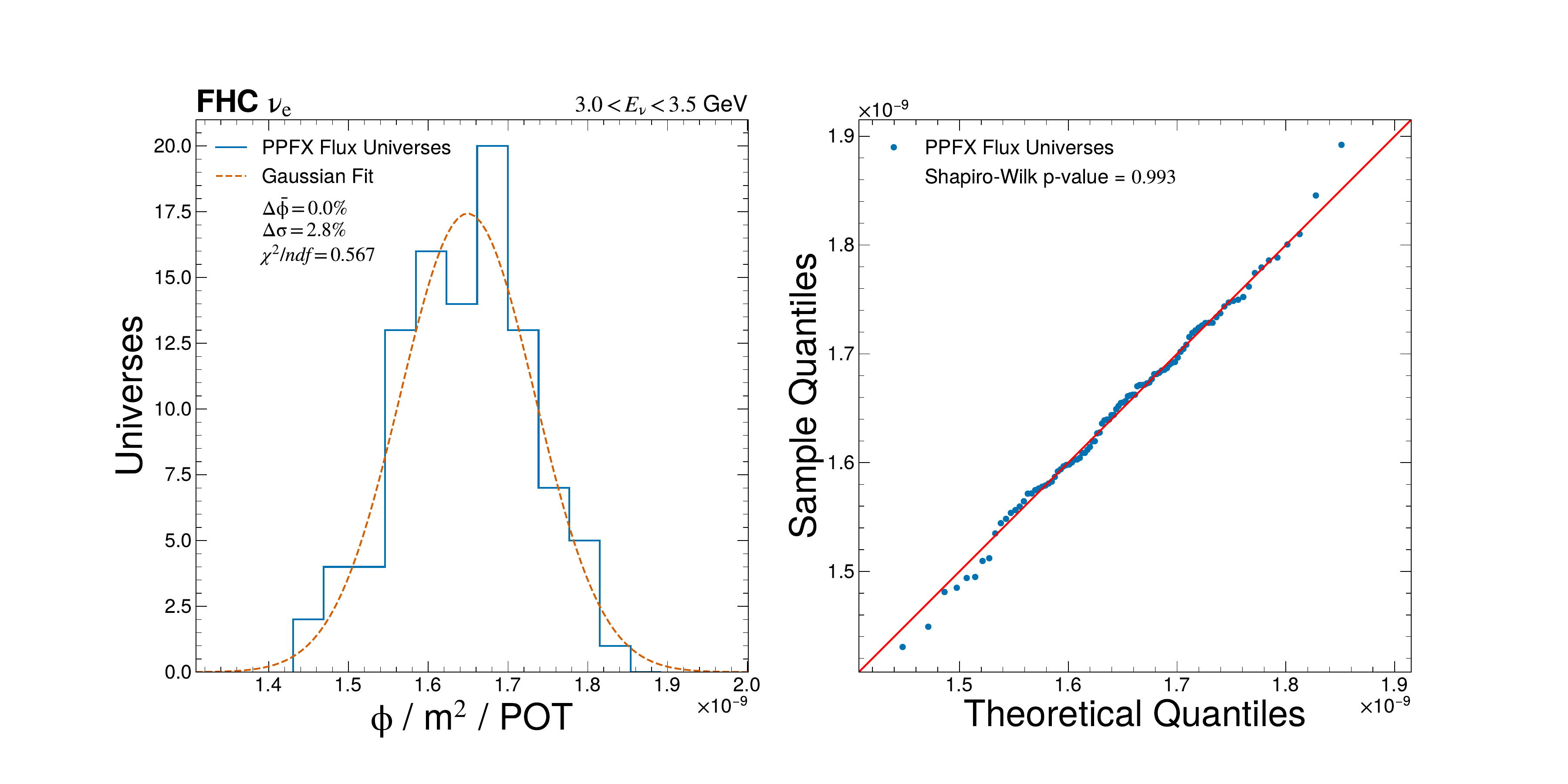}
    \includegraphics[width=0.3\textwidth]{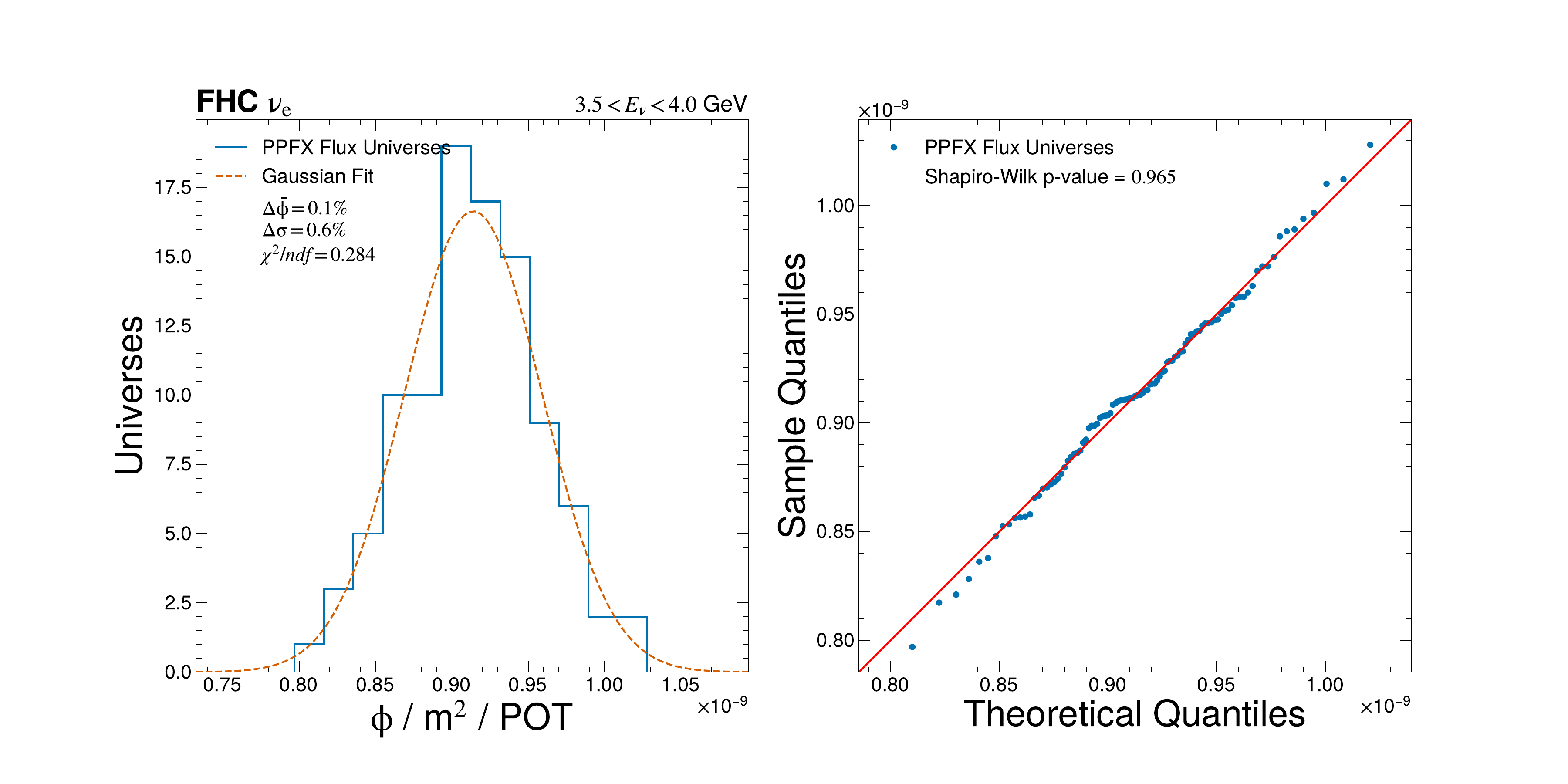}
    \includegraphics[width=0.3\textwidth]{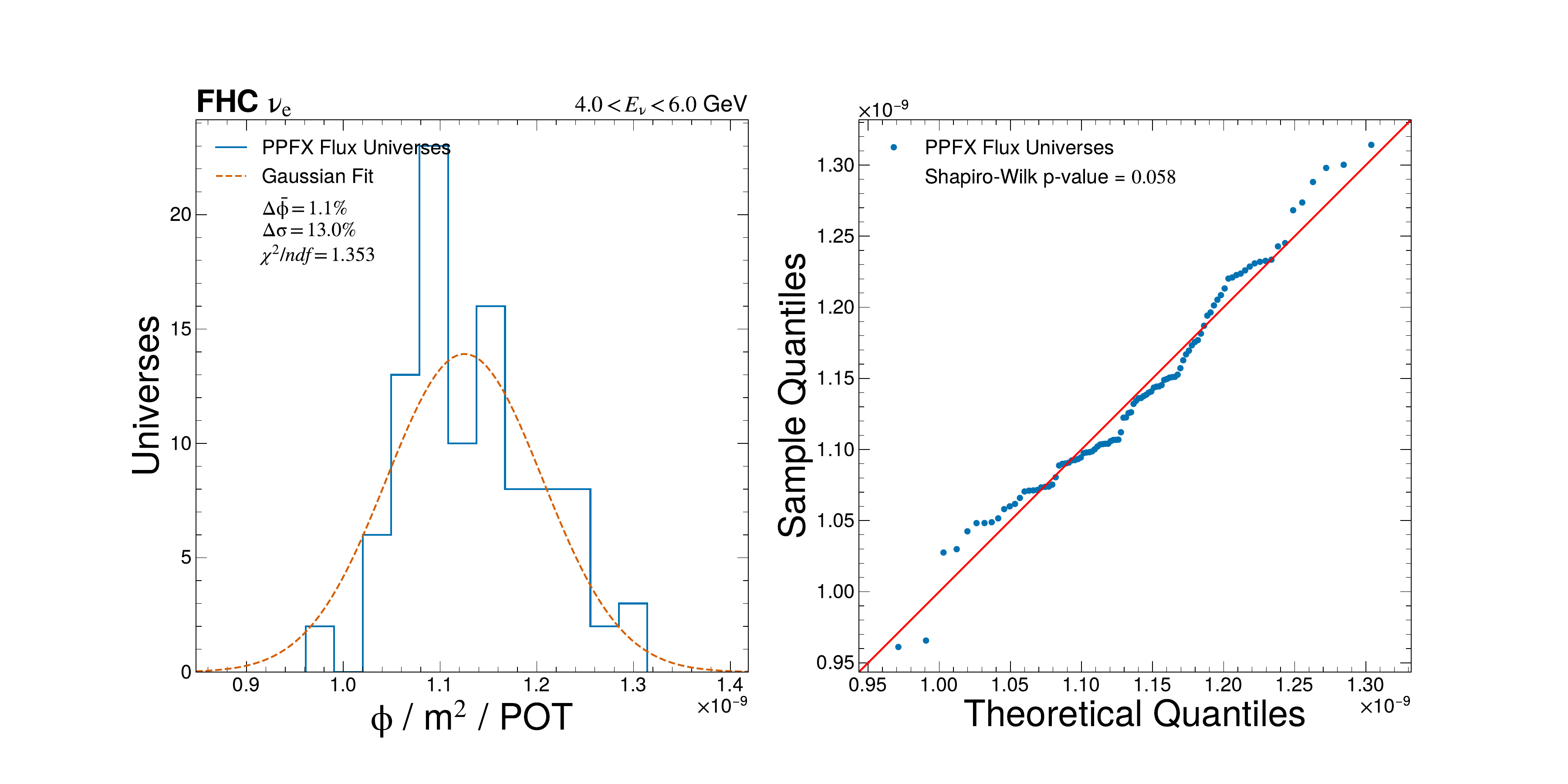}
    \includegraphics[width=0.3\textwidth]{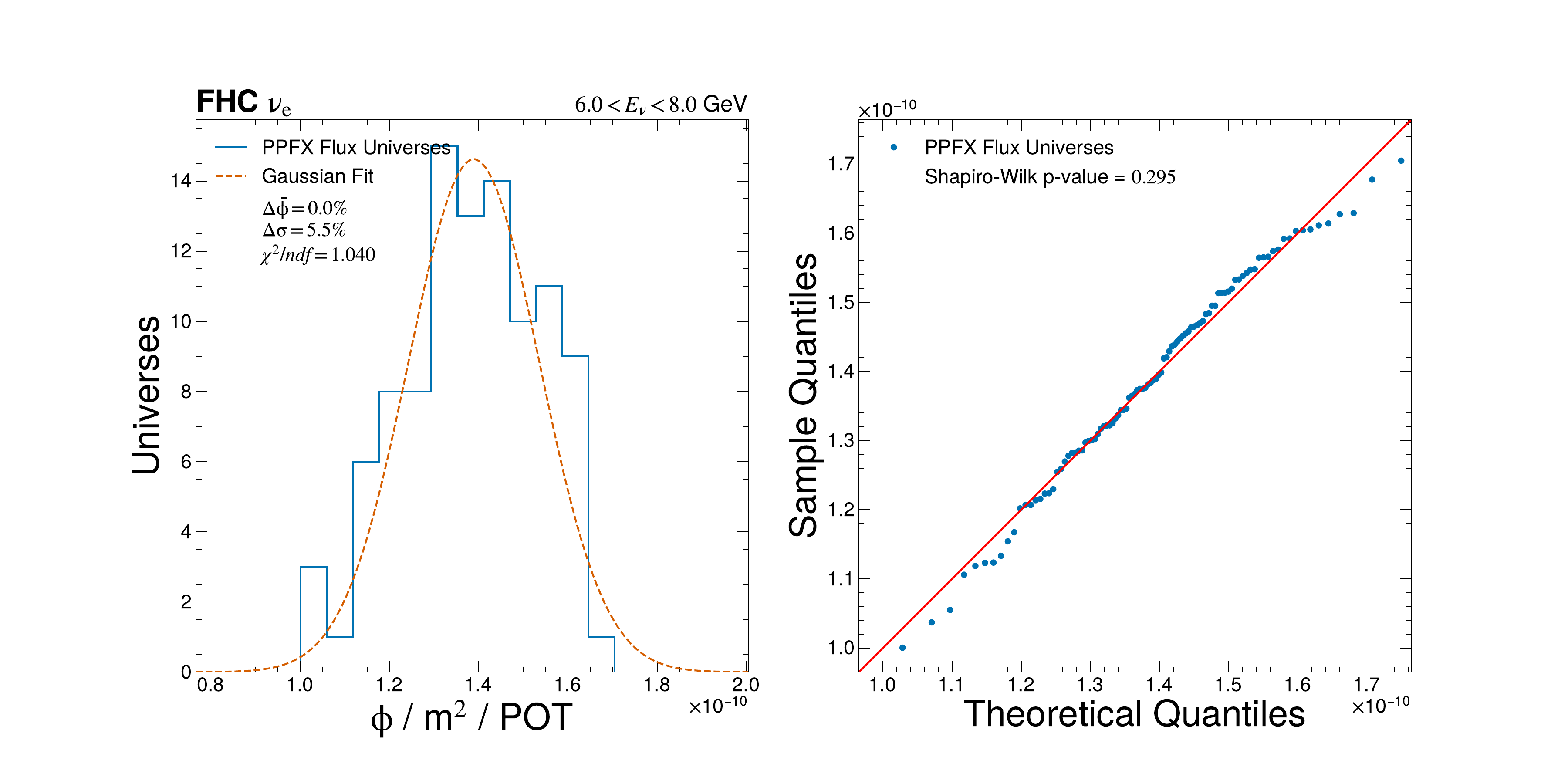}
    \includegraphics[width=0.3\textwidth]{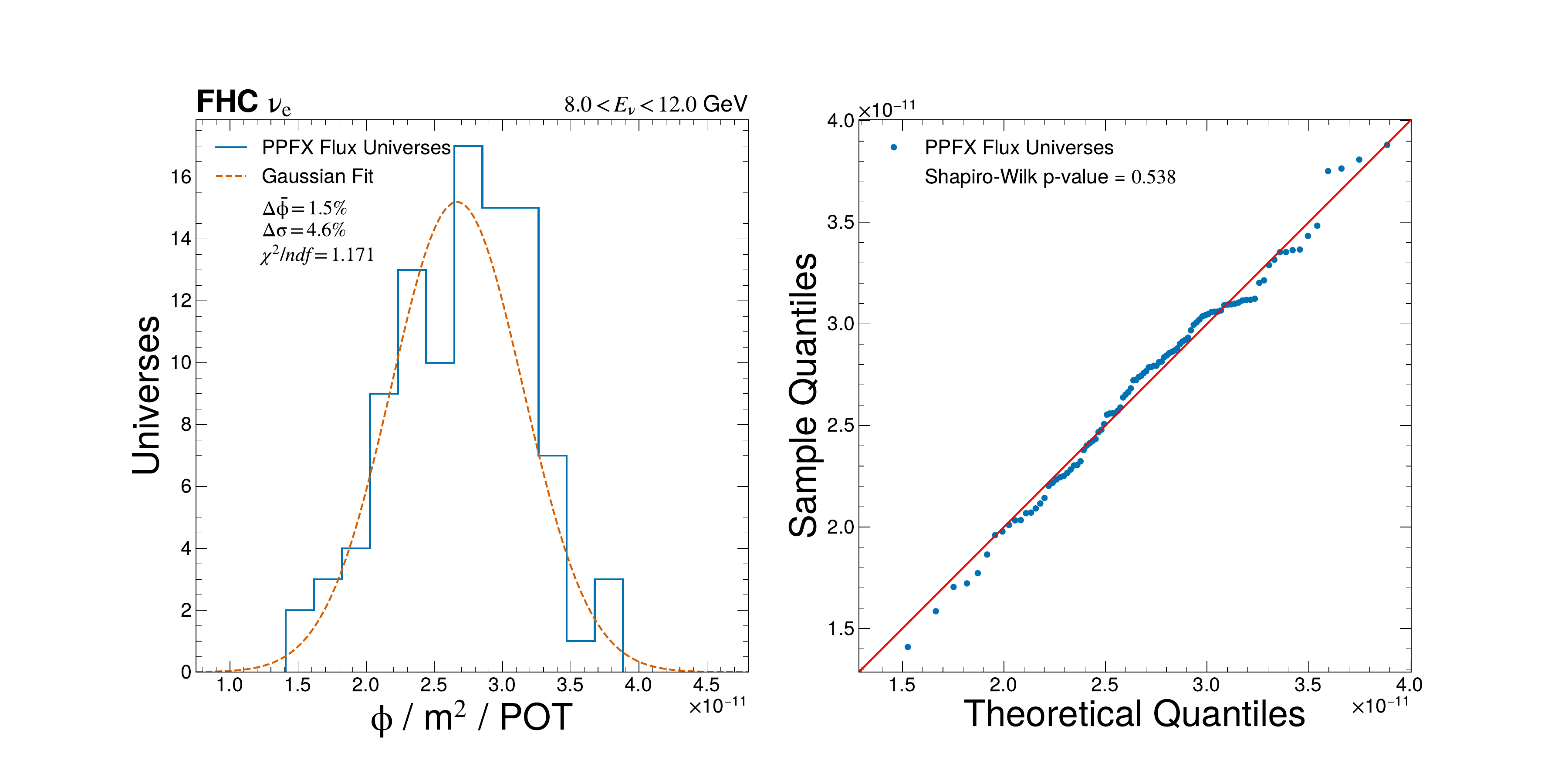}
    \caption[Distribution of PPFX universes for \nue\ (FHC).]{Distribution of PPFX universes for \nue.}
\end{figure}
\begin{figure}[!ht]
    \centering
    \includegraphics[width=0.3\textwidth]{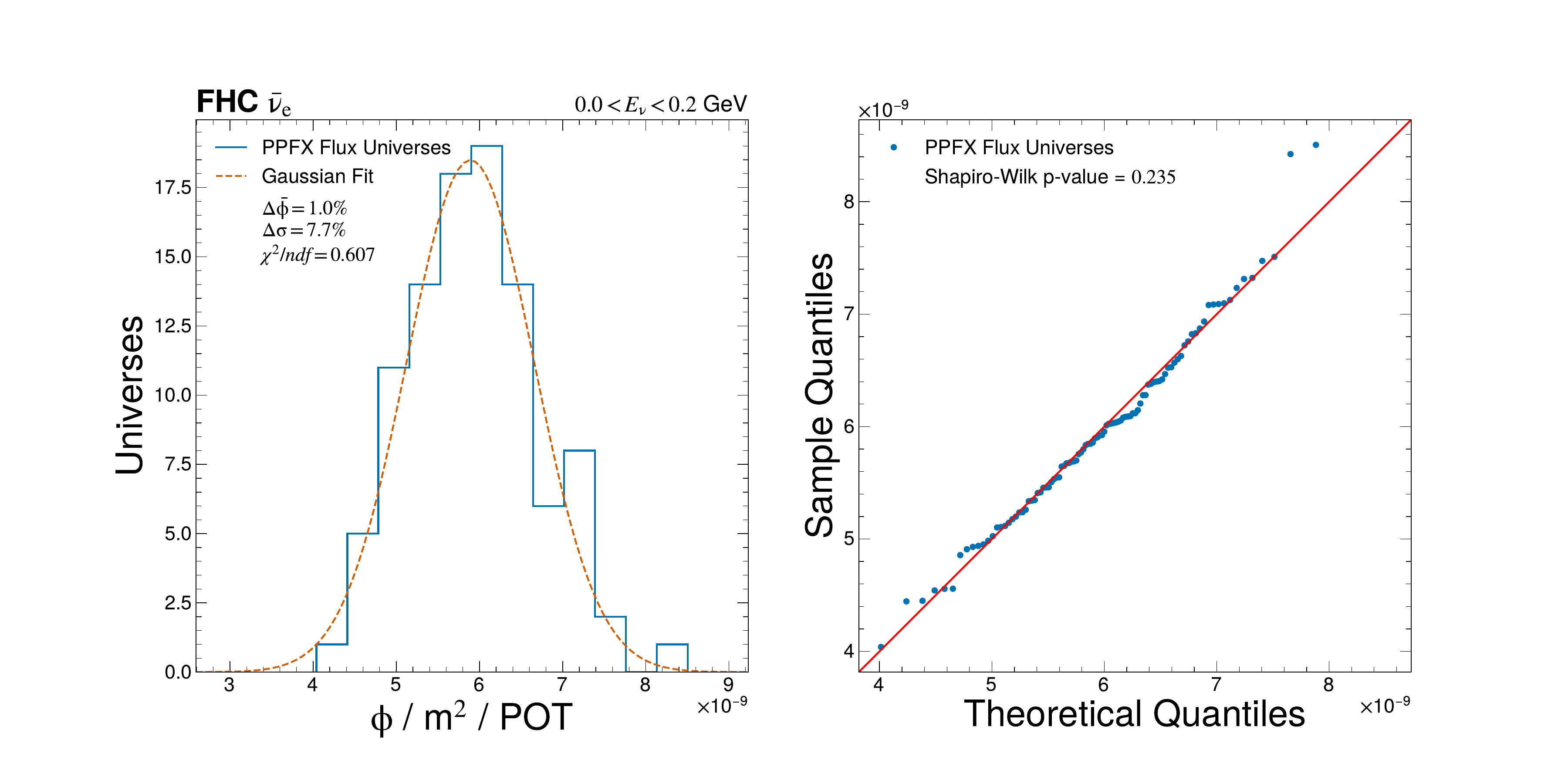}
    \includegraphics[width=0.3\textwidth]{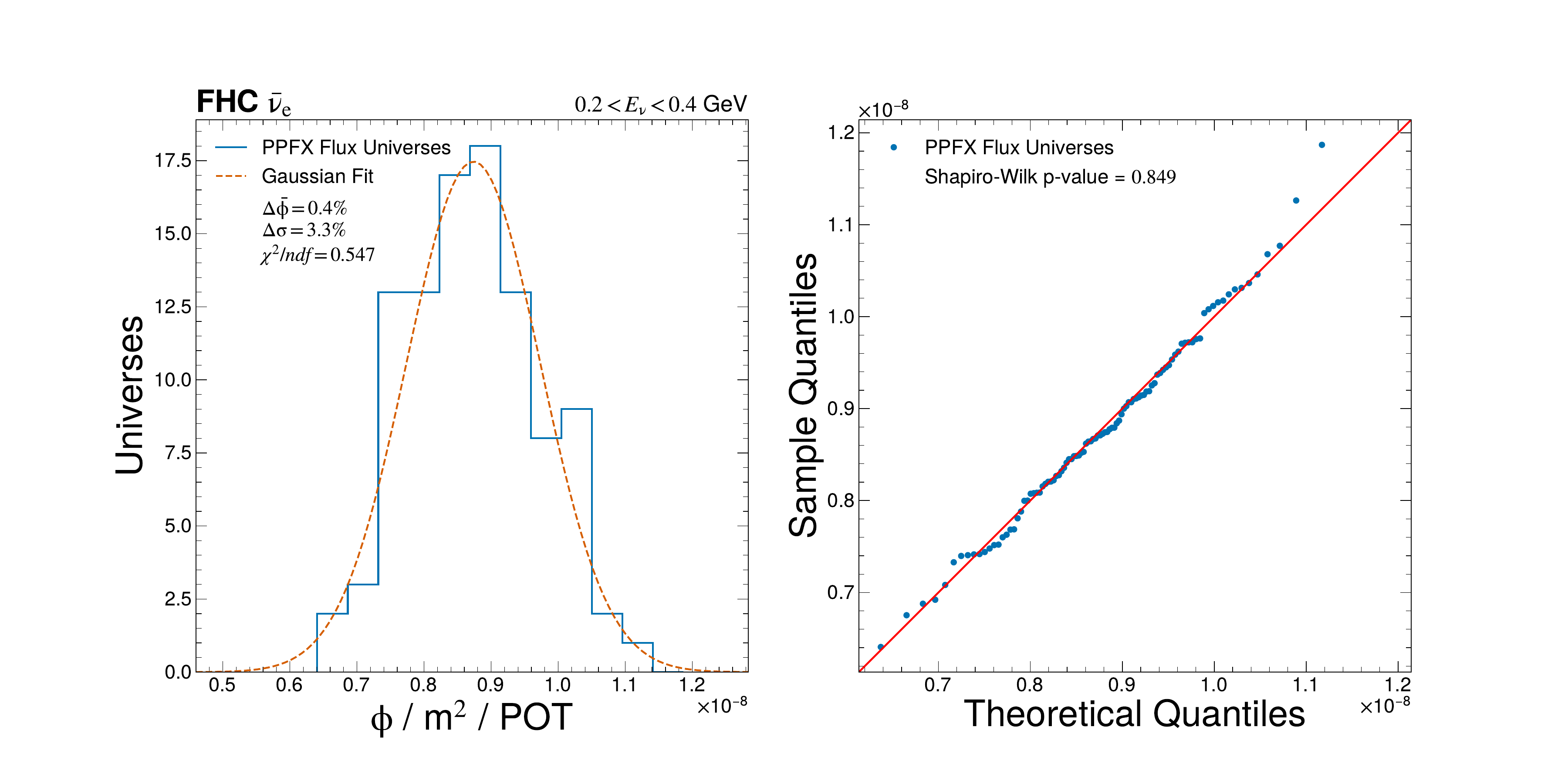}
    \includegraphics[width=0.3\textwidth]{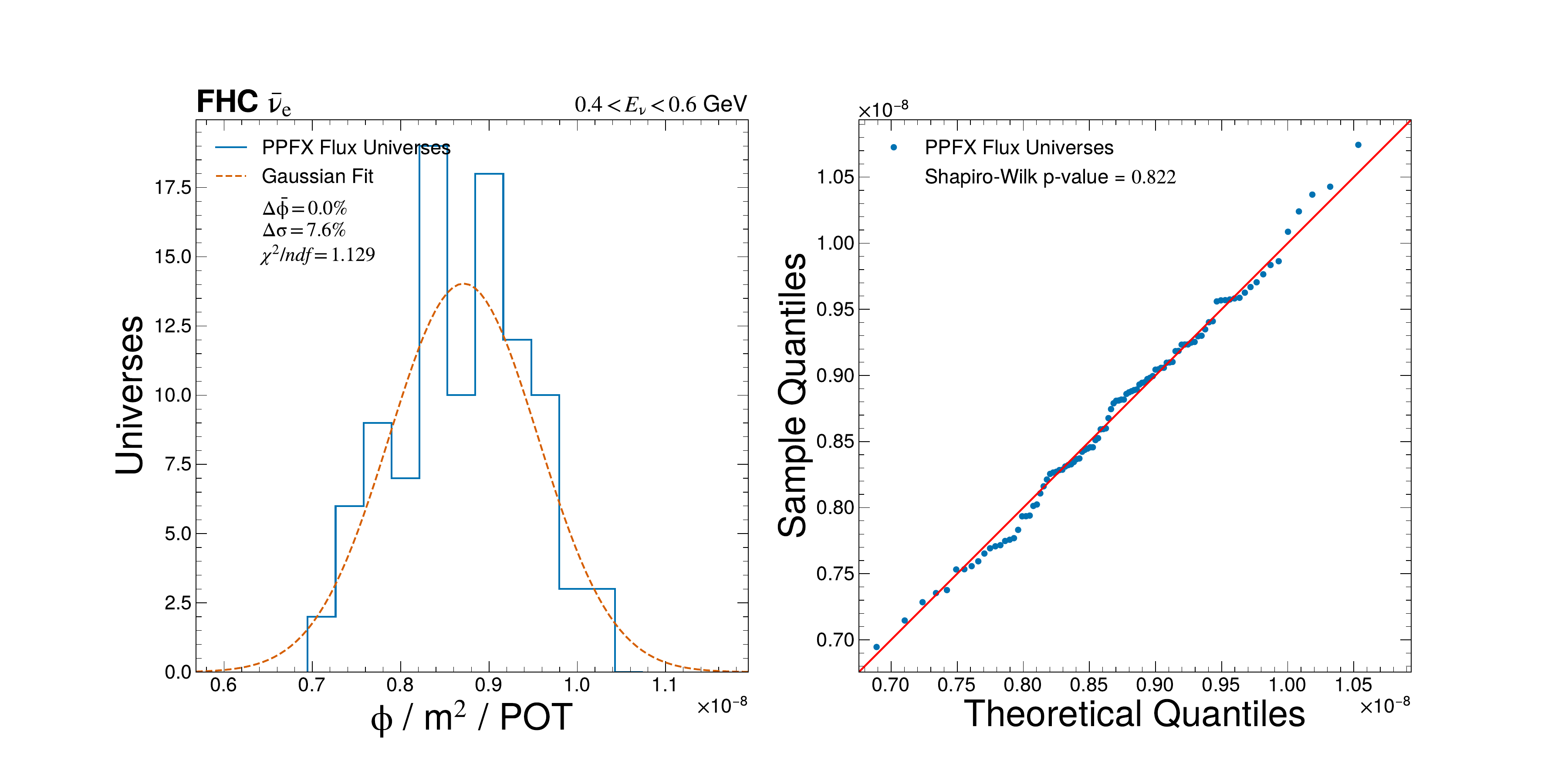}
    \includegraphics[width=0.3\textwidth]{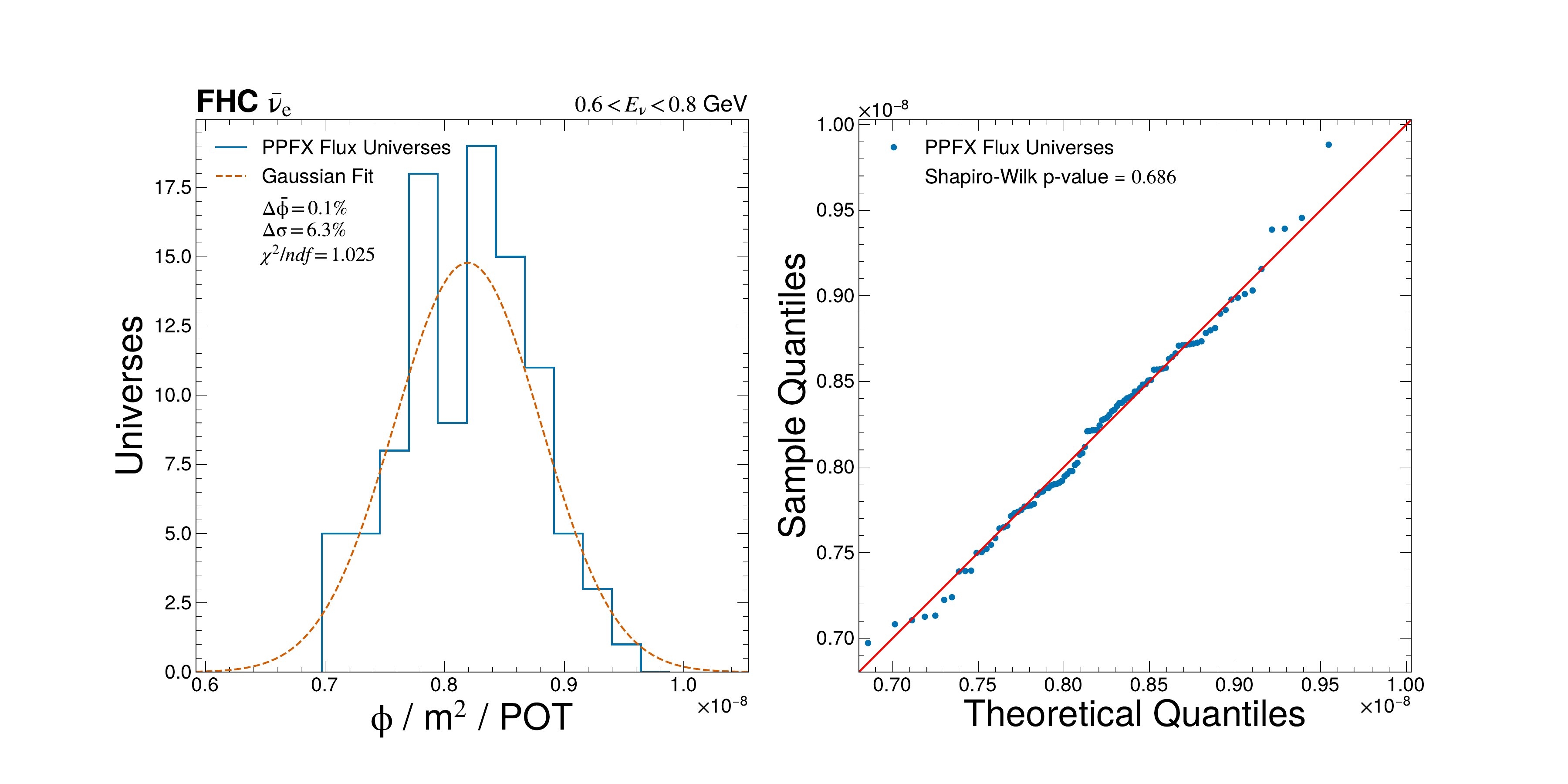}
    \includegraphics[width=0.3\textwidth]{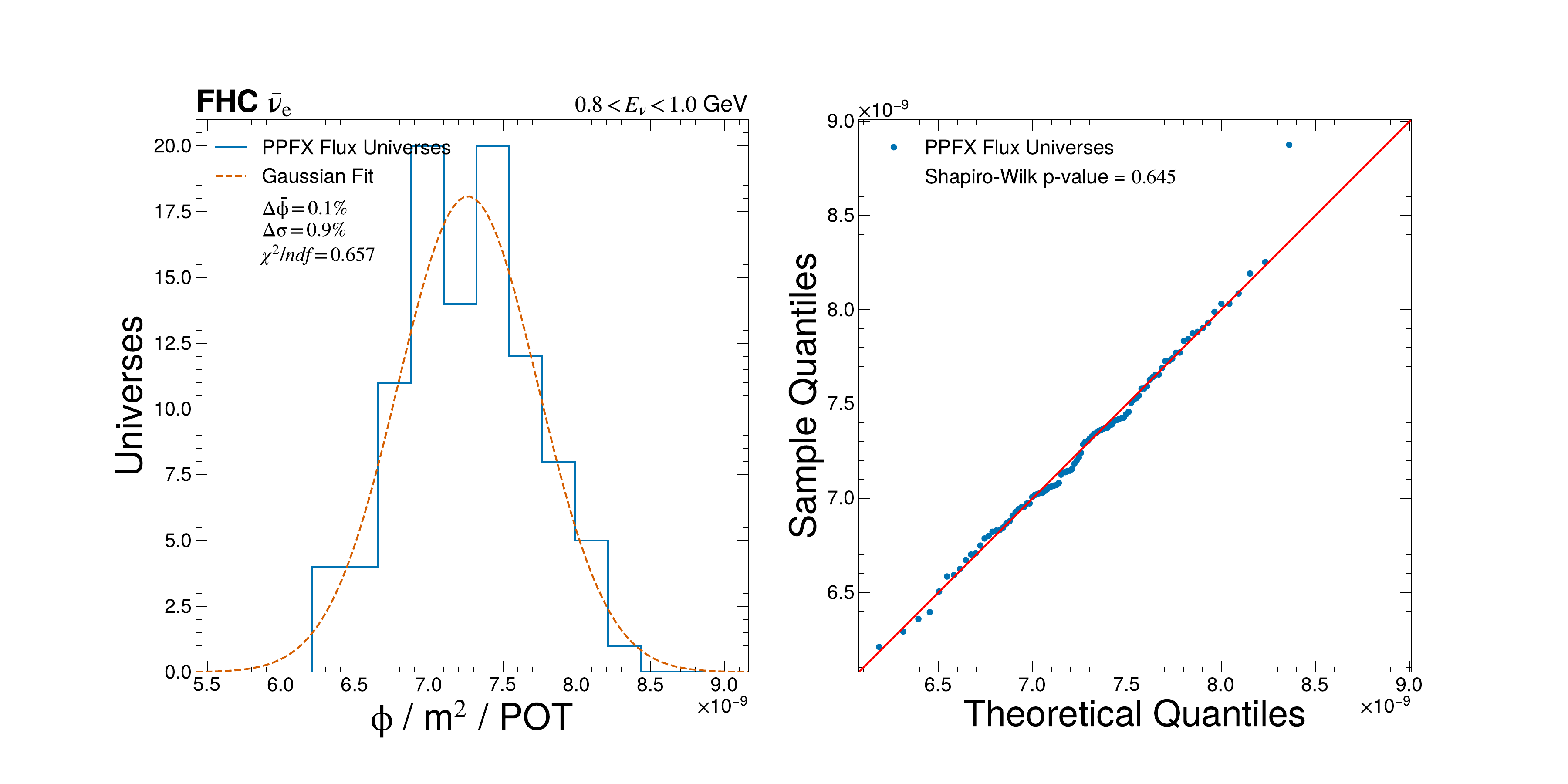}
    \includegraphics[width=0.3\textwidth]{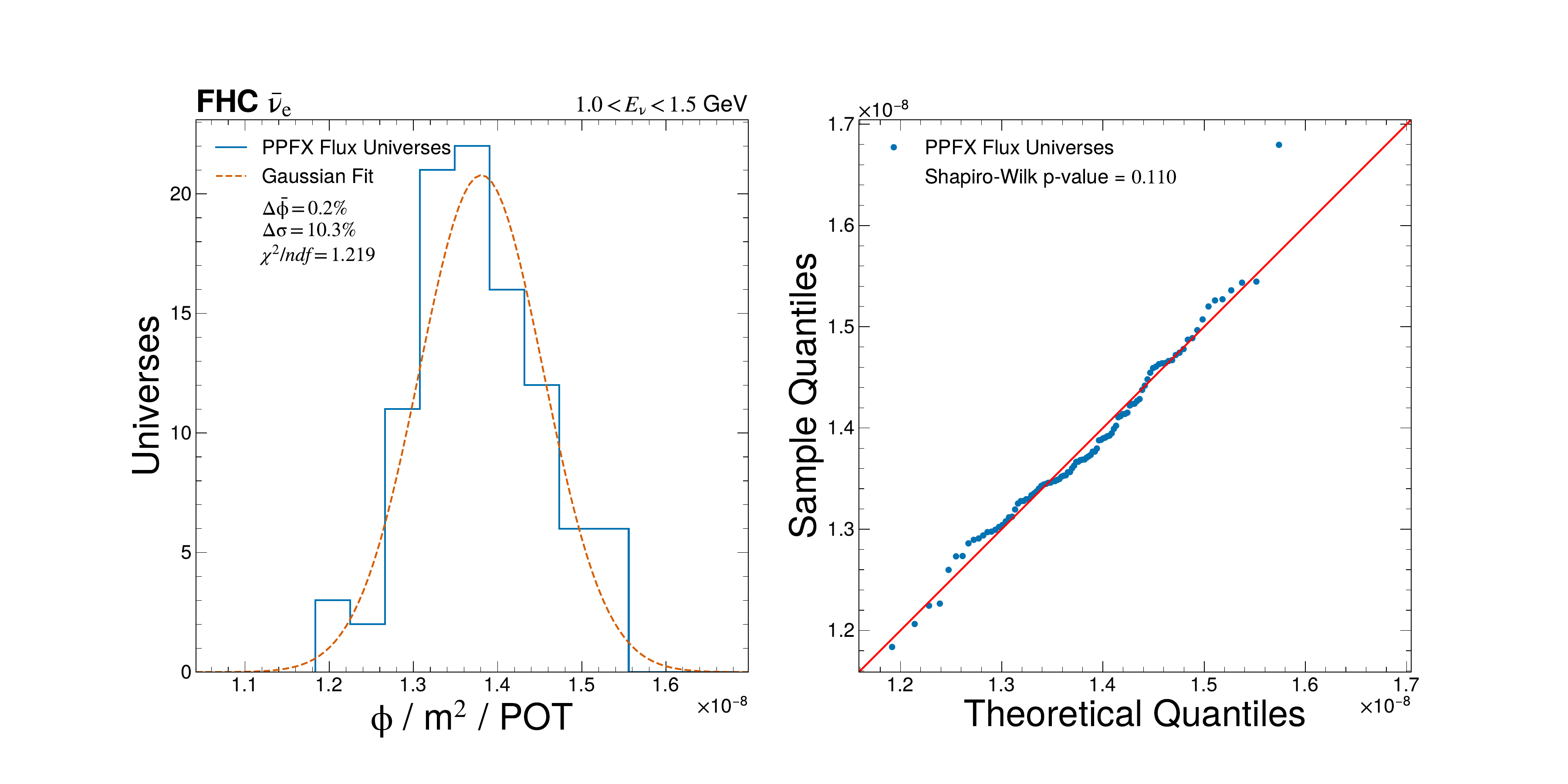}
    \includegraphics[width=0.3\textwidth]{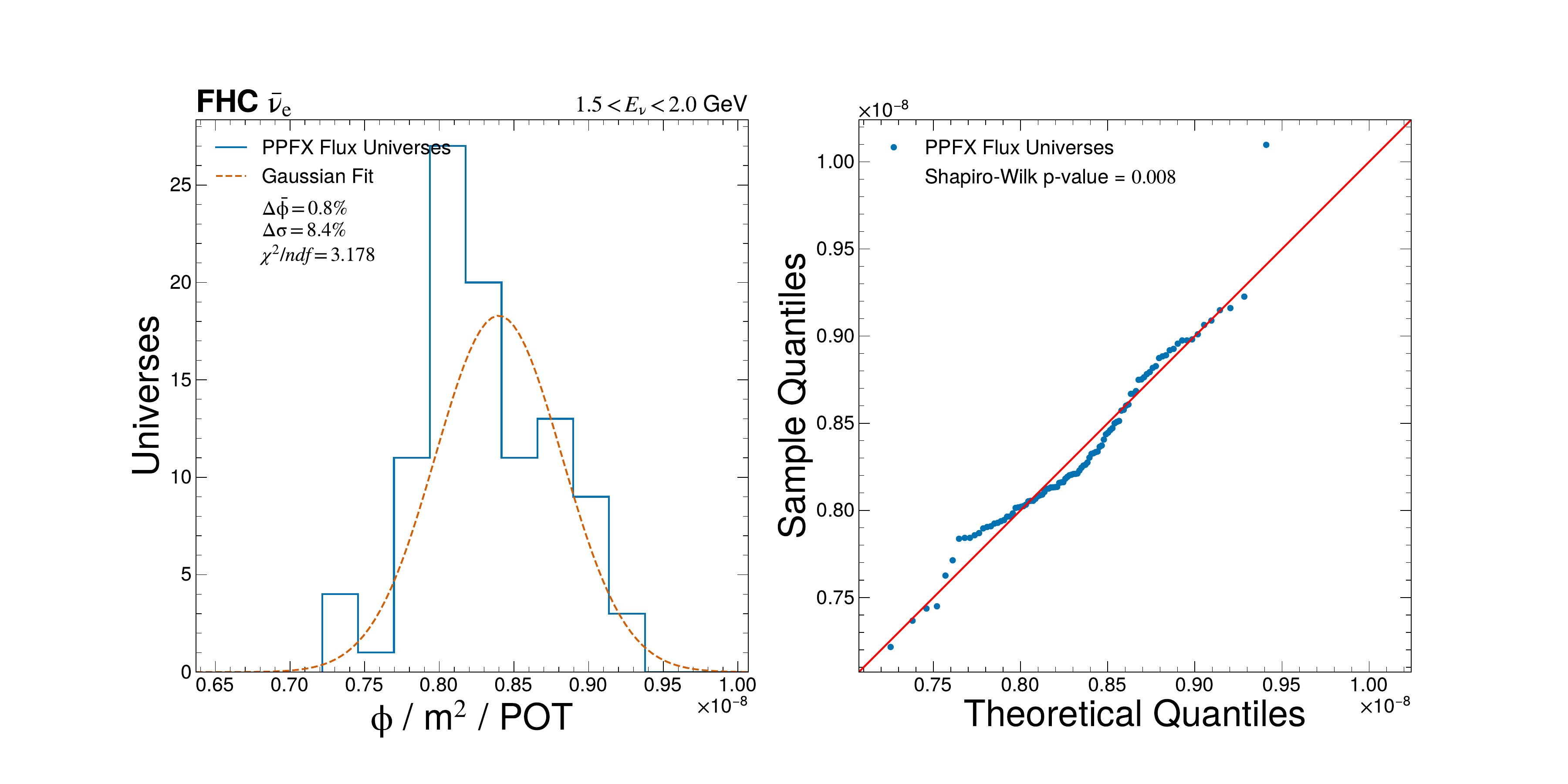}
    \includegraphics[width=0.3\textwidth]{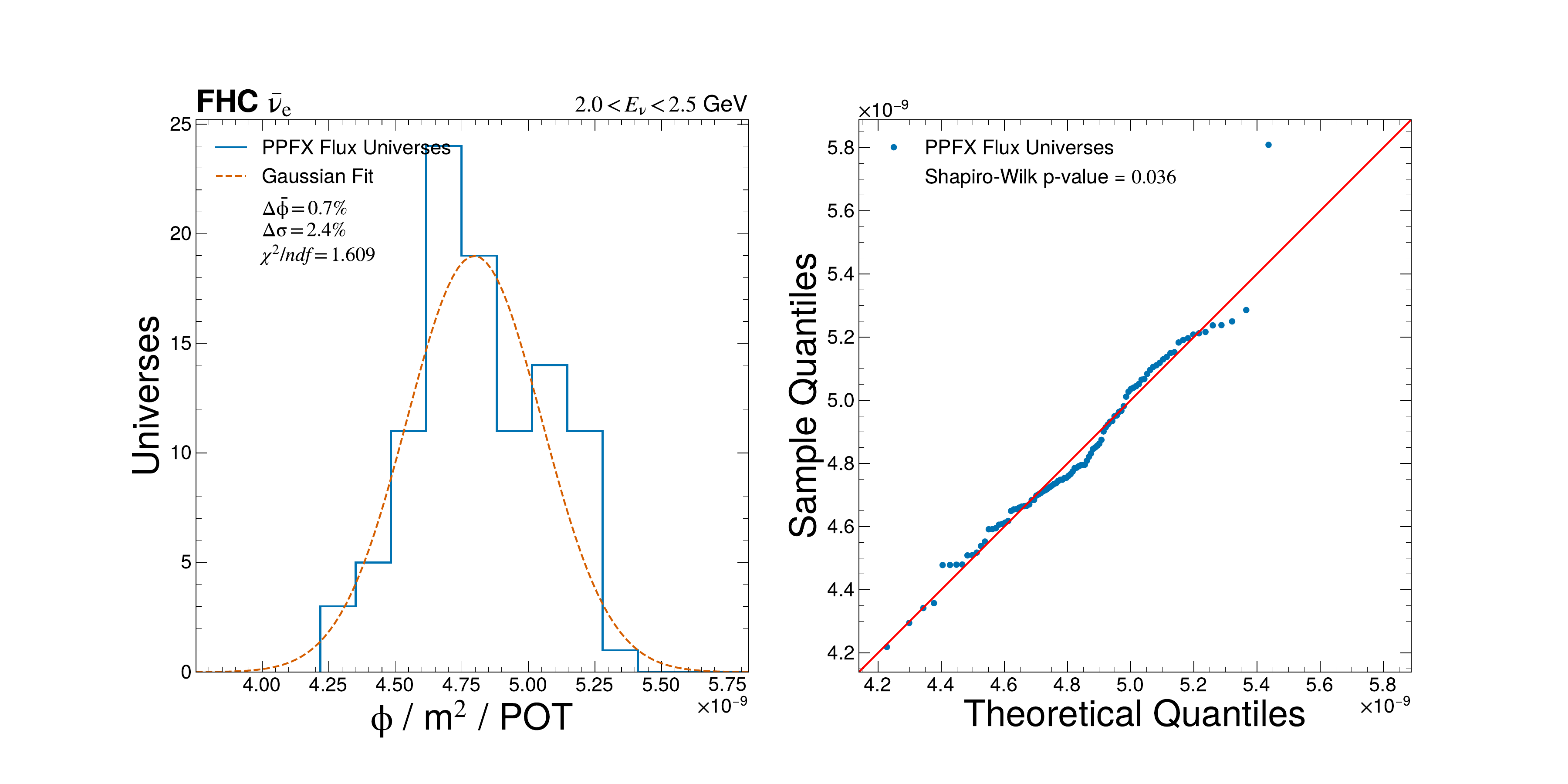}
    \includegraphics[width=0.3\textwidth]{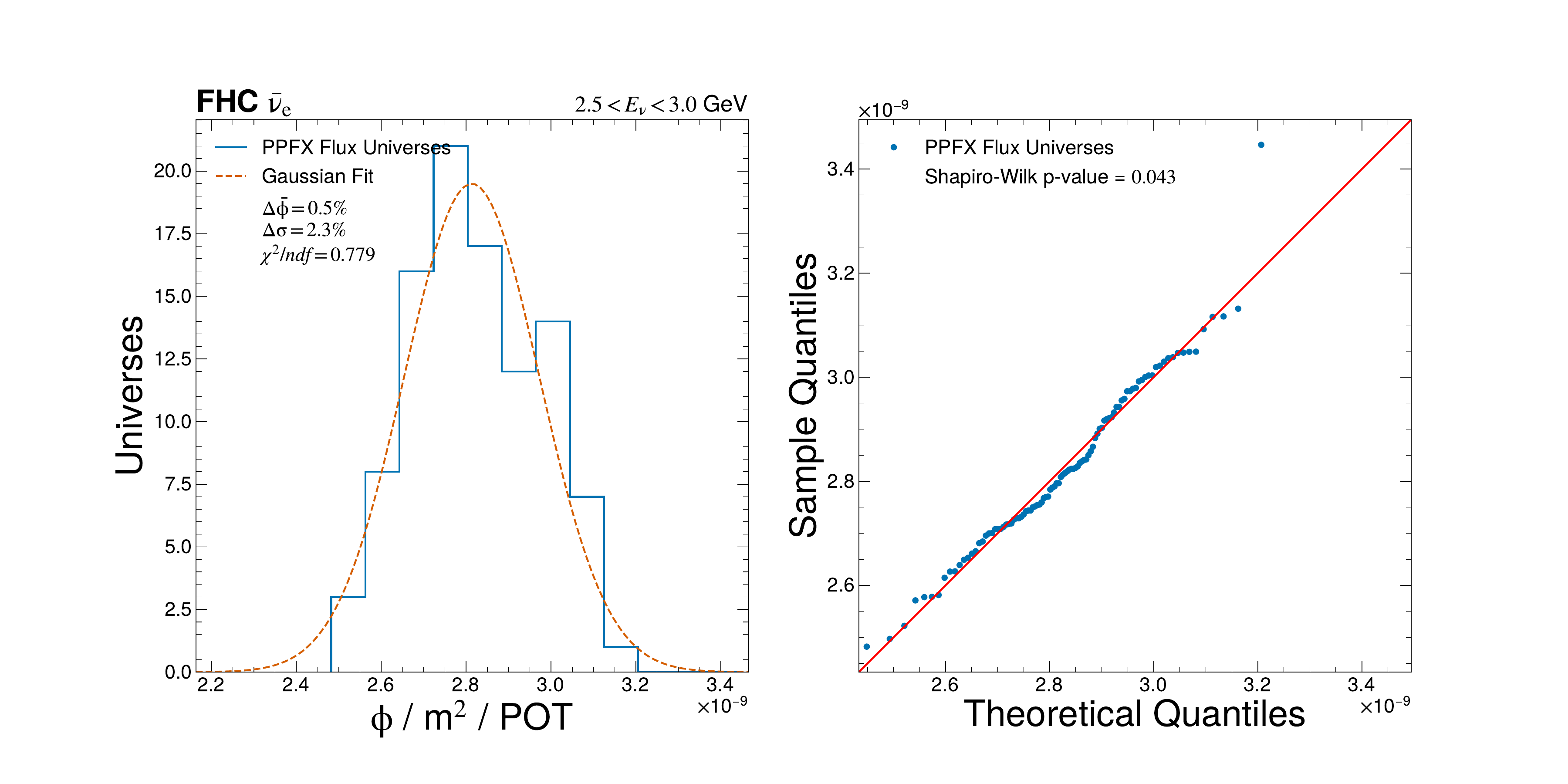}
    \includegraphics[width=0.3\textwidth]{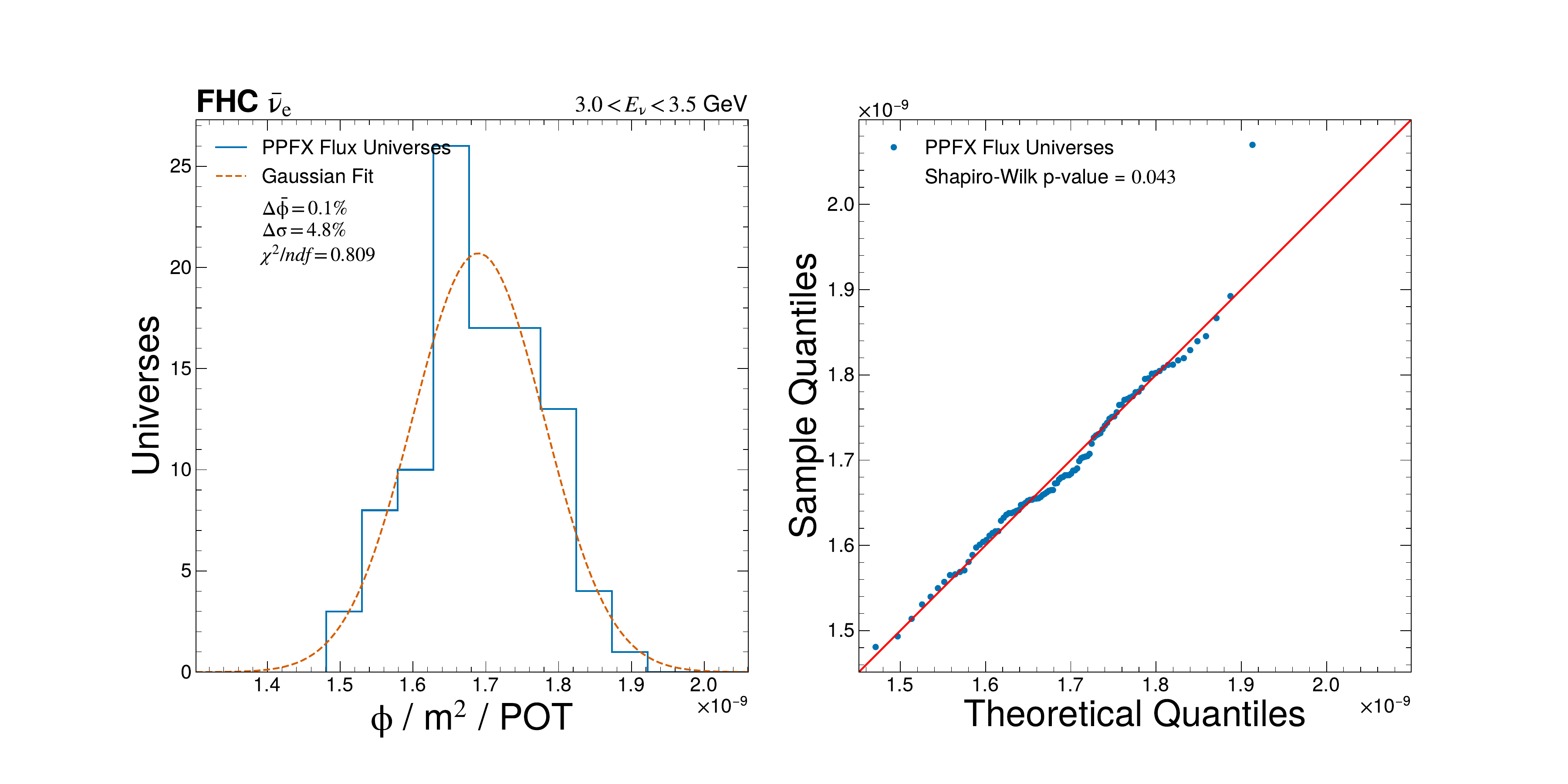}
    \includegraphics[width=0.3\textwidth]{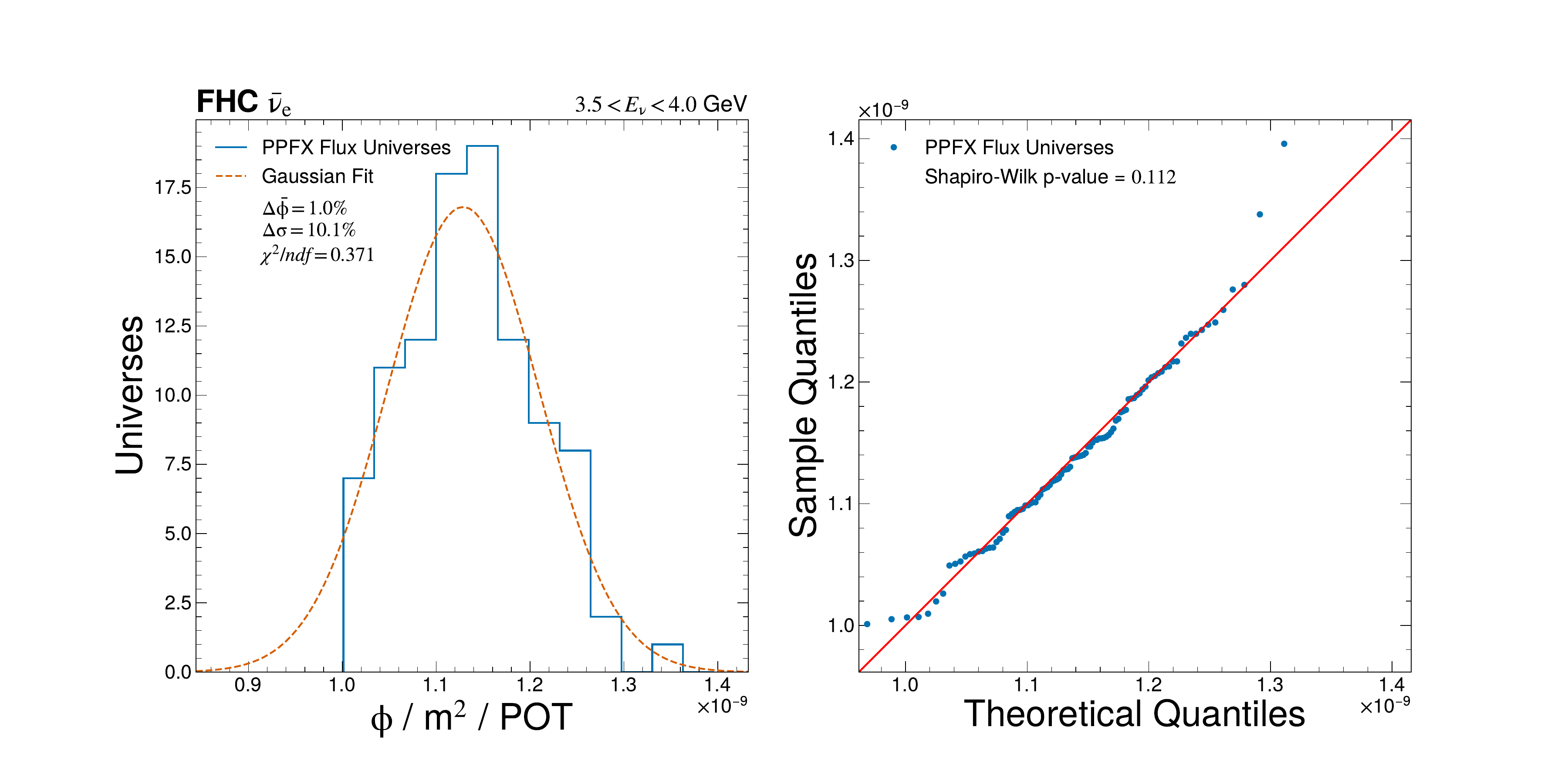}
    \includegraphics[width=0.3\textwidth]{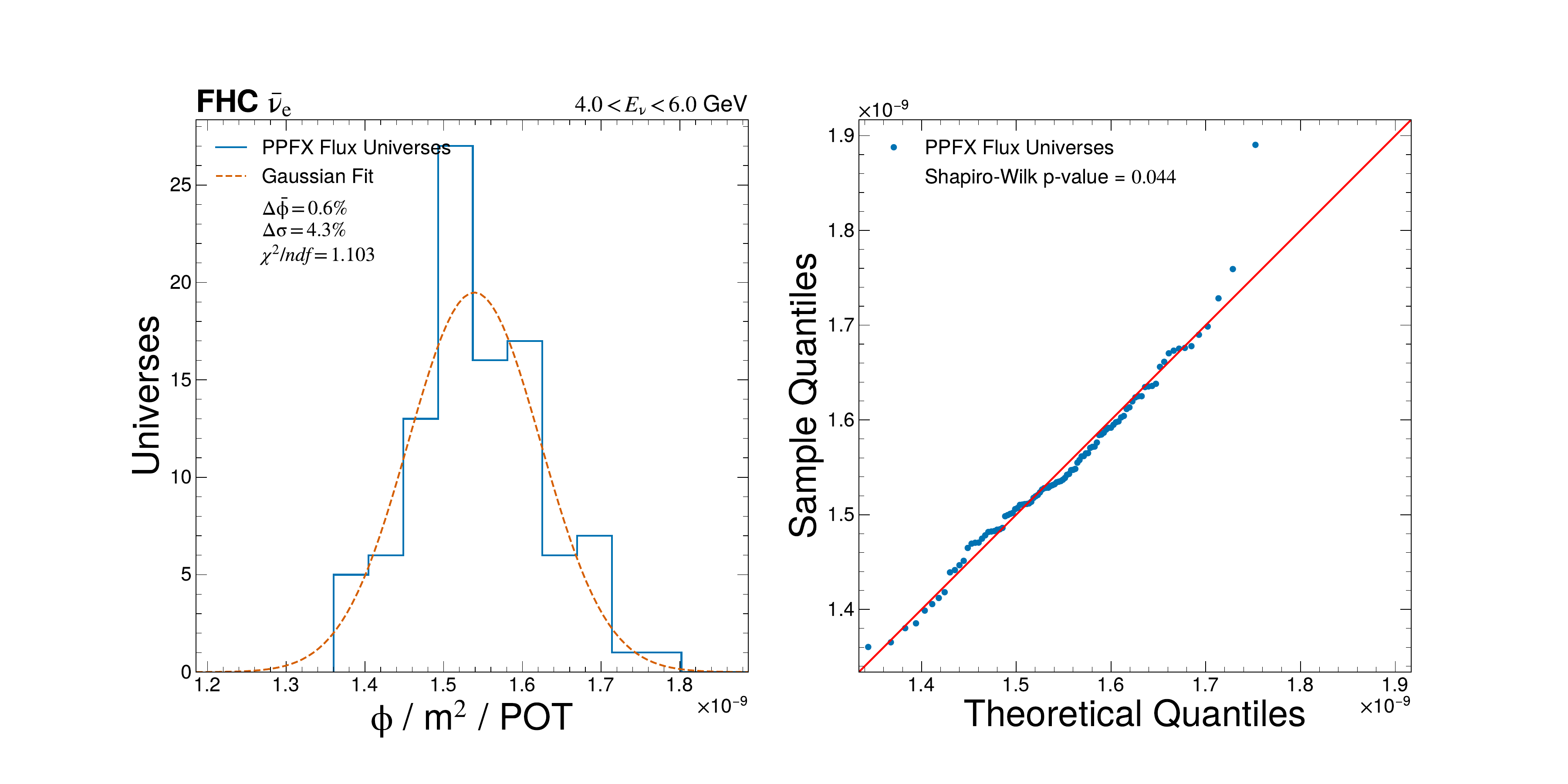}
    \includegraphics[width=0.3\textwidth]{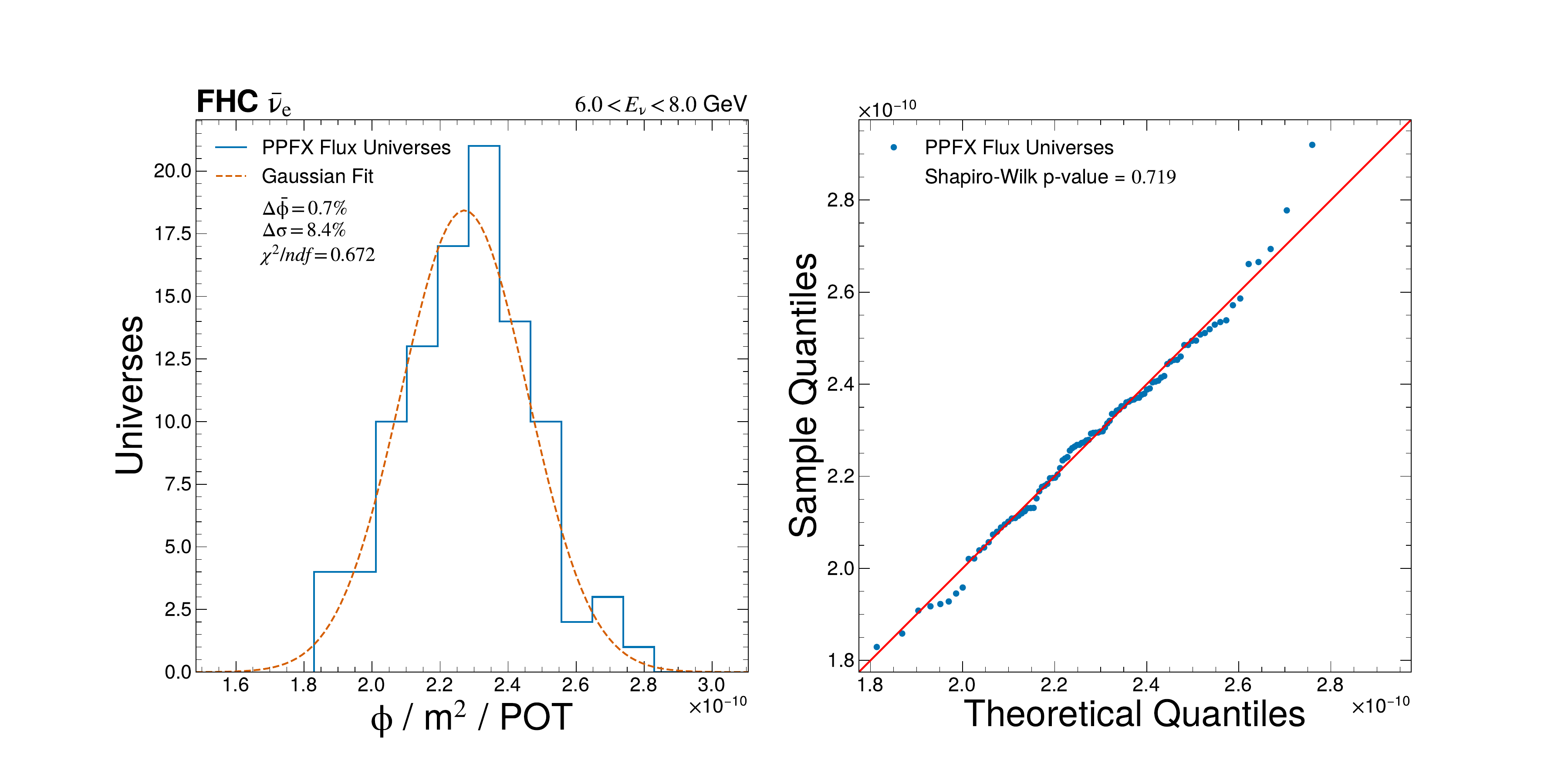}
    \includegraphics[width=0.3\textwidth]{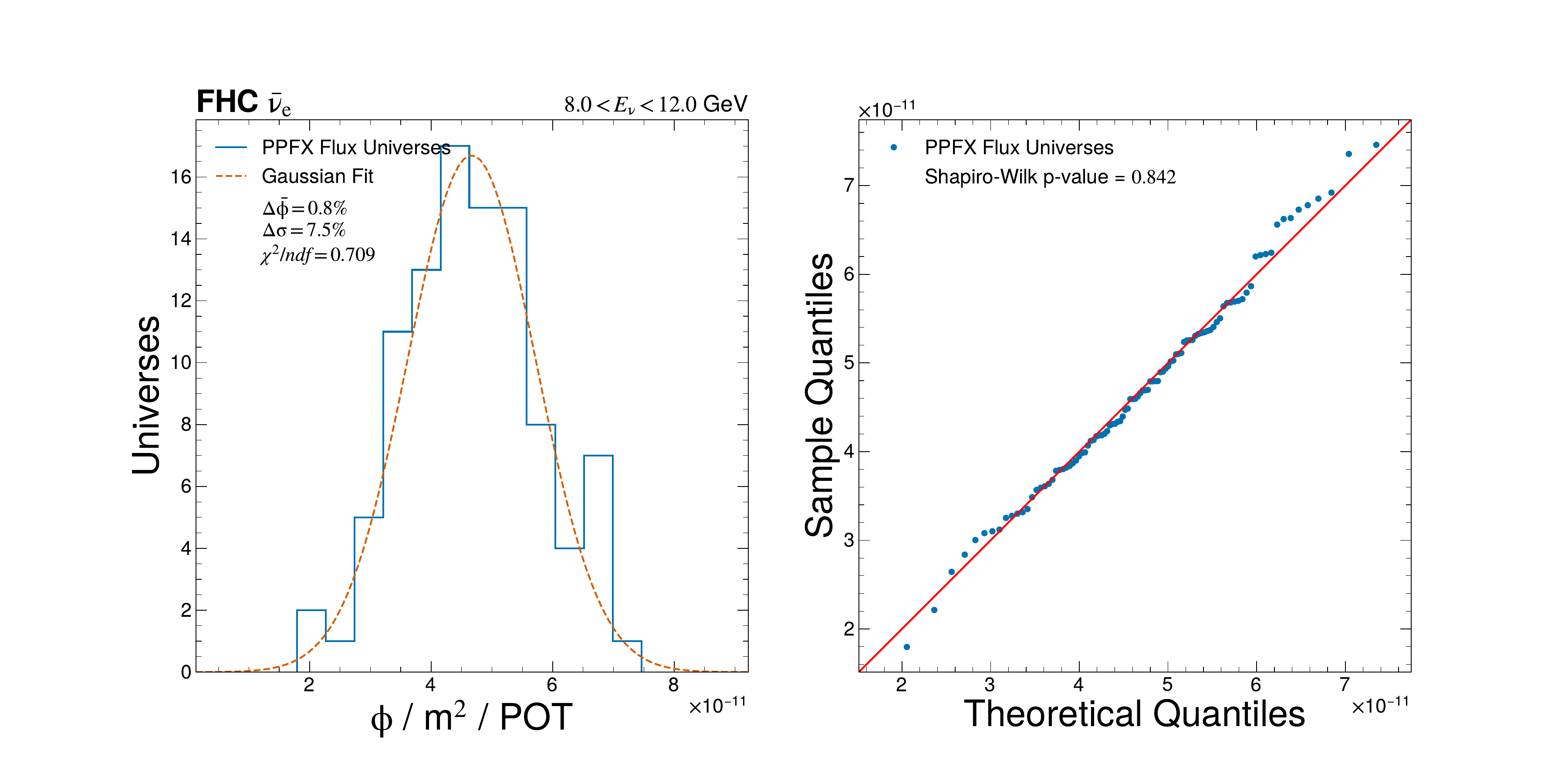}
    \caption[Distribution of PPFX universes for \nueb\ (FHC).]{Distribution of PPFX universes for \nueb.}
\end{figure}
\begin{figure}[!ht]
    \centering
    \includegraphics[width=0.3\textwidth]{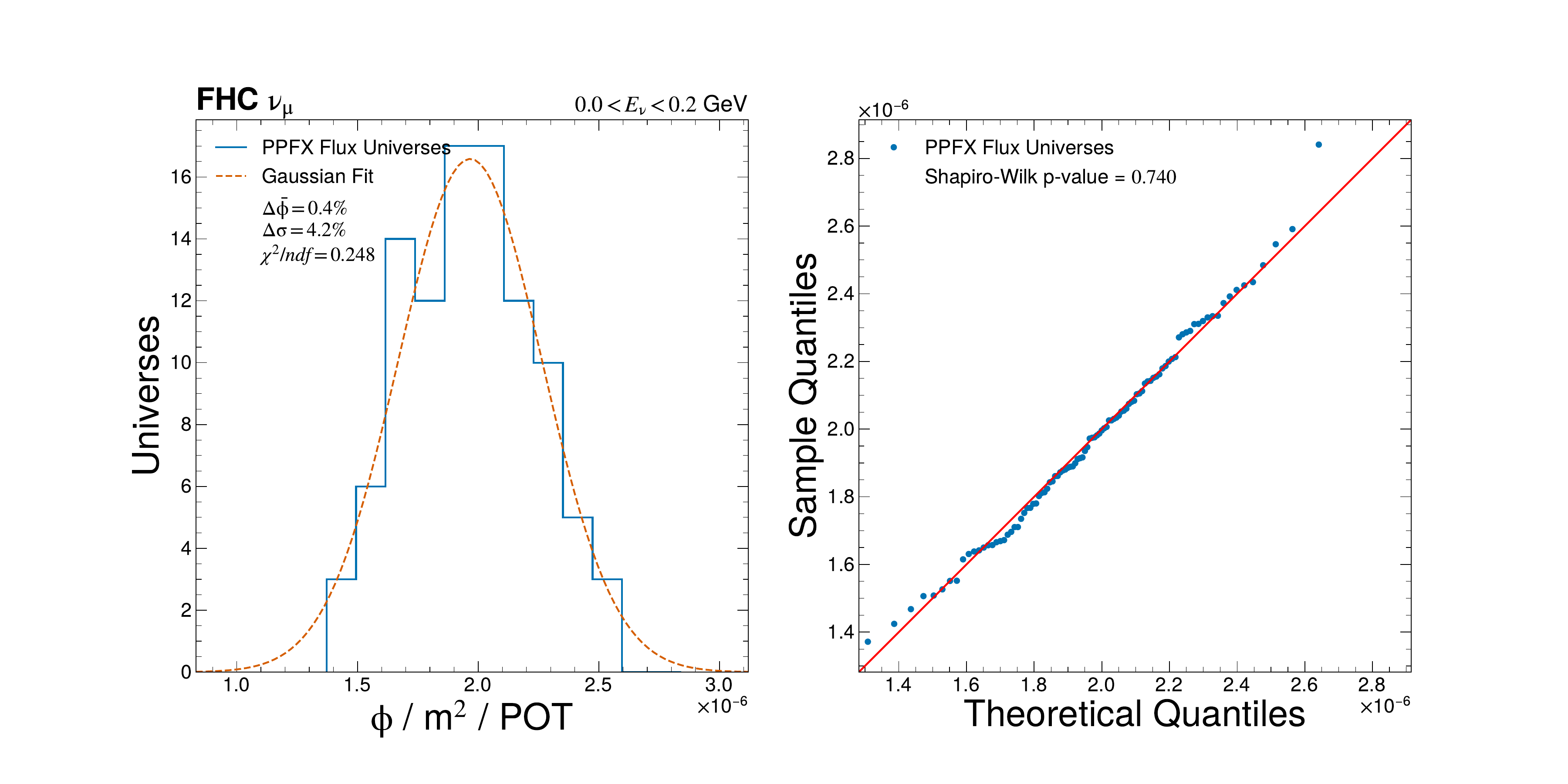}
    \includegraphics[width=0.3\textwidth]{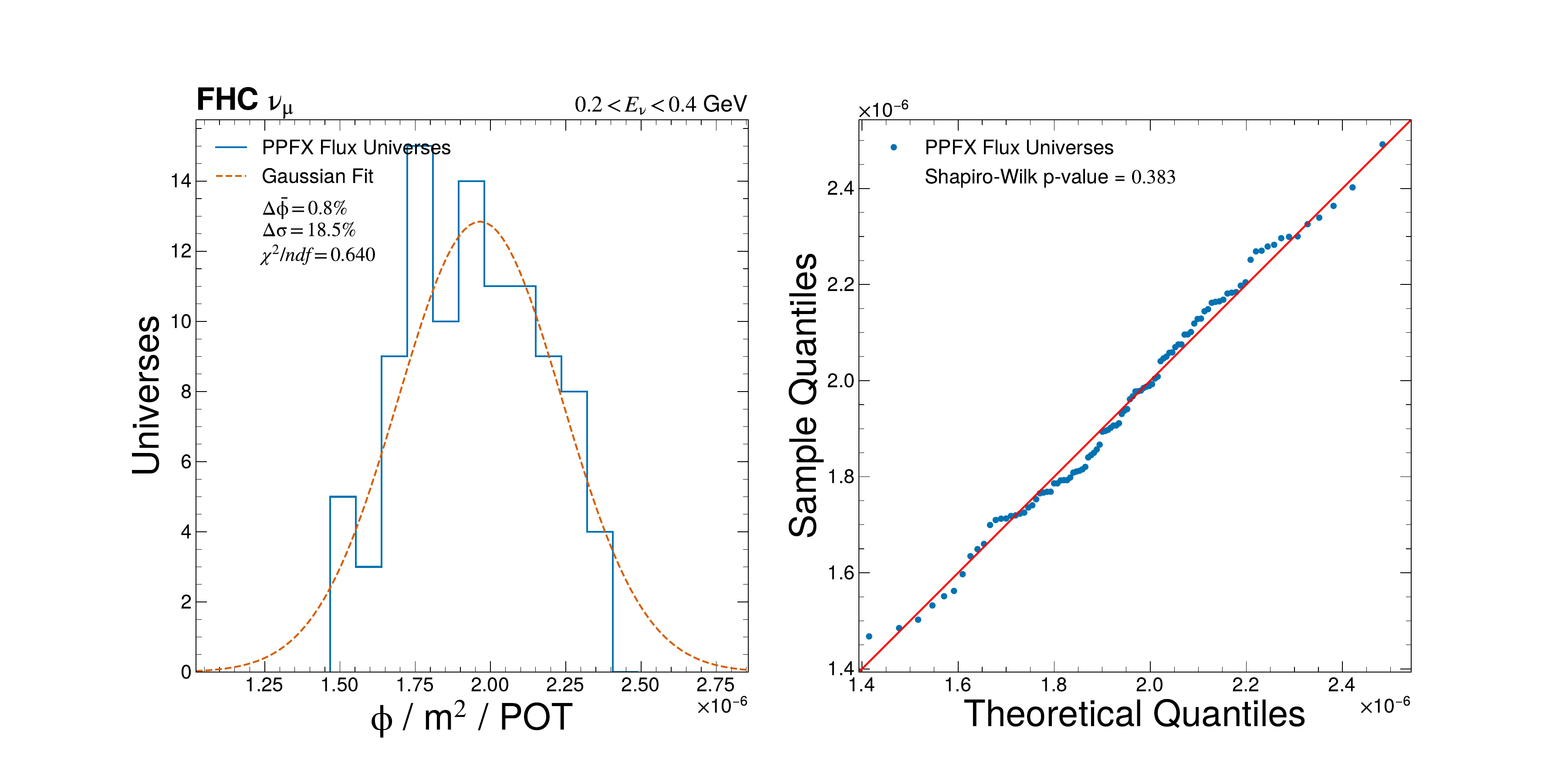}
    \includegraphics[width=0.3\textwidth]{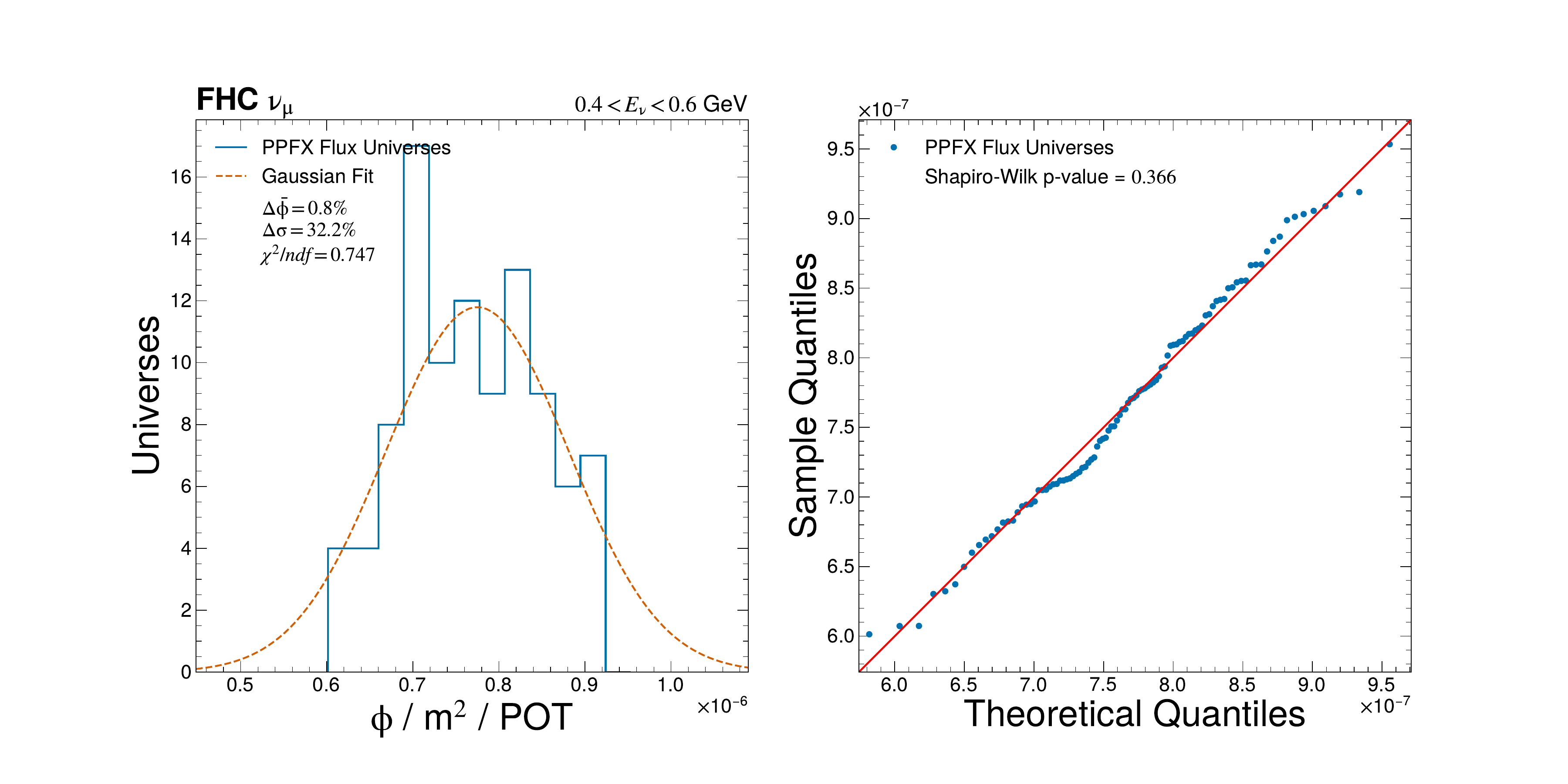}
    \includegraphics[width=0.3\textwidth]{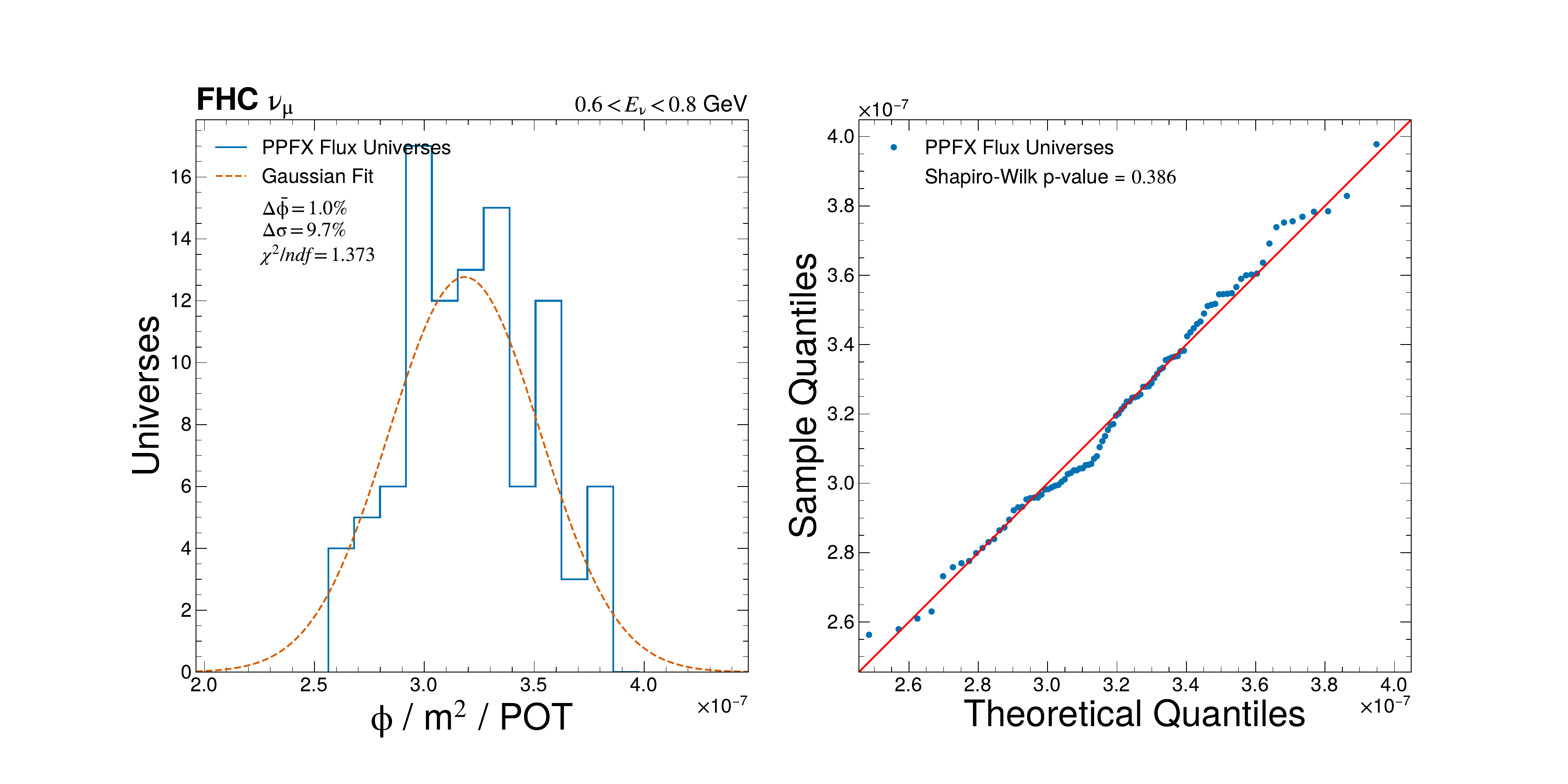}
    \includegraphics[width=0.3\textwidth]{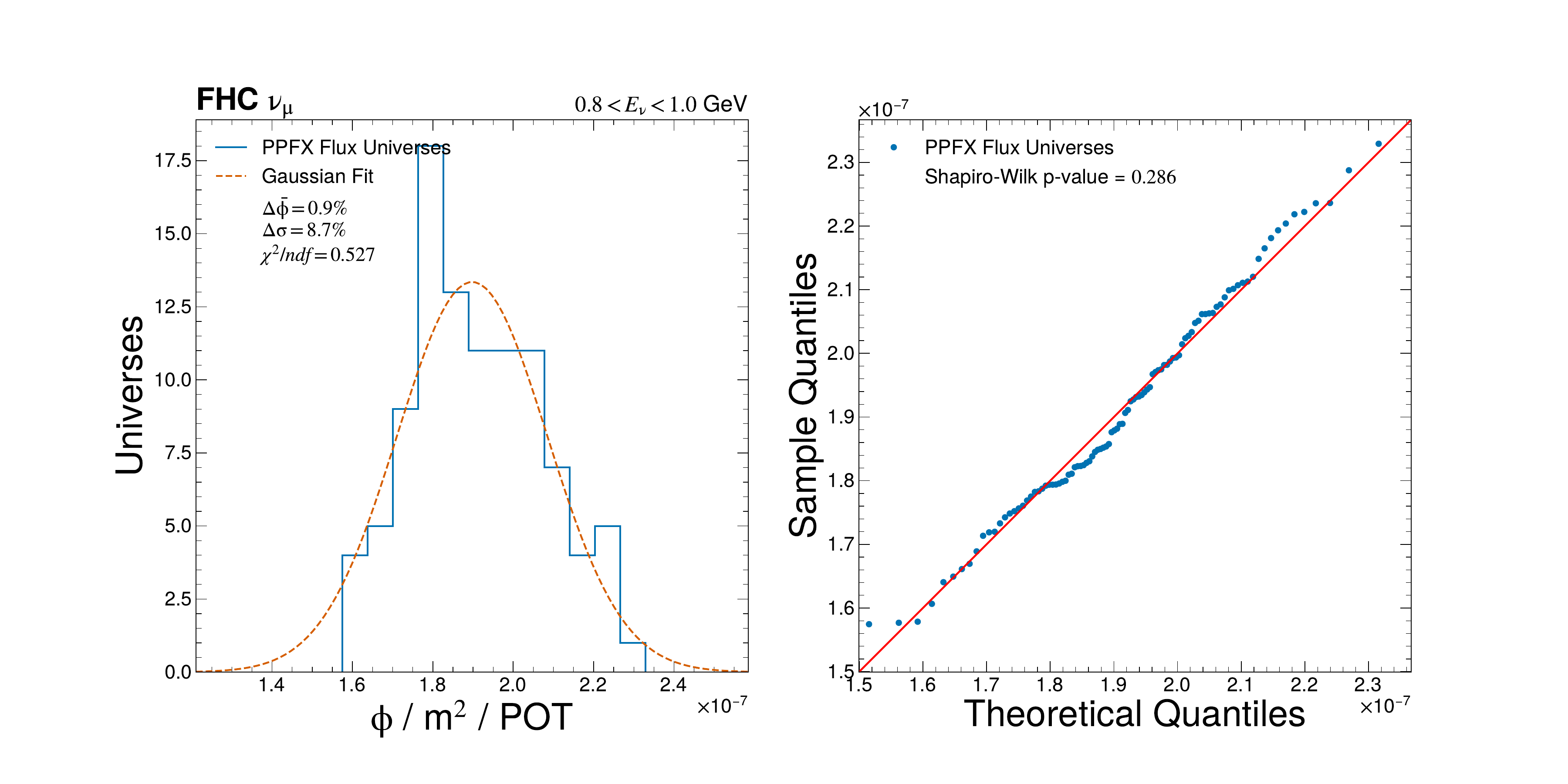}
    \includegraphics[width=0.3\textwidth]{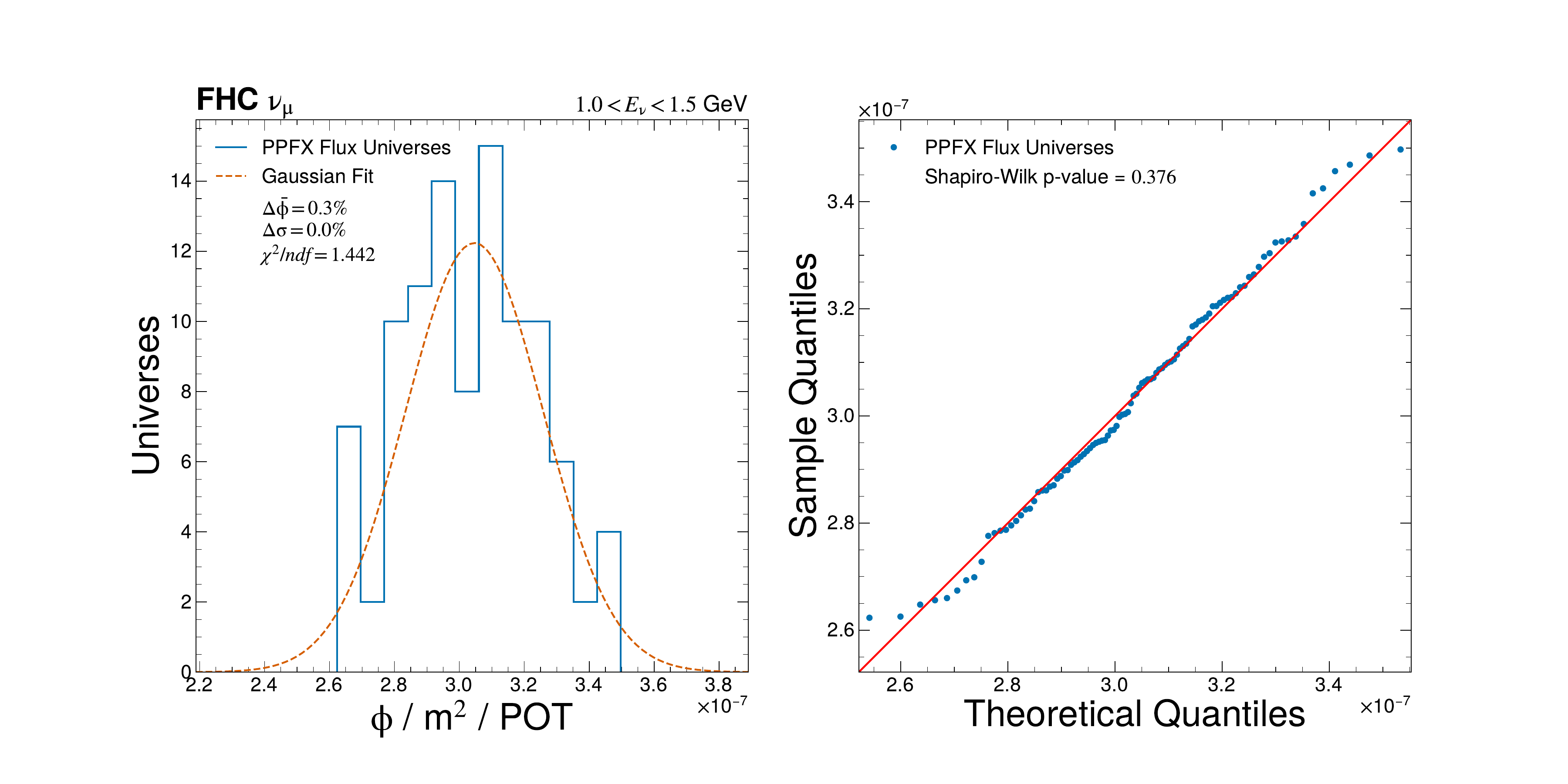}
    \includegraphics[width=0.3\textwidth]{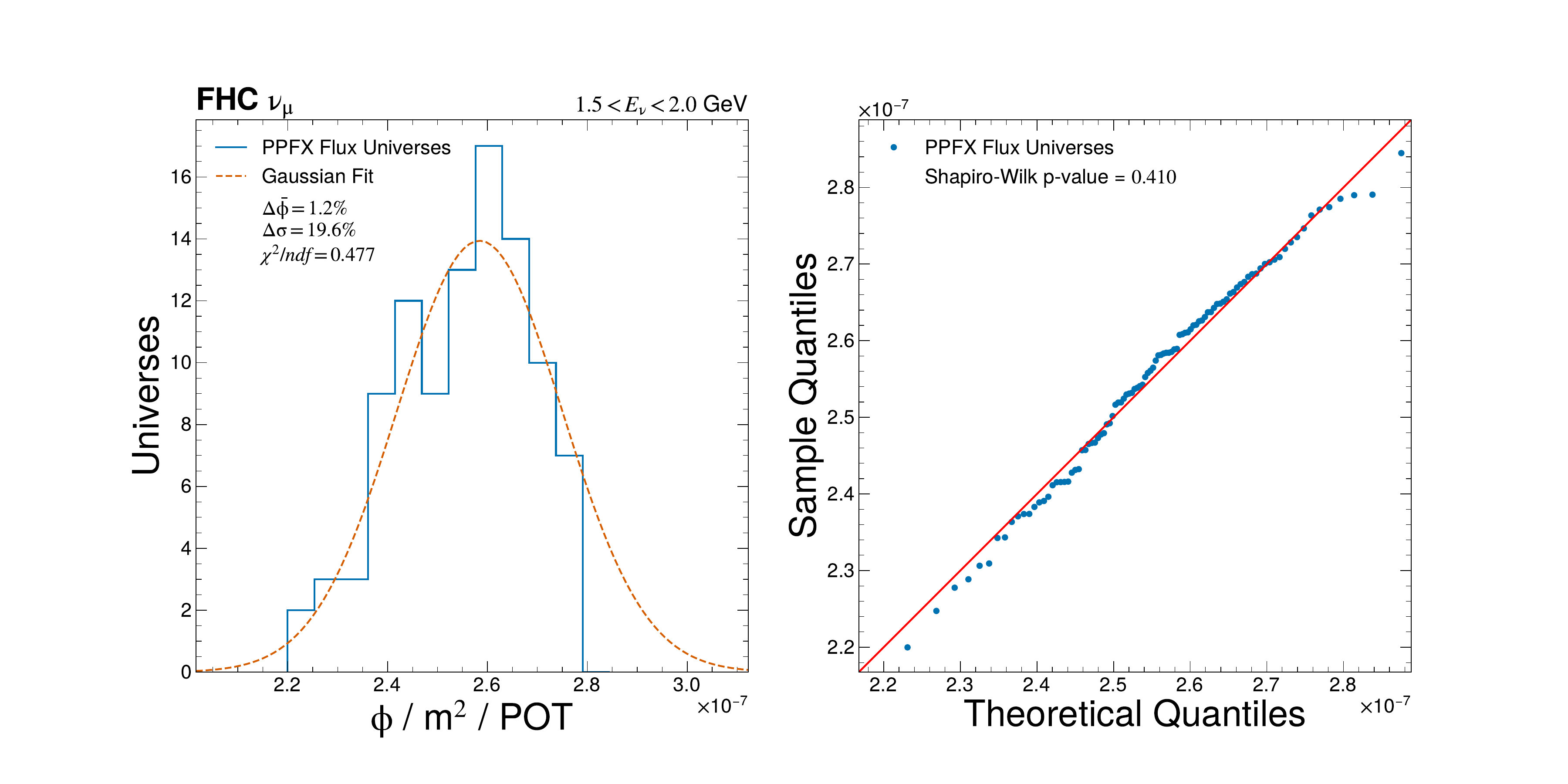}
    \includegraphics[width=0.3\textwidth]{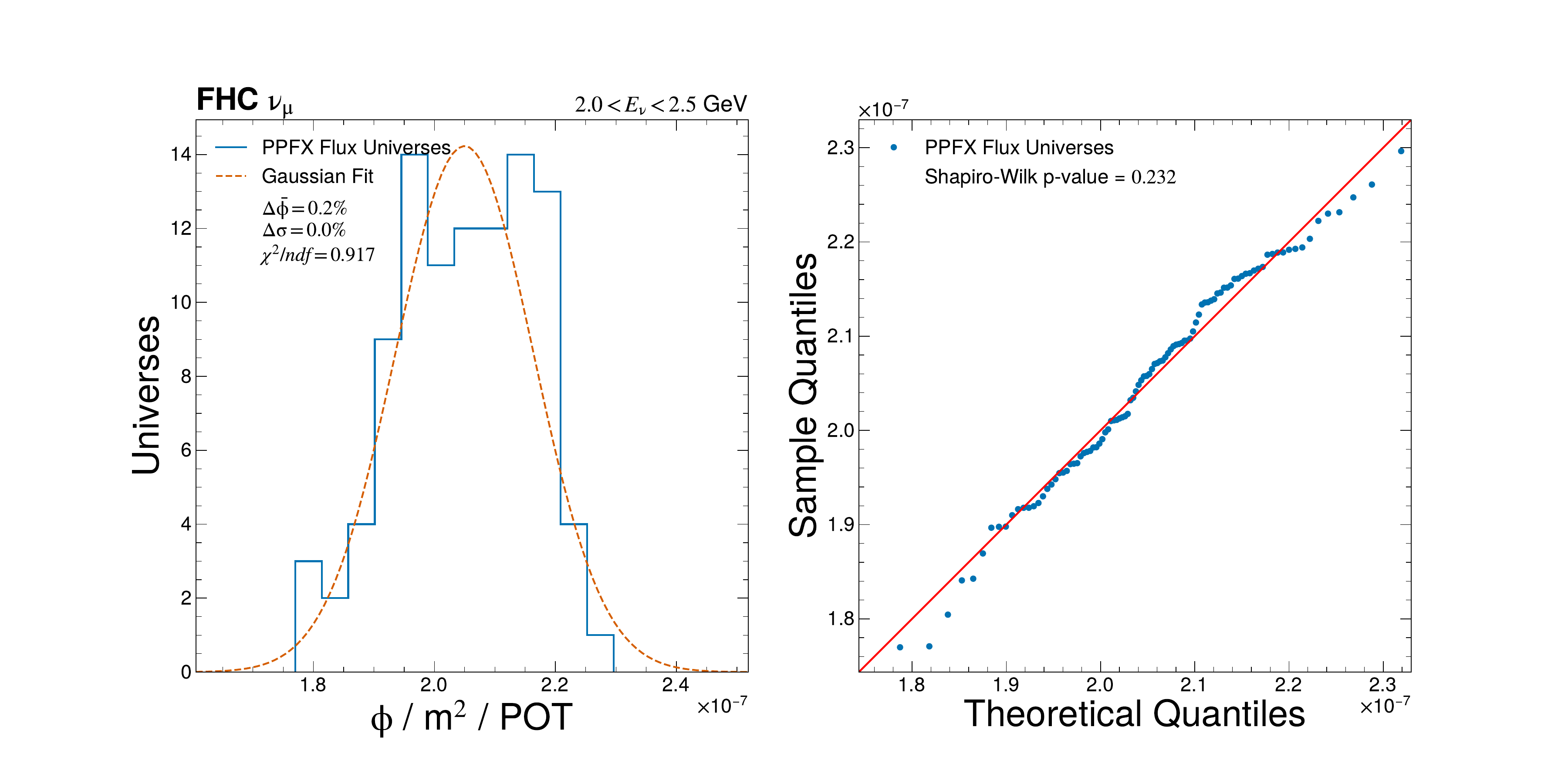}
    \includegraphics[width=0.3\textwidth]{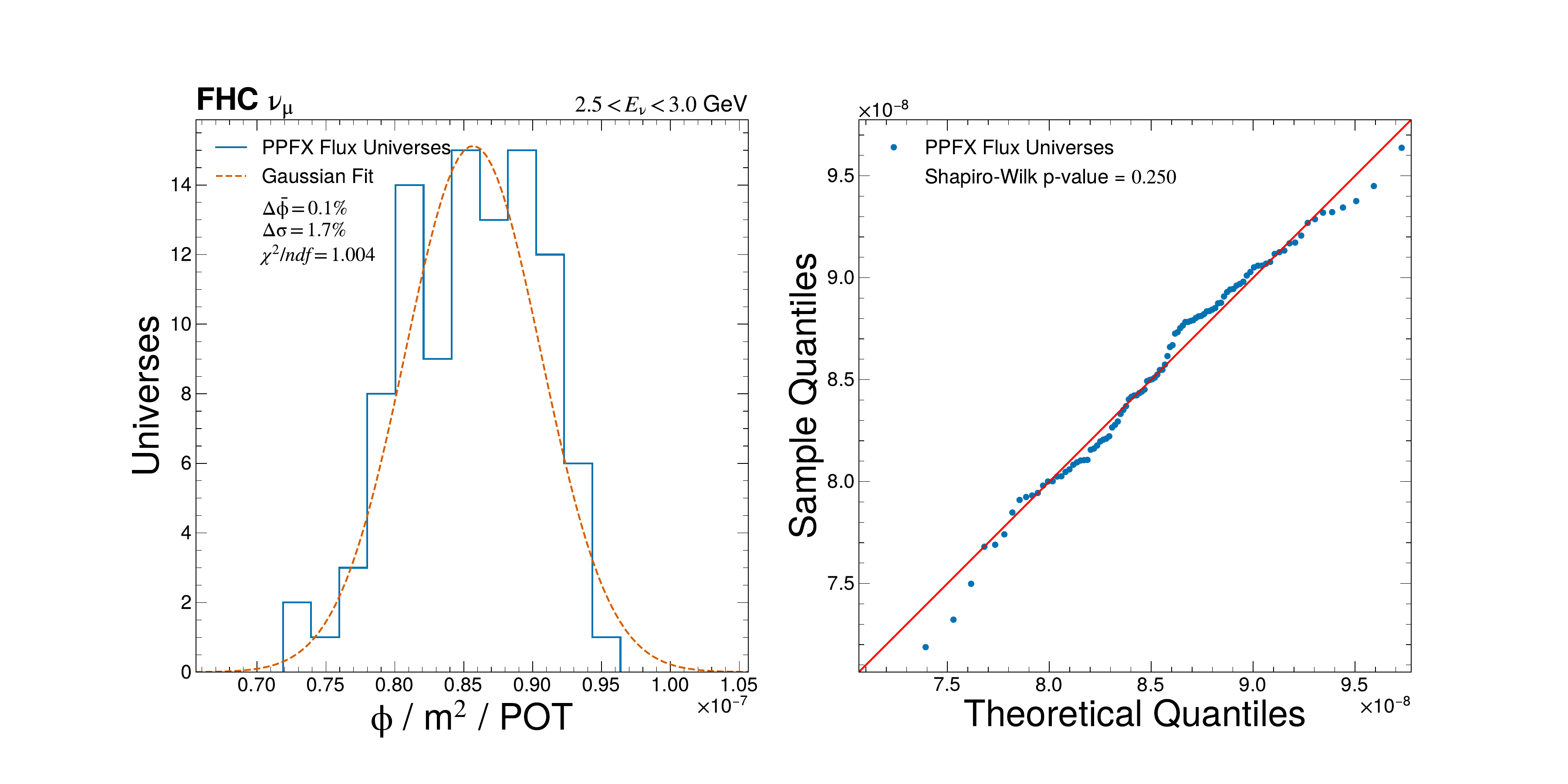}
    \includegraphics[width=0.3\textwidth]{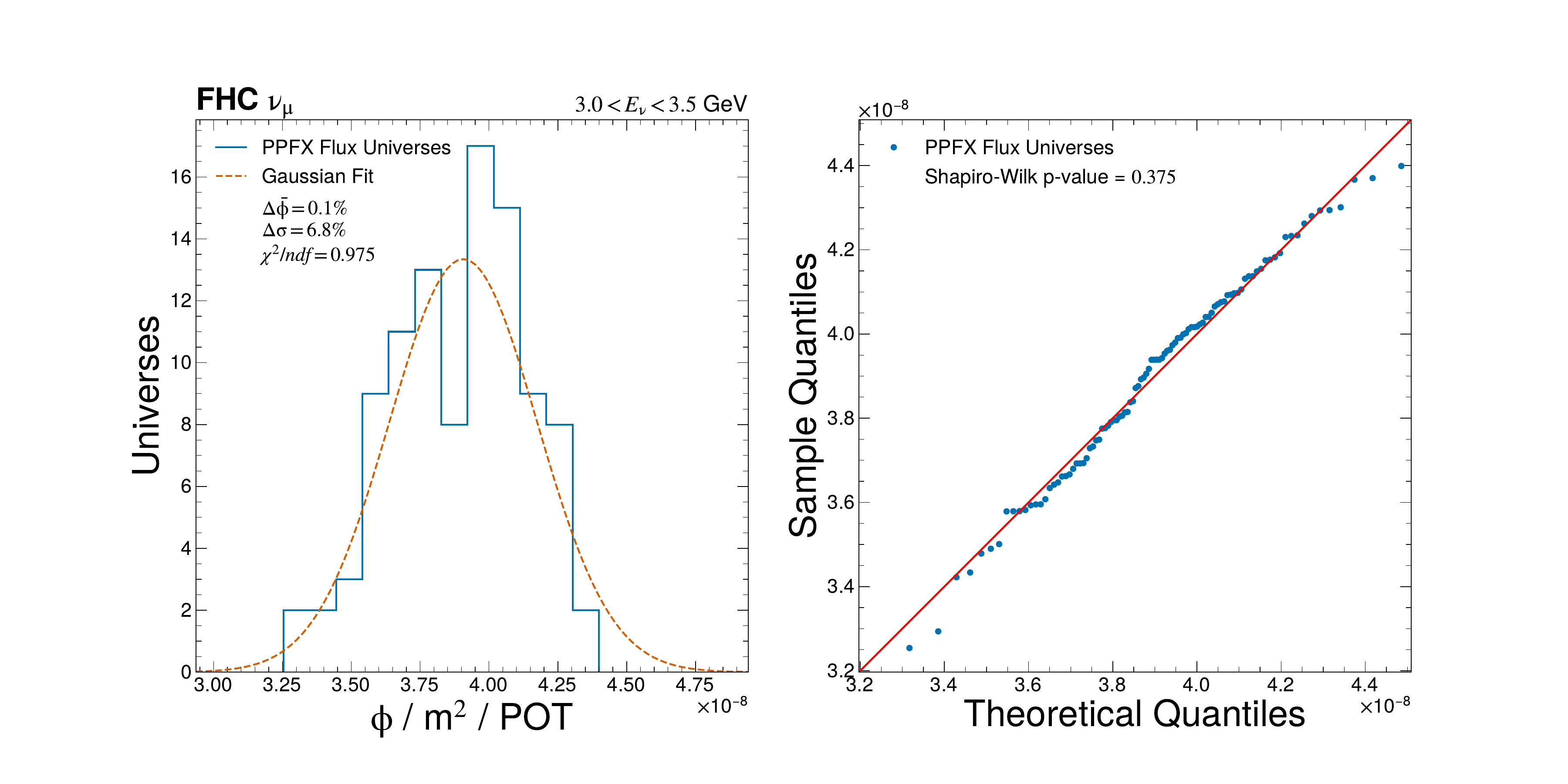}
    \includegraphics[width=0.3\textwidth]{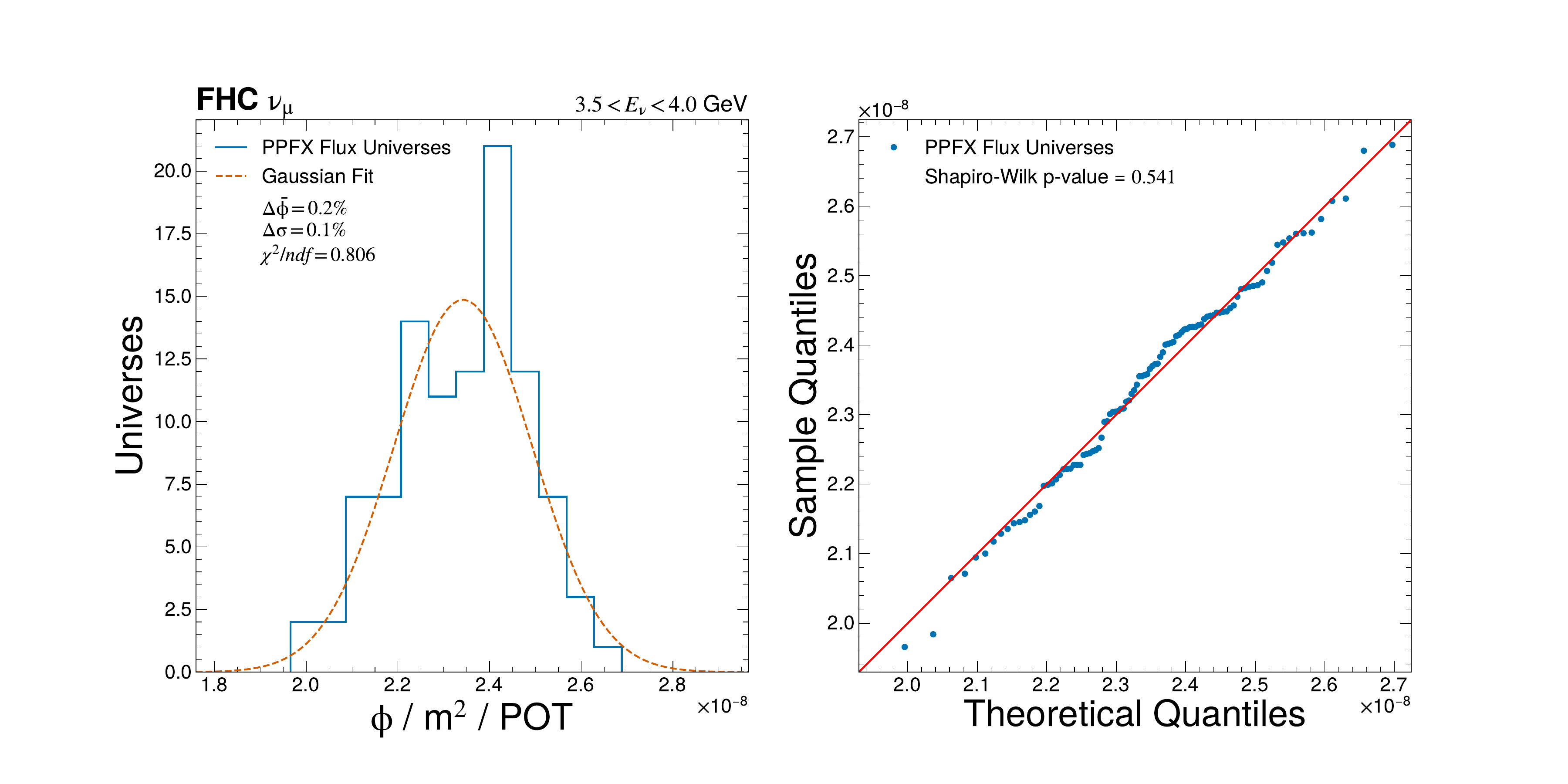}
    \includegraphics[width=0.3\textwidth]{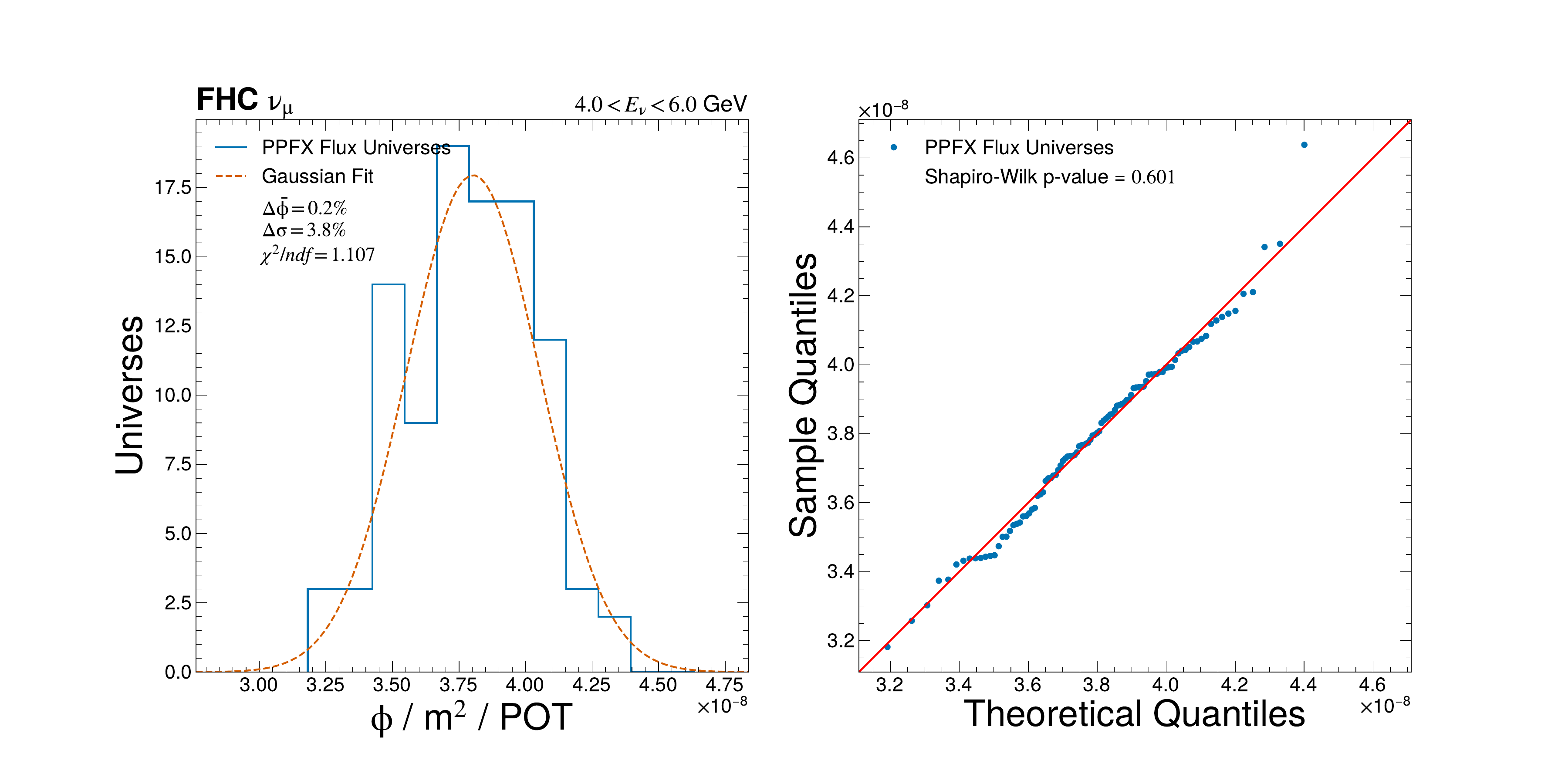}
    \includegraphics[width=0.3\textwidth]{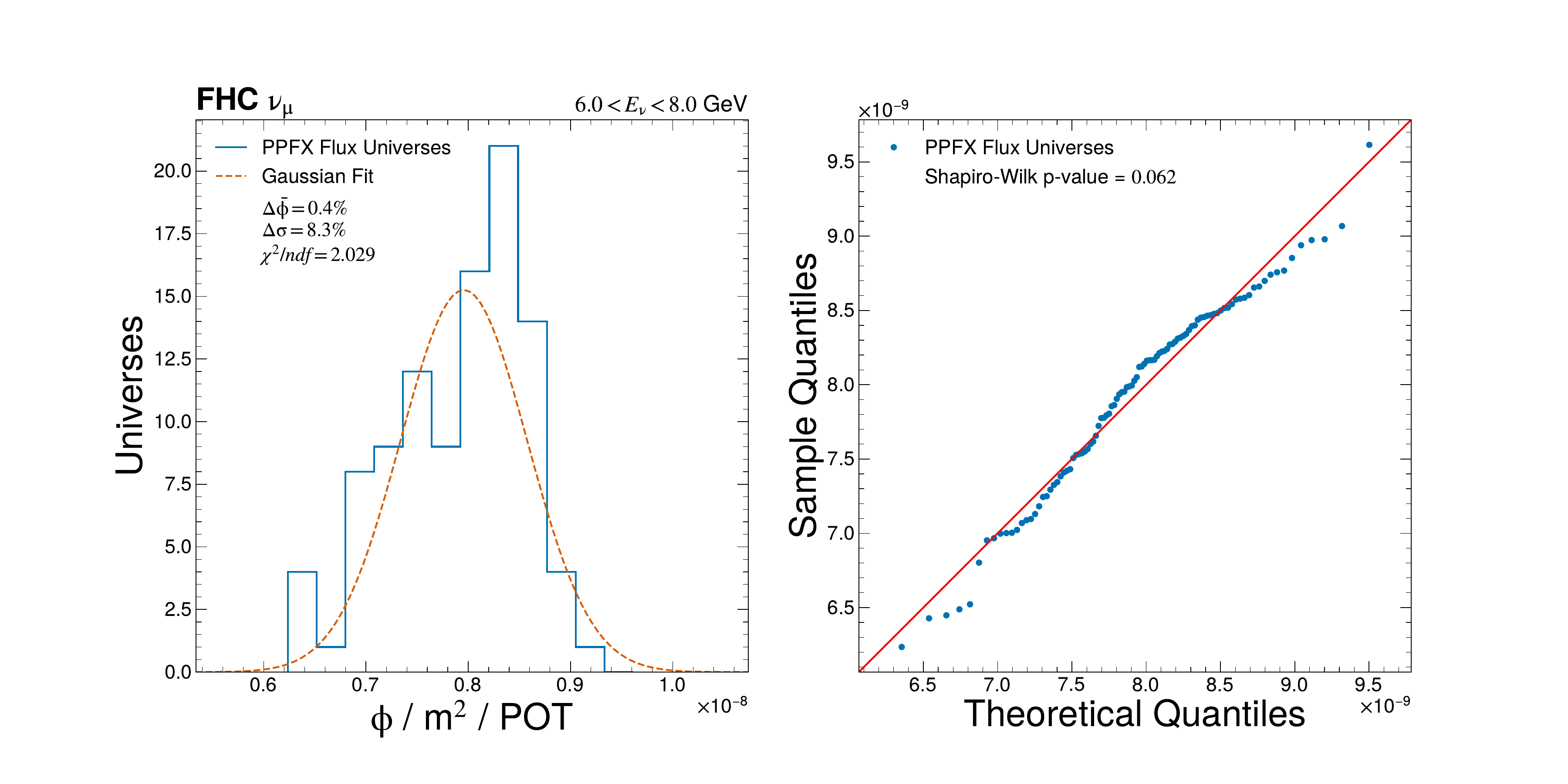}
    \includegraphics[width=0.3\textwidth]{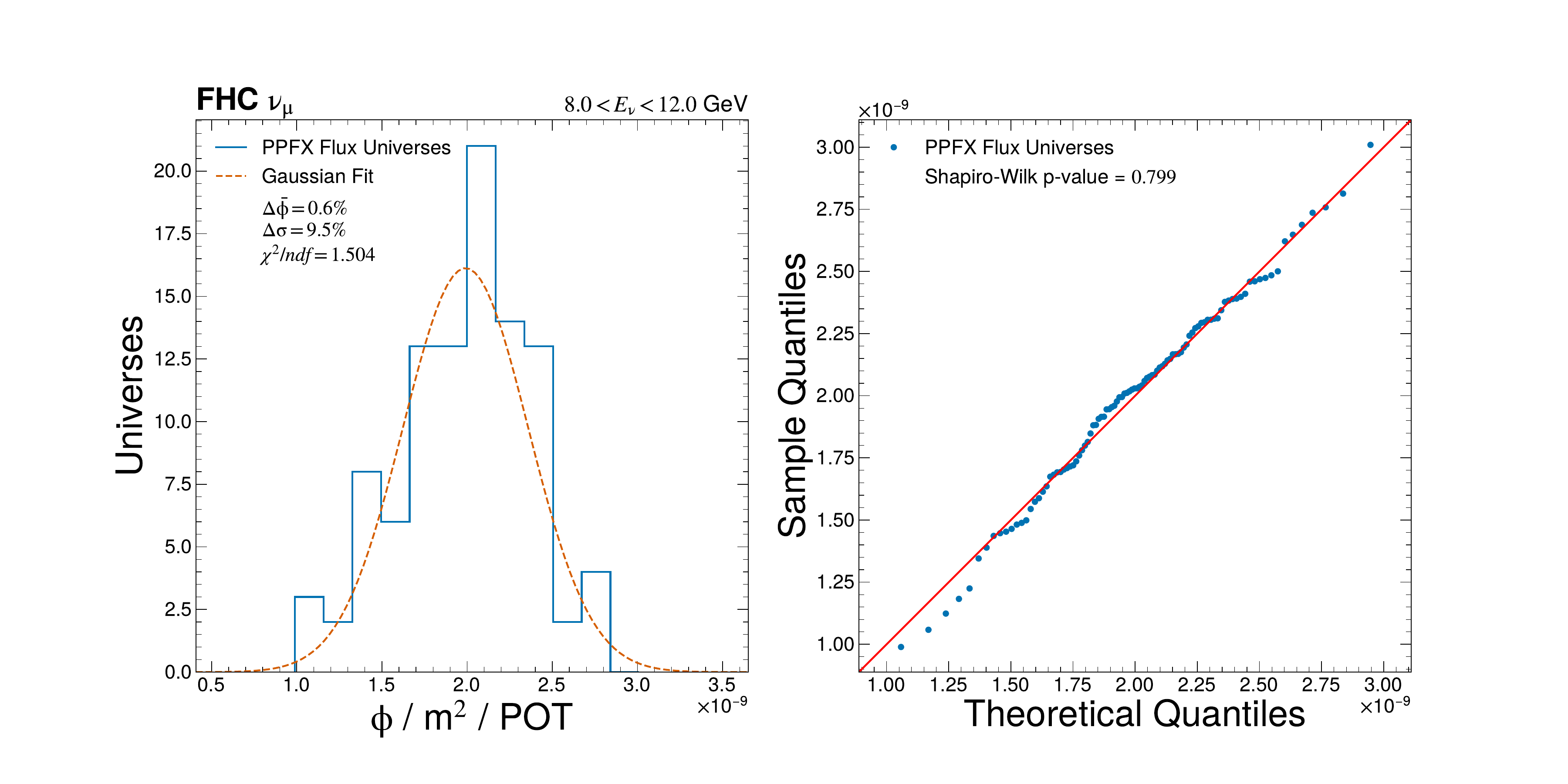}
    \caption[Distribution of PPFX universes for \numu\ (FHC).]{Distribution of PPFX universes for \numu.}
\end{figure}
\begin{figure}[!ht]
    \centering
    \includegraphics[width=0.3\textwidth]{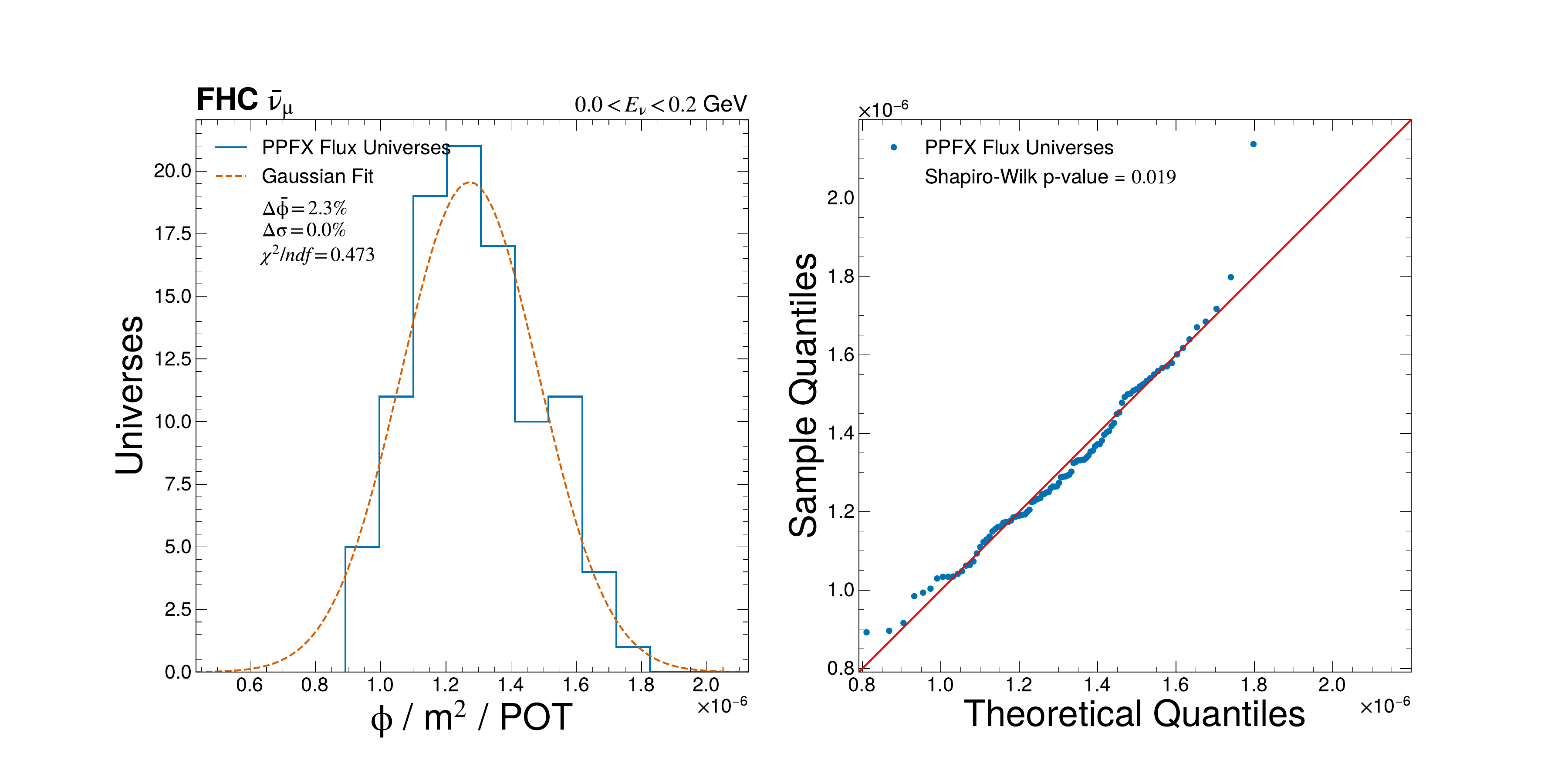}
    \includegraphics[width=0.3\textwidth]{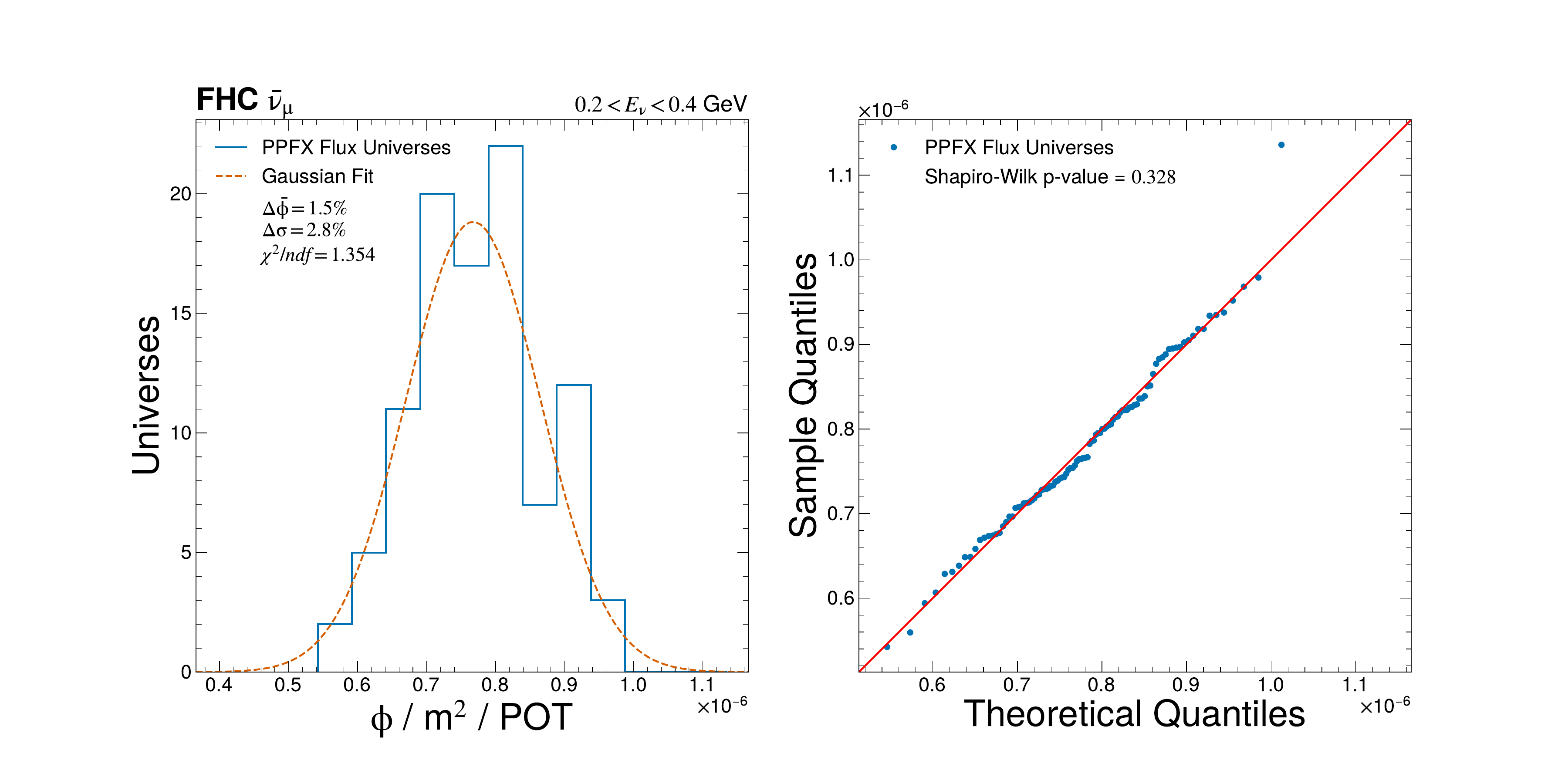}
    \includegraphics[width=0.3\textwidth]{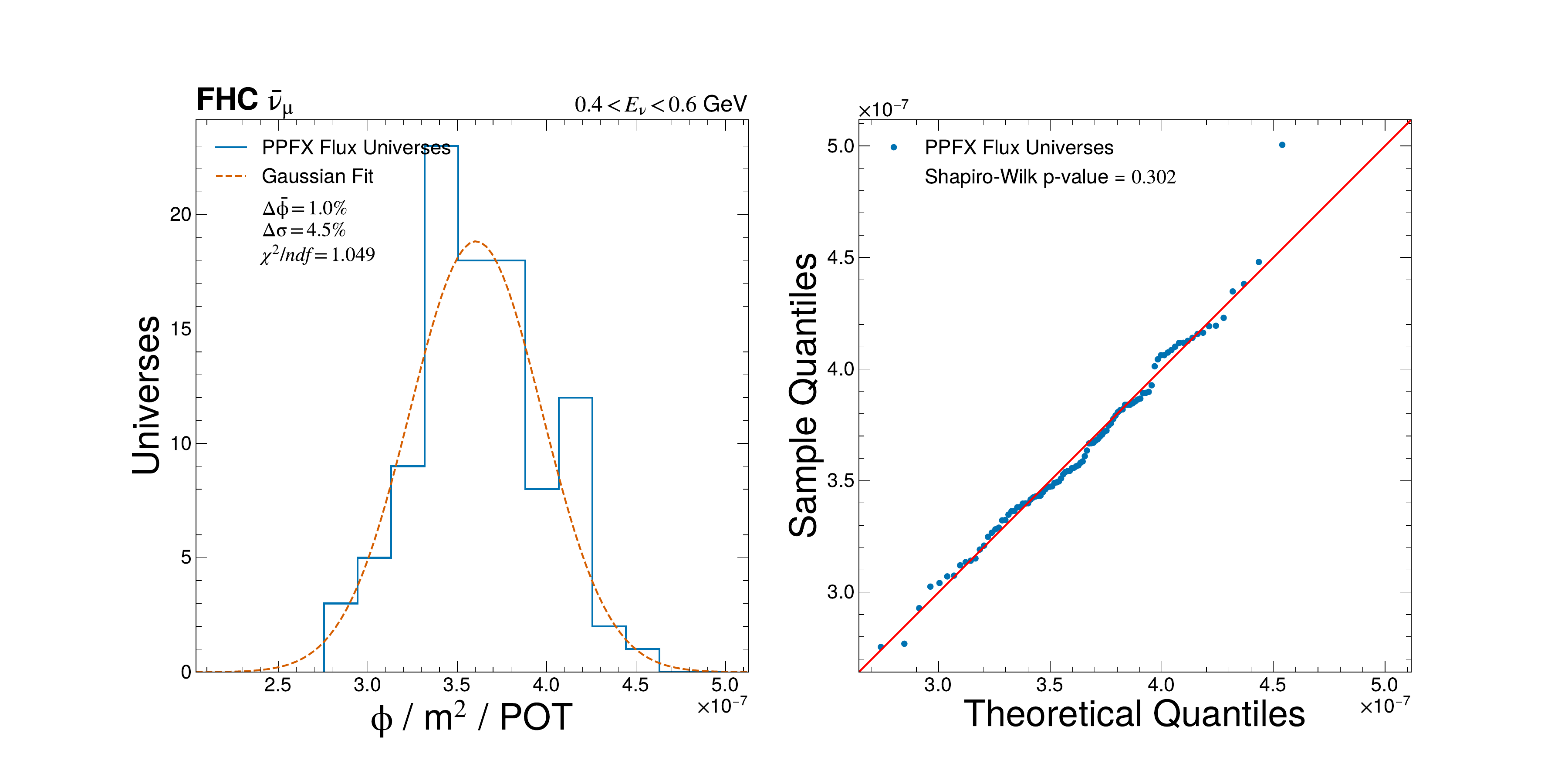}
    \includegraphics[width=0.3\textwidth]{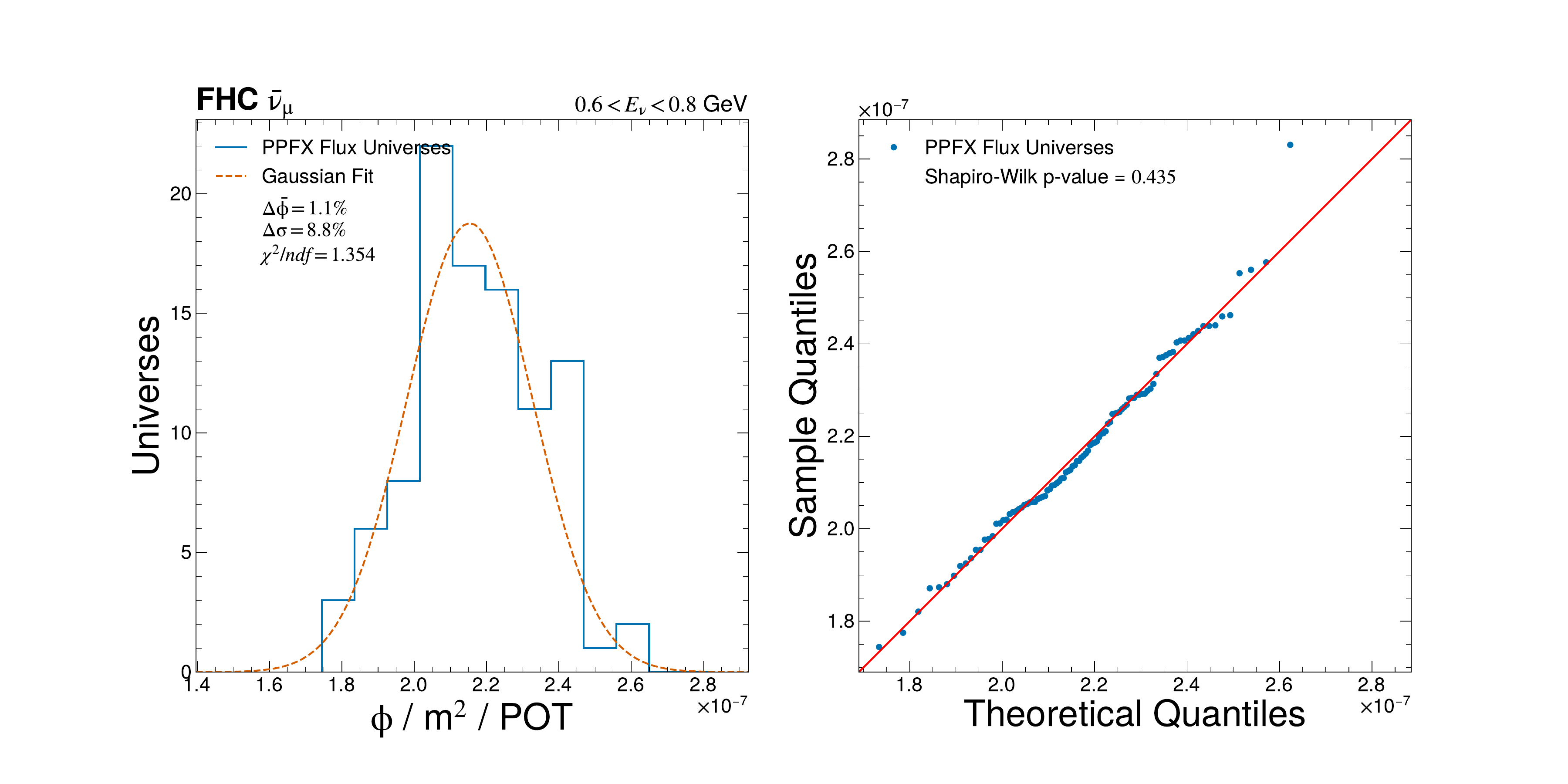}
    \includegraphics[width=0.3\textwidth]{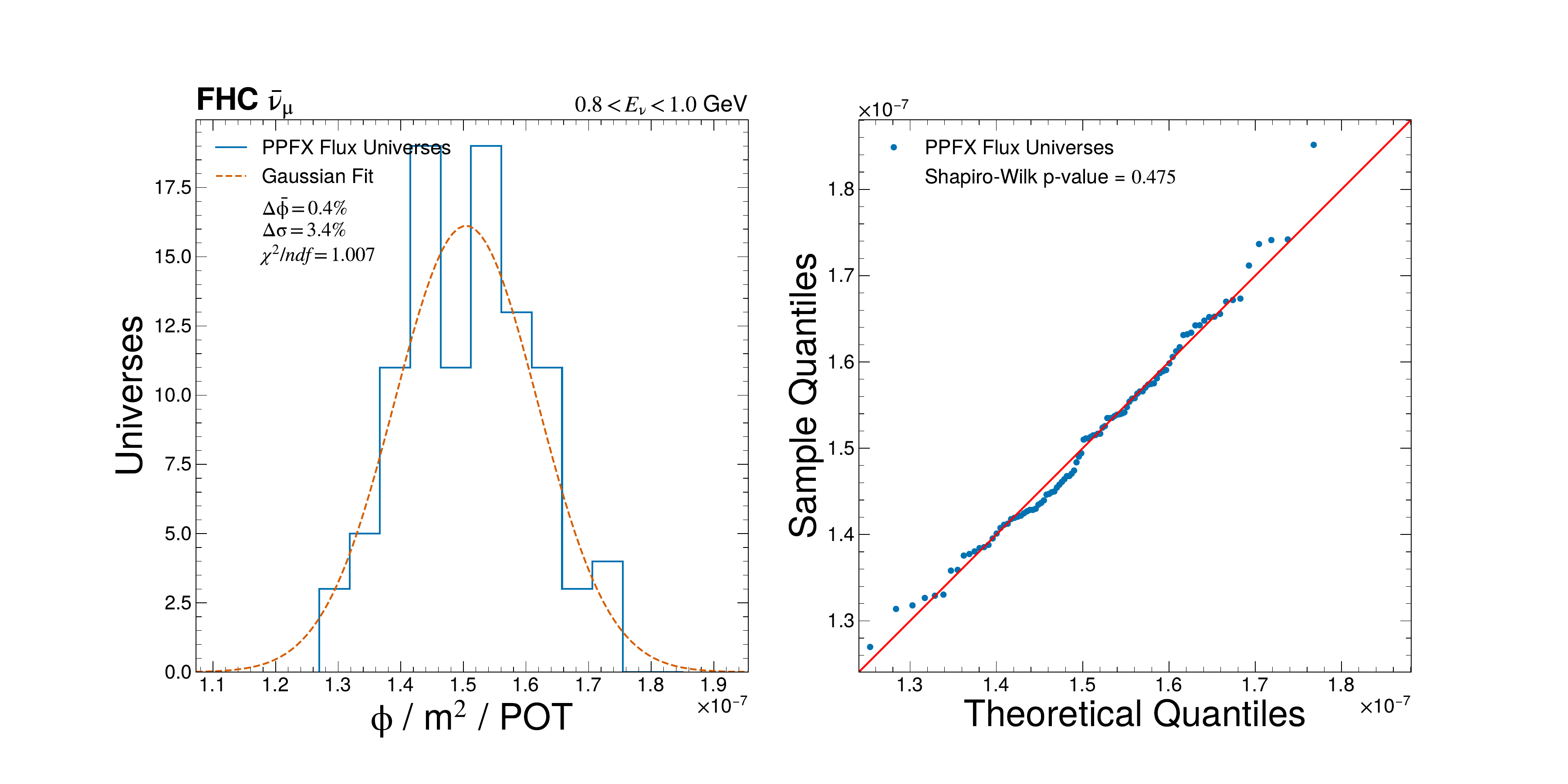}
    \includegraphics[width=0.3\textwidth]{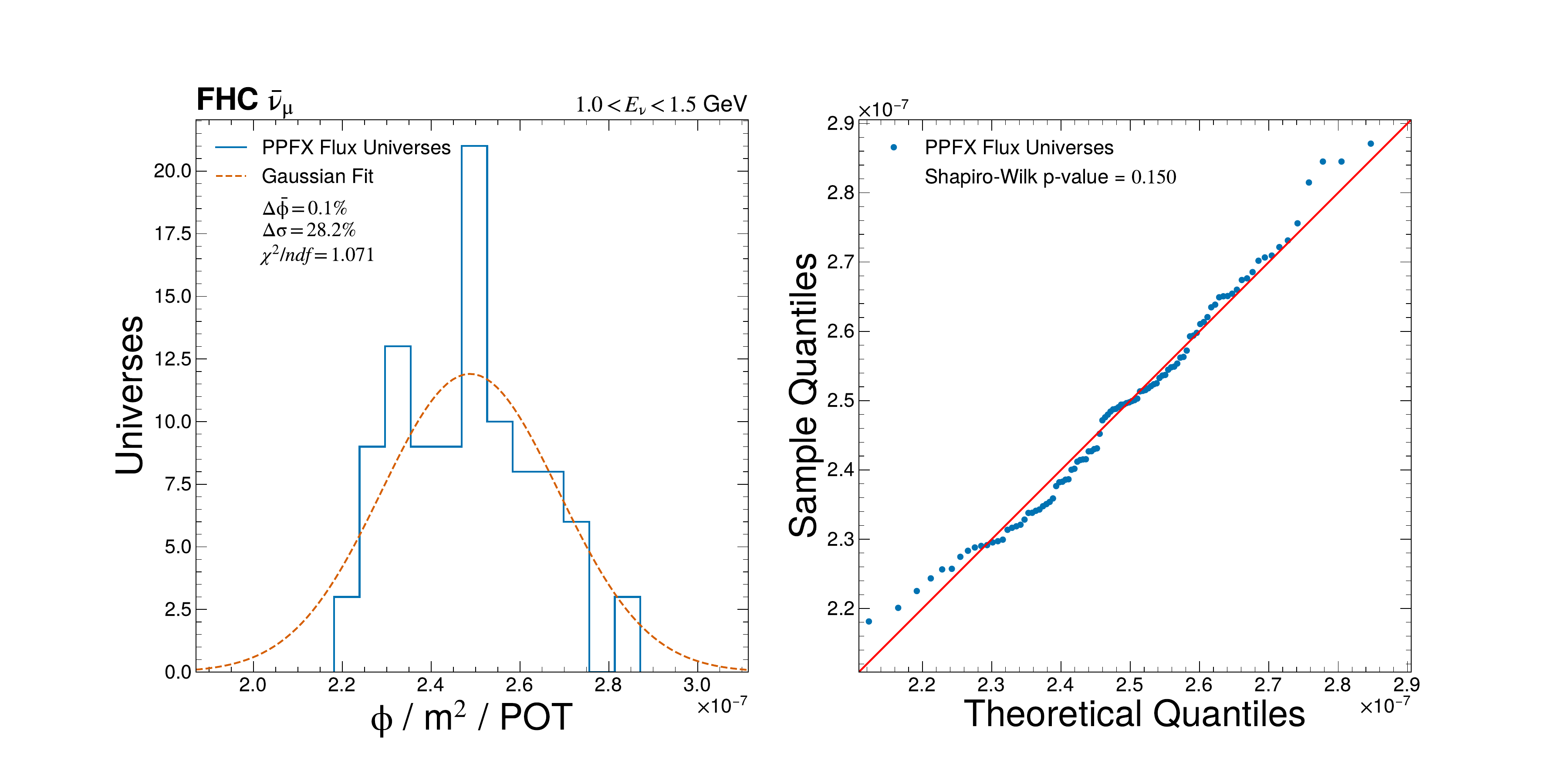}
    \includegraphics[width=0.3\textwidth]{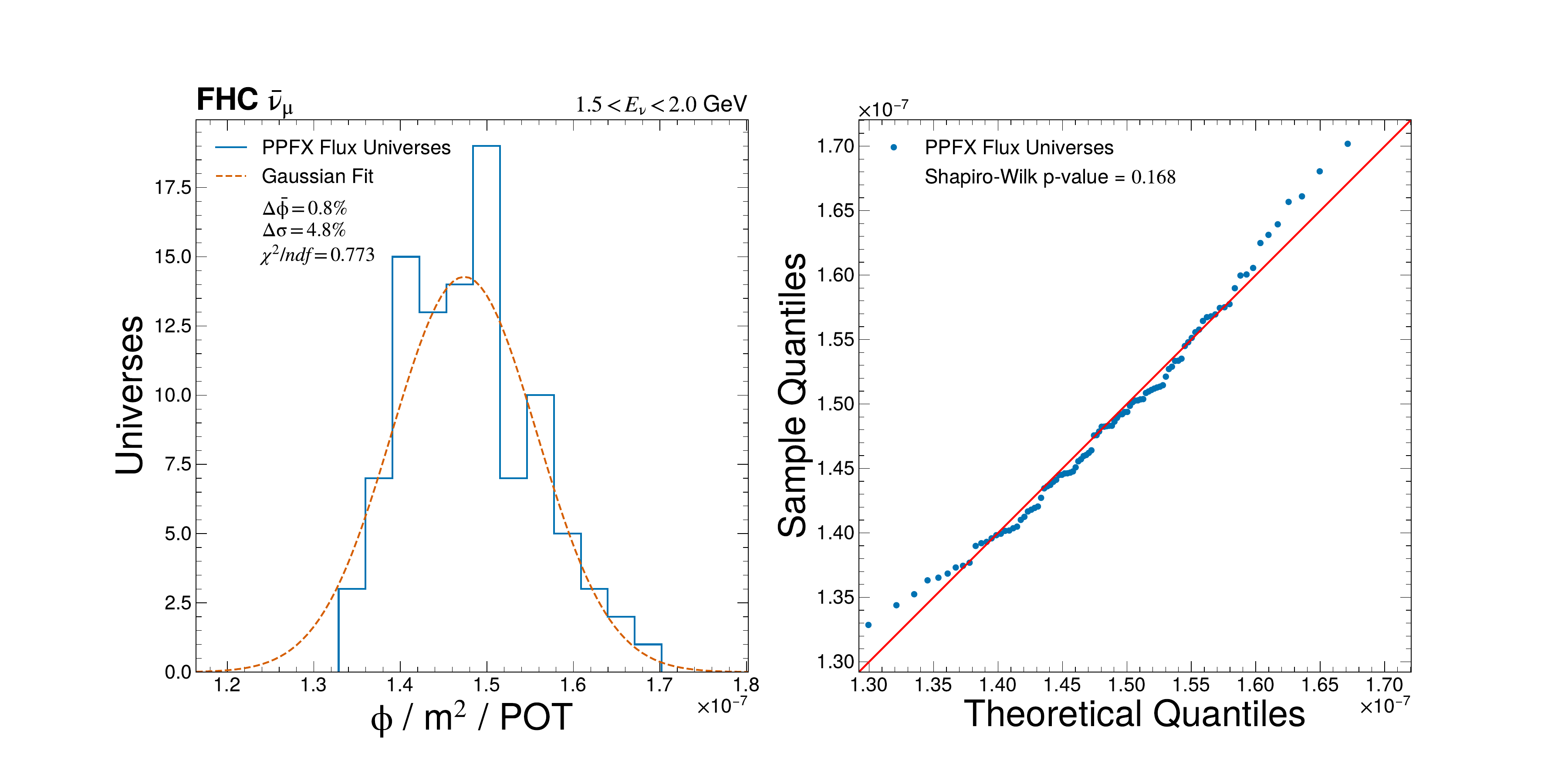}
    \includegraphics[width=0.3\textwidth]{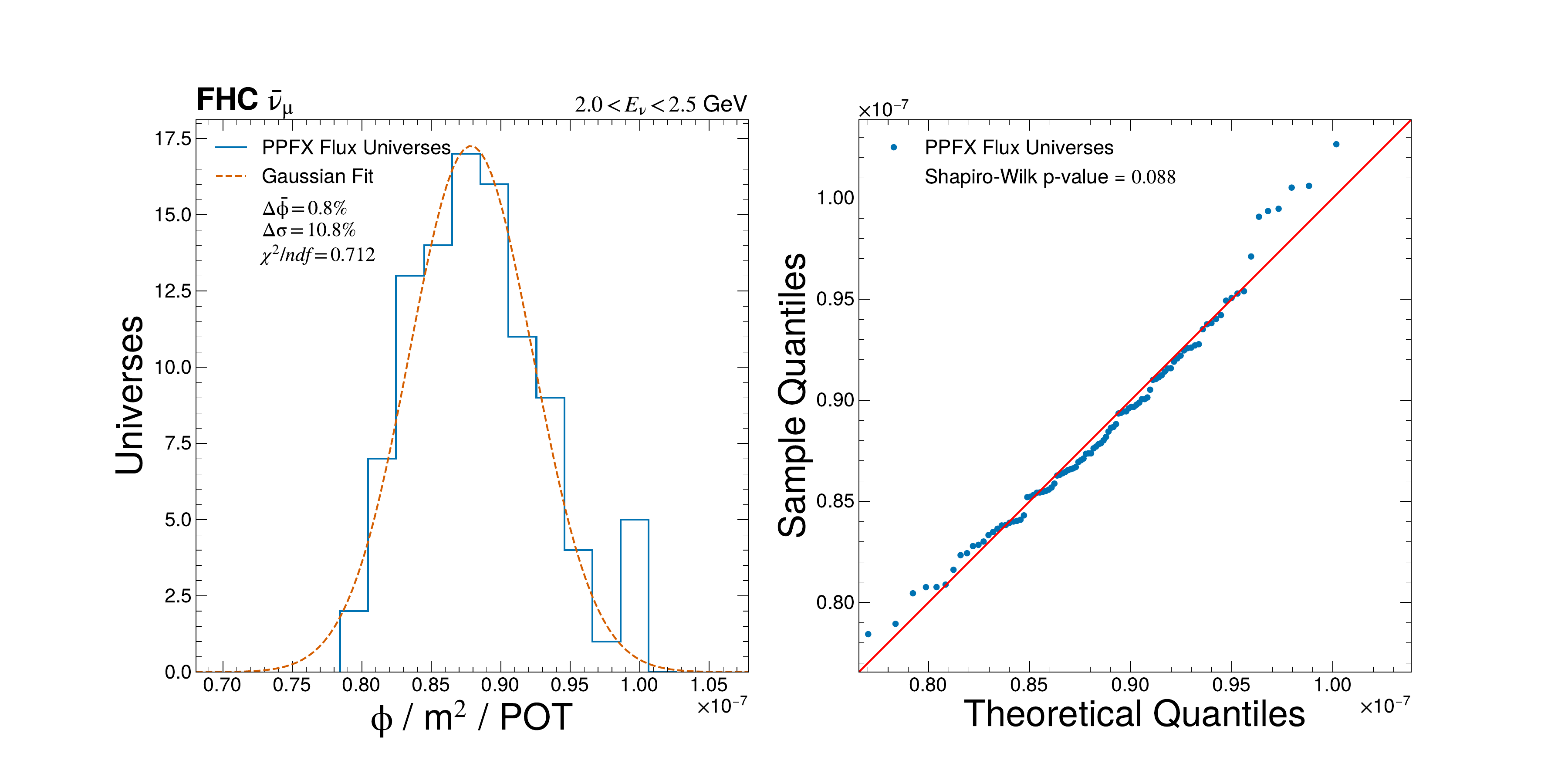}
    \includegraphics[width=0.3\textwidth]{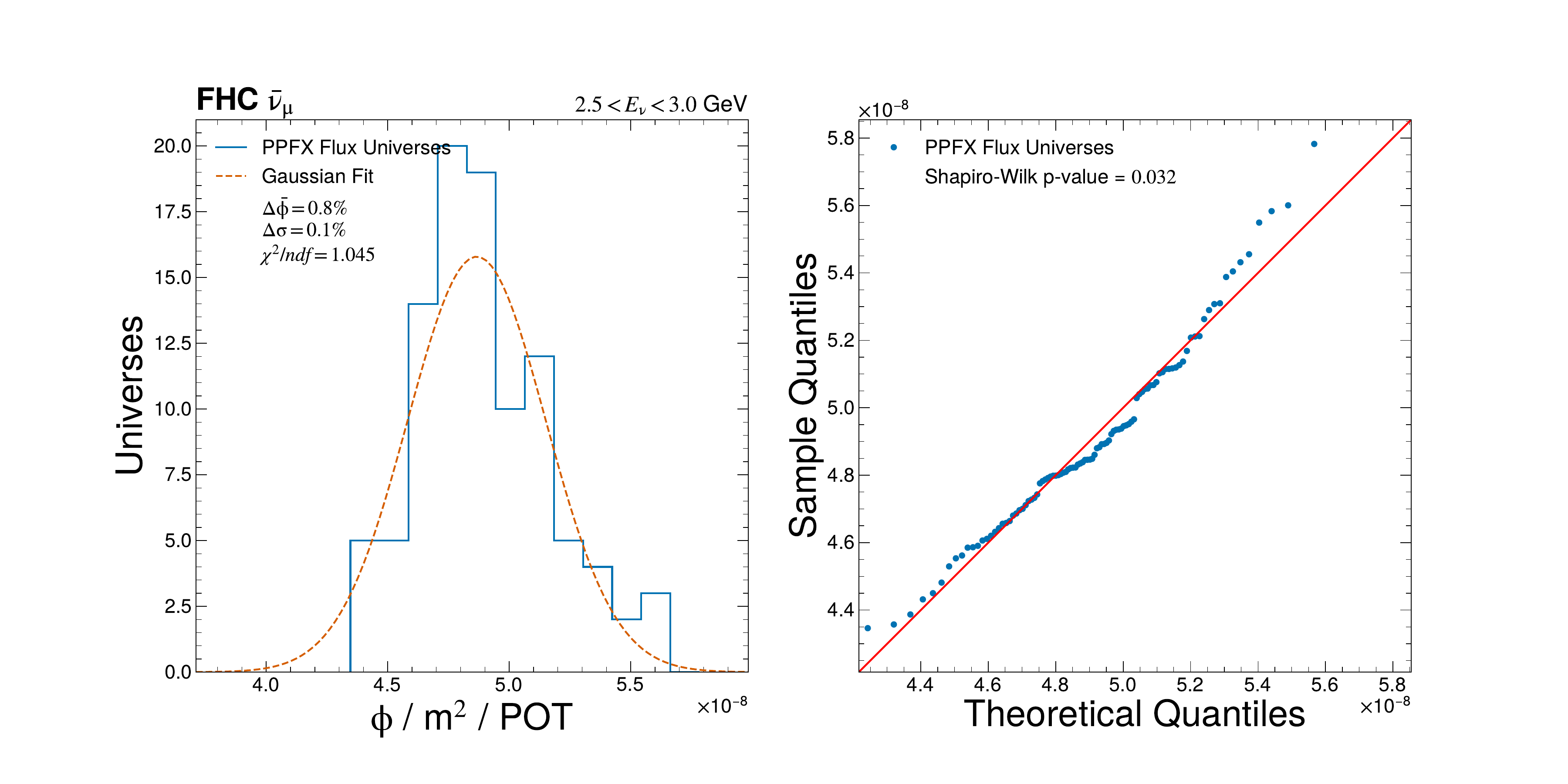}
    \includegraphics[width=0.3\textwidth]{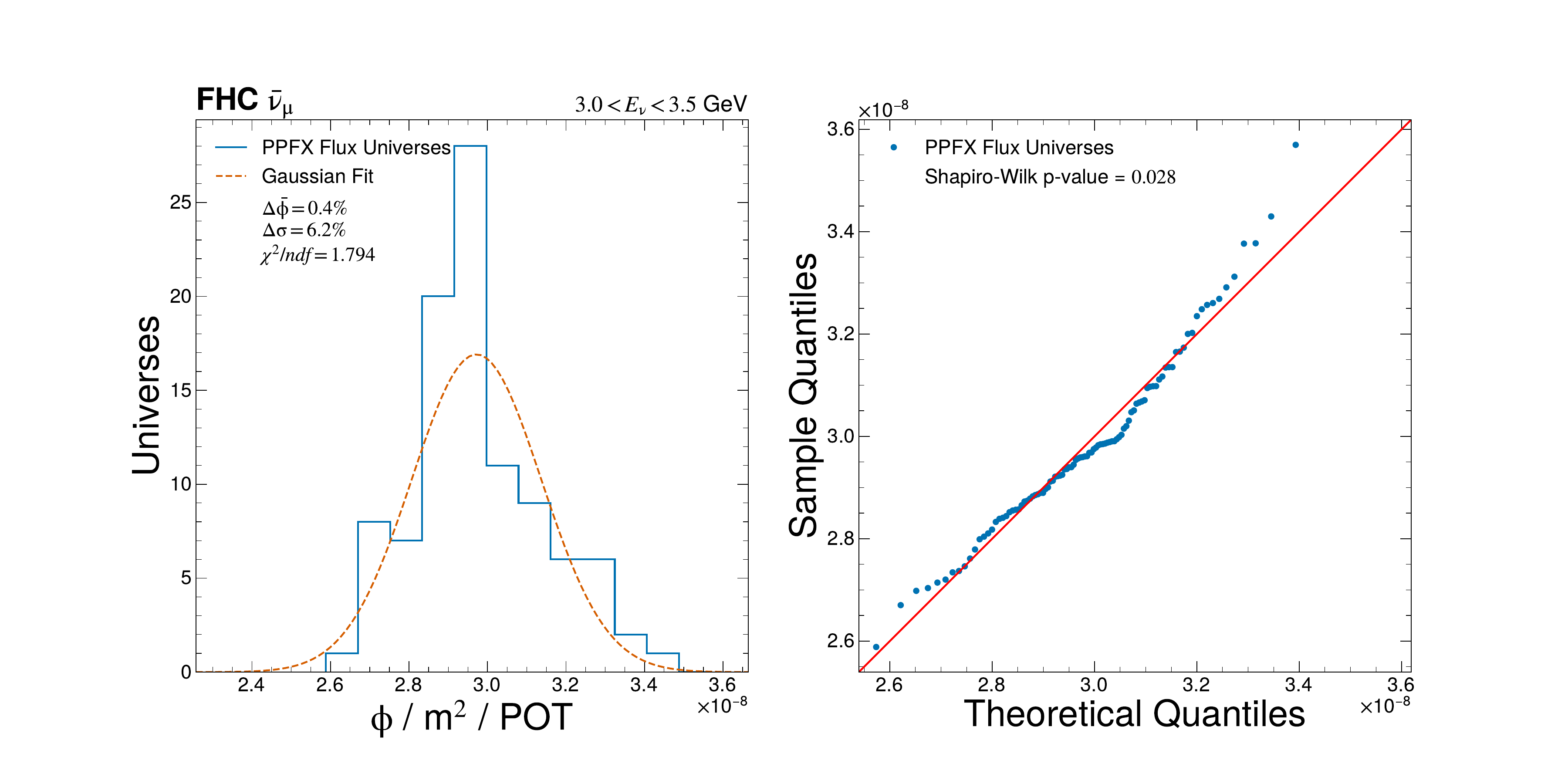}
    \includegraphics[width=0.3\textwidth]{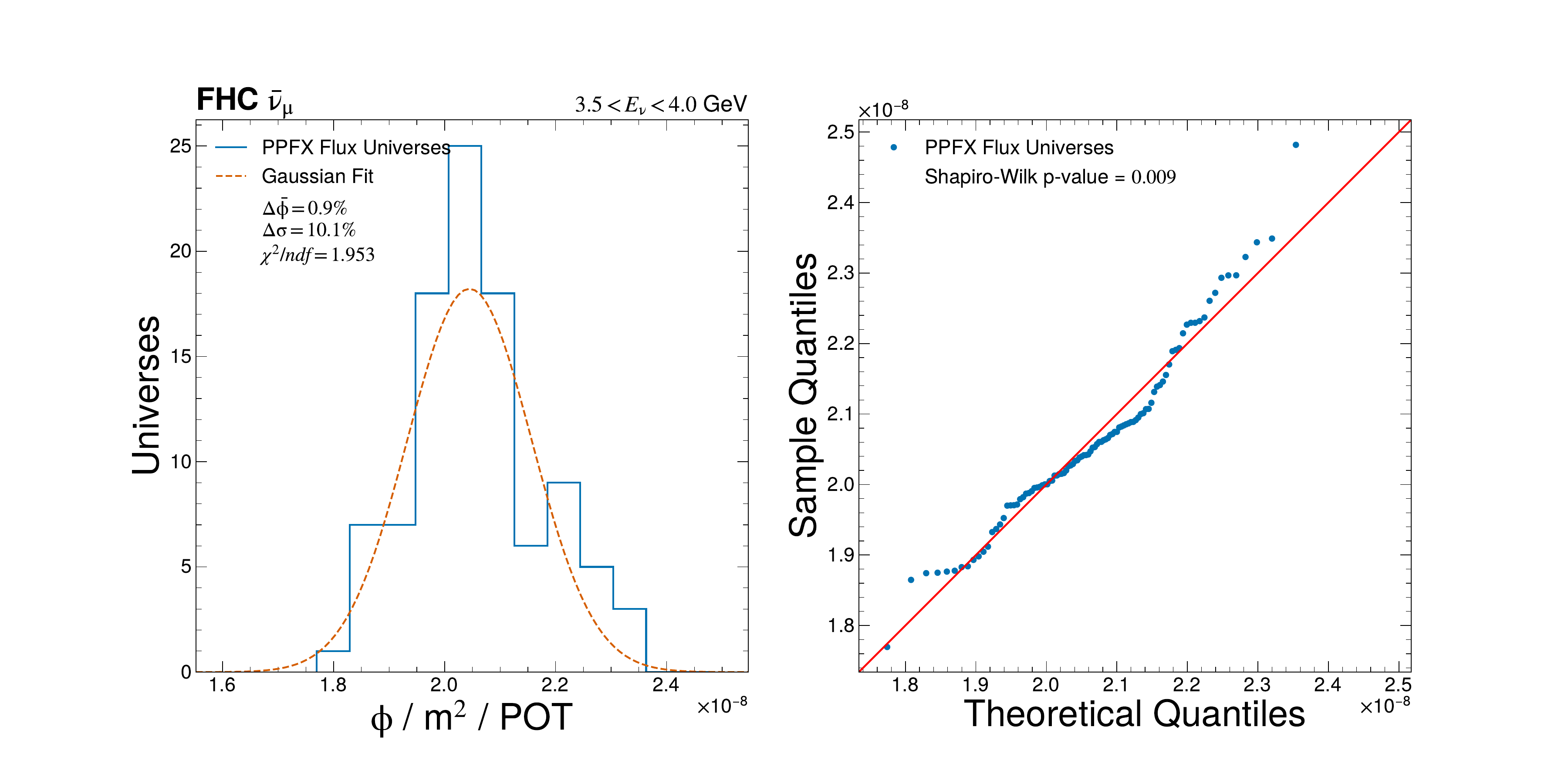}
    \includegraphics[width=0.3\textwidth]{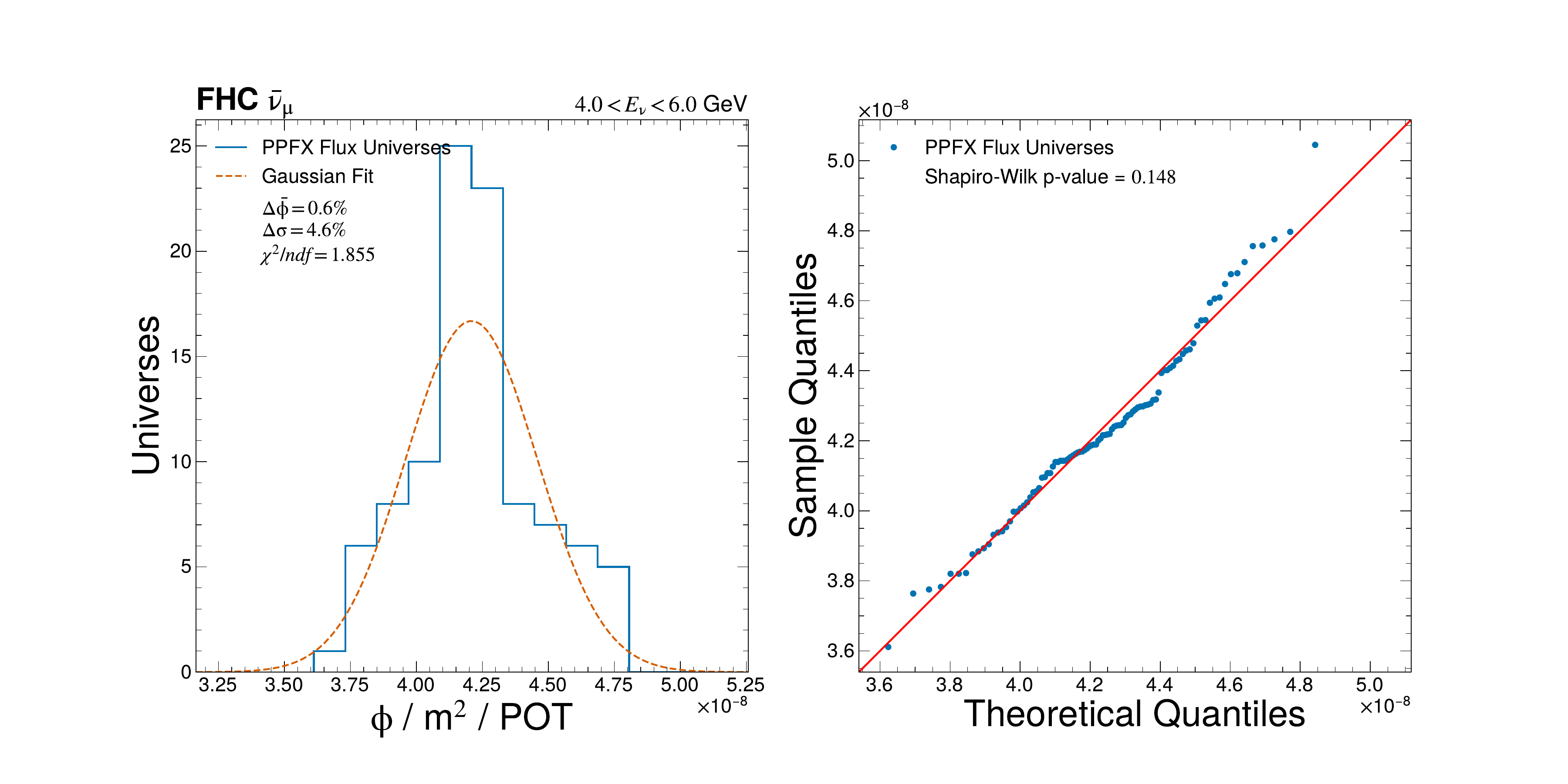}
    \includegraphics[width=0.3\textwidth]{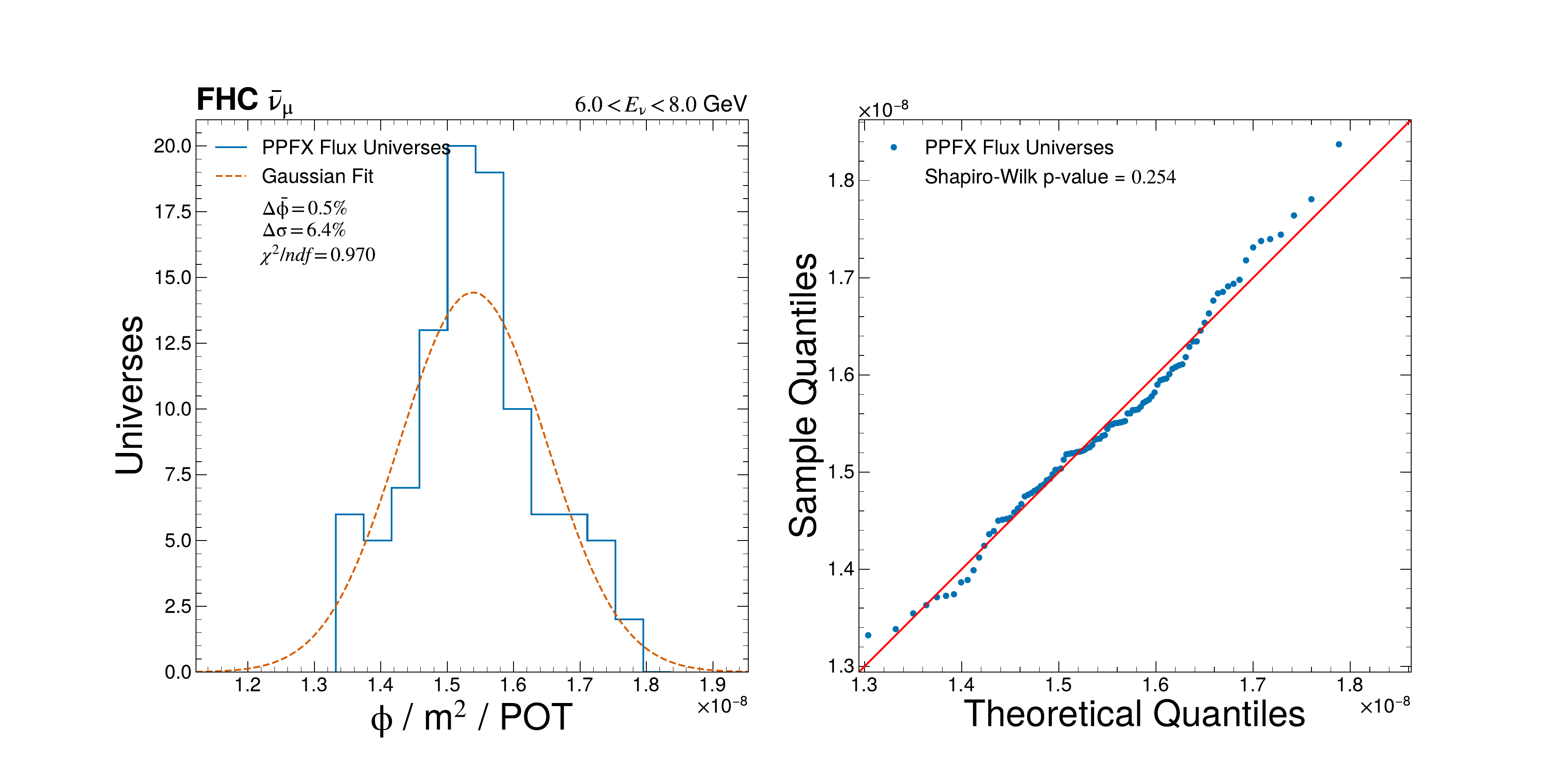}
    \includegraphics[width=0.3\textwidth]{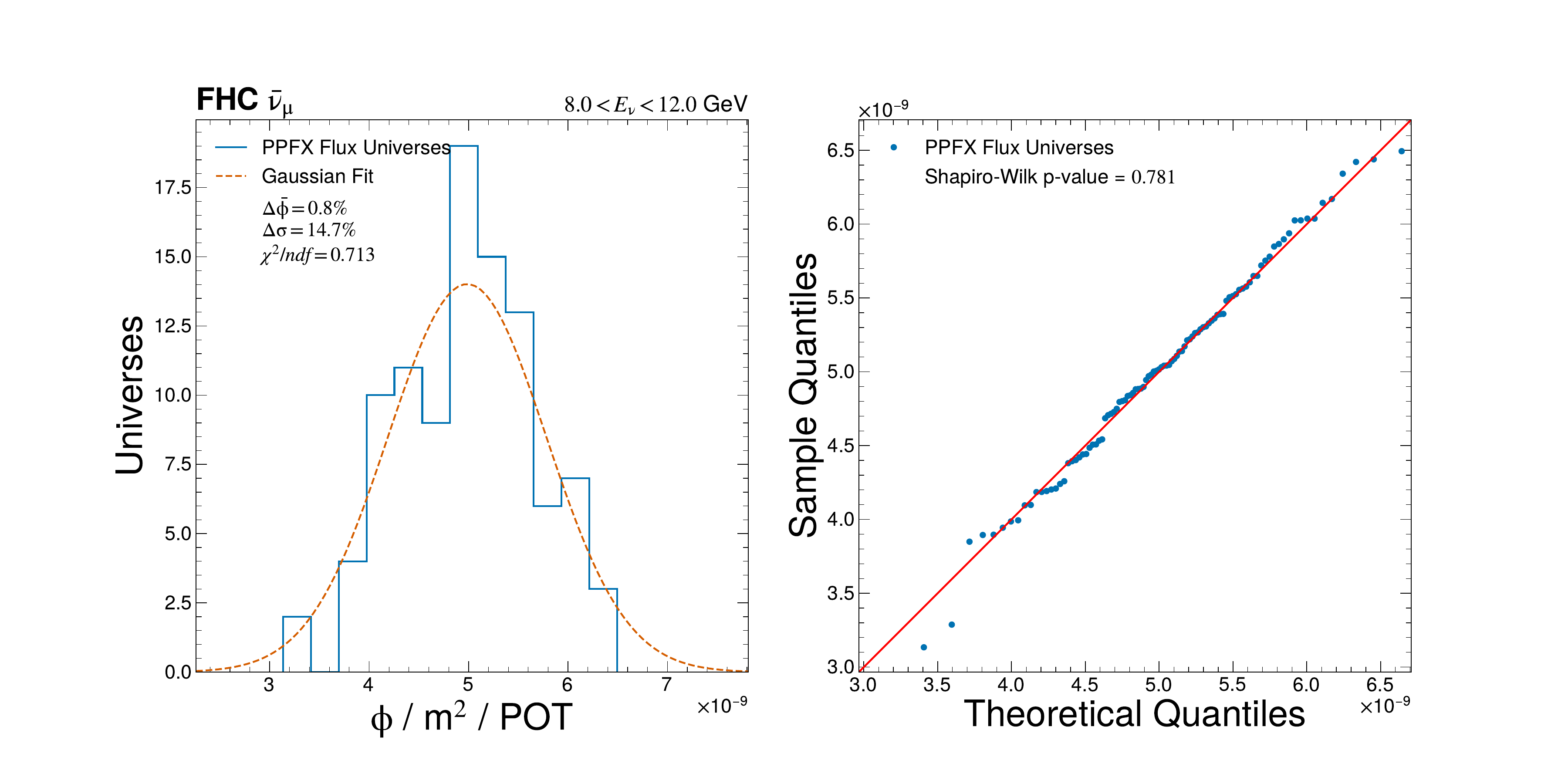}
    \caption[Distribution of PPFX universes for \numub\ (FHC).]{Distribution of PPFX universes for \numub.}
\end{figure}

\clearpage
\section{Reverse Horn Current}
\begin{figure}[!ht]
    \centering
    \includegraphics[width=0.3\textwidth]{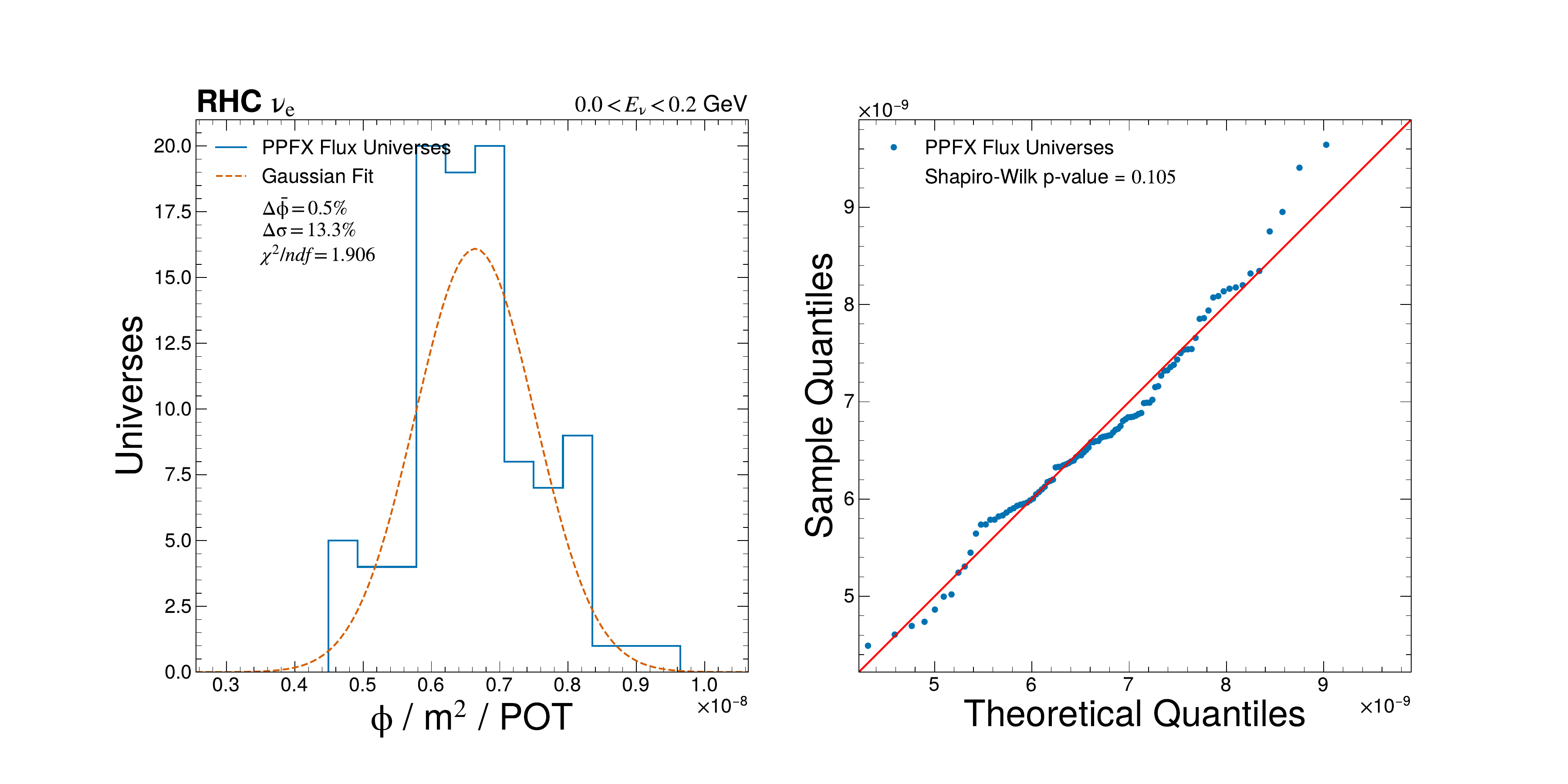}
    \includegraphics[width=0.3\textwidth]{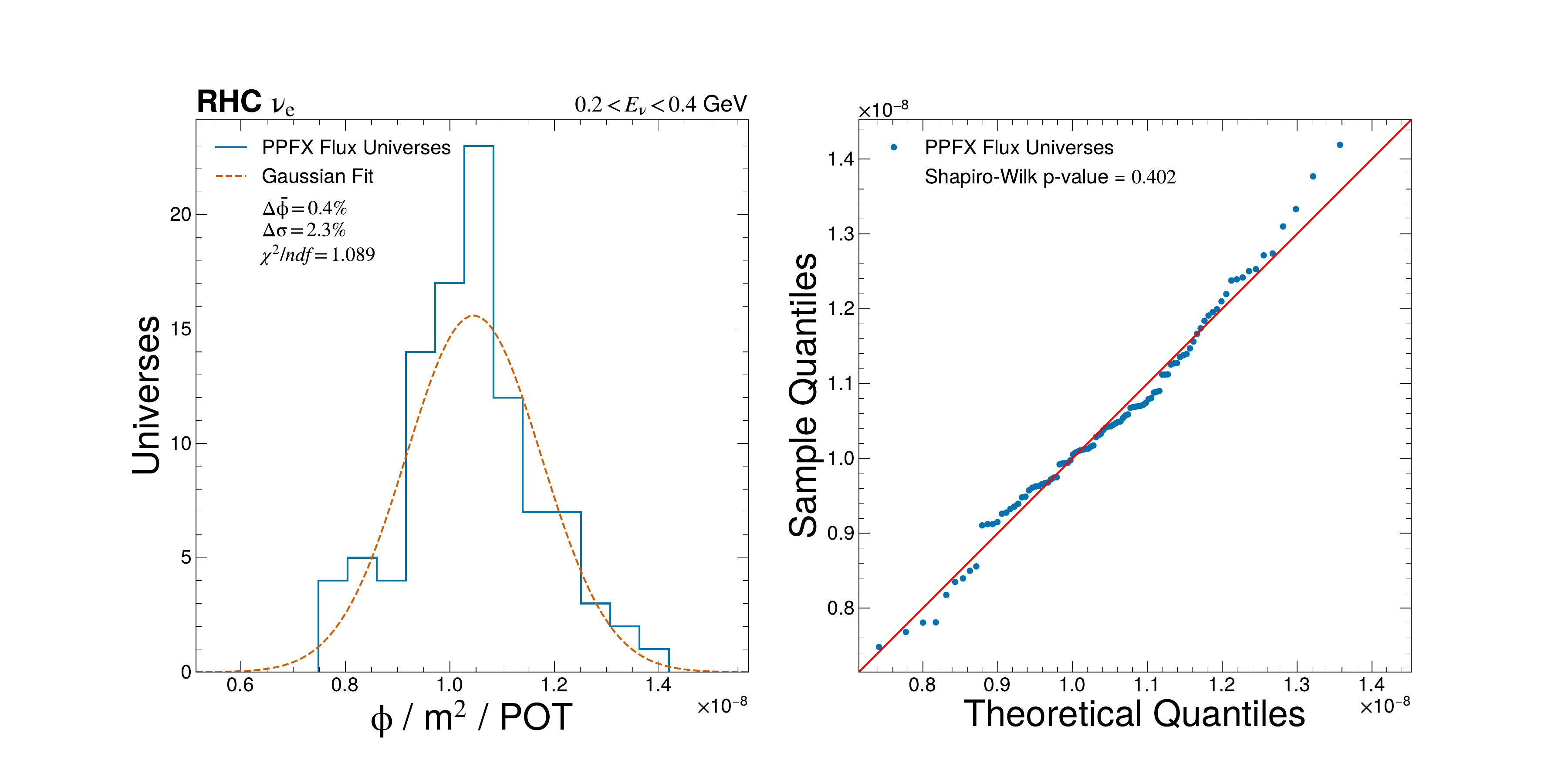}
    \includegraphics[width=0.3\textwidth]{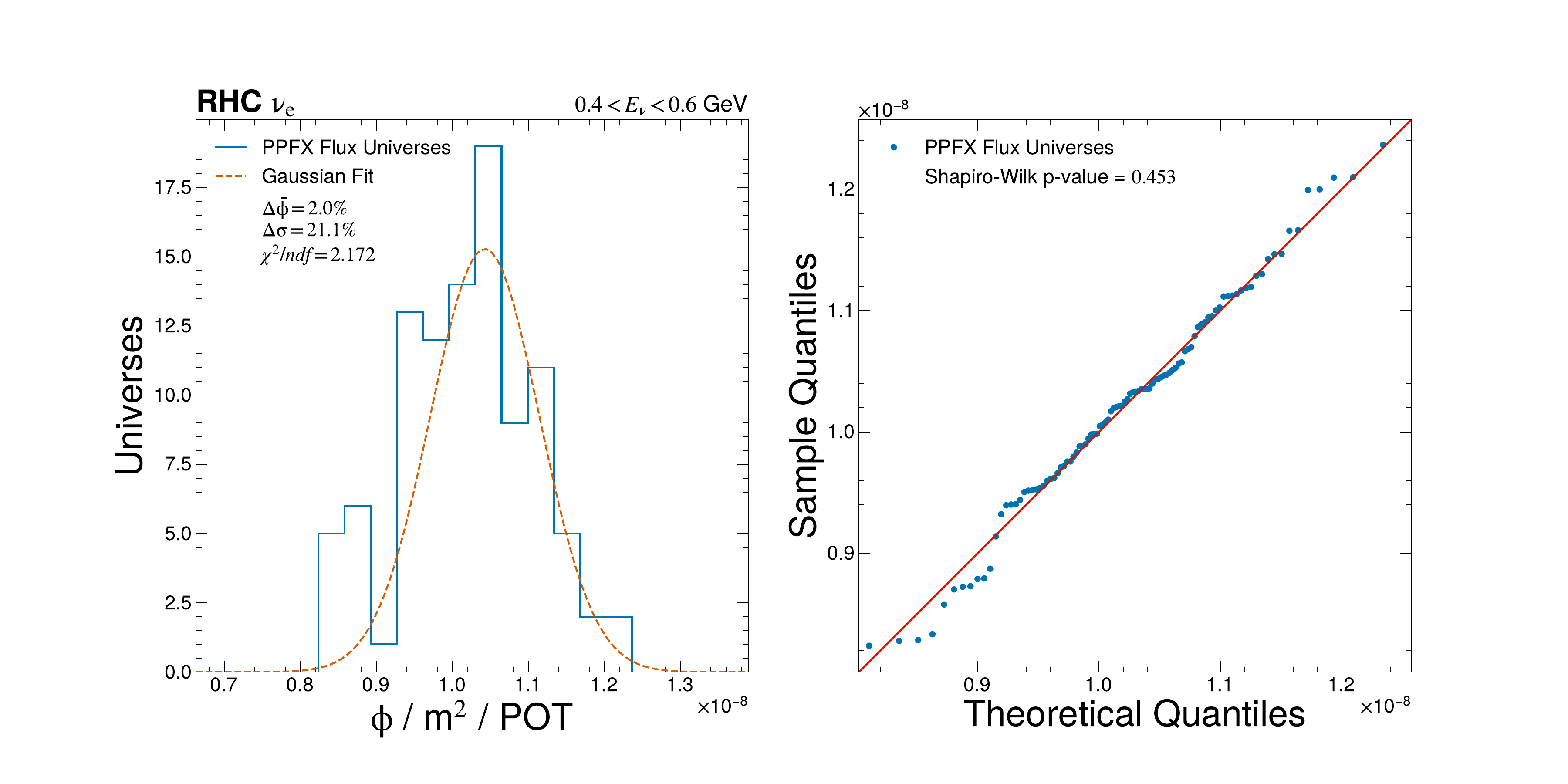}
    \includegraphics[width=0.3\textwidth]{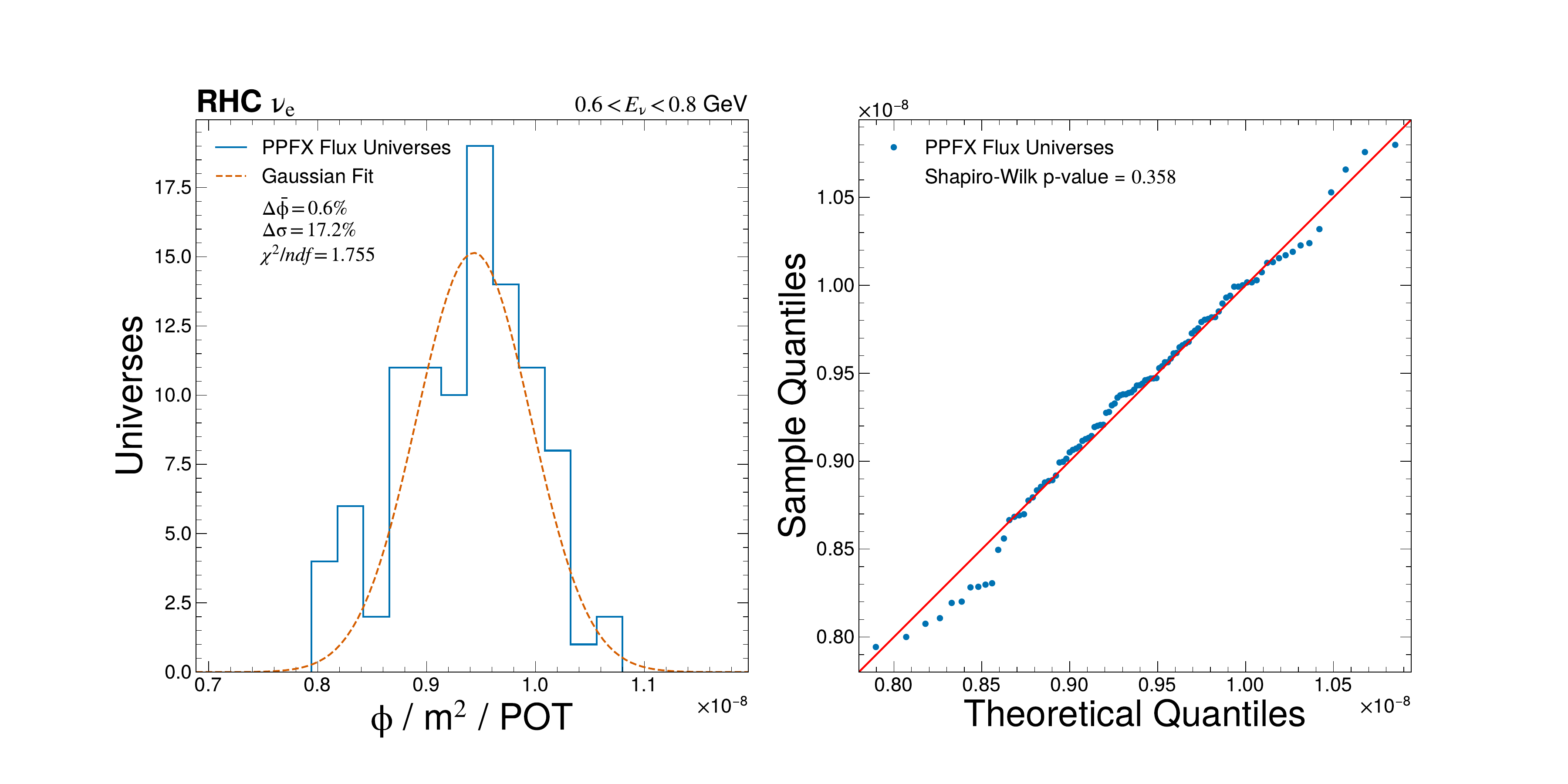}
    \includegraphics[width=0.3\textwidth]{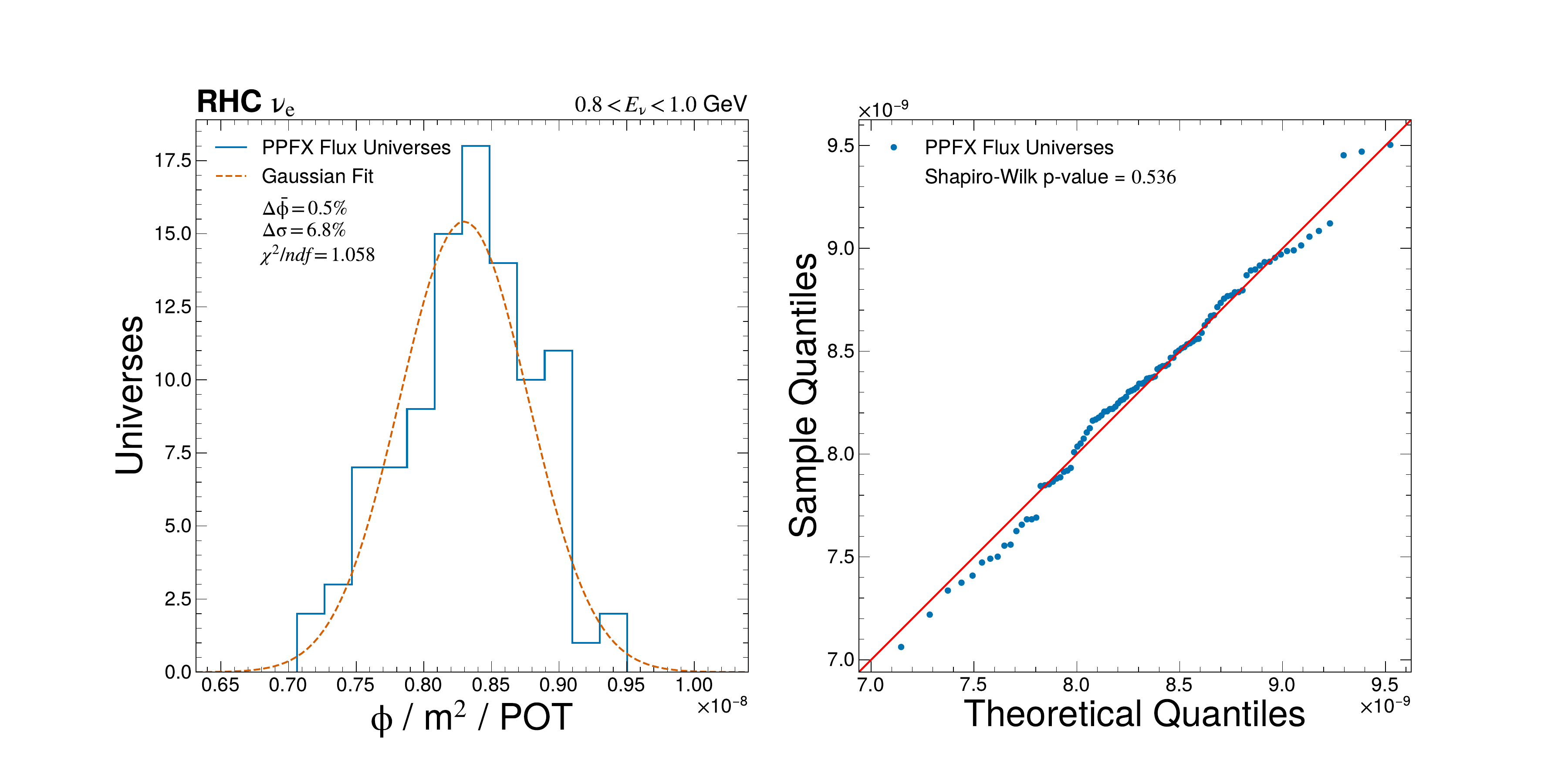}
    \includegraphics[width=0.3\textwidth]{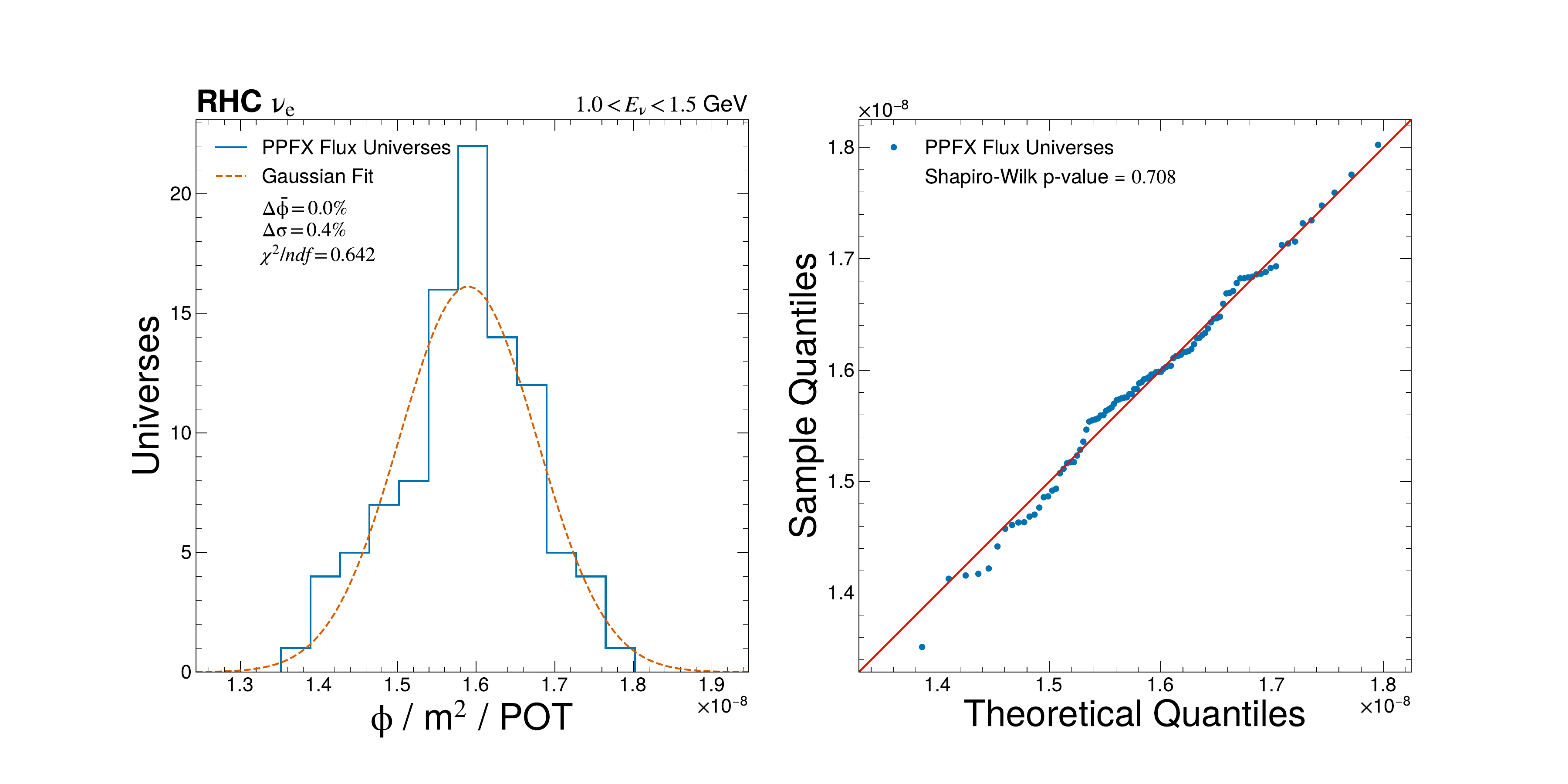}
    \includegraphics[width=0.3\textwidth]{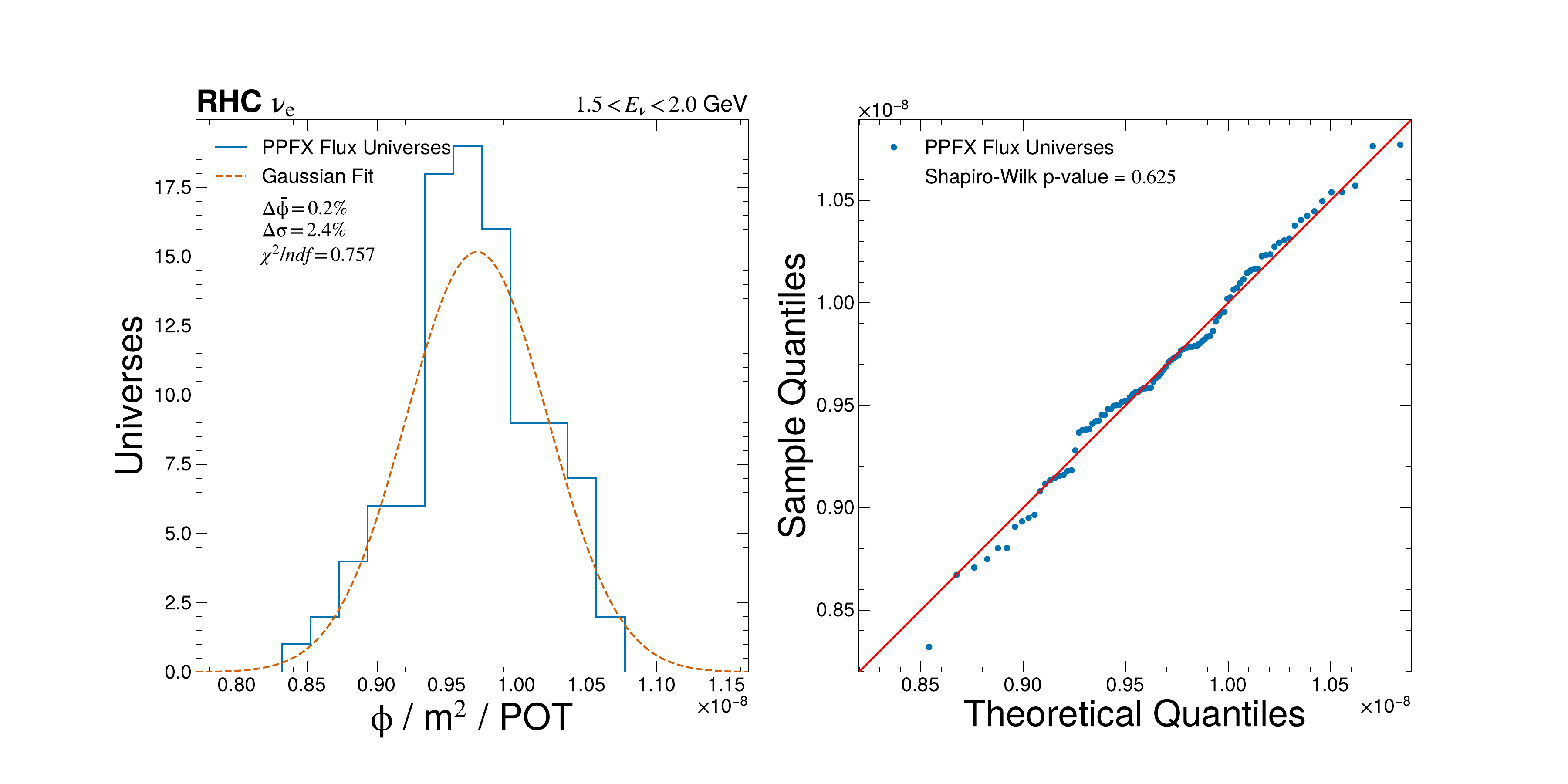}
    \includegraphics[width=0.3\textwidth]{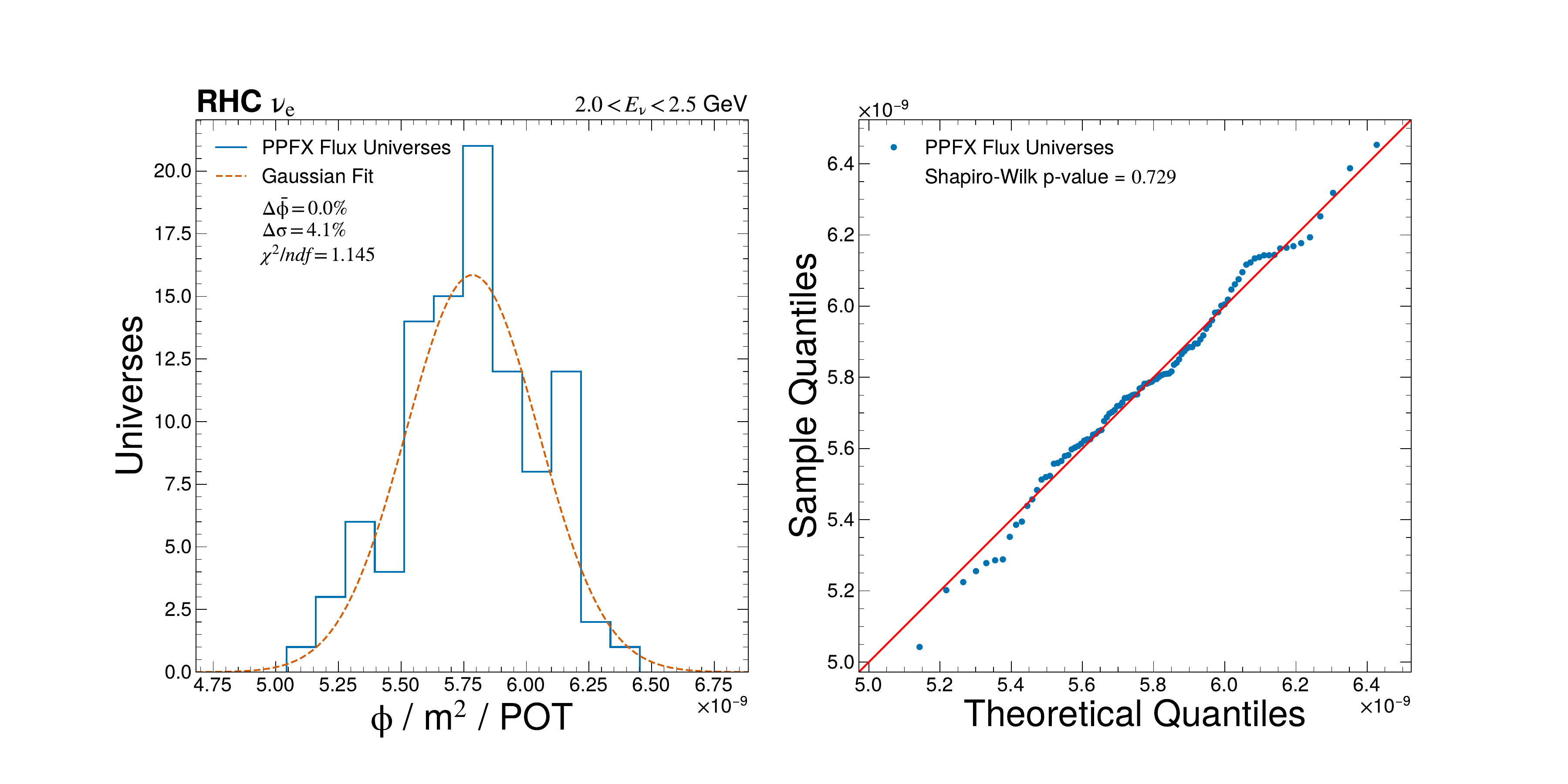}
    \includegraphics[width=0.3\textwidth]{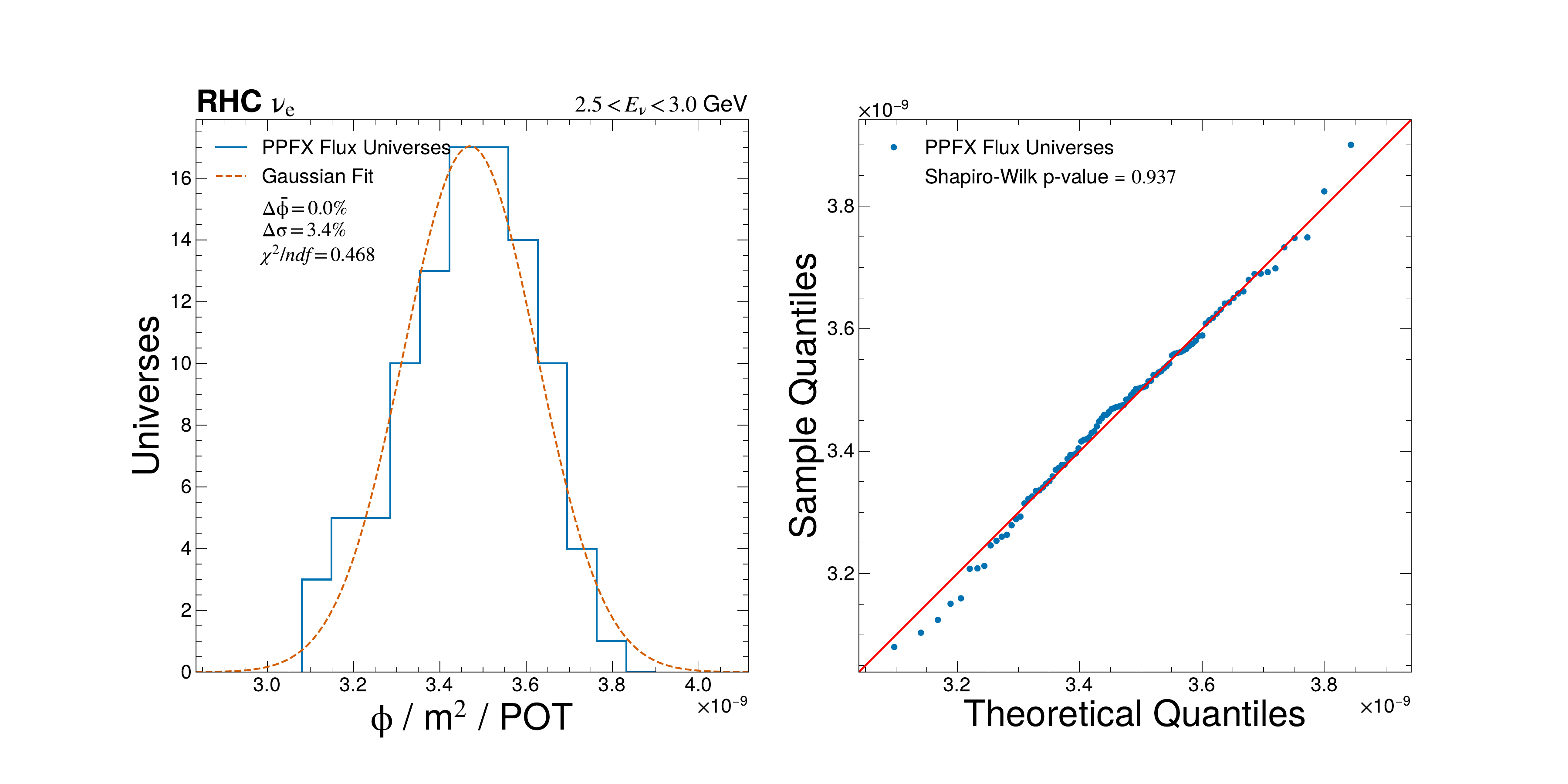}
    \includegraphics[width=0.3\textwidth]{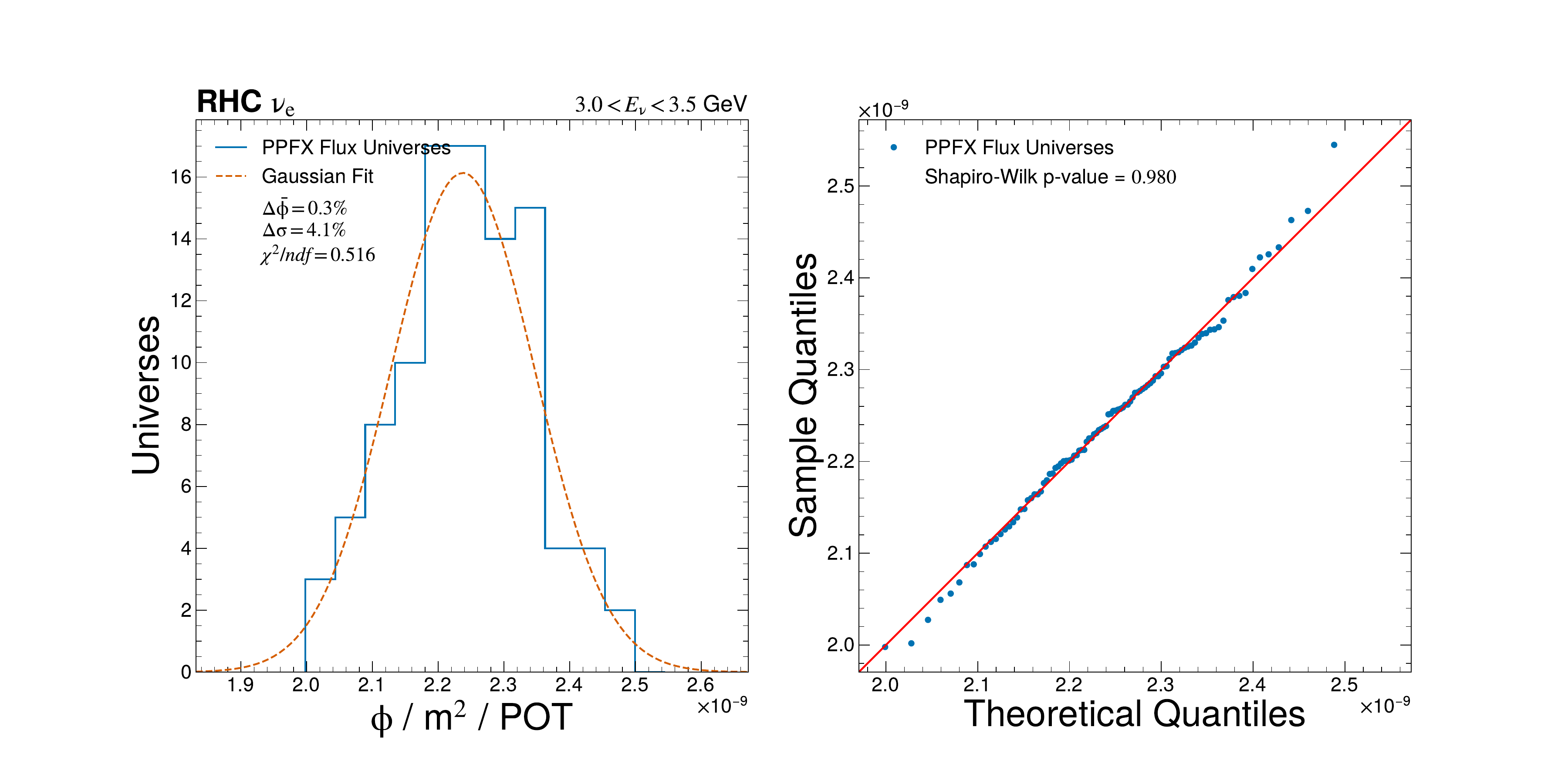}
    \includegraphics[width=0.3\textwidth]{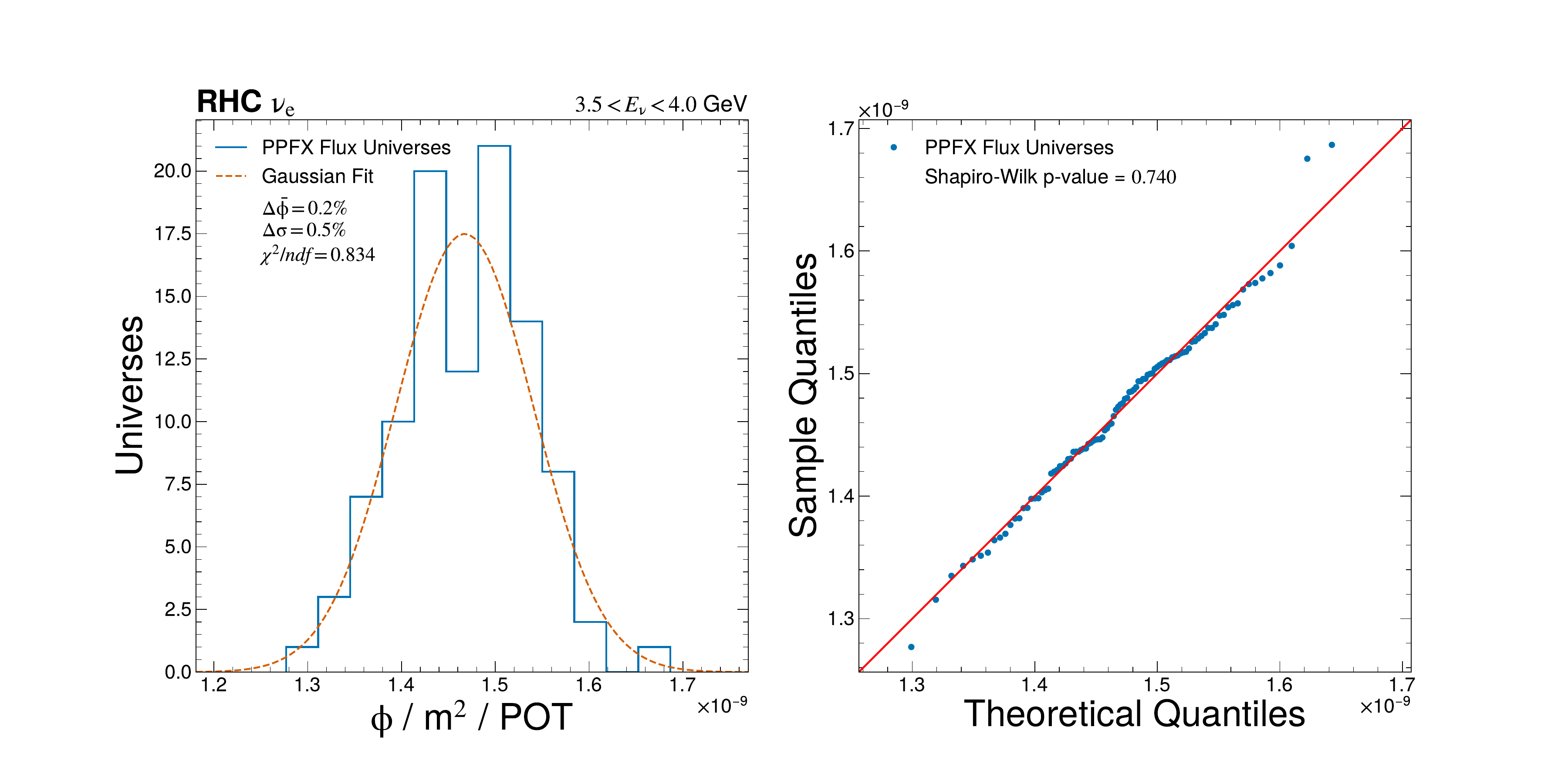}
    \includegraphics[width=0.3\textwidth]{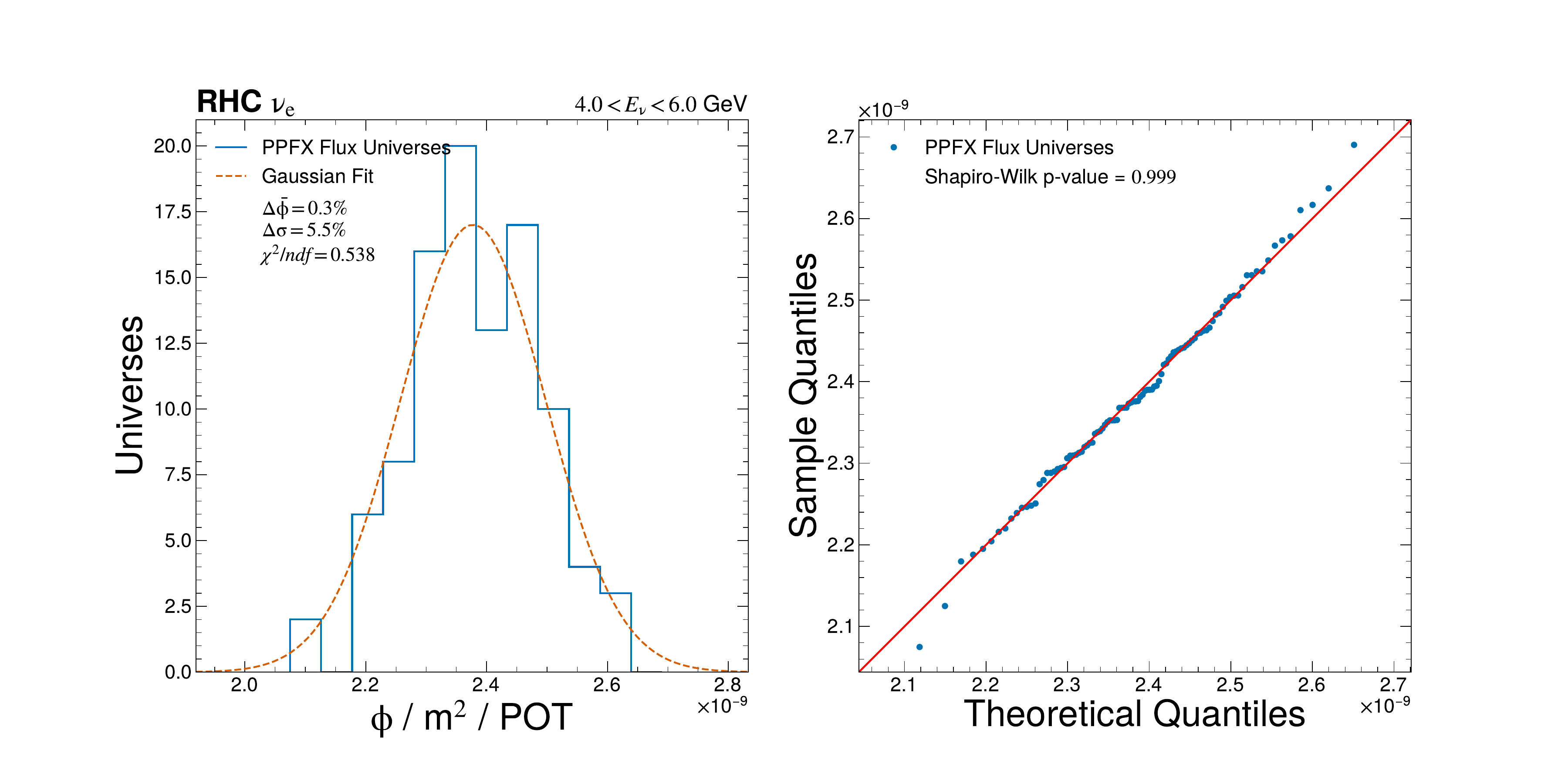}
    \includegraphics[width=0.3\textwidth]{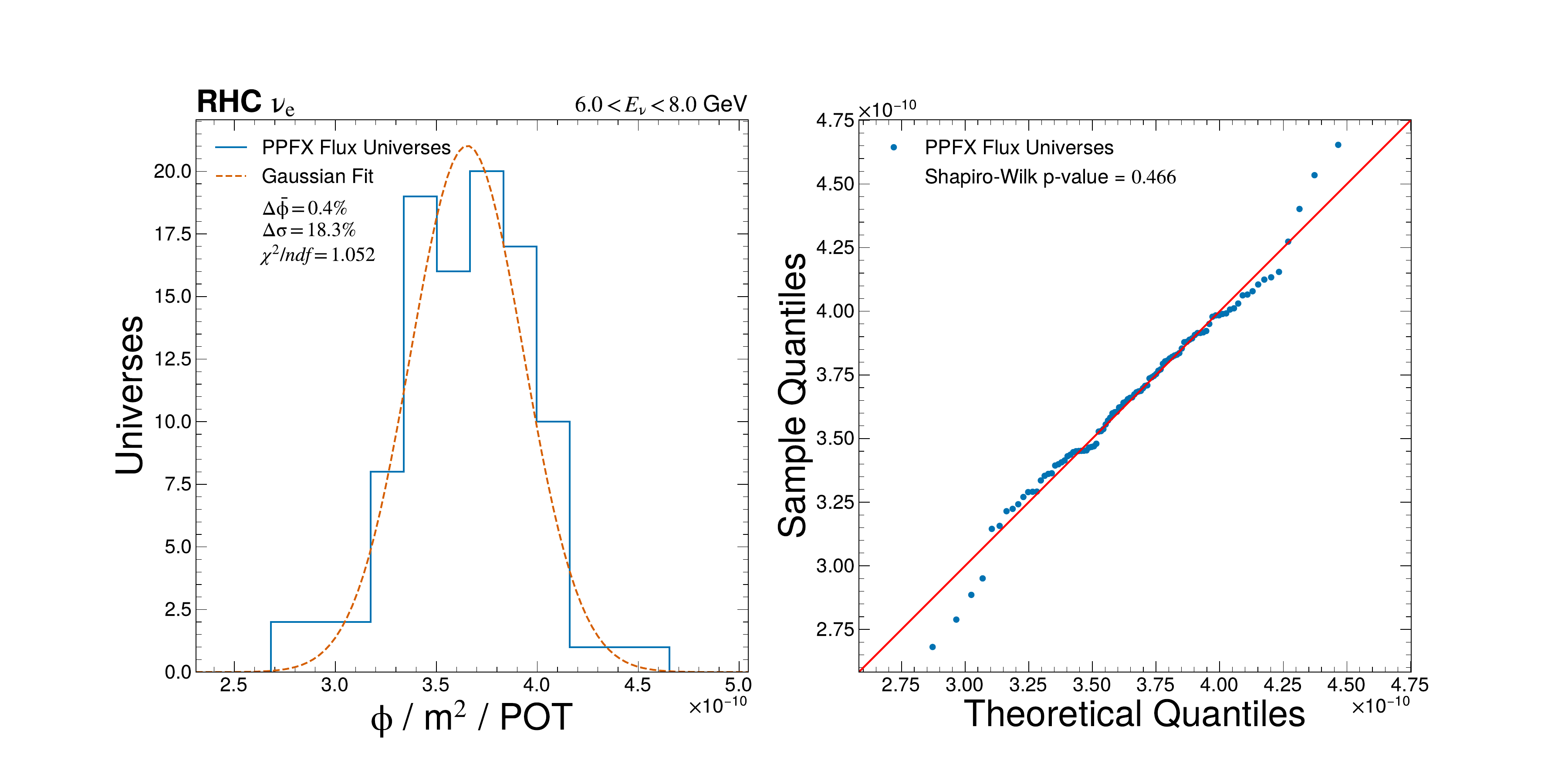}
    \includegraphics[width=0.3\textwidth]{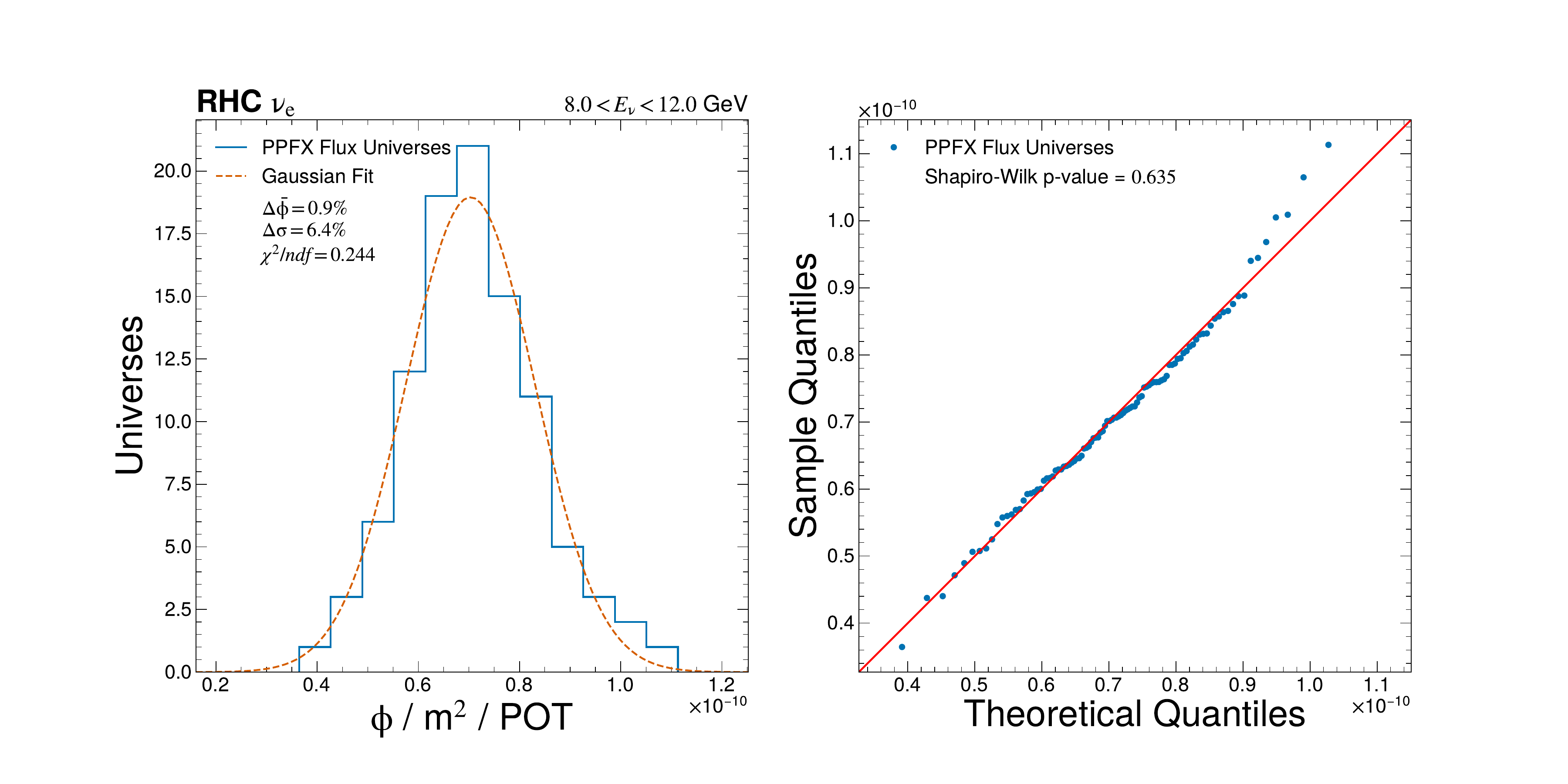}
    \caption[Distribution of PPFX universes for \nue\ (RHC).]{Distribution of PPFX universes for \nue.}
\end{figure}
\begin{figure}[!ht]
    \centering
    \includegraphics[width=0.3\textwidth]{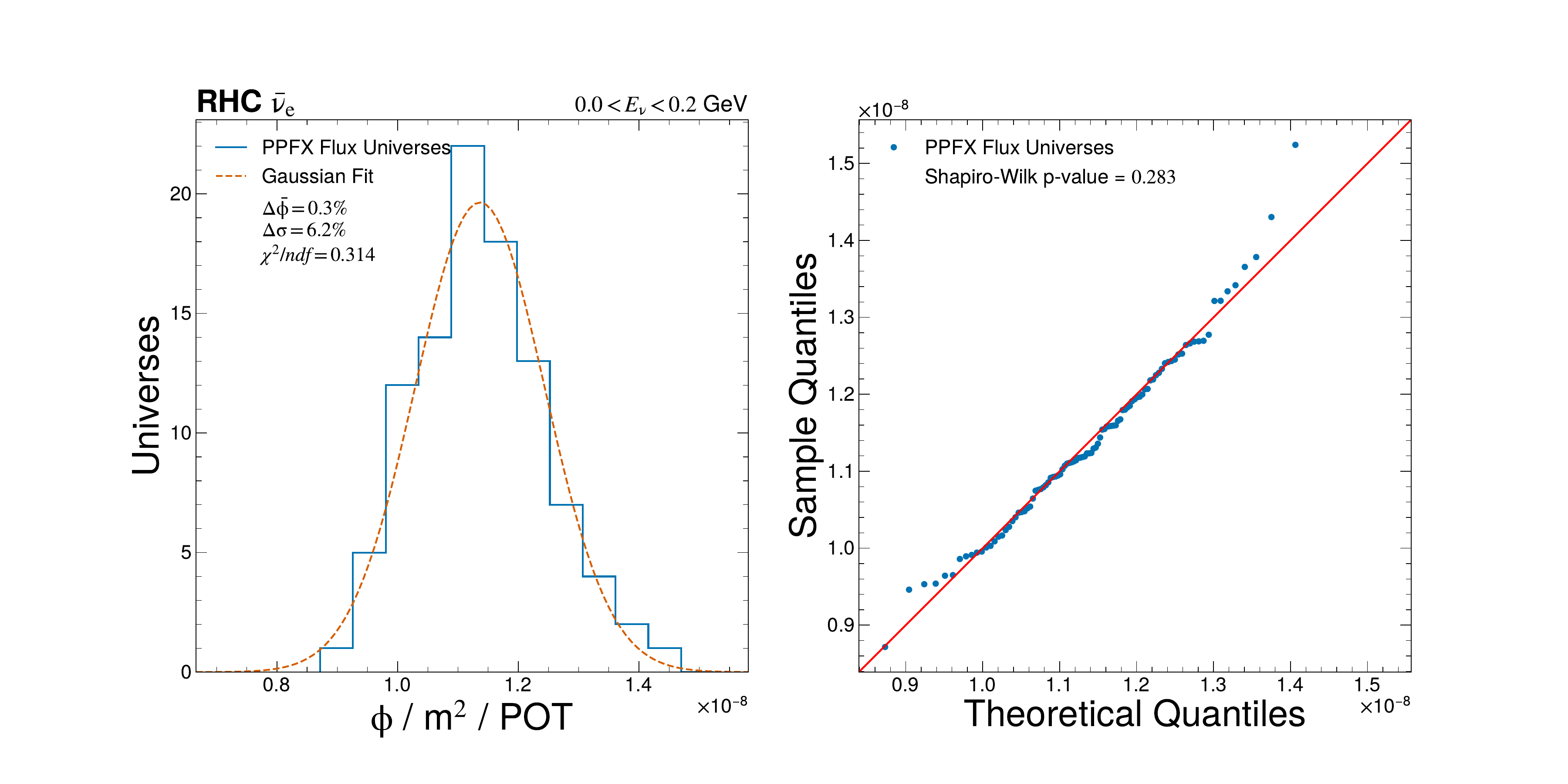}
    \includegraphics[width=0.3\textwidth]{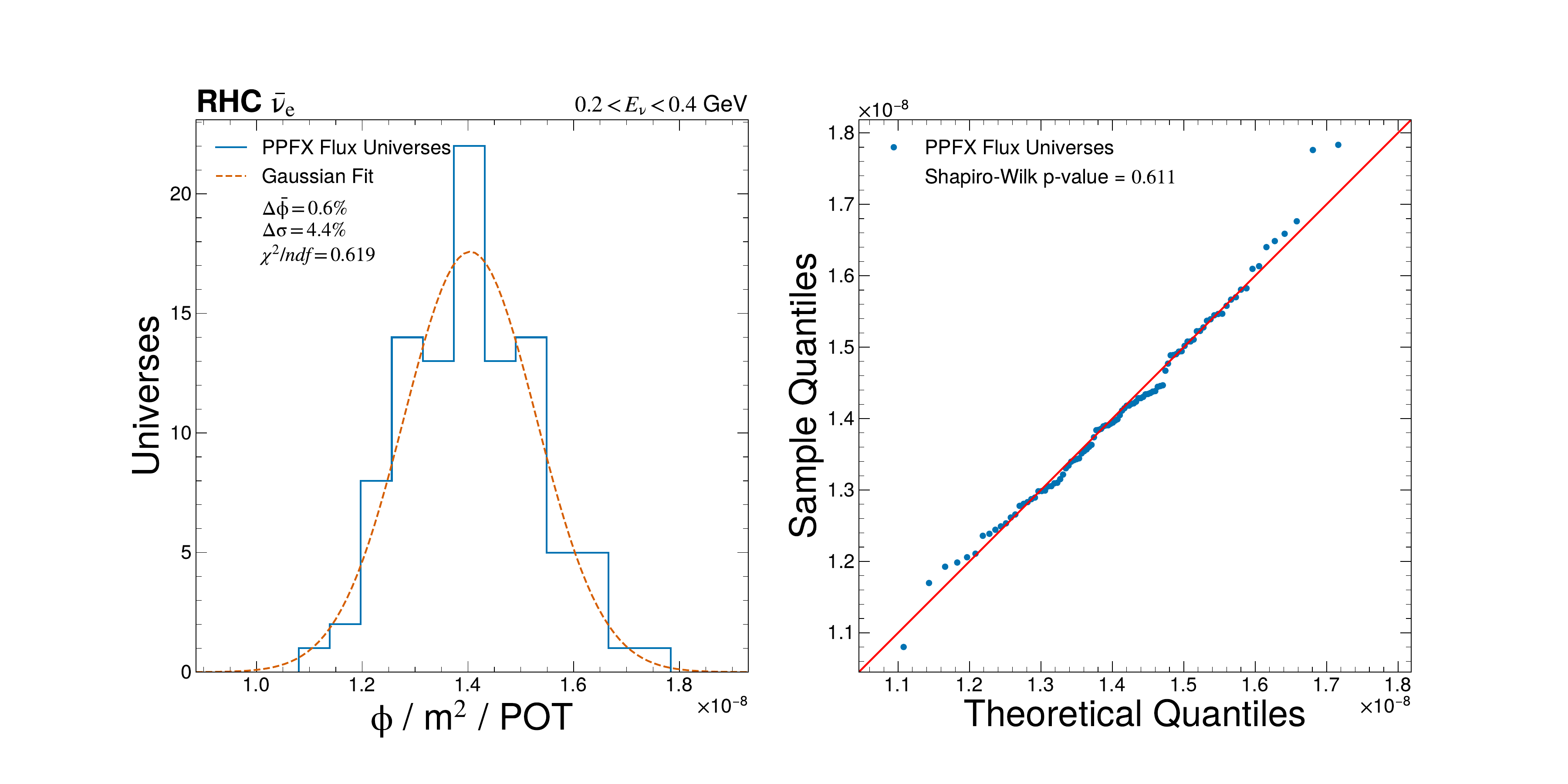}
    \includegraphics[width=0.3\textwidth]{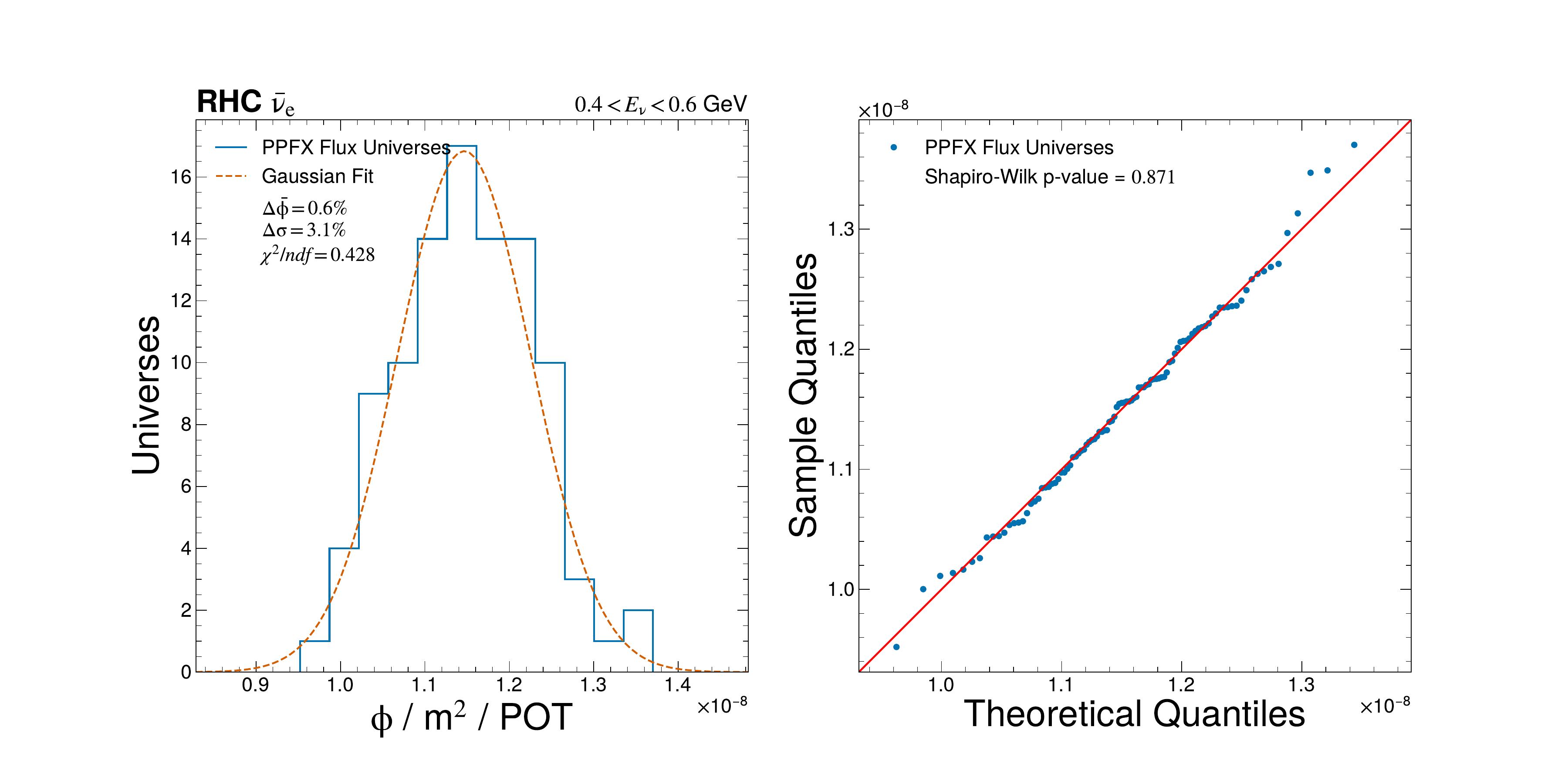}
    \includegraphics[width=0.3\textwidth]{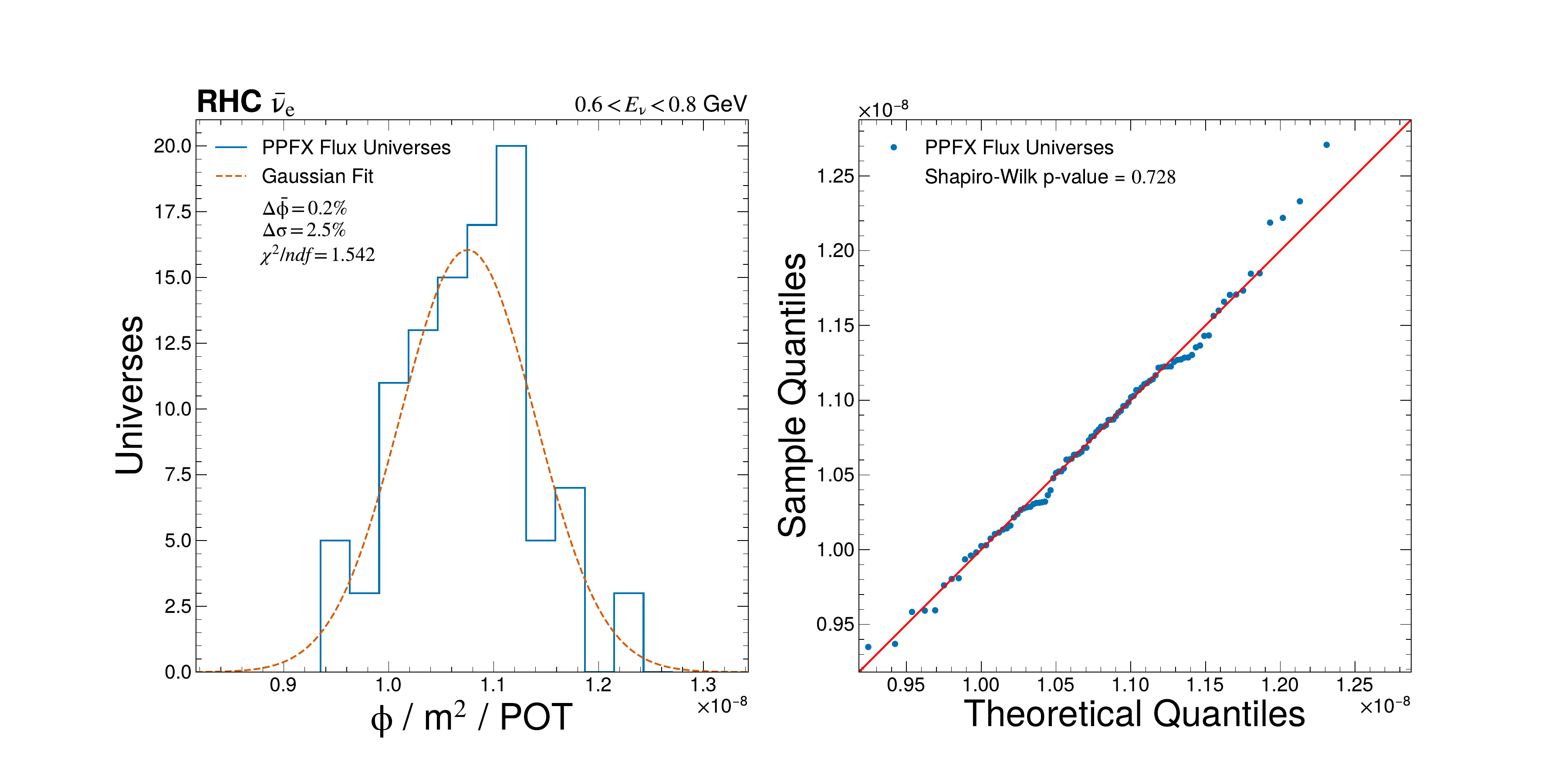}
    \includegraphics[width=0.3\textwidth]{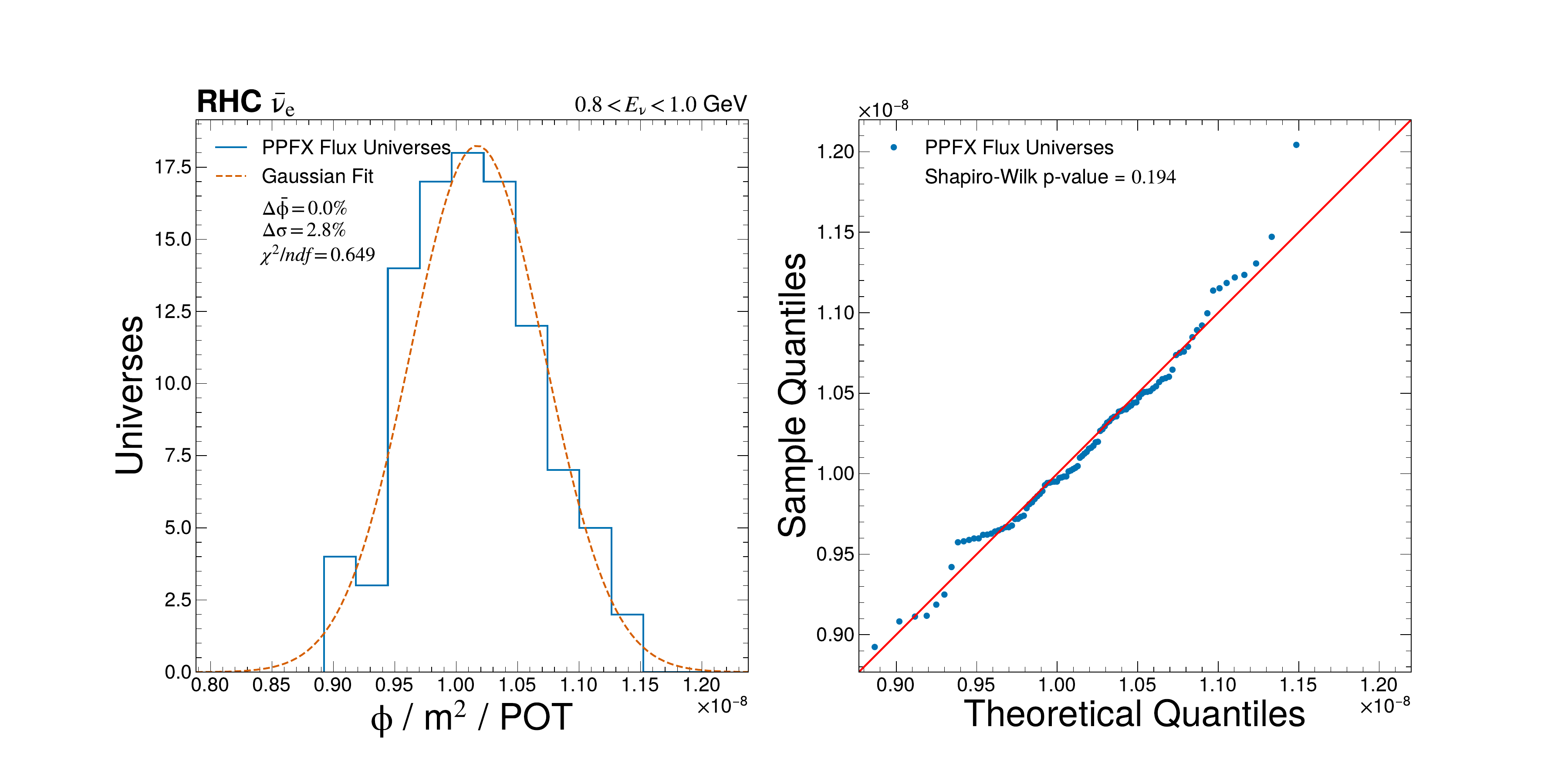}
    \includegraphics[width=0.3\textwidth]{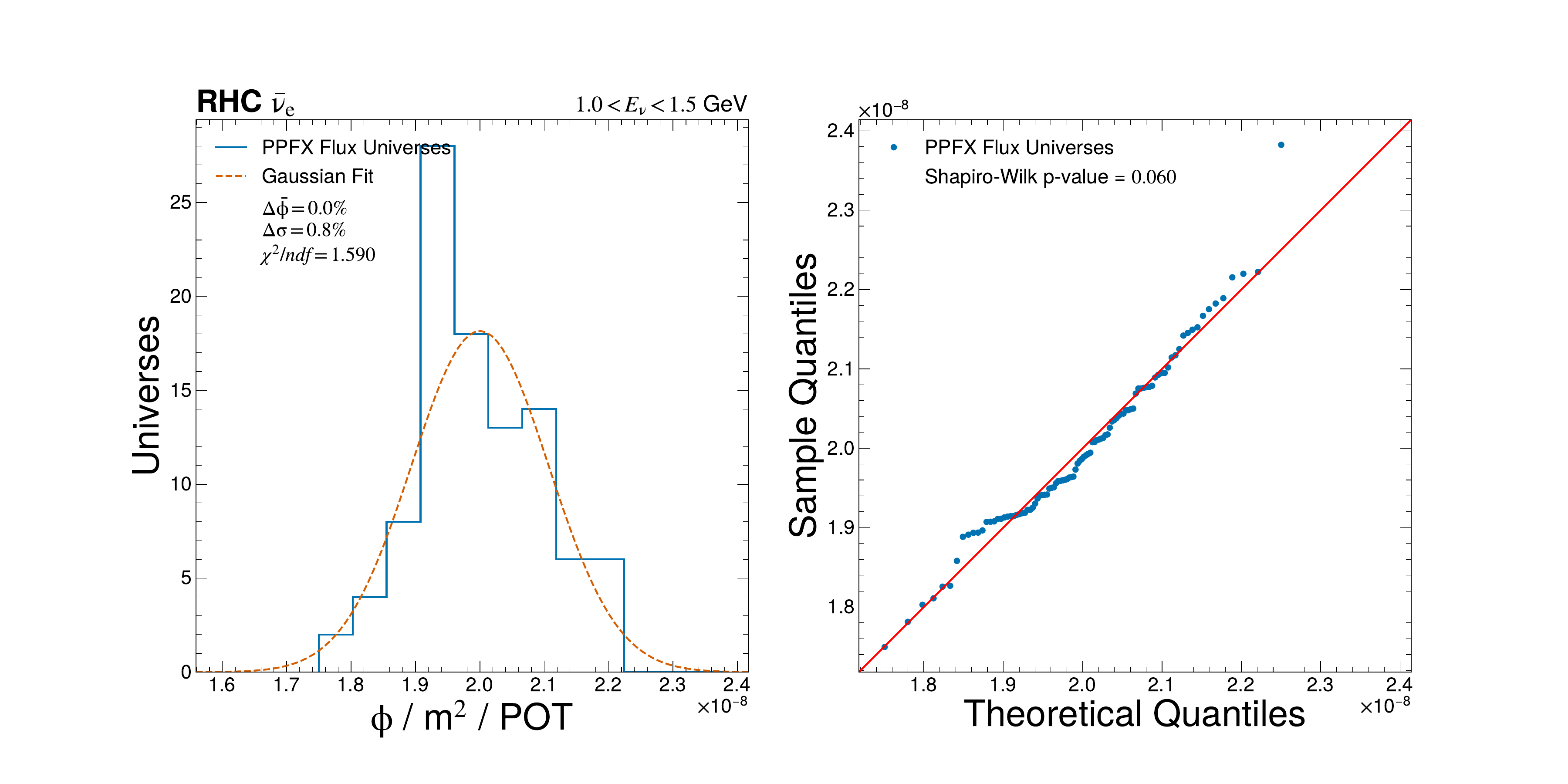}
    \includegraphics[width=0.3\textwidth]{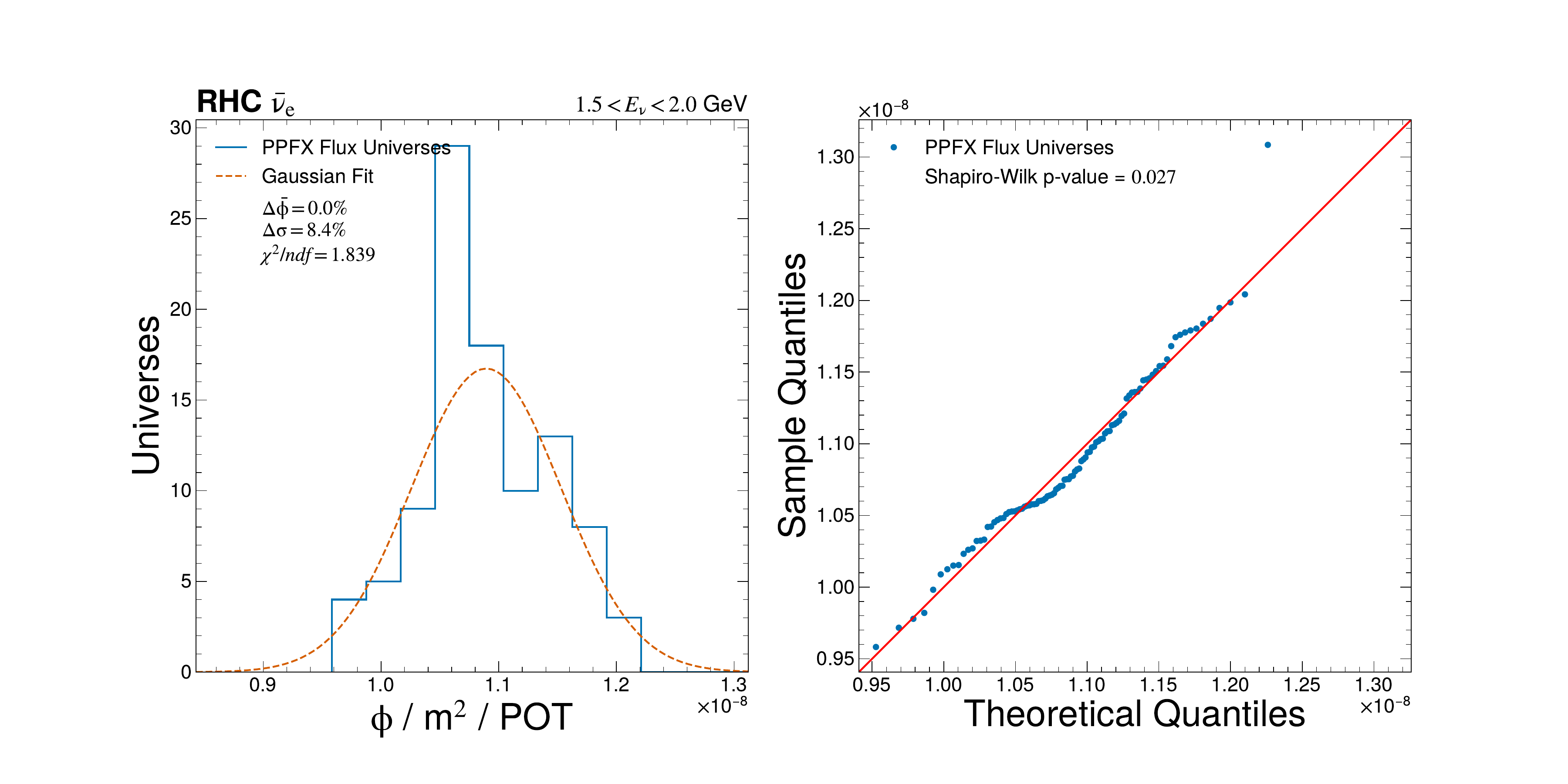}
    \includegraphics[width=0.3\textwidth]{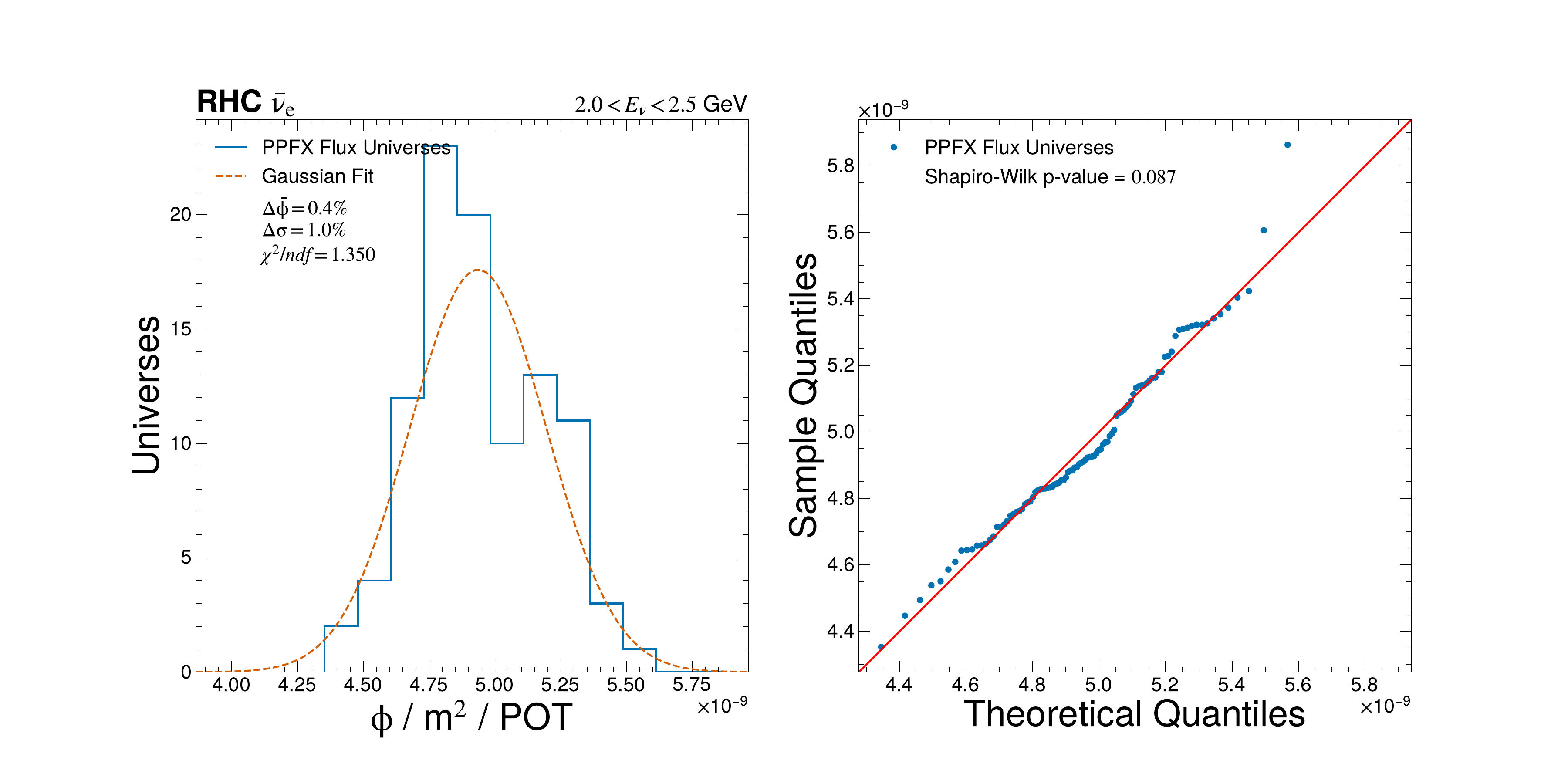}
    \includegraphics[width=0.3\textwidth]{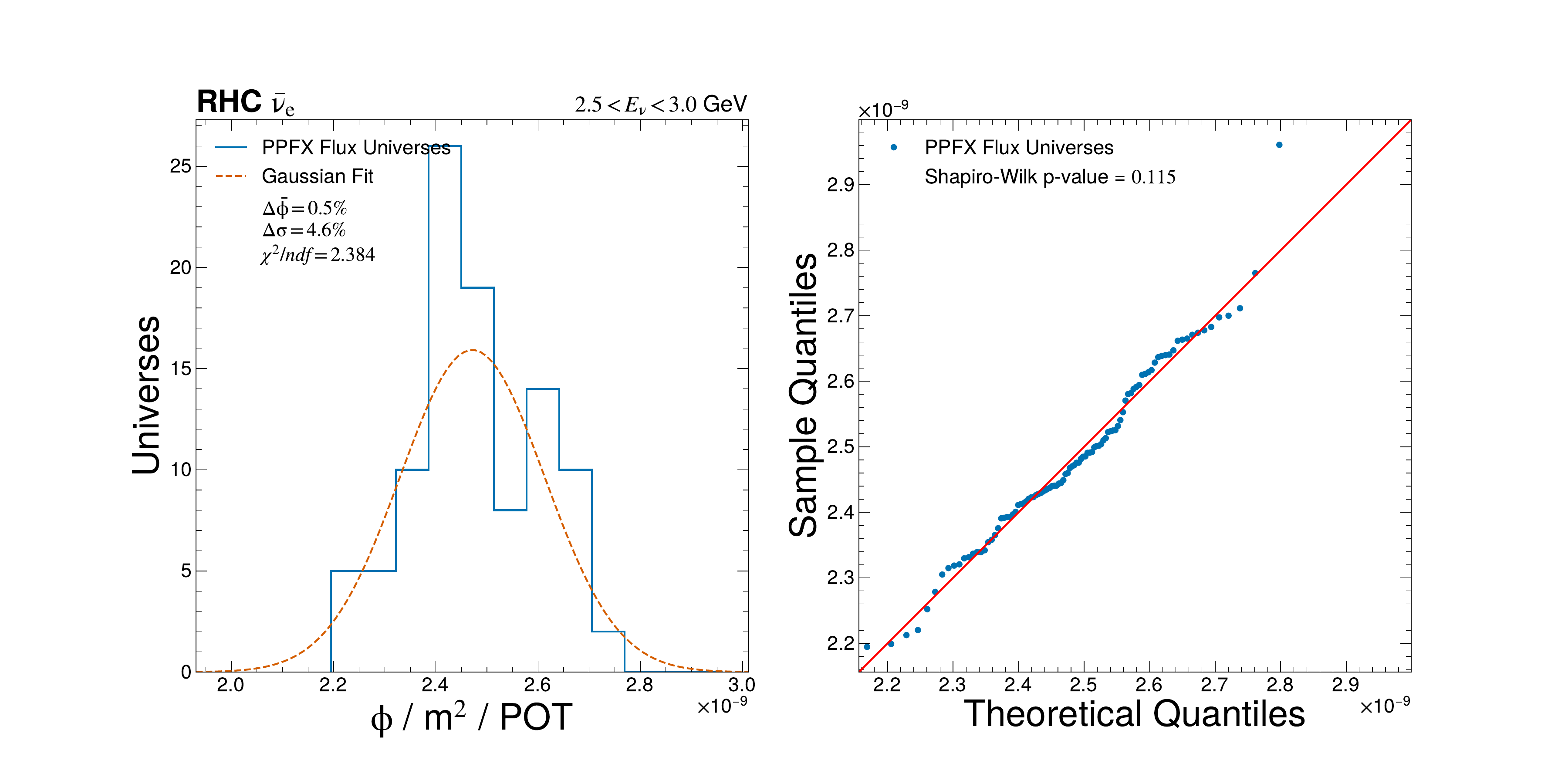}
    \includegraphics[width=0.3\textwidth]{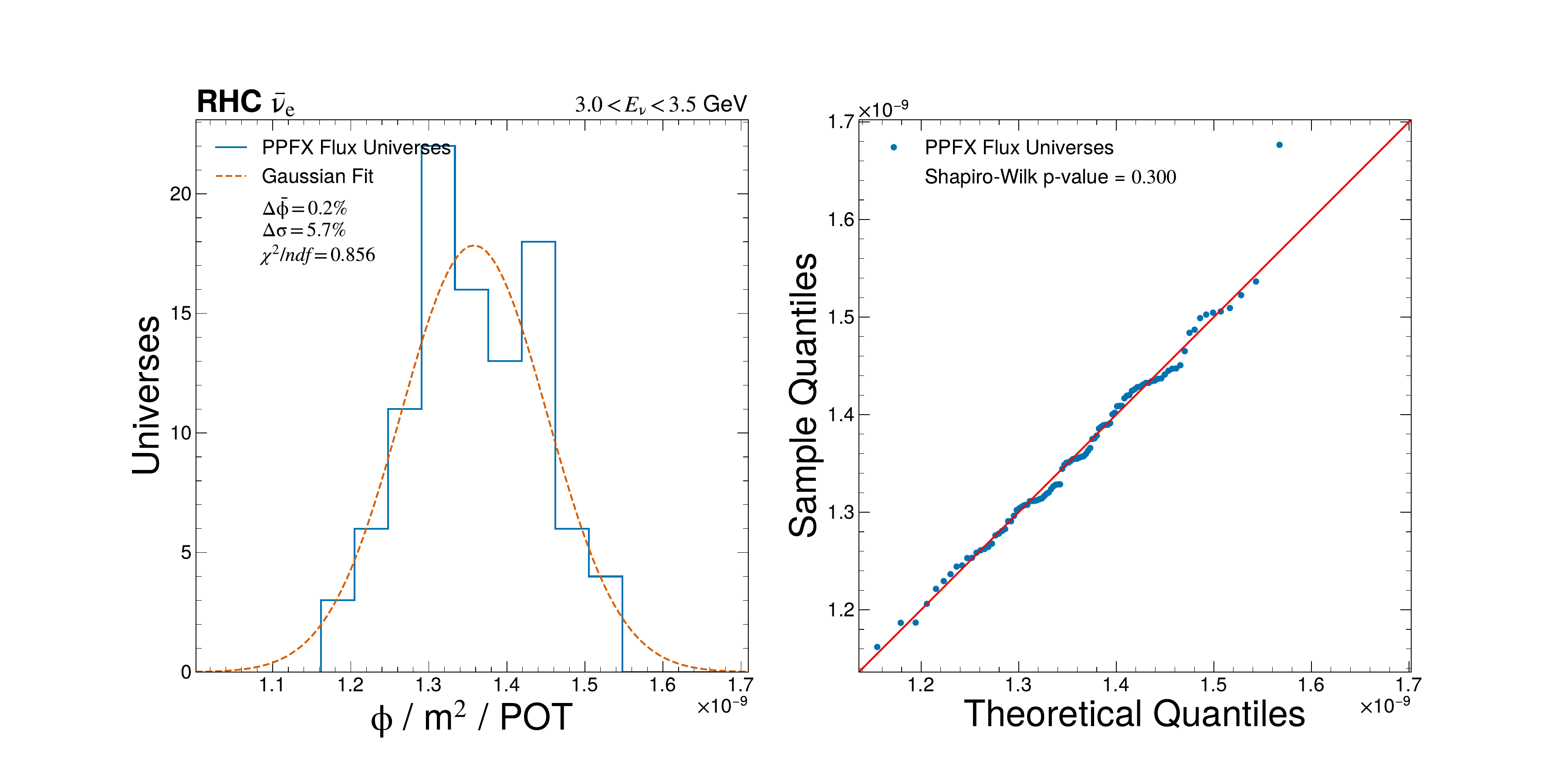}
    \includegraphics[width=0.3\textwidth]{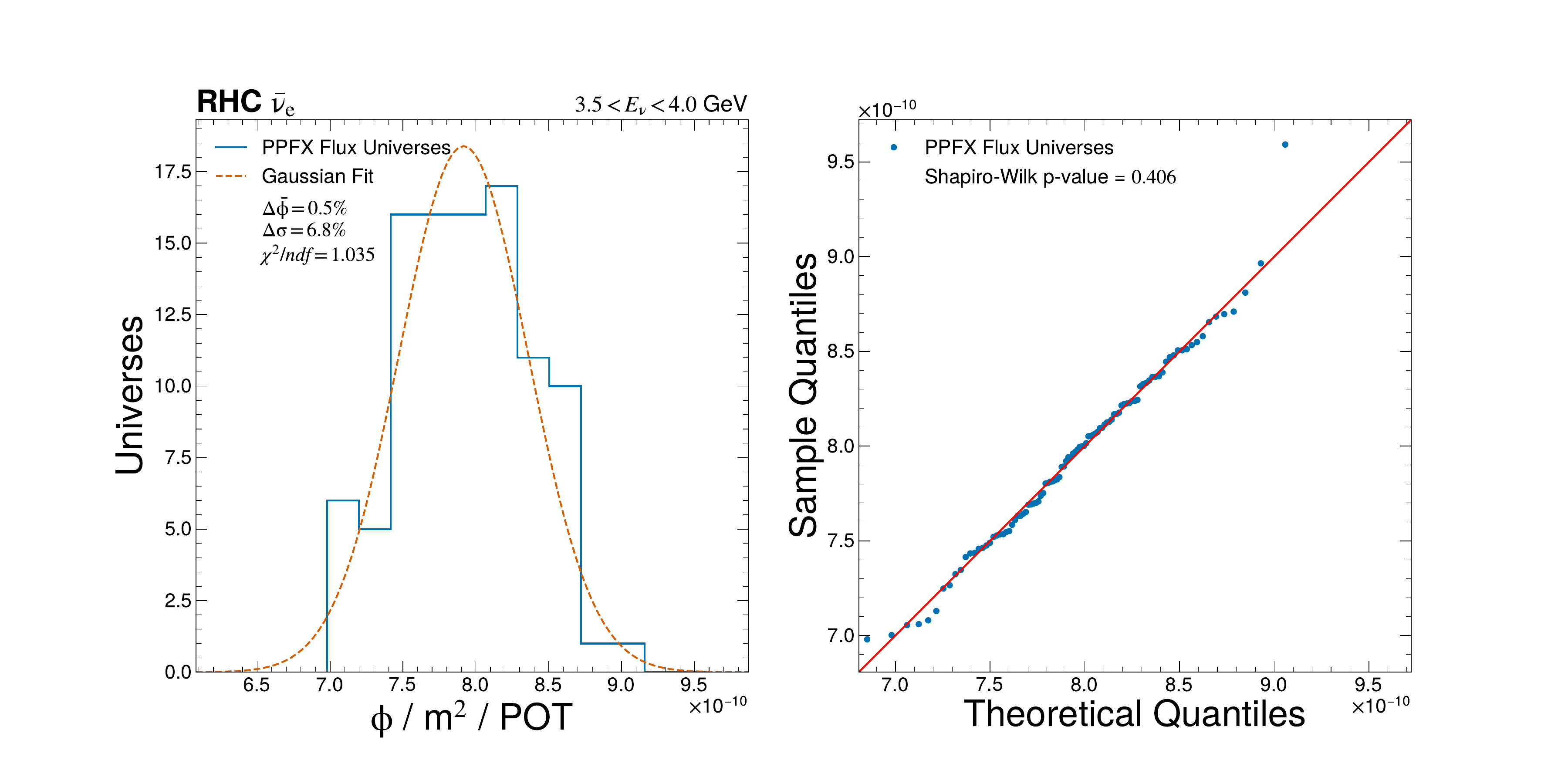}
    \includegraphics[width=0.3\textwidth]{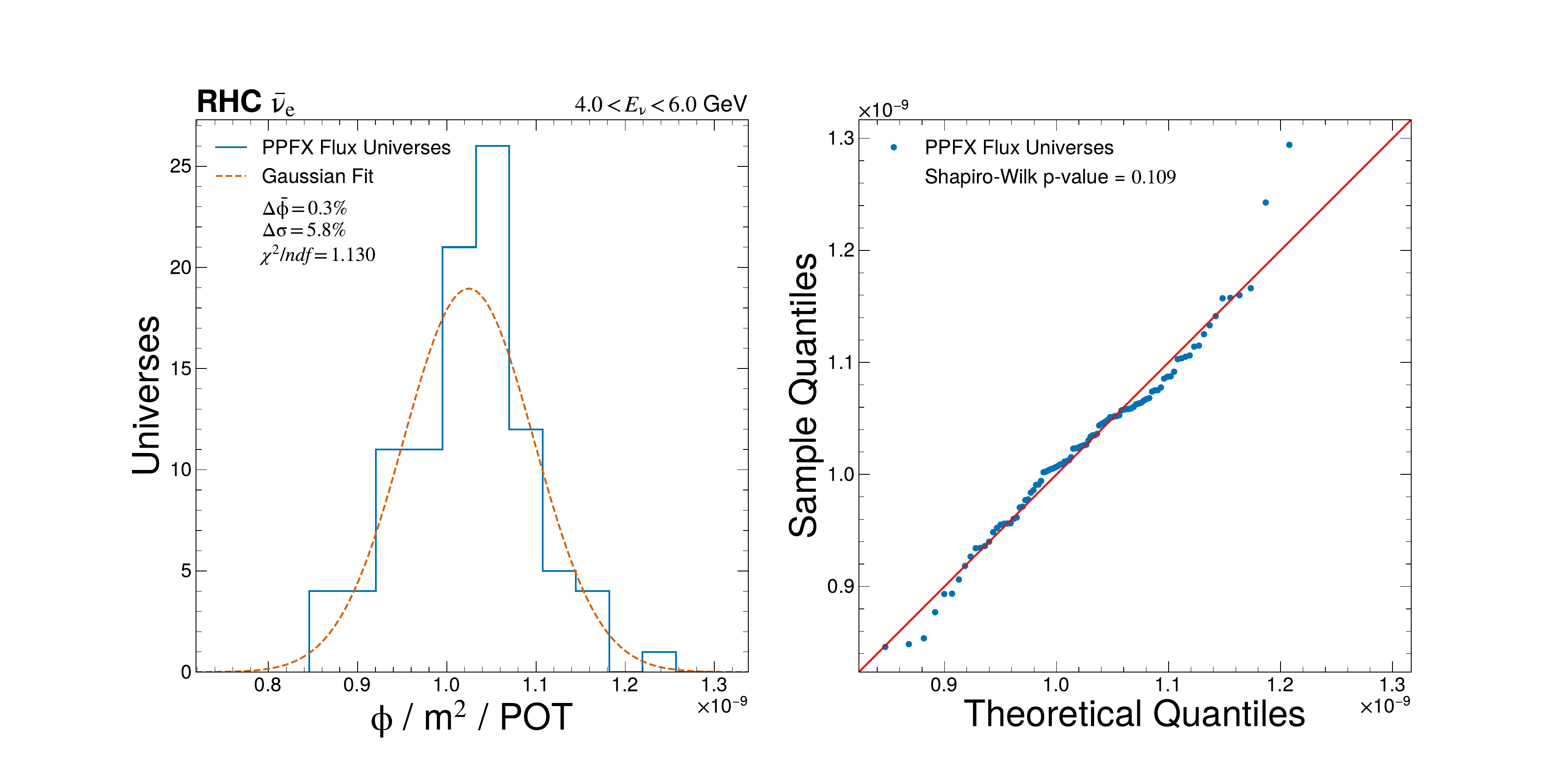}
    \includegraphics[width=0.3\textwidth]{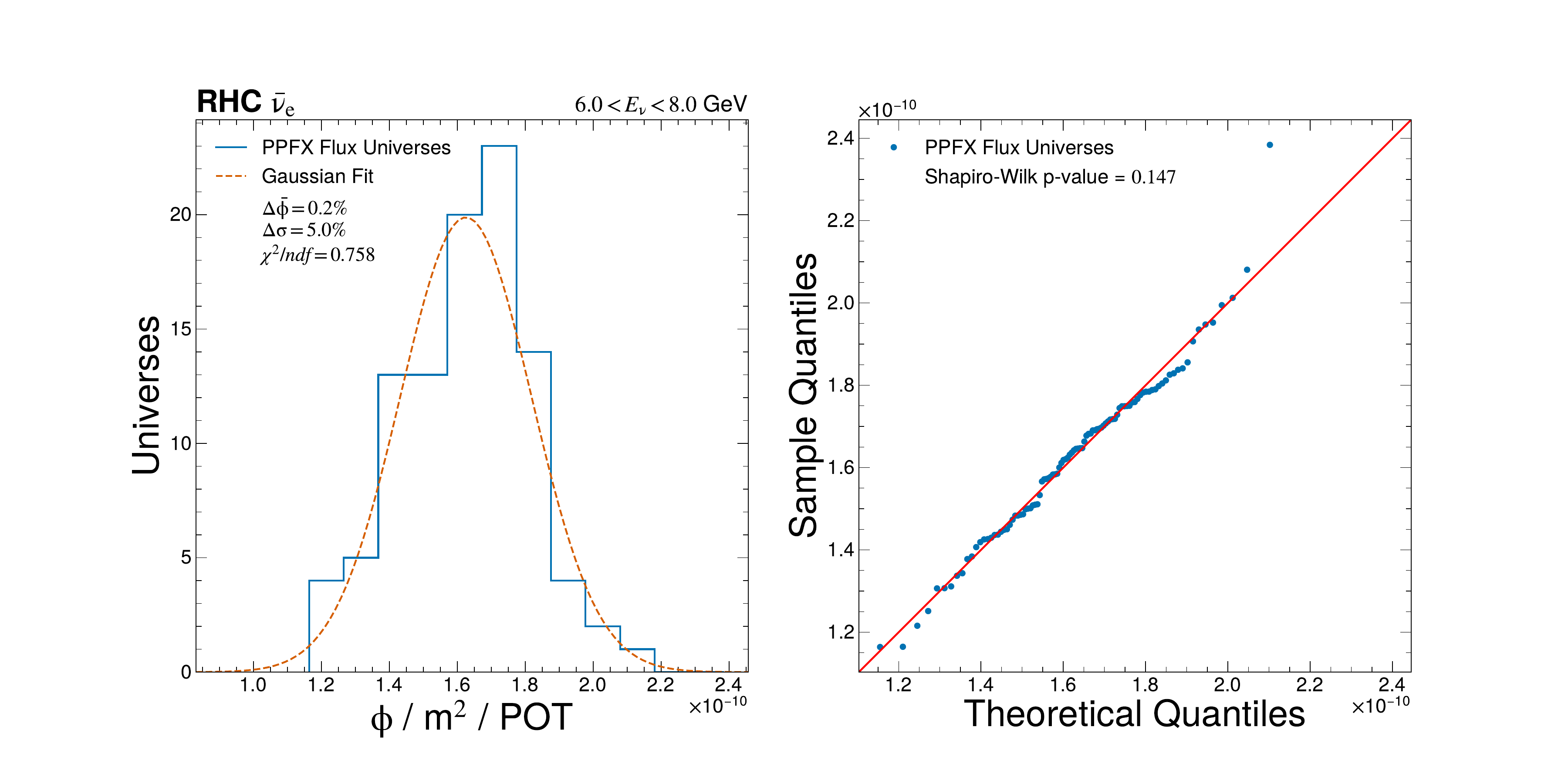}
    \includegraphics[width=0.3\textwidth]{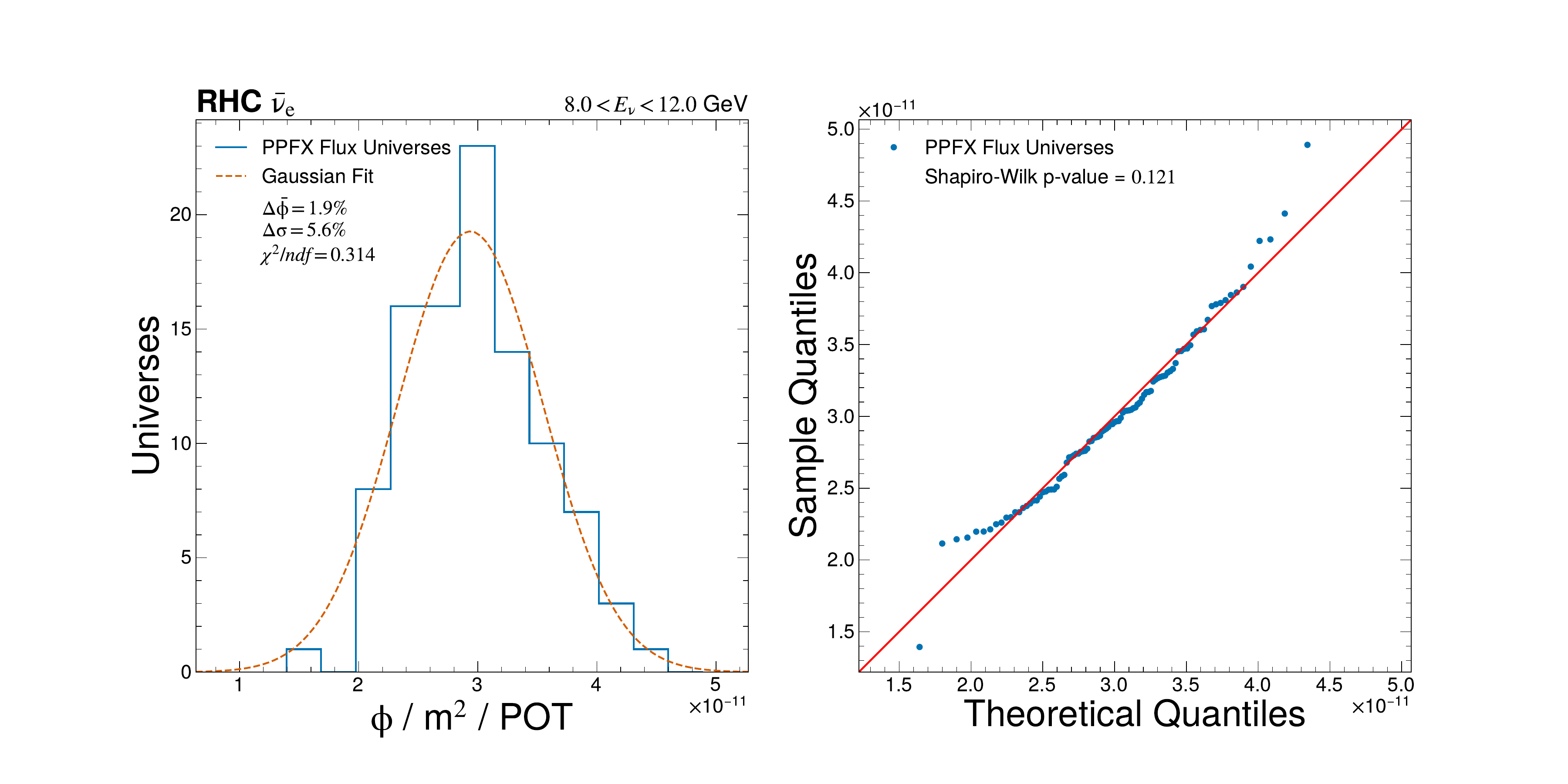}
    \caption[Distribution of PPFX universes for \nueb\ (RHC).]{Distribution of PPFX universes for \nueb.}
\end{figure}
\begin{figure}[!ht]
    \centering
    \includegraphics[width=0.3\textwidth]{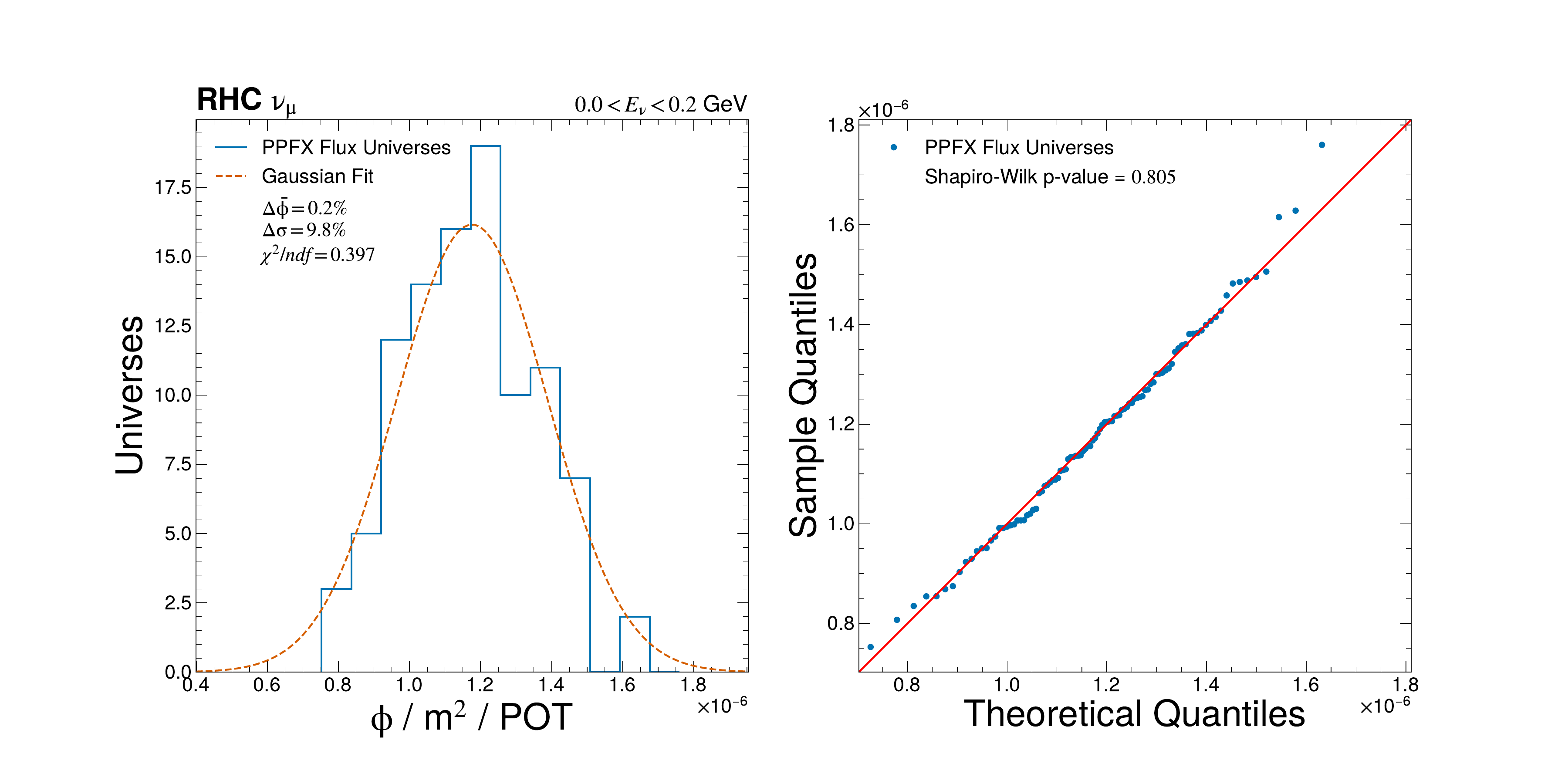}
    \includegraphics[width=0.3\textwidth]{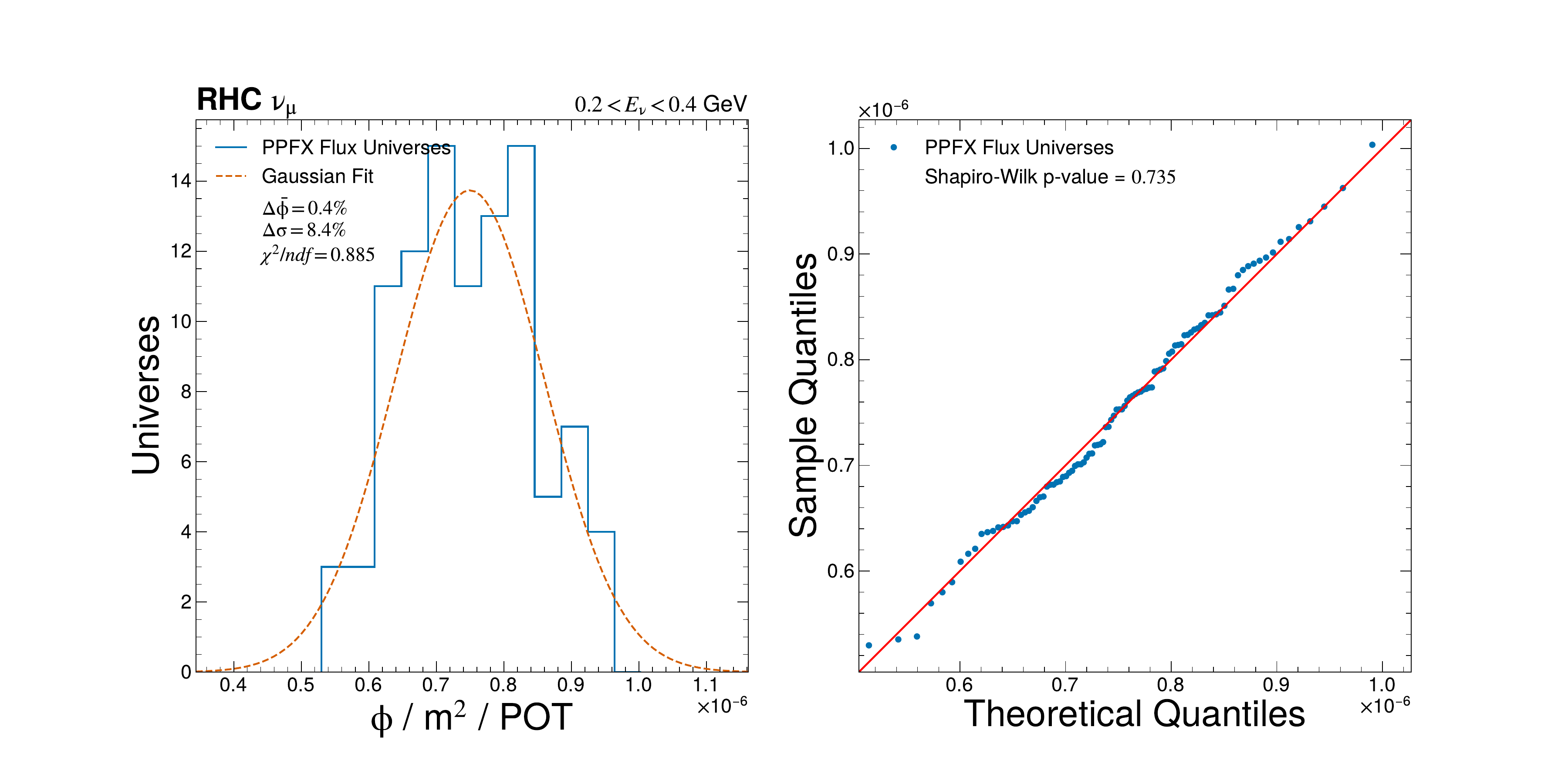}
    \includegraphics[width=0.3\textwidth]{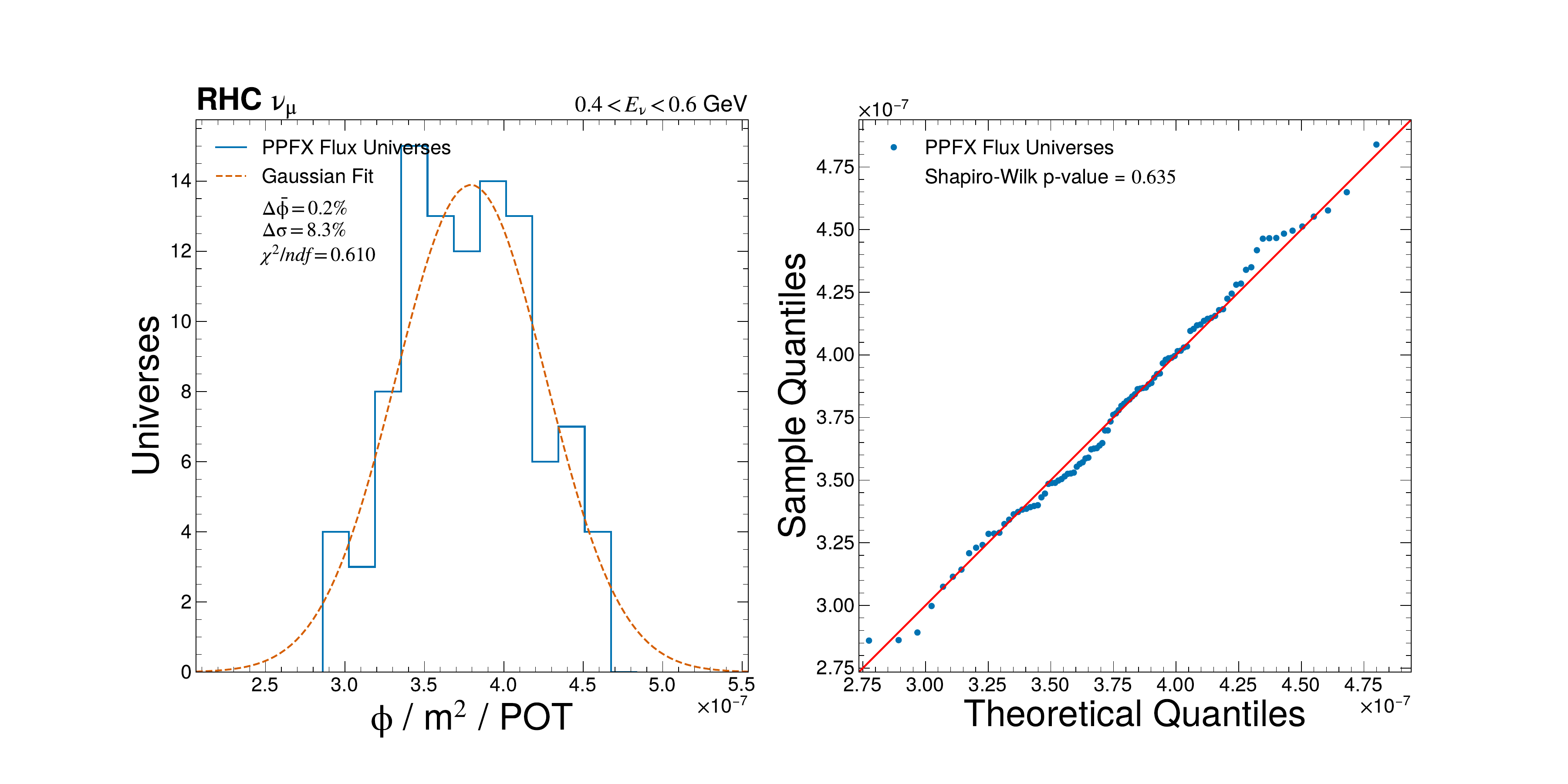}
    \includegraphics[width=0.3\textwidth]{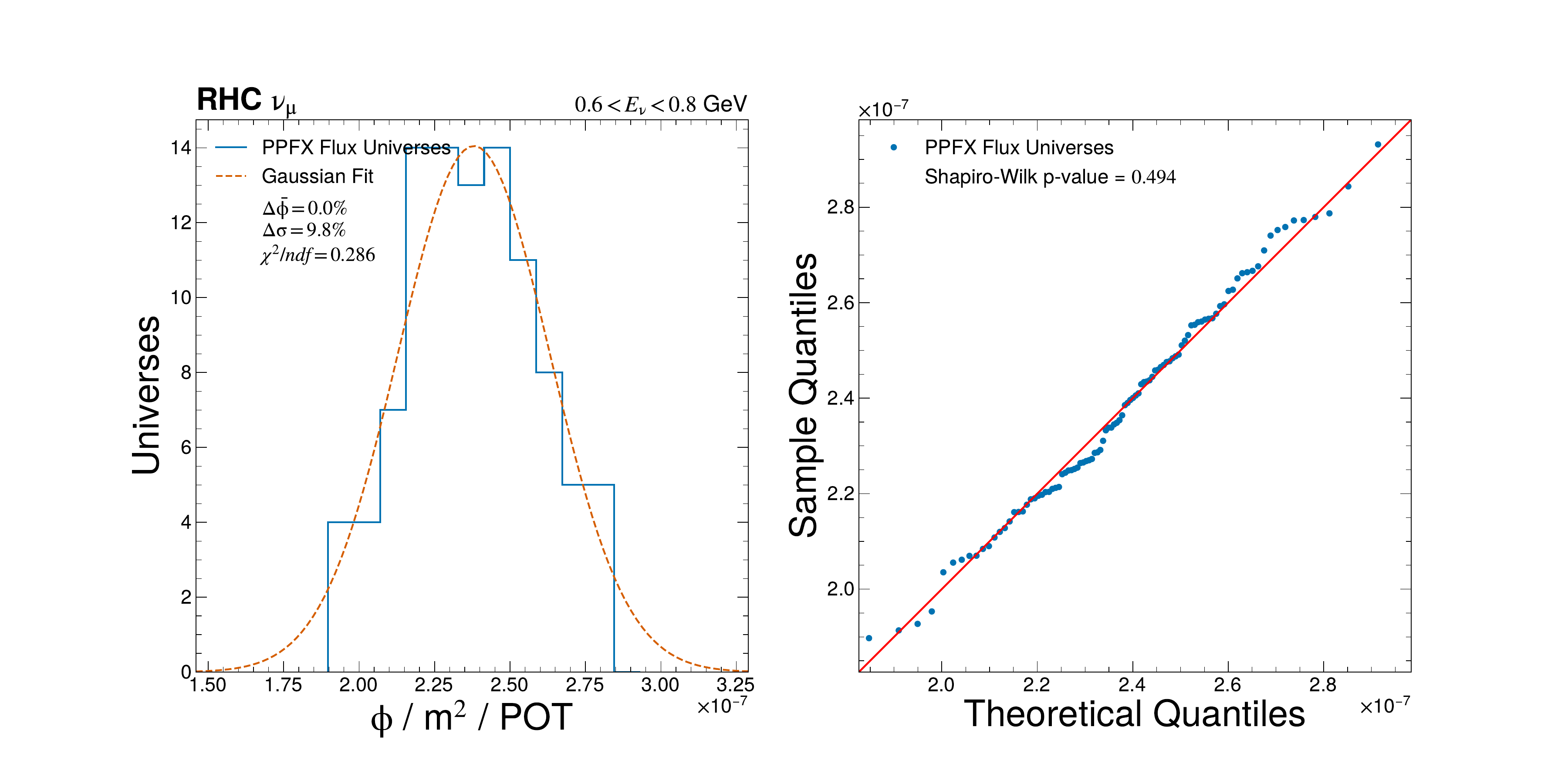}
    \includegraphics[width=0.3\textwidth]{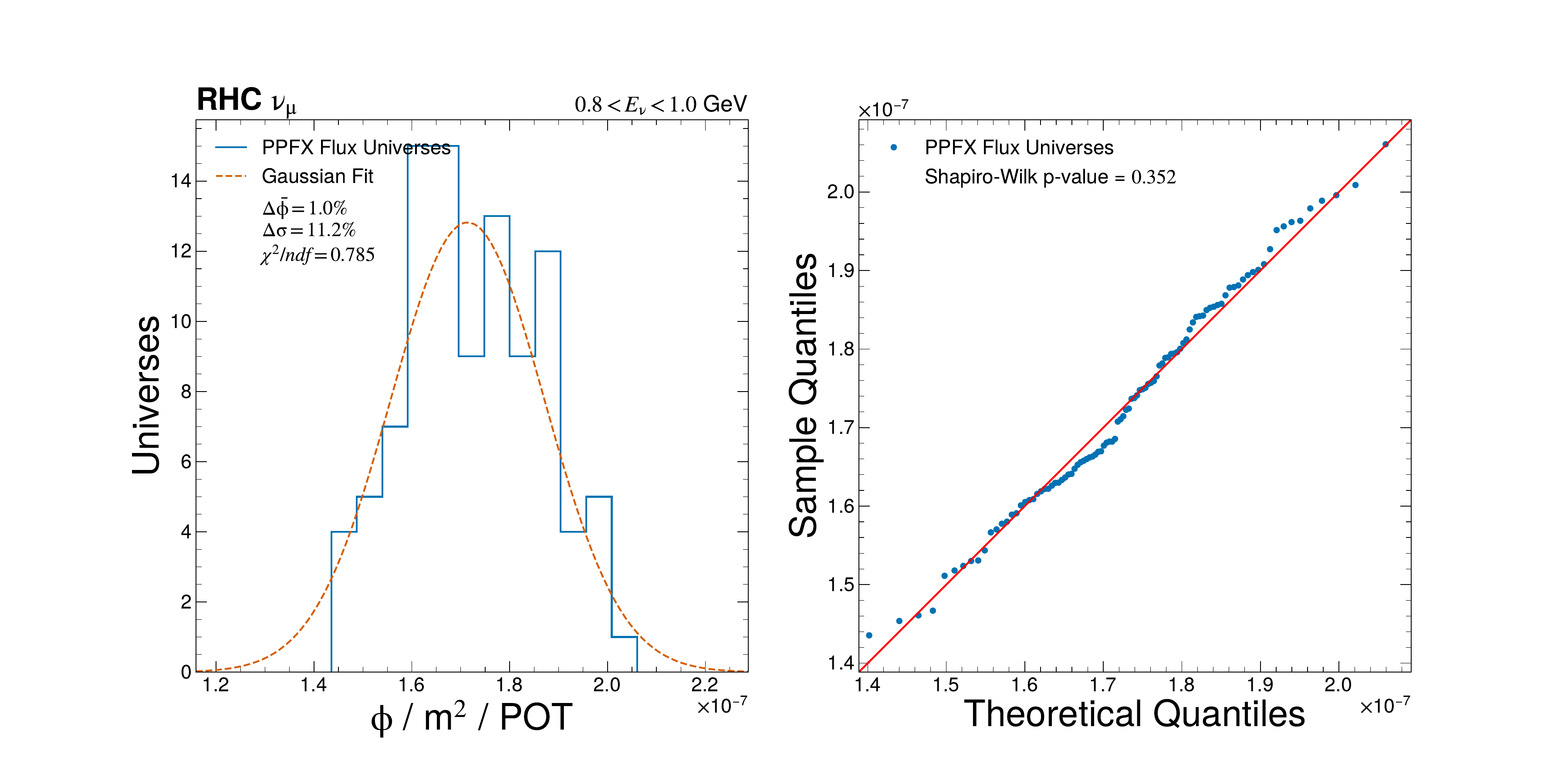}
    \includegraphics[width=0.3\textwidth]{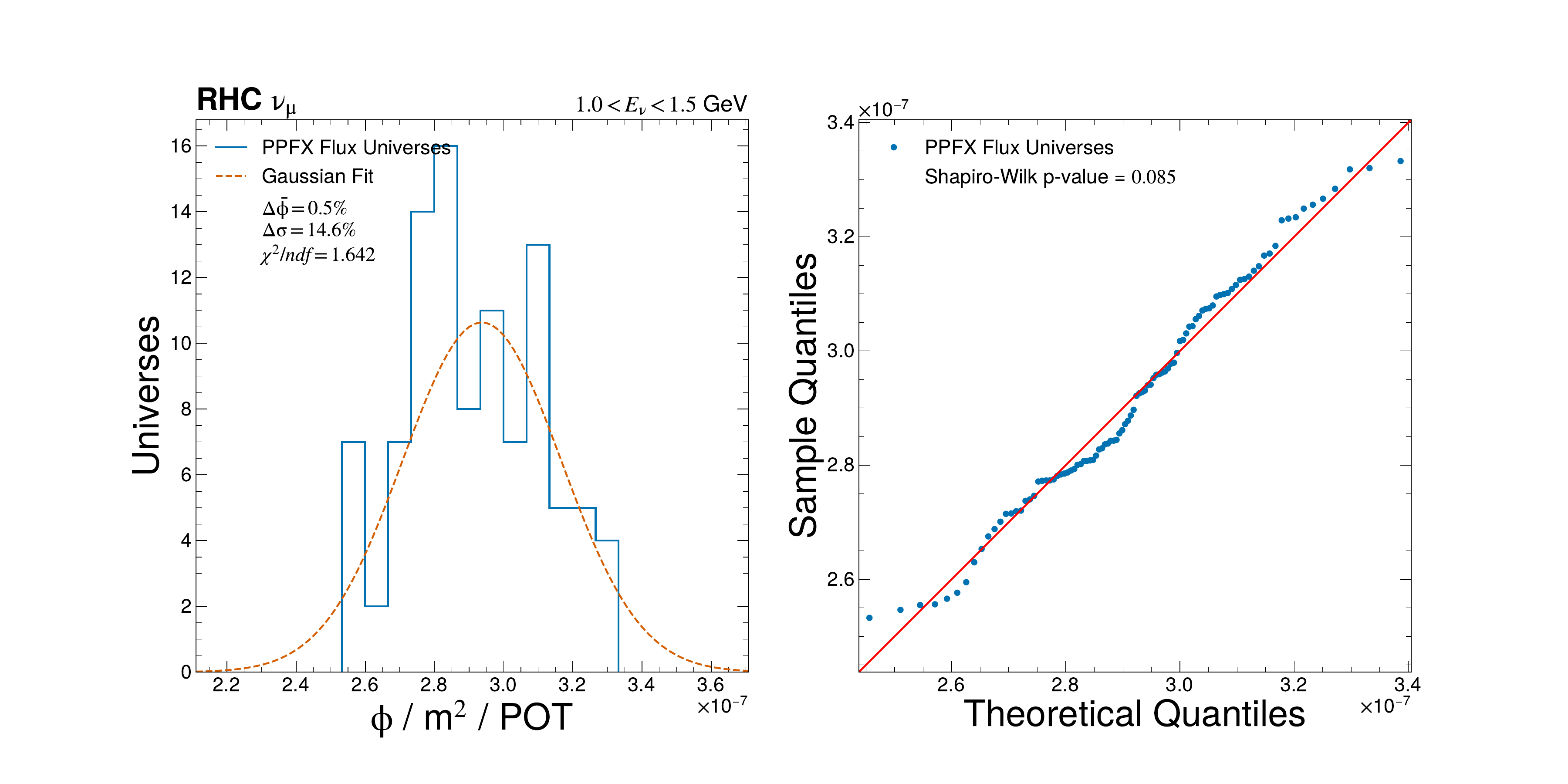}
    \includegraphics[width=0.3\textwidth]{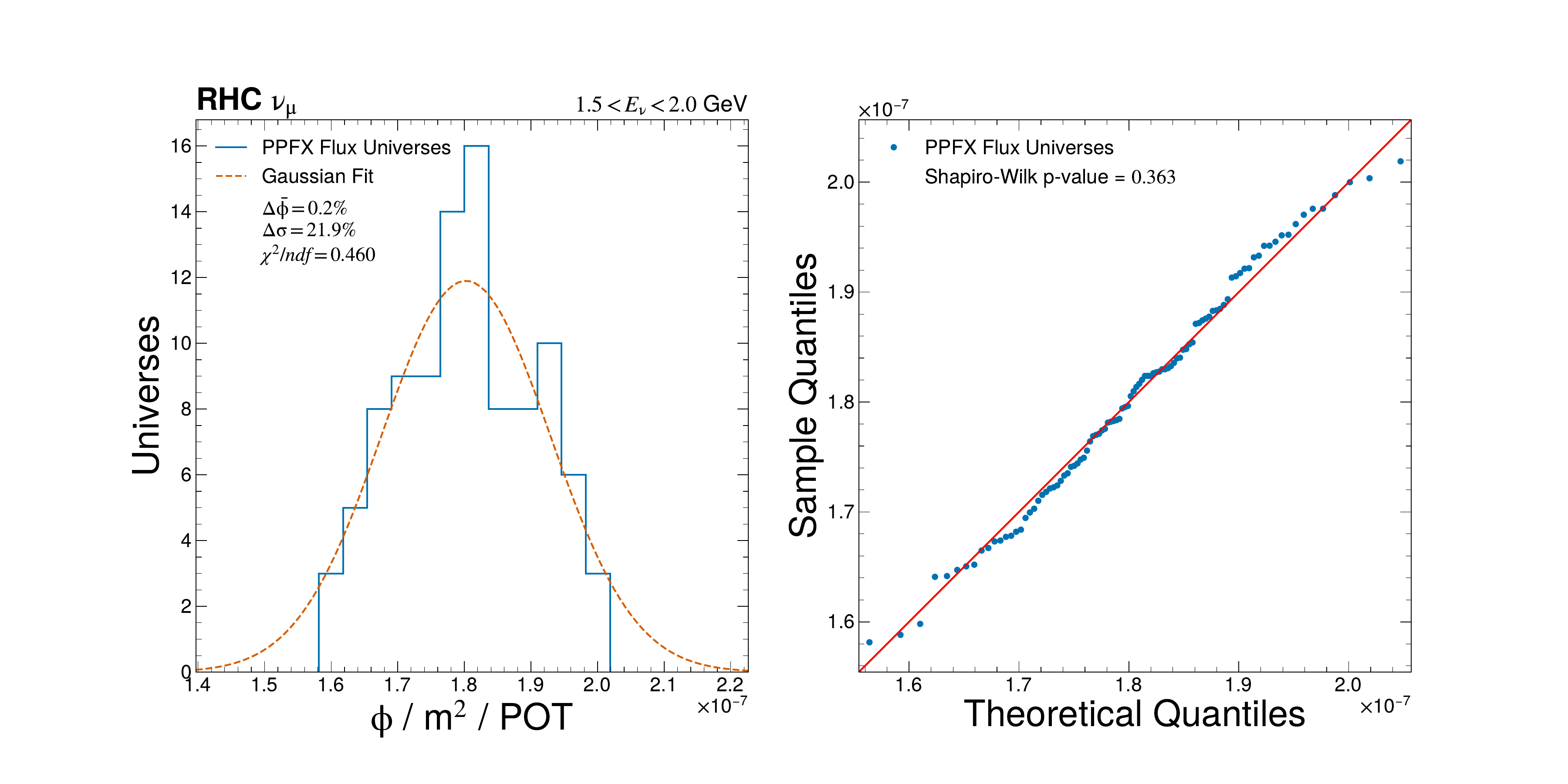}
    \includegraphics[width=0.3\textwidth]{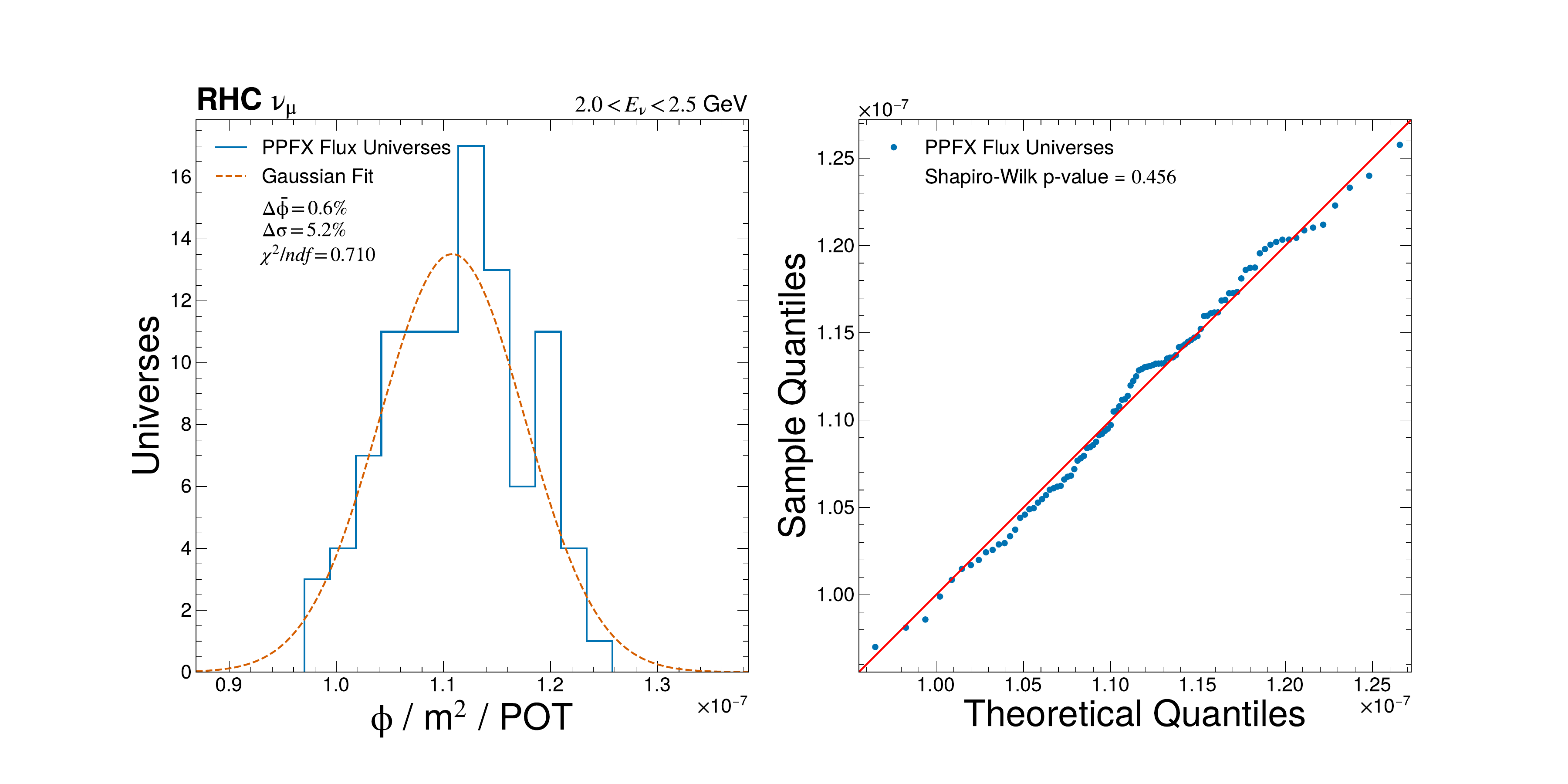}
    \includegraphics[width=0.3\textwidth]{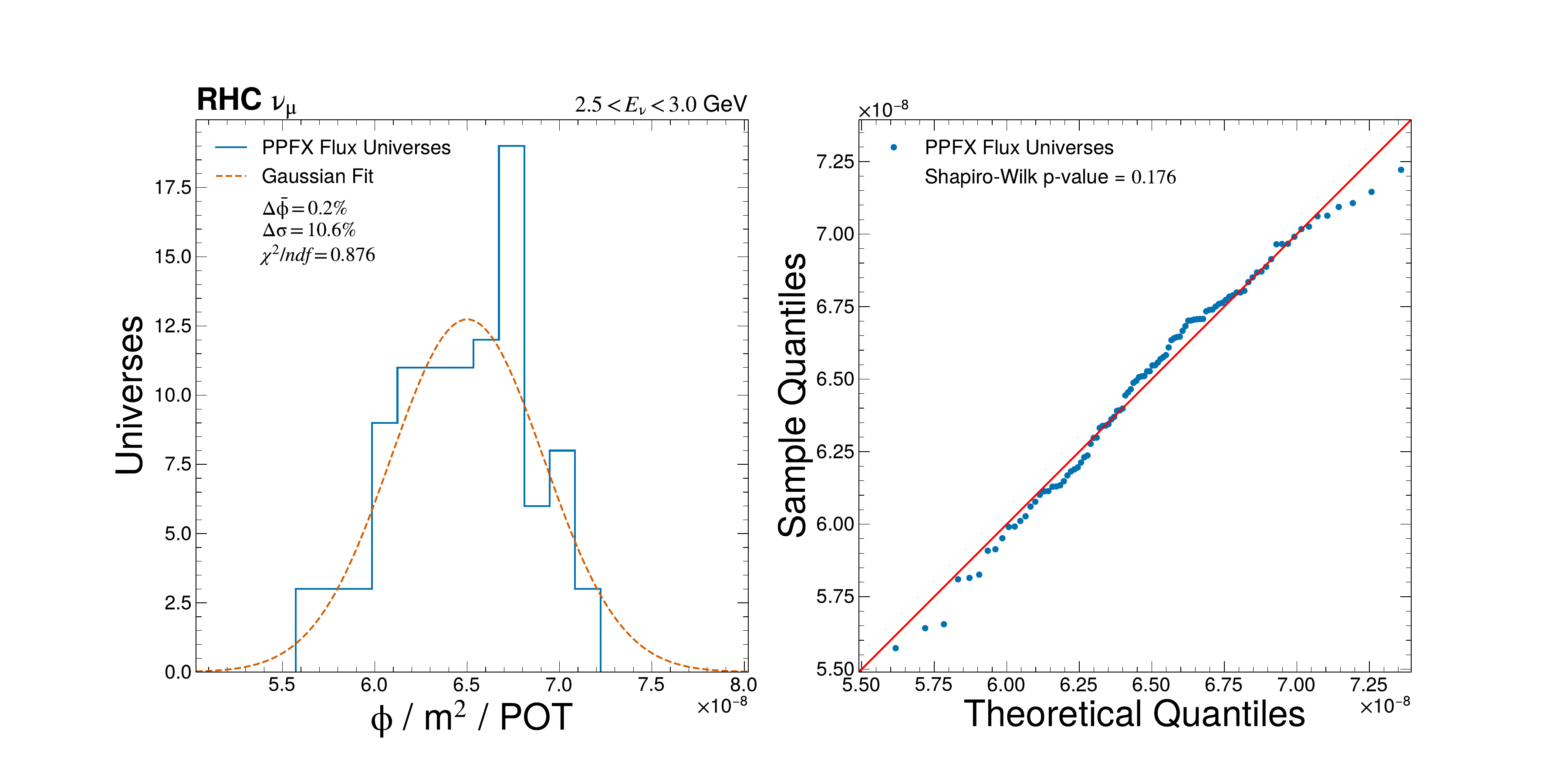}
    \includegraphics[width=0.3\textwidth]{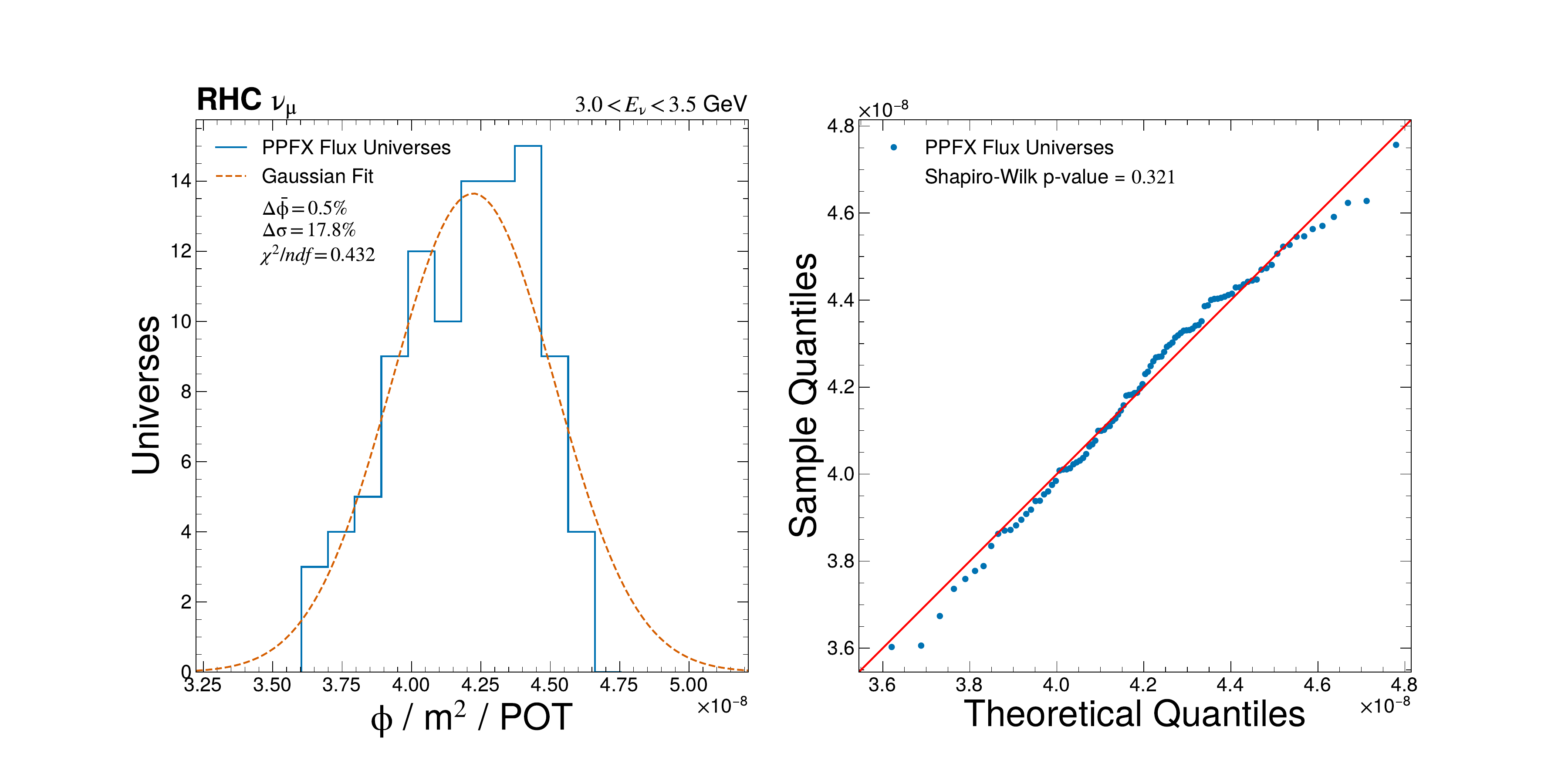}
    \includegraphics[width=0.3\textwidth]{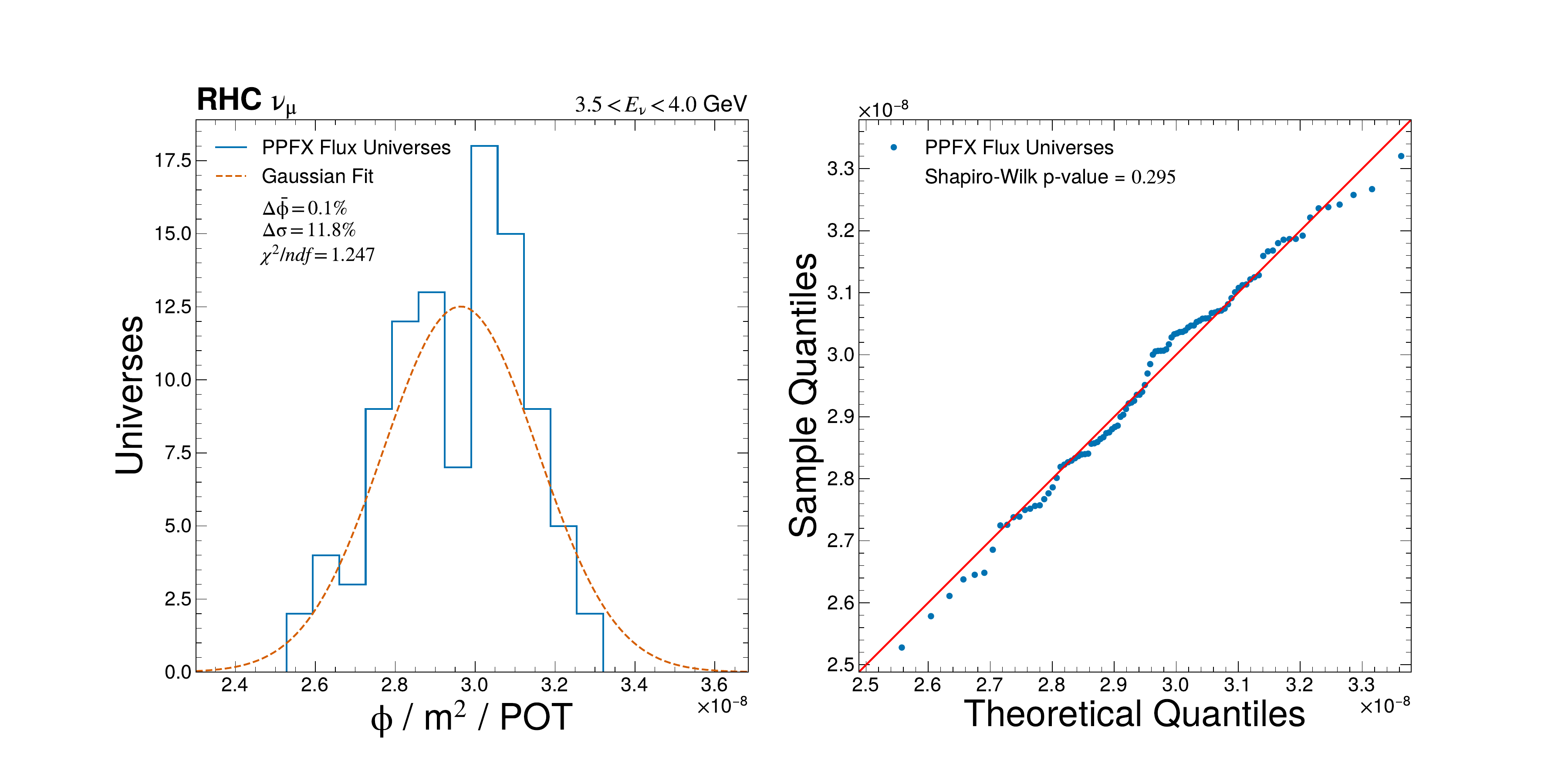}
    \includegraphics[width=0.3\textwidth]{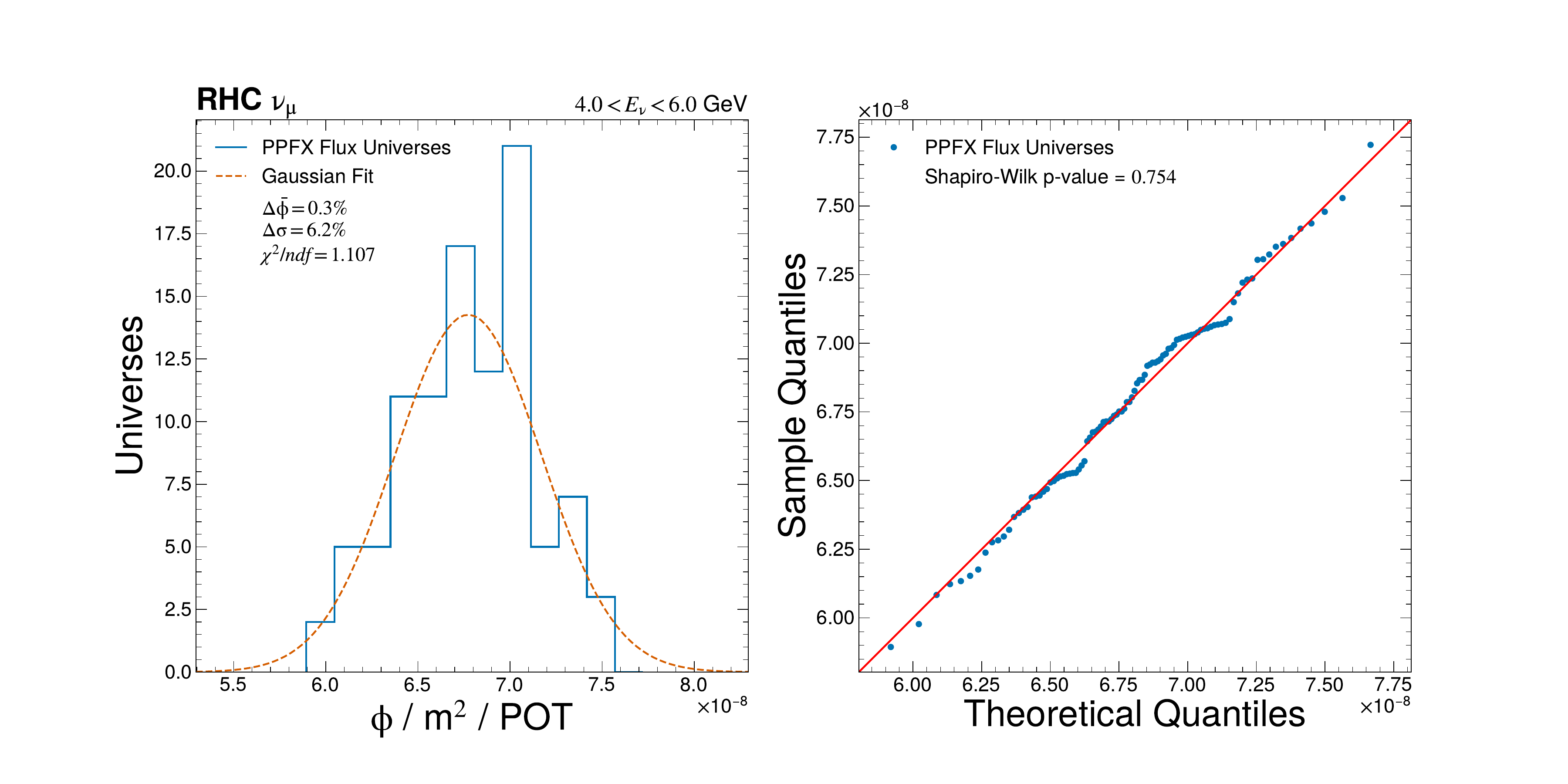}
    \includegraphics[width=0.3\textwidth]{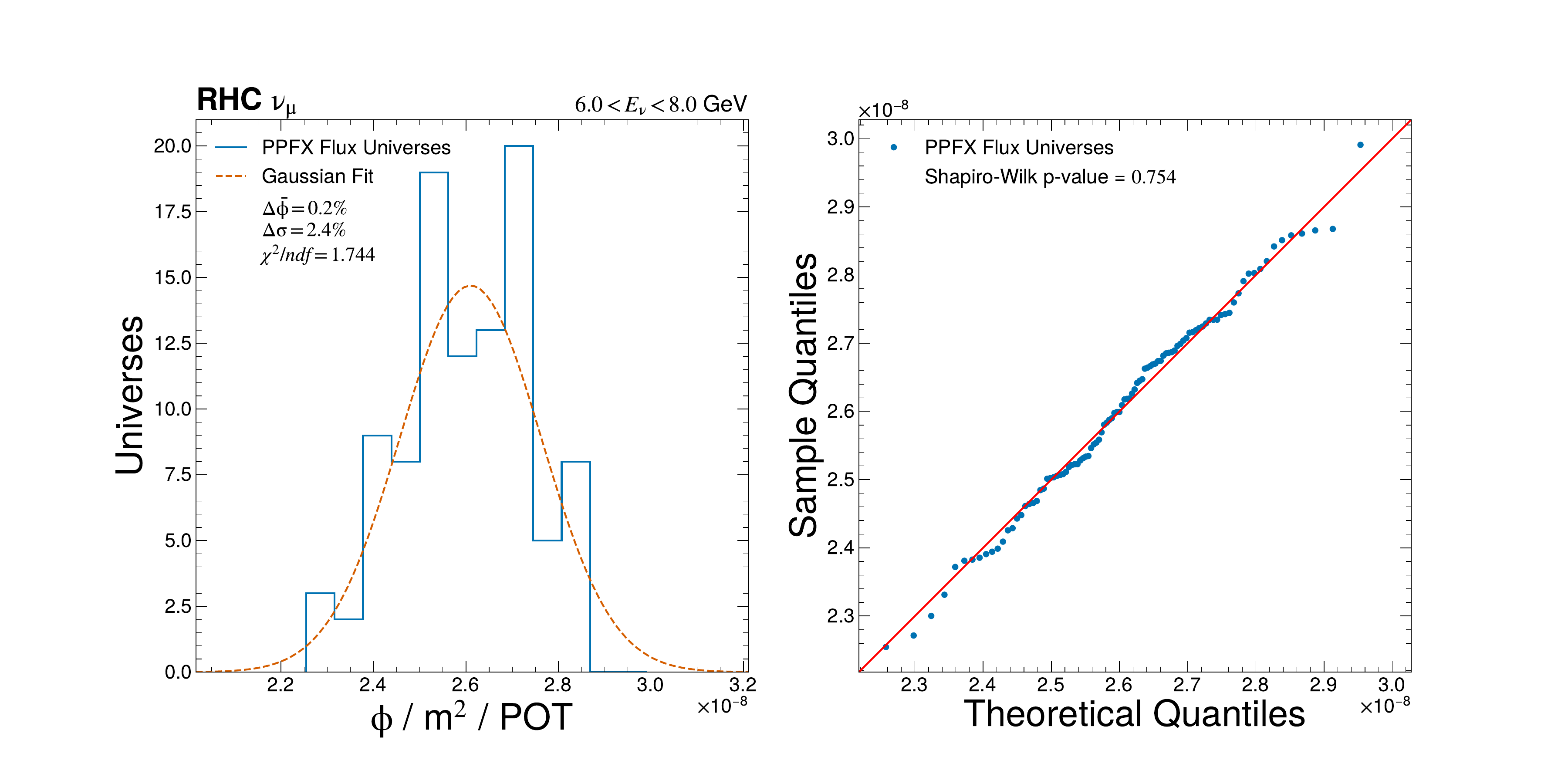}
    \includegraphics[width=0.3\textwidth]{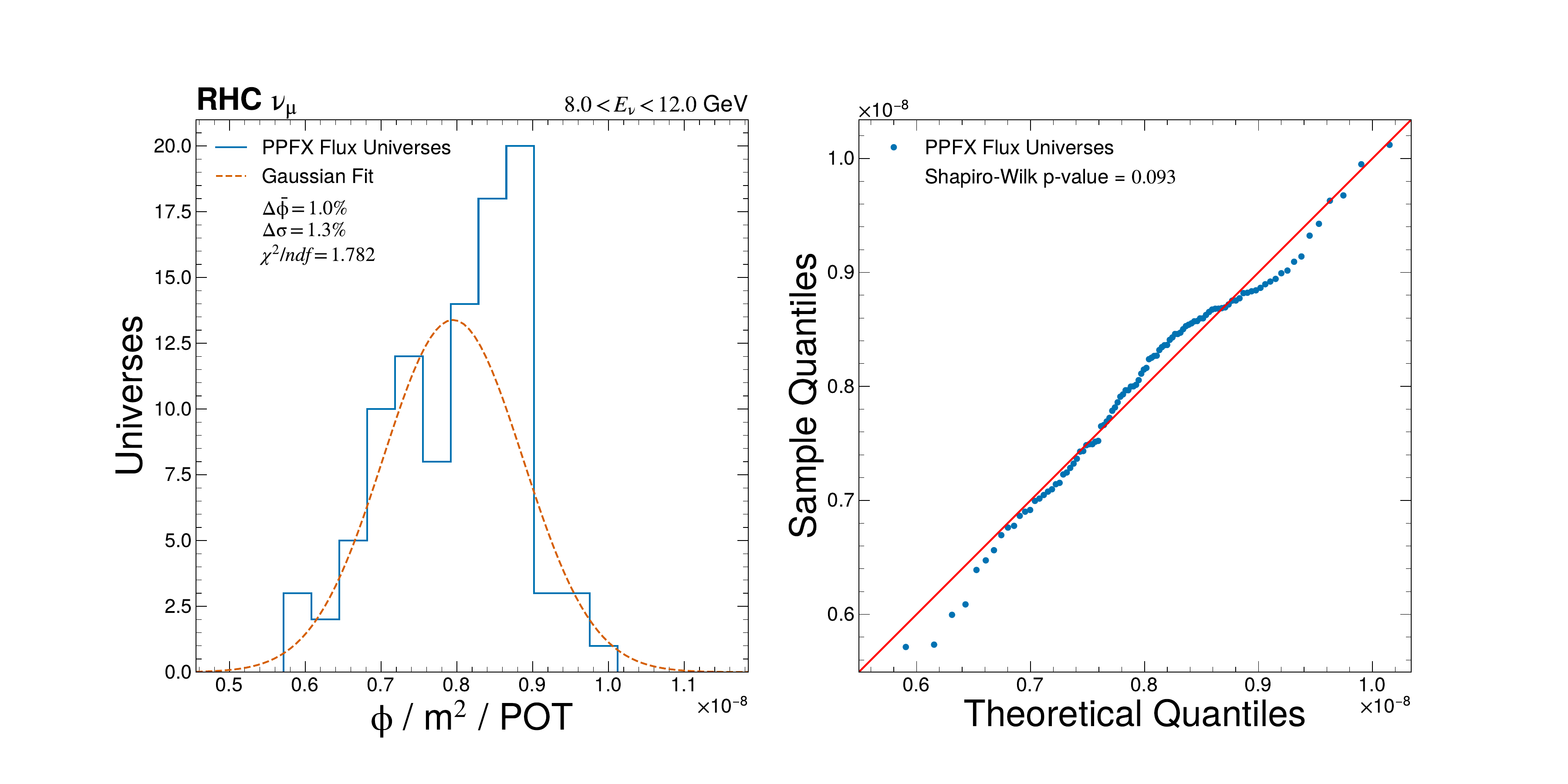}
    \caption[Distribution of PPFX universes for \numu\ (RHC).]{Distribution of PPFX universes for \numu.}
\end{figure}
\begin{figure}[!ht]
    \centering
    \includegraphics[width=0.3\textwidth]{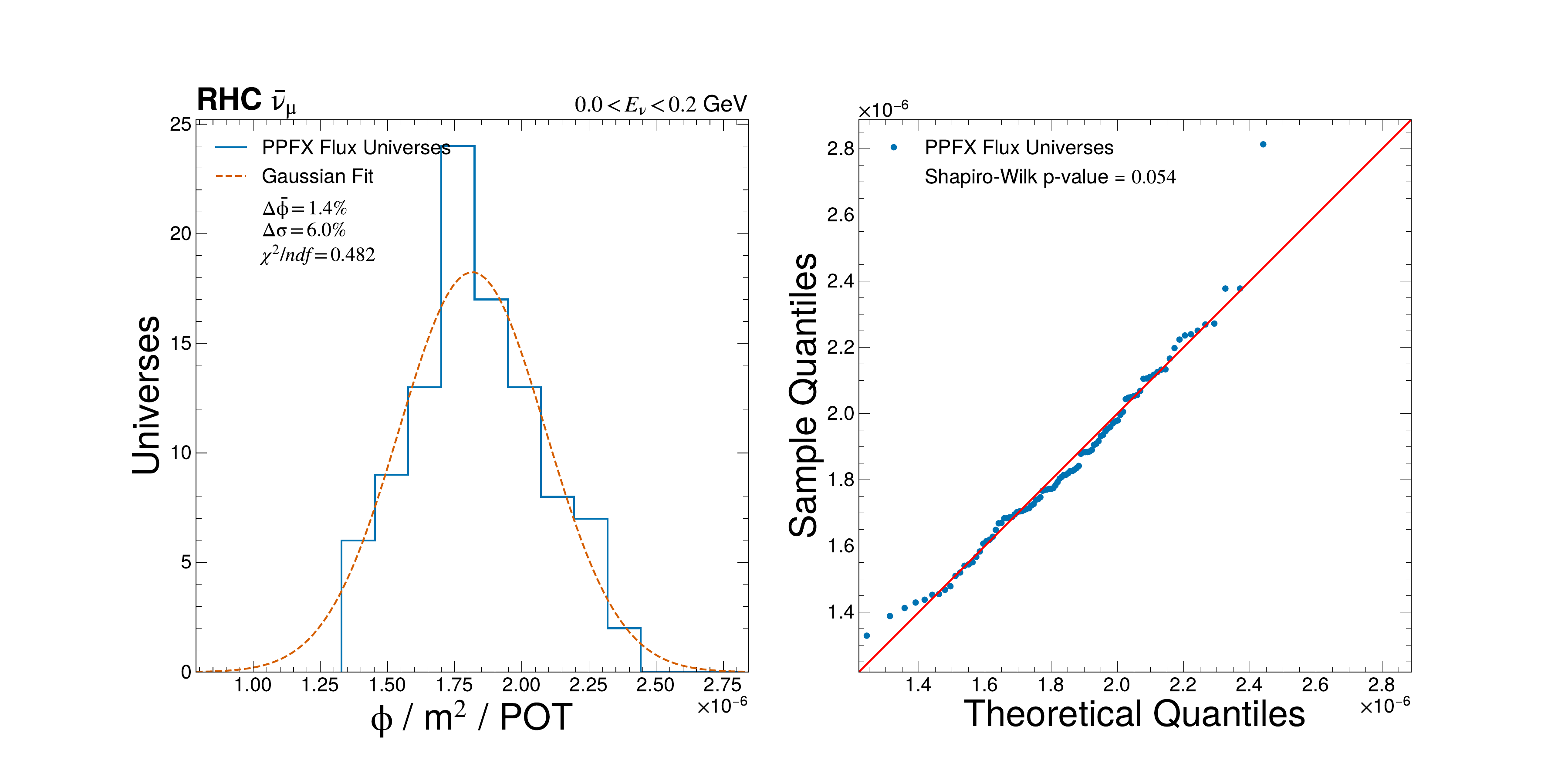}
    \includegraphics[width=0.3\textwidth]{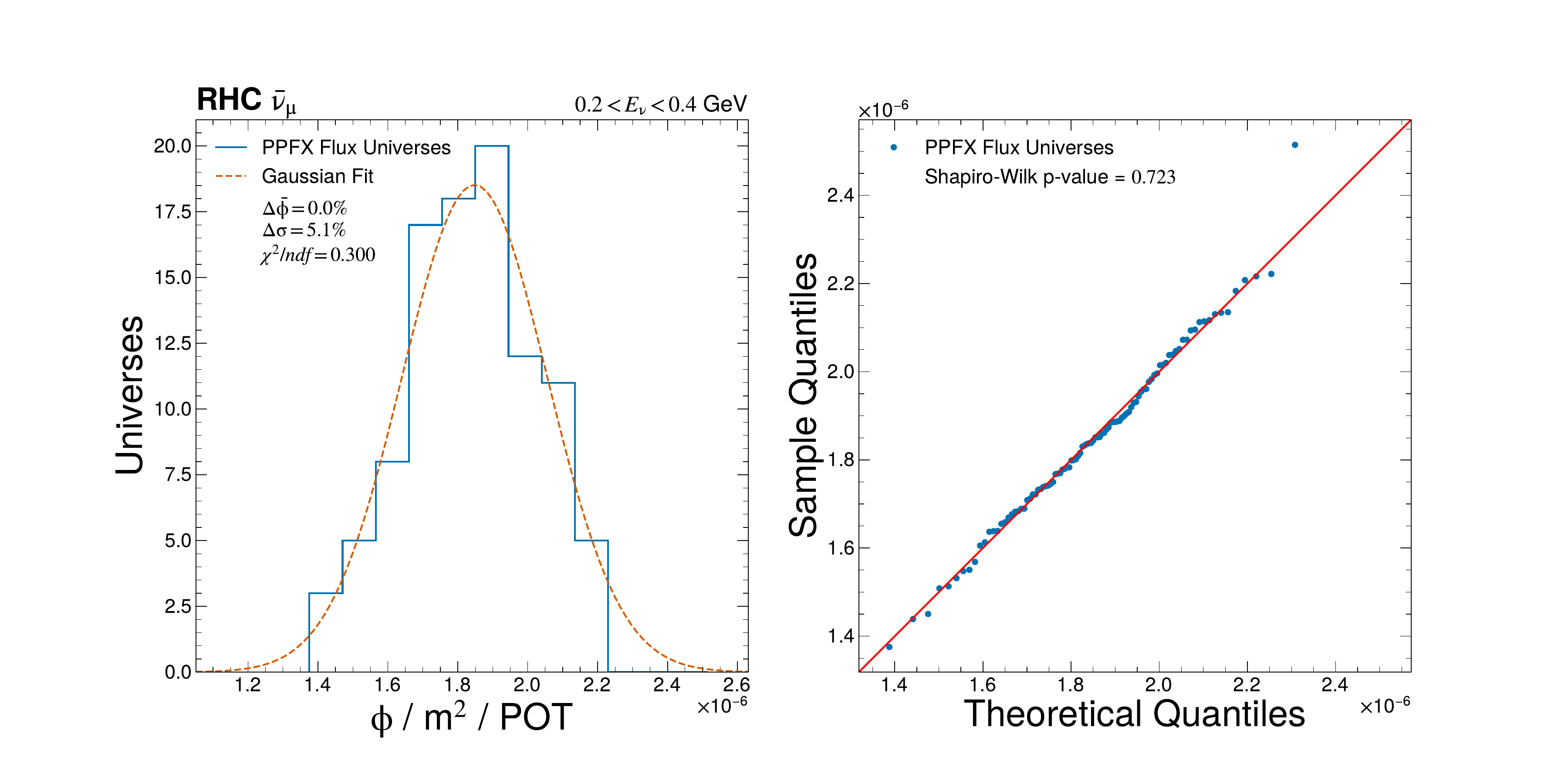}
    \includegraphics[width=0.3\textwidth]{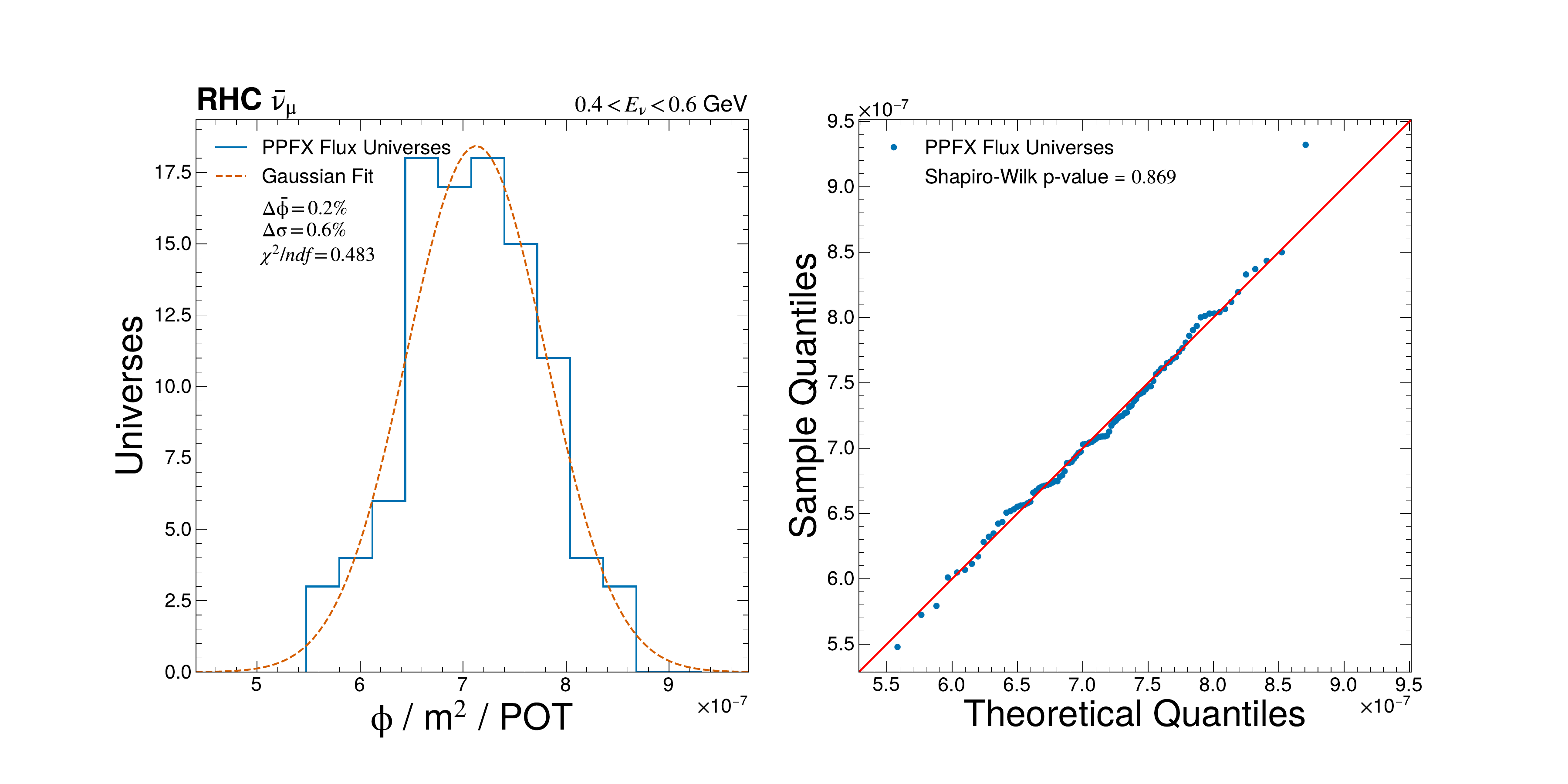}
    \includegraphics[width=0.3\textwidth]{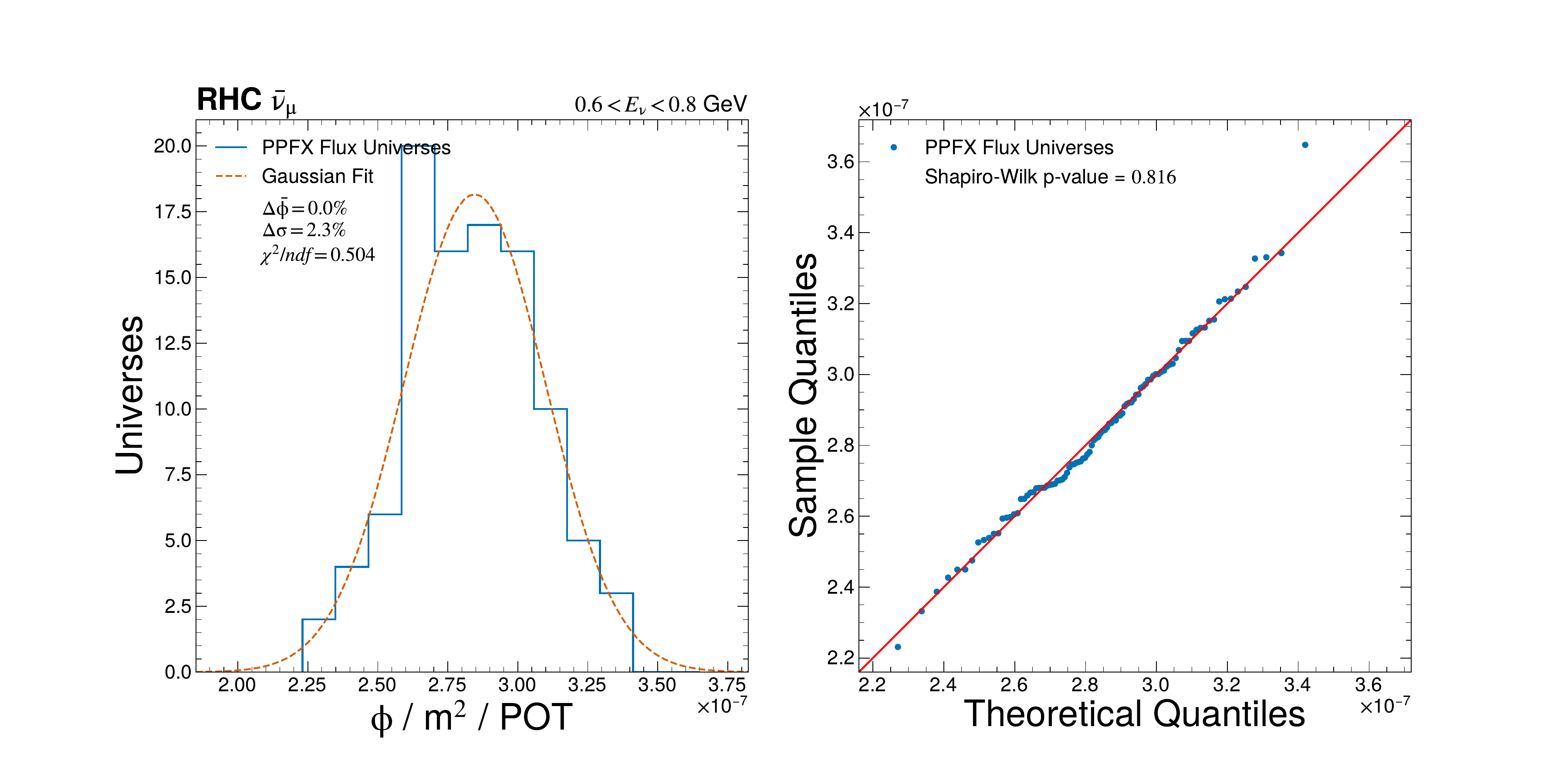}
    \includegraphics[width=0.3\textwidth]{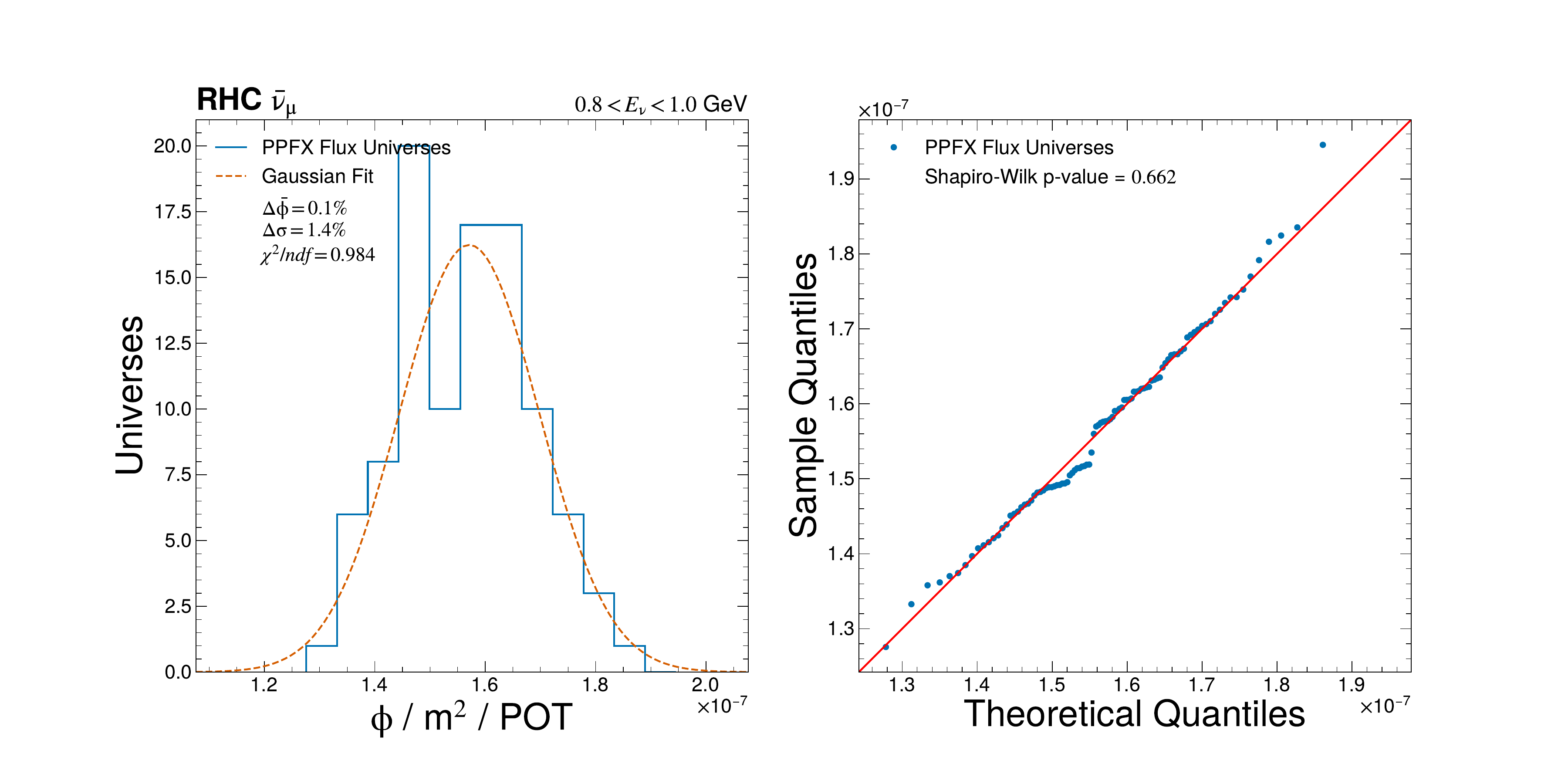}
    \includegraphics[width=0.3\textwidth]{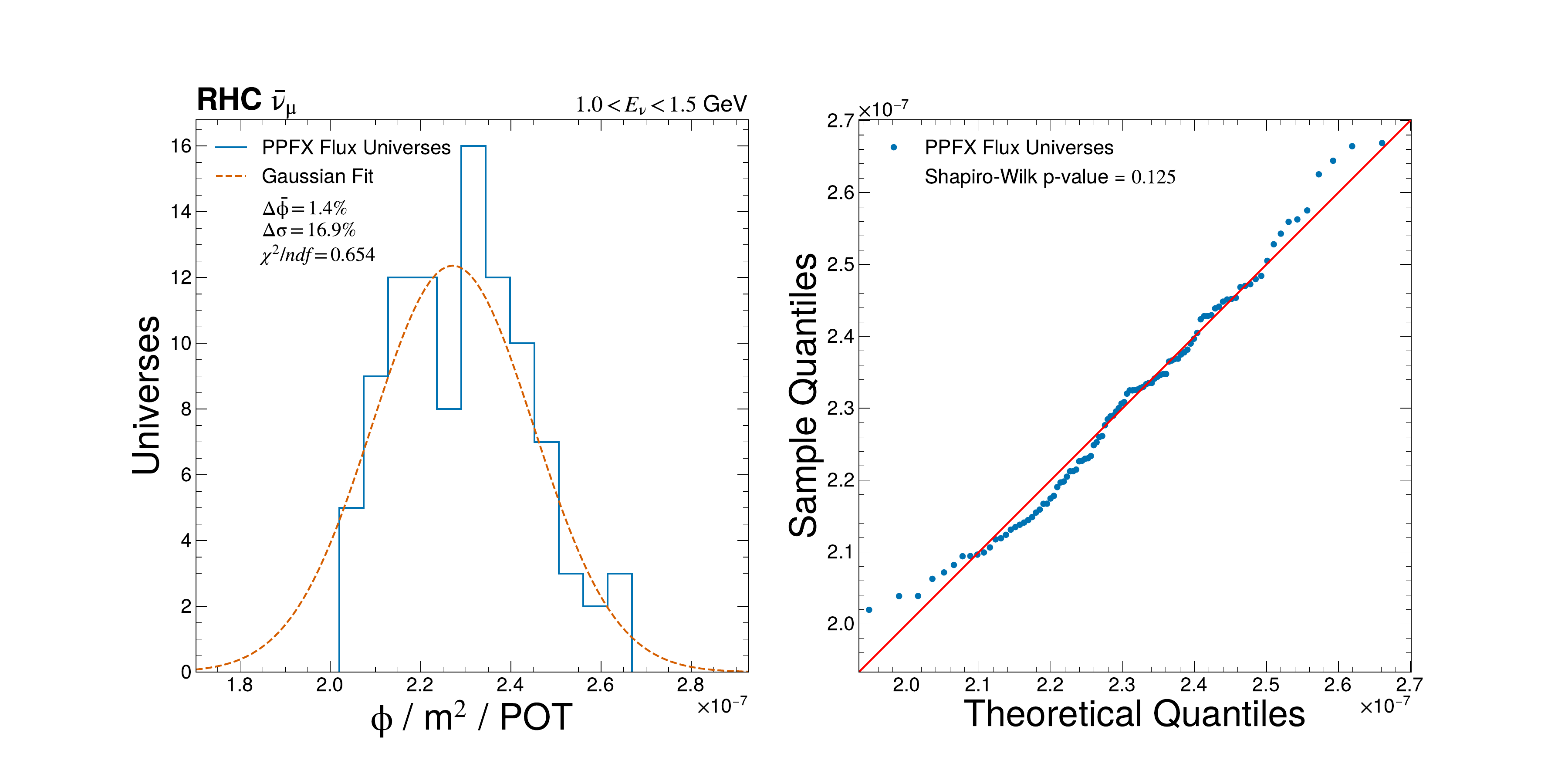}
    \includegraphics[width=0.3\textwidth]{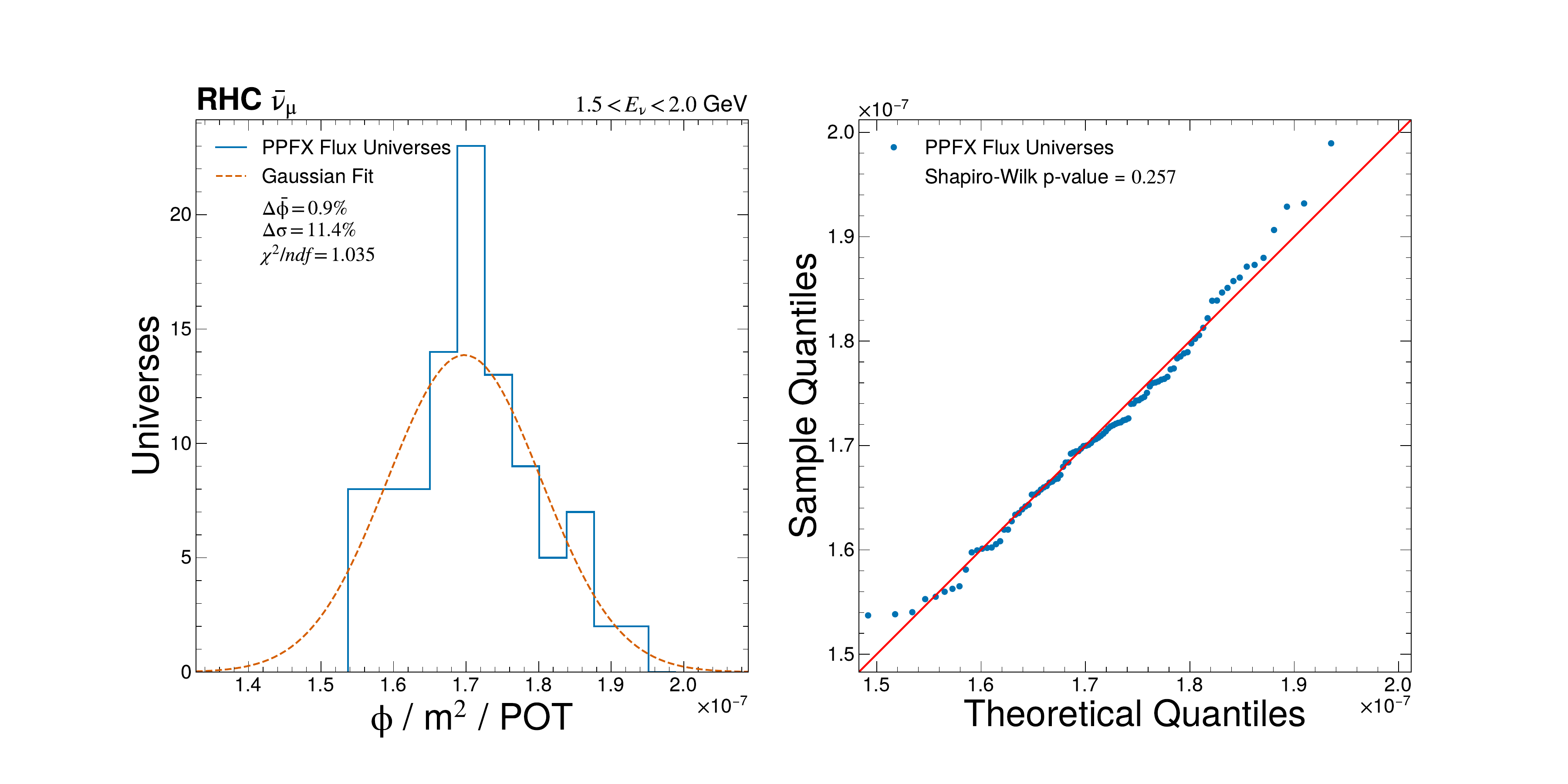}
    \includegraphics[width=0.3\textwidth]{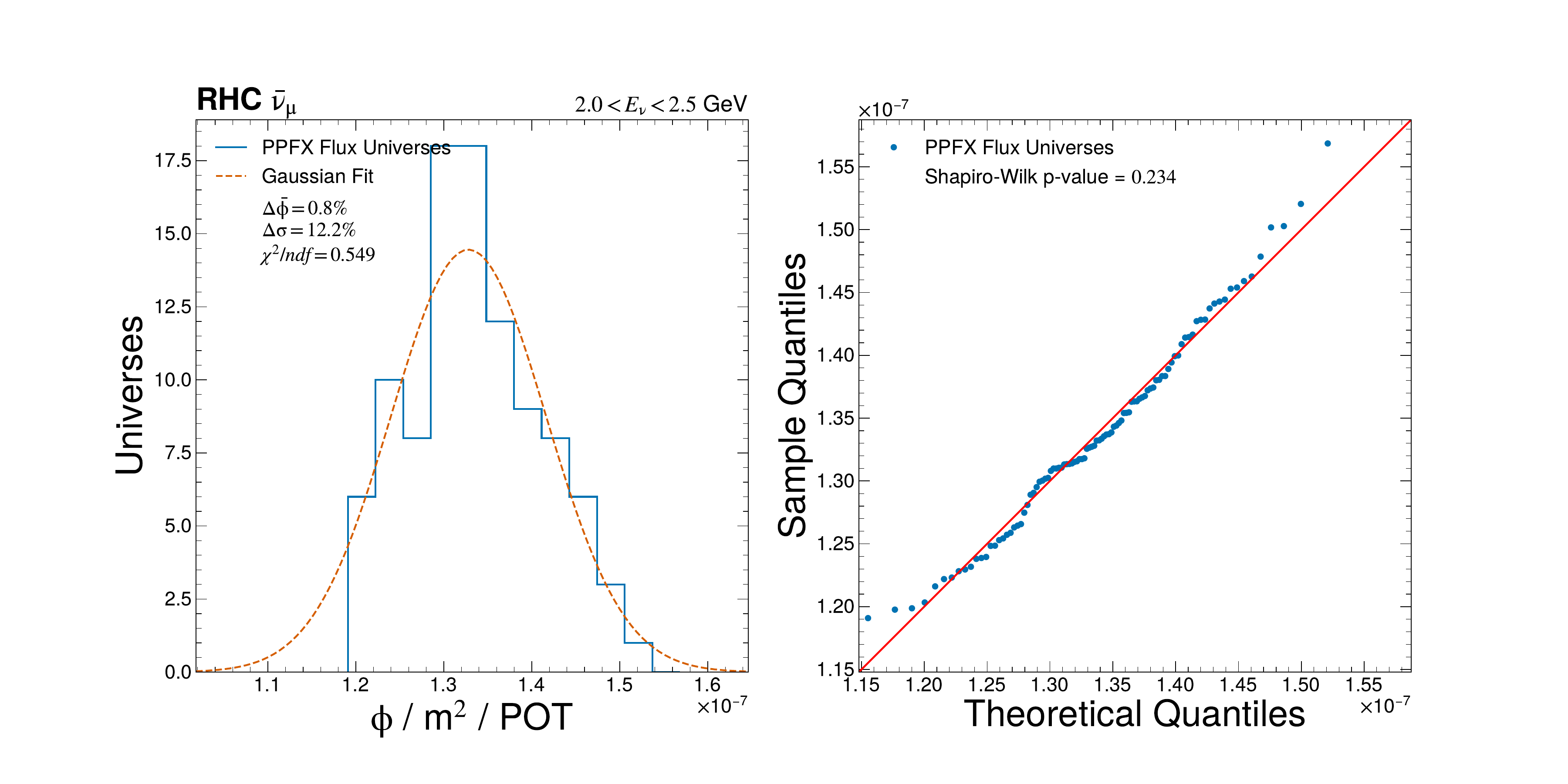}
    \includegraphics[width=0.3\textwidth]{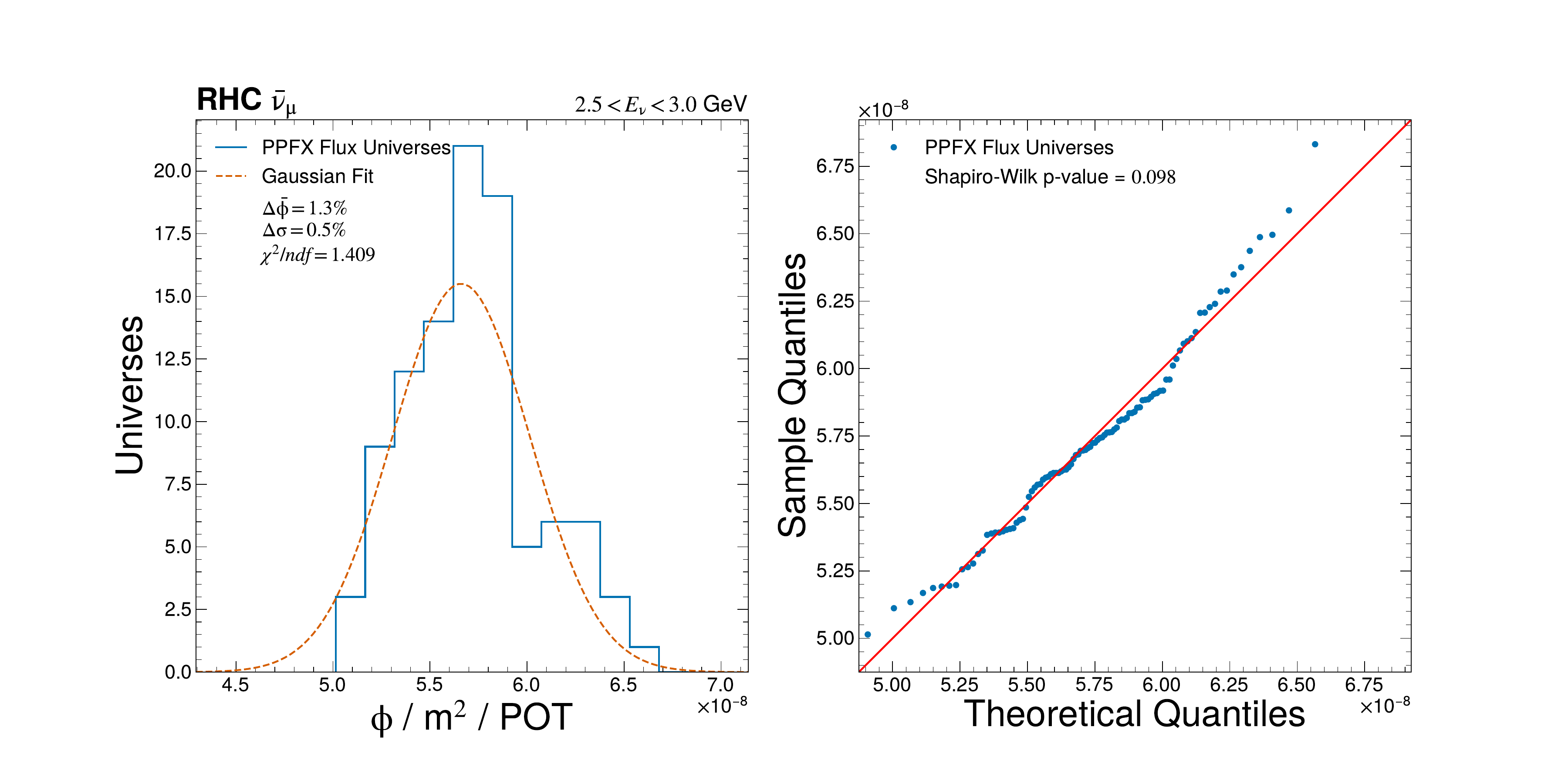}
    \includegraphics[width=0.3\textwidth]{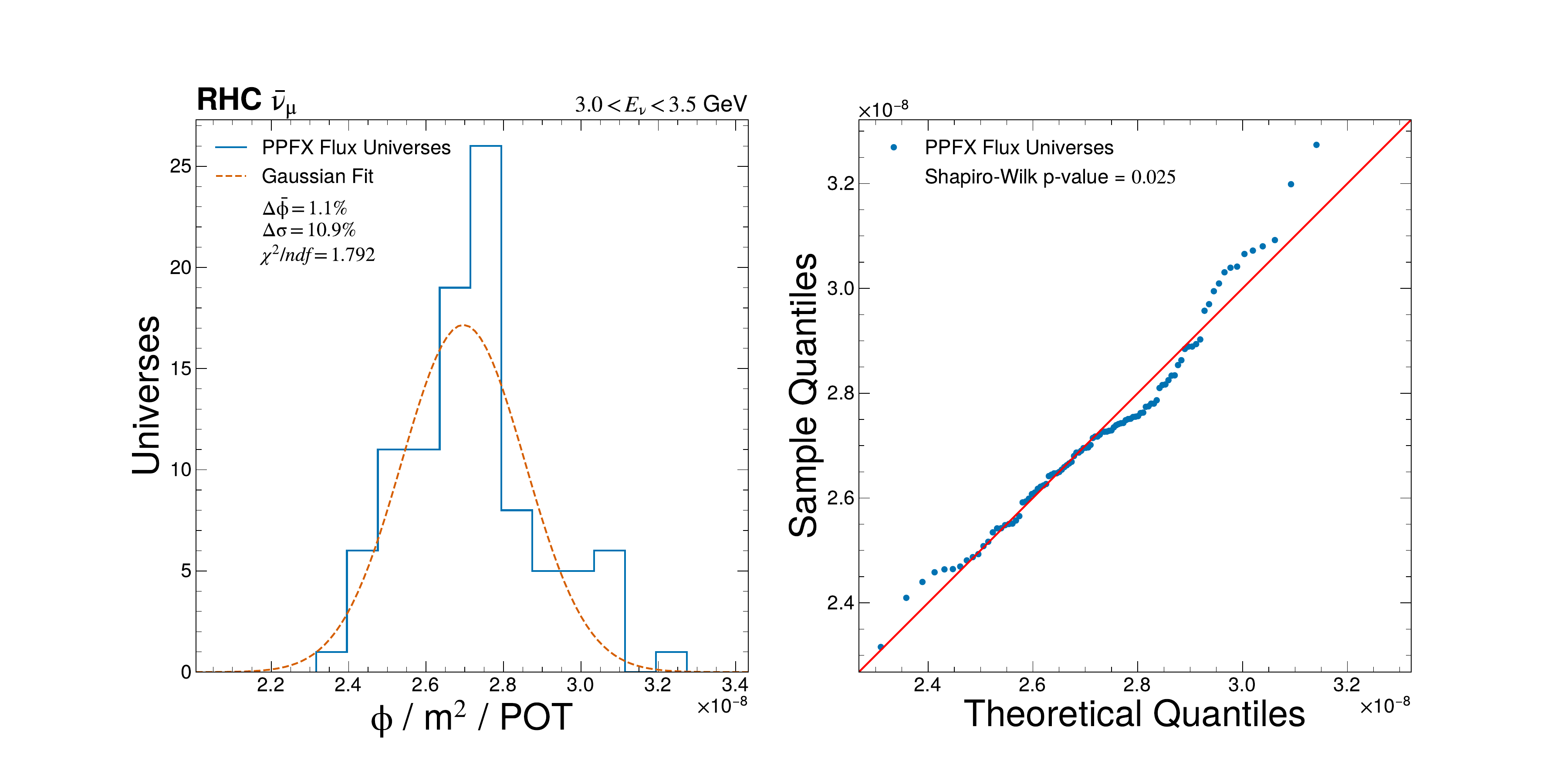}
    \includegraphics[width=0.3\textwidth]{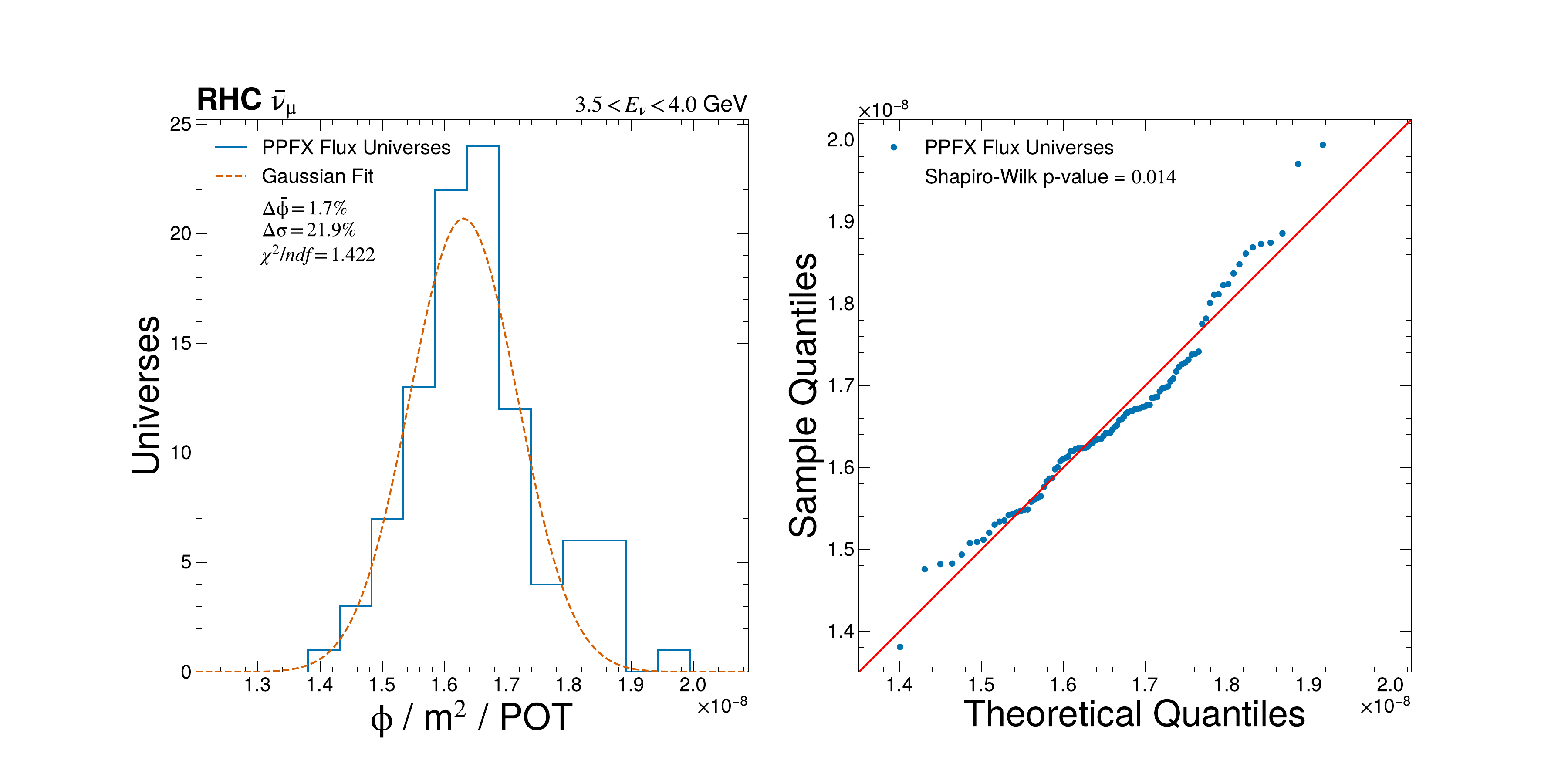}
    \includegraphics[width=0.3\textwidth]{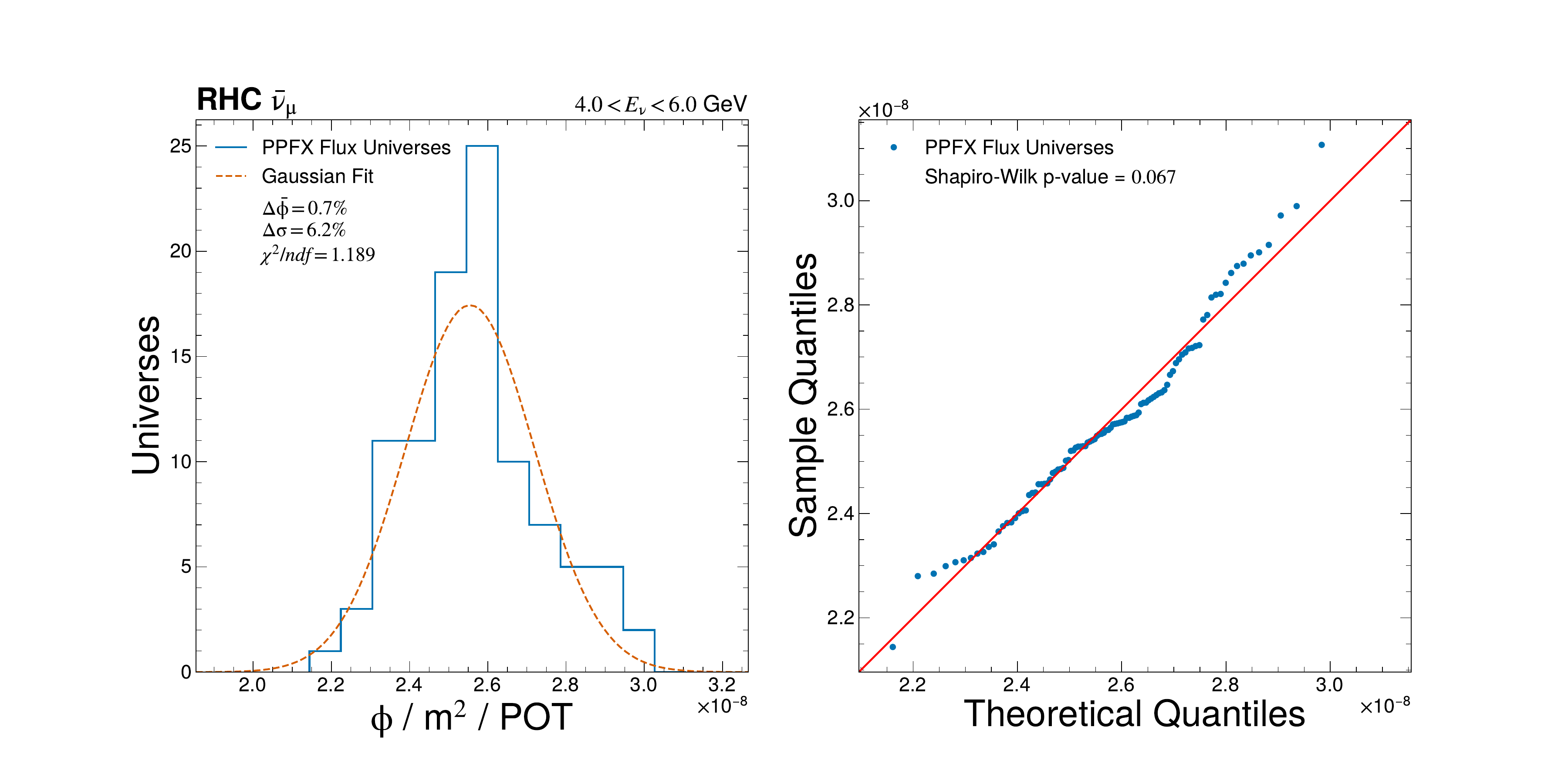}
    \includegraphics[width=0.3\textwidth]{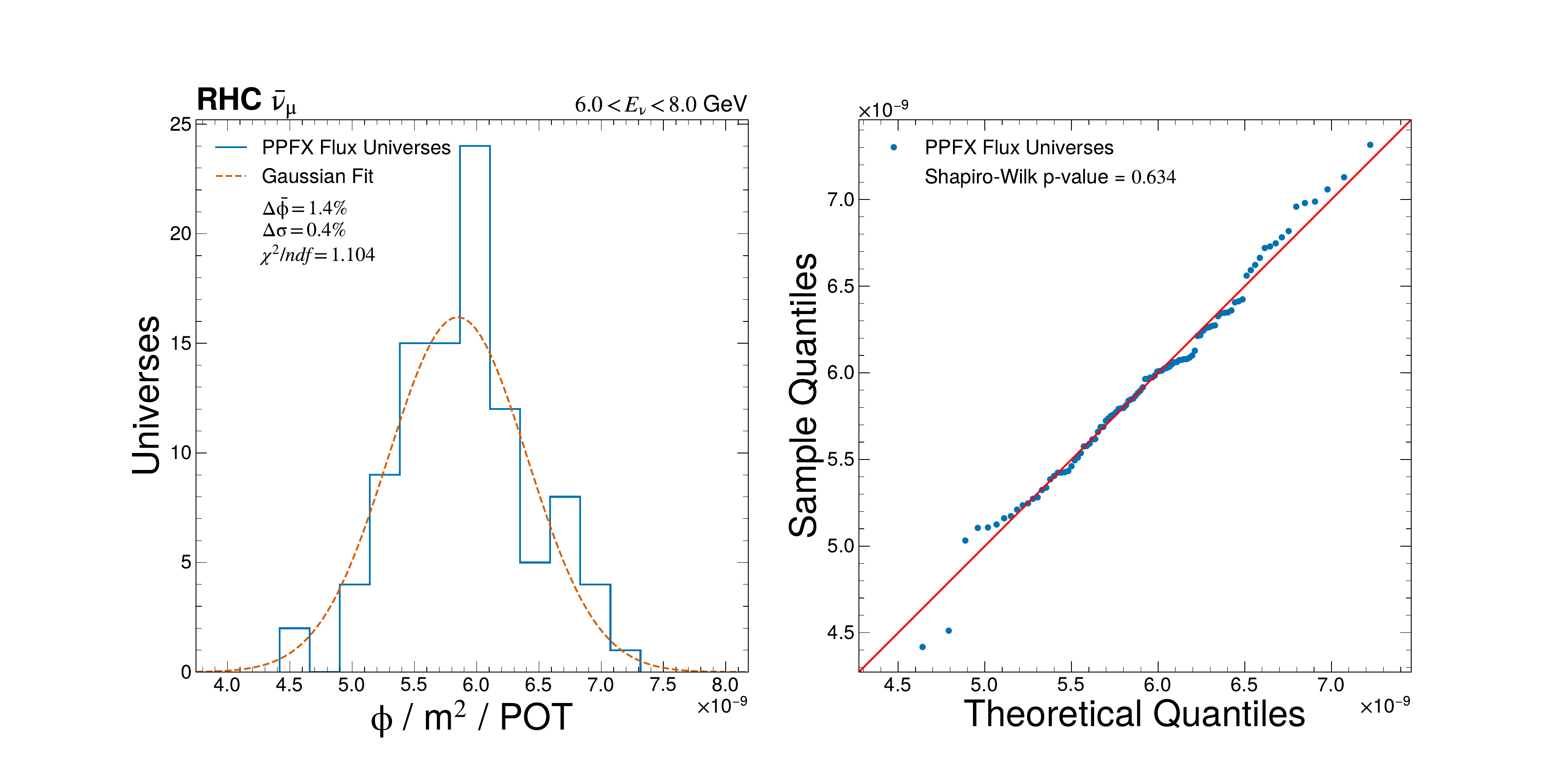}
    \includegraphics[width=0.3\textwidth]{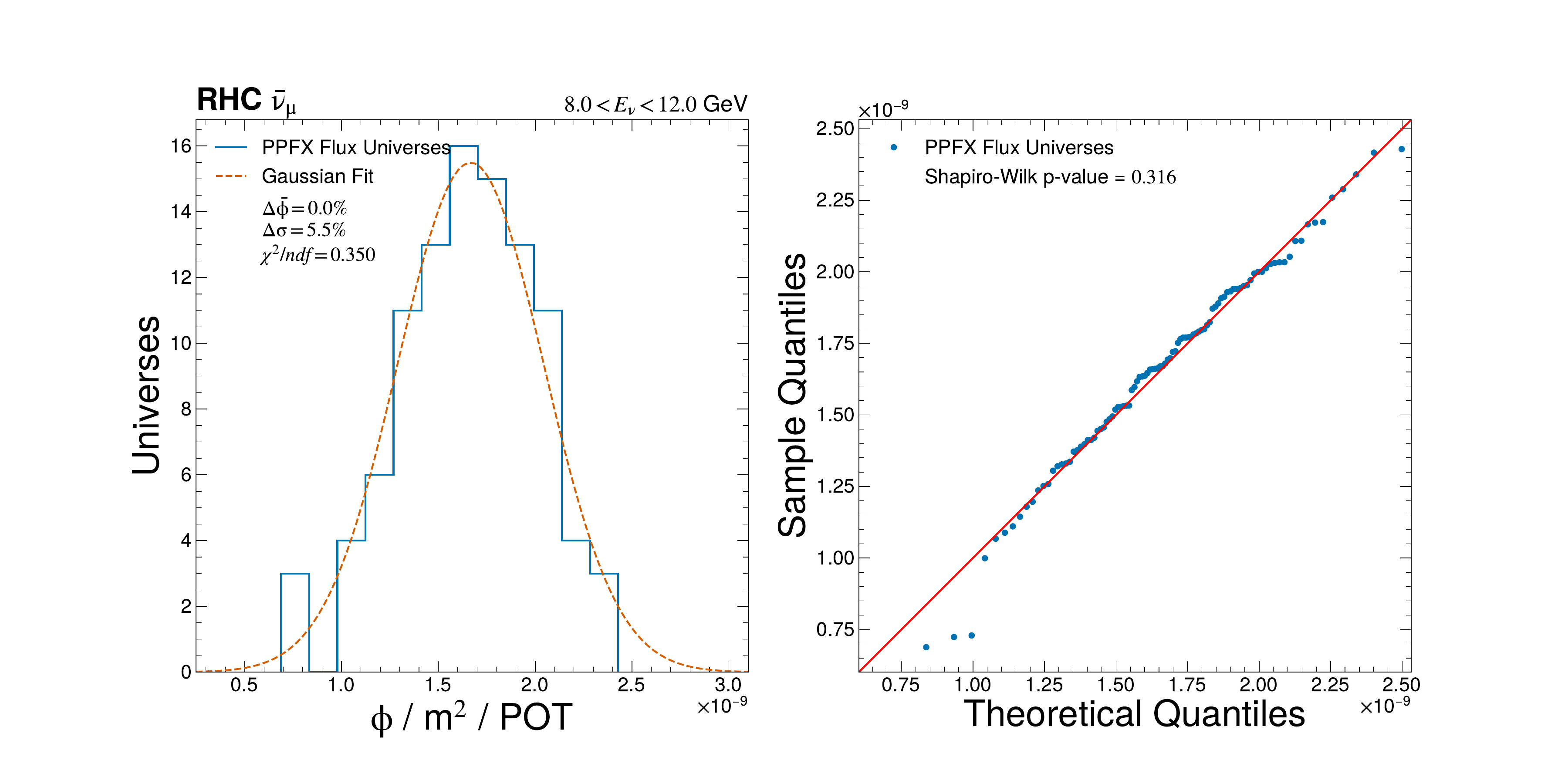}
    \caption[Distribution of PPFX universes for \numub\ (RHC).]{Distribution of PPFX universes for \numub.}
\end{figure}

%% file: flux_prediction.tex
\begin{figure}[!ht]
    \centering
    \begin{subfigure}{0.49\textwidth}
    \includegraphics[width=\textwidth]{UPDATED_fhc_numu_flux_prediction.pdf}
        \caption{FHC \numu\ and \numub}
    \end{subfigure}
    \begin{subfigure}{0.49\textwidth}
    \includegraphics[width=\textwidth]{UPDATED_fhc_nue_flux_prediction.pdf}
    \caption{FHC \nue\ and \nueb}
    \end{subfigure}
    \begin{subfigure}{0.49\textwidth}
    \includegraphics[width=\textwidth]{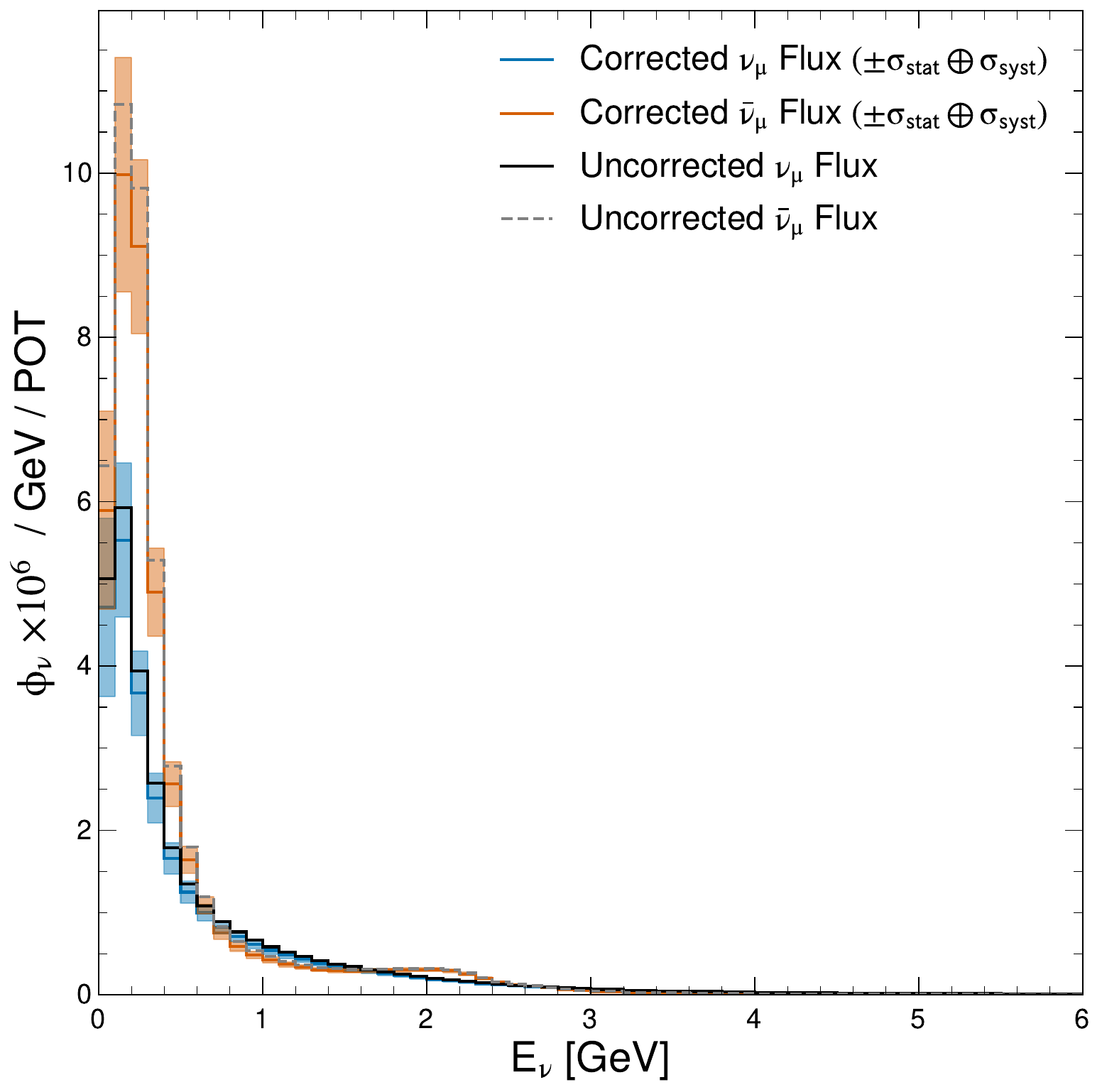}
    \caption{RHC \numu\ and \numub}
    \end{subfigure}
    \begin{subfigure}{0.49\textwidth}
    \includegraphics[width=\textwidth]{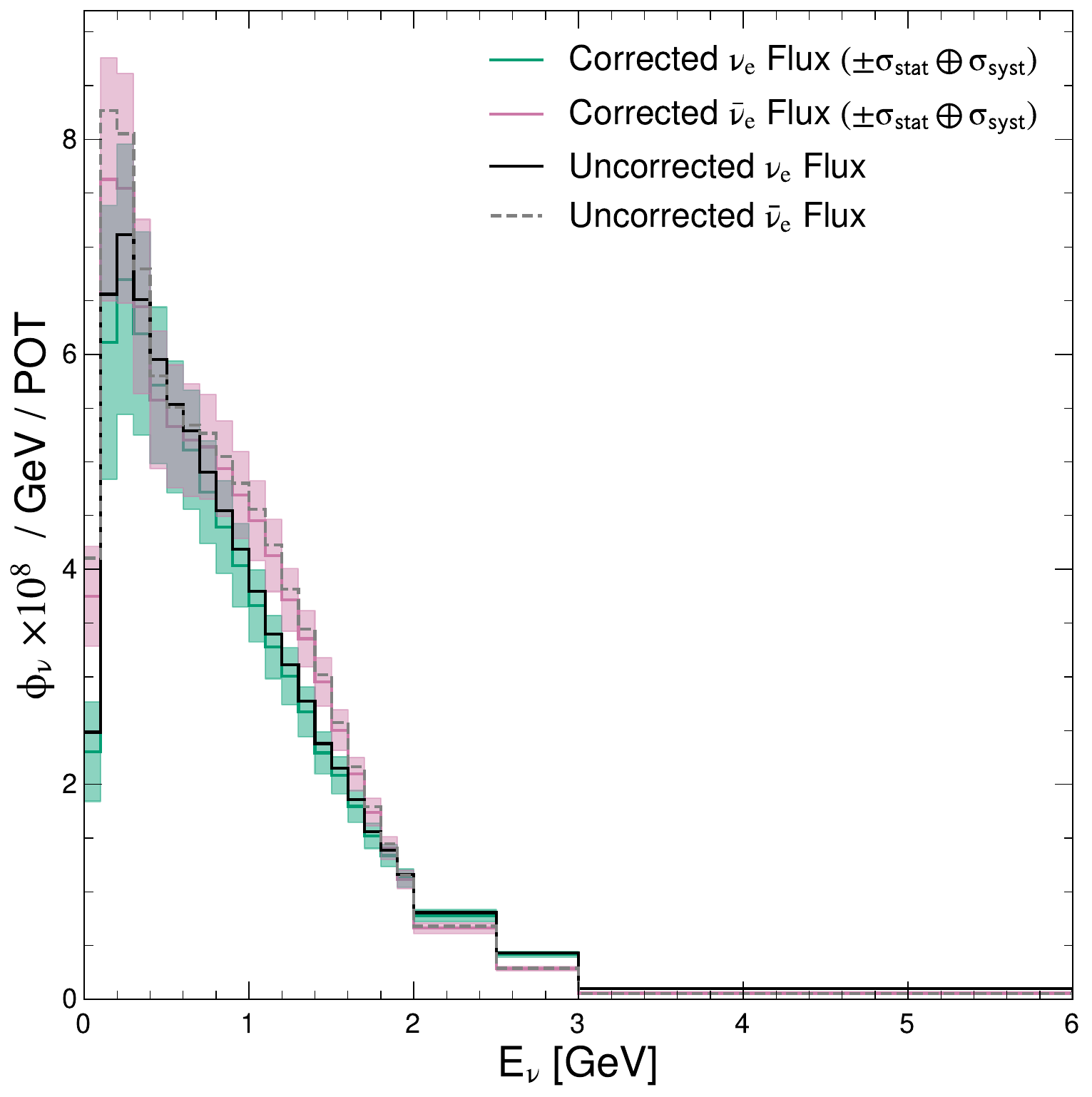}
    \caption{RHC \nue\ and \nueb}
    \end{subfigure}
    \caption[ICARUS Flux Predictions (FHC and RHC)]{ICARUS flux predictions for the FHC and RHC modes.}
\end{figure}

%% file: hp_uncertainties.tex
\clearpage
\section{Forward Horn Current}
\begin{figure}[!ht]
    \centering
    \includegraphics[width=0.23\textwidth,]{fhc_numu_hadron_fractional_uncertainties.pdf}
    \includegraphics[width=0.23\textwidth,]{fhc_nue_hadron_fractional_uncertainties.pdf}
    \includegraphics[width=0.23\textwidth,]{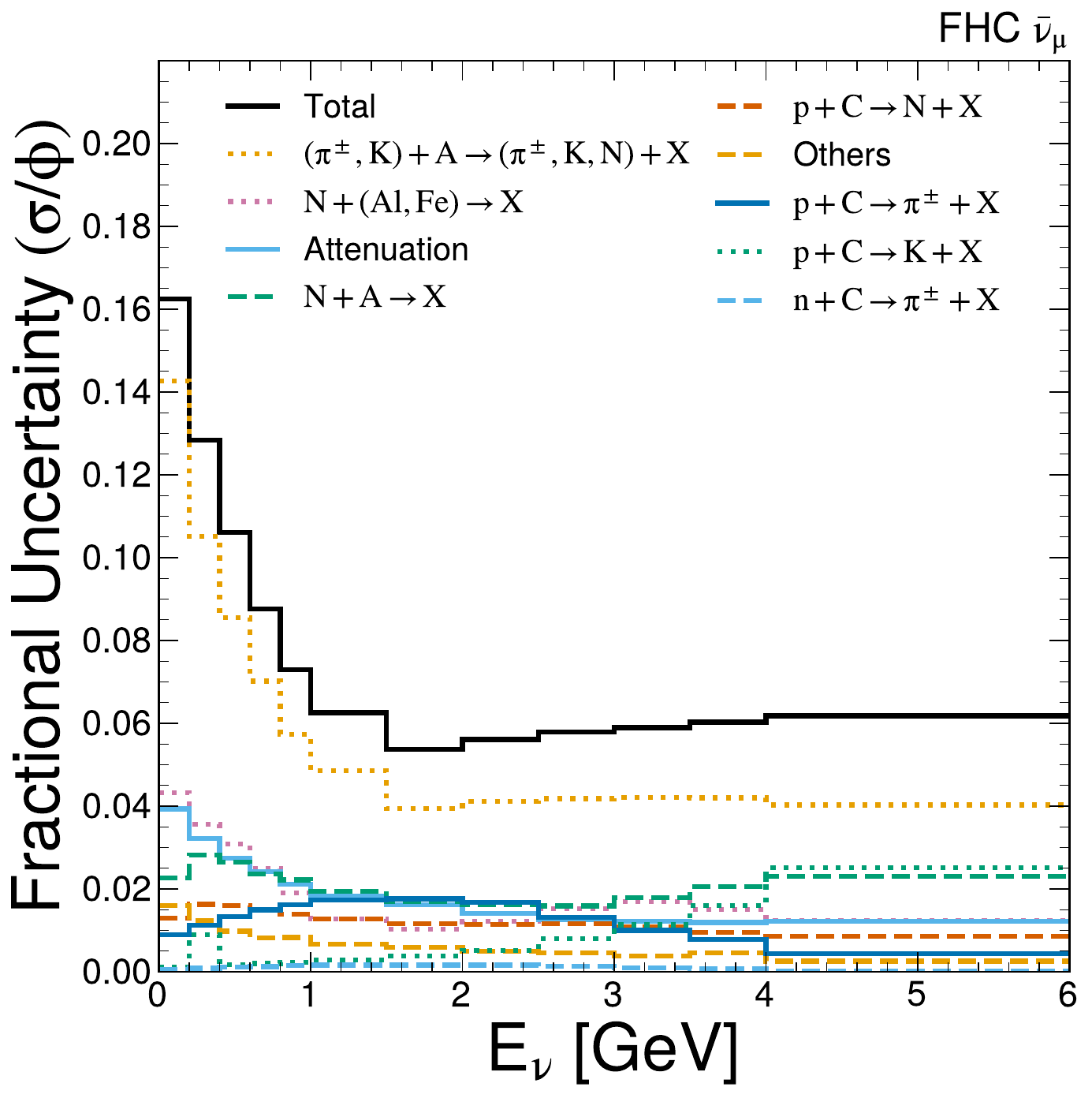}
    \includegraphics[width=0.23\textwidth,]{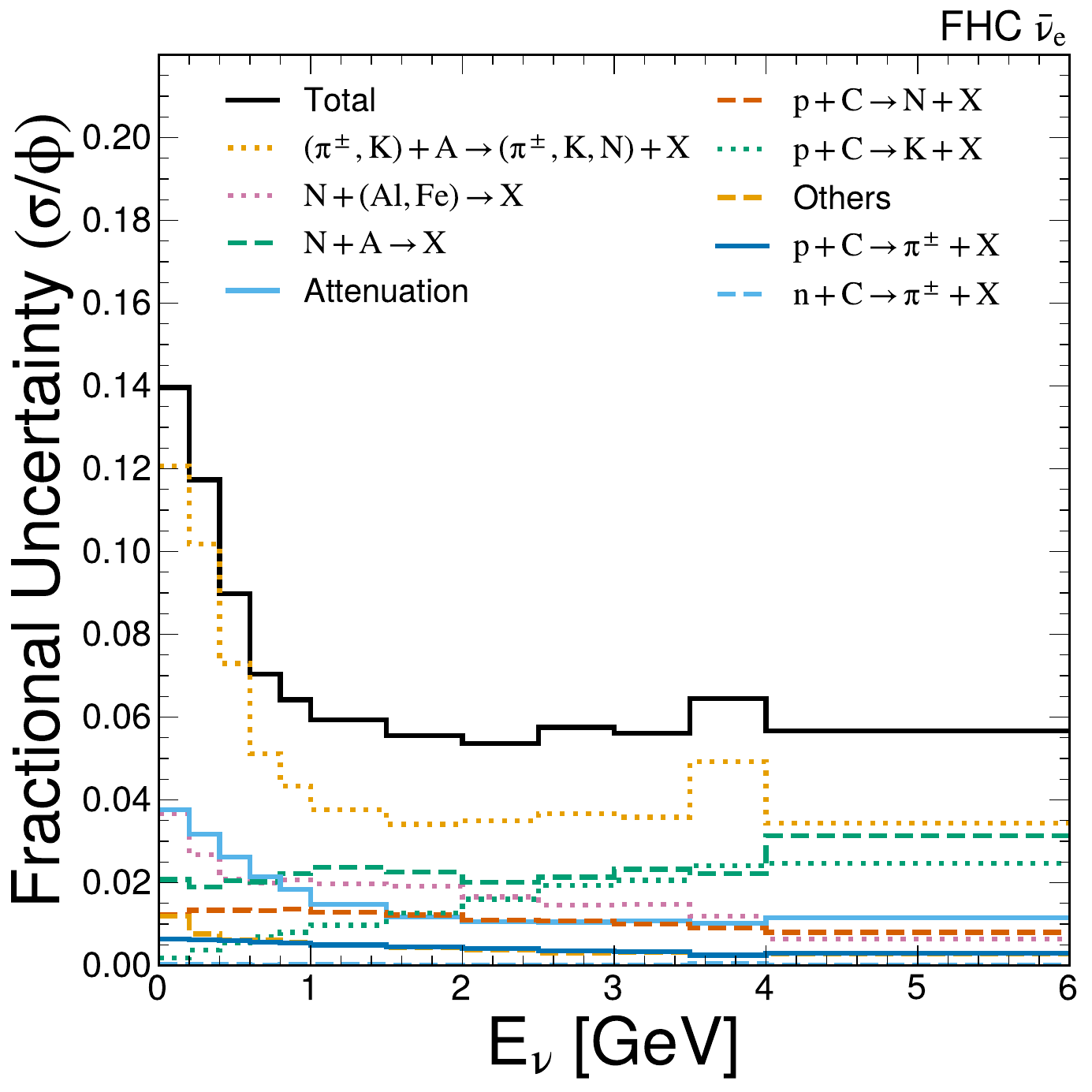}
    \caption[HP Fractional Uncertainties (FHC)]{Hadron interaction systematic uncertainties for all neutrino modes in the forward horn current beam configuration.}
\end{figure}

\begin{figure}[!ht]
    \centering
    \includegraphics[width=0.23\textwidth]{fhc_numu_incoming_hadron_fractional_uncertainties.pdf}
    \includegraphics[width=0.23\textwidth]{fhc_nue_incoming_hadron_fractional_uncertainties.pdf}
    \includegraphics[width=0.23\textwidth]{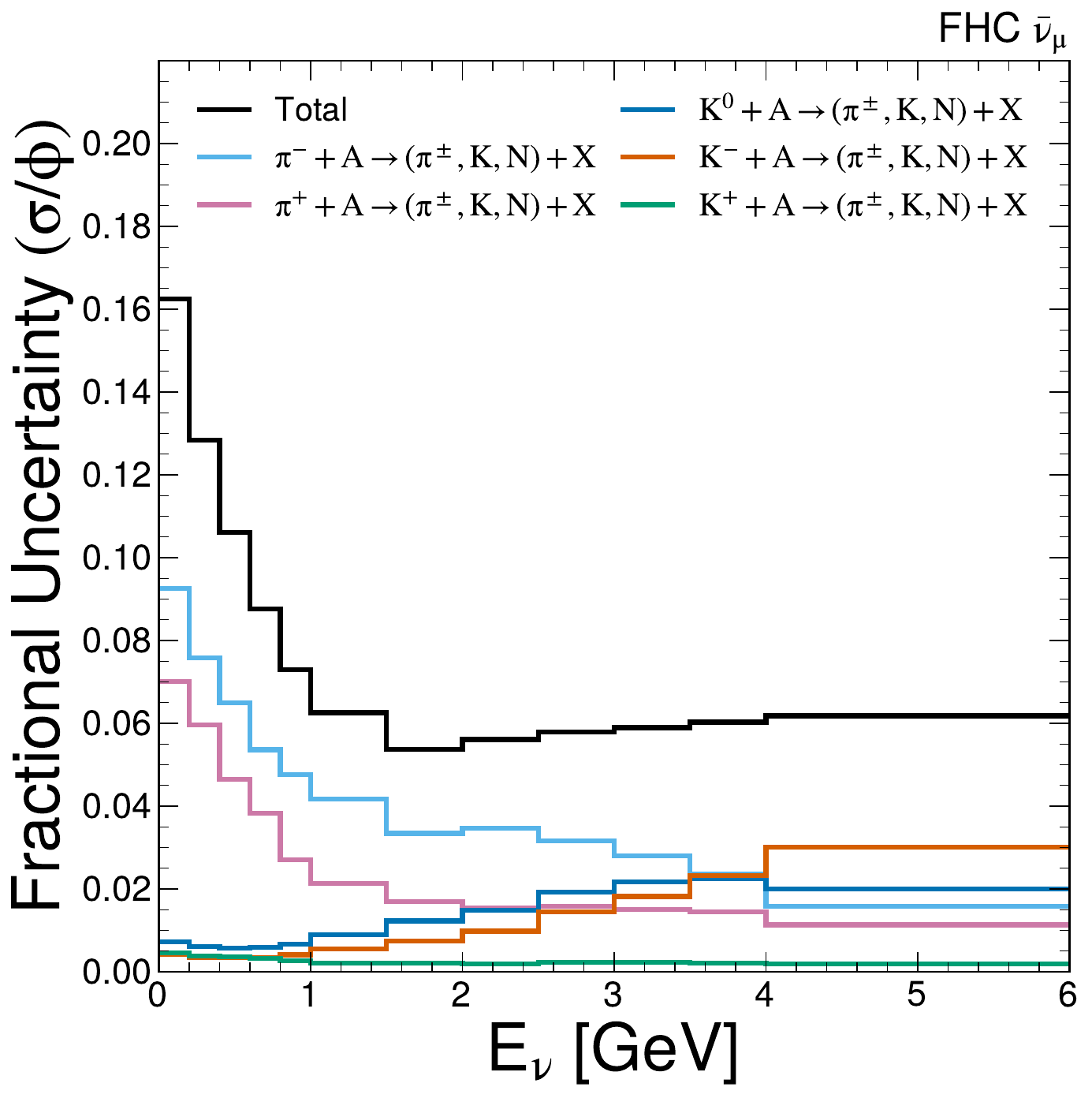}
    \includegraphics[width=0.23\textwidth]{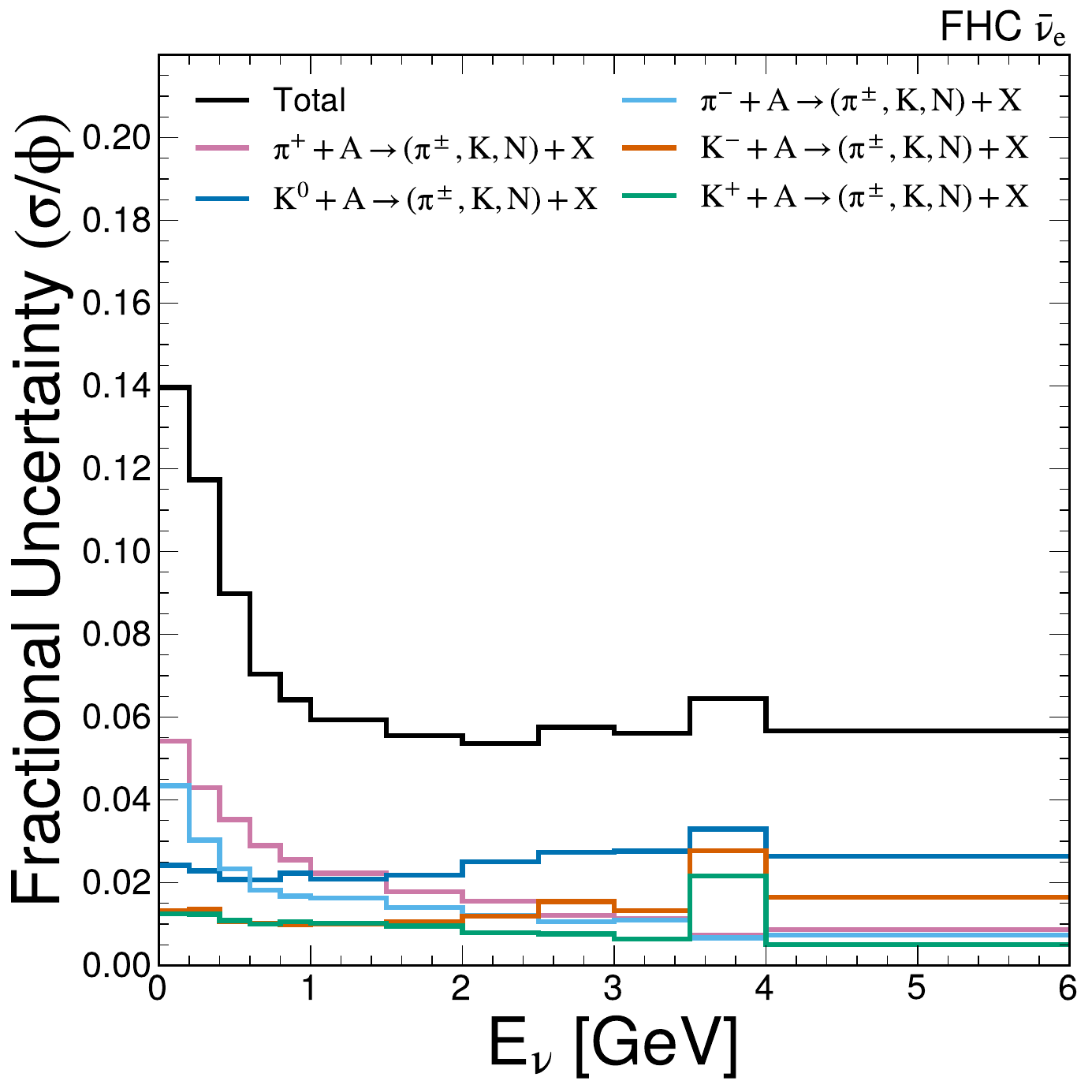}
    \caption[HP Fractional Uncertainties by Incoming Meson (FHC)]{Contribution to the uncertainty by incoming meson.}
\end{figure}

\begin{figure}[!ht]
    \centering
    \includegraphics[width=0.23\textwidth]{fhc_numu_outgoing_hadron_fractional_uncertainties.pdf}
    \includegraphics[width=0.23\textwidth]{fhc_nue_outgoing_hadron_fractional_uncertainties.pdf}
    \includegraphics[width=0.23\textwidth]{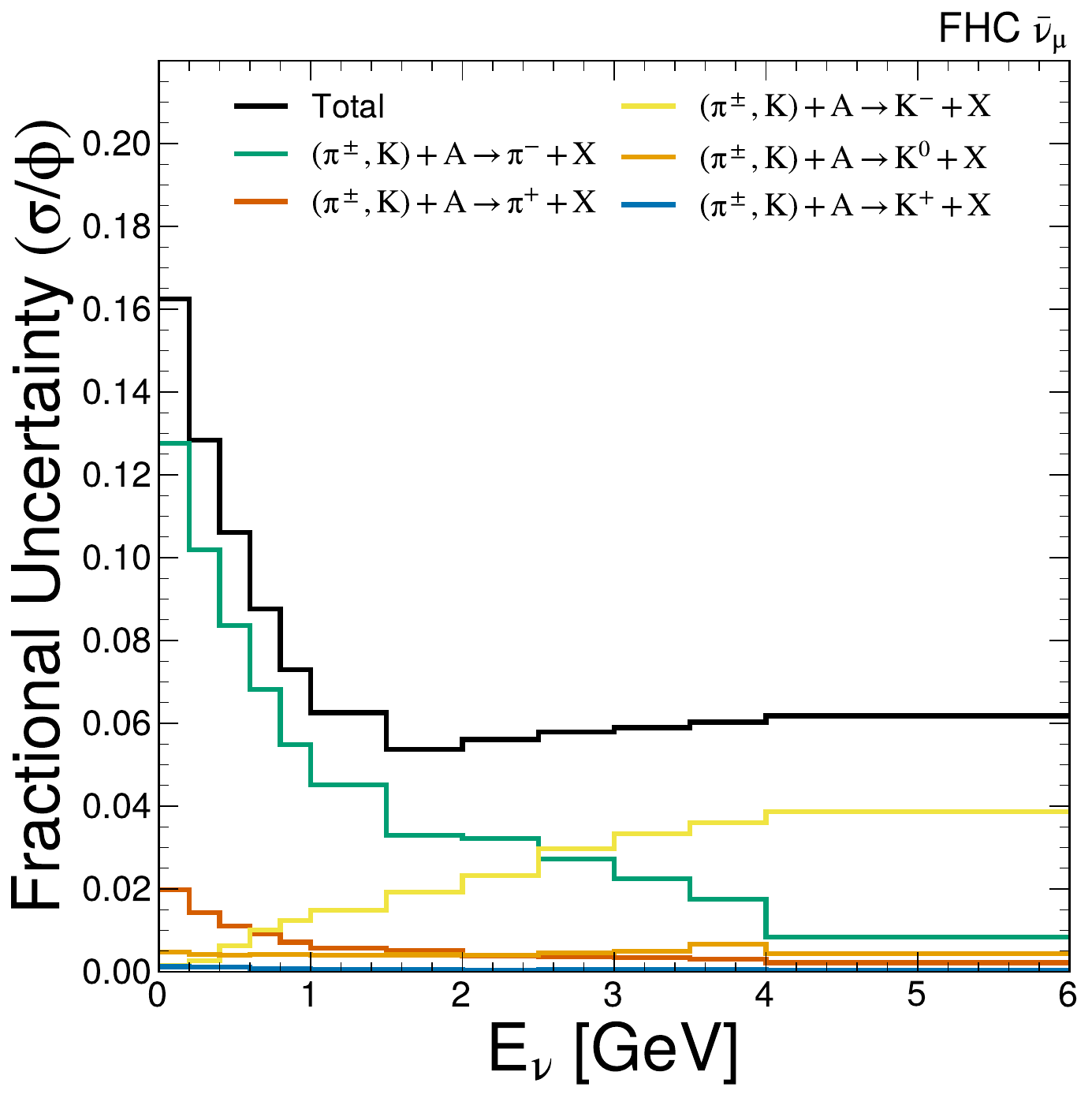}
    \includegraphics[width=0.23\textwidth]{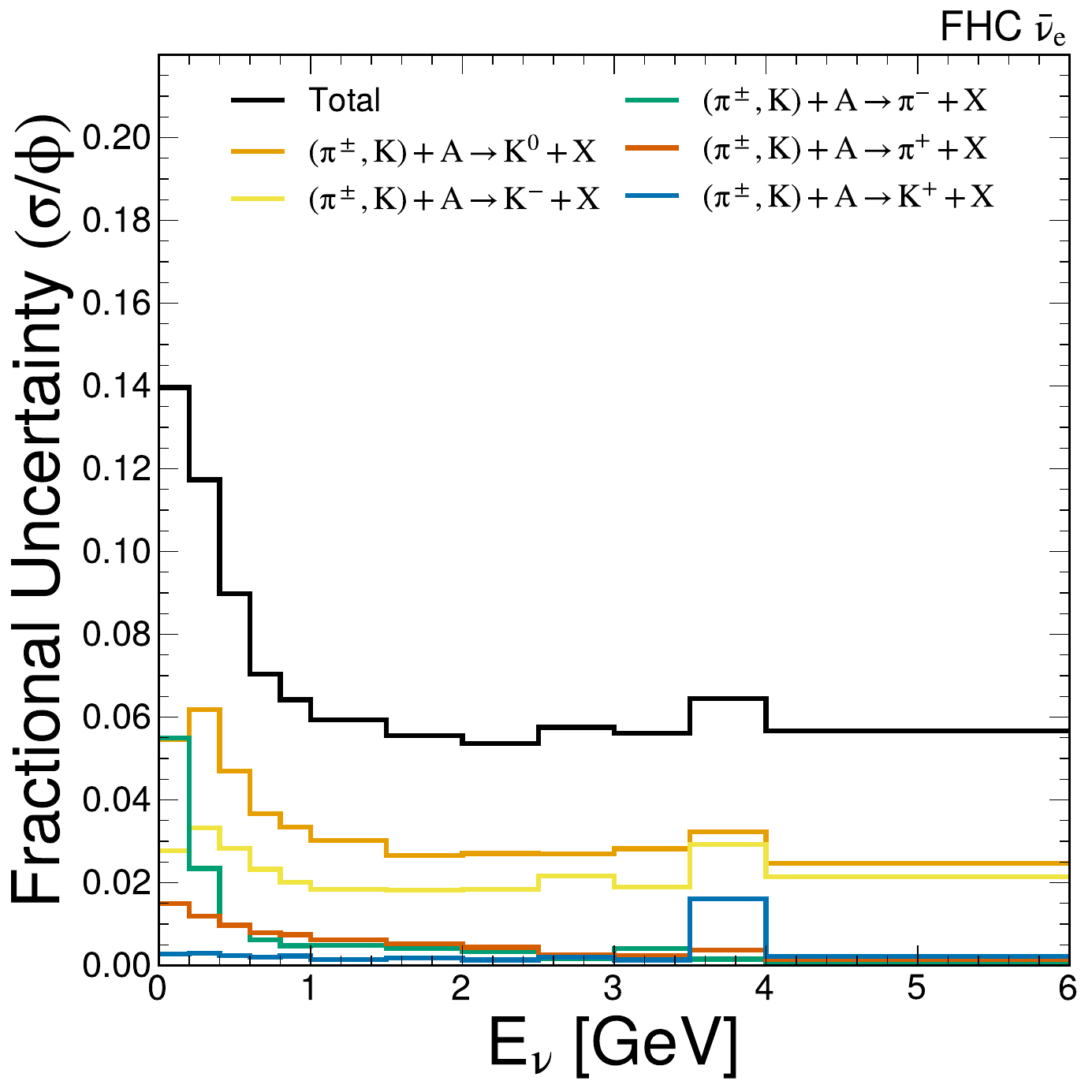}
    \caption[HP Fractional Uncertainties by Outgoing Meson (FHC)]{Contribution to the uncertainty by outgoing meson.}
\end{figure}

\clearpage
\section{Reverse Horn Current}

\begin{figure}[!ht]
    \centering
    \includegraphics[width=0.23\textwidth]{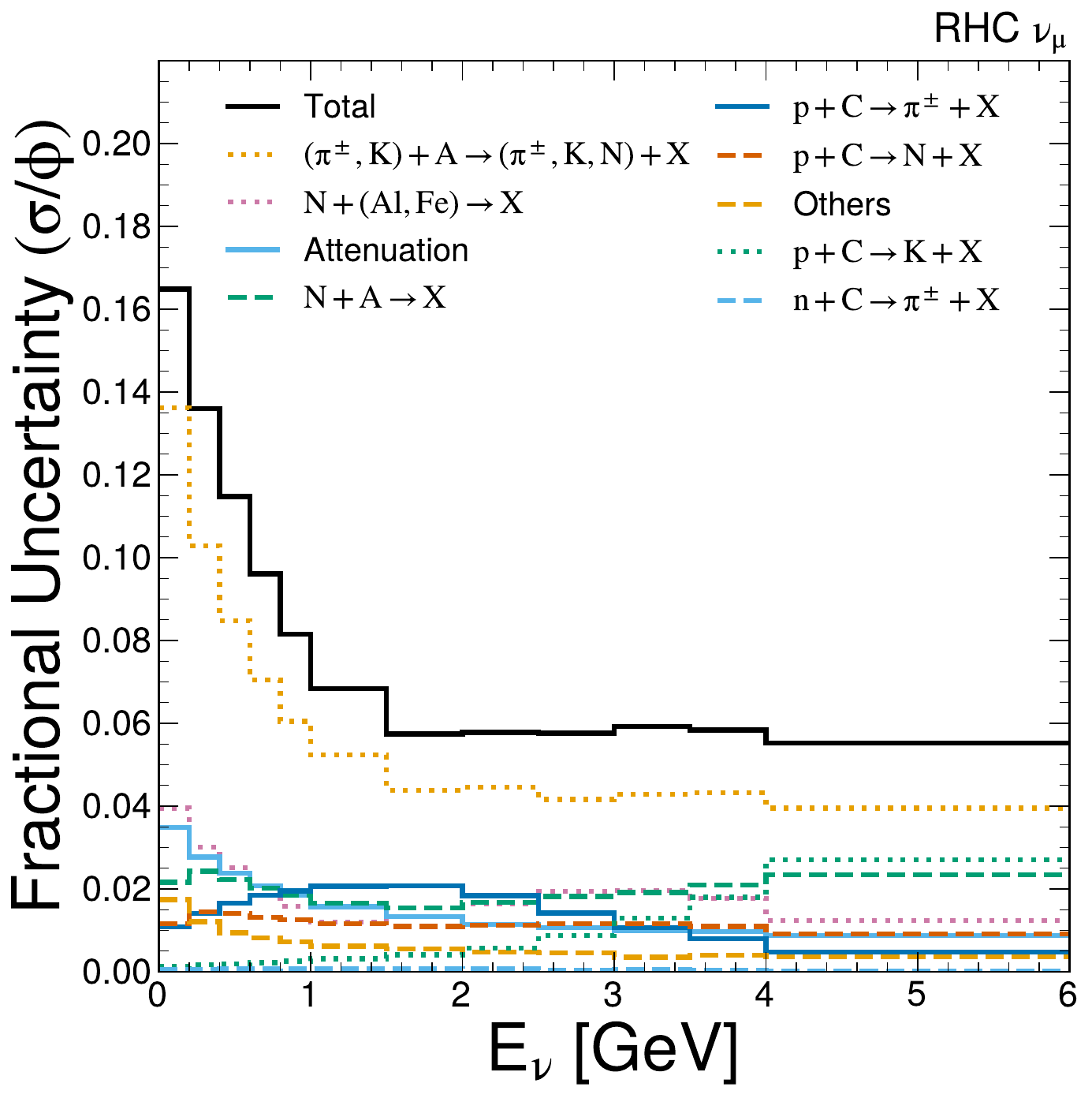}
    \includegraphics[width=0.23\textwidth]{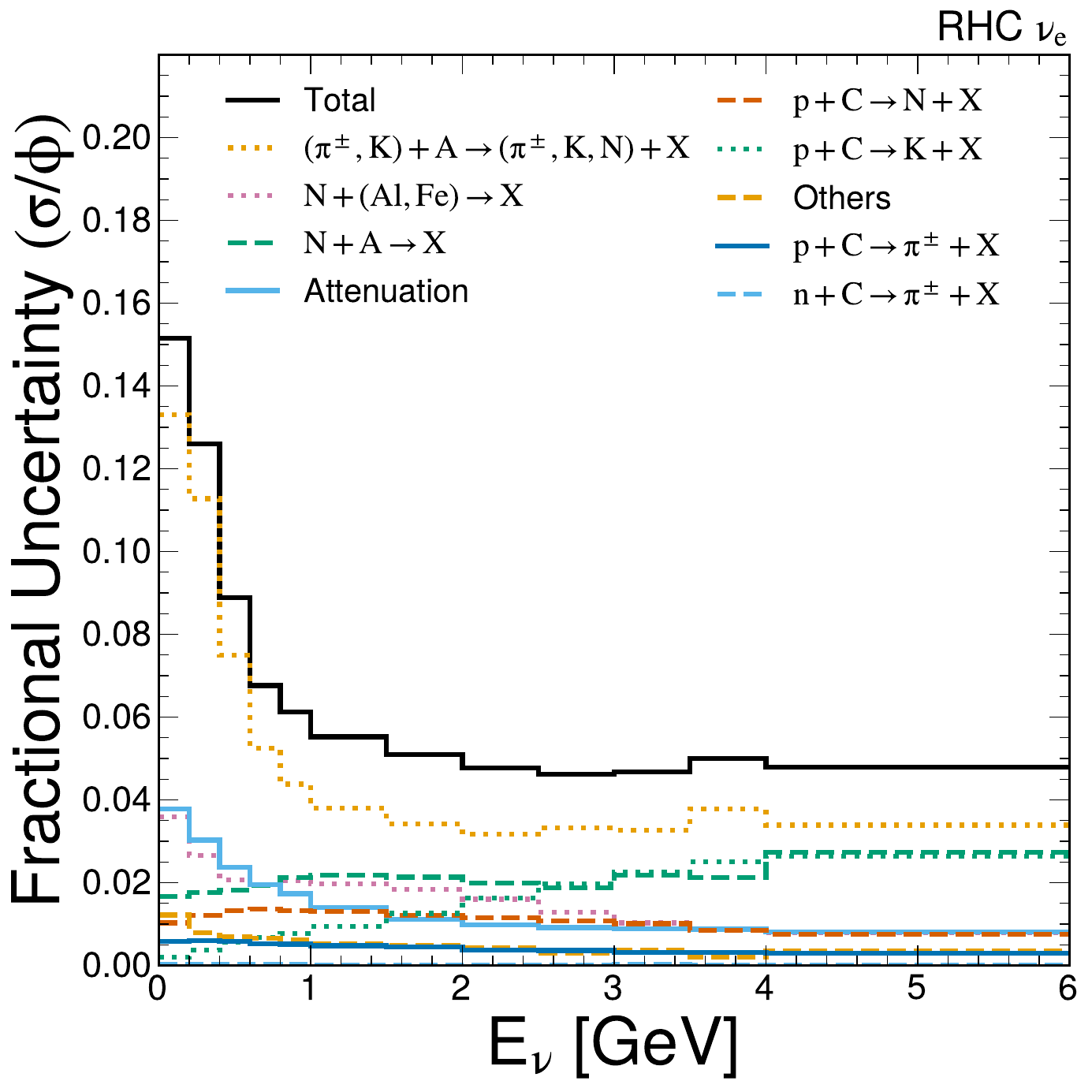}
    \includegraphics[width=0.23\textwidth]{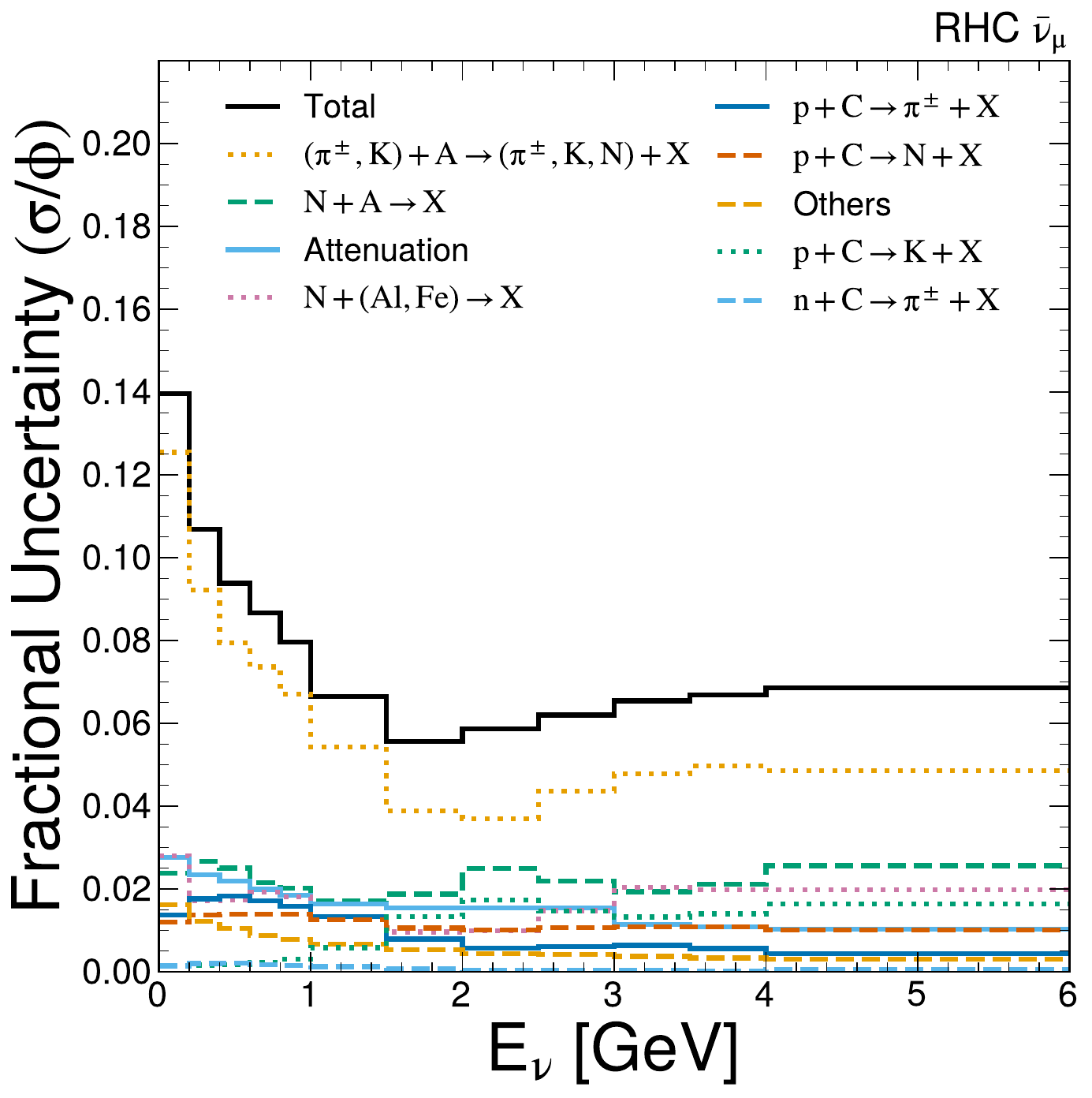}
    \includegraphics[width=0.23\textwidth]{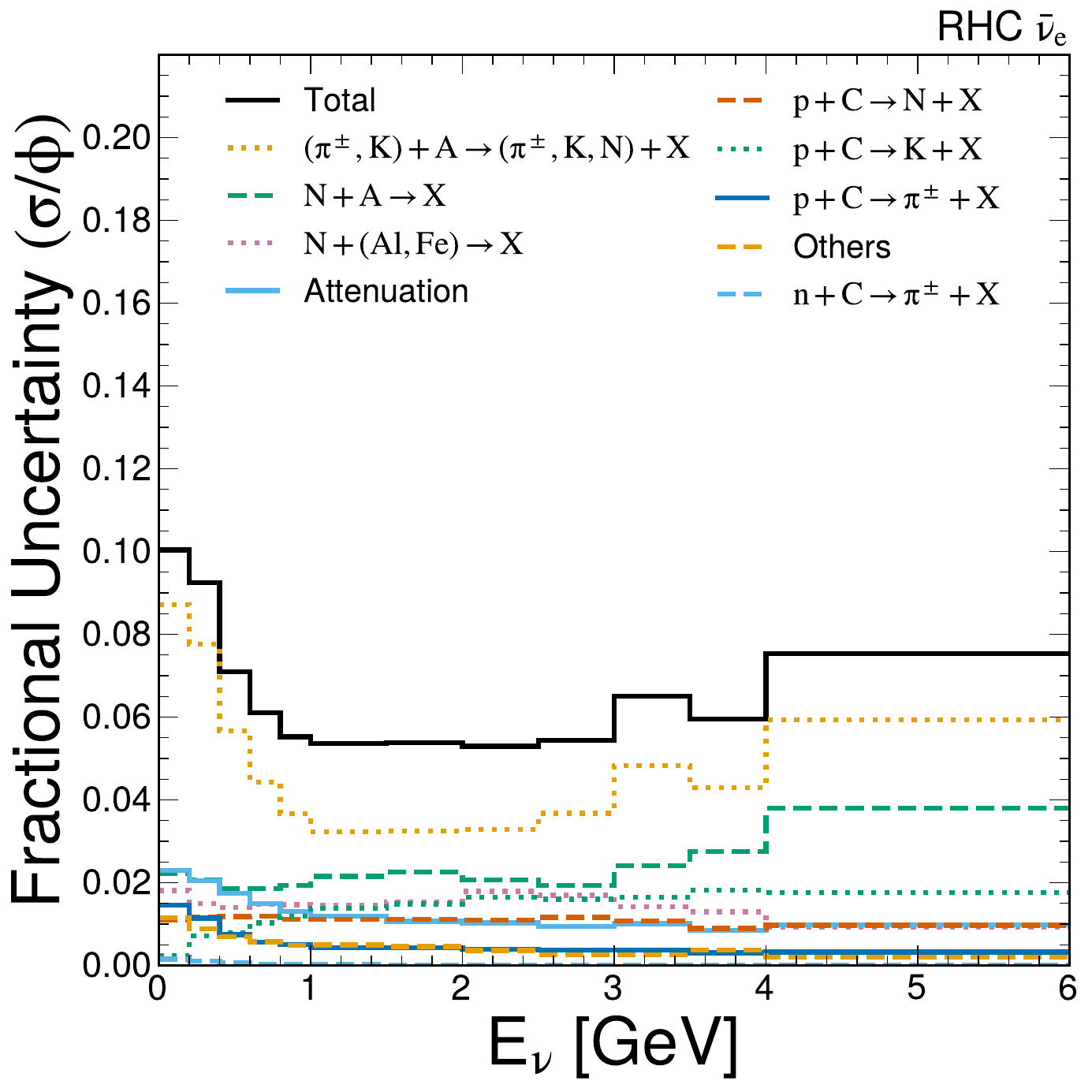}
    \caption[HP Fractional Uncertainties (RHC)]{Hadron interaction systematic uncertainties for all neutrino modes in the forward horn current beam configuration.}
\end{figure}

\begin{figure}[!ht]
    \centering
    \includegraphics[width=0.23\textwidth]{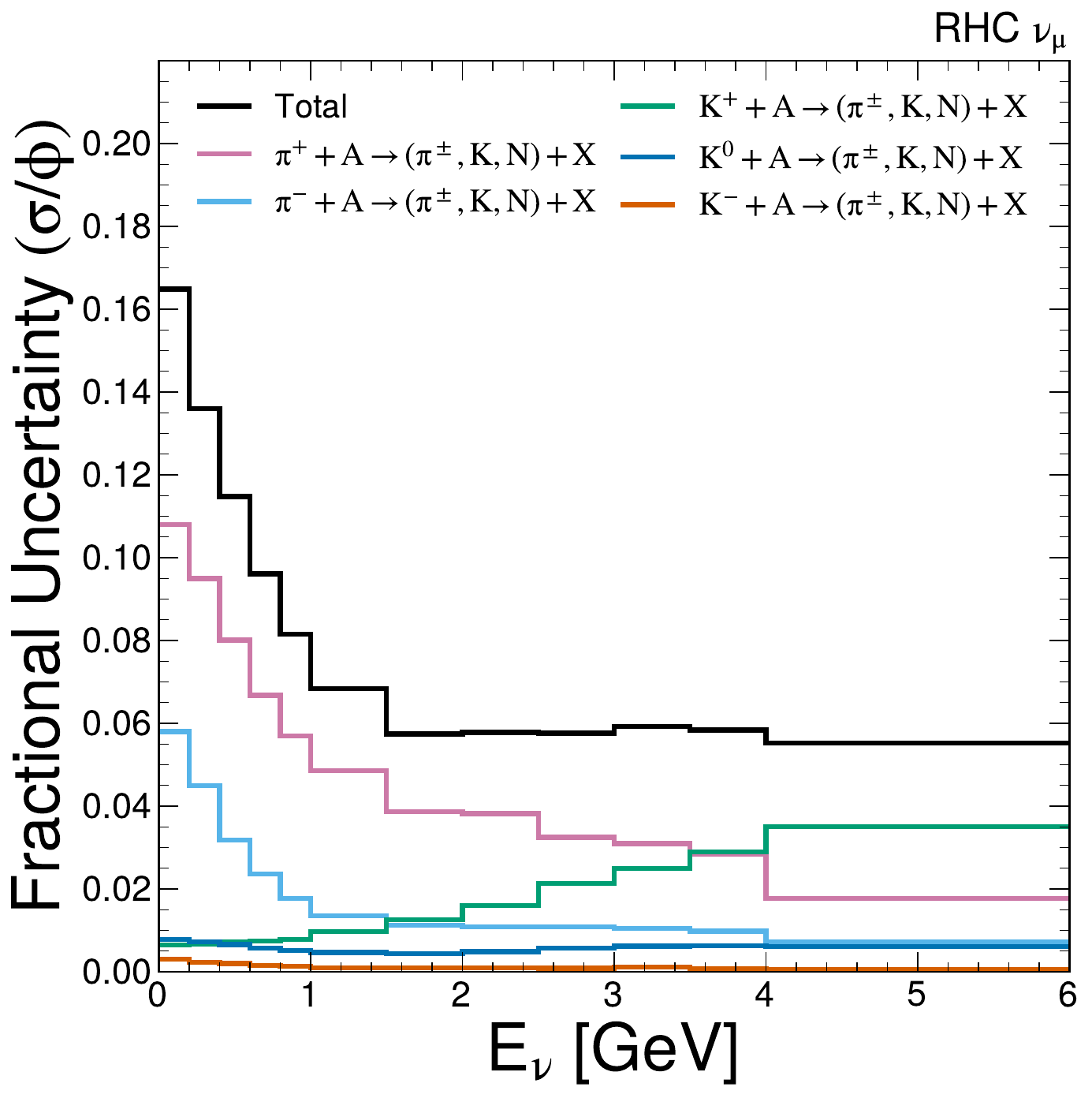}
    \includegraphics[width=0.23\textwidth]{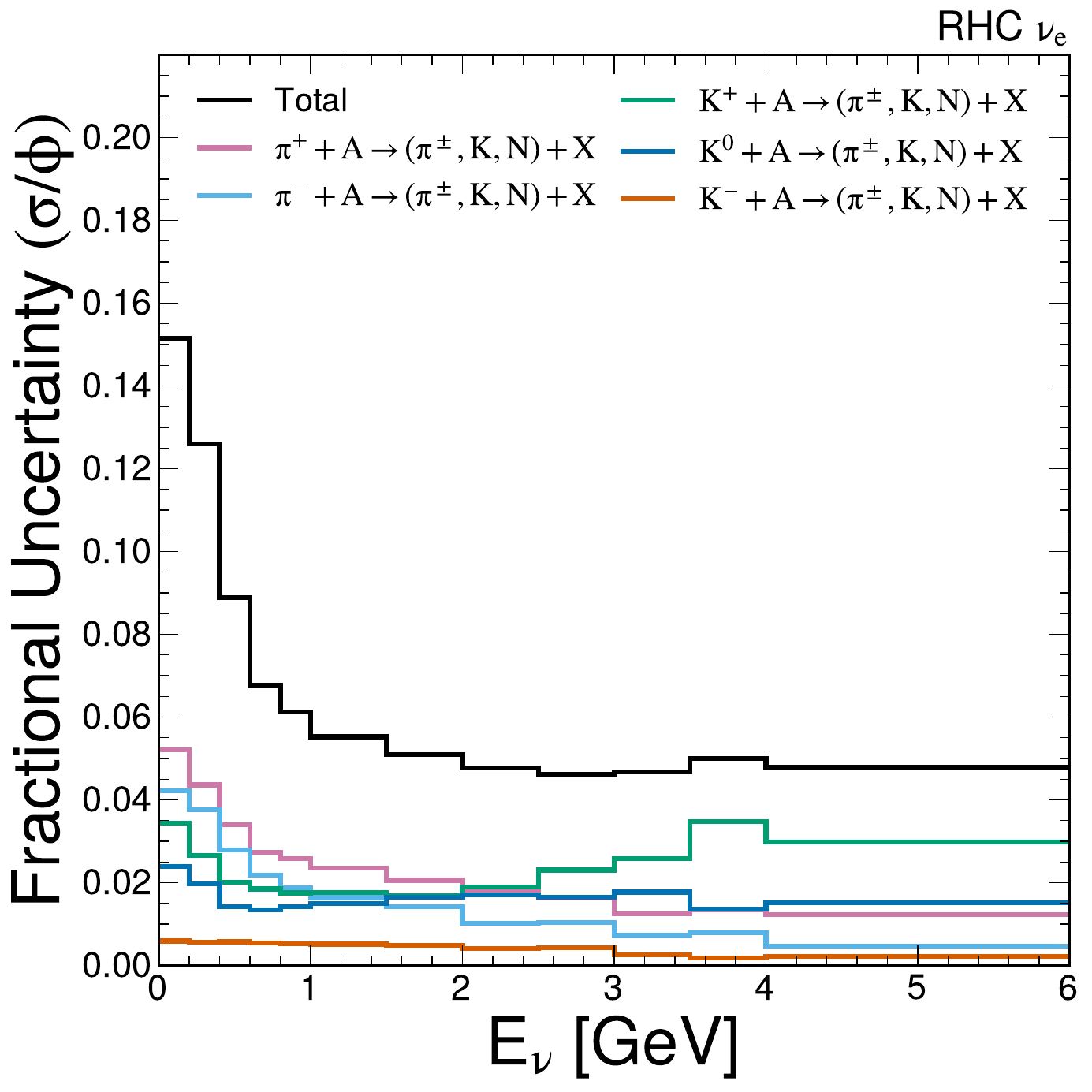}
    \includegraphics[width=0.23\textwidth]{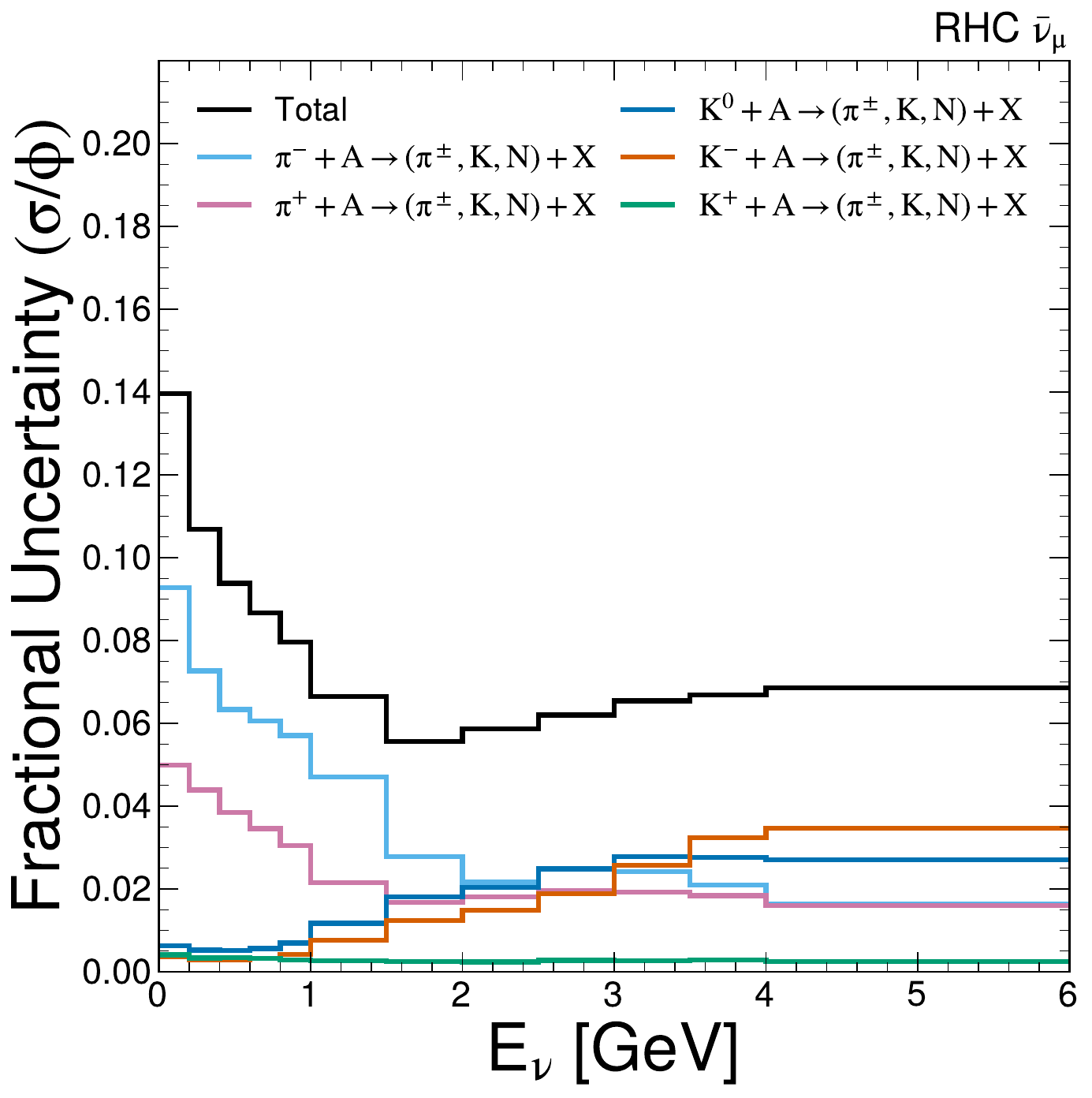}
    \includegraphics[width=0.23\textwidth]{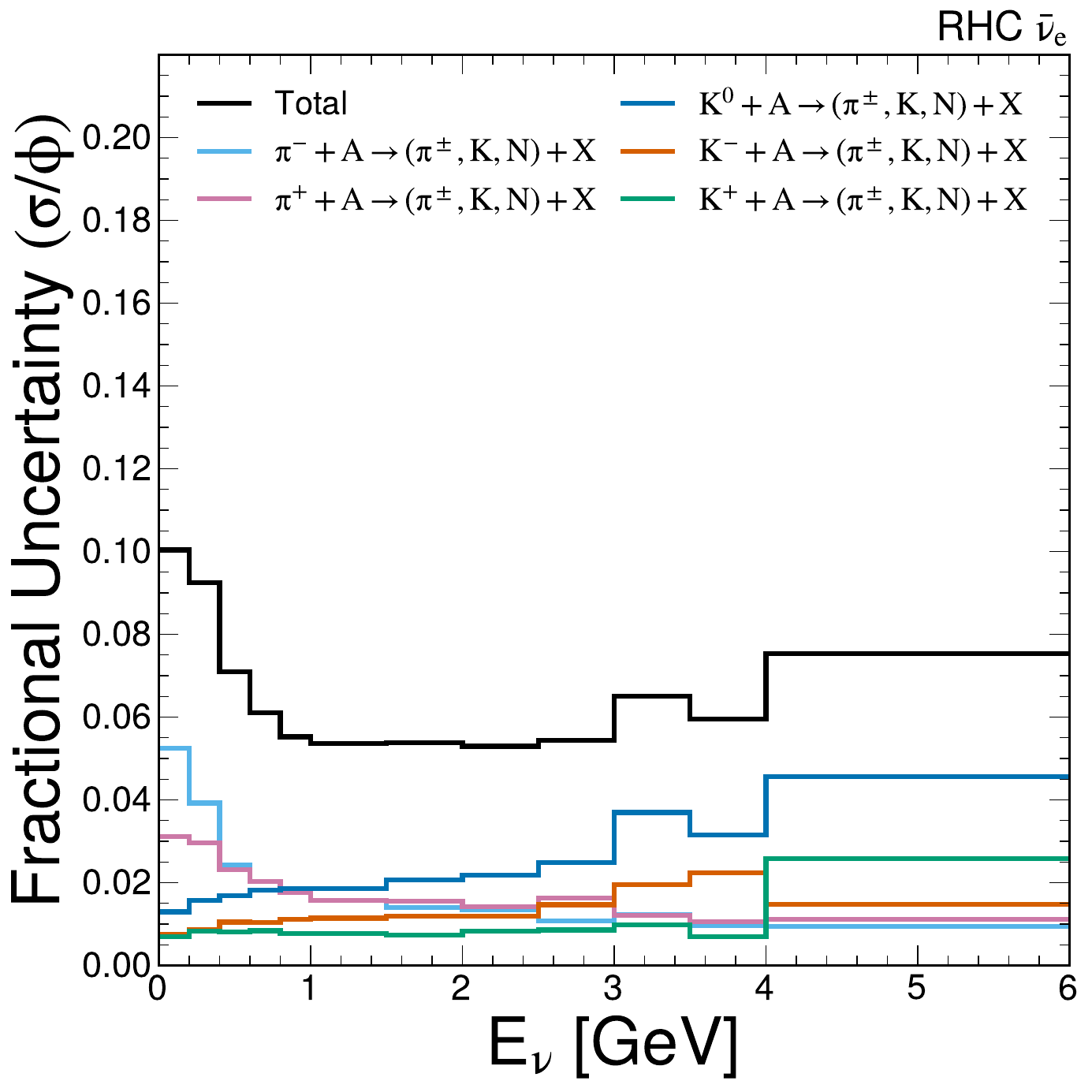}
    \caption[HP Fractional Uncertainties by Incoming Meson (RHC)]{Contribution to the uncertainty by incoming meson.}
\end{figure}

\begin{figure}[!ht]
    \centering
    \includegraphics[width=0.23\textwidth]{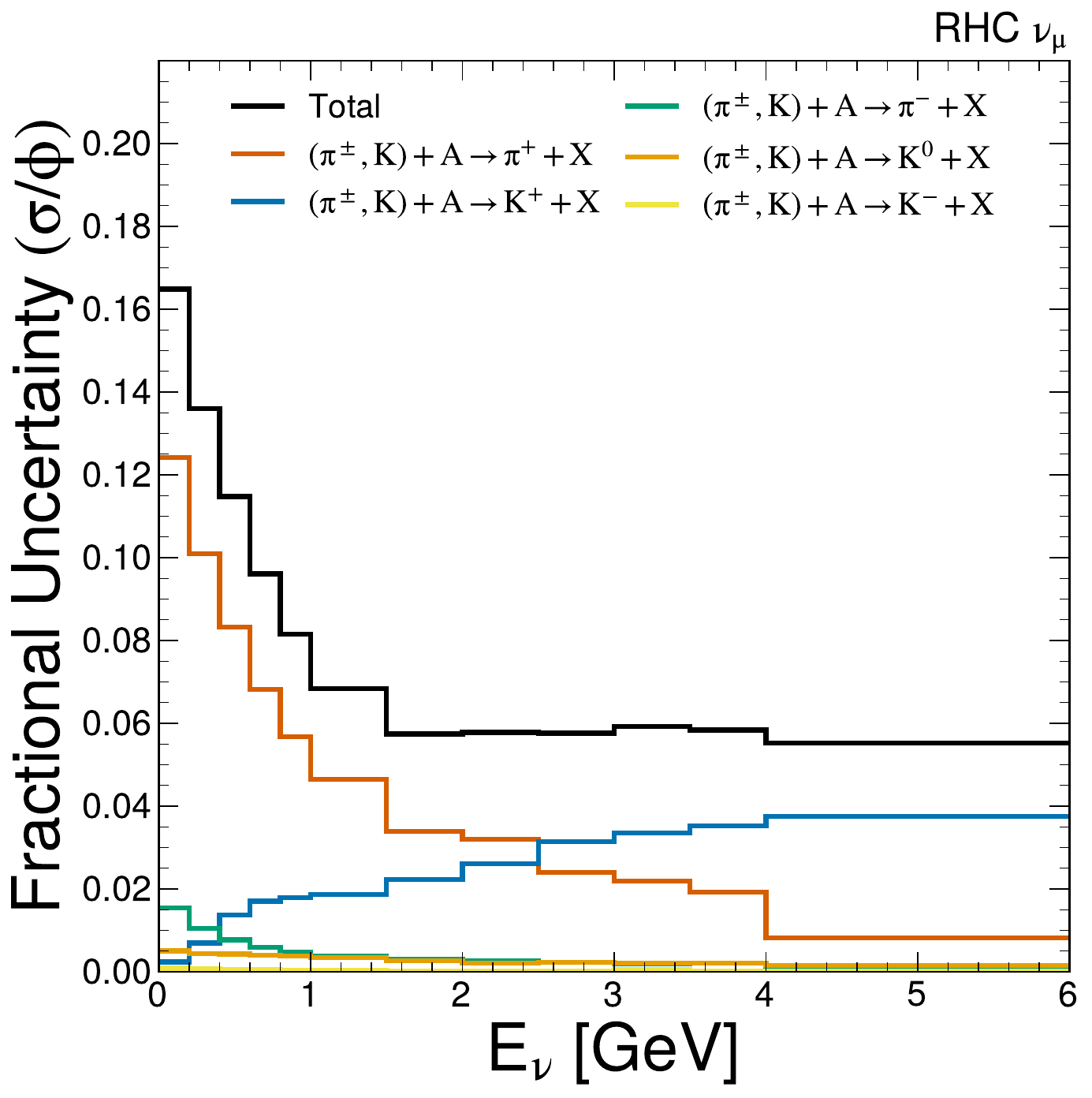}
    \includegraphics[width=0.23\textwidth]{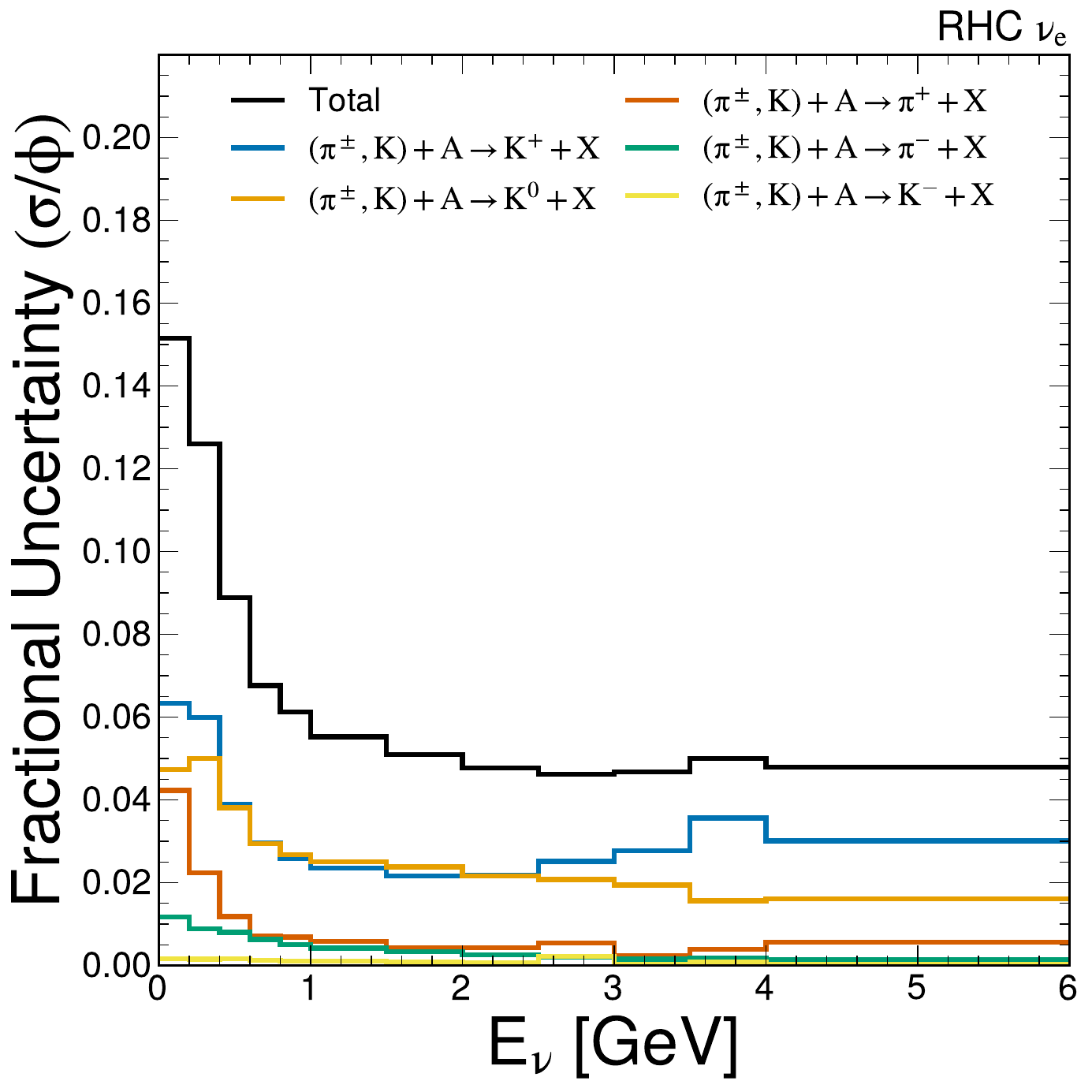}
    \includegraphics[width=0.23\textwidth]{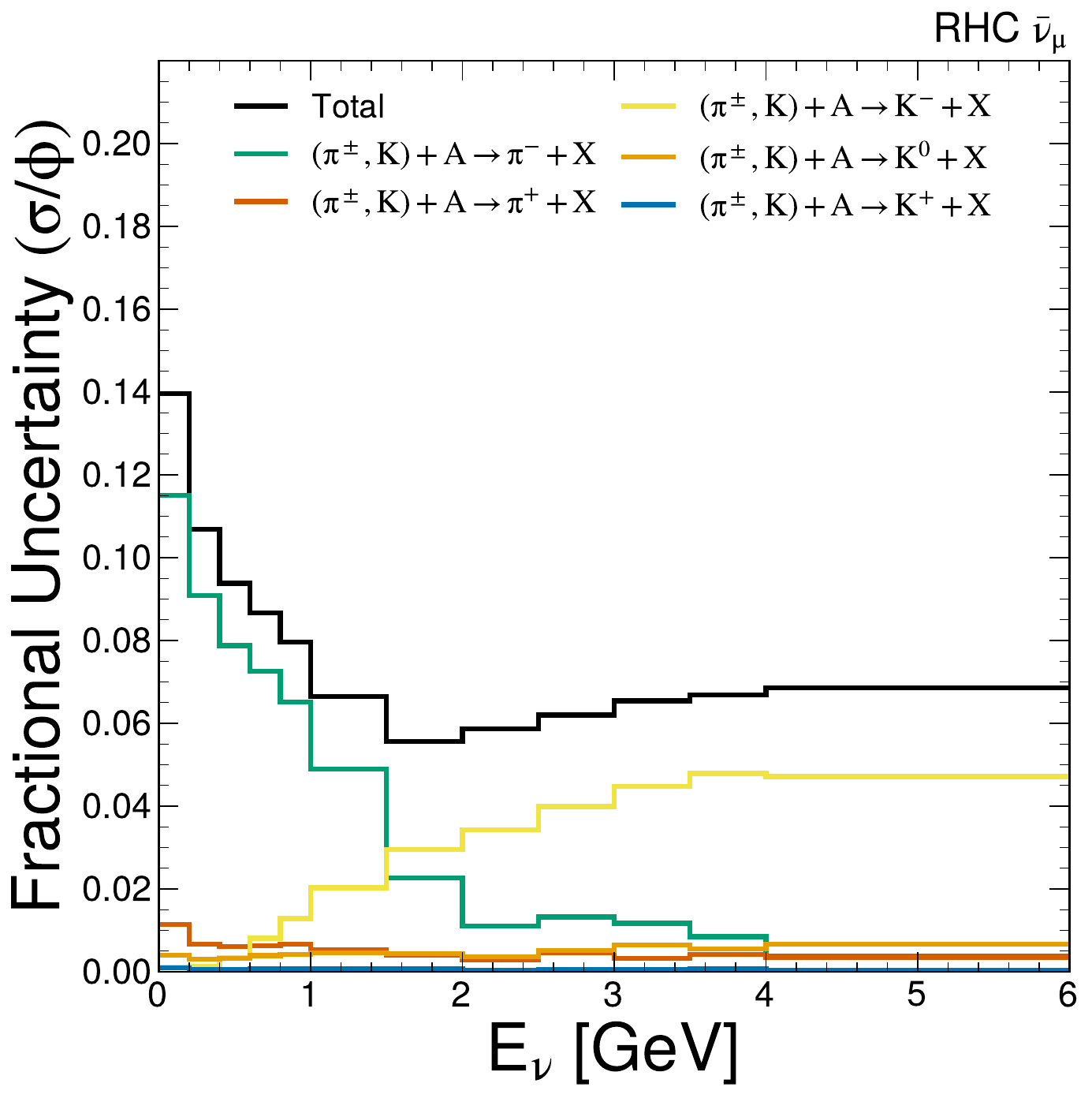}
    \includegraphics[width=0.23\textwidth]{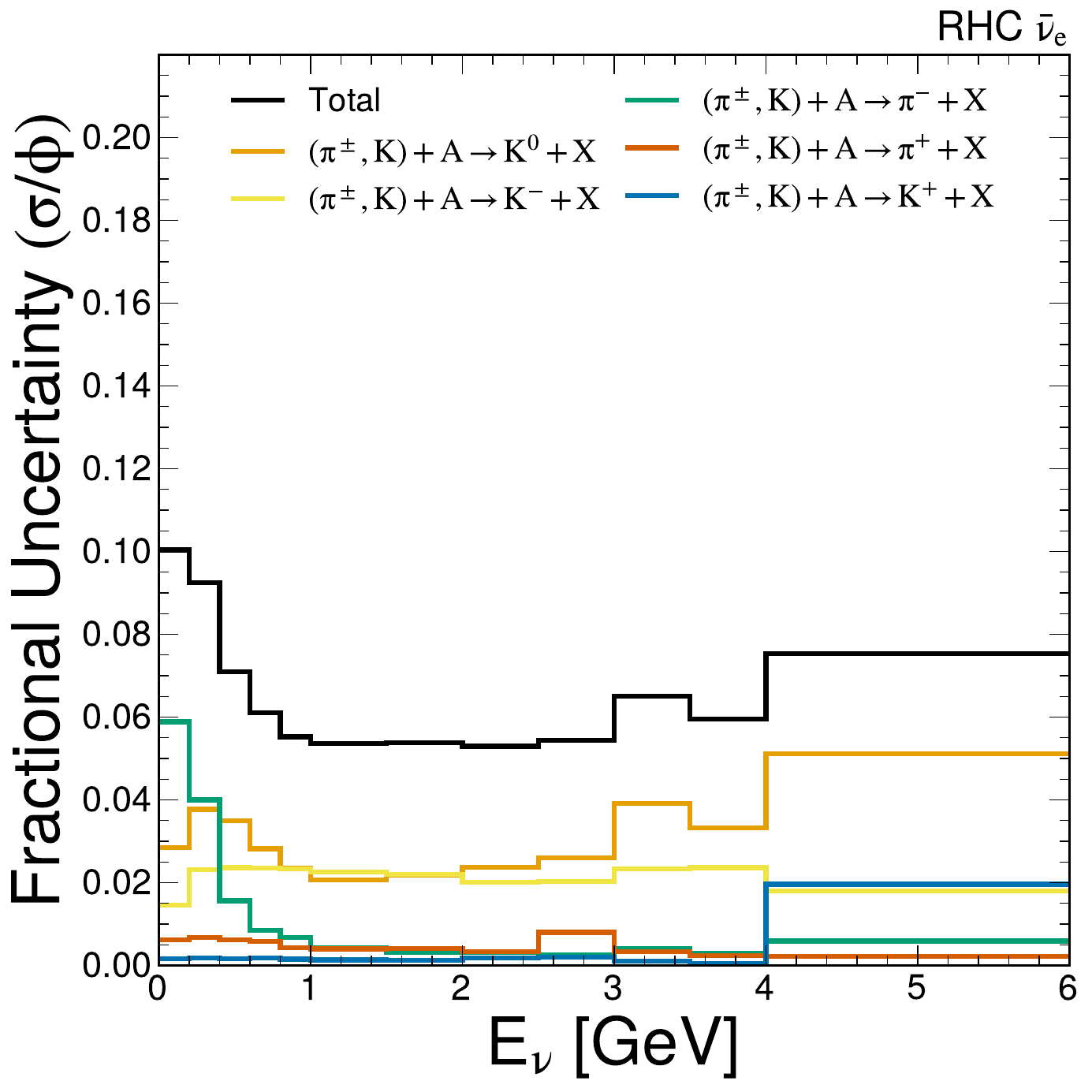}
    \caption[HP Fractional Uncertainties by Outgoing Meson (RHC)]{Contribution to the uncertainty by outgoing meson.}
\end{figure}

%% file: hp_covariance_matrices.tex
\clearpage
\section{Covariance Matrices}
\begin{figure}[!ht]
    \centering
    \begin{subfigure}[]{0.27\textwidth}
    \includegraphics[width=\textwidth]{hcov_hadron_total_covariance_matrix.pdf}
    \caption{Total}
    \end{subfigure}
    \begin{subfigure}[]{0.27\textwidth}
    \includegraphics[width=\textwidth]{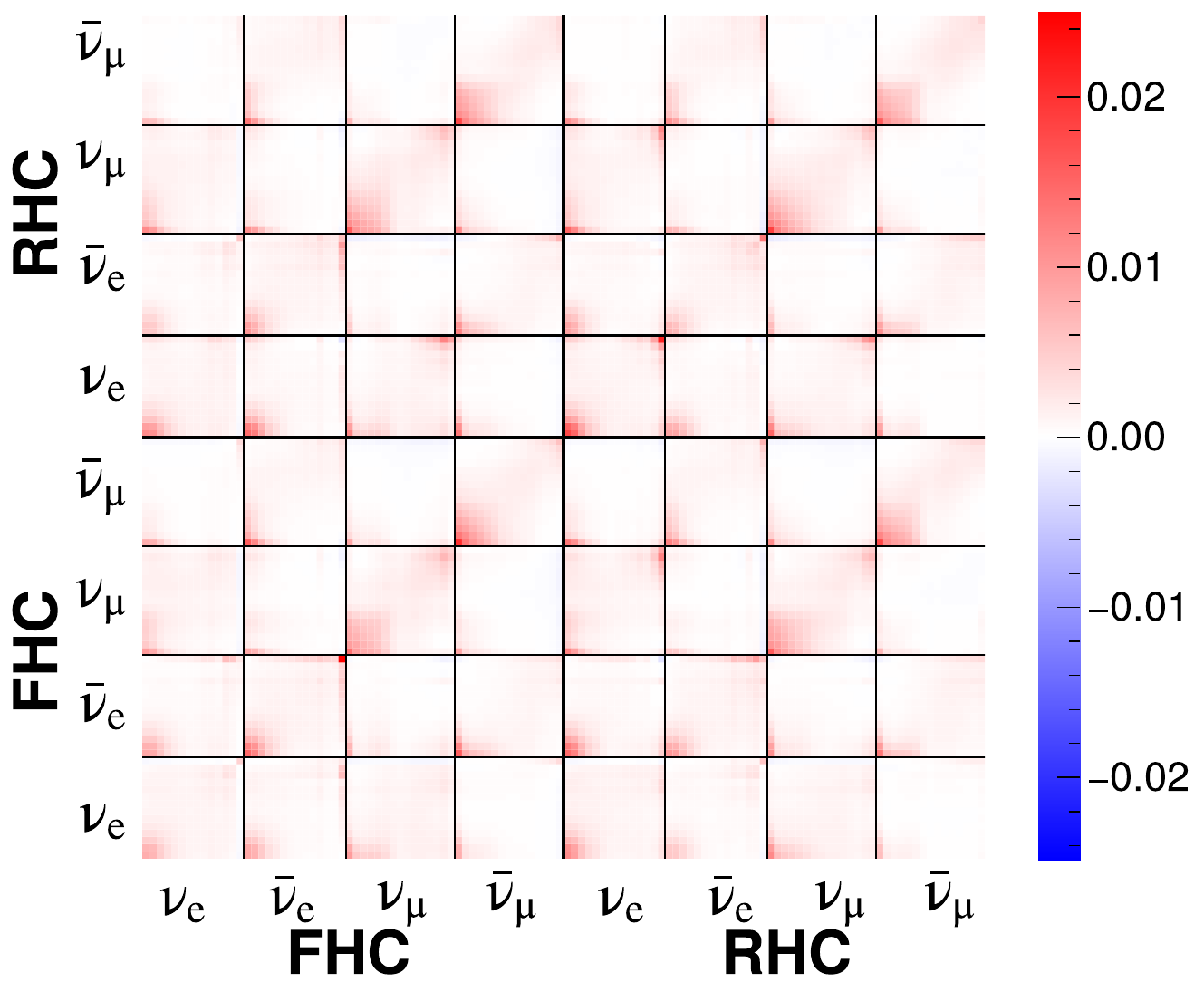}
        \caption{mesinc}
    \end{subfigure}
    \begin{subfigure}[]{0.27\textwidth}
    \includegraphics[width=\textwidth]{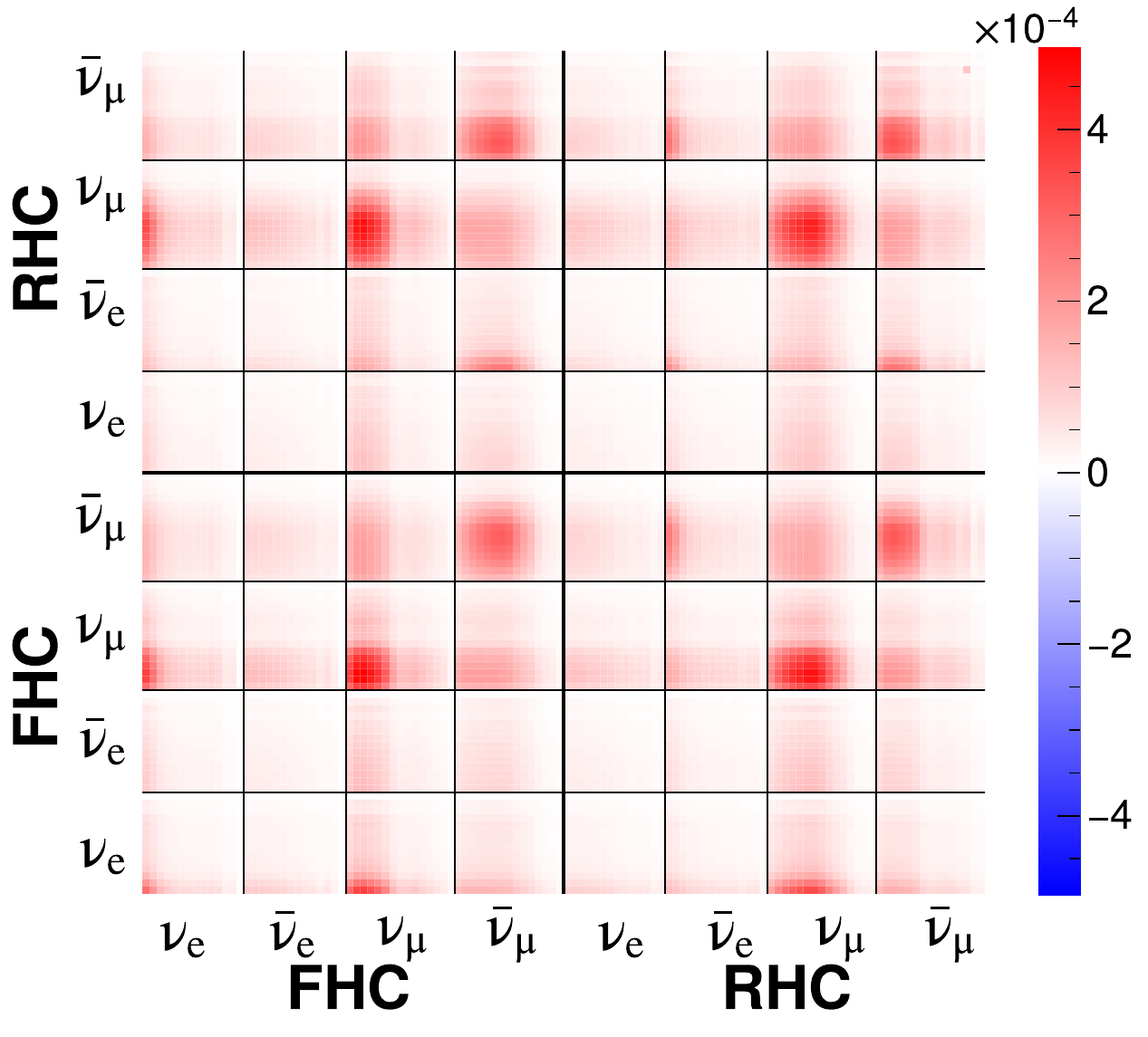}
        \caption{pCpi}
    \end{subfigure}
    \begin{subfigure}[]{0.27\textwidth}
    \includegraphics[width=\textwidth]{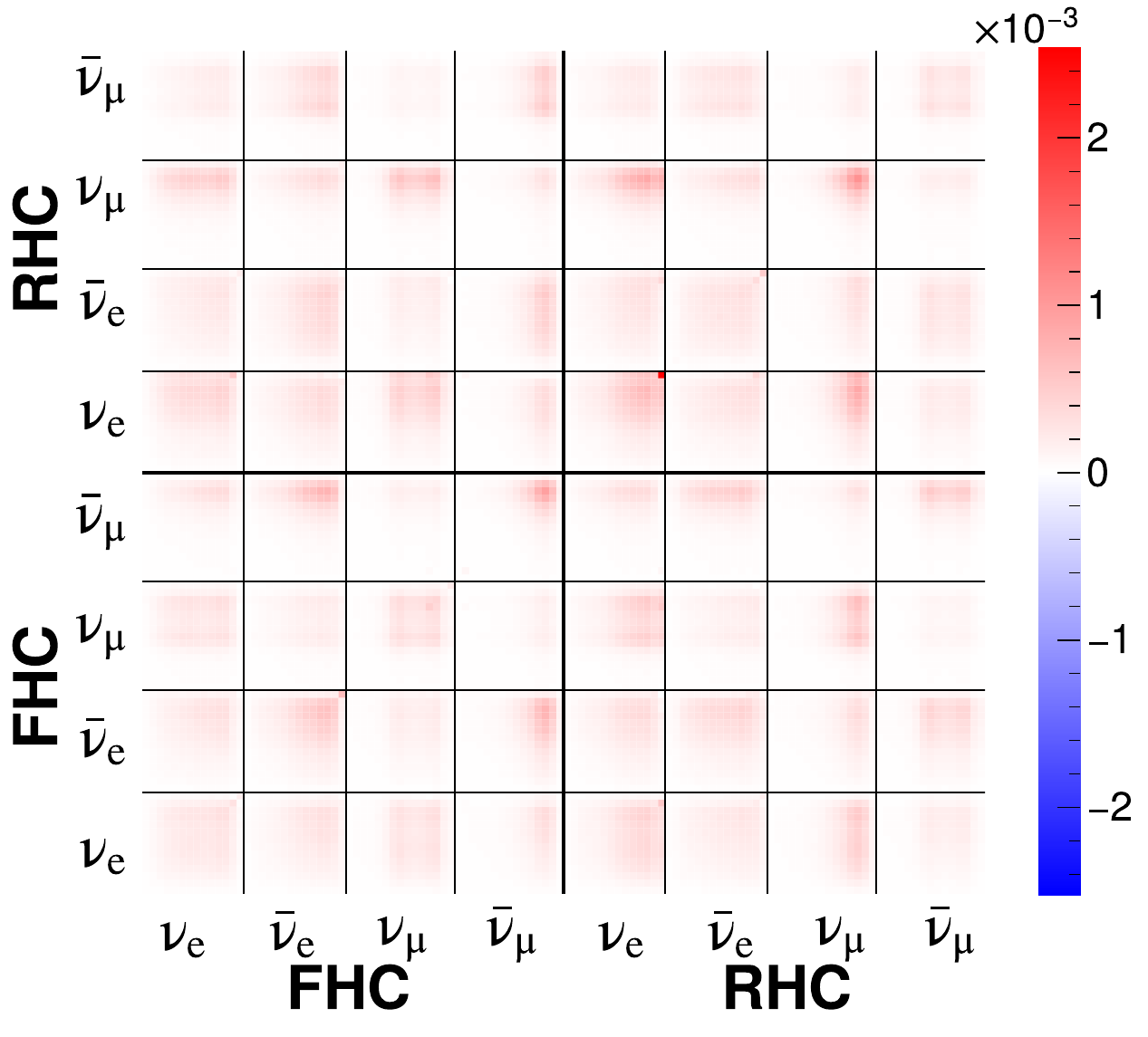}
        \caption{pCk}
    \end{subfigure}
    \begin{subfigure}[]{0.27\textwidth}
    \includegraphics[width=\textwidth]{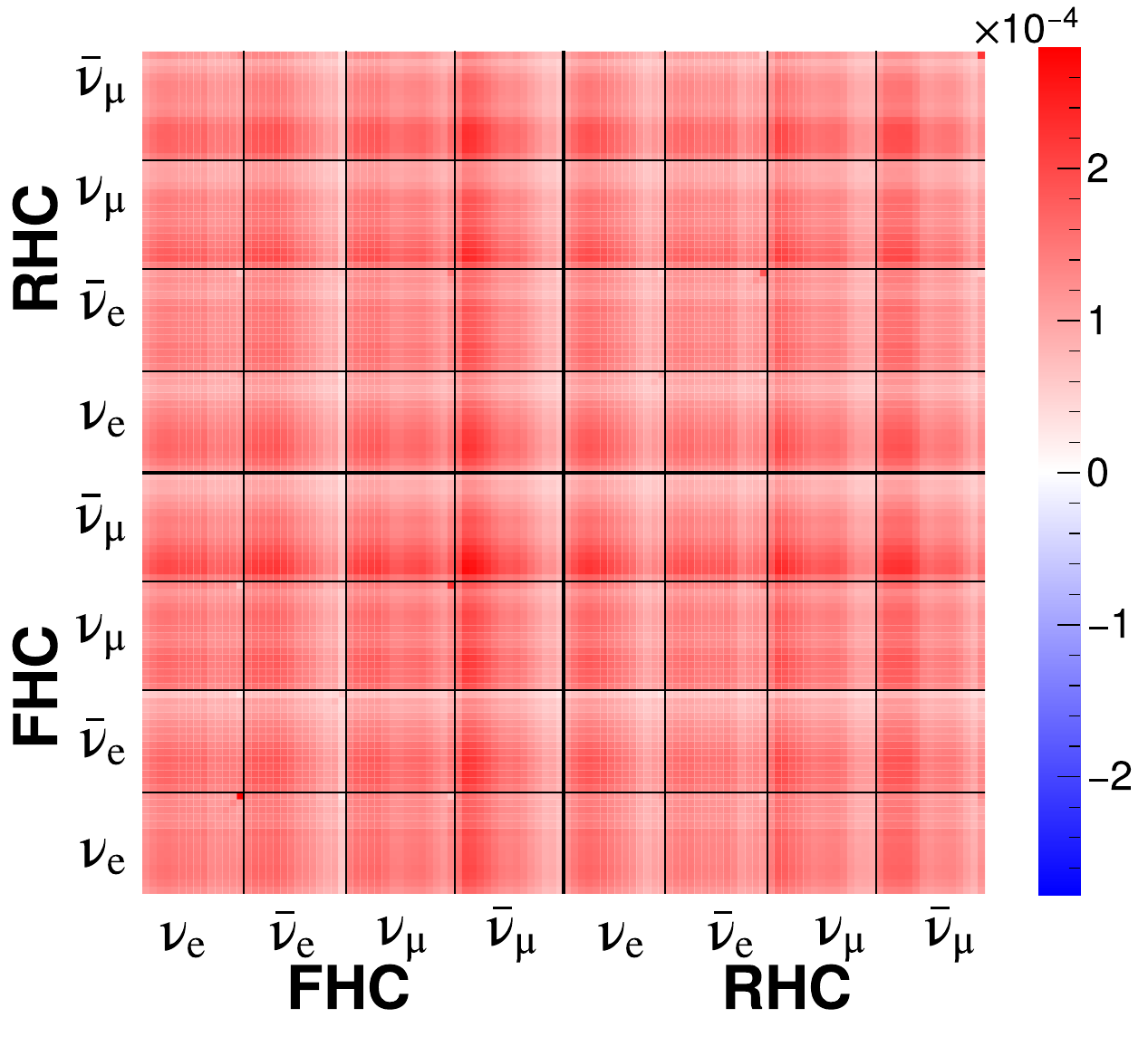}
        \caption{pCnu}
    \end{subfigure}
    \begin{subfigure}[]{0.27\textwidth}
    \includegraphics[width=\textwidth]{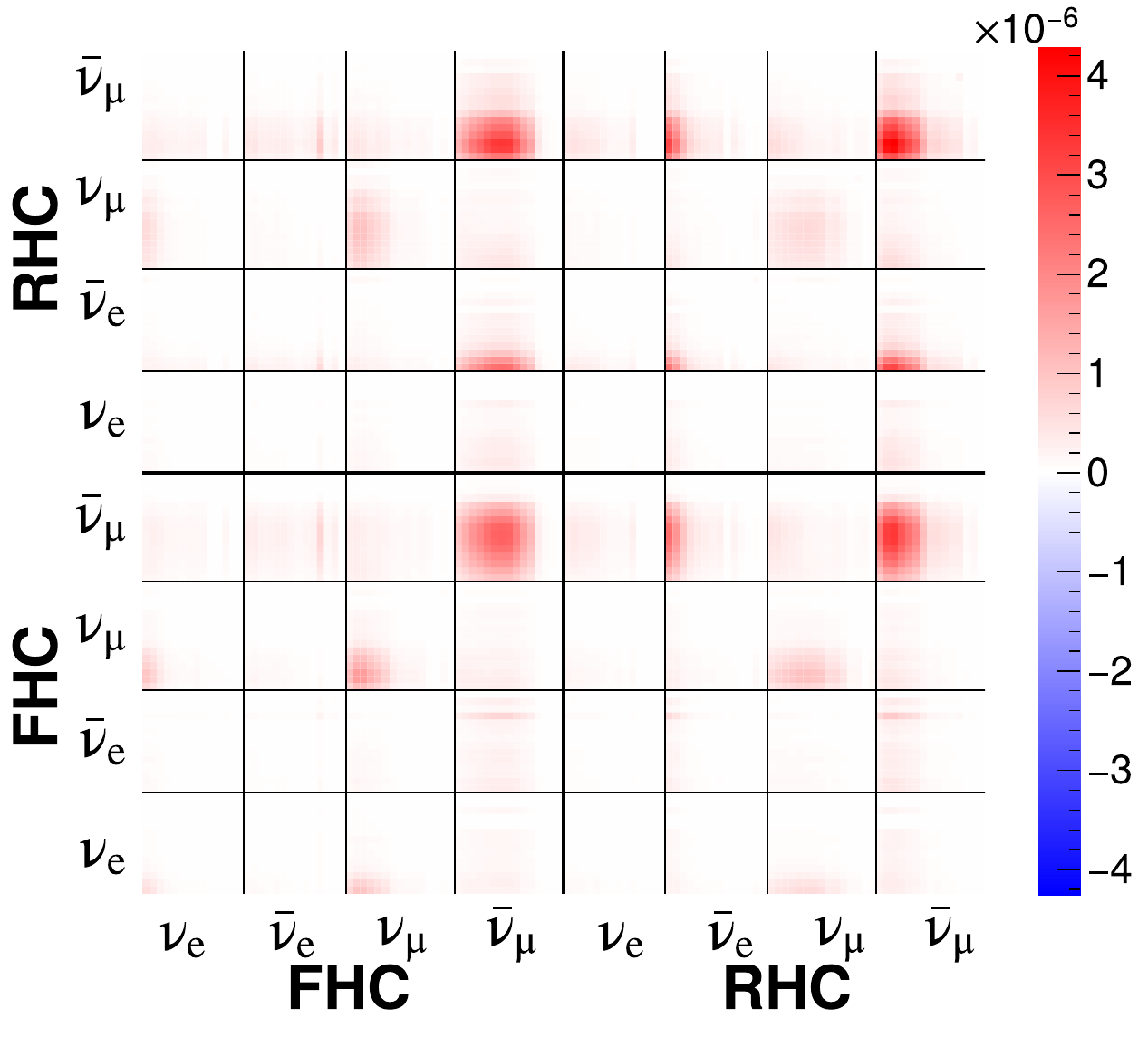}
        \caption{nCpi}
    \end{subfigure}
    \begin{subfigure}[]{0.27\textwidth}
    \includegraphics[width=\textwidth]{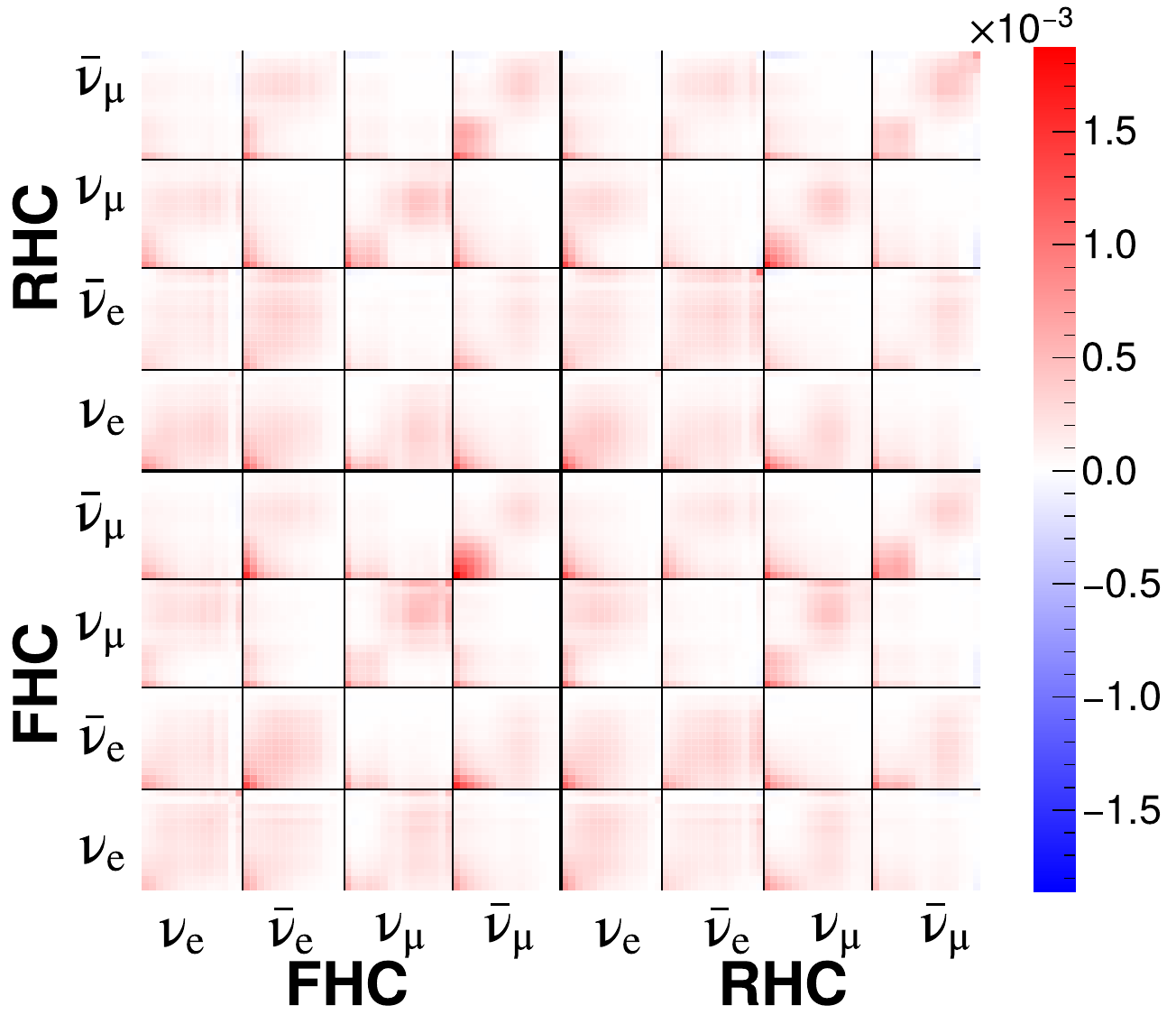}
        \caption{nuAlFe}
    \end{subfigure}
    \begin{subfigure}[]{0.27\textwidth}
    \includegraphics[width=\textwidth]{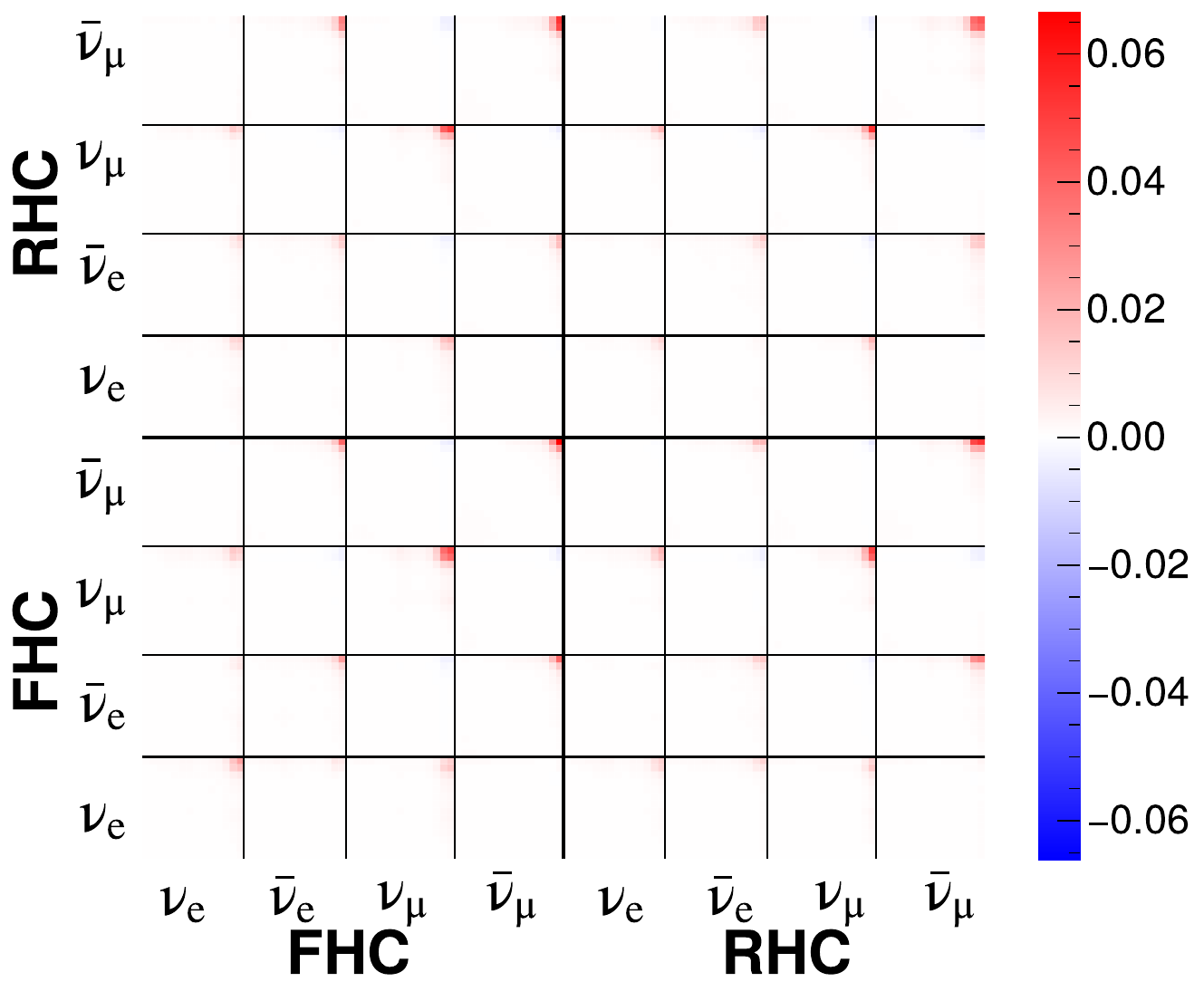}
        \caption{nua}
    \end{subfigure}
    \begin{subfigure}[]{0.27\textwidth}
        \includegraphics[width=\textwidth]{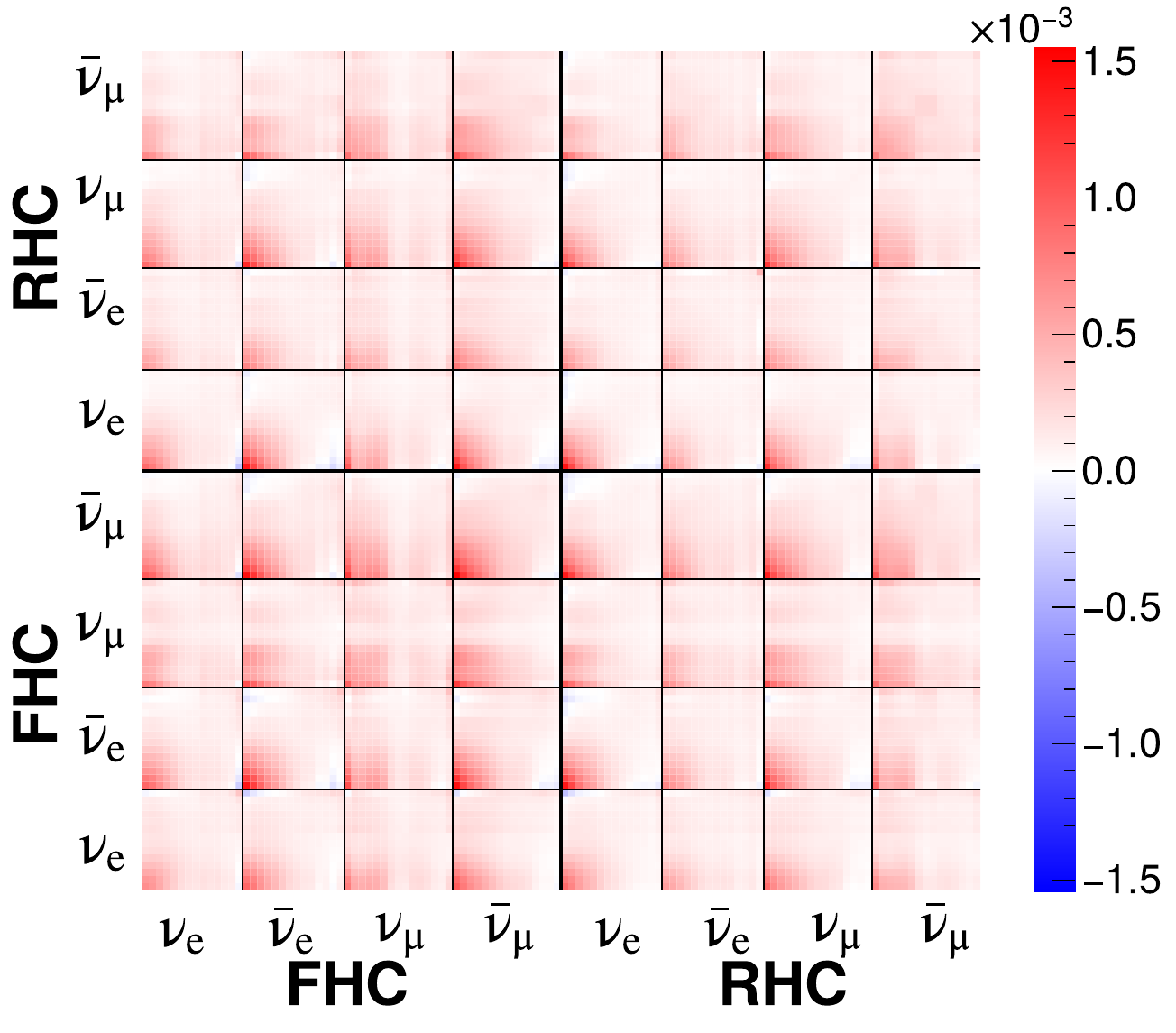}
        \caption{Attenuation}
    \end{subfigure}
    \begin{subfigure}[]{0.27\textwidth}
    \includegraphics[width=\textwidth]{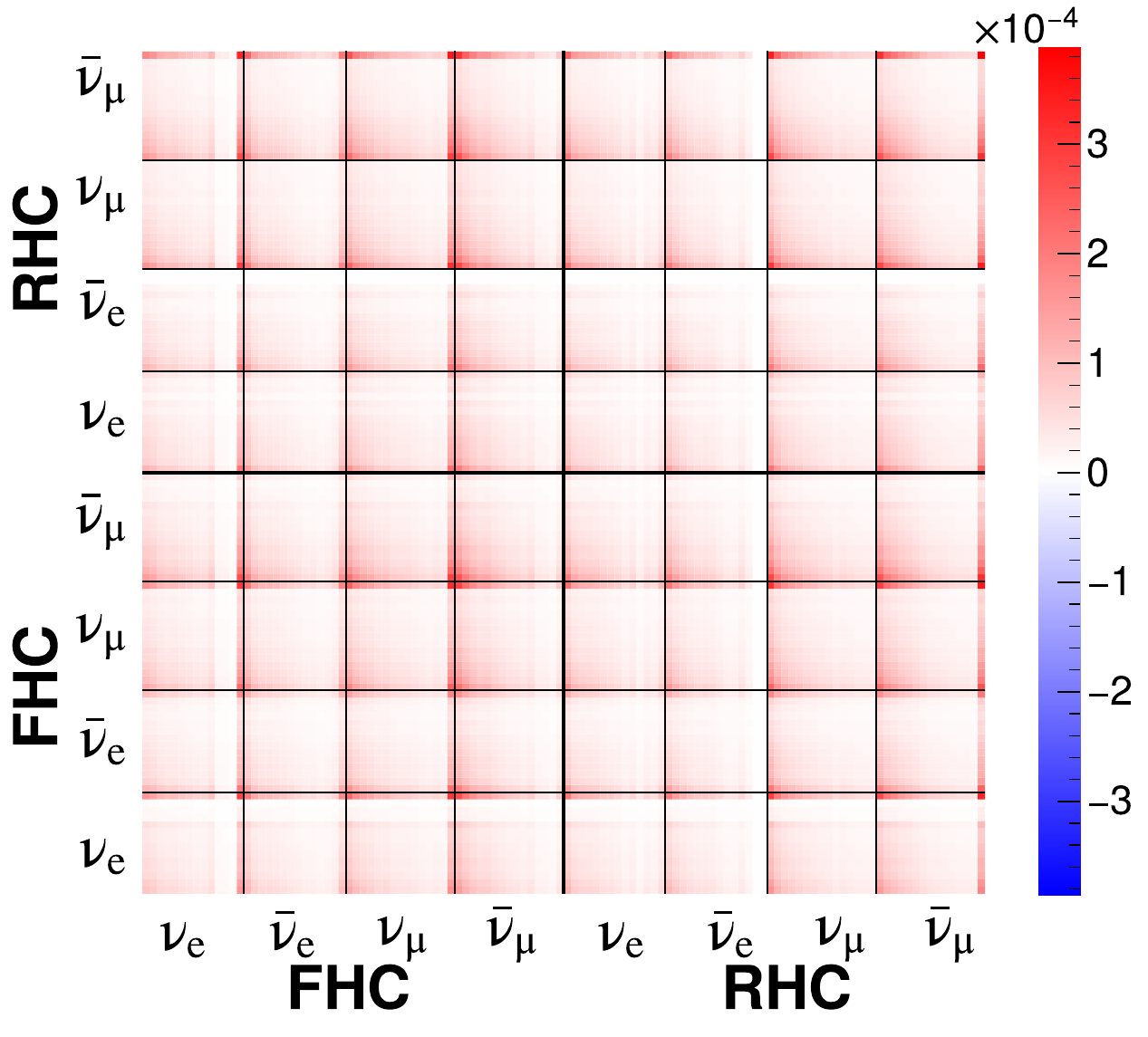}
        \caption{Others}
    \end{subfigure}
    \caption[Individual HP Covariance Matrices]{All hadron production covariance matrices.}
\end{figure}

\clearpage
\section{Correlation Matrices}
\begin{figure}[!ht]
    \centering
    \begin{subfigure}[]{0.27\textwidth}
        \includegraphics[width=\textwidth]{hcorr_hadron_total_correlation_matrix.pdf}
        \caption{Total}
    \end{subfigure}
    \begin{subfigure}[]{0.27\textwidth}
        \includegraphics[width=\textwidth]{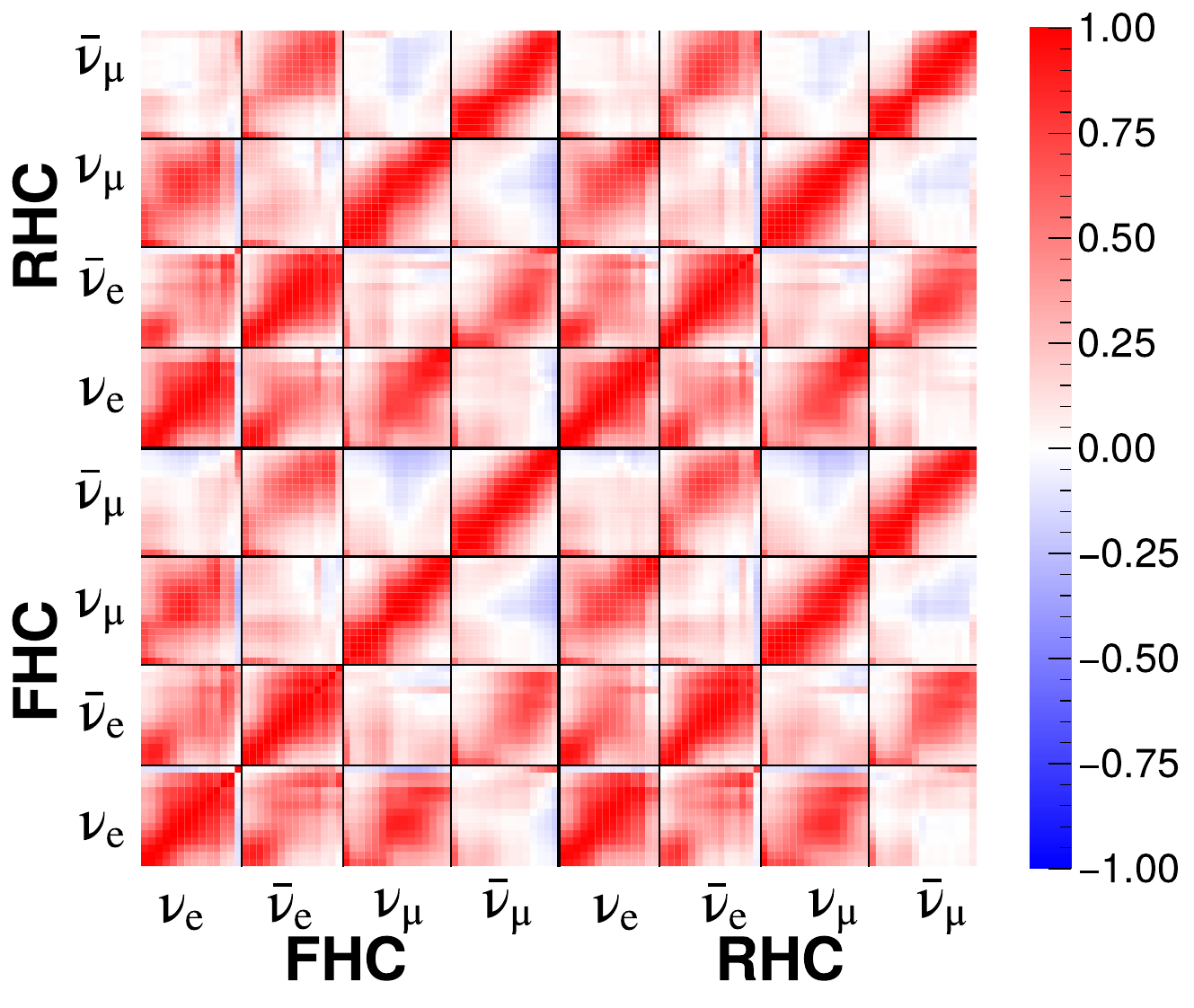}
        \caption{mesinc}
    \end{subfigure}
    \begin{subfigure}[]{0.27\textwidth}
        \includegraphics[width=\textwidth]{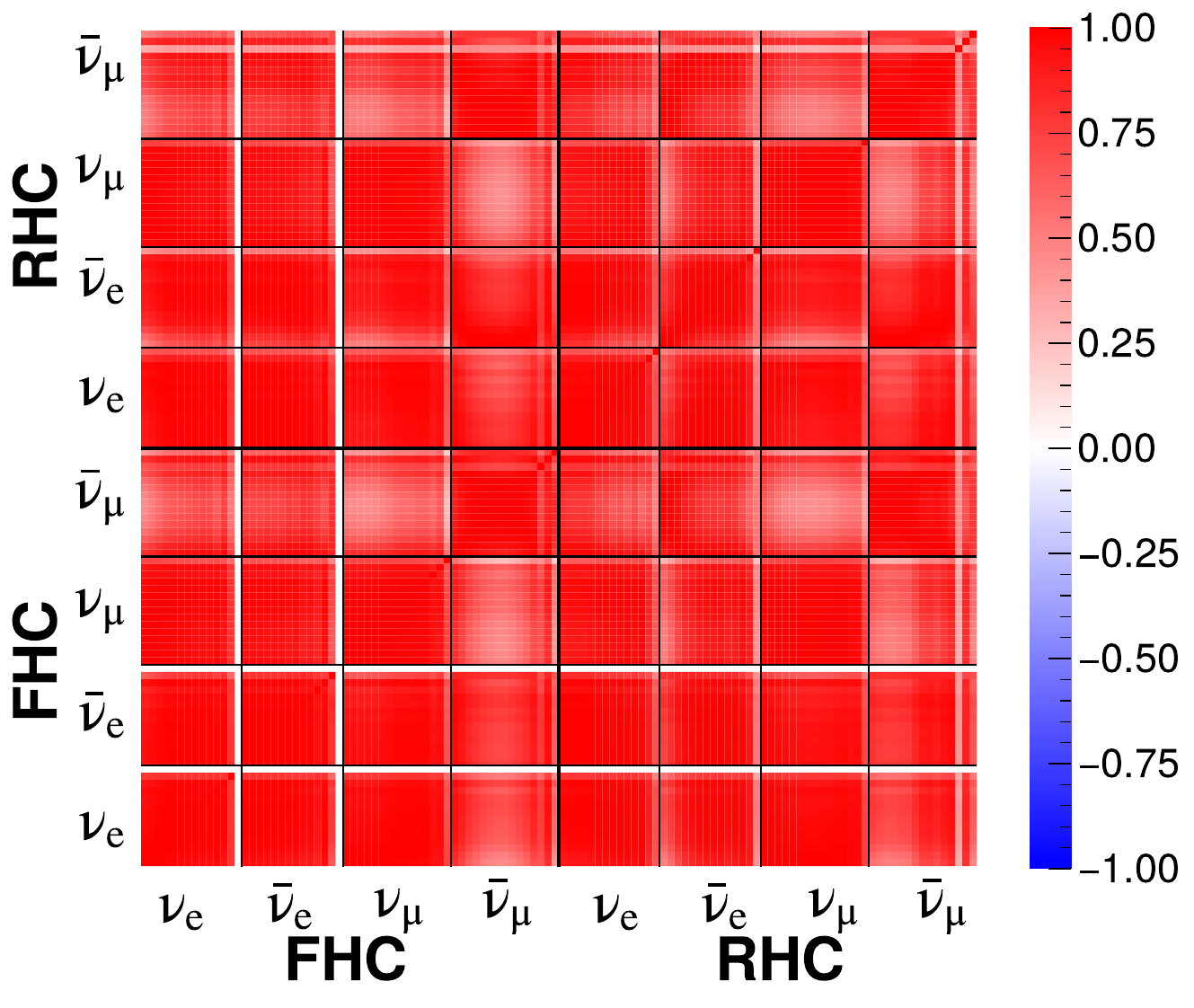}
        \caption{pCpi}
    \end{subfigure}
    \begin{subfigure}[]{0.27\textwidth}
        \includegraphics[width=\textwidth]{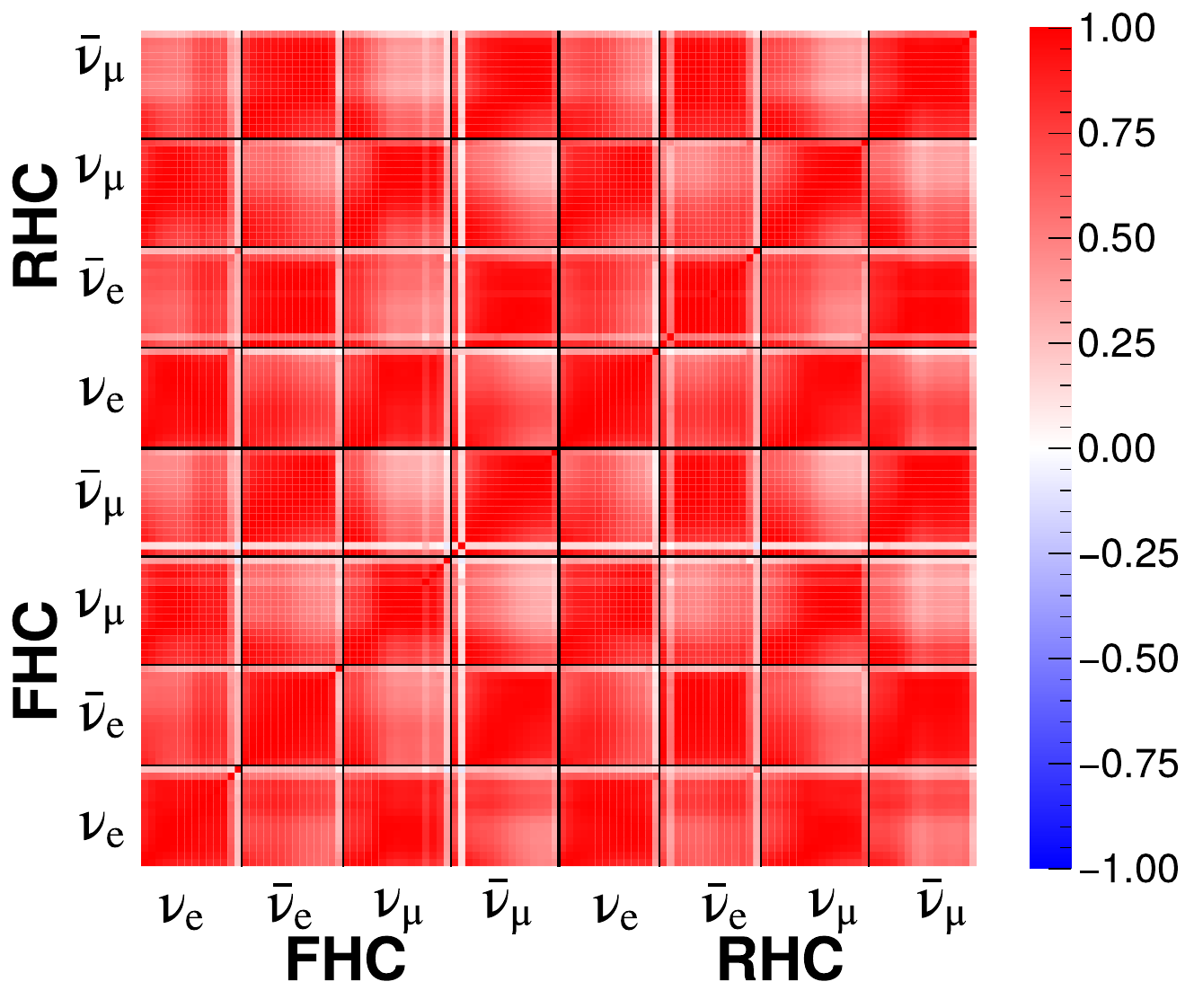}
        \caption{pCk}
    \end{subfigure}
    \begin{subfigure}[]{0.27\textwidth}
        \includegraphics[width=\textwidth]{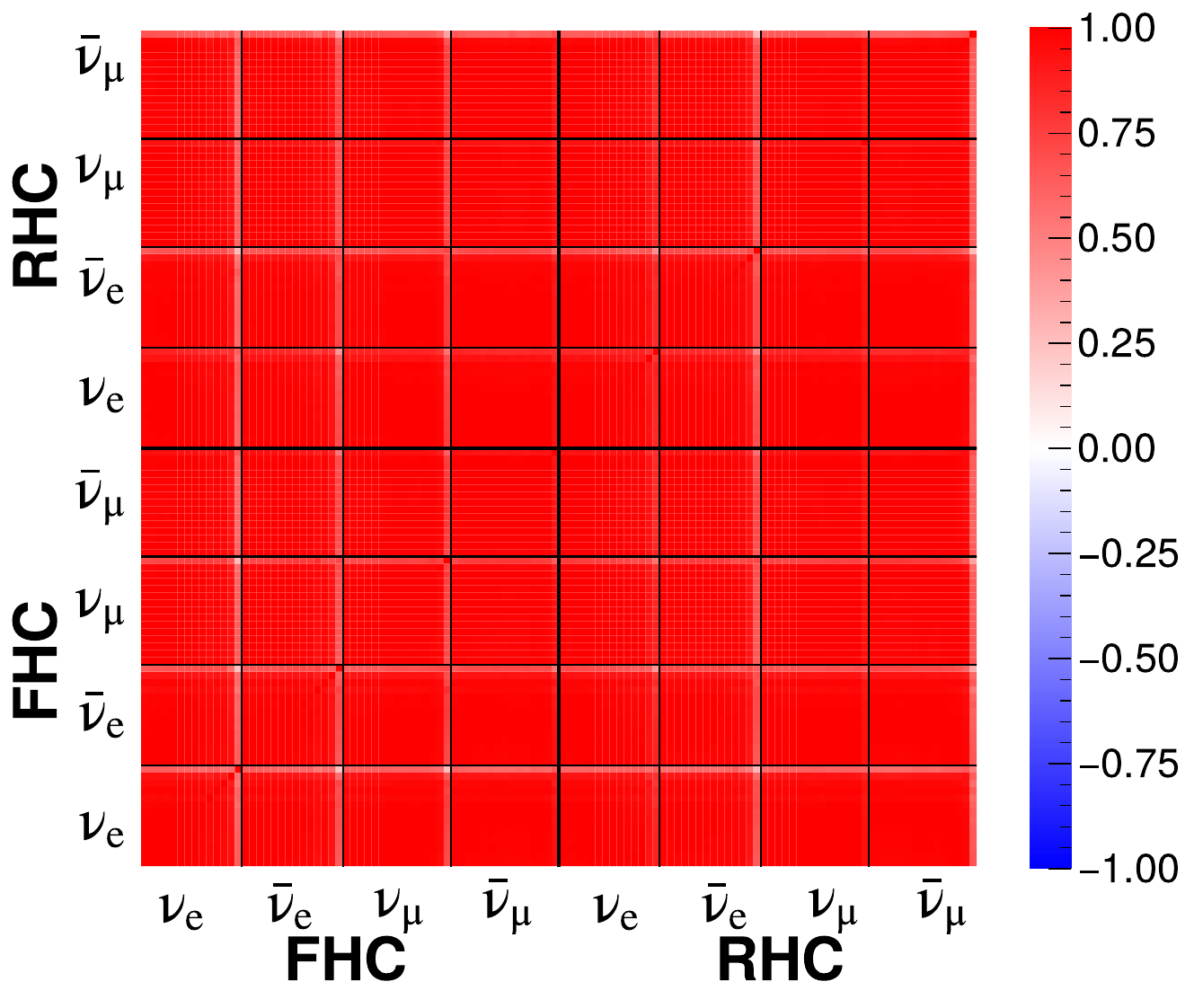}
        \caption{pCnu}
    \end{subfigure}
    \begin{subfigure}[]{0.27\textwidth}
        \includegraphics[width=\textwidth]{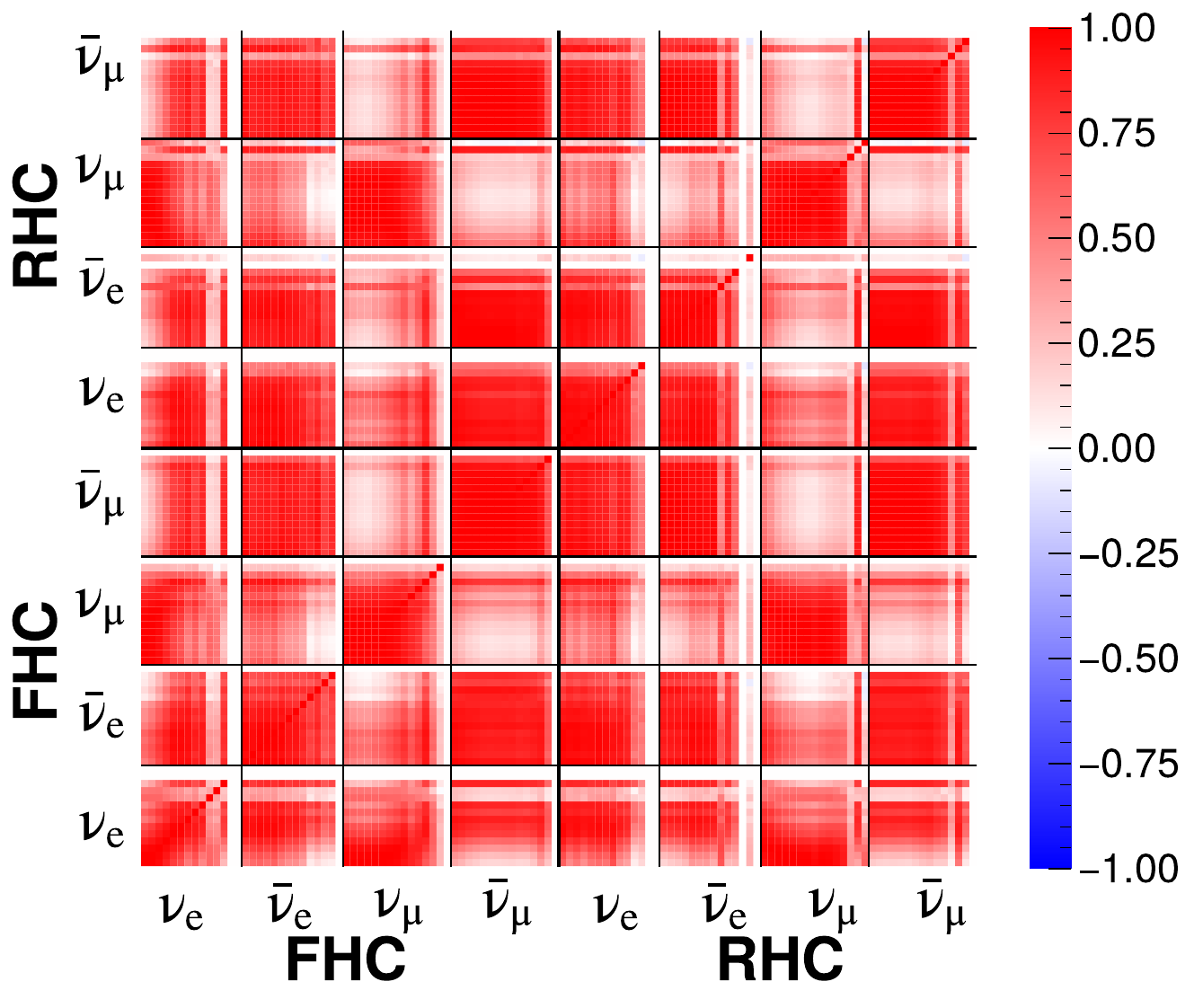}
        \caption{nCpi}
    \end{subfigure}
    \begin{subfigure}[]{0.27\textwidth}
        \includegraphics[width=\textwidth]{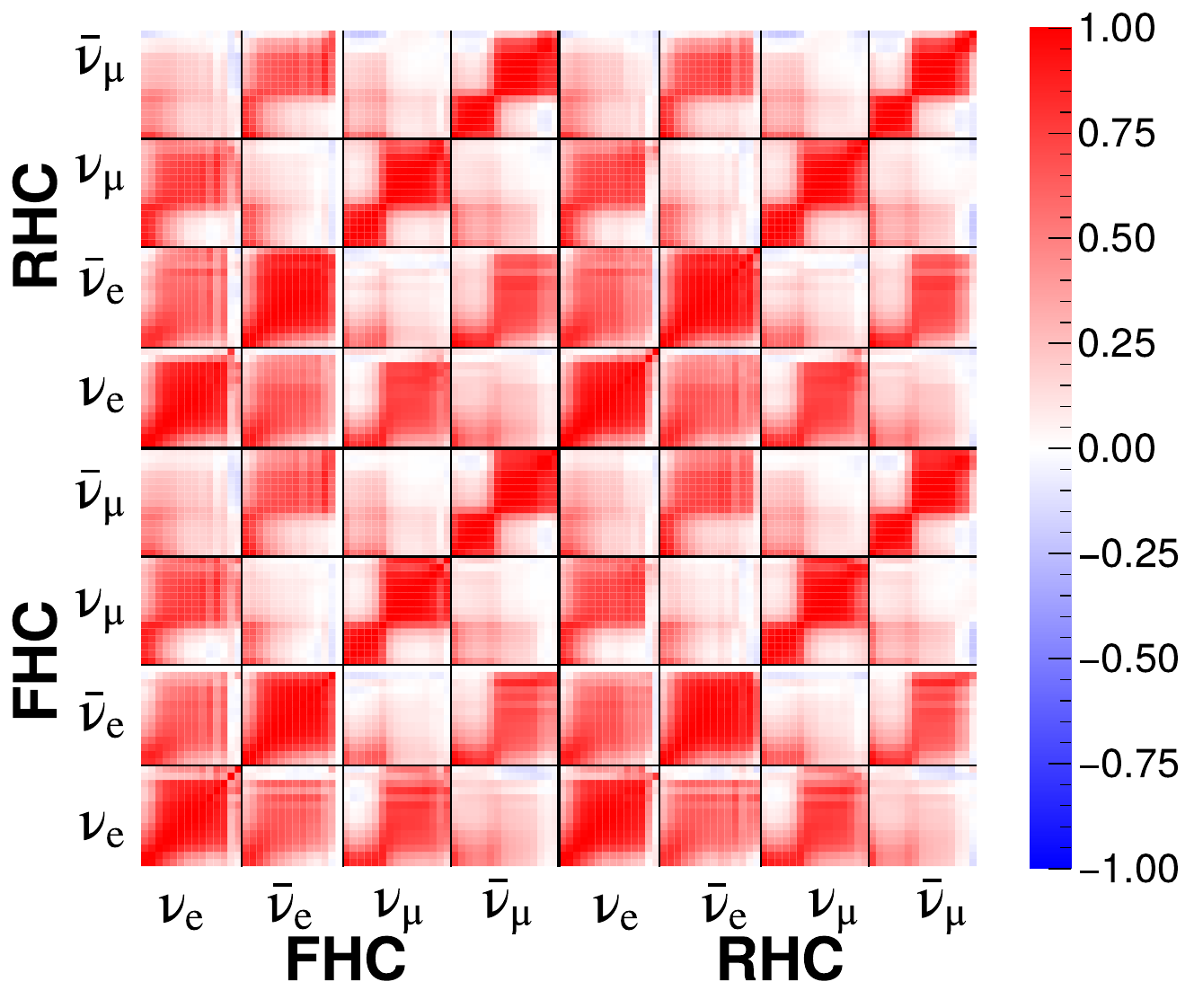}
        \caption{nuAlFe}
    \end{subfigure}
    \begin{subfigure}[]{0.27\textwidth}
        \includegraphics[width=\textwidth]{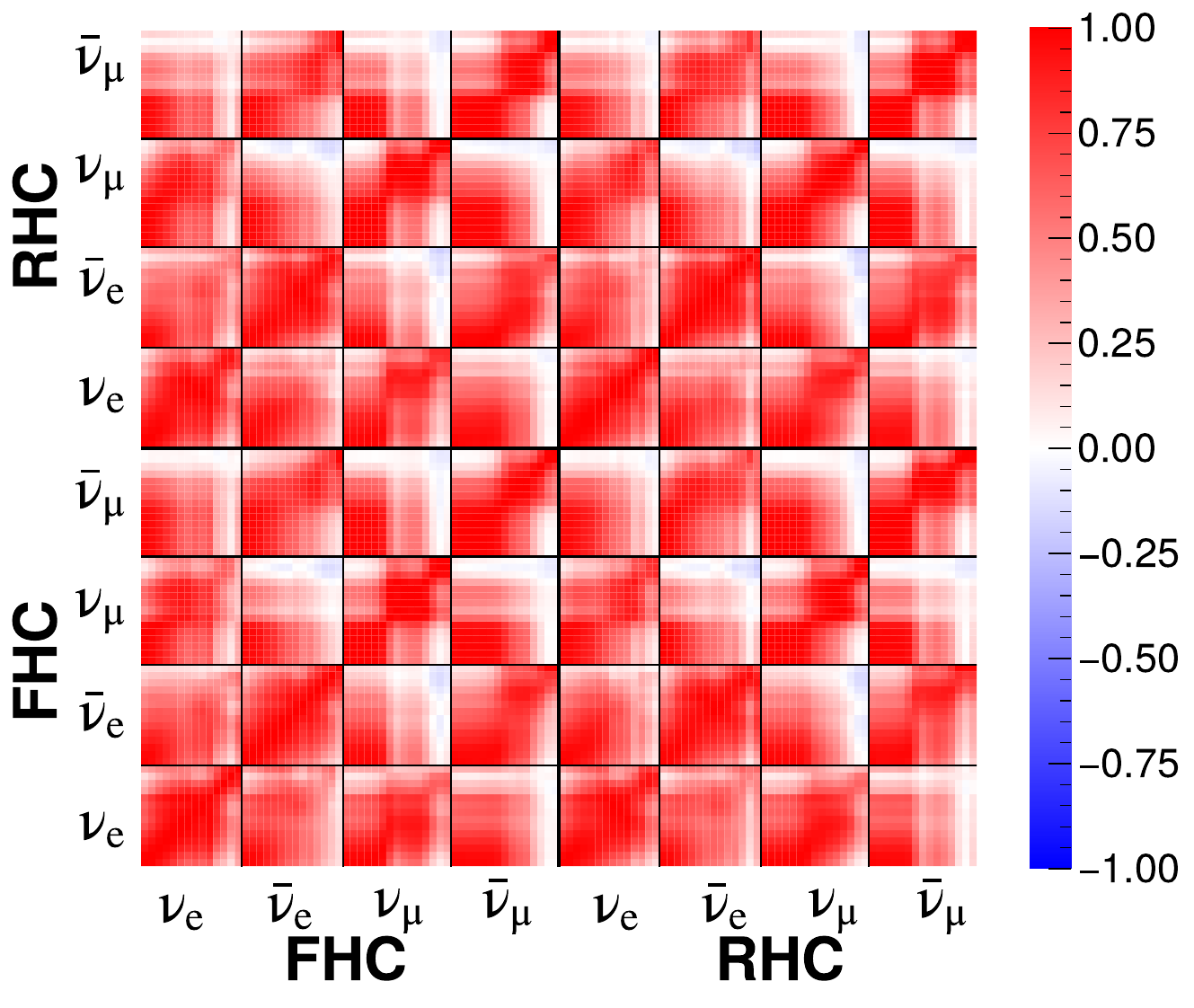}
        \caption{nua}
    \end{subfigure}
    \begin{subfigure}[]{0.27\textwidth}
        \includegraphics[width=\textwidth]{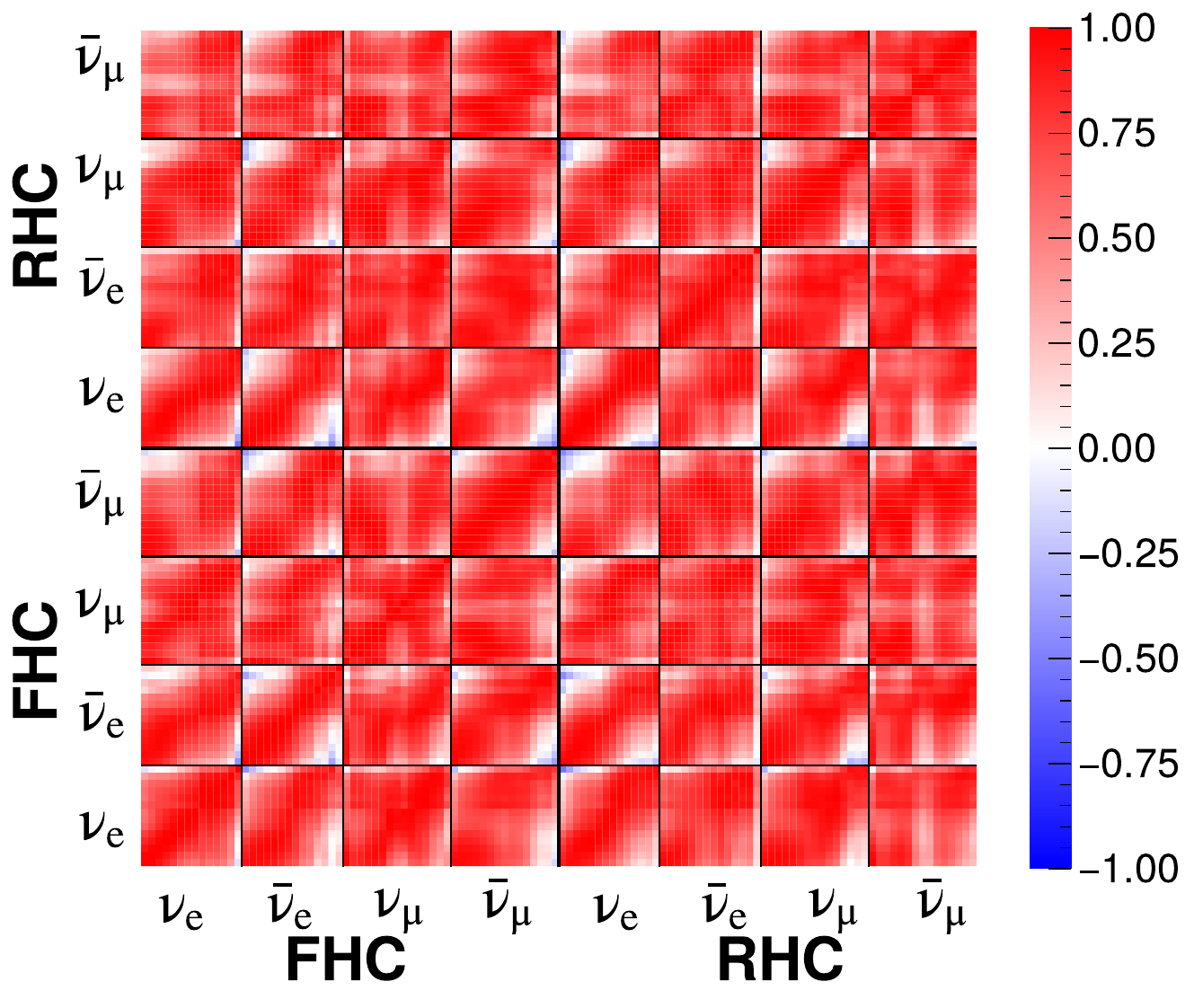}
        \caption{Attenuation}
    \end{subfigure}
    \begin{subfigure}[]{0.27\textwidth}
        \includegraphics[width=\textwidth]{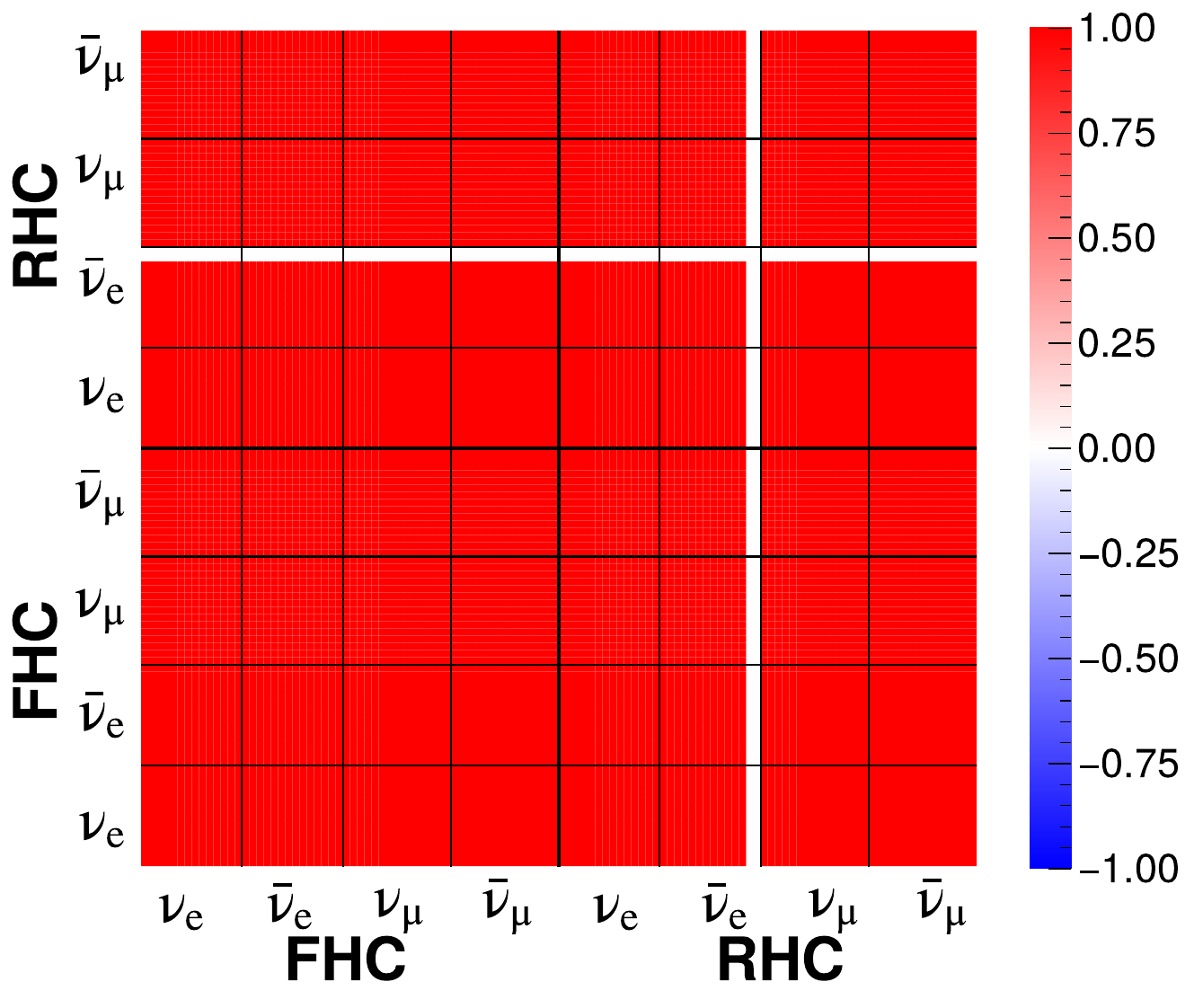}
        \caption{Others}
    \end{subfigure}
    \caption[Individual HP Correlation Matrices]{All hadron production correlation matrices.}
\end{figure}

%% file: pca_variance_plots.tex
\clearpage
\section{Top Four Principal Components}
\begin{figure}[!ht]
    \centering
    \includegraphics[width=\textwidth]{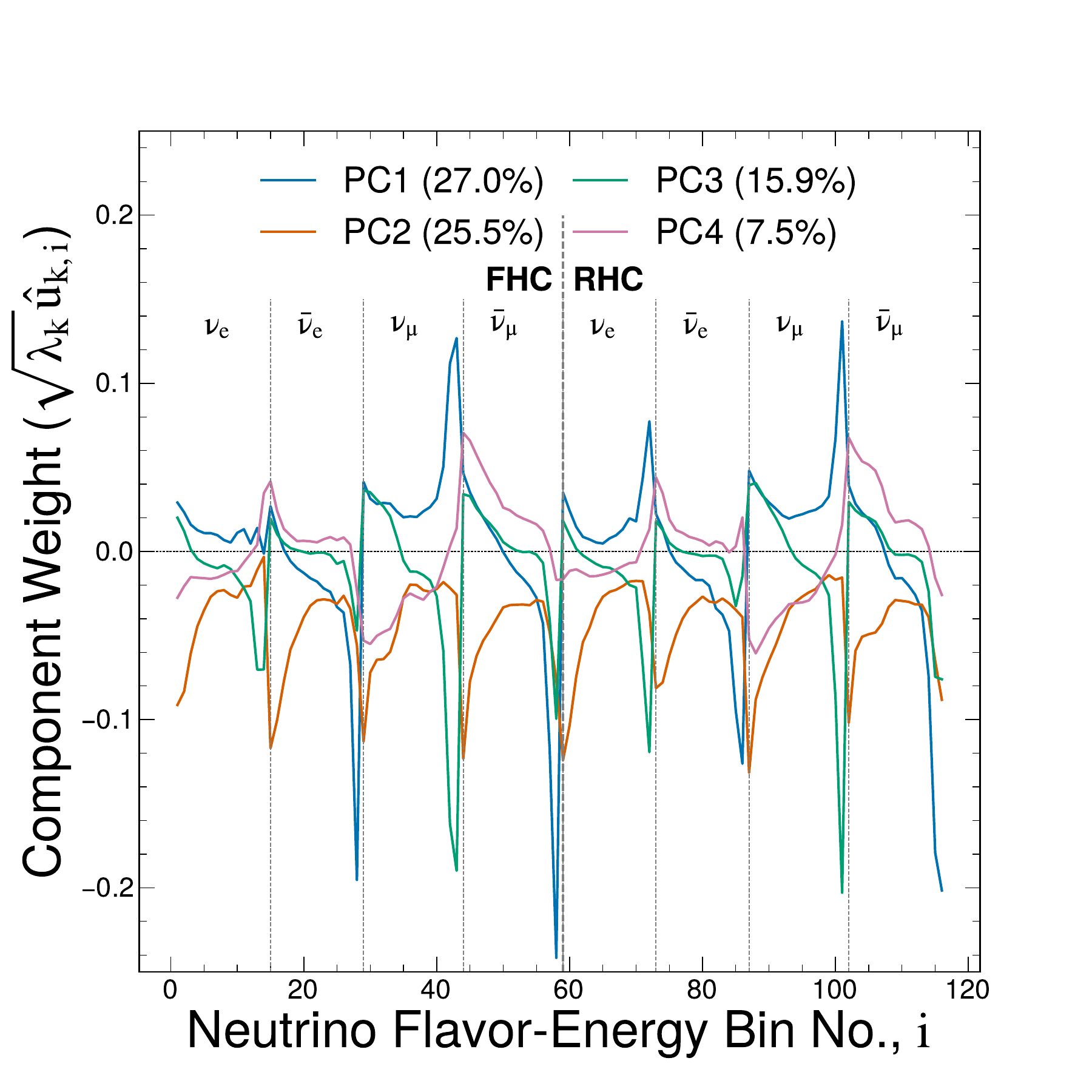}
    \caption[PCA Top Four Principal Components]{The top four principal components with the largest contributions to the total variance of the Hadron Production Covariance Matrix. The first two PCs contribute comparable amounts to the total variance, with PC1 describing more of the uncertainty in the HE flux tail, and PC2 the LE regions of the flux.}
    \label{fig:top-pca-comps}
\end{figure}

\clearpage
\section{Physics vs. PCA Variance Comparison}
\subsection{Forward Horn Current}
\begin{figure}[!ht]
    \centering
    \includegraphics[width=0.98\textwidth]{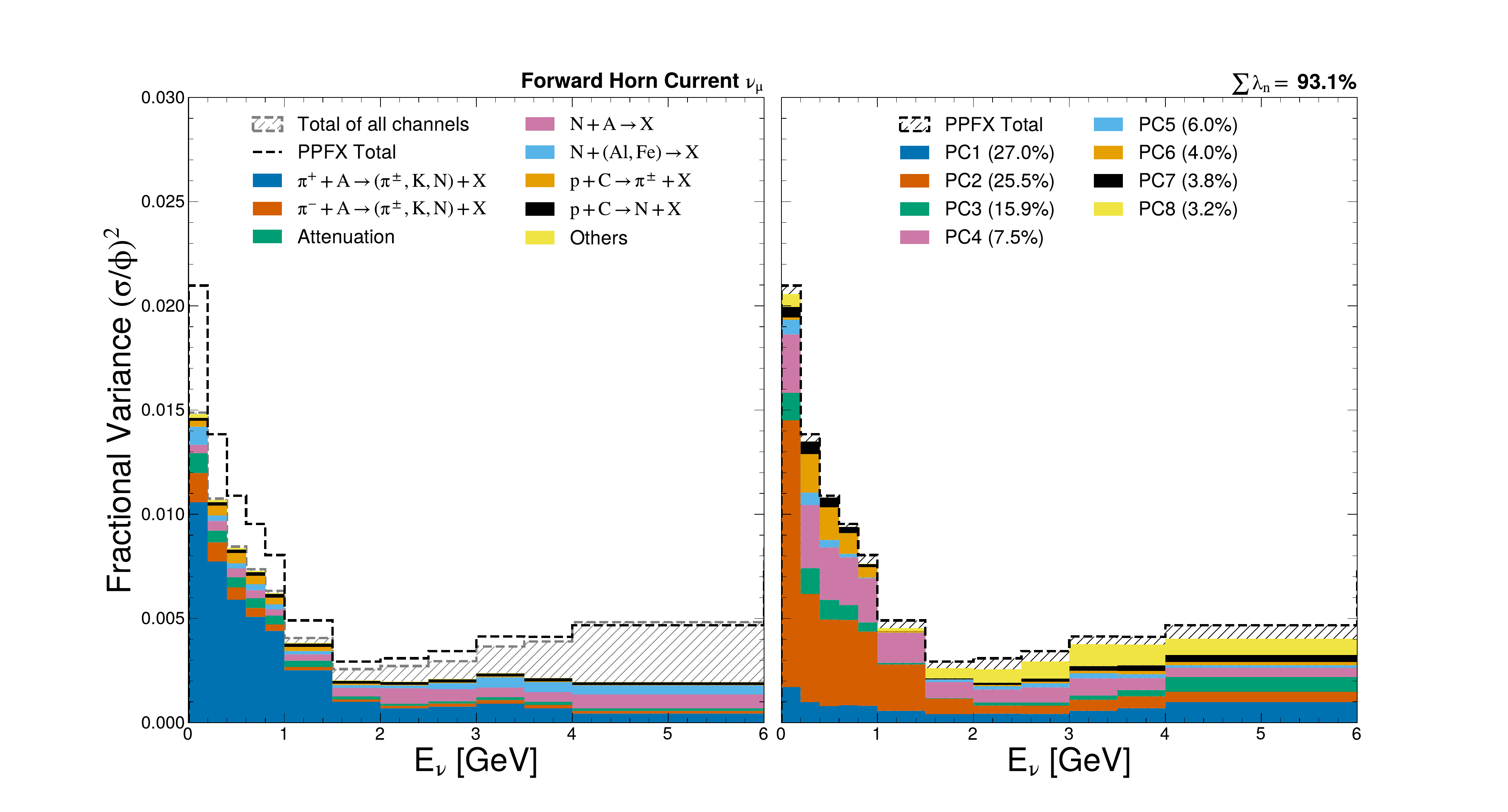}
    \includegraphics[width=0.98\textwidth]{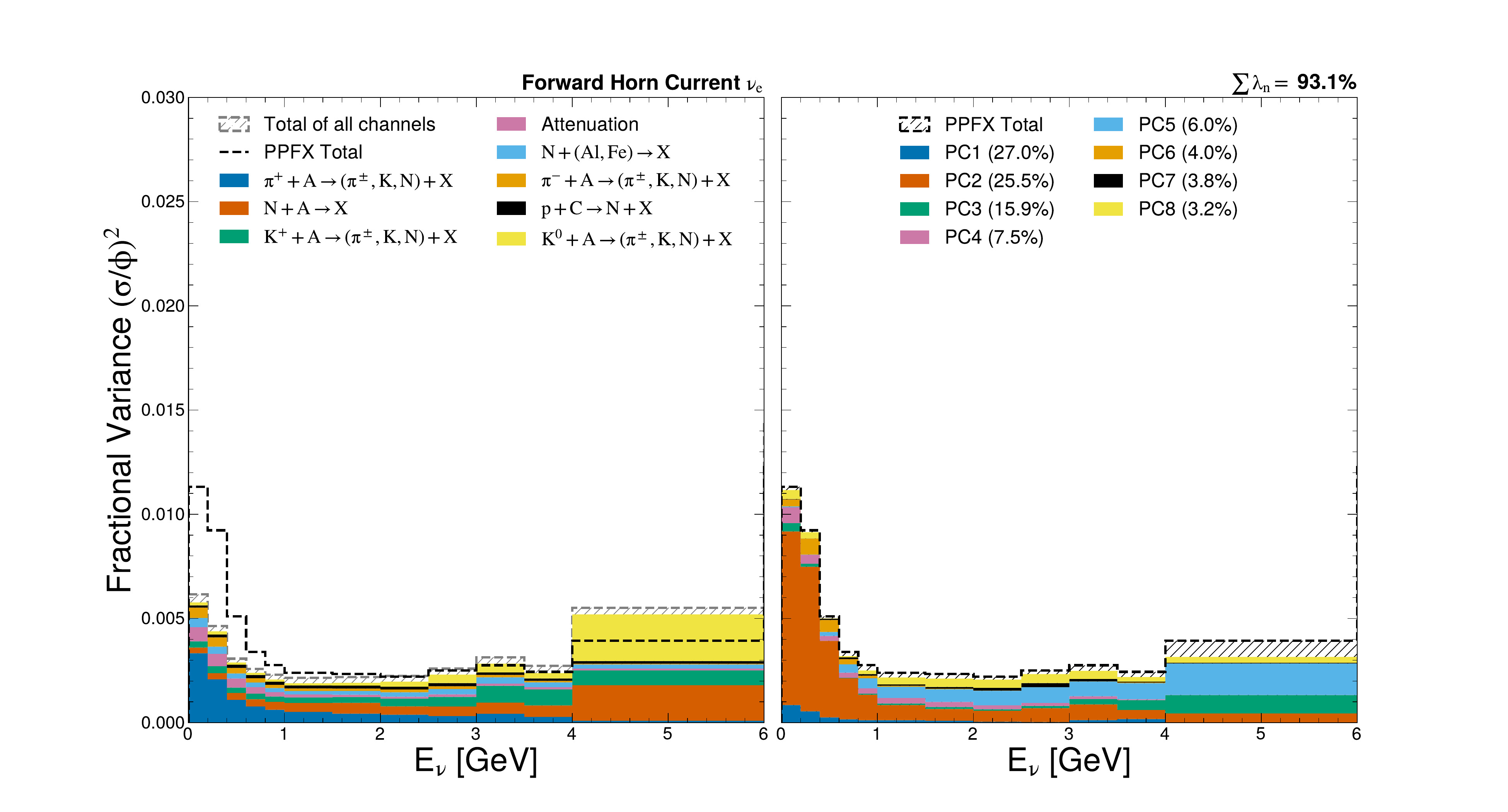}
    \caption[PCA Variance by Incoming Meson (FHC, $\nu$)]{Fractional variance comparison between physics and PCA descriptions by incoming meson (FHC, $\nu$).}
\end{figure}
\begin{figure}[!ht]
    \centering
    \includegraphics[width=0.98\textwidth]{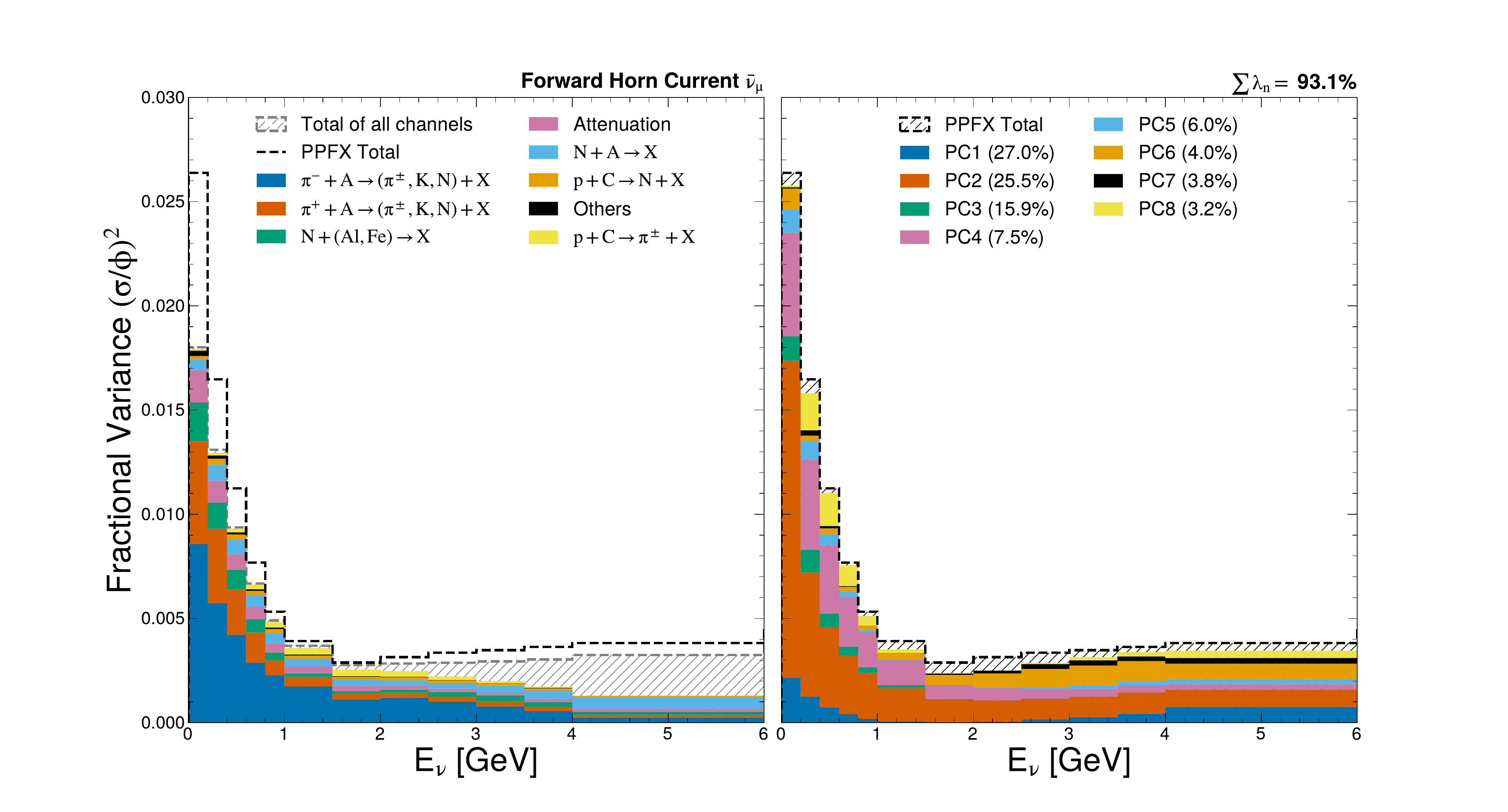}
    \includegraphics[width=0.98\textwidth]{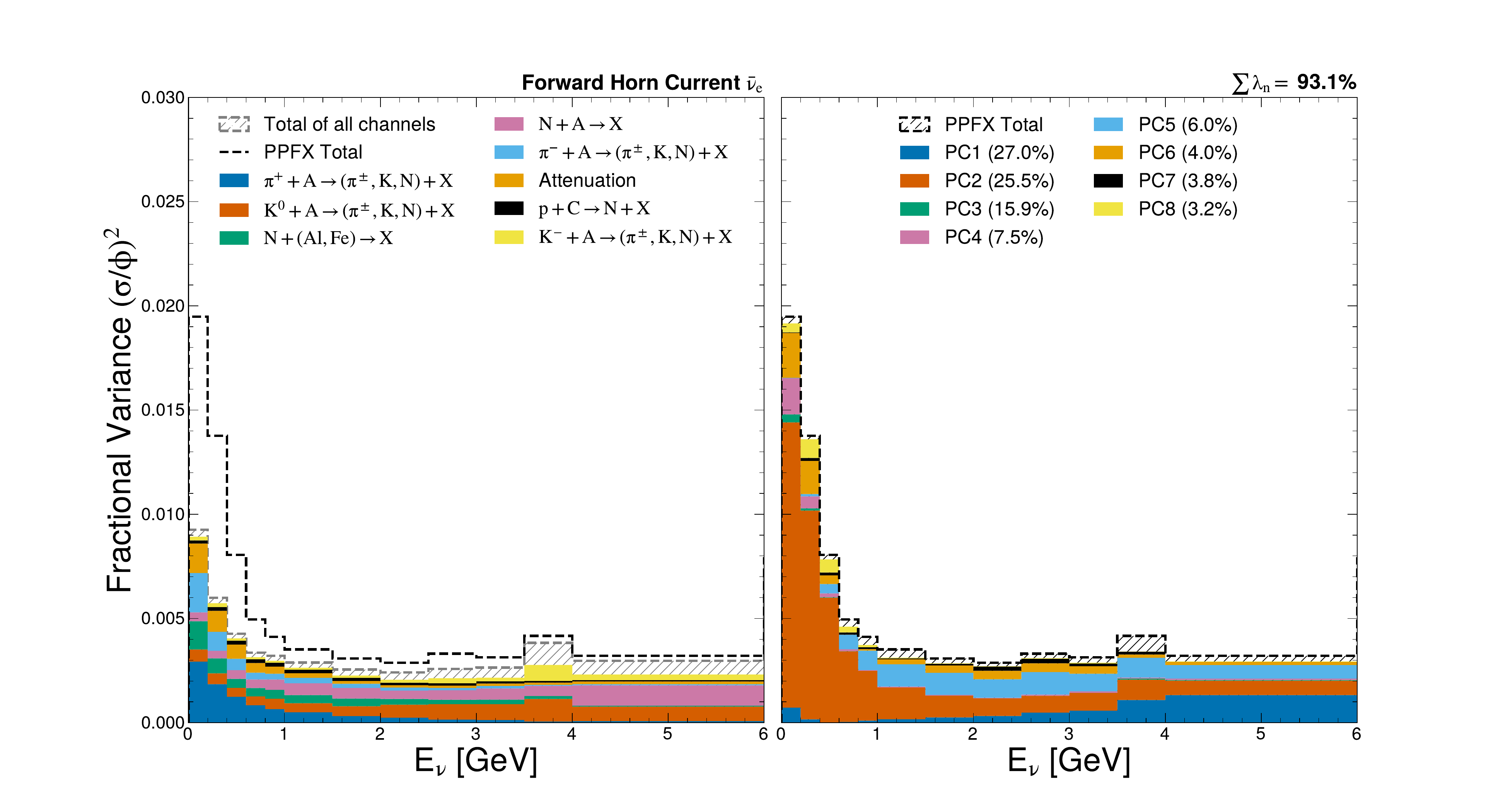}
    \caption[PCA Variance by Incoming Meson (FHC, $\bar{\nu}$)]{Fractional variance comparison between physics and PCA descriptions by incoming meson (FHC, $\bar{\nu}$).}
\end{figure}

\begin{figure}[!ht]
    \centering
    \includegraphics[width=0.98\textwidth]{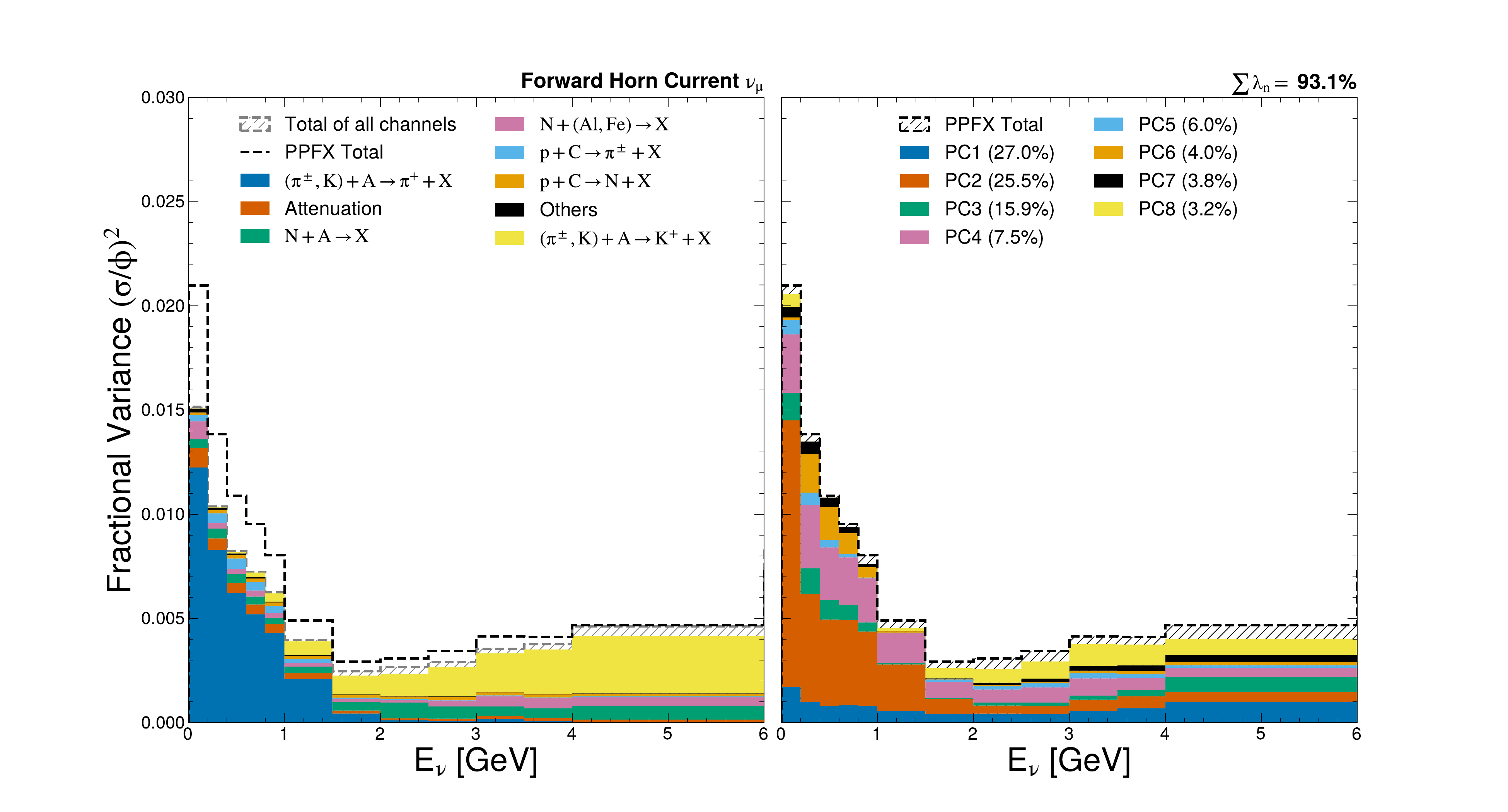}
    \includegraphics[width=0.98\textwidth]{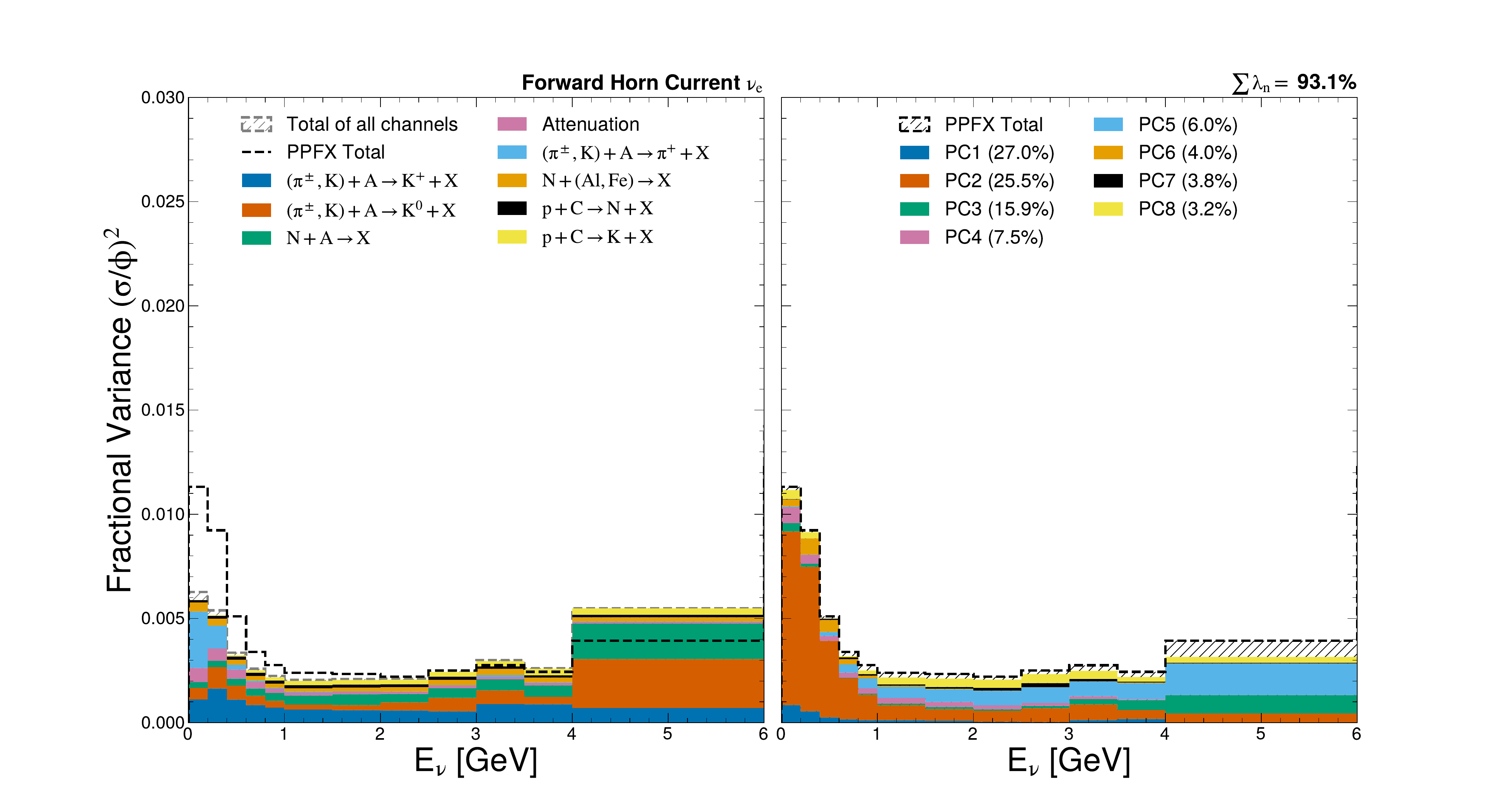}
    \caption[PCA Variance by Outgoing Meson (FHC, $\nu$)]{Fractional variance comparison between physics and PCA descriptions by outgoing meson (FHC, $\nu$).}
\end{figure}
\begin{figure}[!ht]
    \centering
    \includegraphics[width=0.98\textwidth]{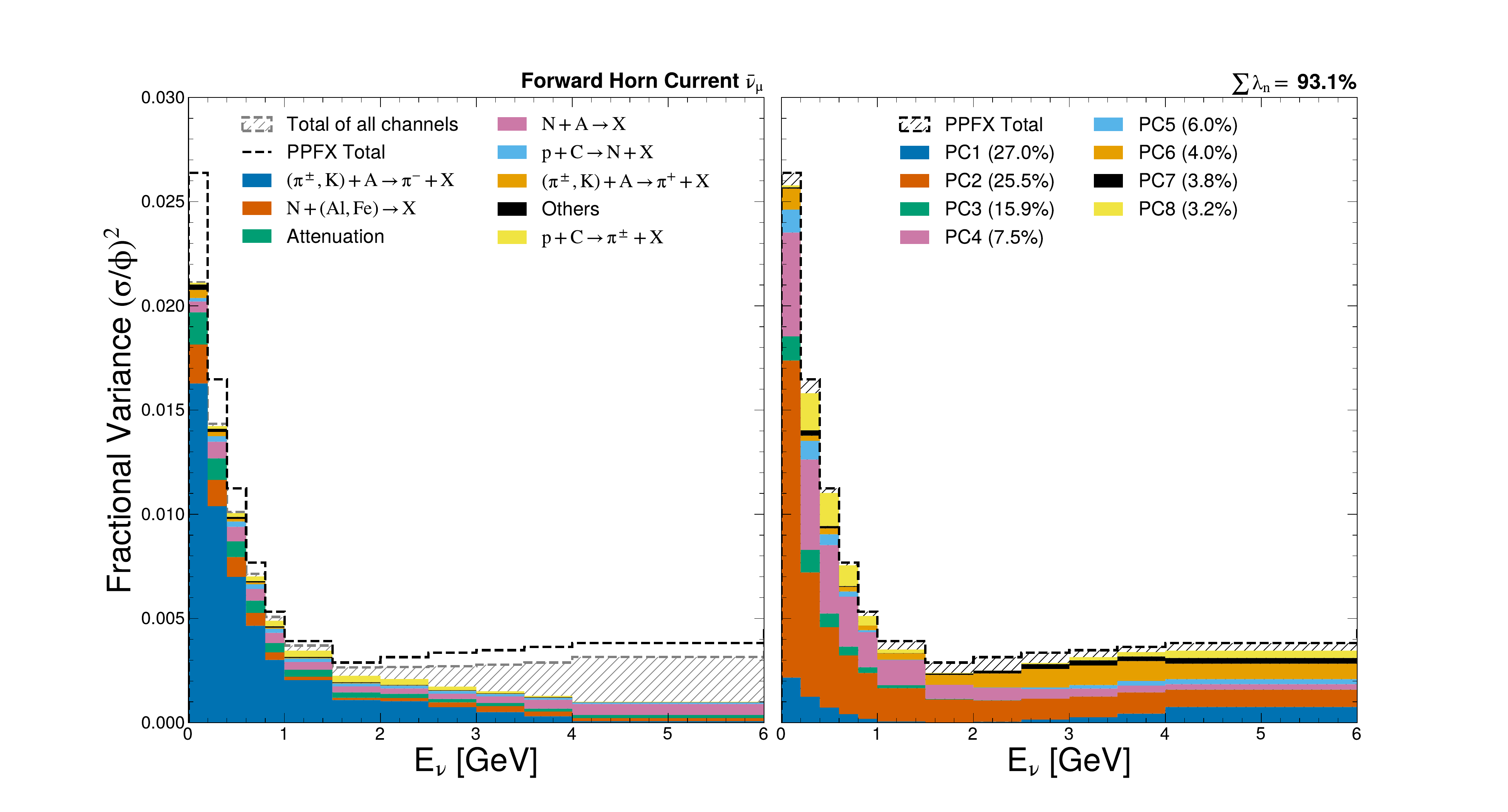}
    \includegraphics[width=0.98\textwidth]{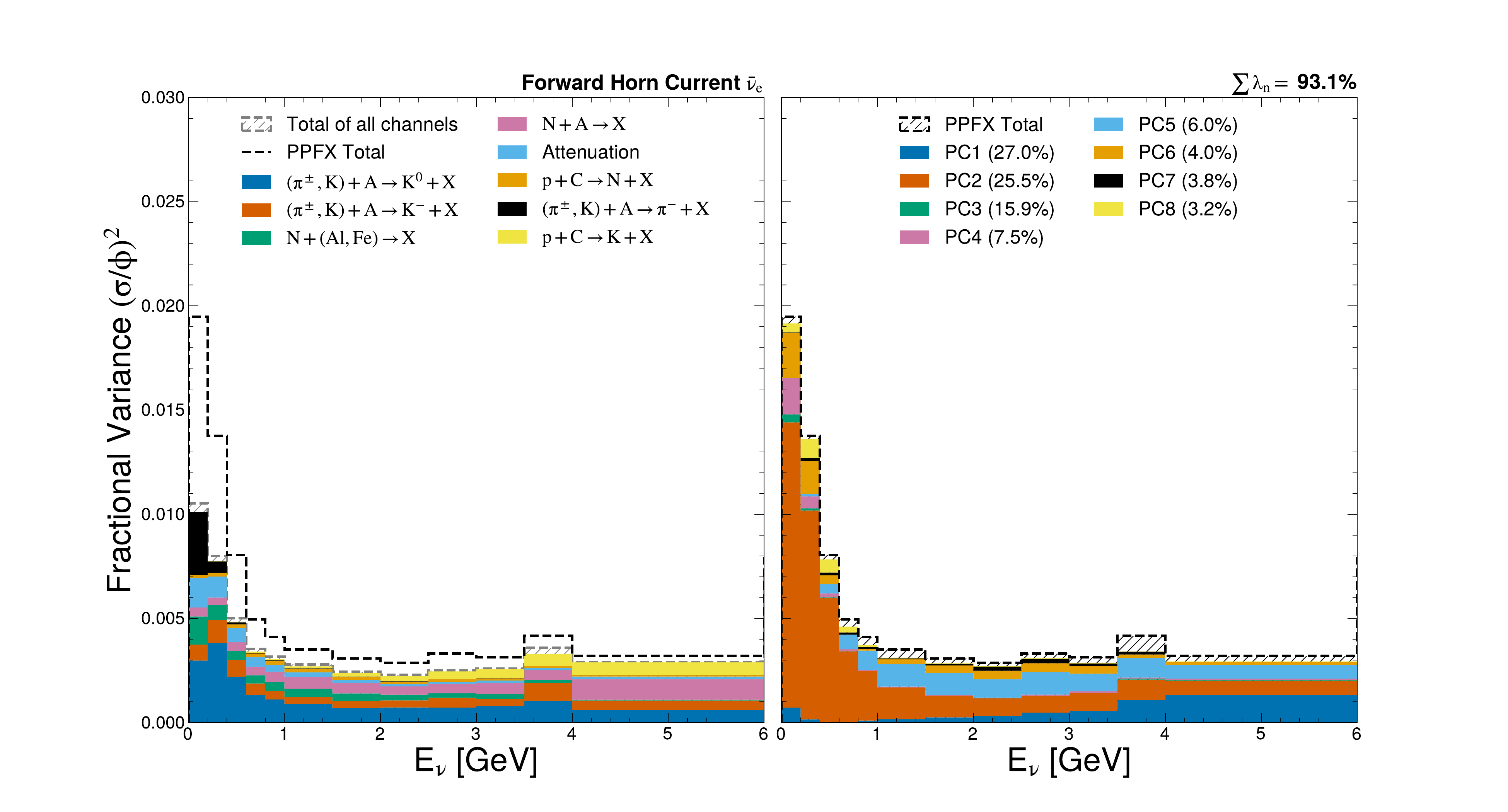}
    \caption[PCA Variance by Outgoing Meson (FHC, $\bar{\nu}$)]{Fractional variance comparison between physics and PCA descriptions by outgoing meson (FHC, $\bar{\nu}$).}
\end{figure}

\clearpage
\subsection{Reverse Horn Current}
\begin{figure}[!ht]
    \centering
    \includegraphics[width=0.98\textwidth]{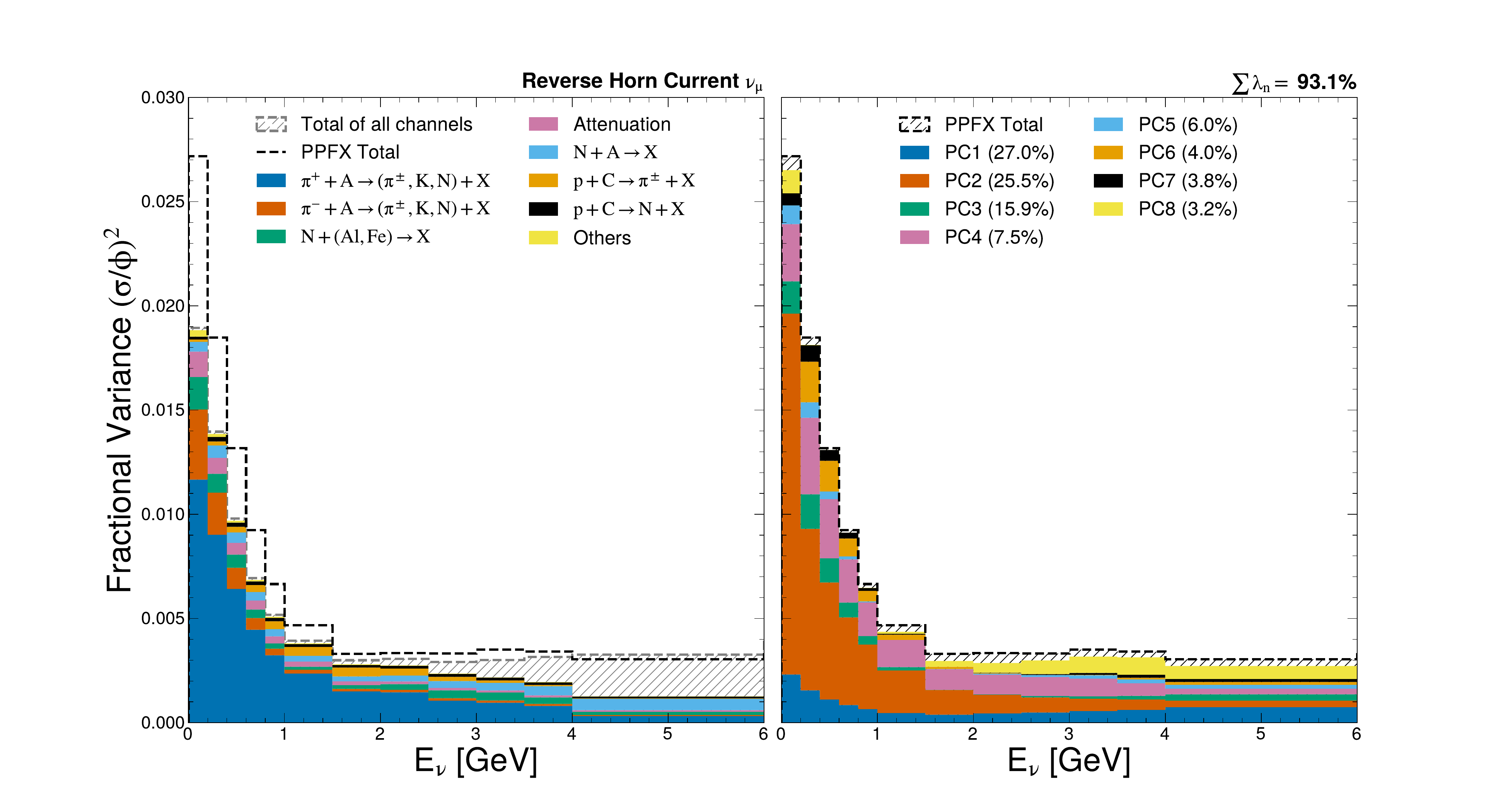}
    \includegraphics[width=0.98\textwidth]{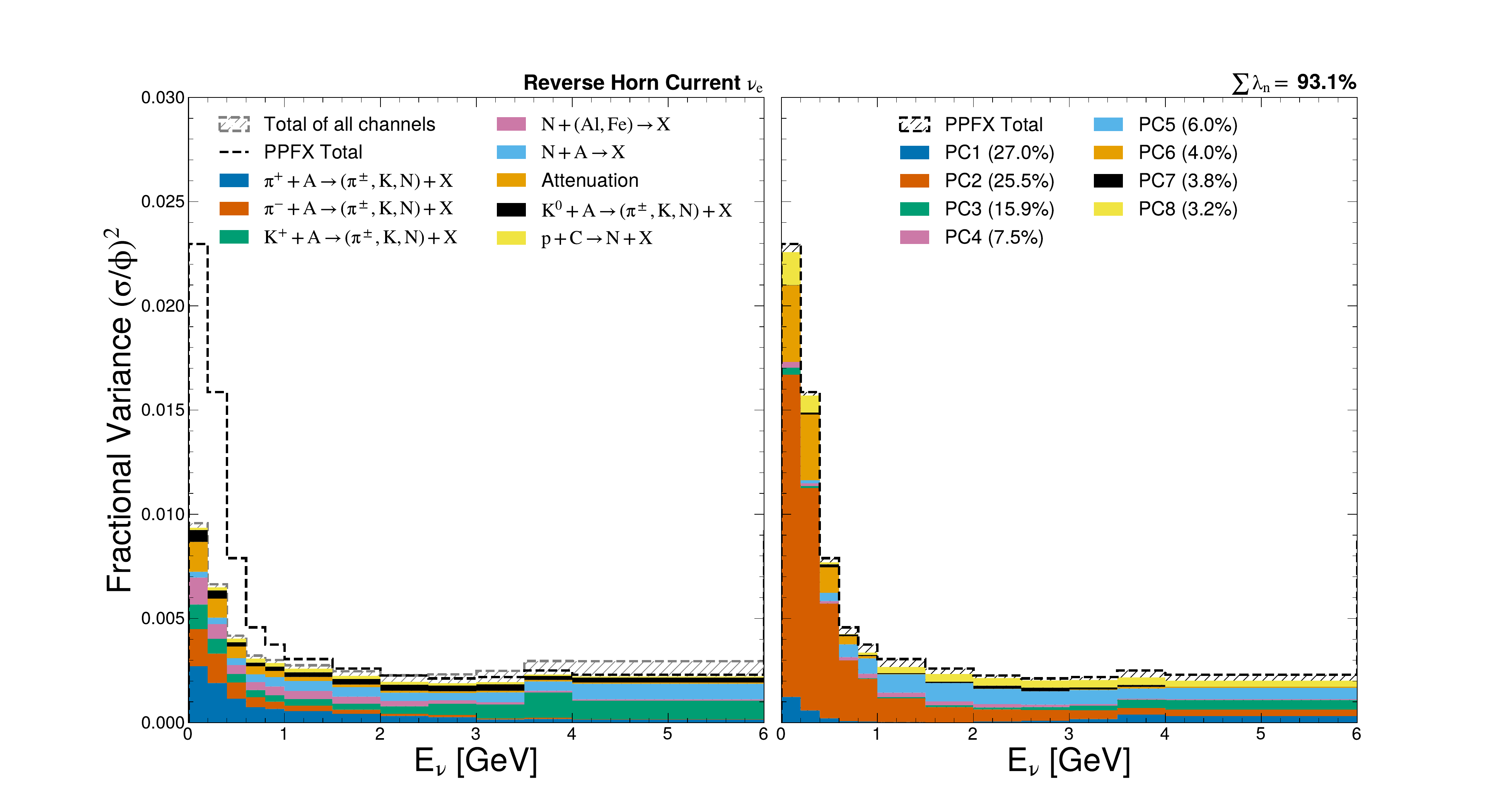}
    \caption[PCA Variance by Incoming Meson (RHC, $\nu$)]{Fractional variance comparison between physics and PCA descriptions by incoming meson (RHC, $\nu$).}
\end{figure}
\begin{figure}[!ht]
    \centering
    \includegraphics[width=0.98\textwidth]{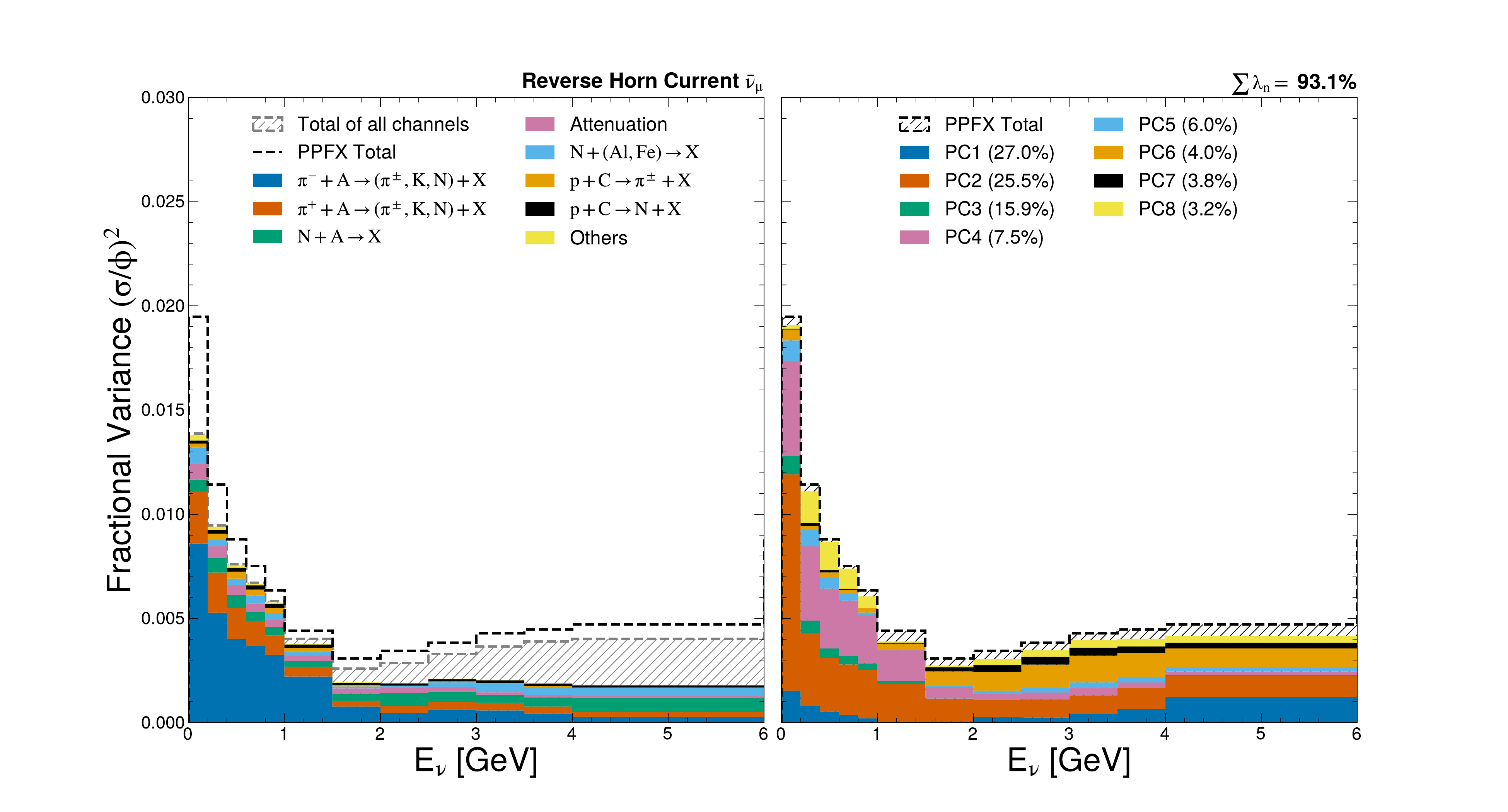}
    \includegraphics[width=0.98\textwidth]{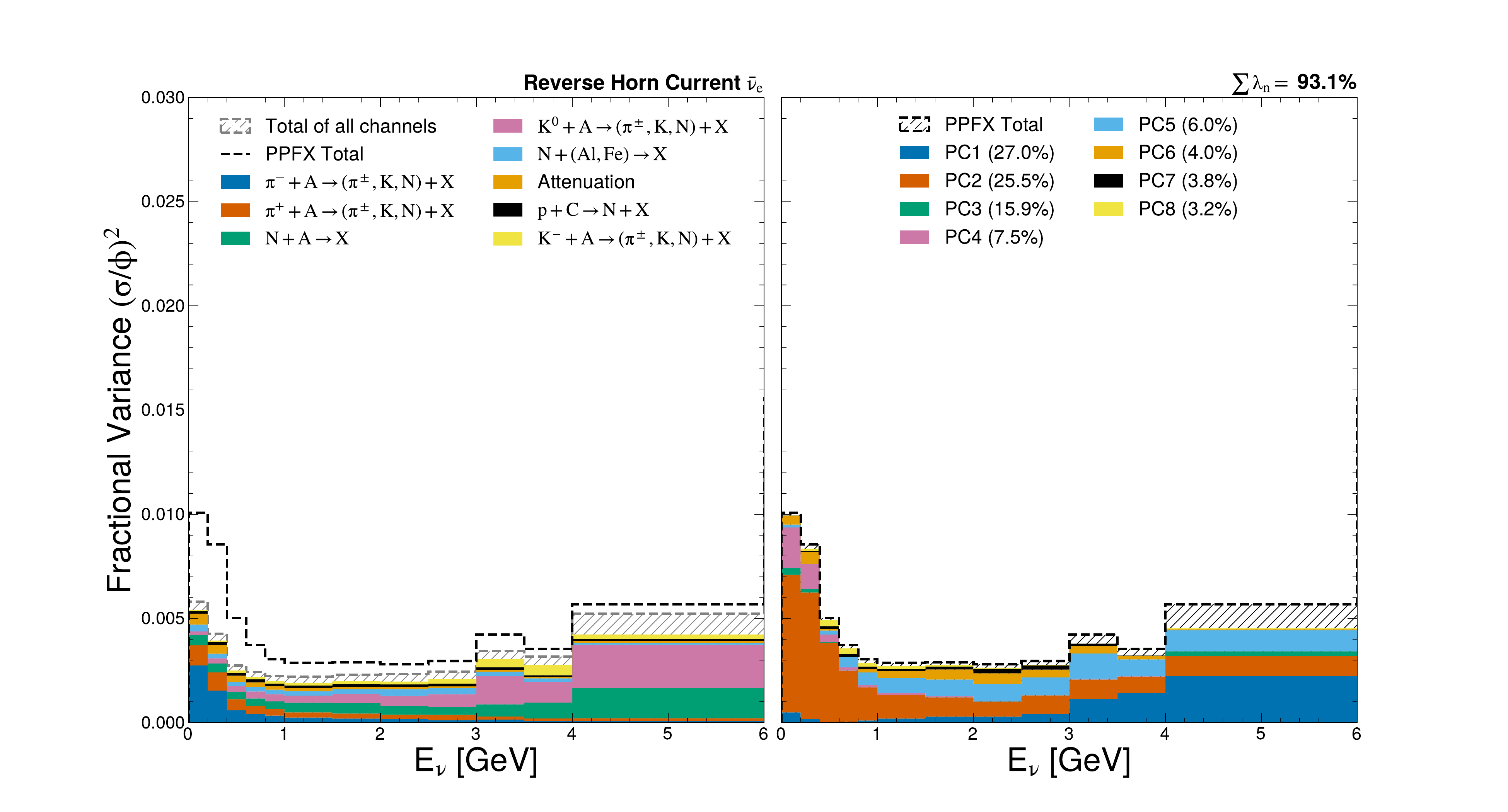}
    \caption[PCA Variance by Incoming Meson (RHC, $\bar{\nu}$)]{Fractional variance comparison between physics and PCA descriptions by incoming meson (RHC, $\bar{\nu}$).}
\end{figure}

\begin{figure}[!ht]
    \centering
    \includegraphics[width=0.98\textwidth]{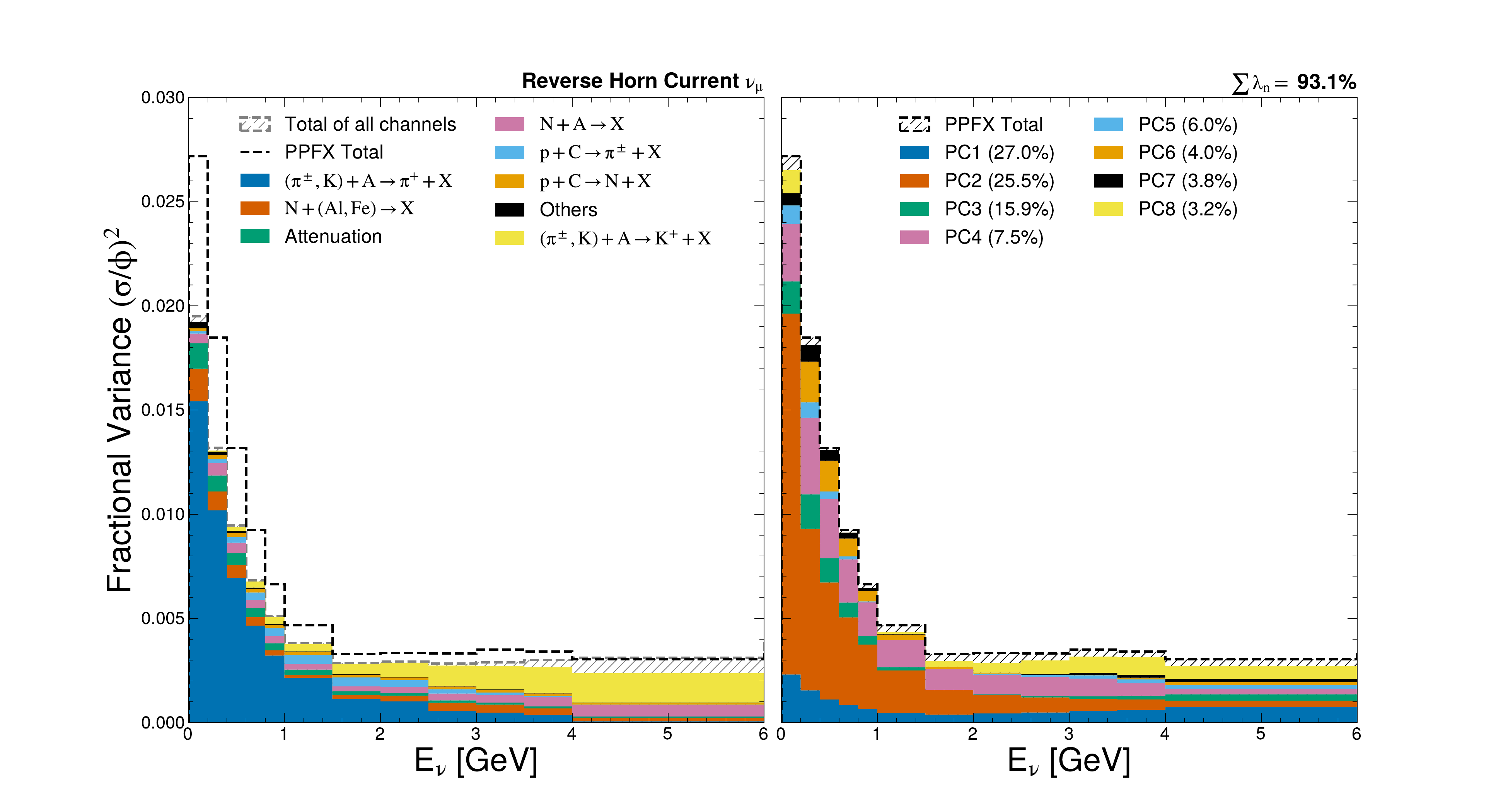}
    \includegraphics[width=0.98\textwidth]{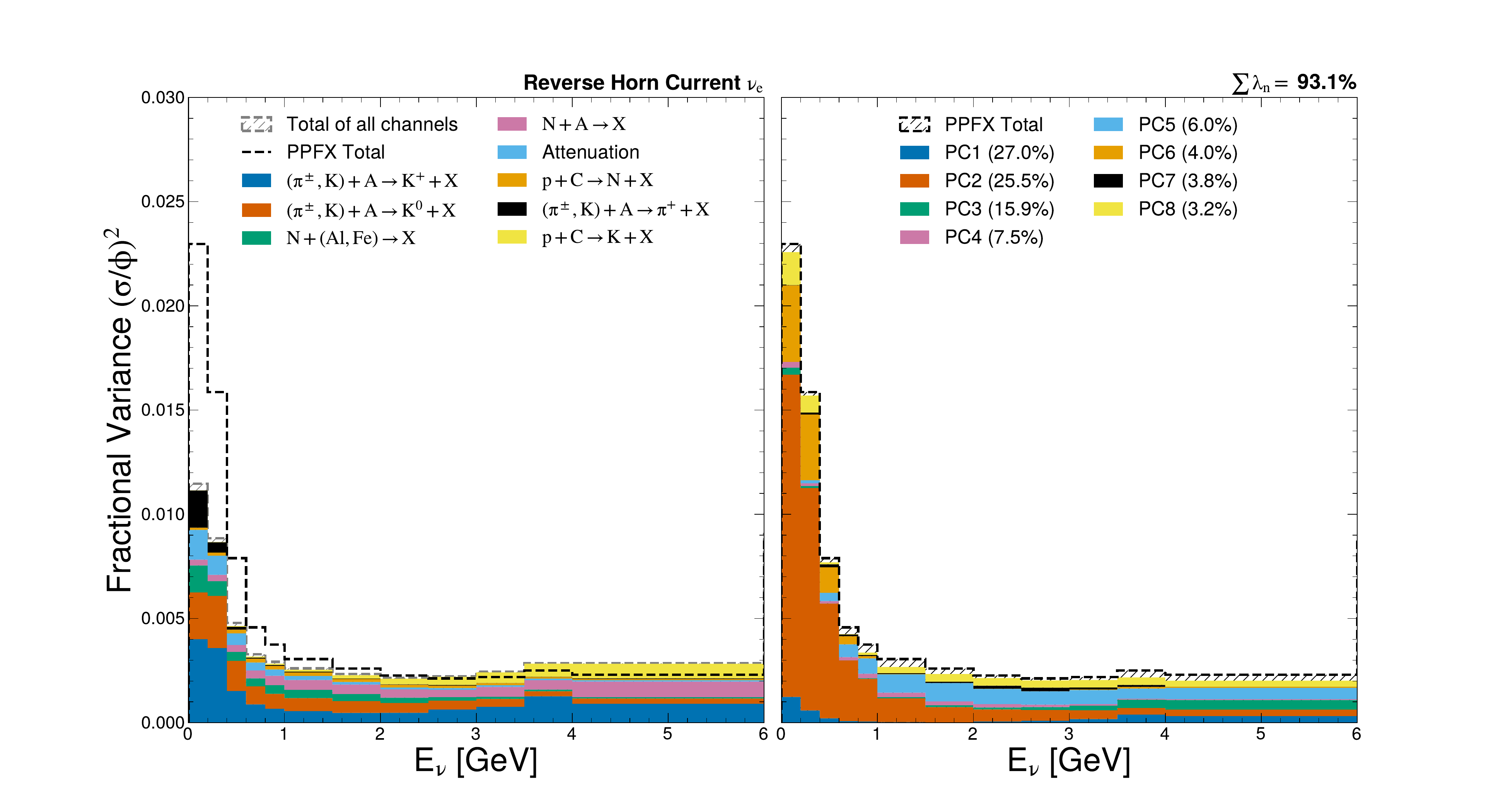}
    \caption[PCA Variance by Outgoing Meson (RHC, $\nu$)]{Fractional variance comparison between physics and PCA descriptions by outgoing meson (RHC, $\nu$).}
\end{figure}
\begin{figure}[!ht]
    \centering
    \includegraphics[width=0.98\textwidth]{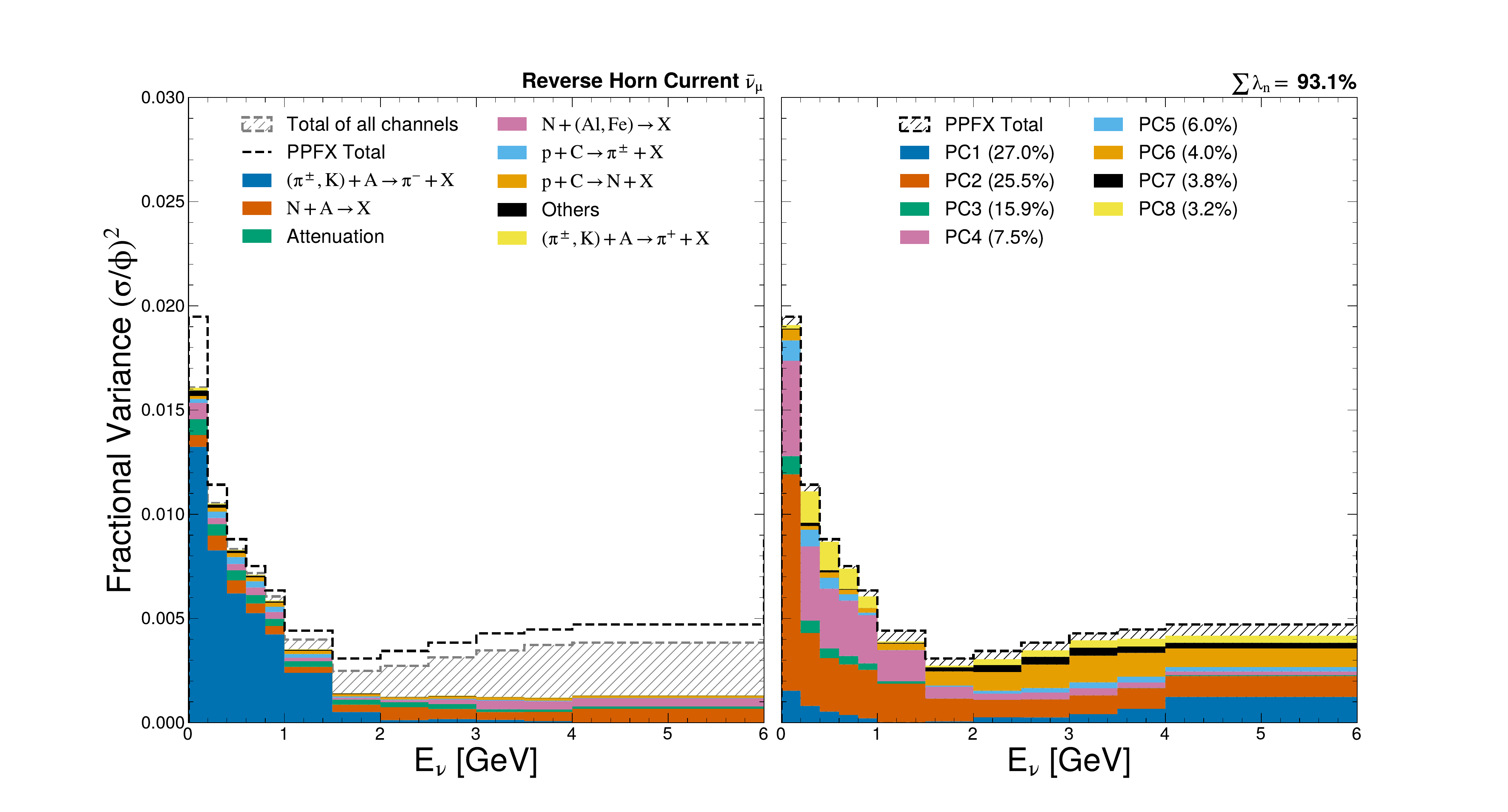}
    \includegraphics[width=0.98\textwidth]{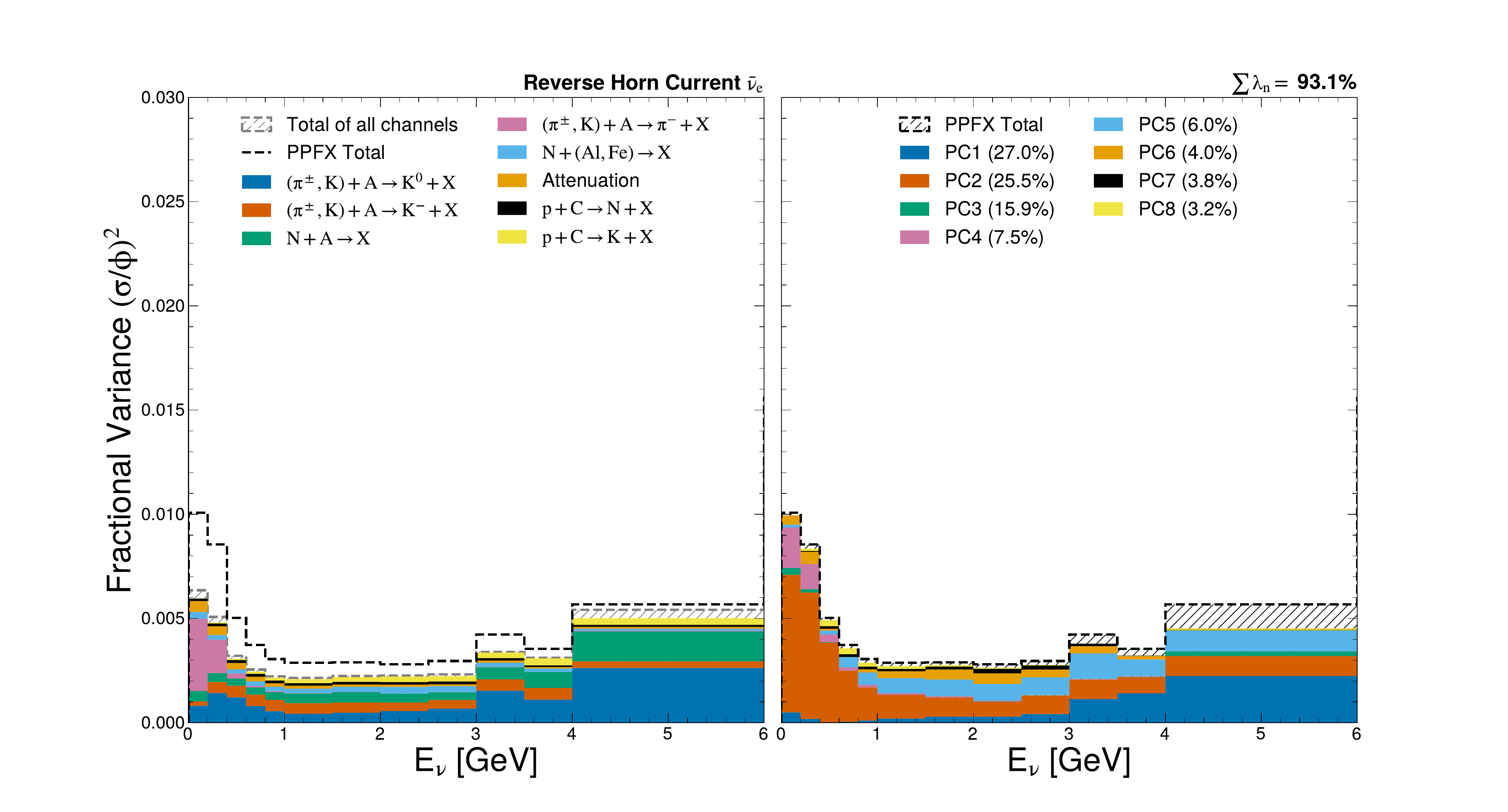}
    \caption[PCA Variance by Outgoing Meson (RHC, $\bar{\nu}$)]{Fractional variance comparison between physics and PCA descriptions by outgoing meson (RHC, $\bar{\nu}$).}
\end{figure}

%% file: beam-samples-descriptions.tex
\clearpage
\begin{table}[!ht]
    \centering
    \caption{NuMI Monte Carlo Simulation Nominal Configuration}
    \begin{tabular}{lr}
      \toprule\toprule
      \textbf{Run 15 (nominal)} & \\
      \midrule
      \textit{Beam spot size}    & \SI{1.3}{\mm} \\
      \textit{Horn current}       & \(\pm\)\SI{200}{\kA}\\
      \textit{Horn water layer}   & \SI{1}{\mm} \\
      \textit{Target z-position}  & \SI{-143.3}{\cm}\\
      \textit{Decay pipe B-field} & none \\
      \bottomrule\bottomrule
    \end{tabular}
    \label{tab:numi-nominal}
\end{table}

\begin{table}[!ht]
    \centering
    \caption[List of NuMI Beamline Focusing Samples]{List of the NuMI beamline focusing samples and their inclusion status in this analysis. Samples marked as ``excluded'' were consistent with statistical fluctuations present in the nominal sample.}
    \begin{tabular}{l l r}
        \toprule\toprule\\
        Run ID & Description & Inclusion Status \\
        \midrule
        8  & horn current $+\SI{2}{\kA}$ & included \\
        9  & horn current $-\SI{2}{\kA}$ & excluded \\
        10 & horn1 position $x+\SI{0.3}{\cm}$ & included \\
        11 & horn1 position $x-\SI{0.3}{\cm}$ & included \\
        12 & horn1 position $y+\SI{0.3}{\cm}$ & included \\
        13 & horn1 position $y-\SI{0.3}{\cm}$ & included \\
        14 & beam spot size $+\SI{0.2}{\cm}$ & included \\
        15 & nominal & included \\
        16 & beam spot size $-\SI{0.2}{\cm}$ & included \\
        17 & horn2 position $x+\SI{0.3}{\cm}$ & excluded \\
        18 & horn2 position $x-\SI{0.3}{\cm}$ & excluded \\
        19 & horn2 position $y+\SI{0.3}{\cm}$ & excluded \\
        20 & horn2 position $y-\SI{0.3}{\cm}$ & excluded \\
        21 & horn water layer $+\SI{1}{\mm}$ & included $\left(E_\nu \leq \SI{1}{\GeV}\right)$ \\
        22 & horn water layer $-\SI{1}{\mm}$ & included $\left(E_\nu \leq \SI{1}{\GeV}\right)$ \\
        24 & beam shift $x +\SI{1}{\mm}$ & included \\
        25 & beam shift $x -\SI{1}{\mm}$ & included \\
        26 & beam shift $y +\SI{1}{\mm}$ & included \\
        27 & beam shift $y -\SI{1}{\mm}$ & included \\
        28 & target position $z +\SI{7}{\mm}$ & excluded \\
        29 & target position $z -\SI{7}{\mm}$ & excluded \\
        30 & B-field in decay pipe & excluded \\
        32 & beam divergence $\SI{54}{\micro\radian}$ & included $\left(E_\nu \geq \SI{1}{\GeV}\right)$ \\
        \bottomrule\bottomrule
    \end{tabular}
    \label{tab:beam-runs}
\end{table}

%% file: beam_run_shifts.tex
\clearpage
\section{Forward Horn Current}
\begin{figure}[!ht]
    \centering
    \includegraphics[width=0.25\textwidth]{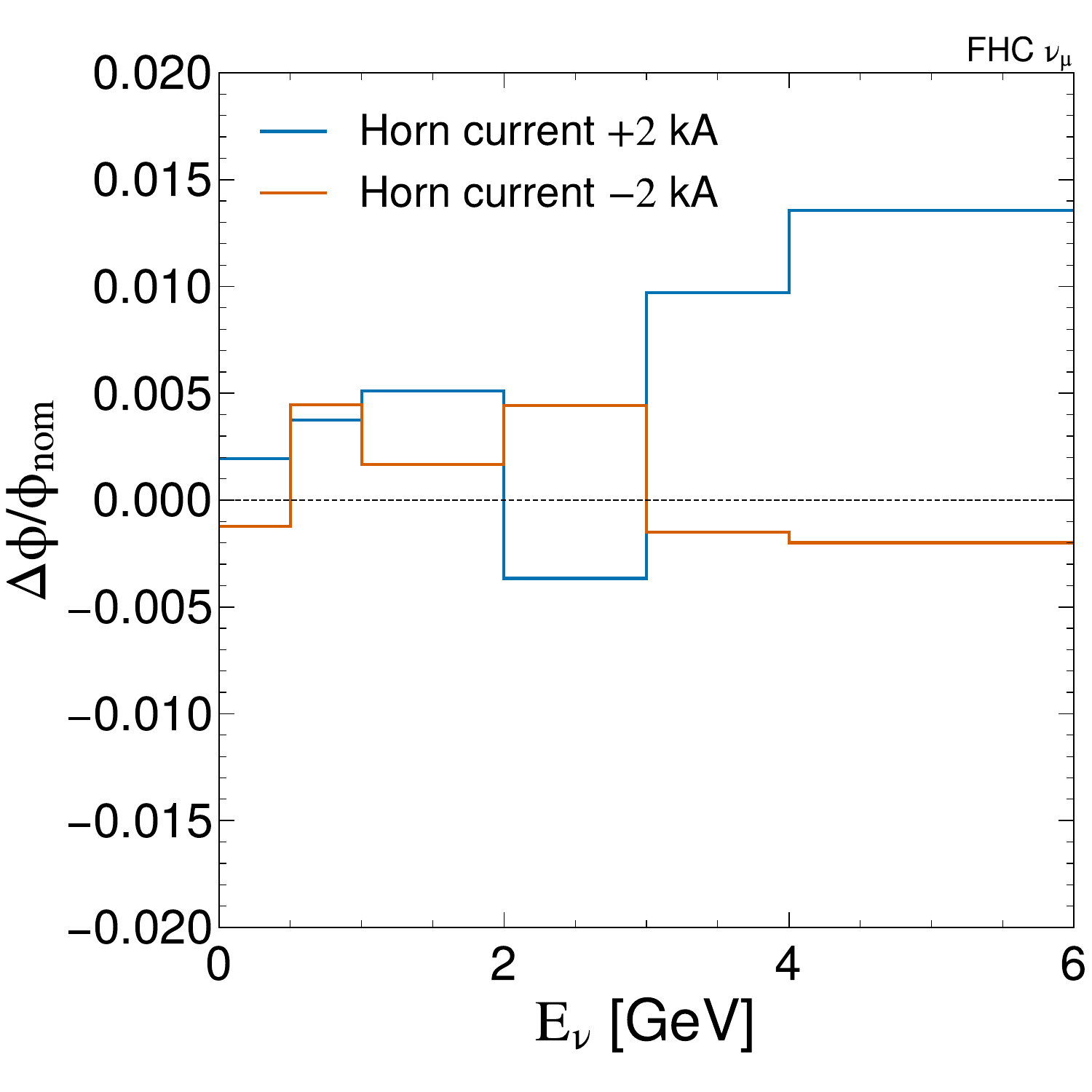}
    \includegraphics[width=0.25\textwidth]{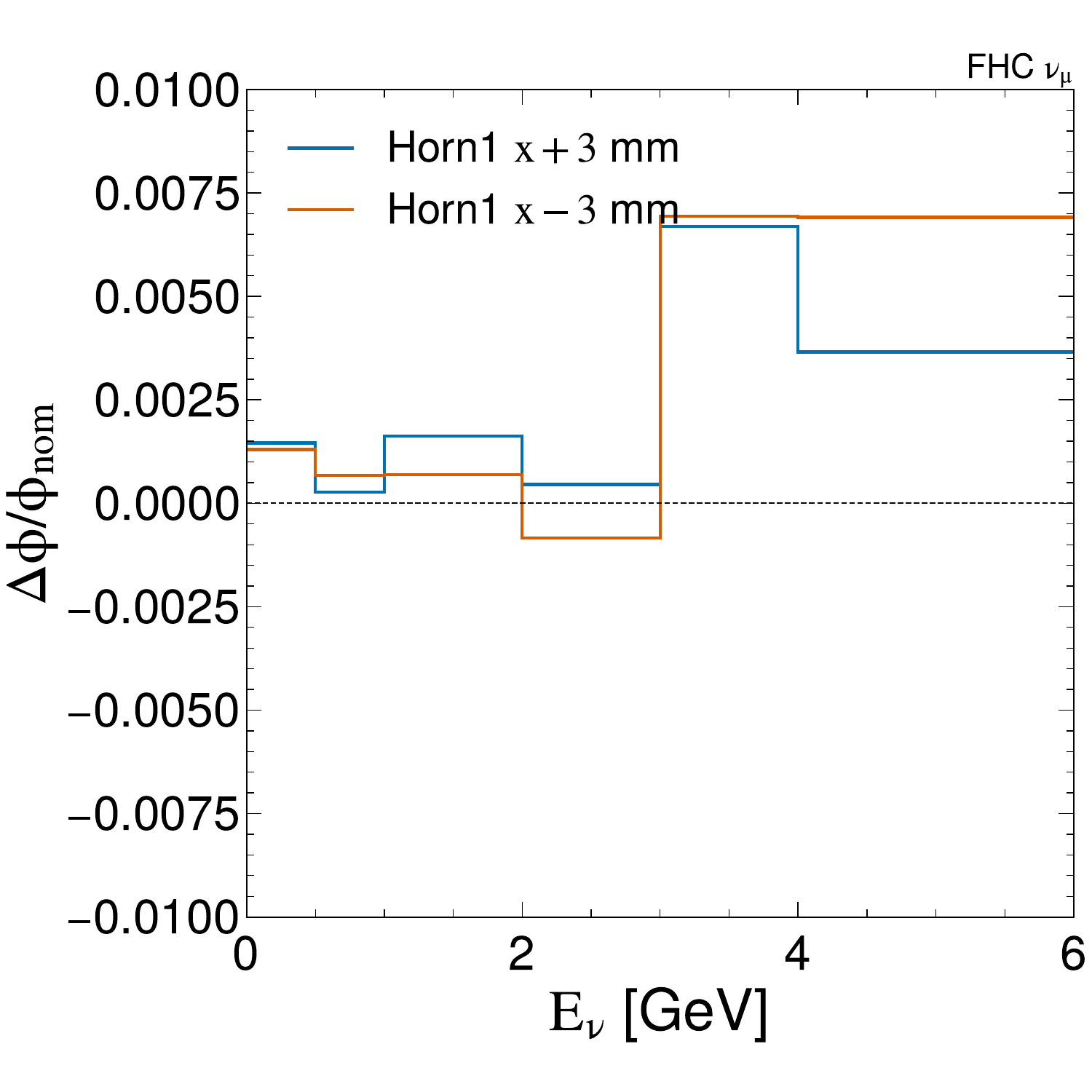}
    \includegraphics[width=0.25\textwidth]{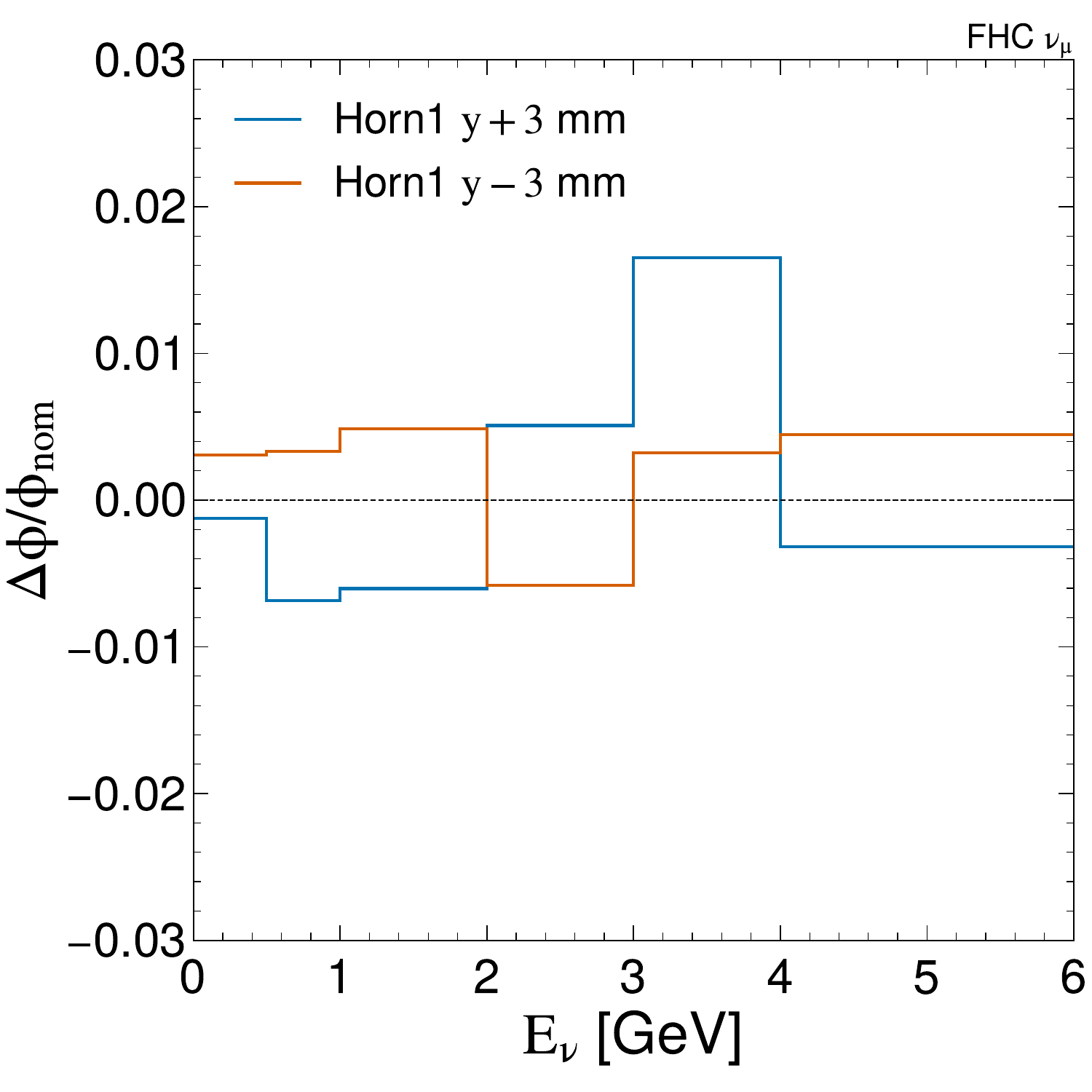}\\
    \includegraphics[width=0.25\textwidth]{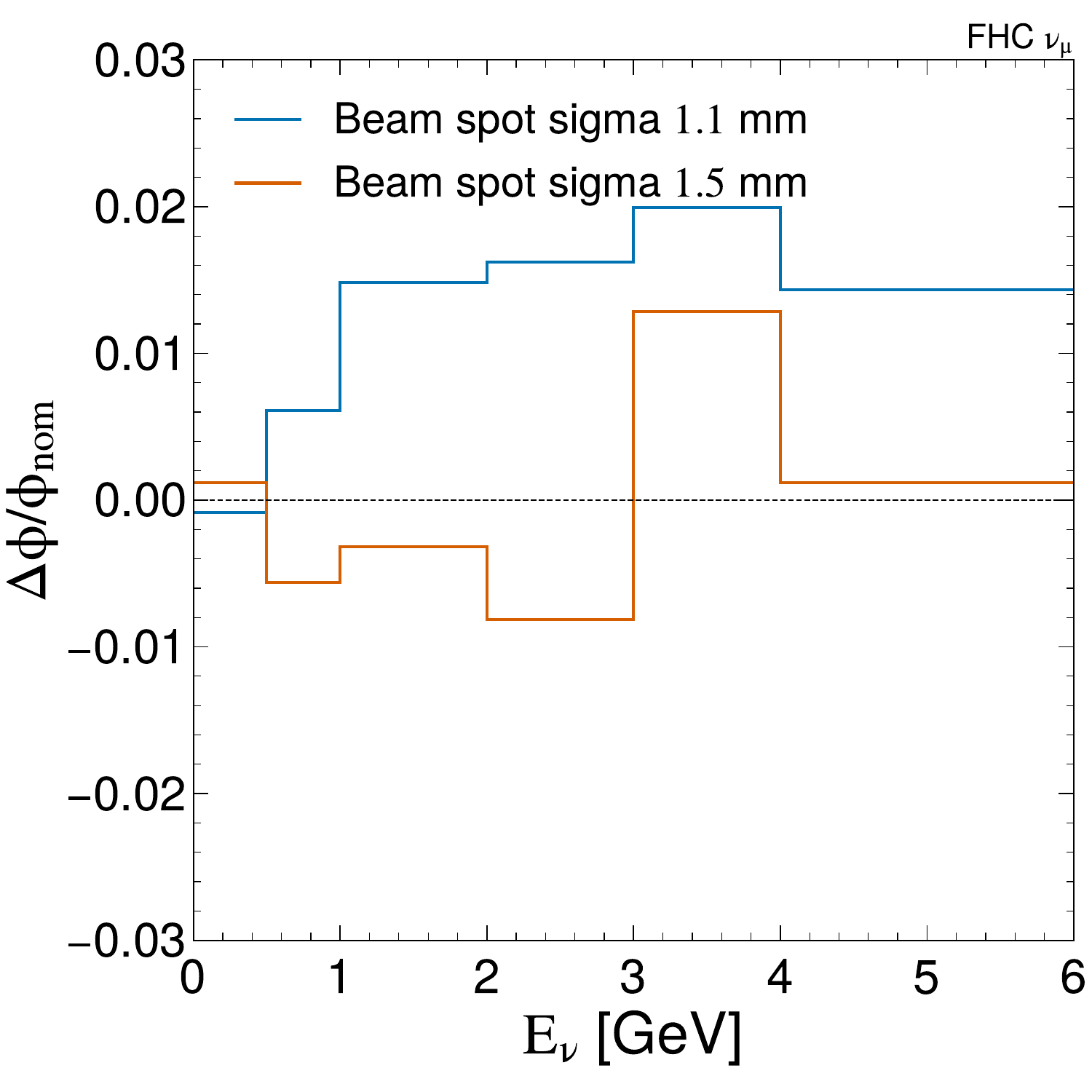}
    \includegraphics[width=0.25\textwidth]{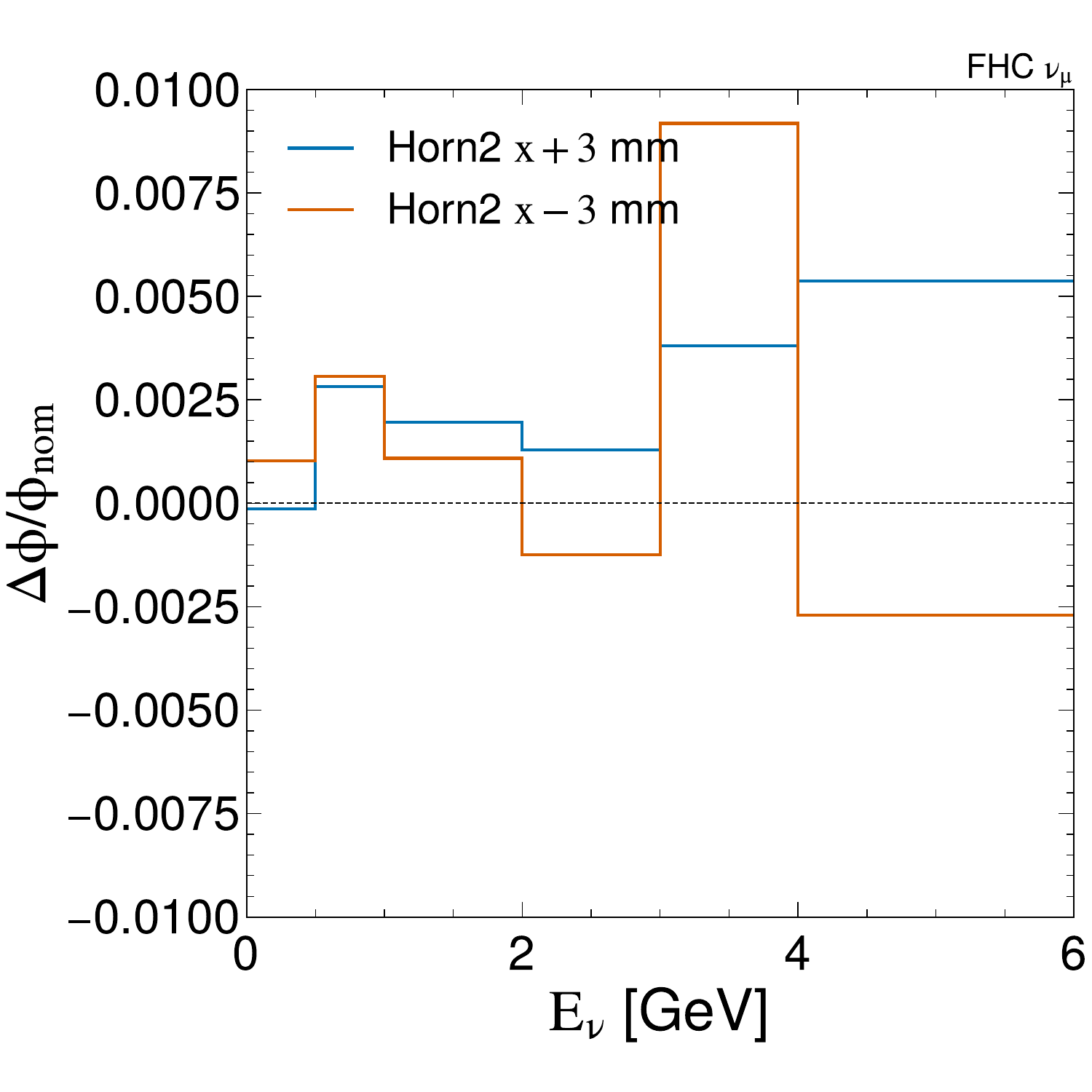}
    \includegraphics[width=0.25\textwidth]{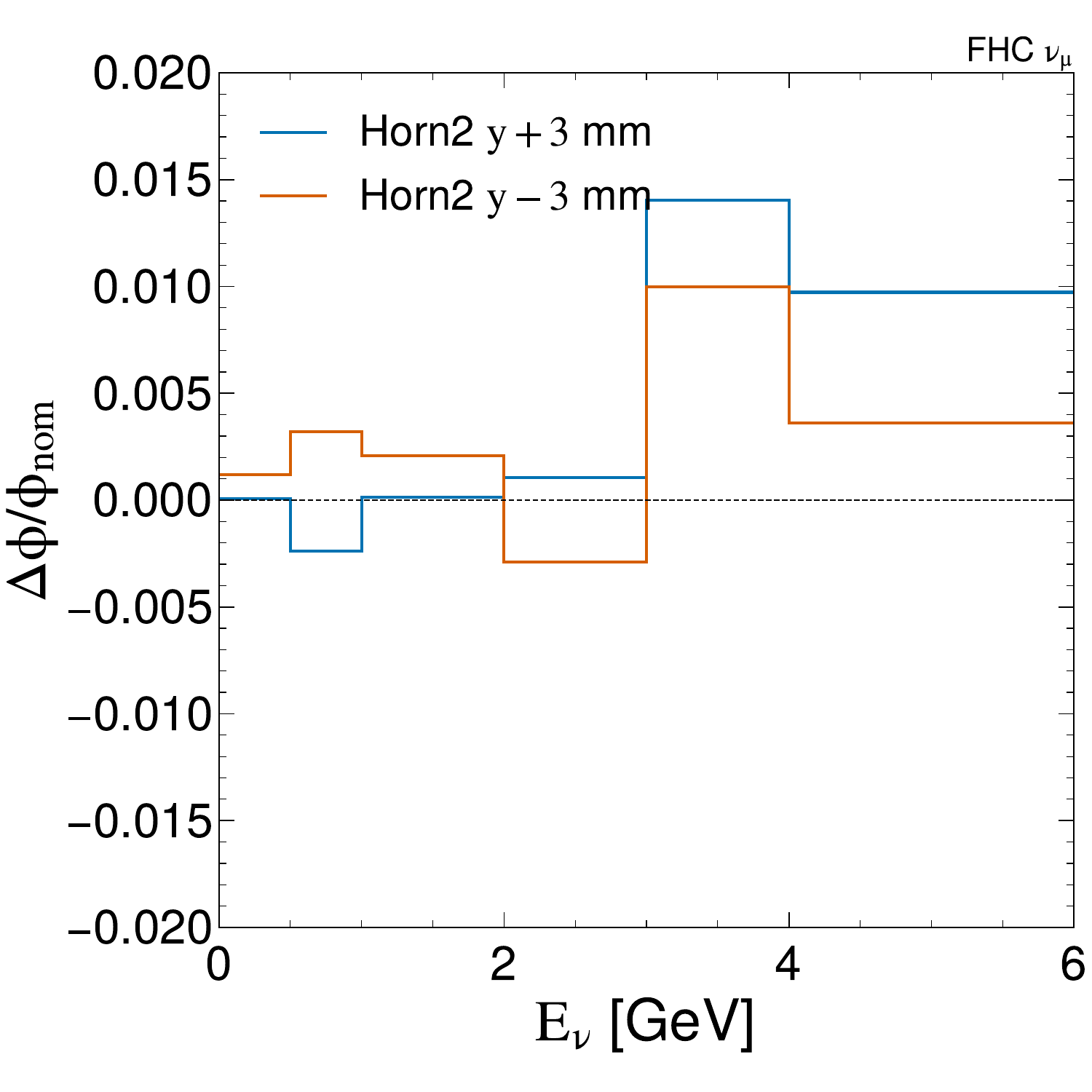}\\
    \includegraphics[width=0.25\textwidth]{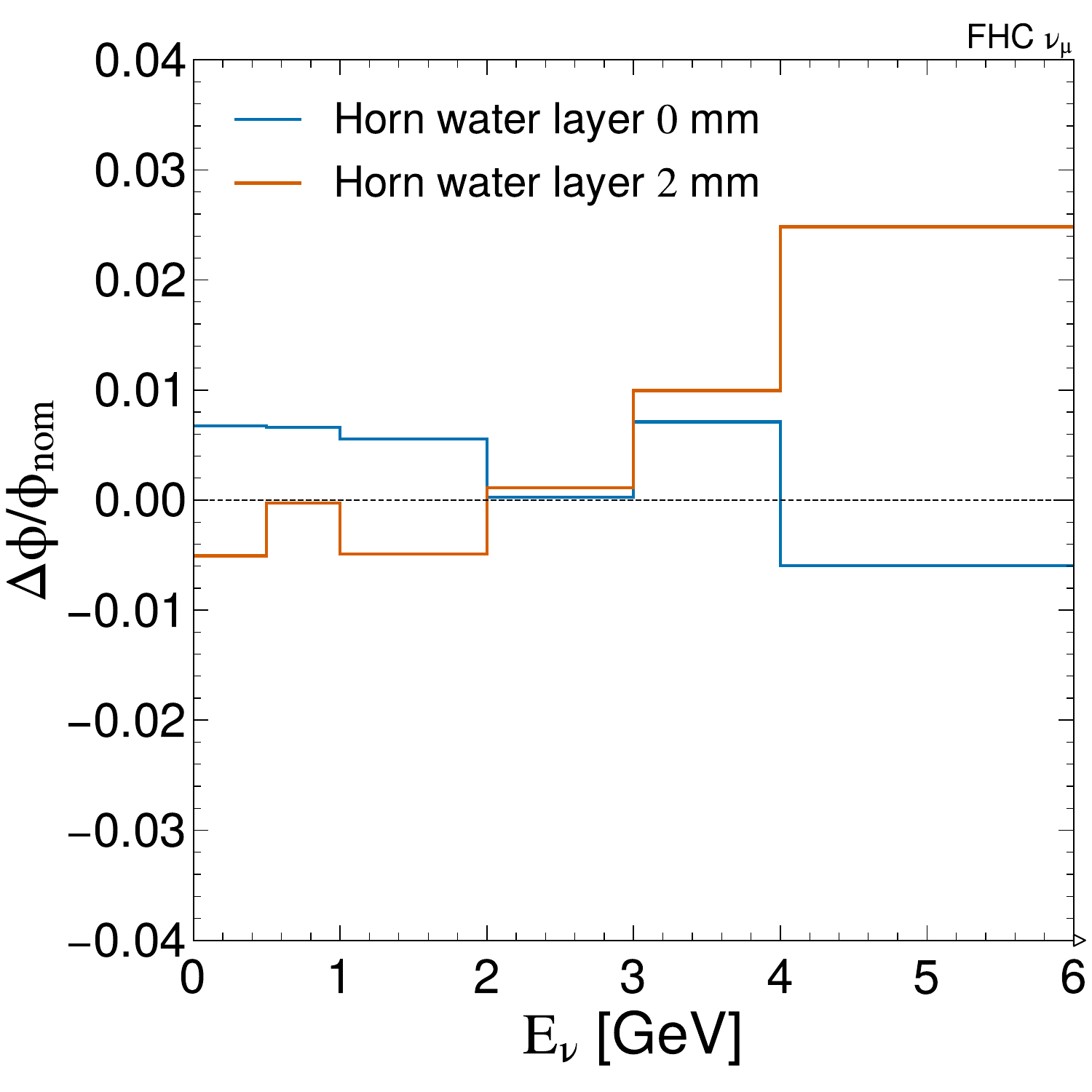}
    \includegraphics[width=0.25\textwidth]{fhc_numu_run15_run16.pdf}
    \includegraphics[width=0.25\textwidth]{fhc_numu_run17_run18.pdf}\\
    \includegraphics[width=0.25\textwidth]{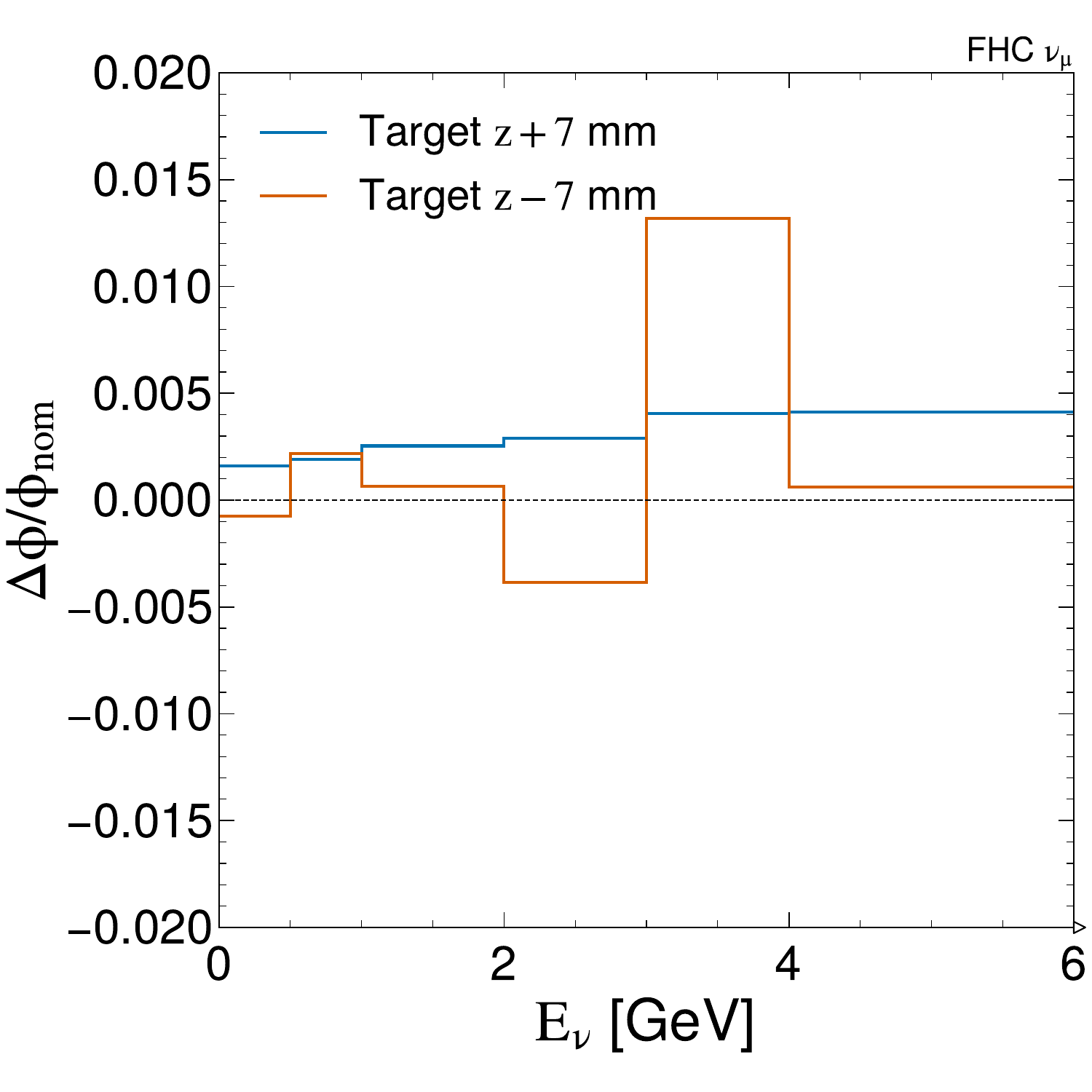}
    \caption[Beam Focusing Systematic Shifts (FHC, \numu)]{Beam focusing systematic shifts in the fractional scale (FHC, \numu).}
\end{figure}
\begin{figure}[!ht]
    \centering
    \includegraphics[width=0.25\textwidth]{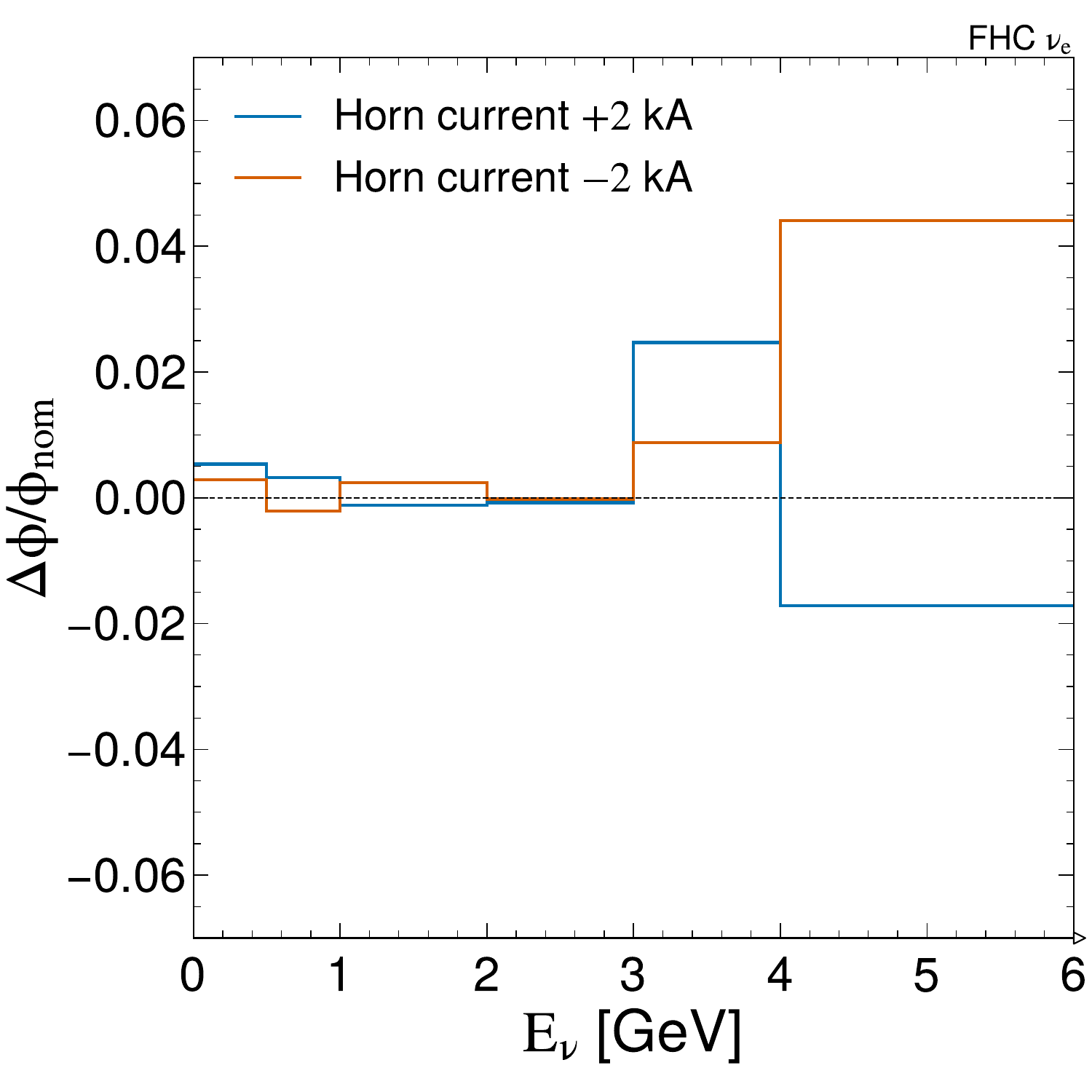}
    \includegraphics[width=0.25\textwidth]{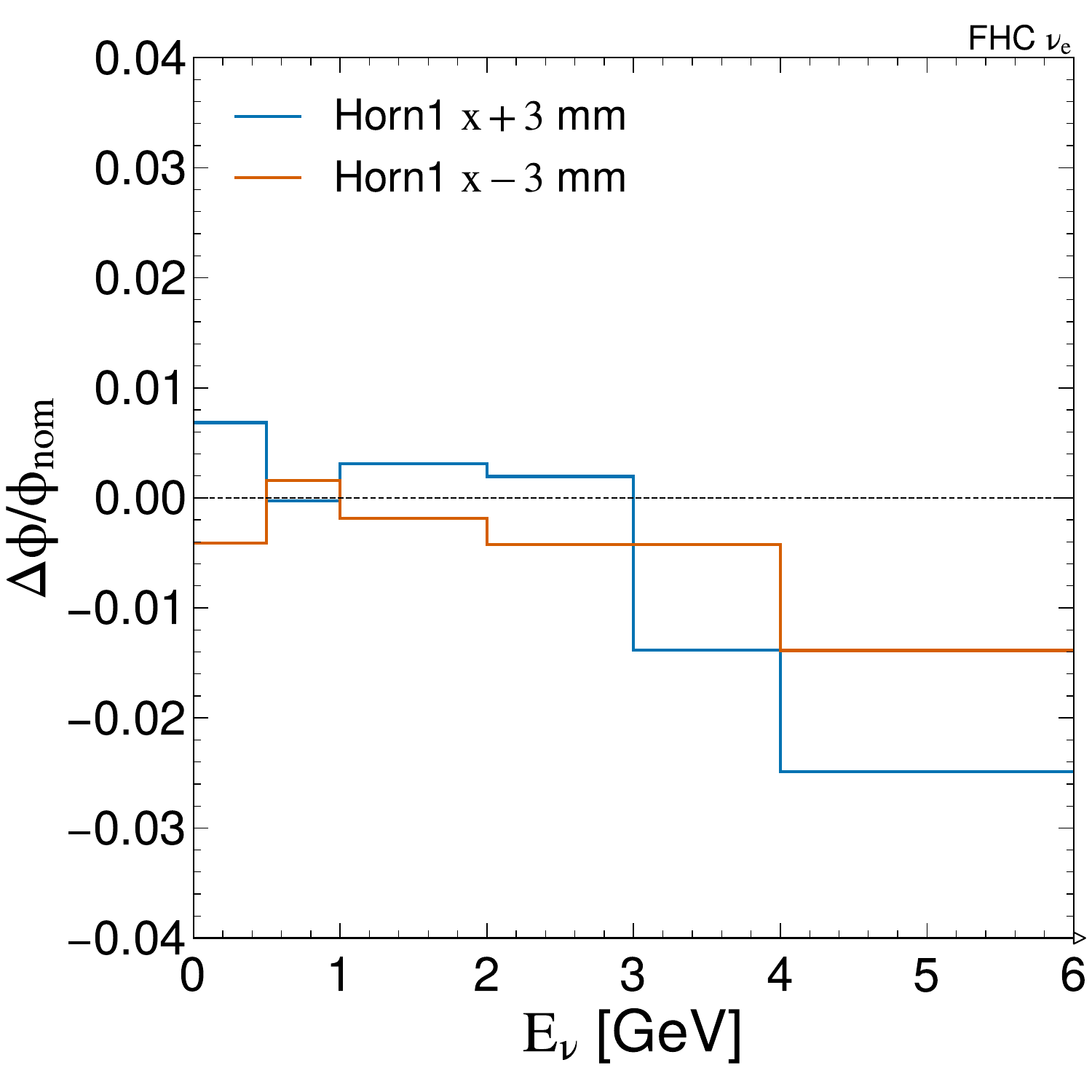}
    \includegraphics[width=0.25\textwidth]{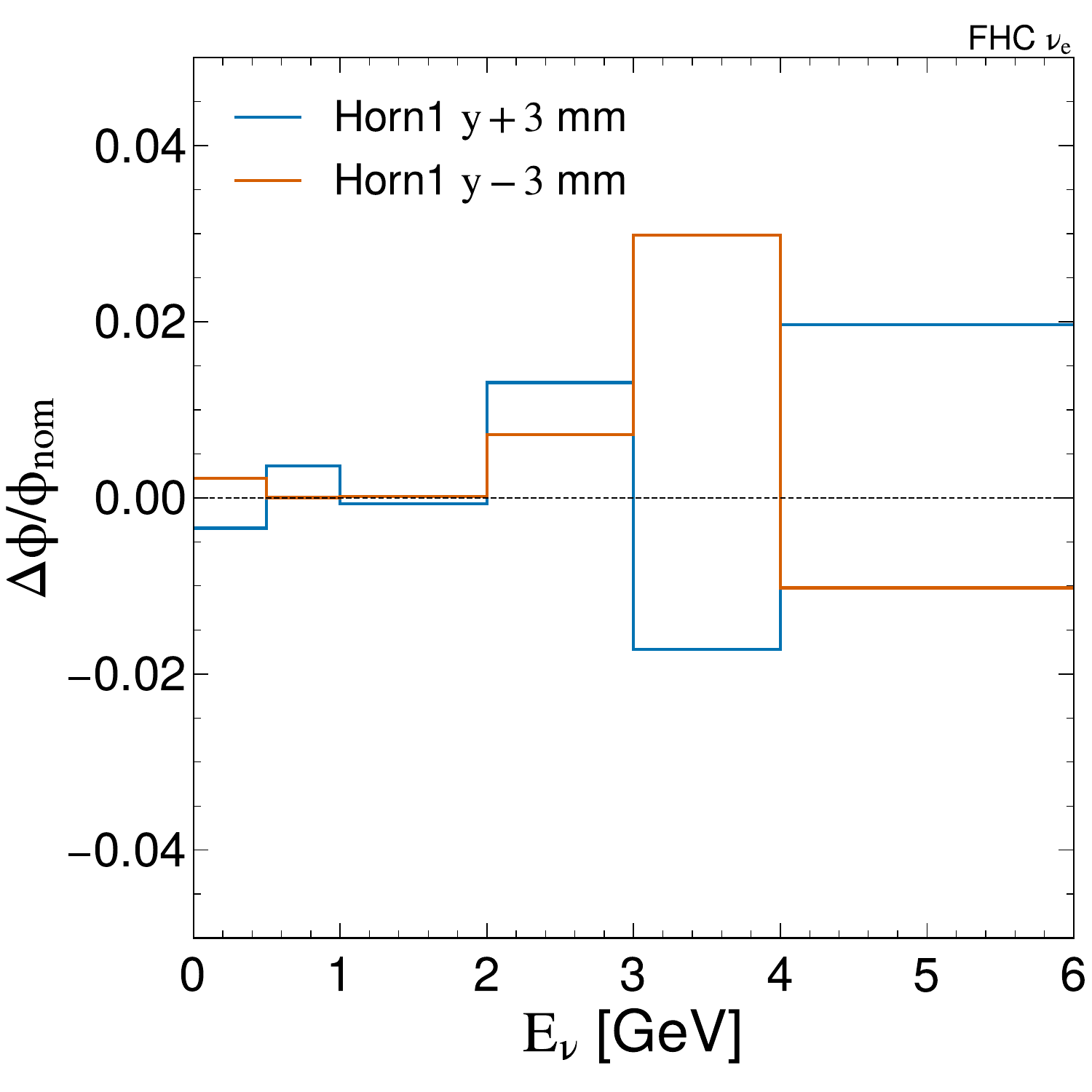}\\
    \includegraphics[width=0.25\textwidth]{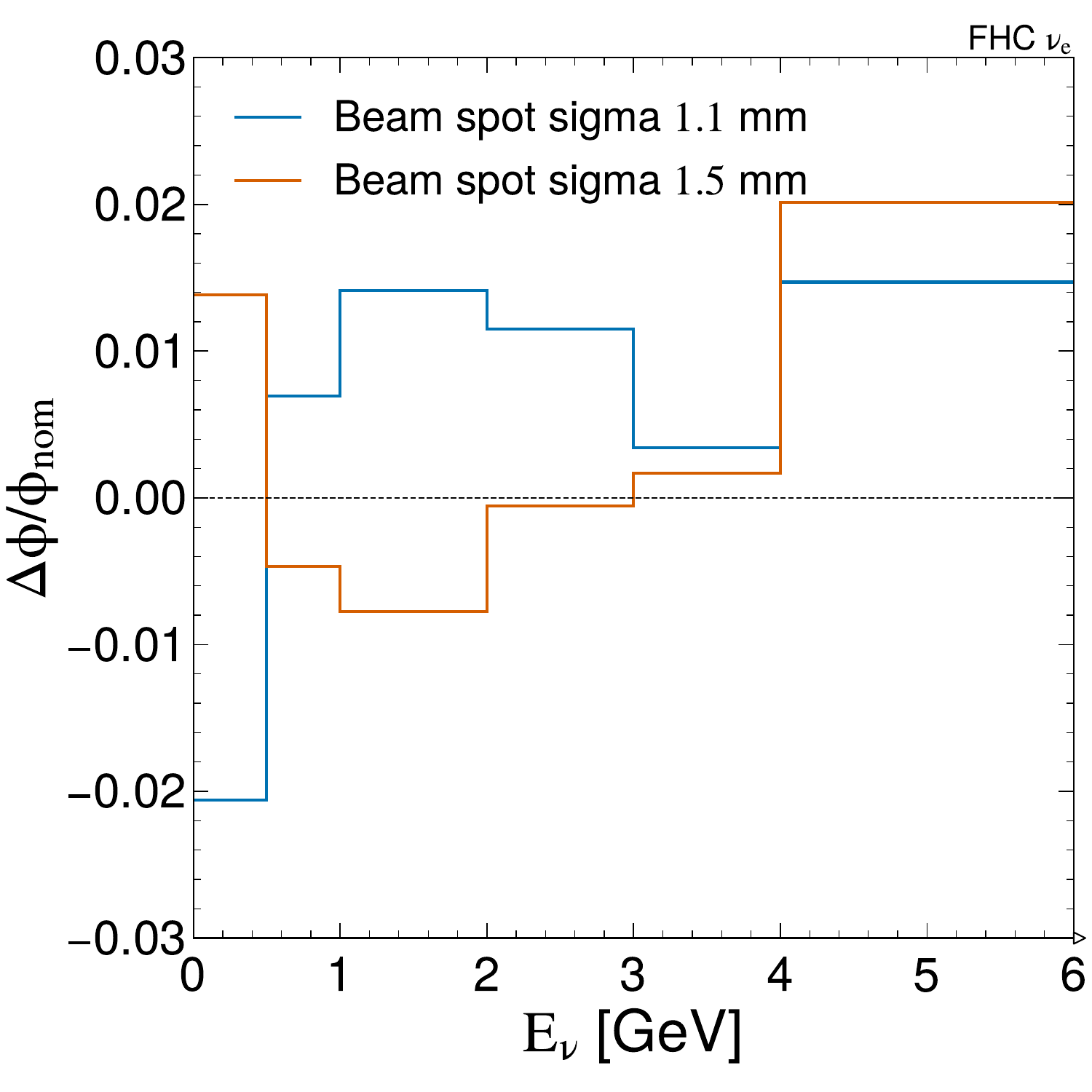}
    \includegraphics[width=0.25\textwidth]{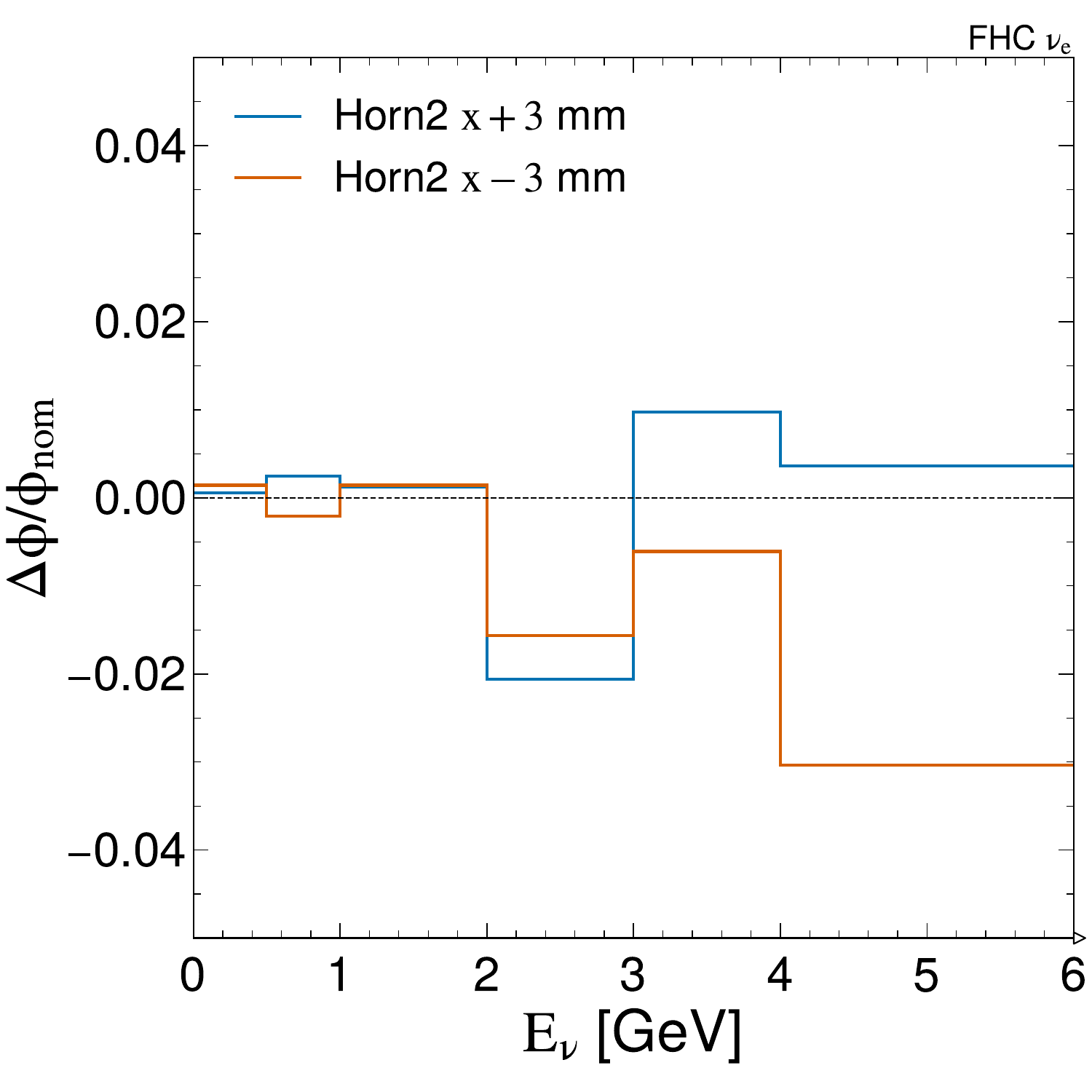}
    \includegraphics[width=0.25\textwidth]{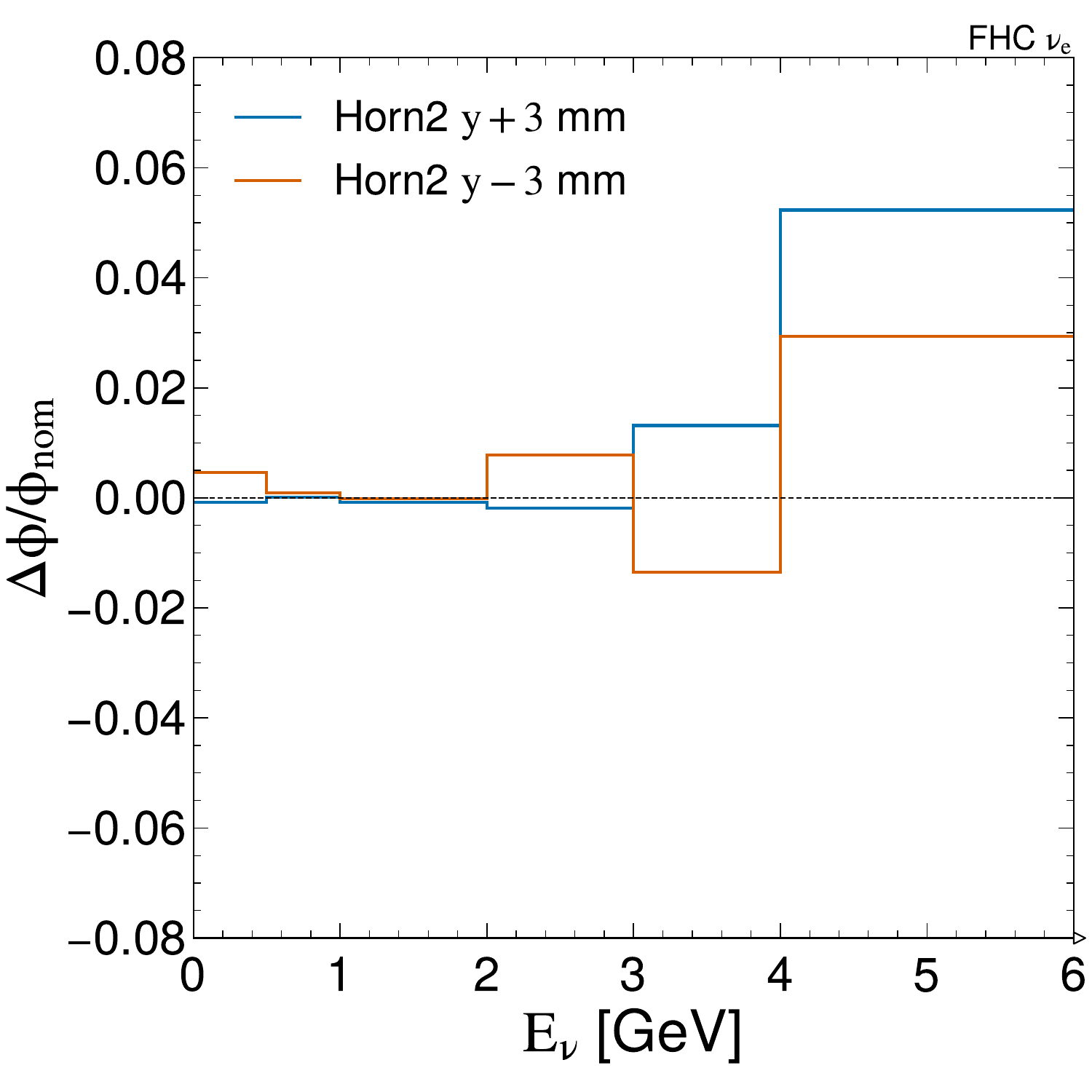}\\
    \includegraphics[width=0.25\textwidth]{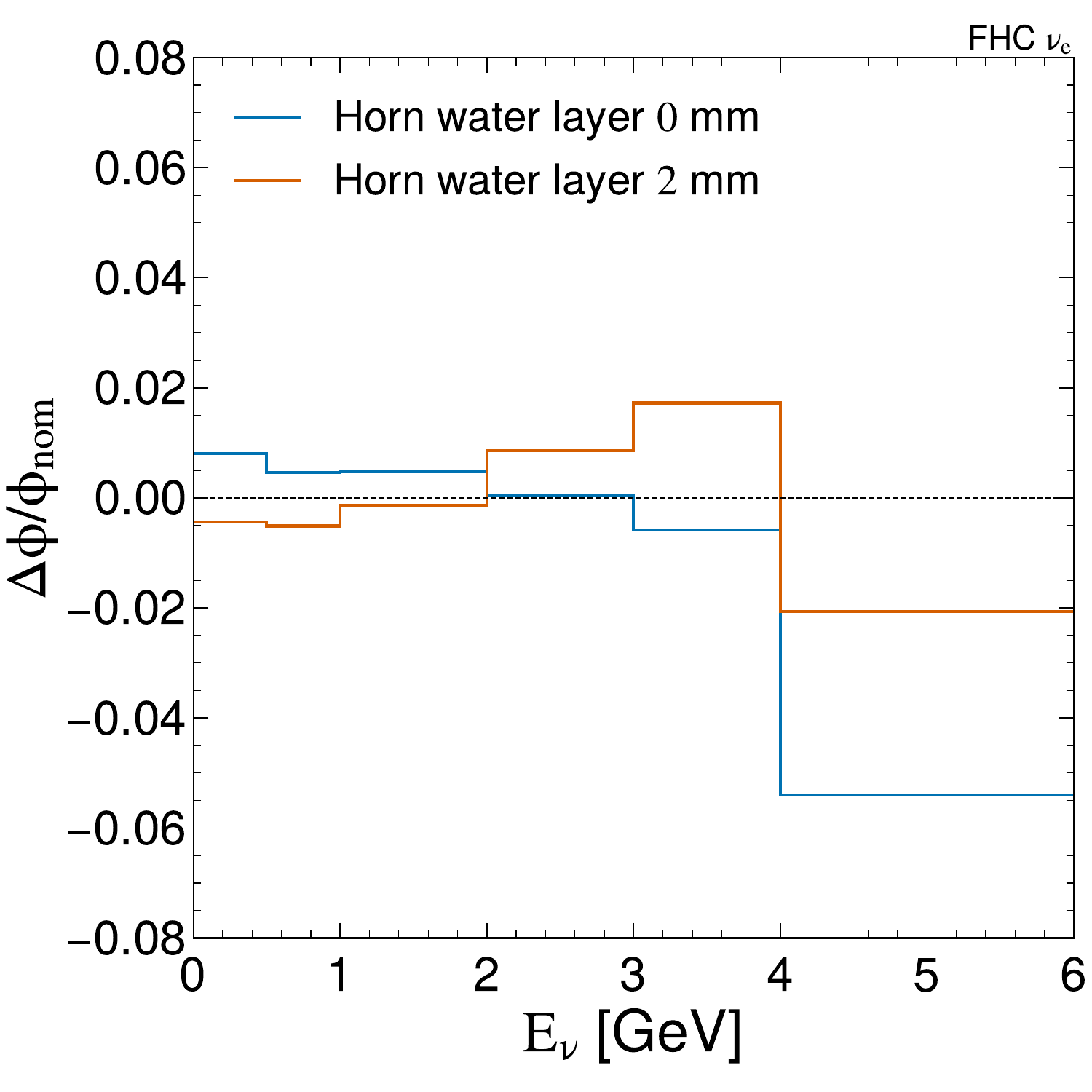}
    \includegraphics[width=0.25\textwidth]{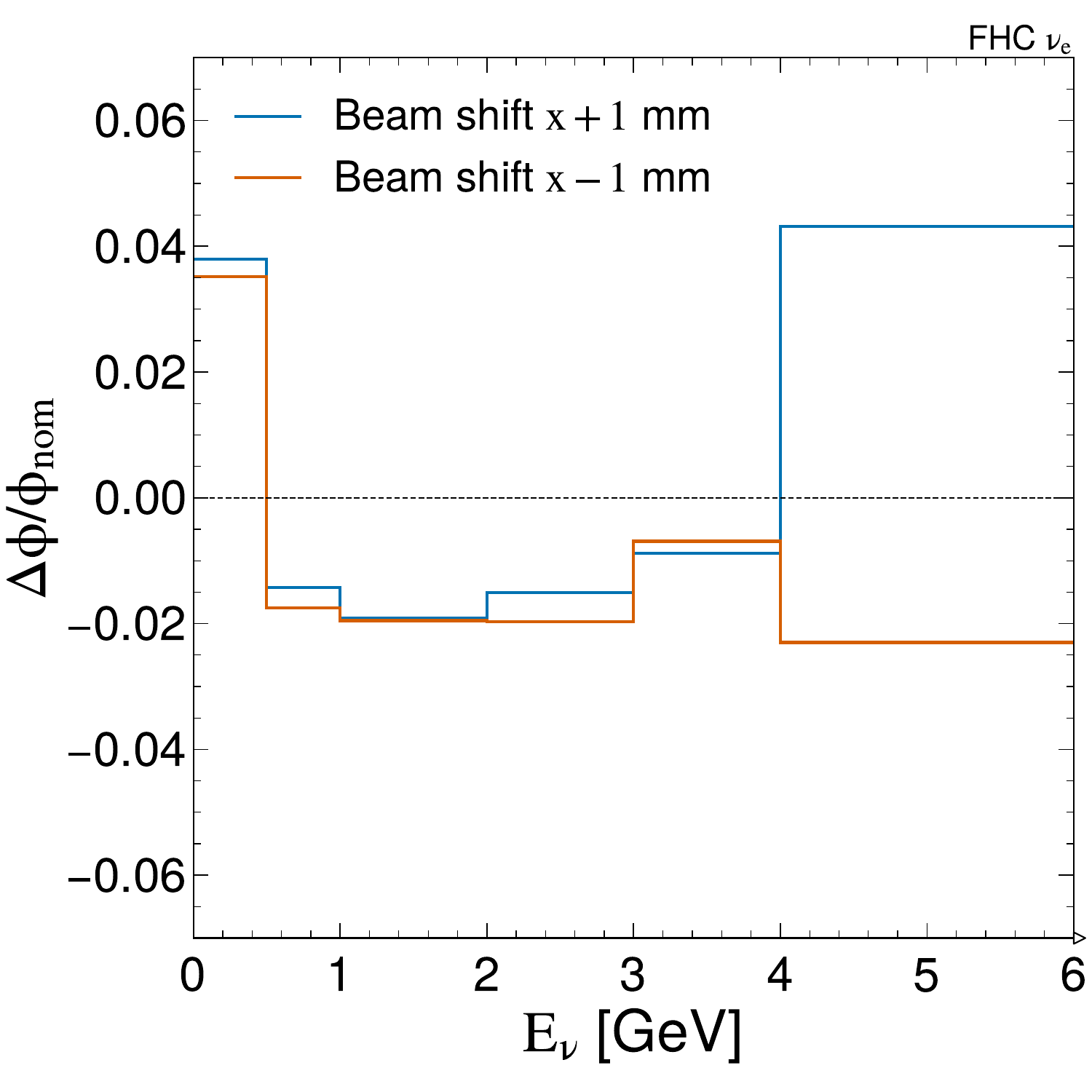}
    \includegraphics[width=0.25\textwidth]{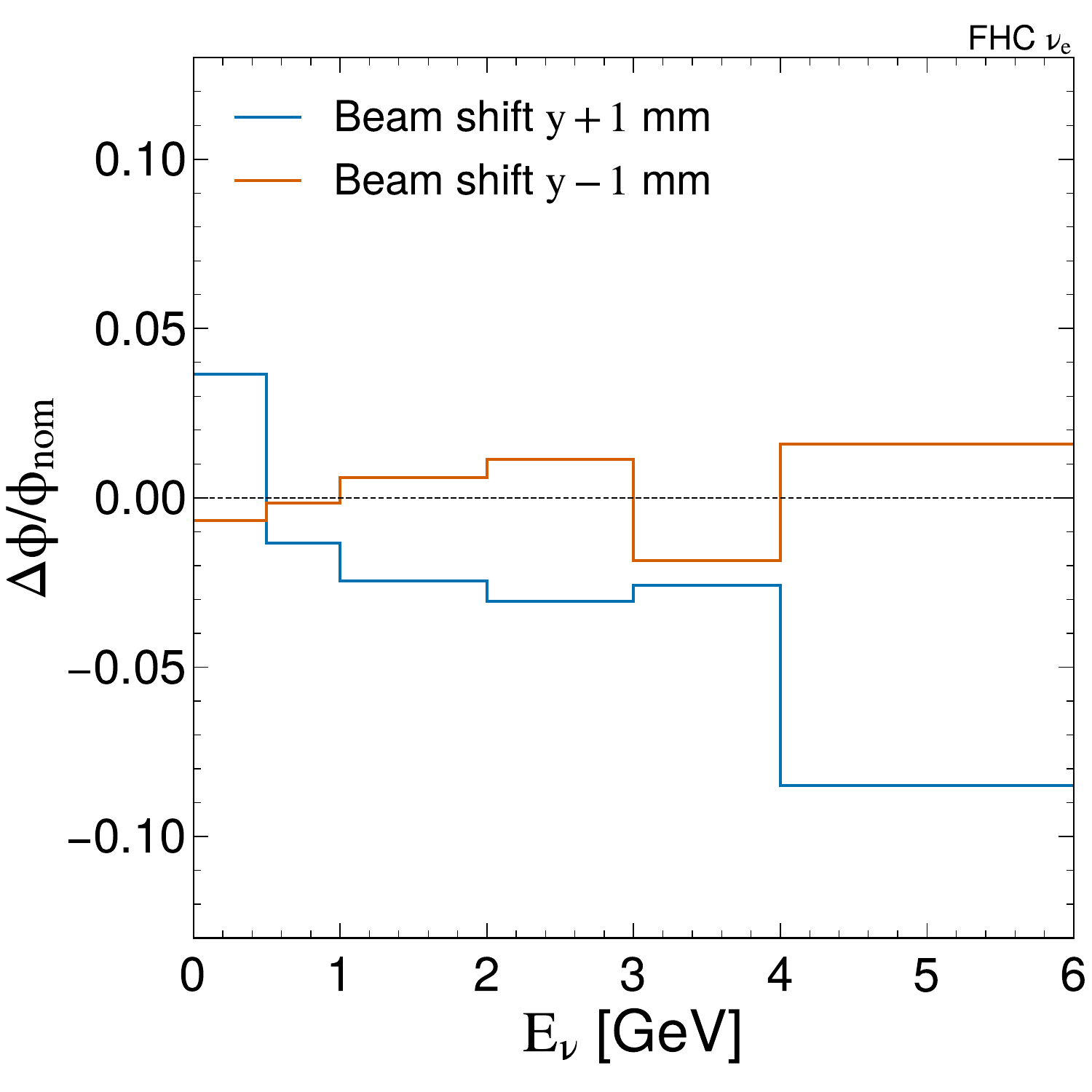}\\
    \includegraphics[width=0.25\textwidth]{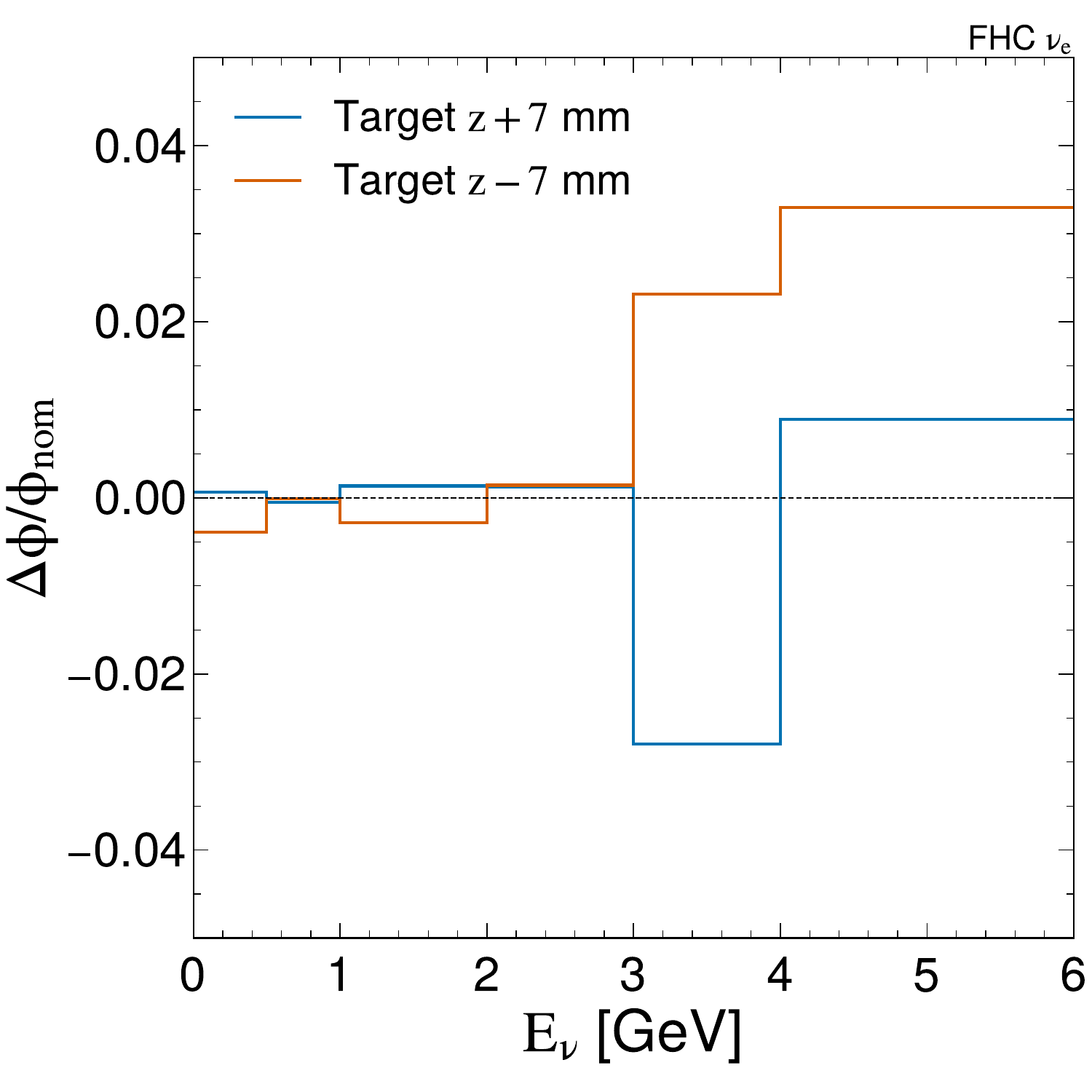}
    \caption[Beam Focusing Systematic Shifts (FHC, \nue)]{Beam focusing systematic shifts in the fractional scale (FHC, \nue).}
\end{figure}
\begin{figure}[!ht]
    \centering
    \includegraphics[width=0.25\textwidth]{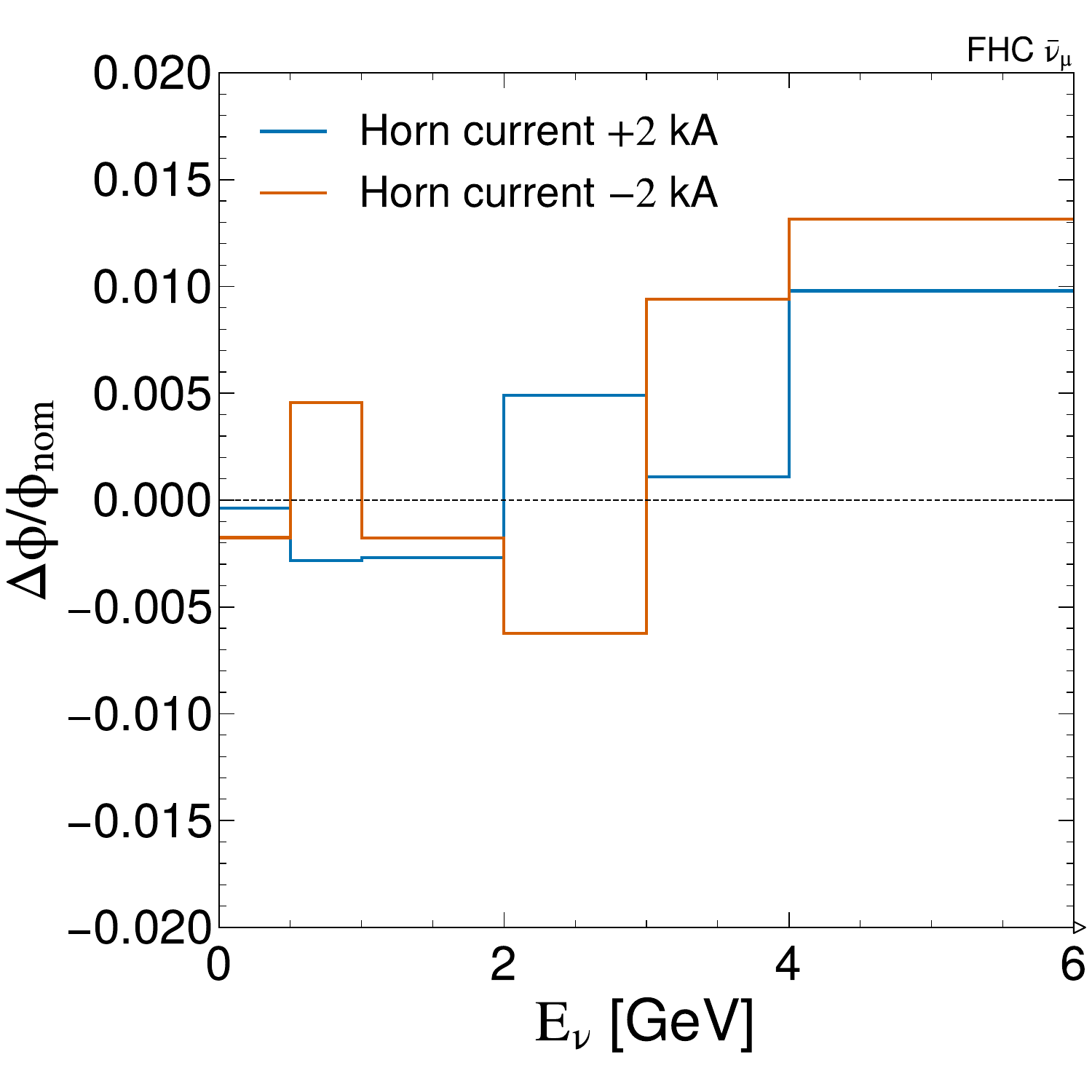}
    \includegraphics[width=0.25\textwidth]{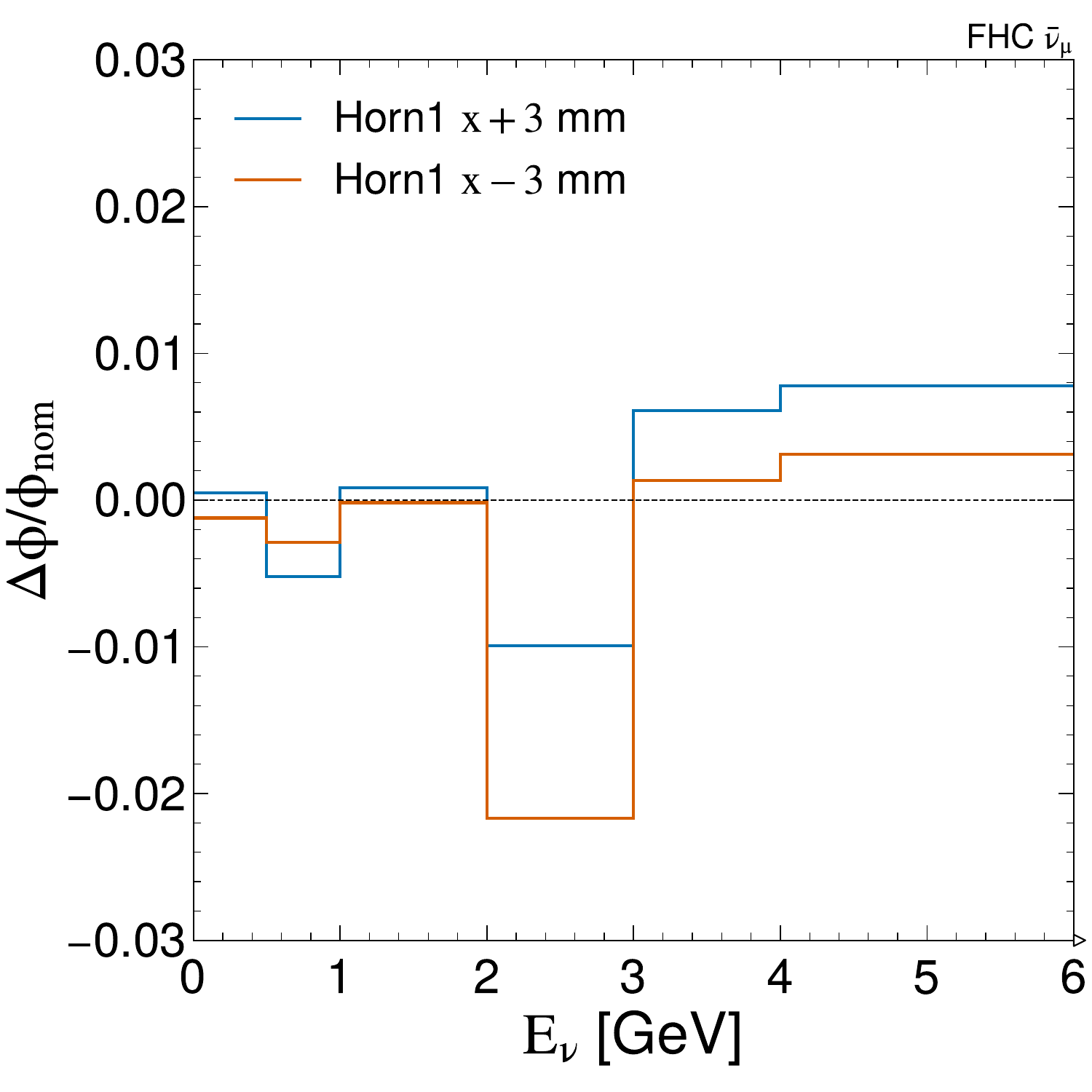}
    \includegraphics[width=0.25\textwidth]{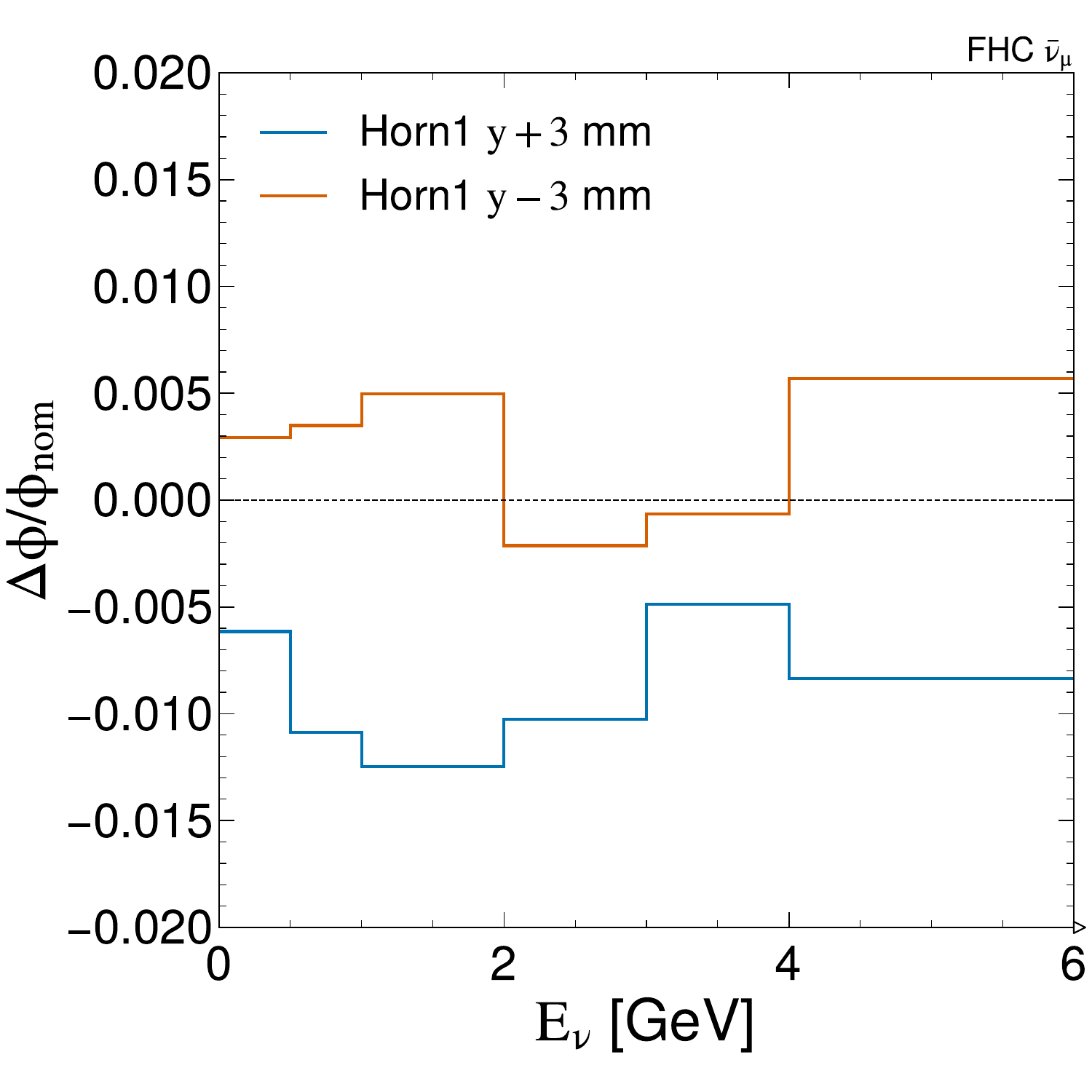}\\
    \includegraphics[width=0.25\textwidth]{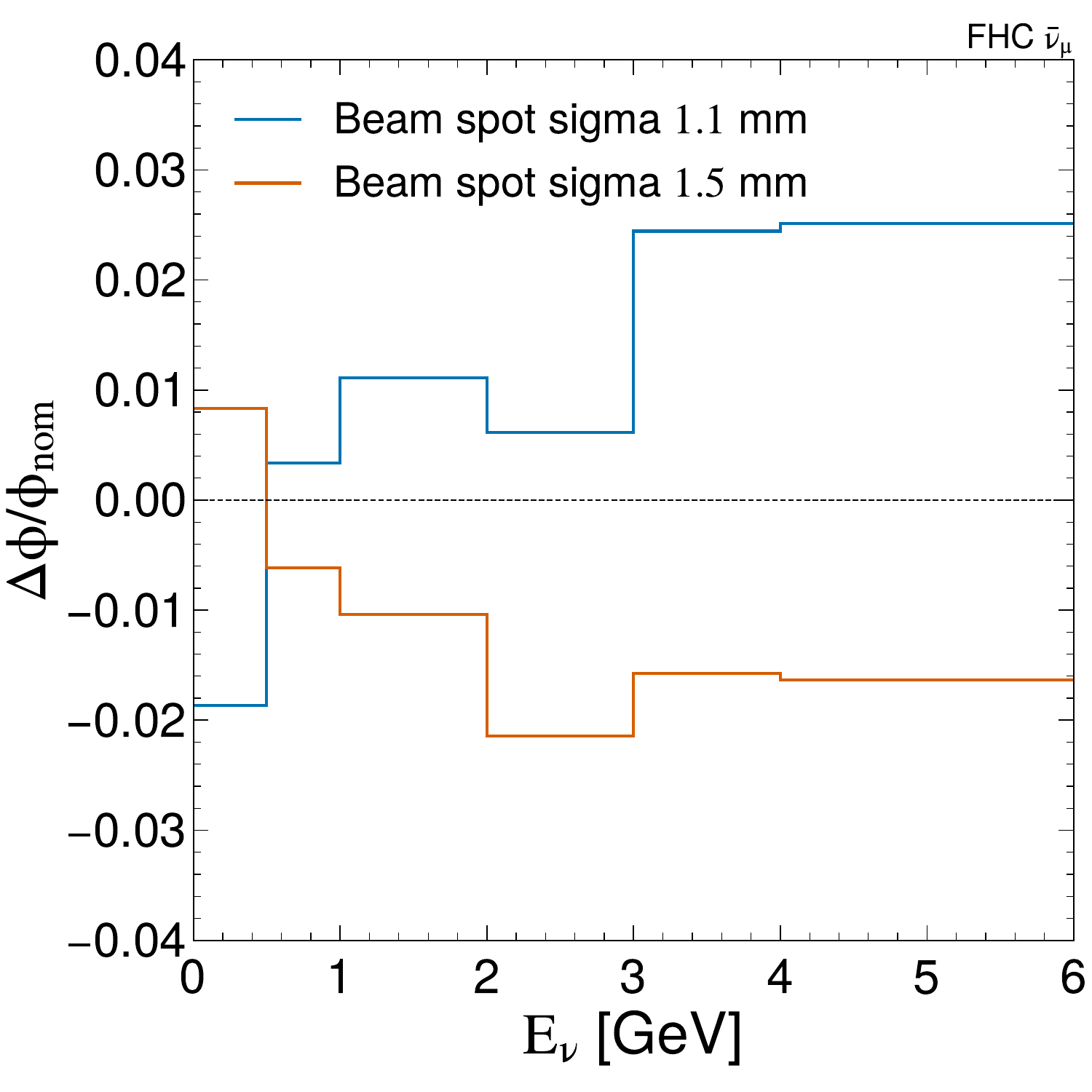}
    \includegraphics[width=0.25\textwidth]{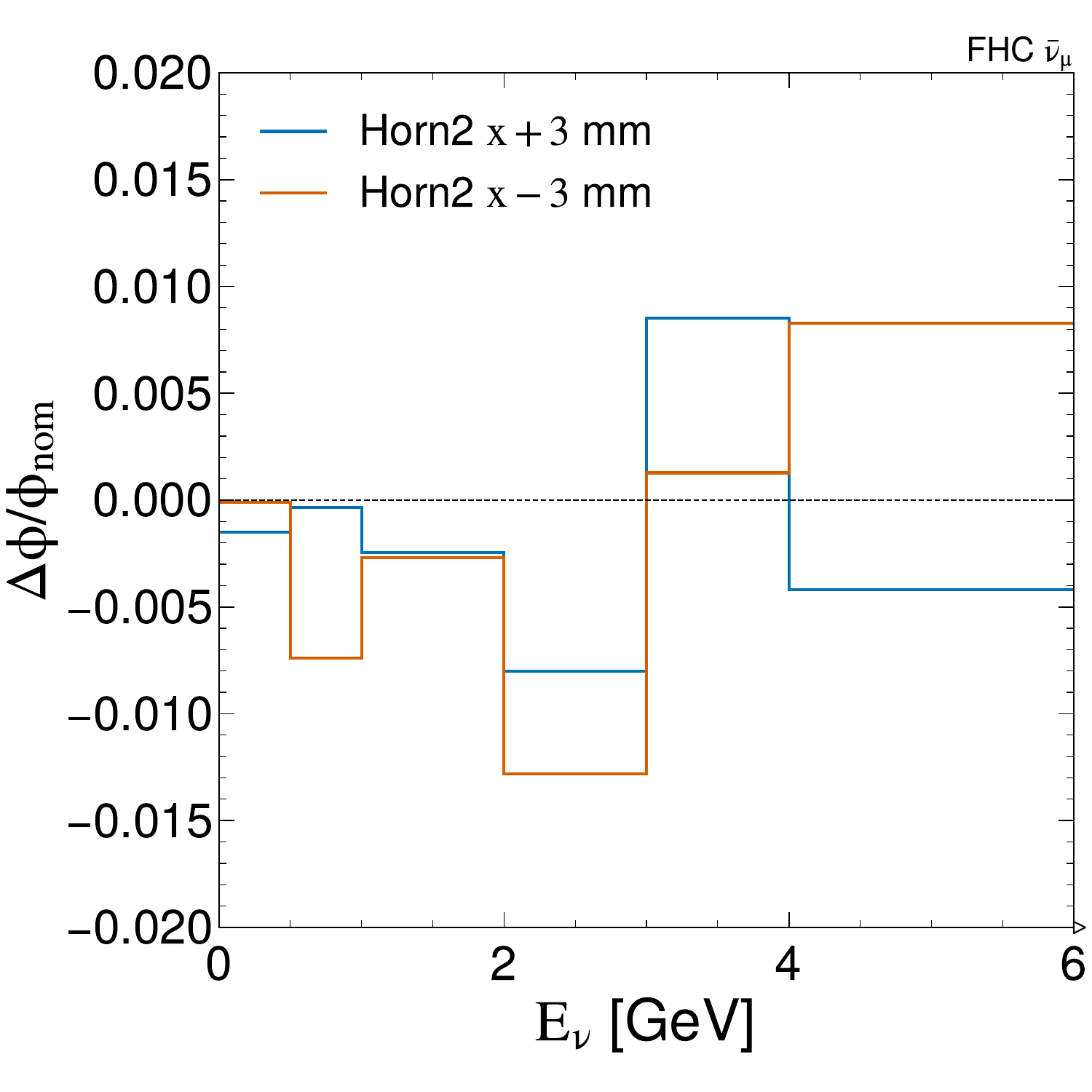}
    \includegraphics[width=0.25\textwidth]{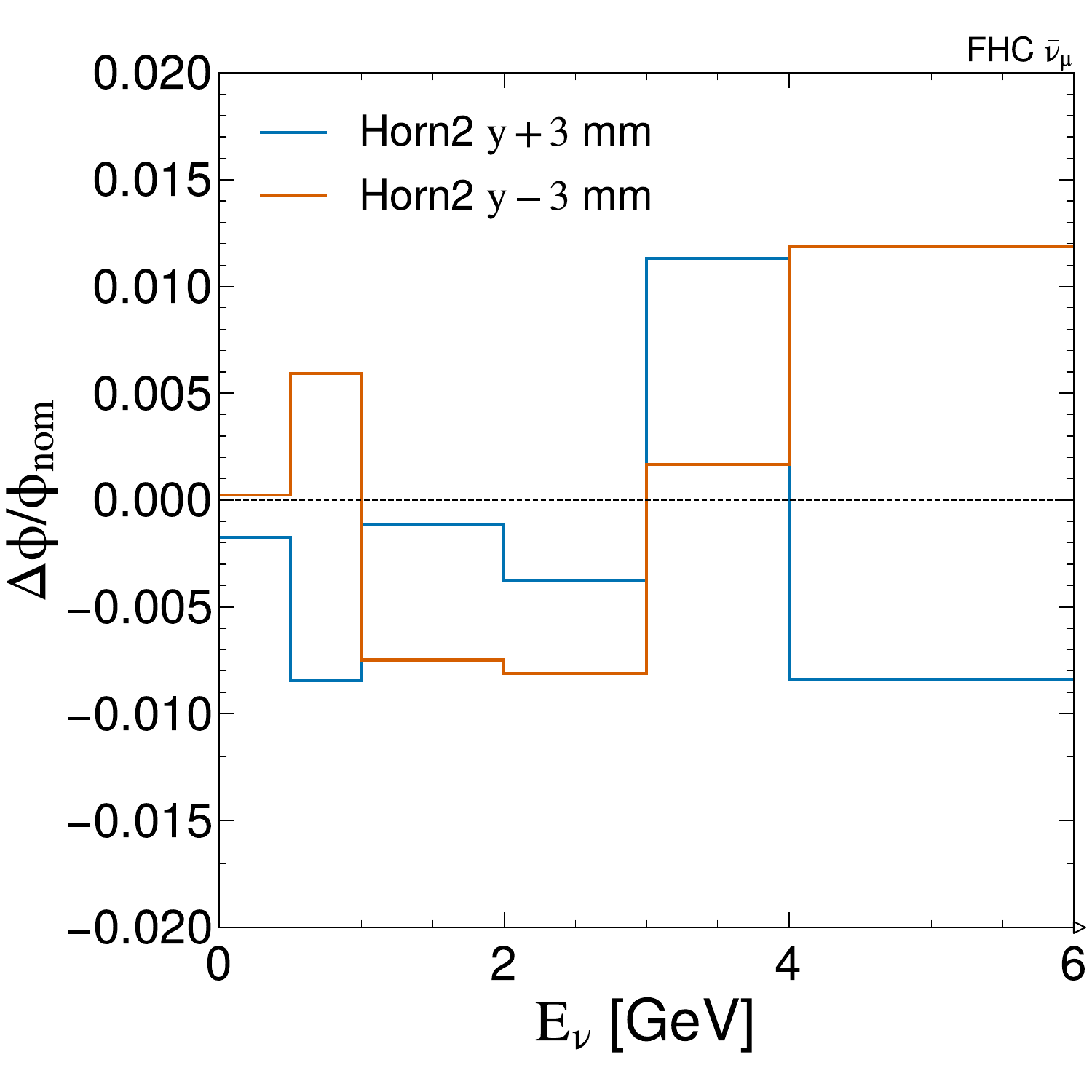}\\
    \includegraphics[width=0.25\textwidth]{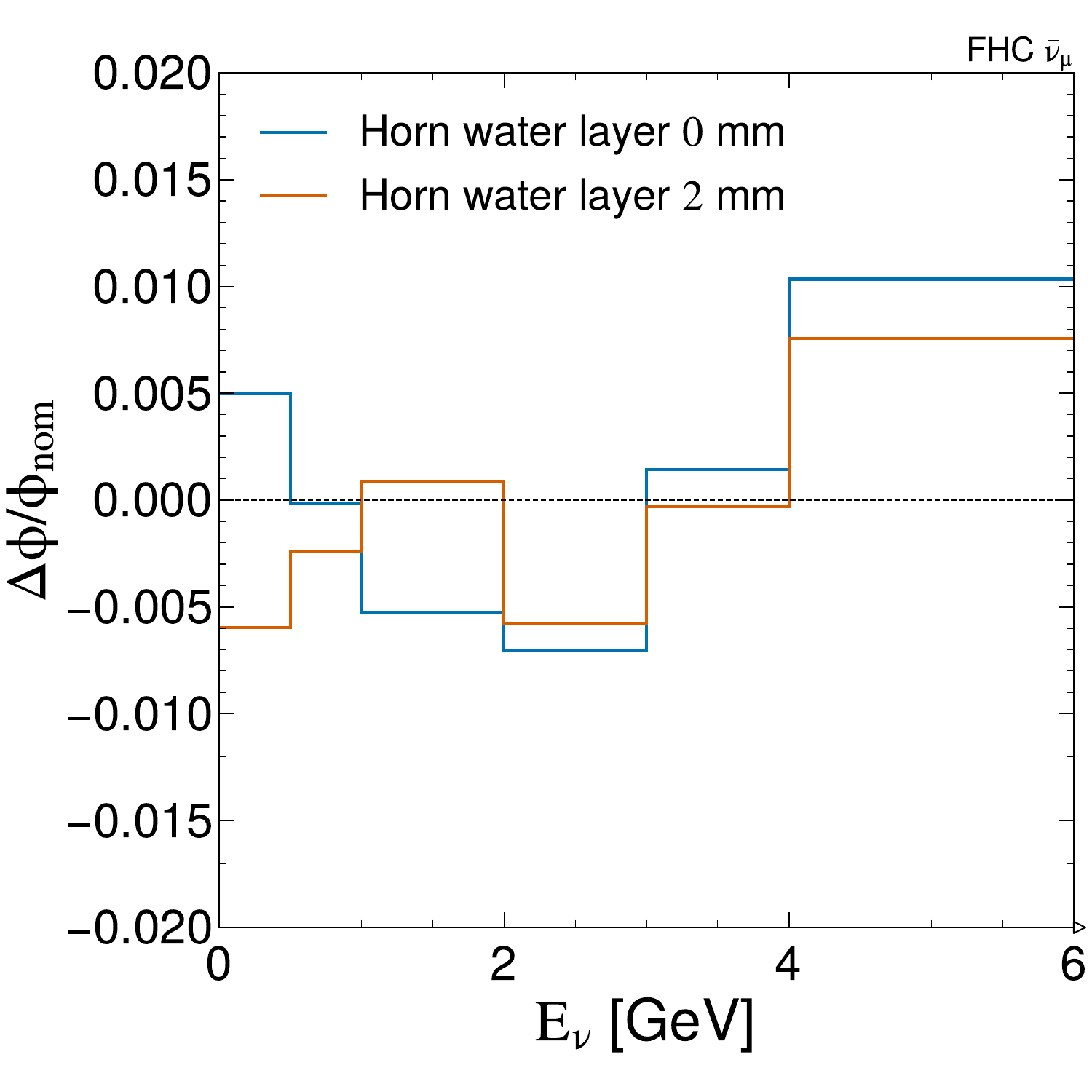}
    \includegraphics[width=0.25\textwidth]{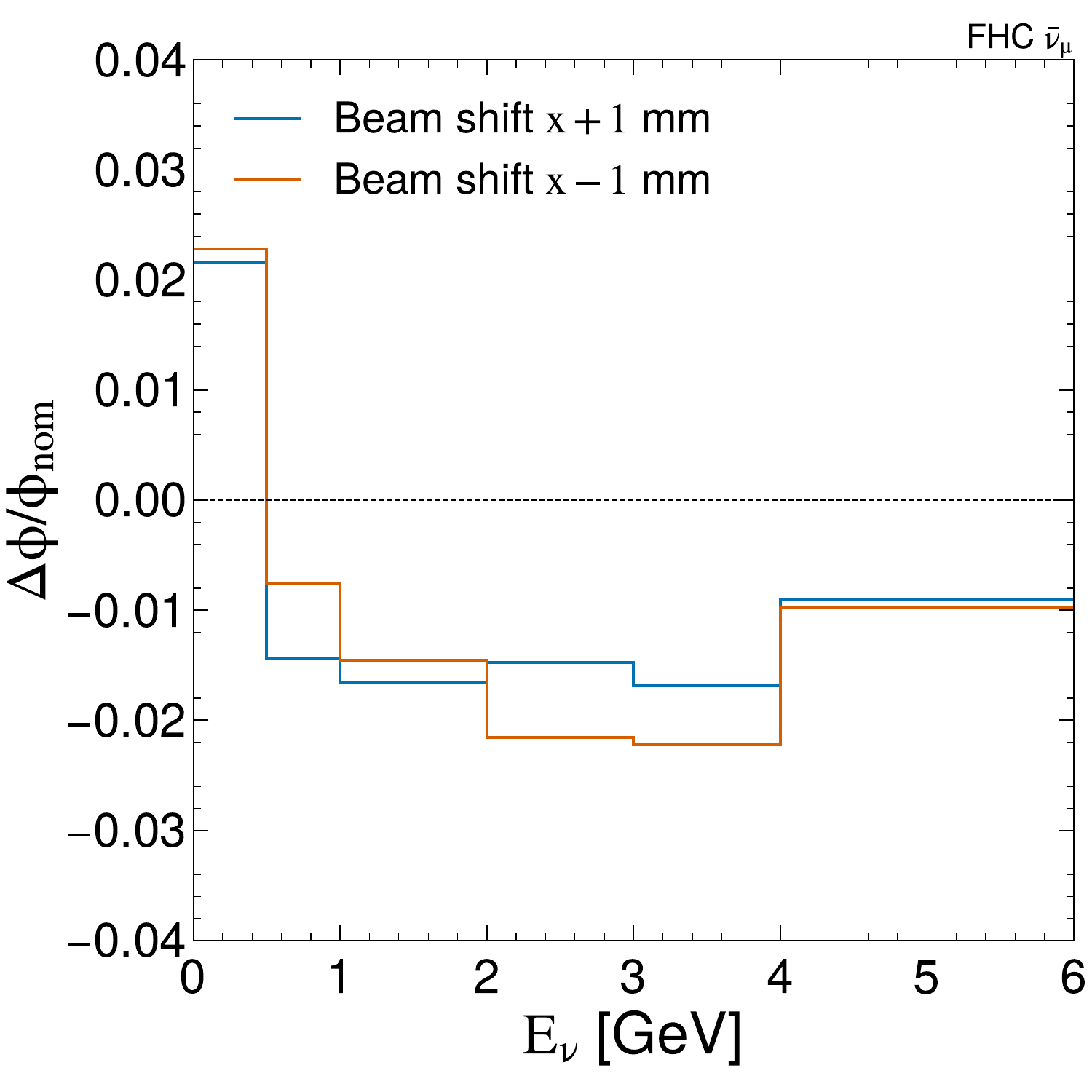}
    \includegraphics[width=0.25\textwidth]{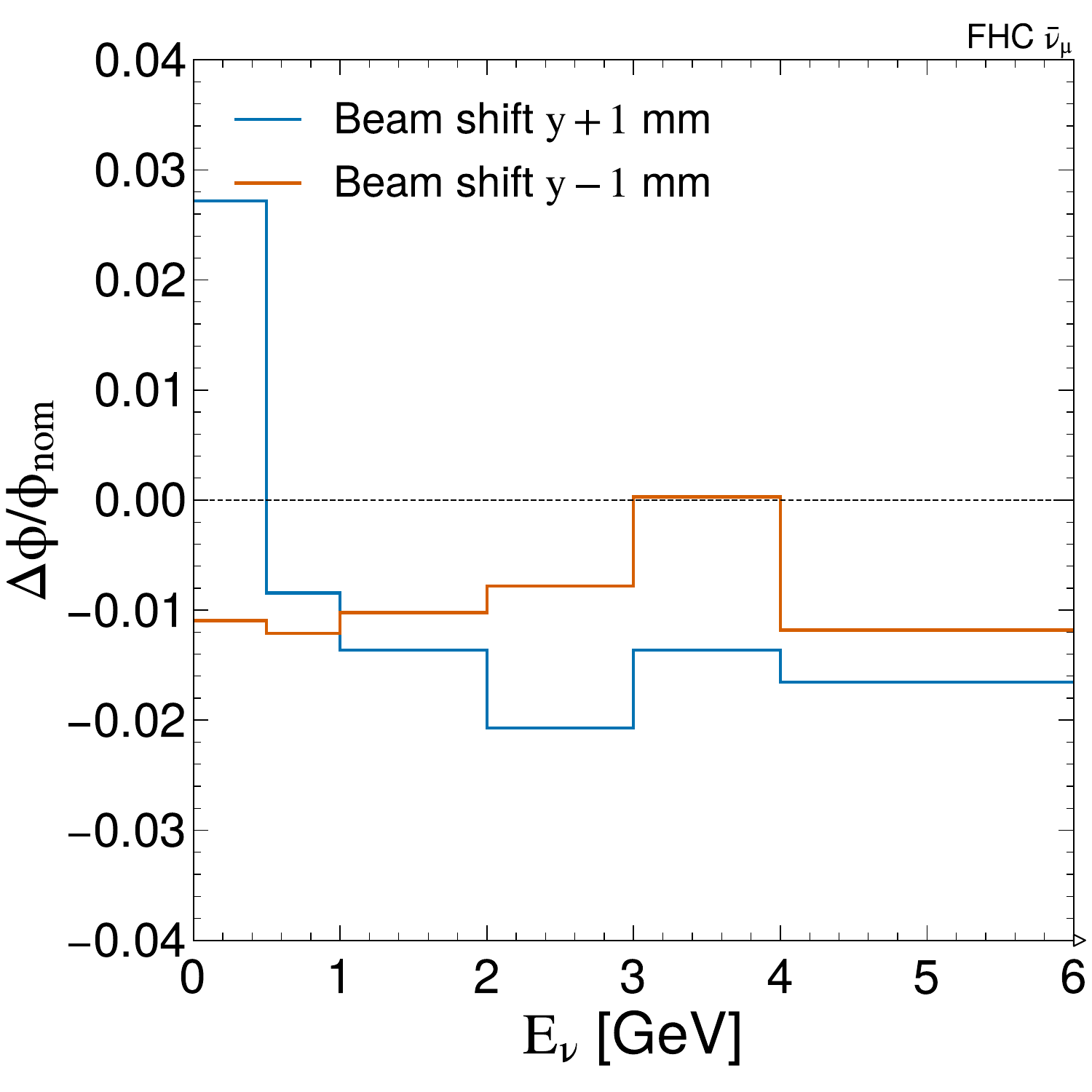}\\
    \includegraphics[width=0.25\textwidth]{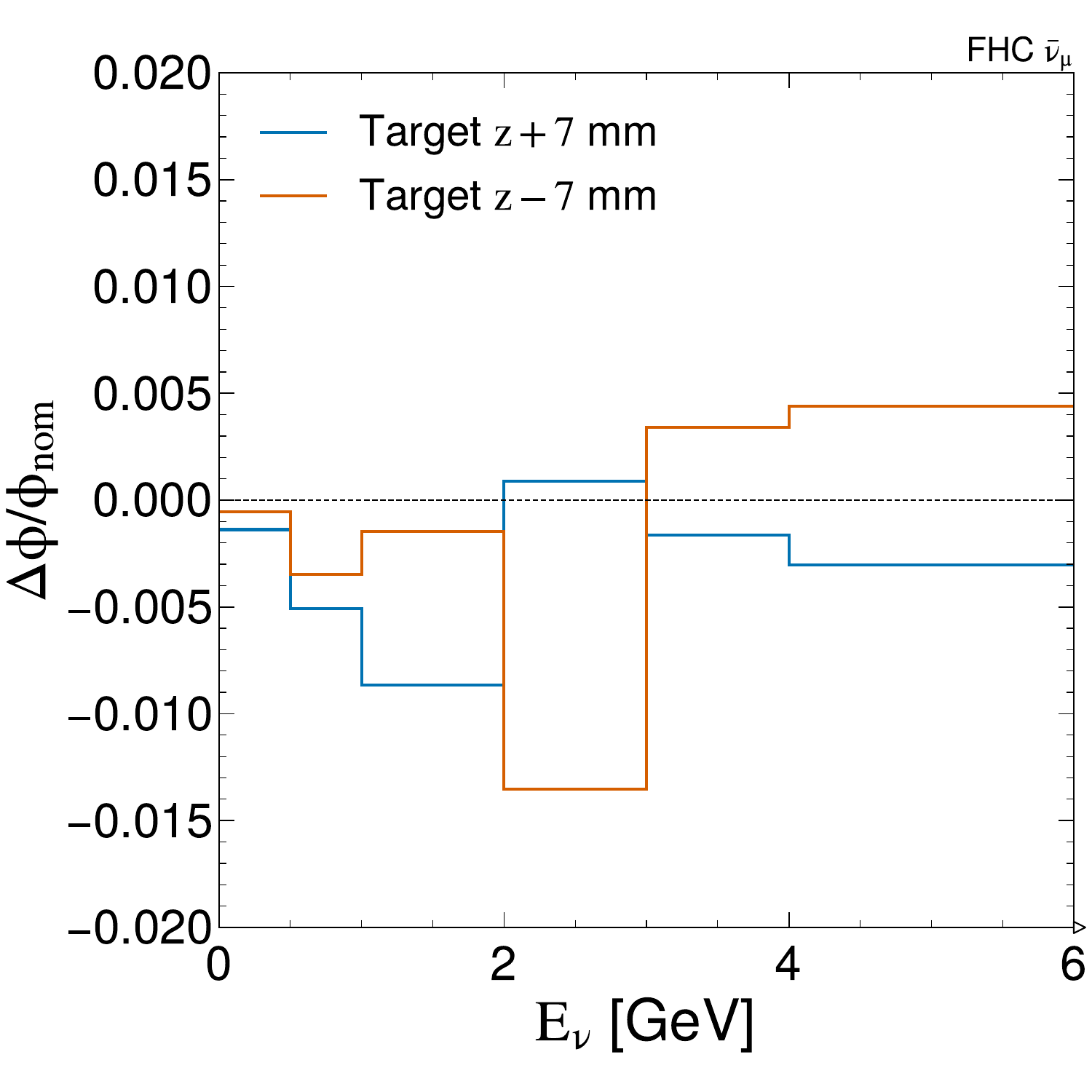}
    \caption[Beam Focusing Systematic Shifts (FHC, \numub)]{Beam focusing systematic shifts in the fractional scale (FHC, \numub).}
\end{figure}
\begin{figure}[!ht]
    \centering
    \includegraphics[width=0.25\textwidth]{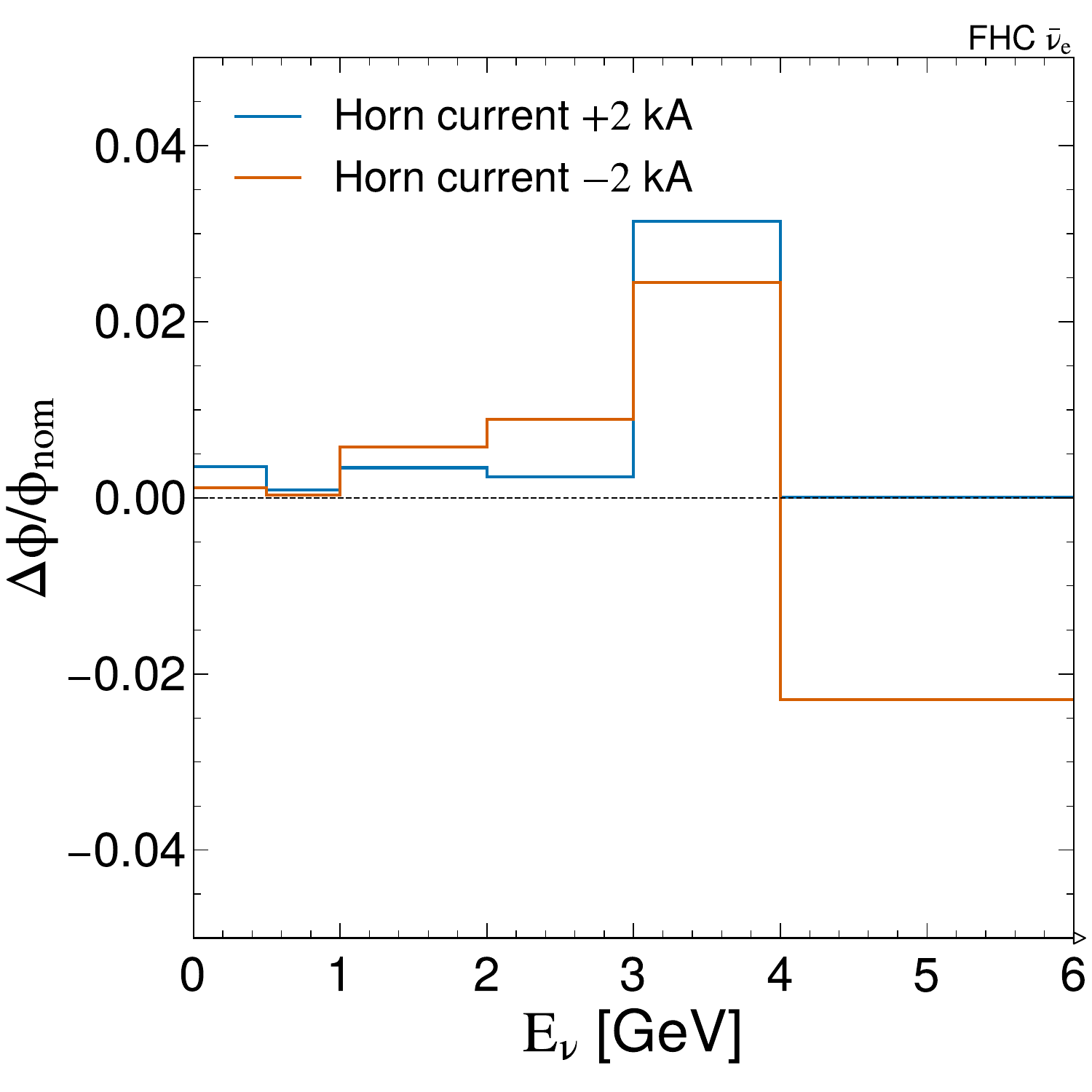}
    \includegraphics[width=0.25\textwidth]{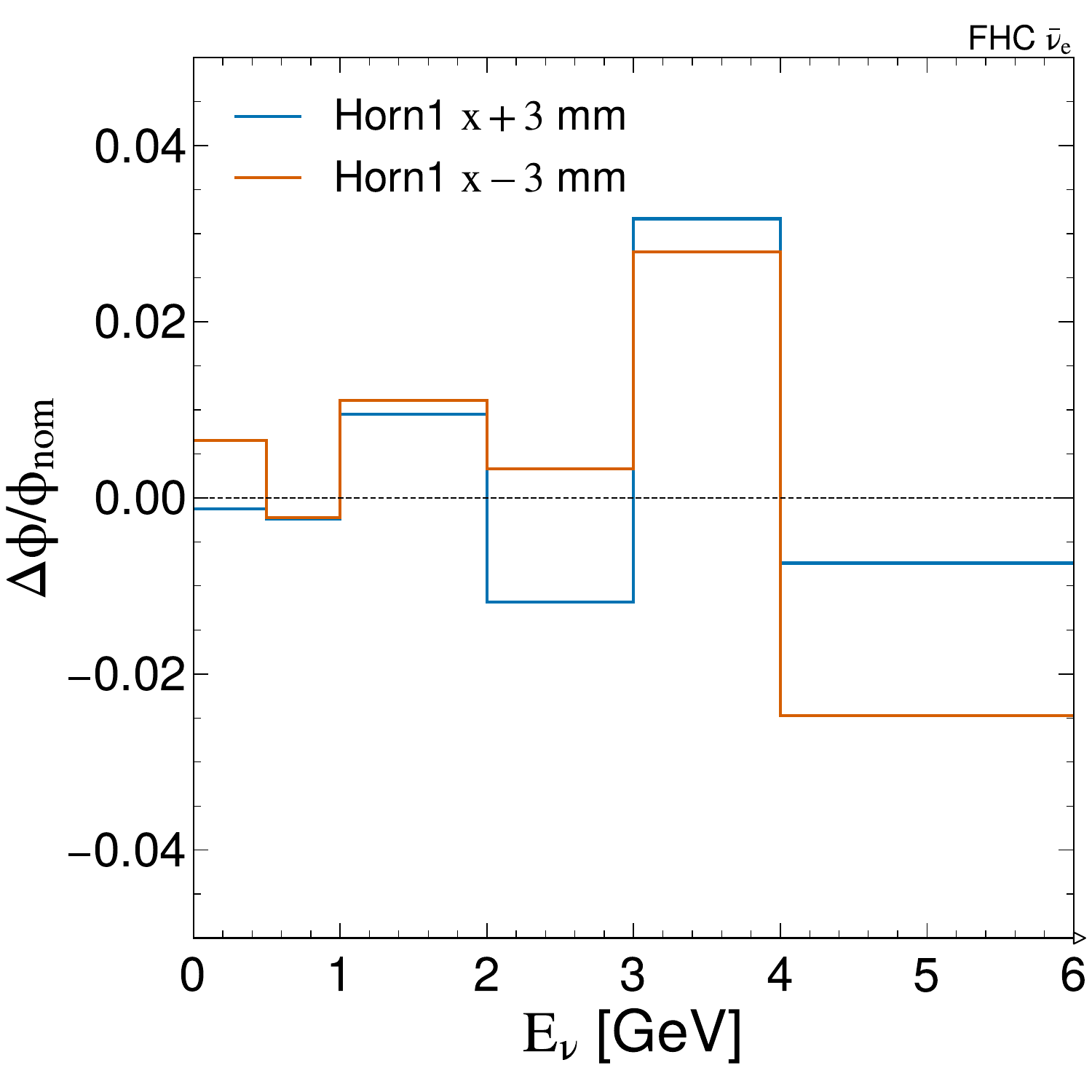}
    \includegraphics[width=0.25\textwidth]{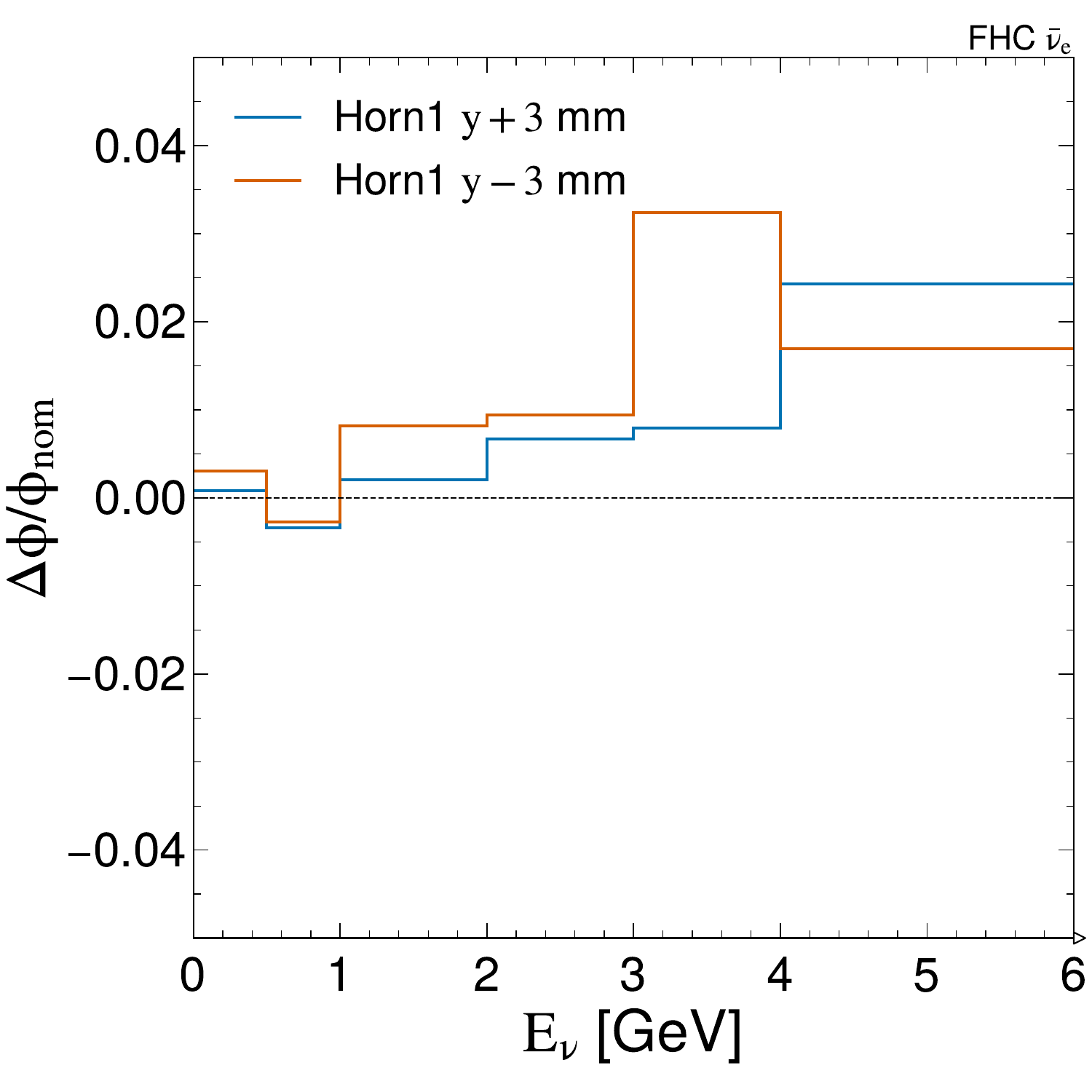}\\
    \includegraphics[width=0.25\textwidth]{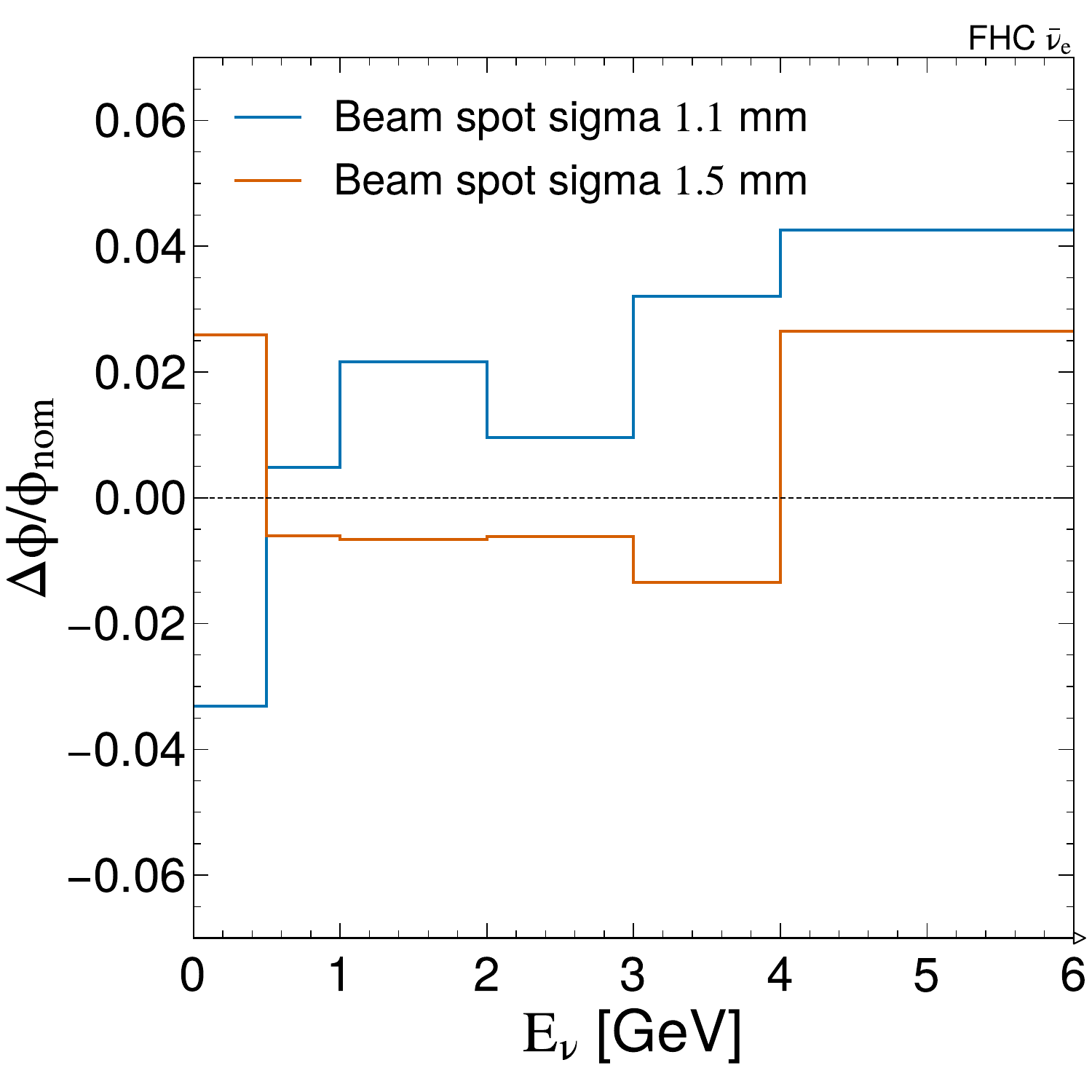}
    \includegraphics[width=0.25\textwidth]{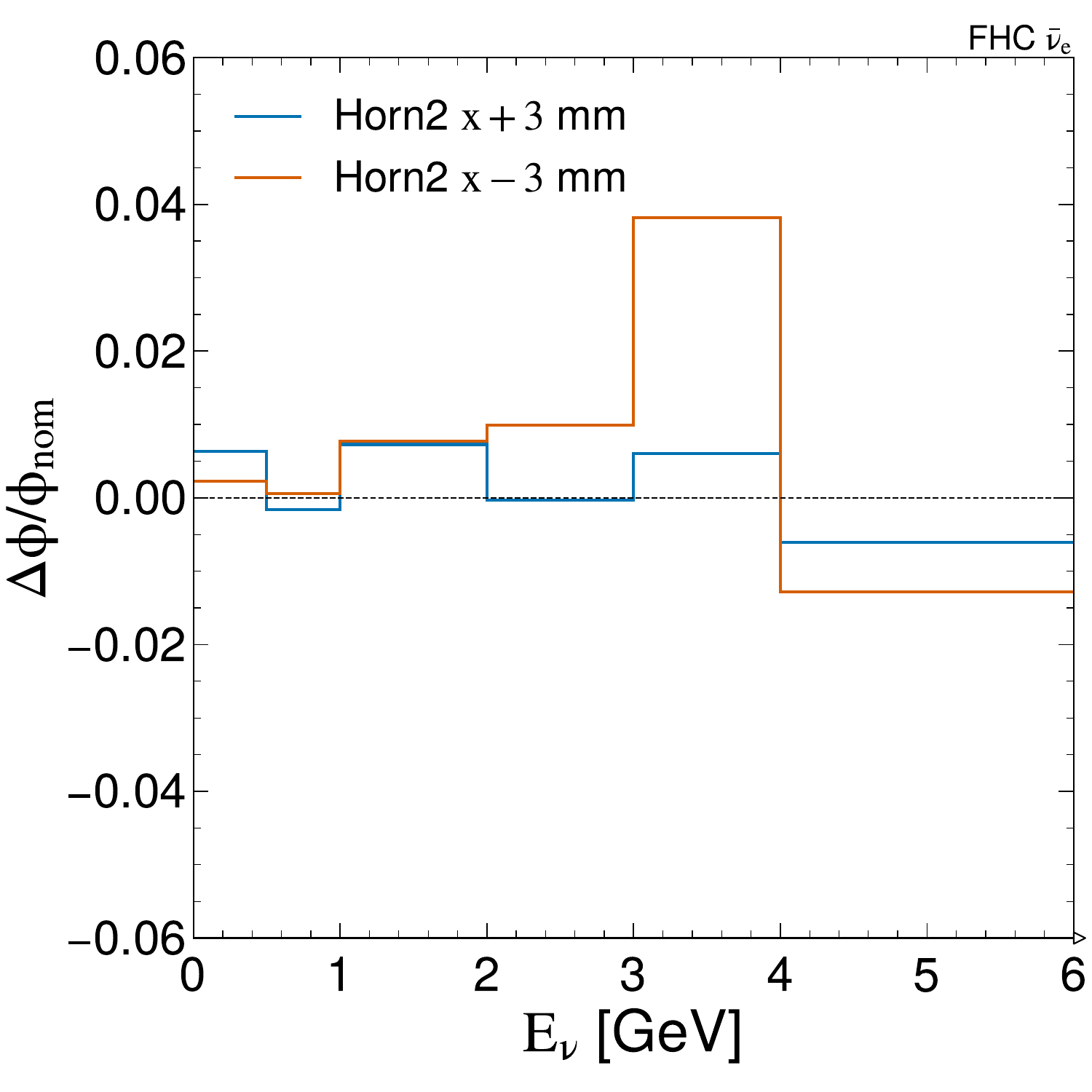}
    \includegraphics[width=0.25\textwidth]{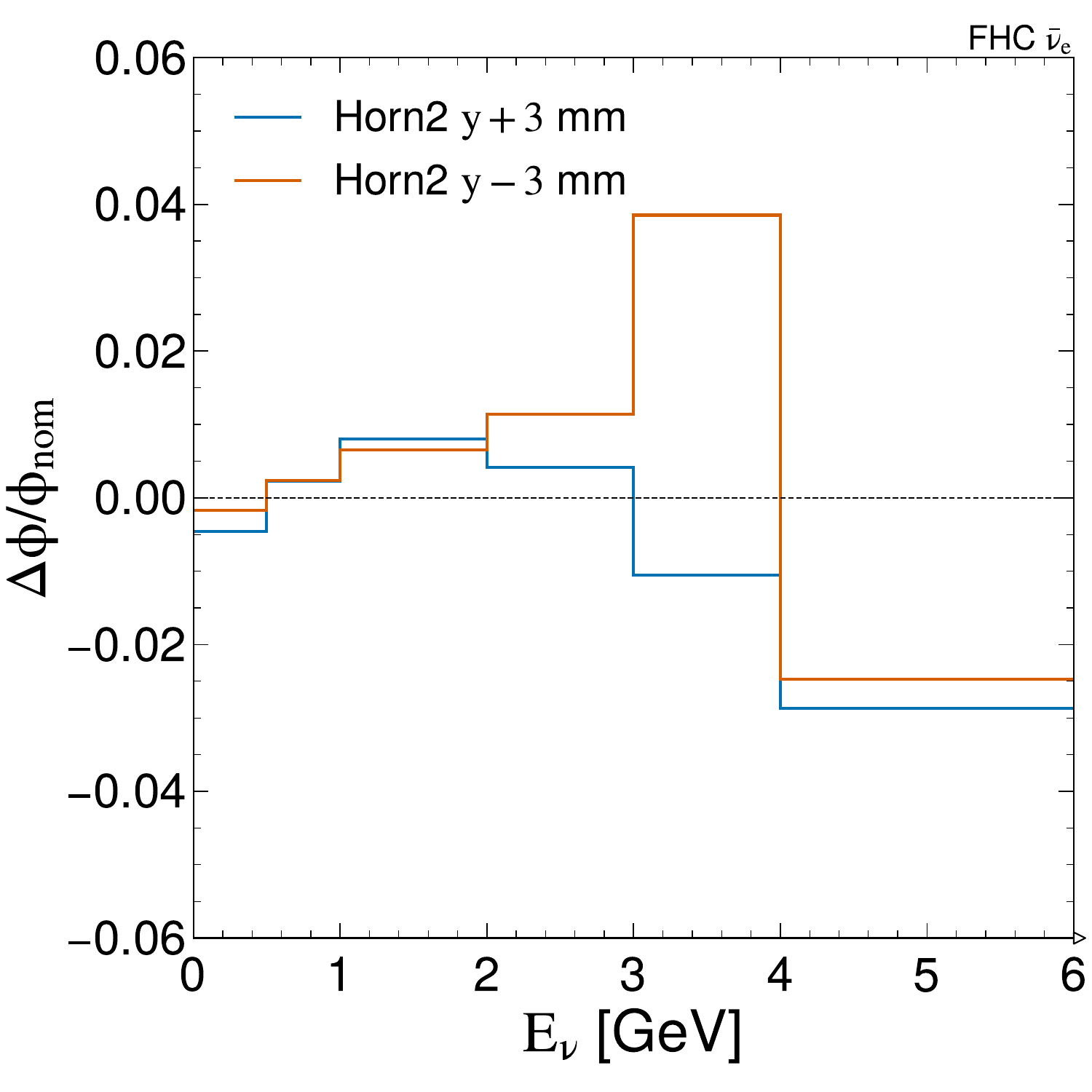}\\
    \includegraphics[width=0.25\textwidth]{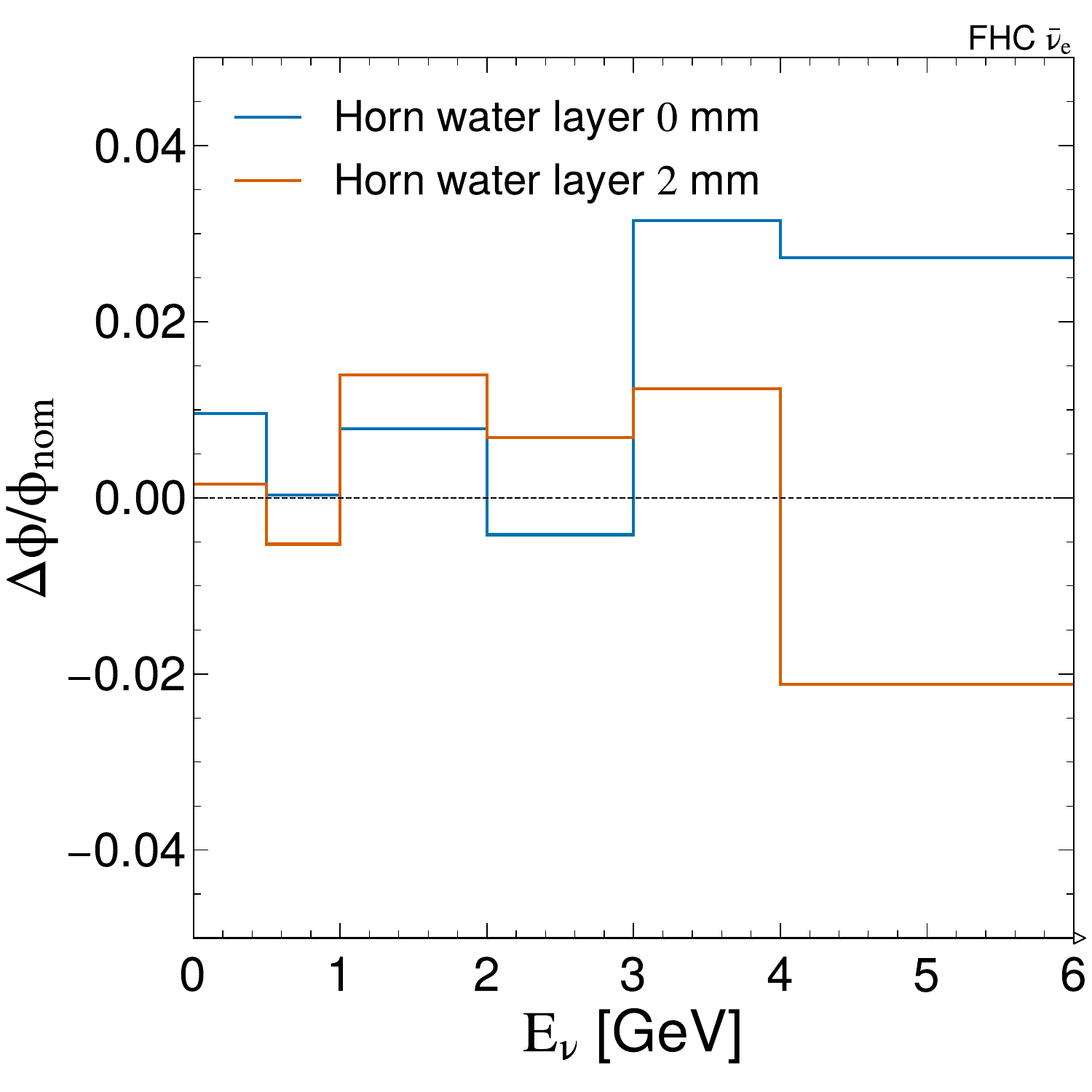}
    \includegraphics[width=0.25\textwidth]{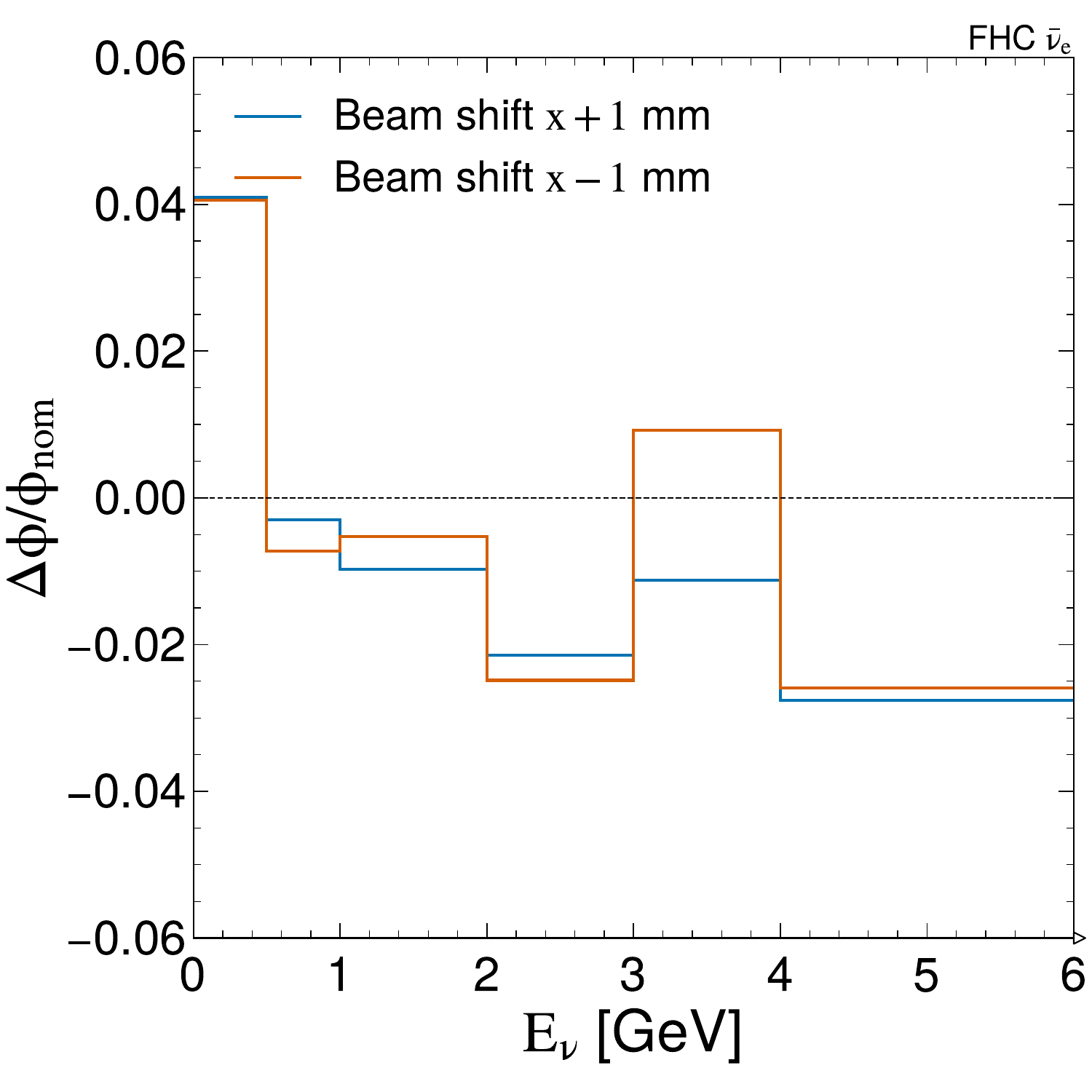}
    \includegraphics[width=0.25\textwidth]{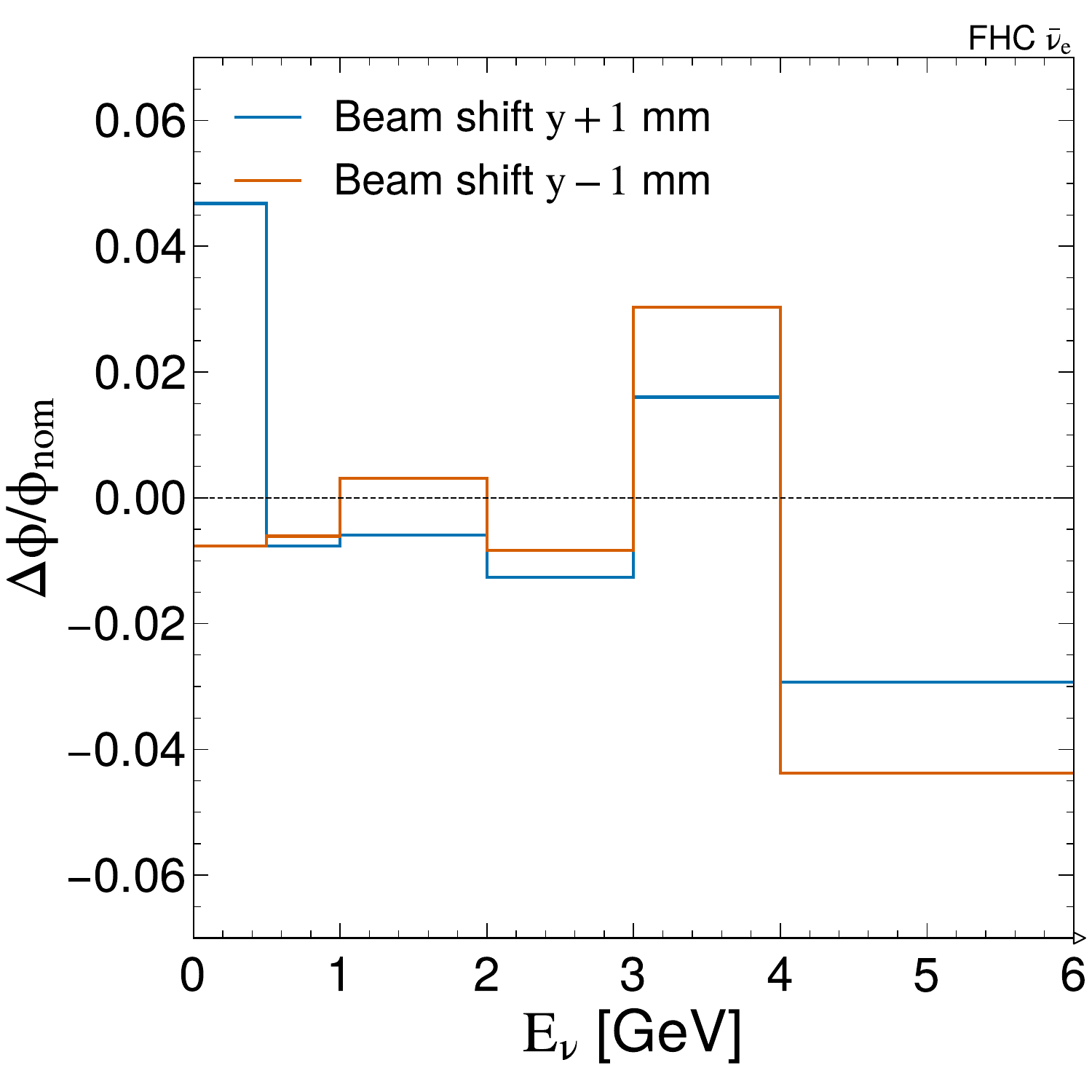}\\
    \includegraphics[width=0.25\textwidth]{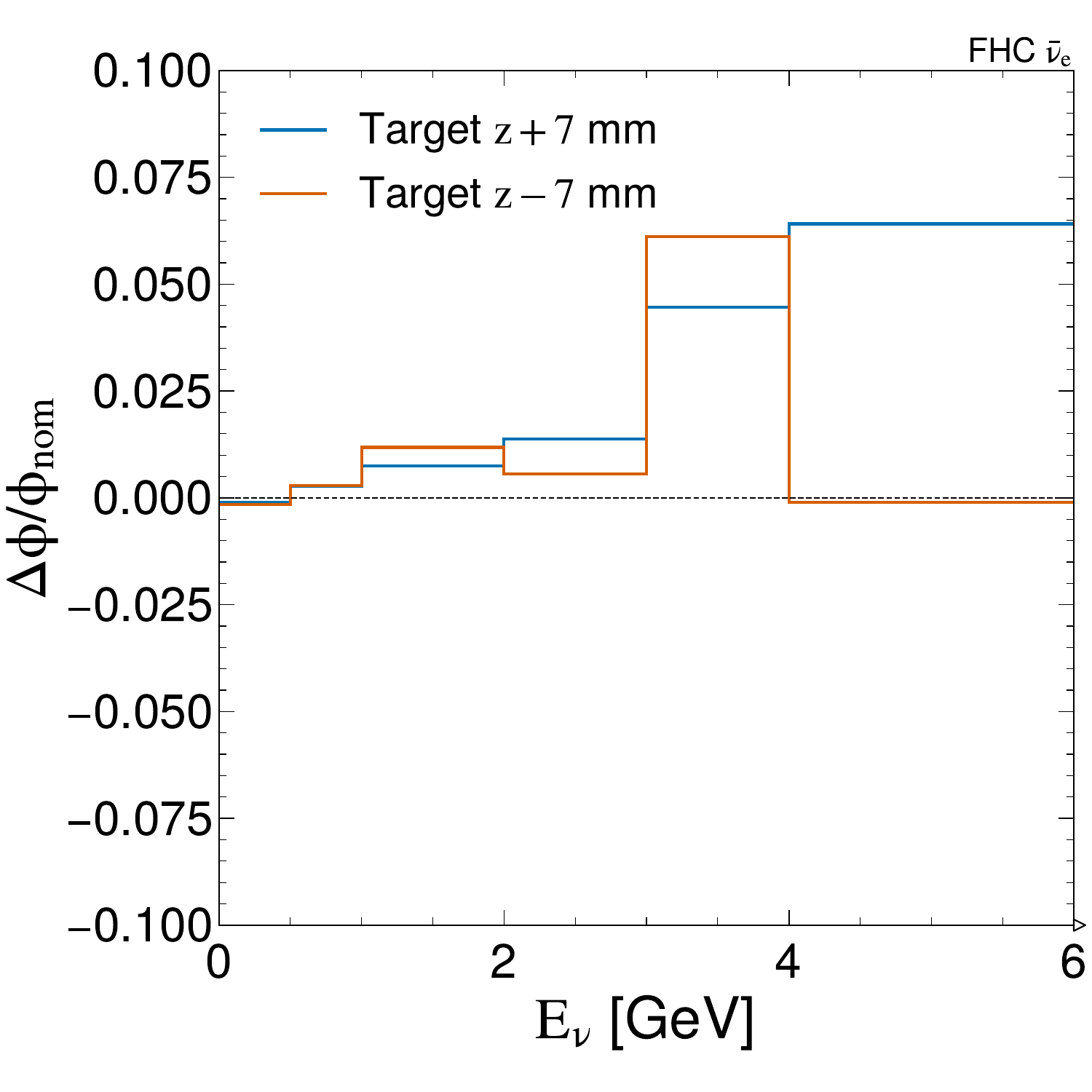}
    \caption[Beam Focusing Systematic Shifts (FHC, \nueb)]{Beam focusing systematic shifts in the fractional scale (FHC, \nueb).}
\end{figure}
\clearpage
\section{Reverse Horn Current}
\begin{figure}[!ht]
    \centering
    \includegraphics[width=0.25\textwidth]{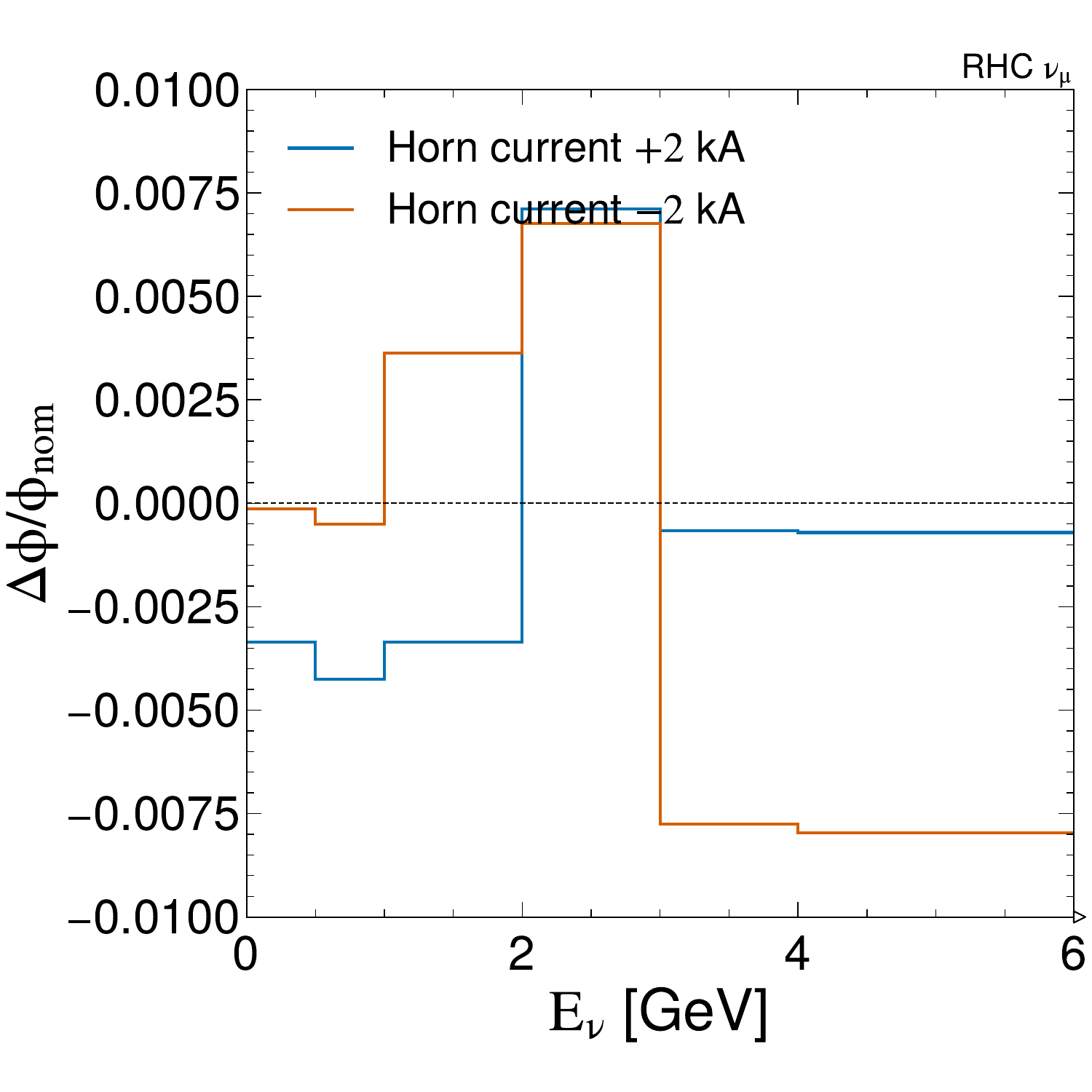}
    \includegraphics[width=0.25\textwidth]{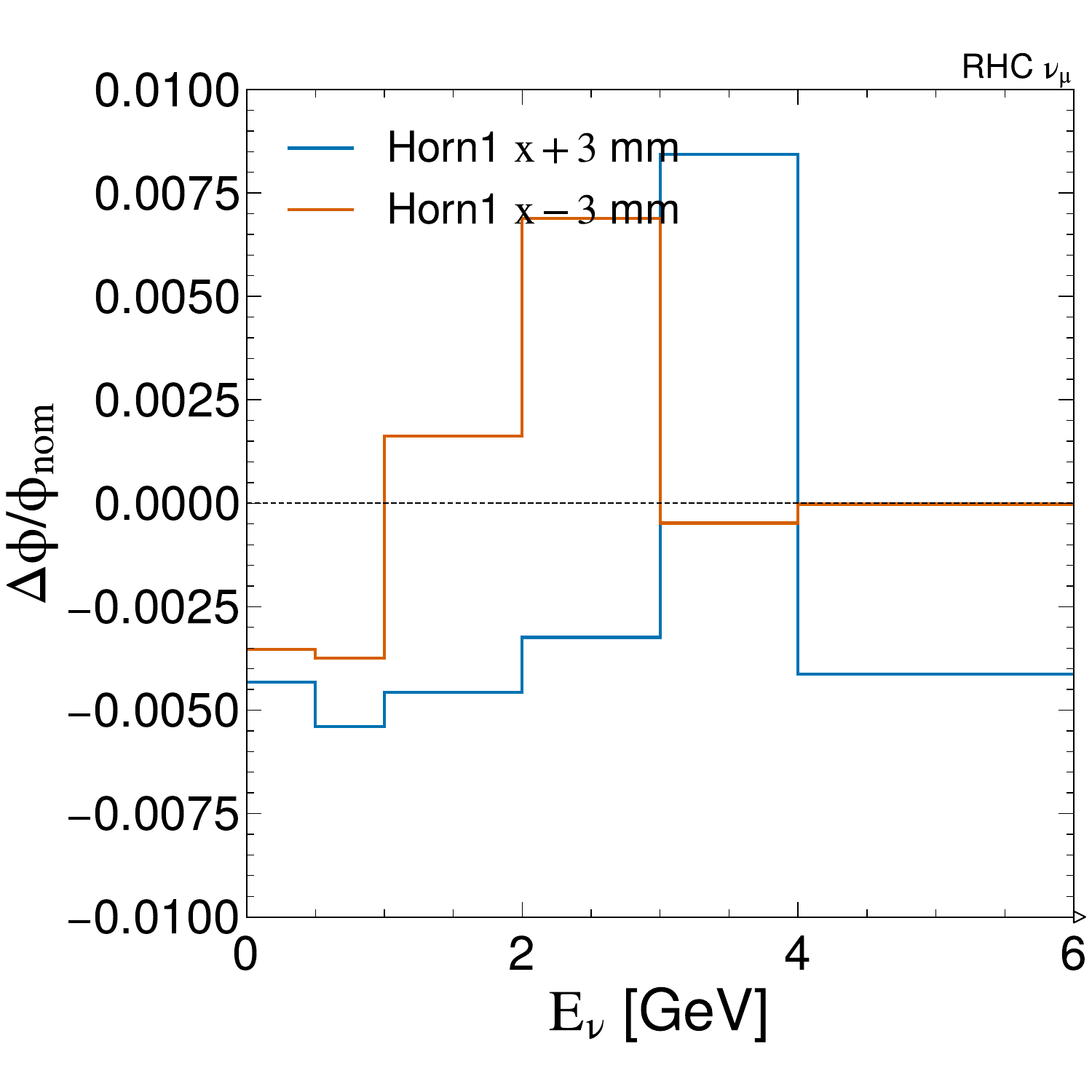}
    \includegraphics[width=0.25\textwidth]{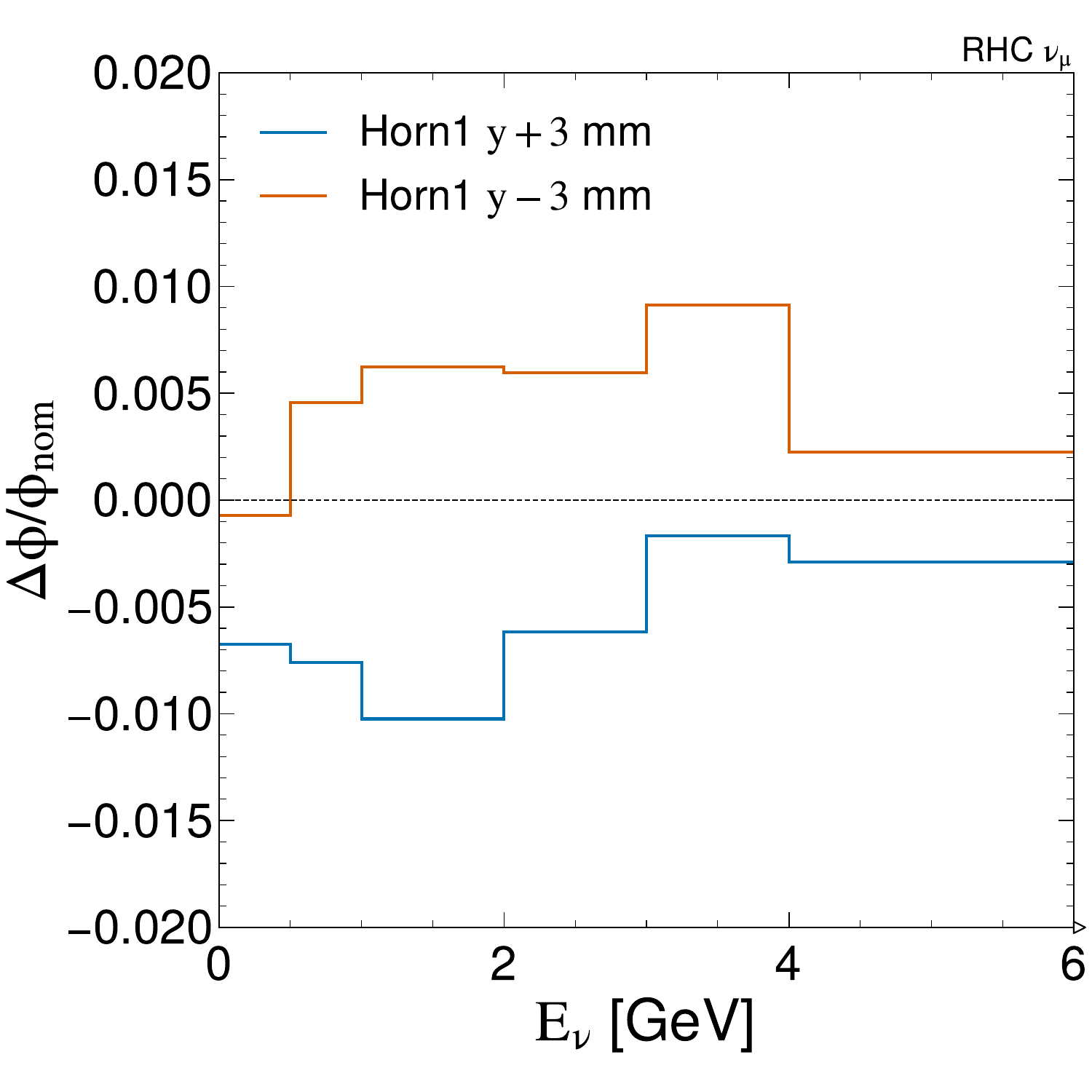}\\
    \includegraphics[width=0.25\textwidth]{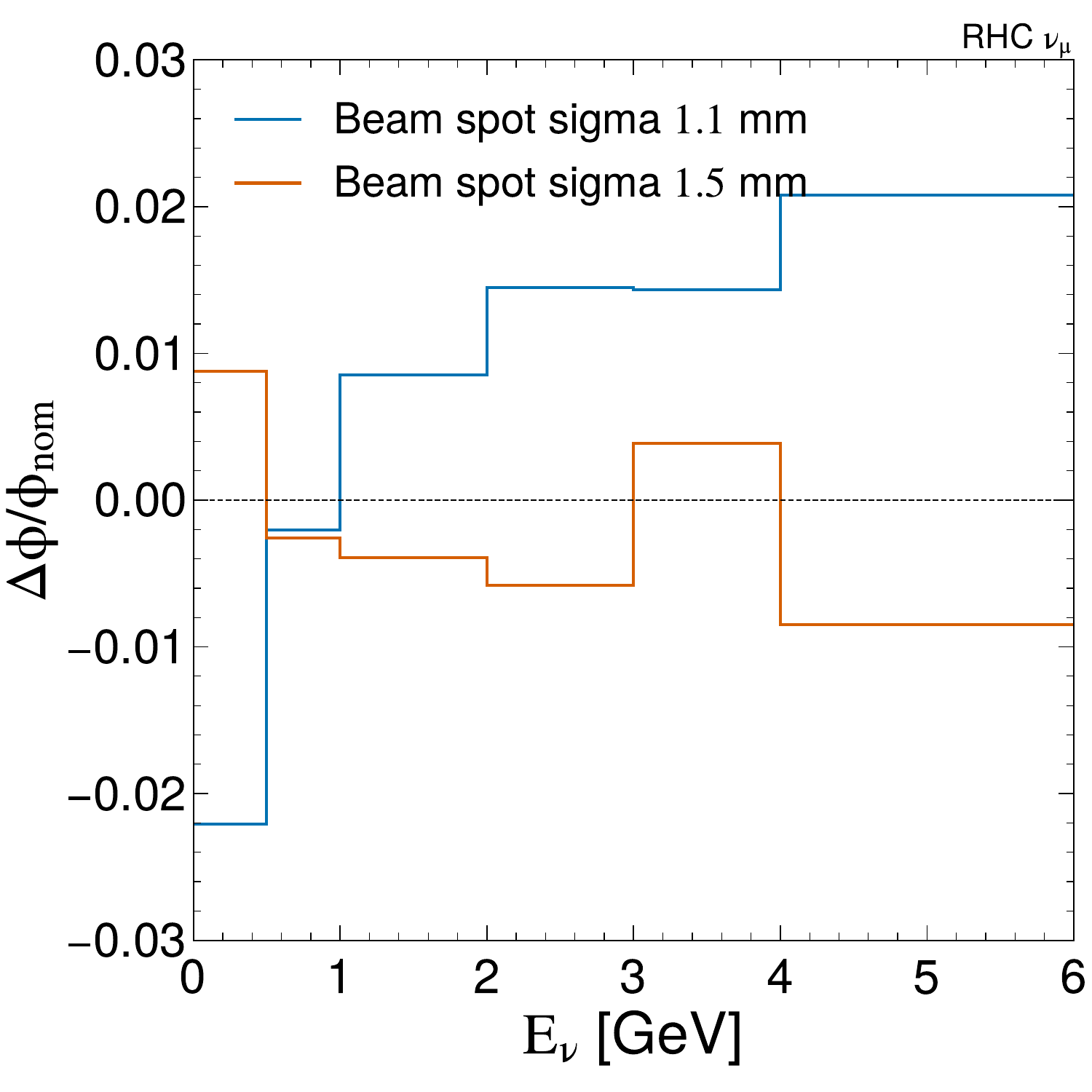}
    \includegraphics[width=0.25\textwidth]{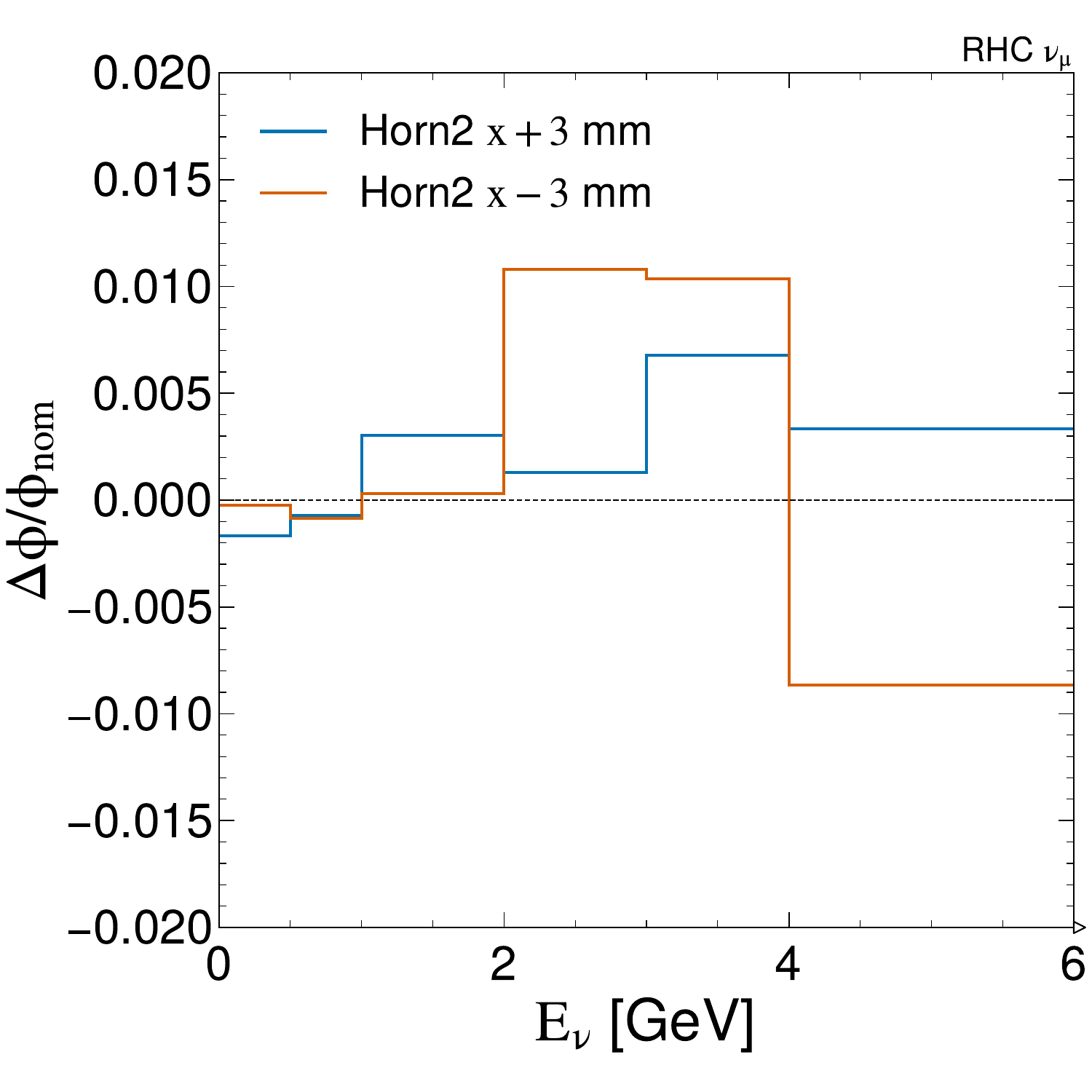}
    \includegraphics[width=0.25\textwidth]{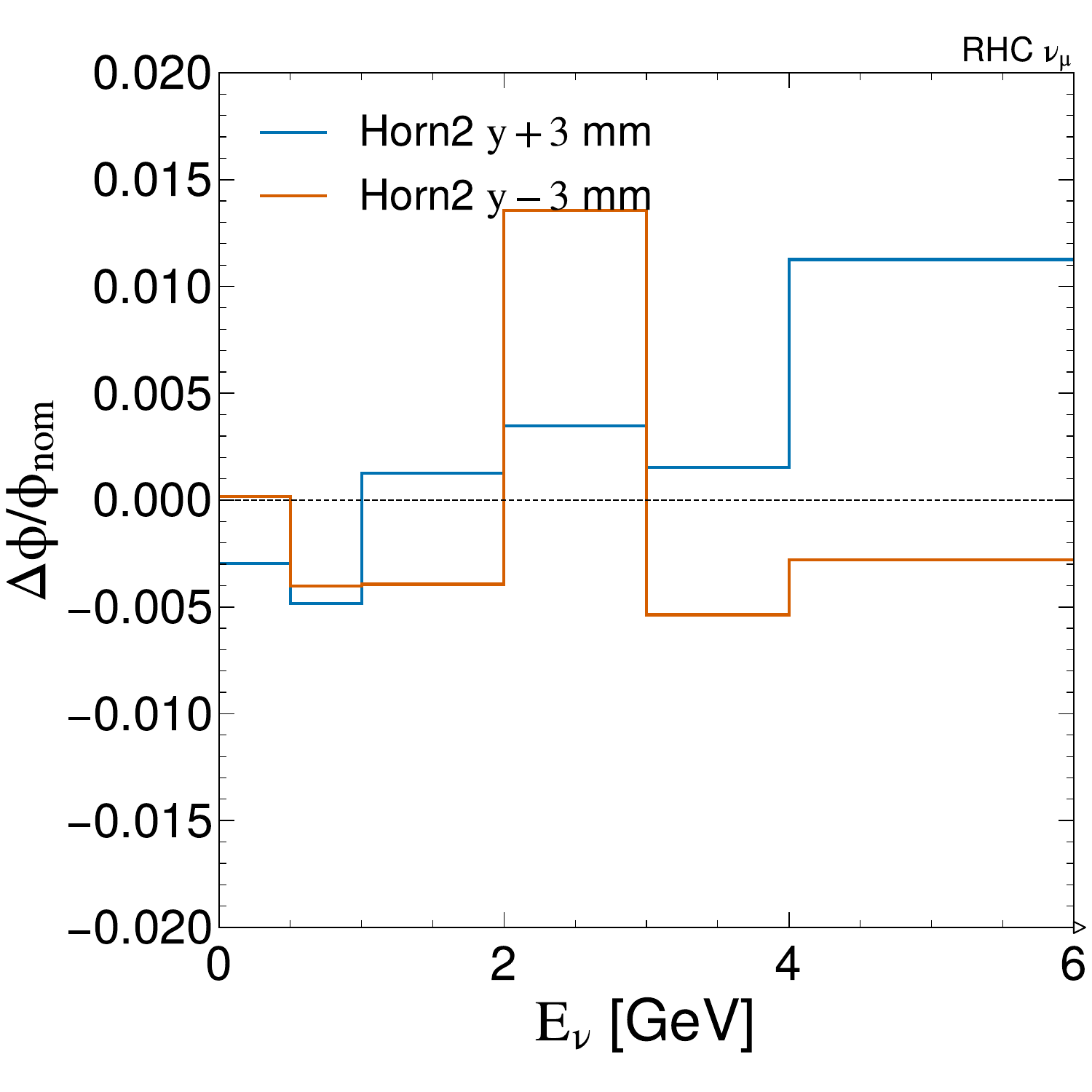}\\
    \includegraphics[width=0.25\textwidth]{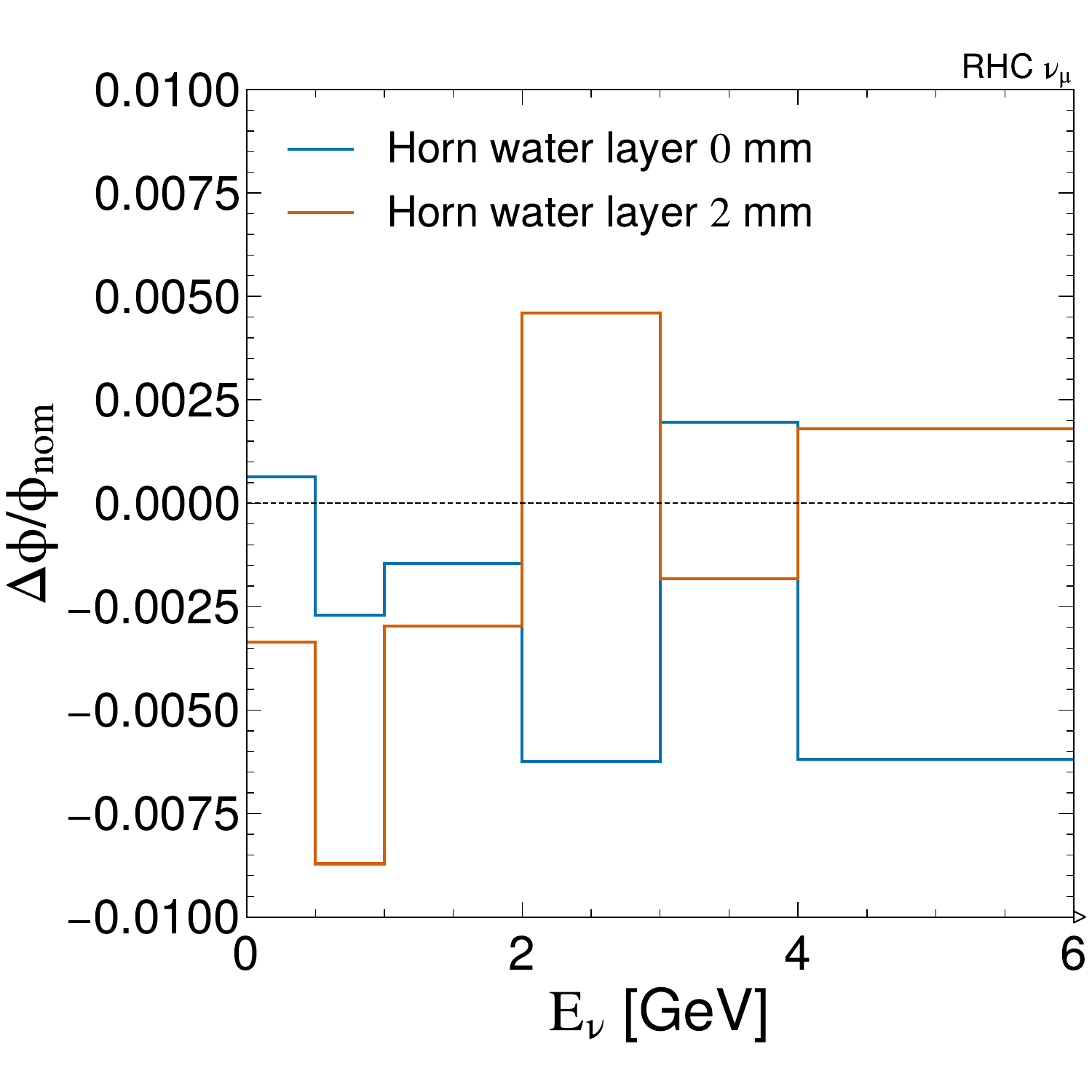}
    \includegraphics[width=0.25\textwidth]{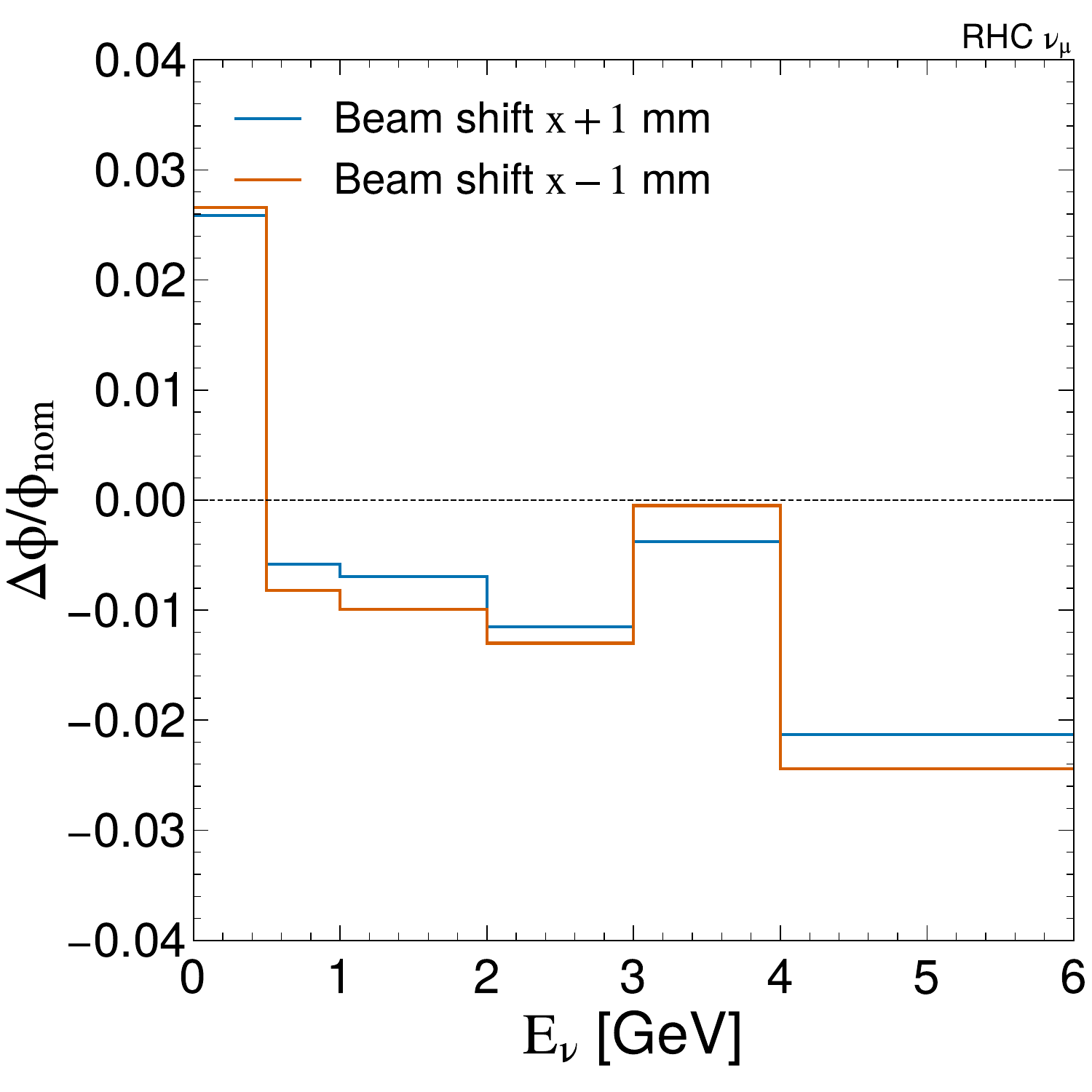}
    \includegraphics[width=0.25\textwidth]{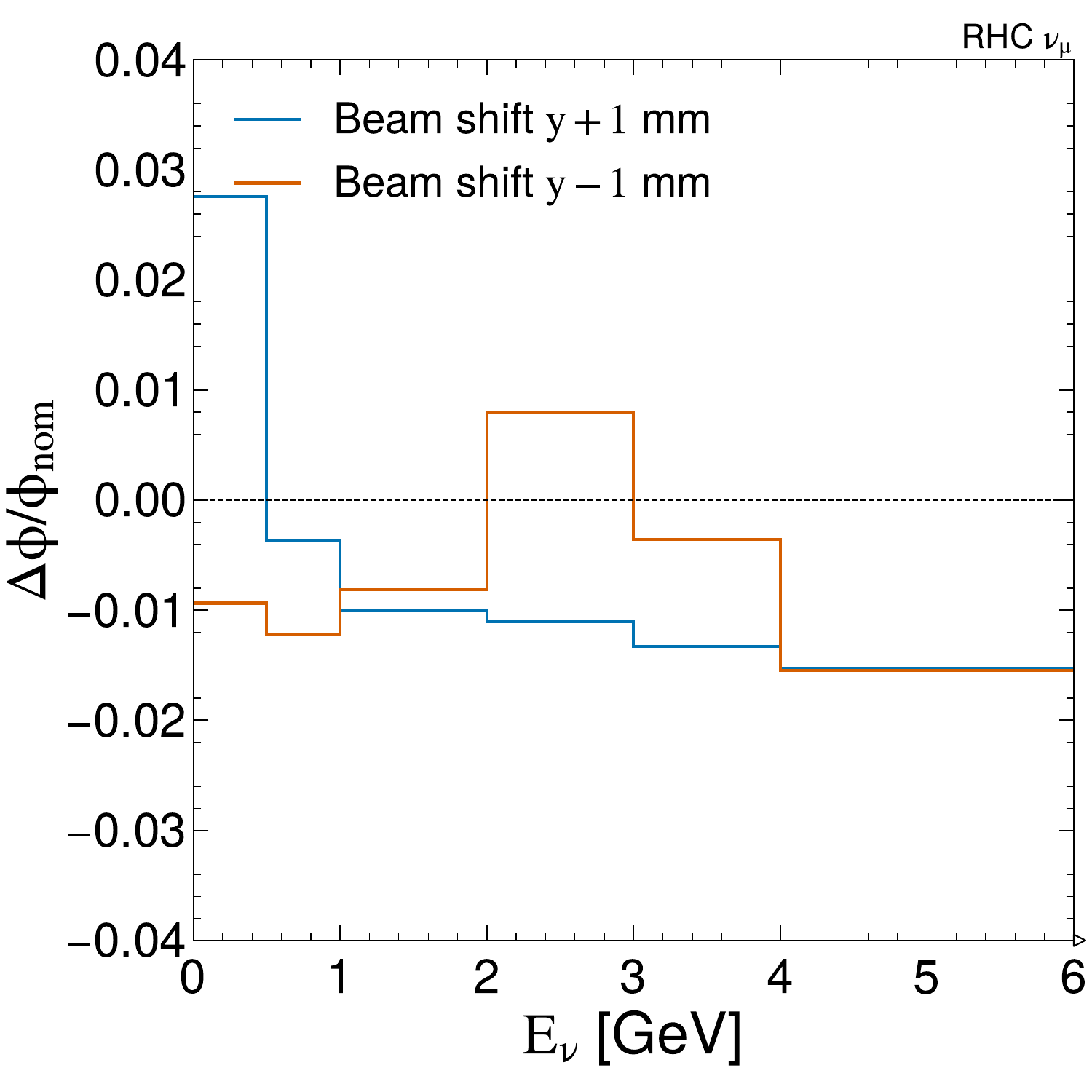}\\
    \includegraphics[width=0.25\textwidth]{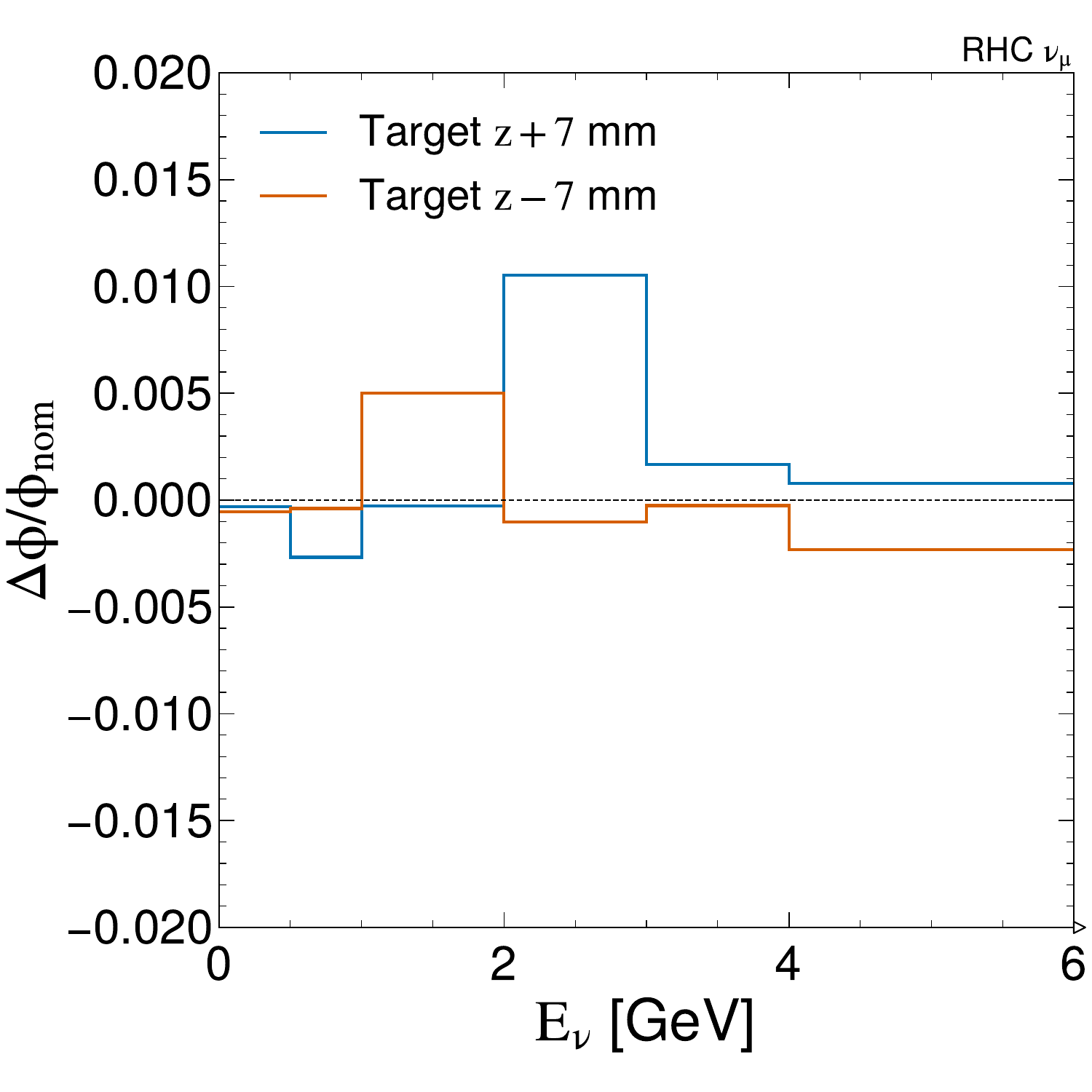}
    \caption[Beam Focusing Systematic Shifts (RHC, \numu)]{Beam focusing systematic shifts in the fractional scale (RHC, \numu).}
\end{figure}
\begin{figure}[!ht]
    \centering
    \includegraphics[width=0.25\textwidth]{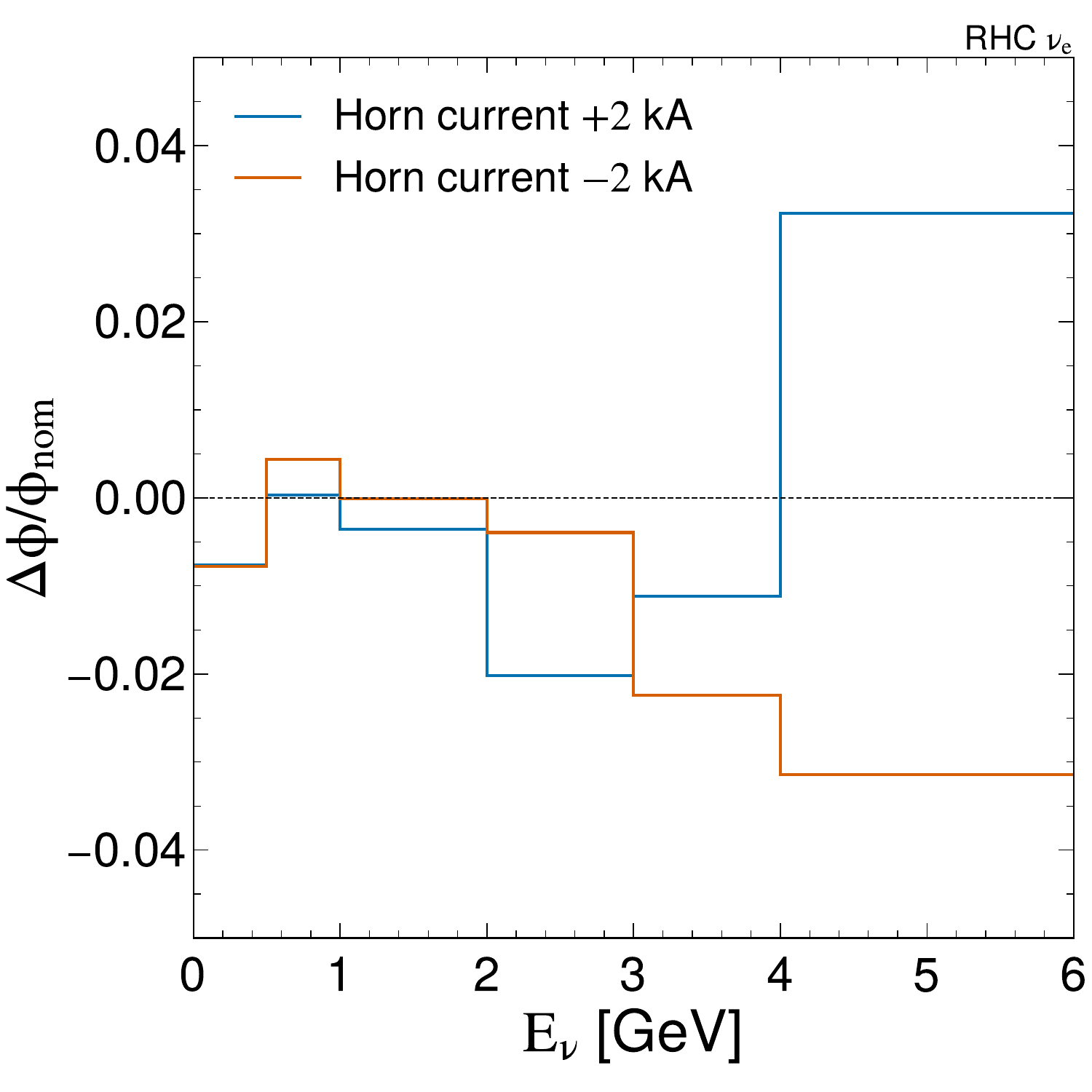}
    \includegraphics[width=0.25\textwidth]{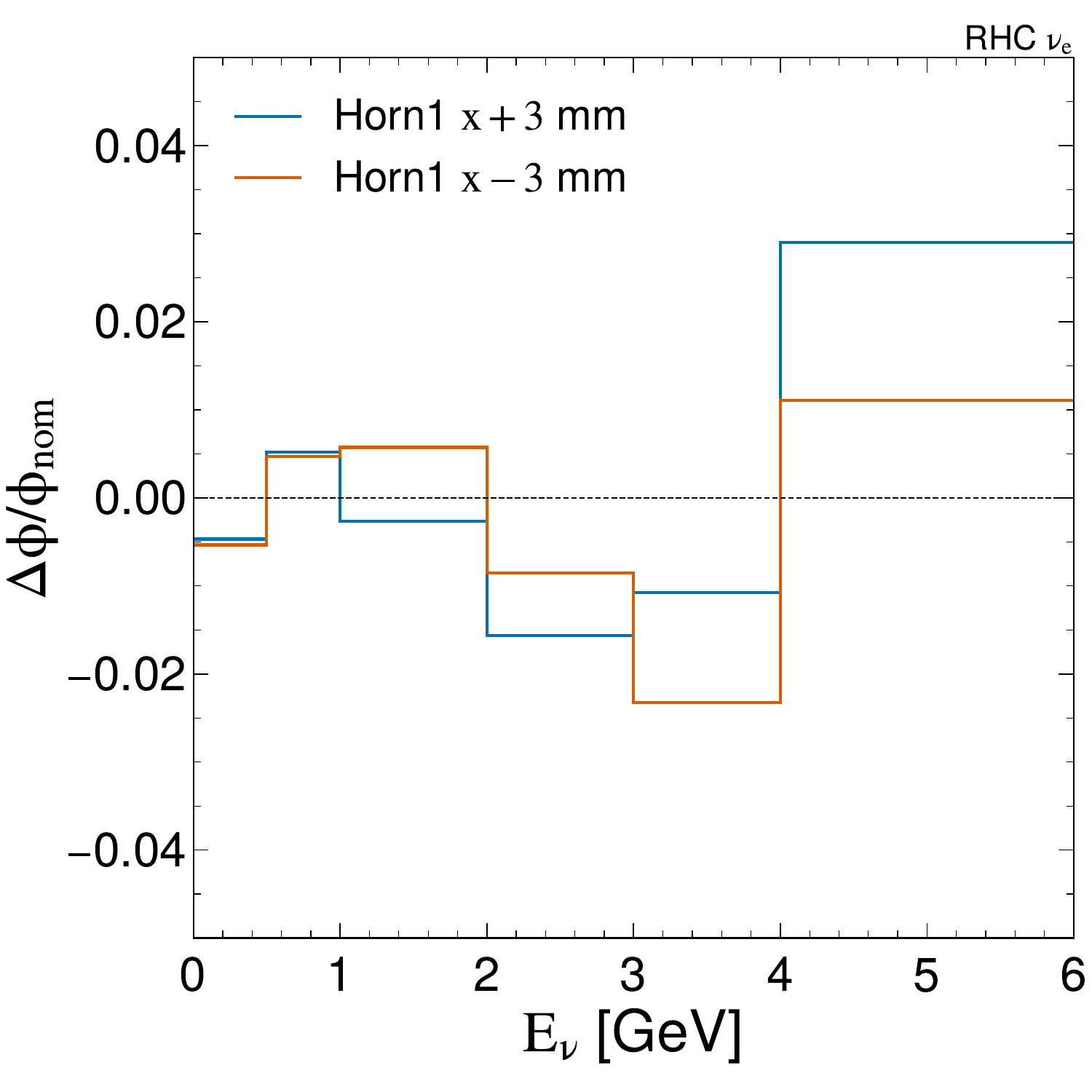}
    \includegraphics[width=0.25\textwidth]{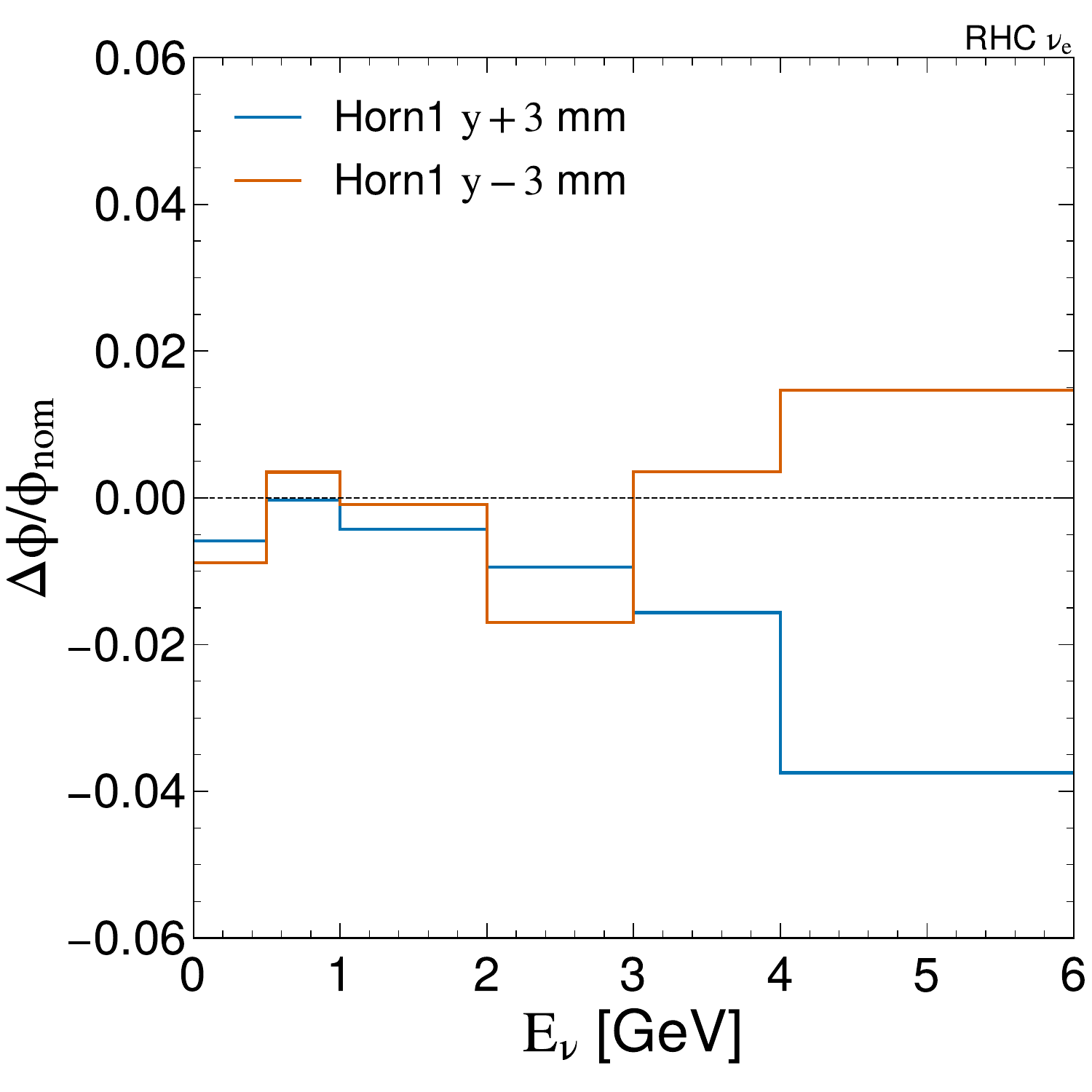}\\
    \includegraphics[width=0.25\textwidth]{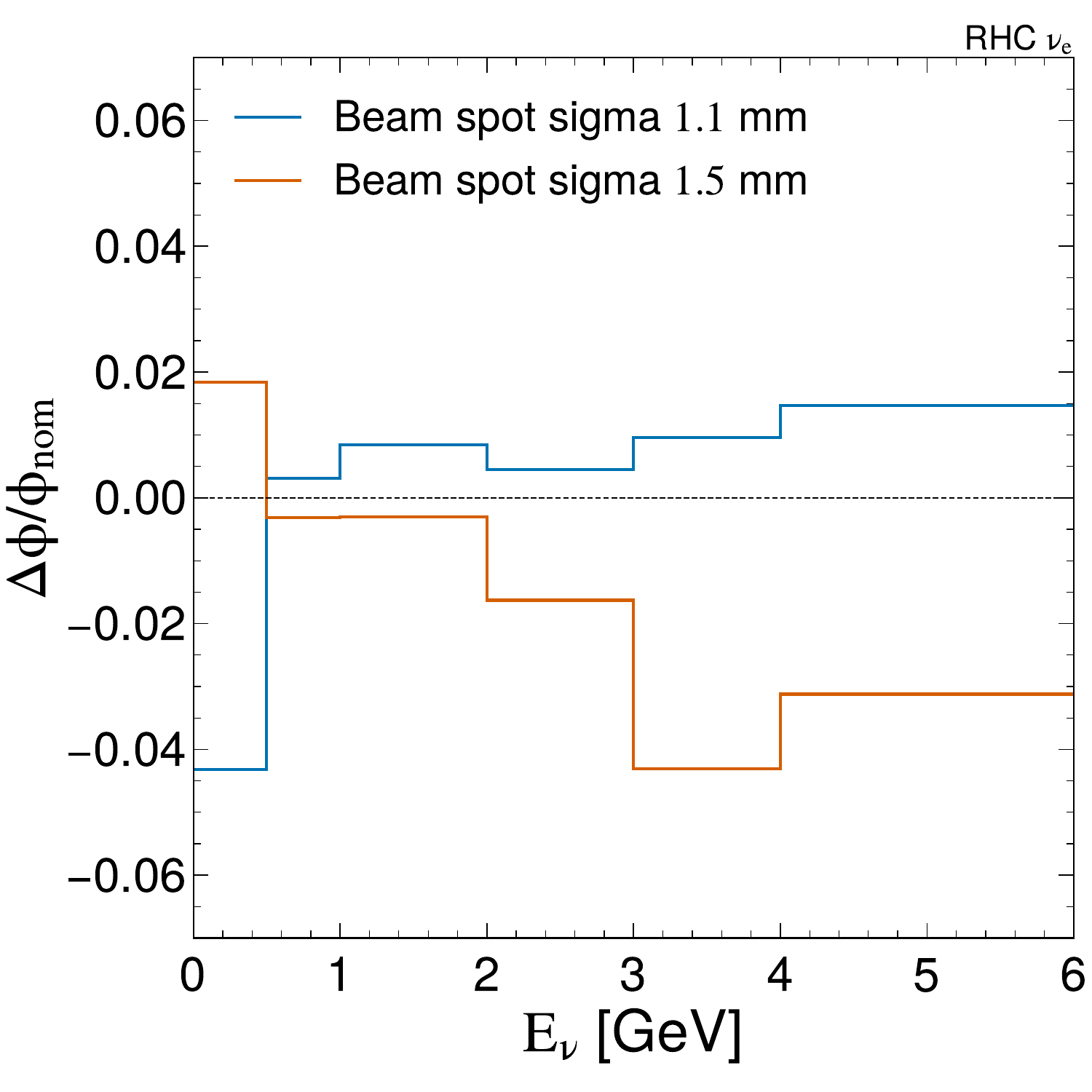}
    \includegraphics[width=0.25\textwidth]{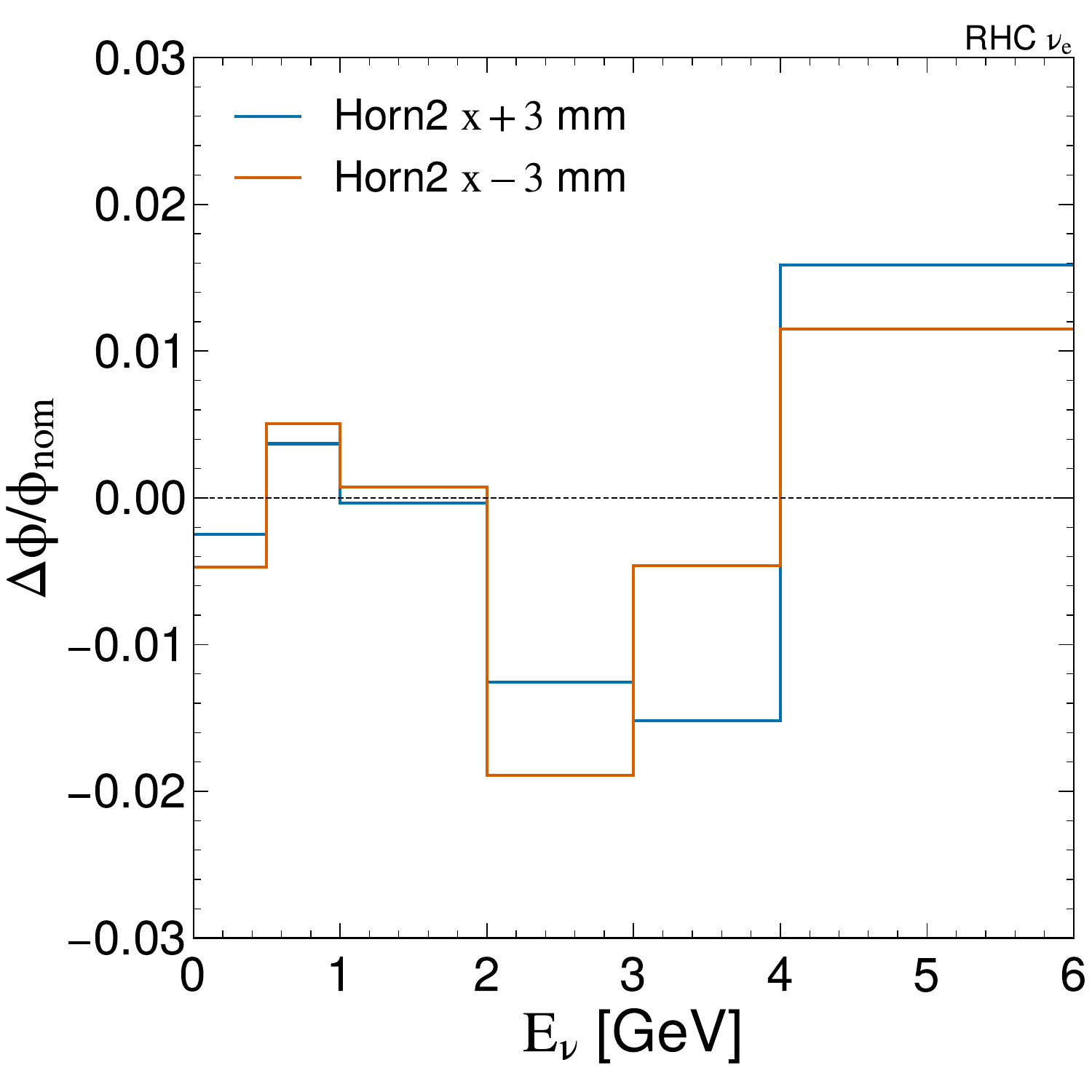}
    \includegraphics[width=0.25\textwidth]{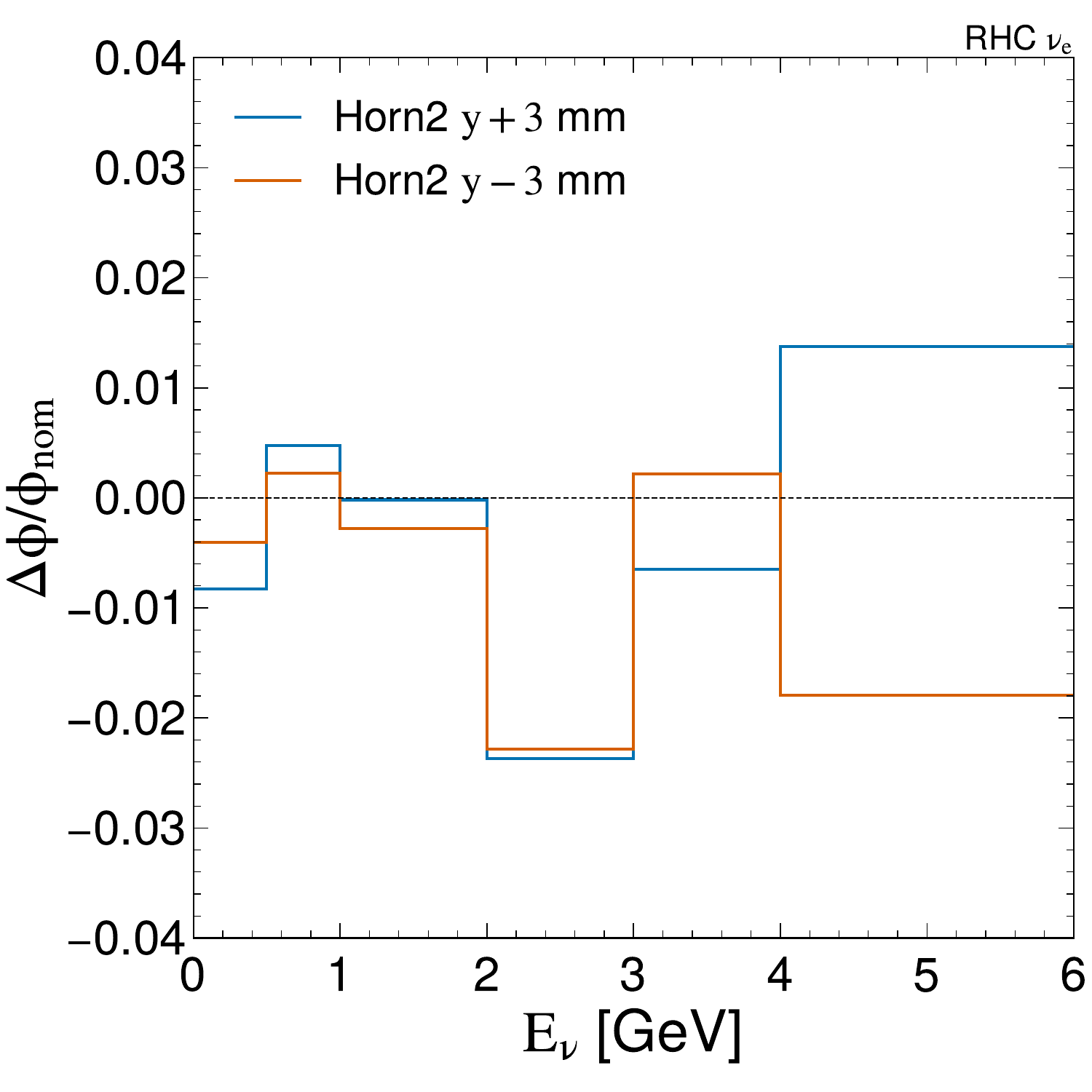}\\
    \includegraphics[width=0.25\textwidth]{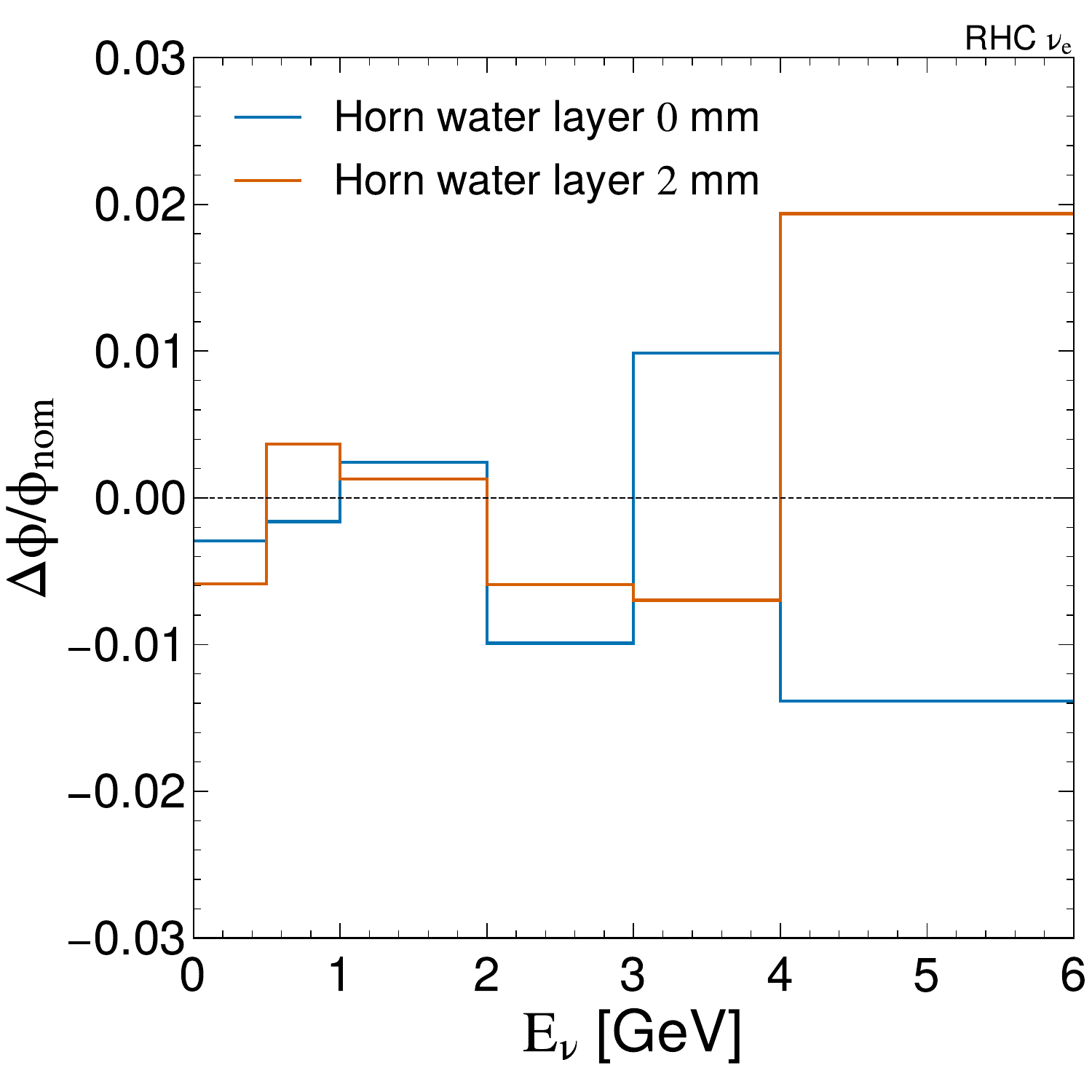}
    \includegraphics[width=0.25\textwidth]{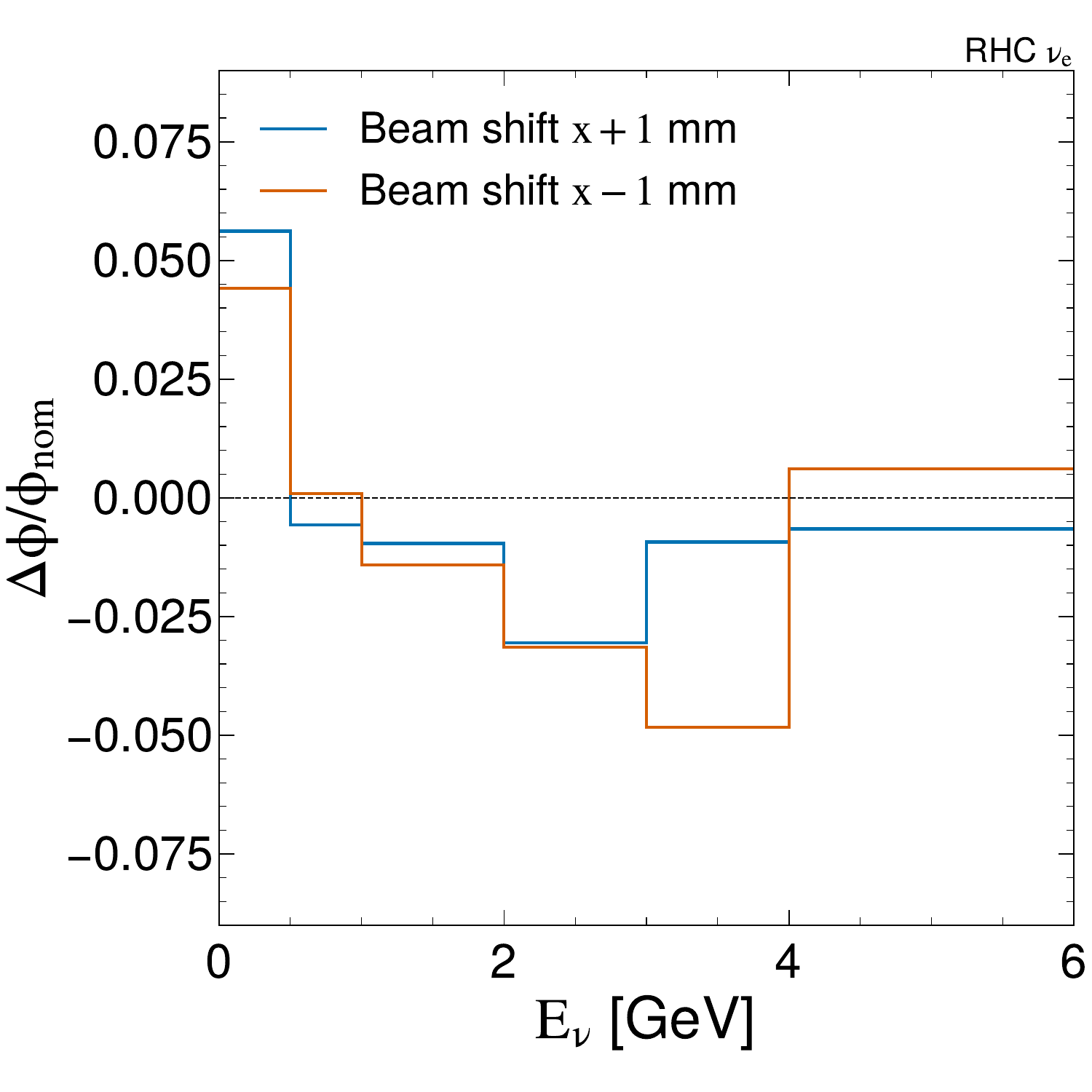}
    \includegraphics[width=0.25\textwidth]{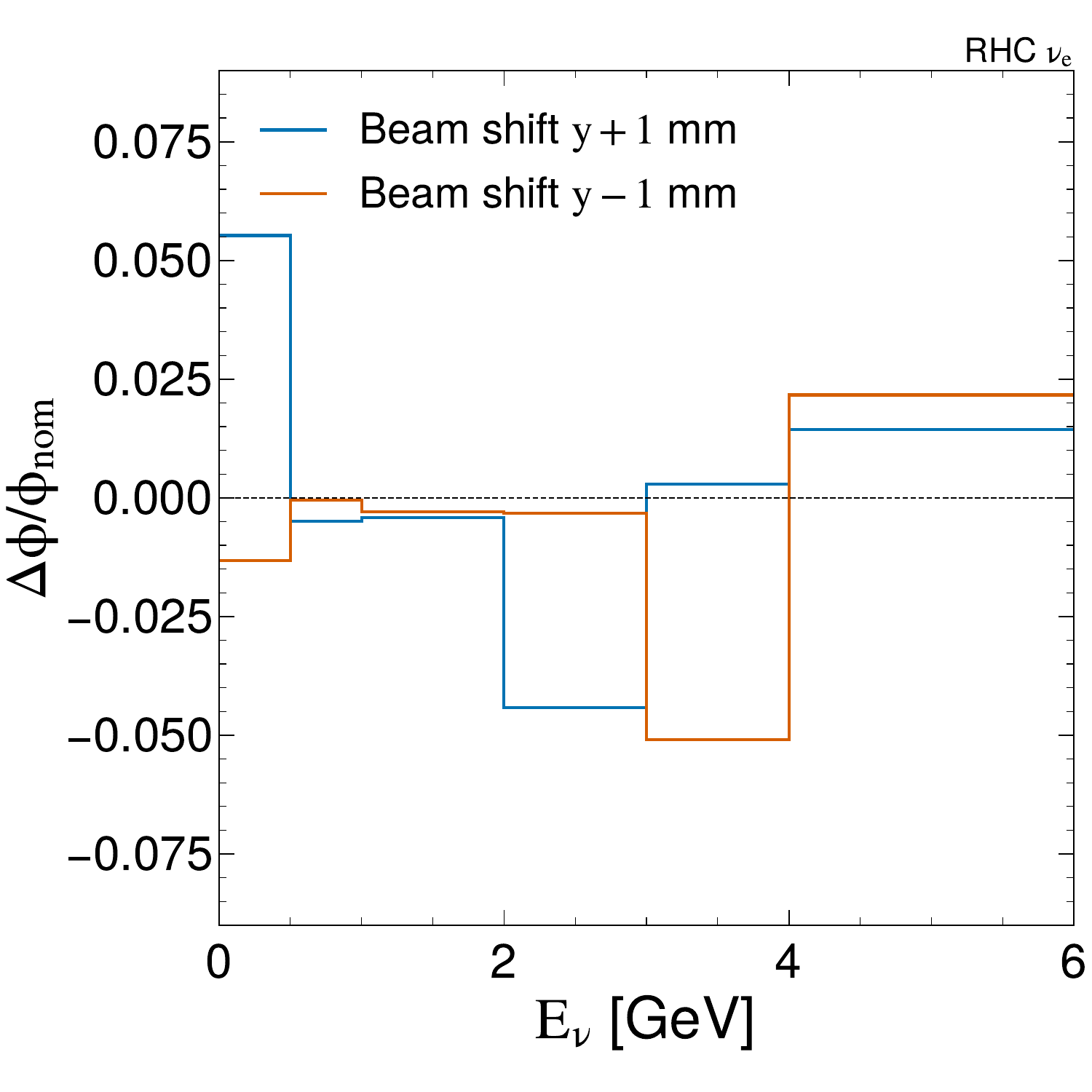}\\
    \includegraphics[width=0.25\textwidth]{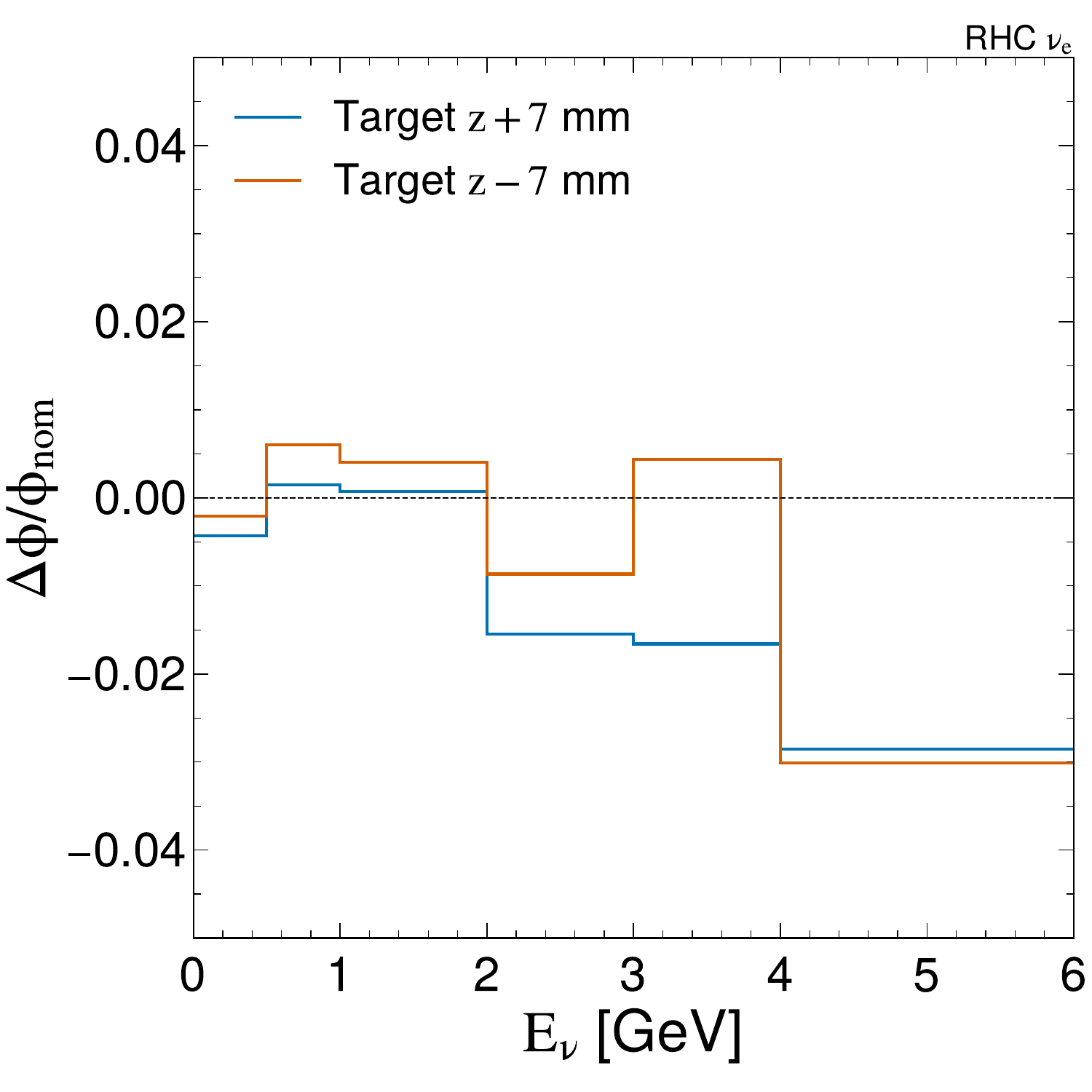}
    \caption[Beam Focusing Systematic Shifts (RHC, \nue)]{Beam focusing systematic shifts in the fractional scale (RHC, \nue).}
\end{figure}
\begin{figure}[!ht]
    \centering
    \includegraphics[width=0.25\textwidth]{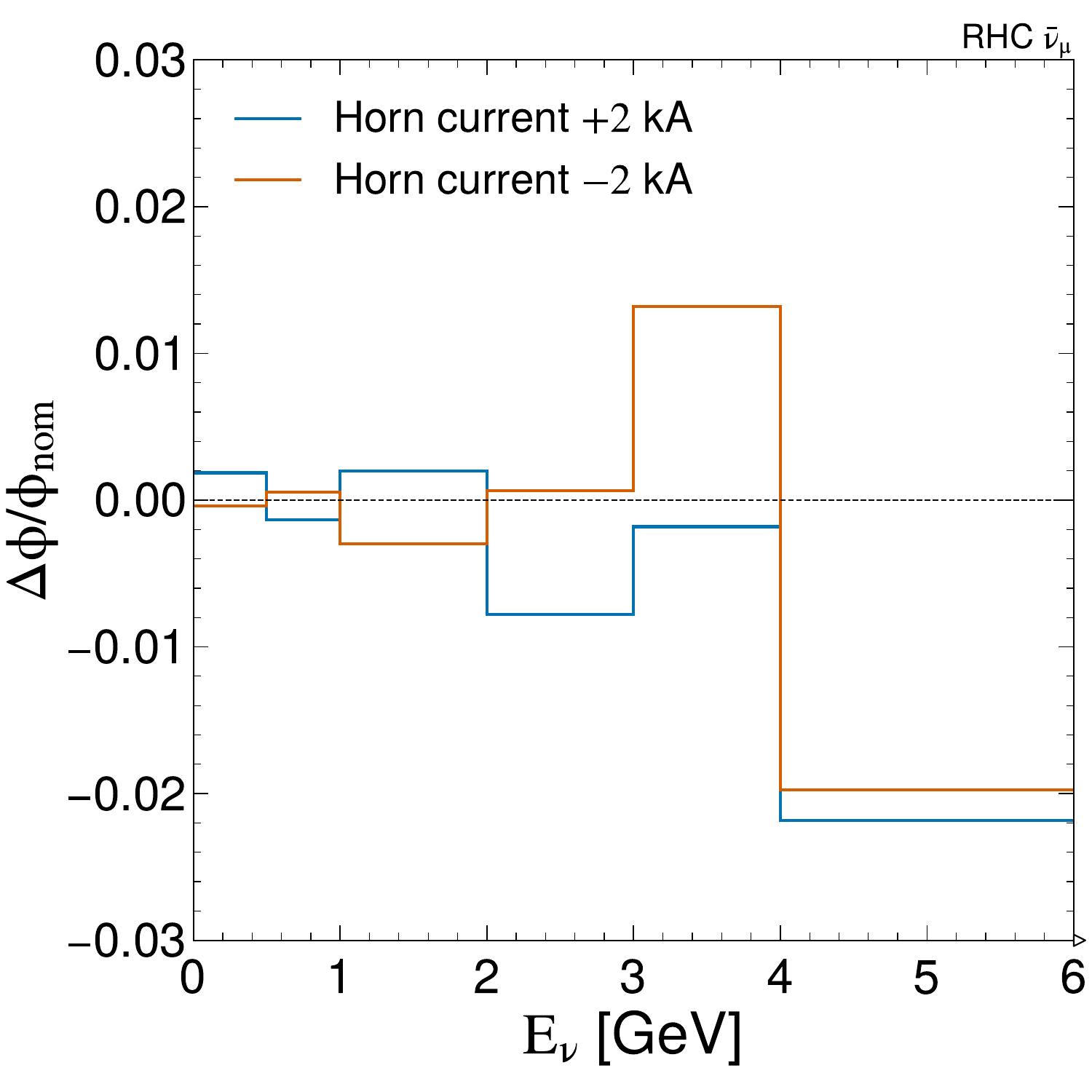}
    \includegraphics[width=0.25\textwidth]{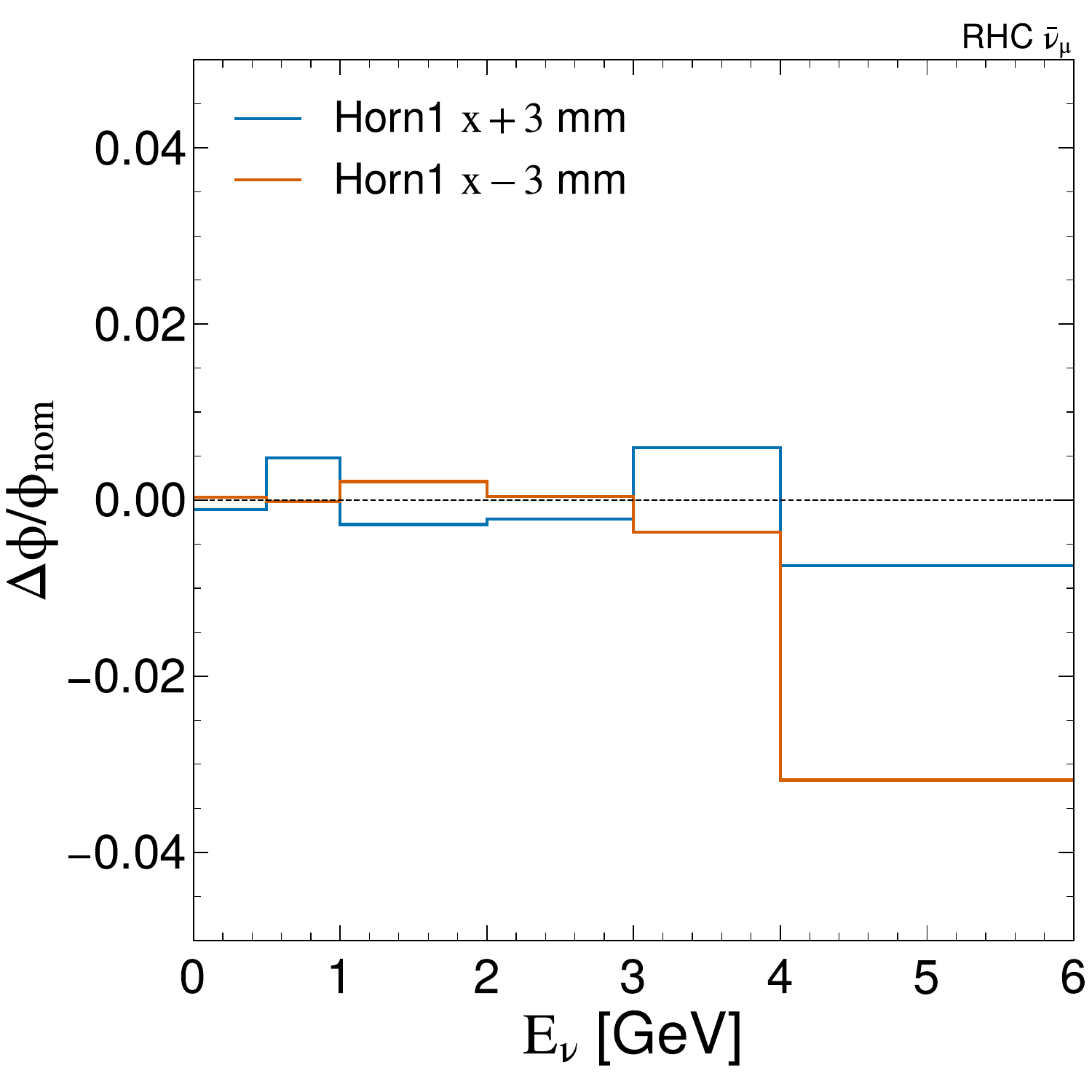}
    \includegraphics[width=0.25\textwidth]{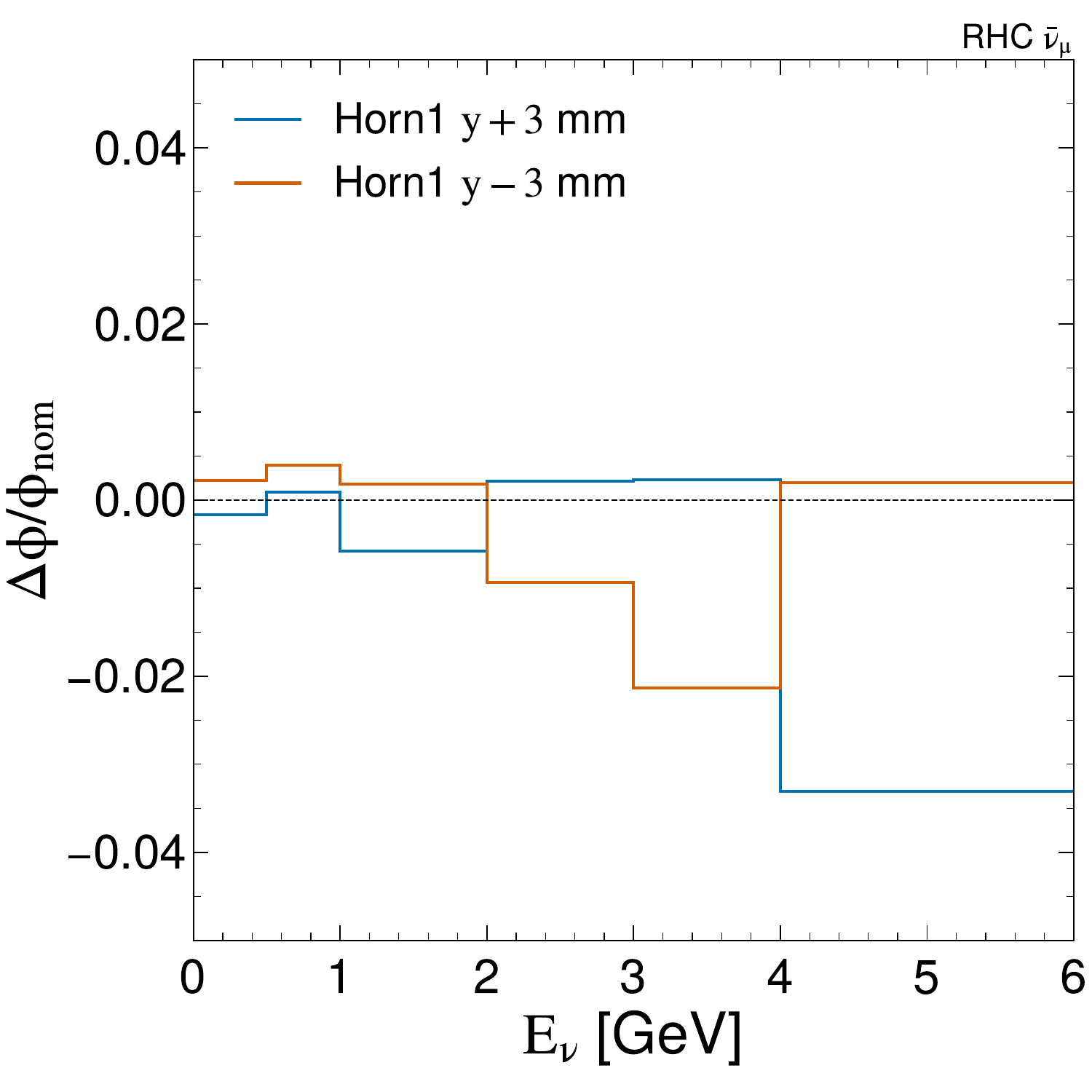}\\
    \includegraphics[width=0.25\textwidth]{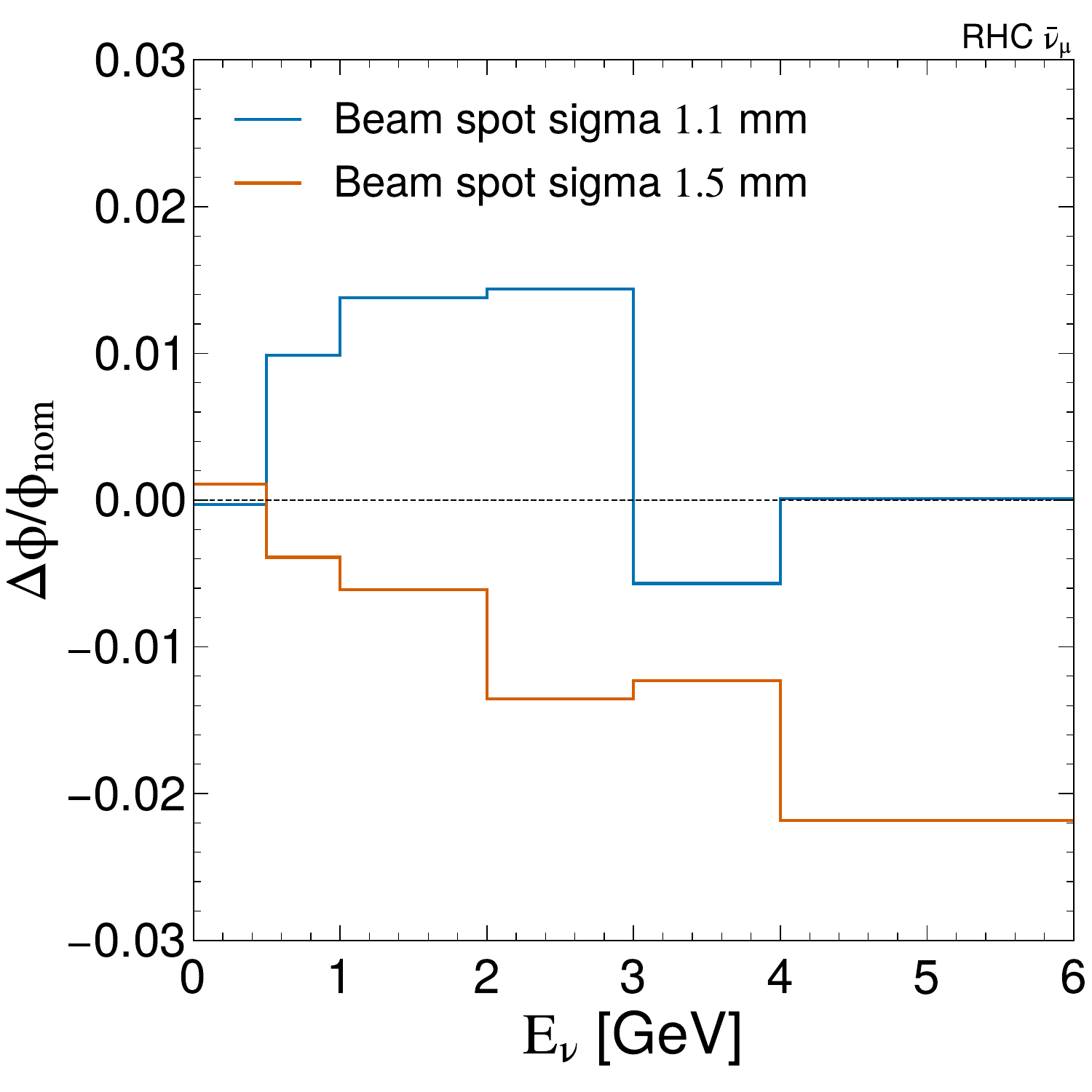}
    \includegraphics[width=0.25\textwidth]{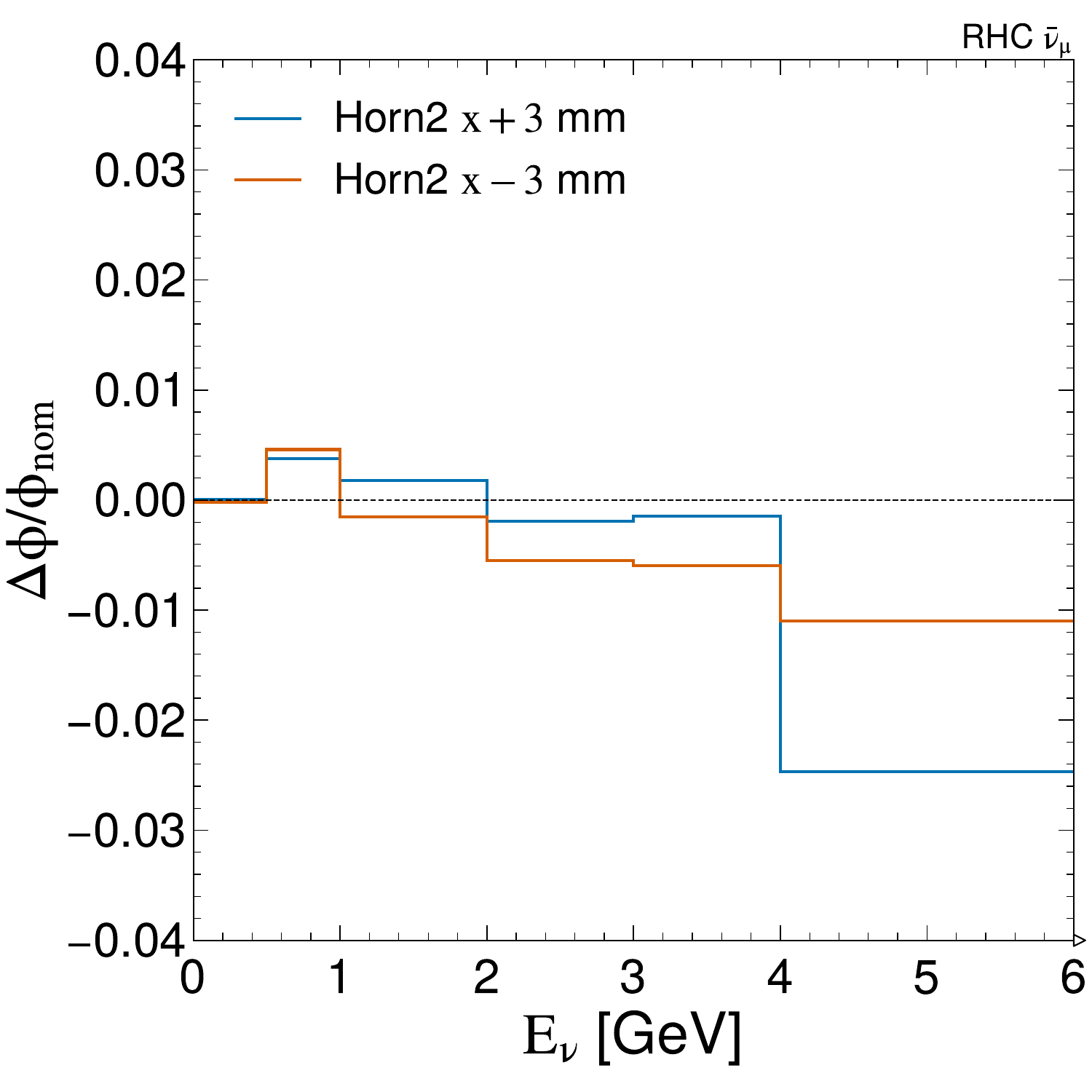}
    \includegraphics[width=0.25\textwidth]{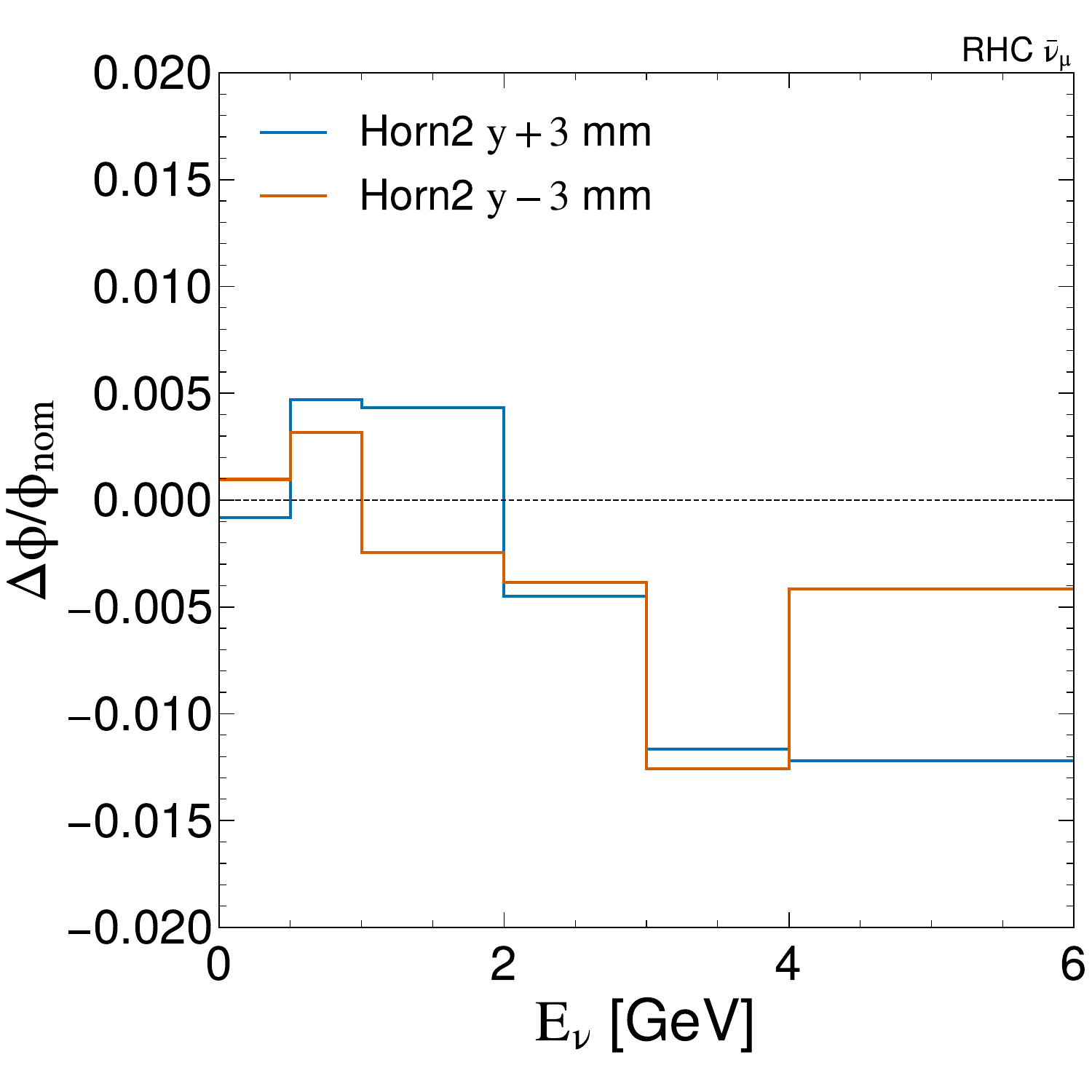}\\
    \includegraphics[width=0.25\textwidth]{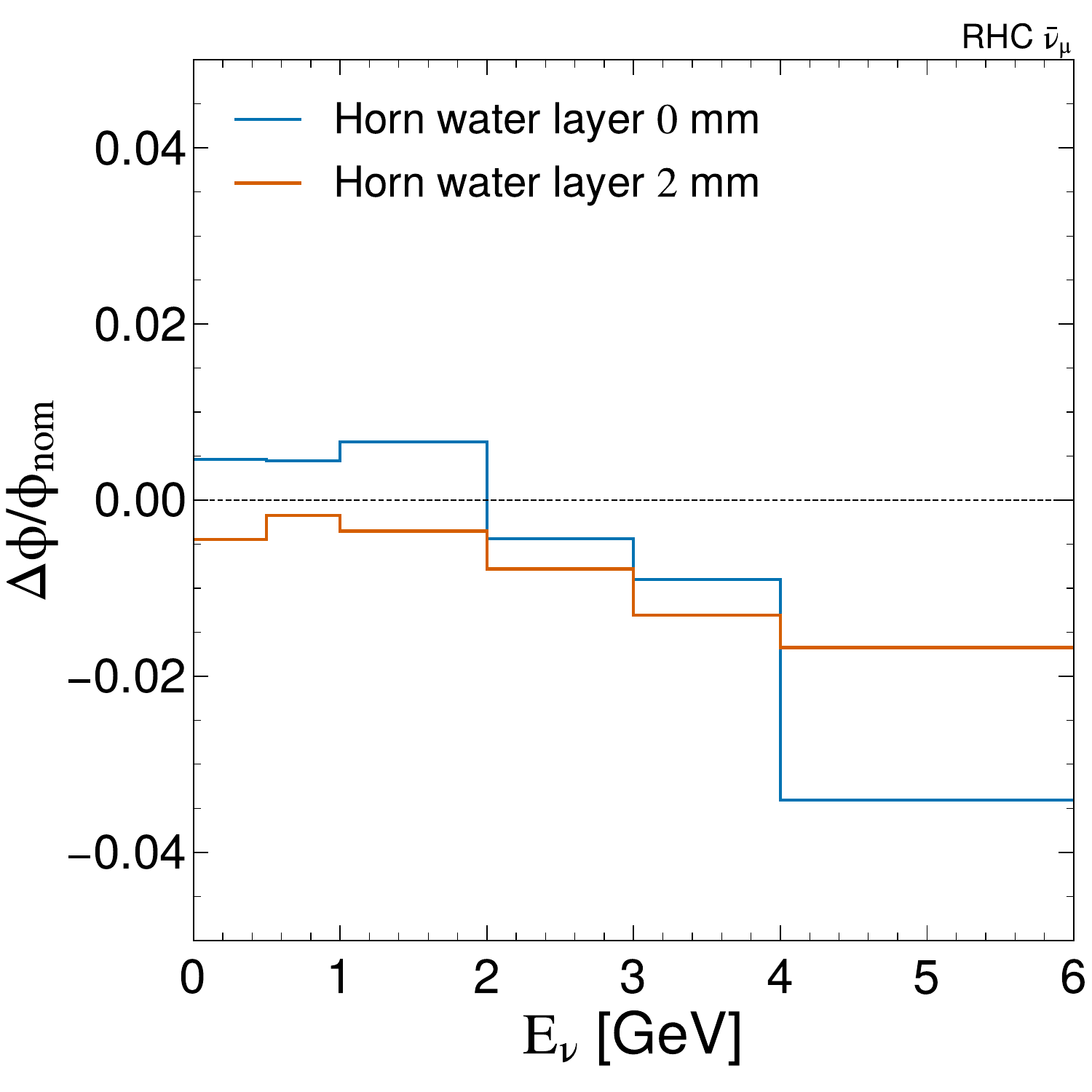}
    \includegraphics[width=0.25\textwidth]{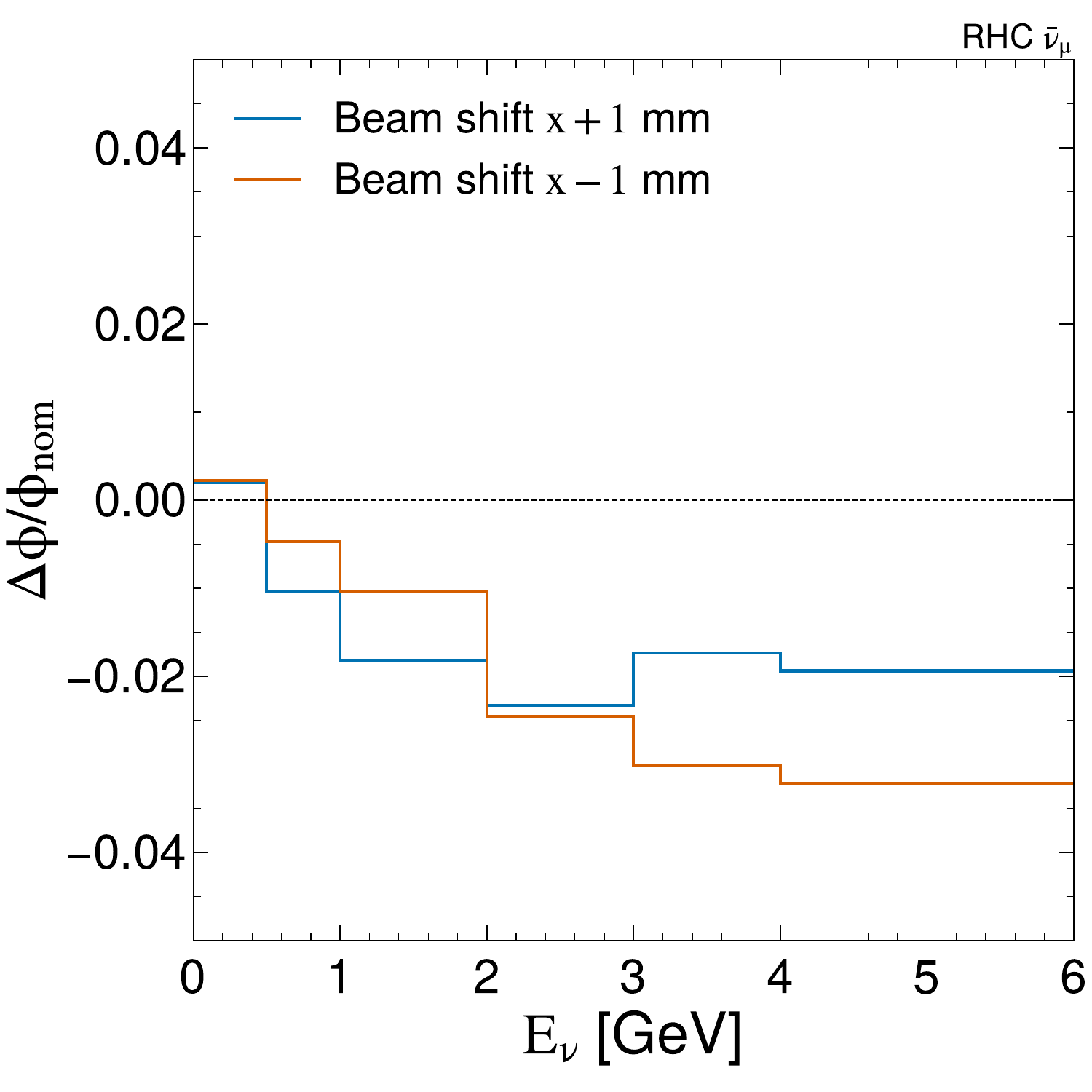}
    \includegraphics[width=0.25\textwidth]{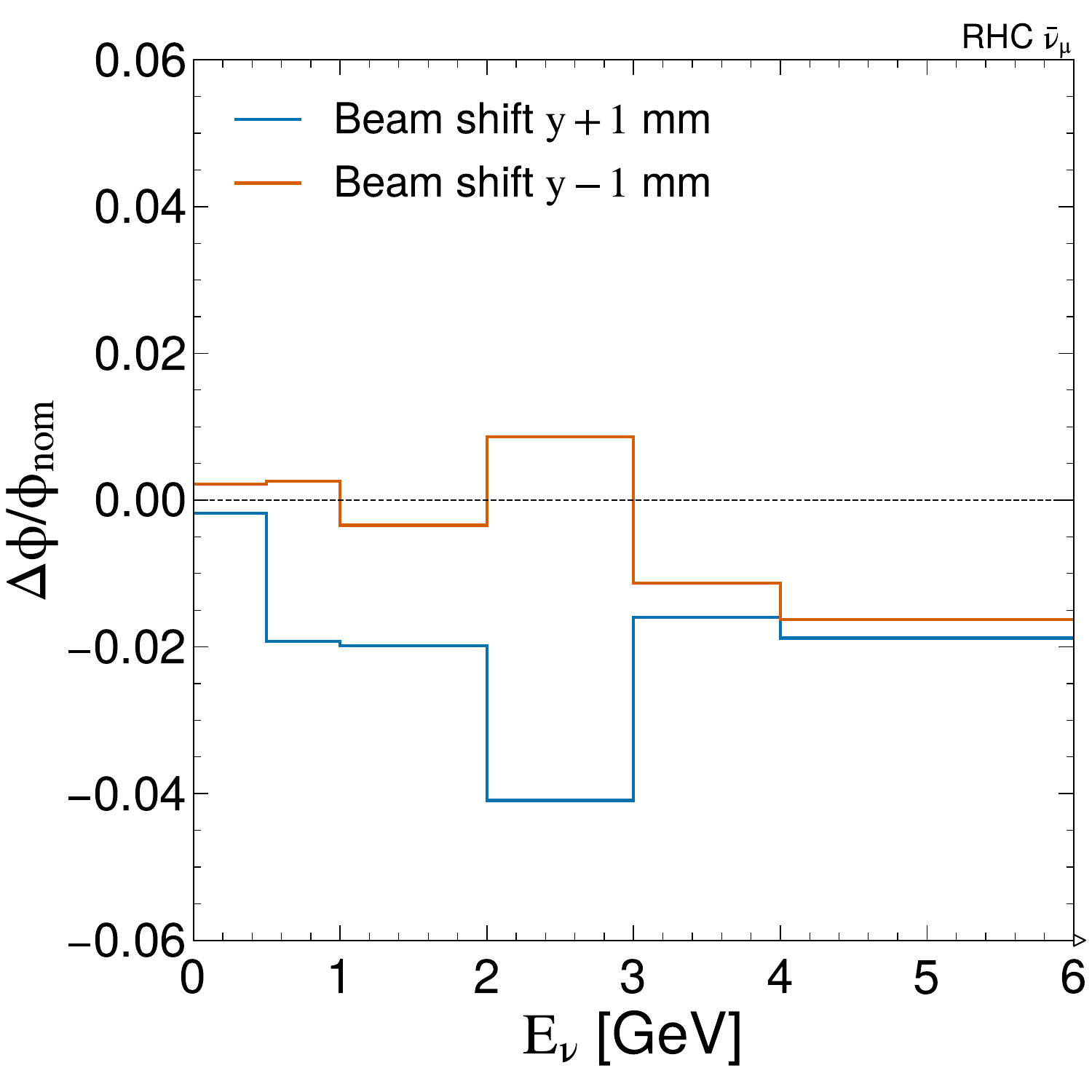}\\
    \includegraphics[width=0.25\textwidth]{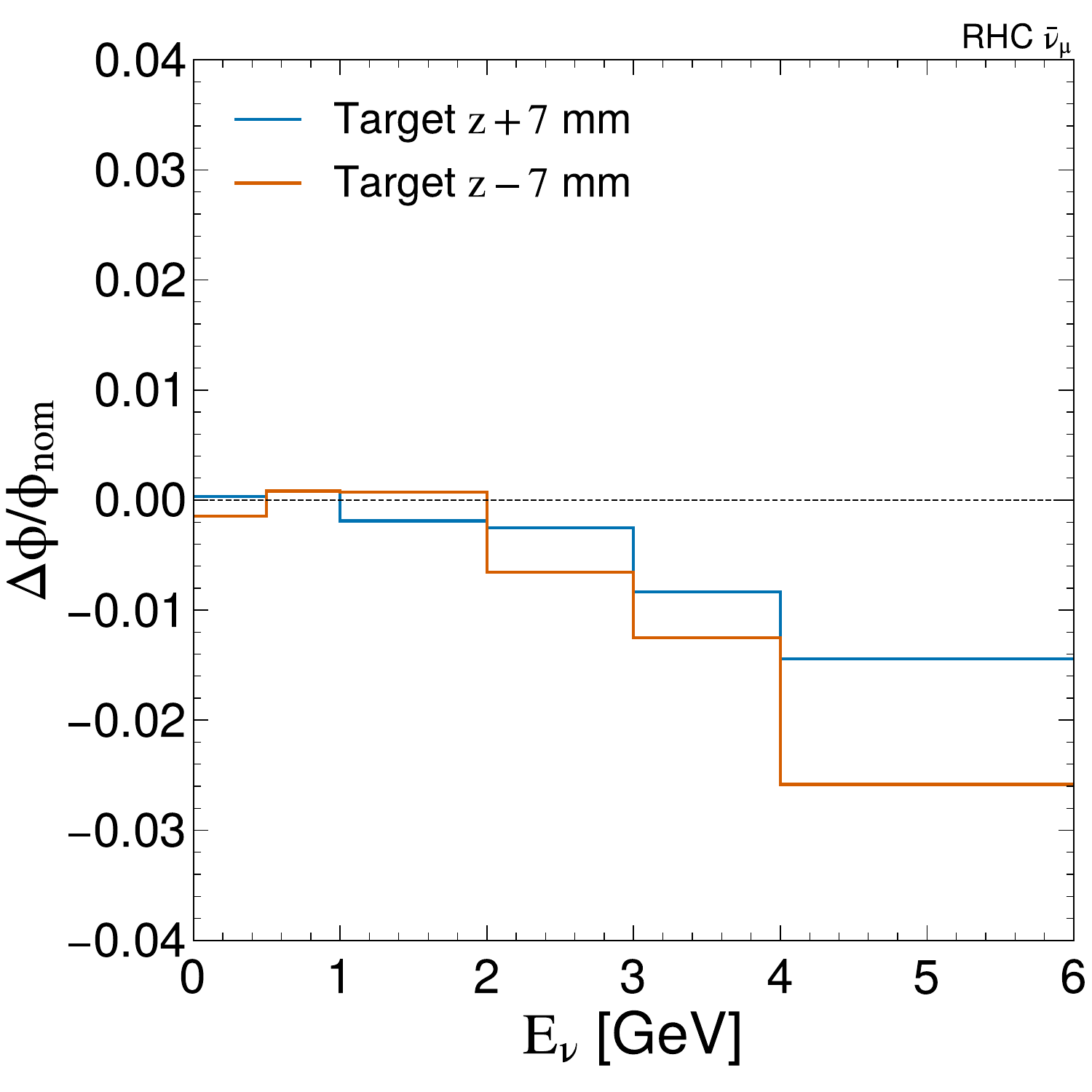}
    \caption[Beam Focusing Systematic Shifts (RHC, \numub)]{Beam focusing systematic shifts in the fractional scale (RHC, \numub).}
\end{figure}
\begin{figure}[!ht]
    \centering
    \includegraphics[width=0.25\textwidth]{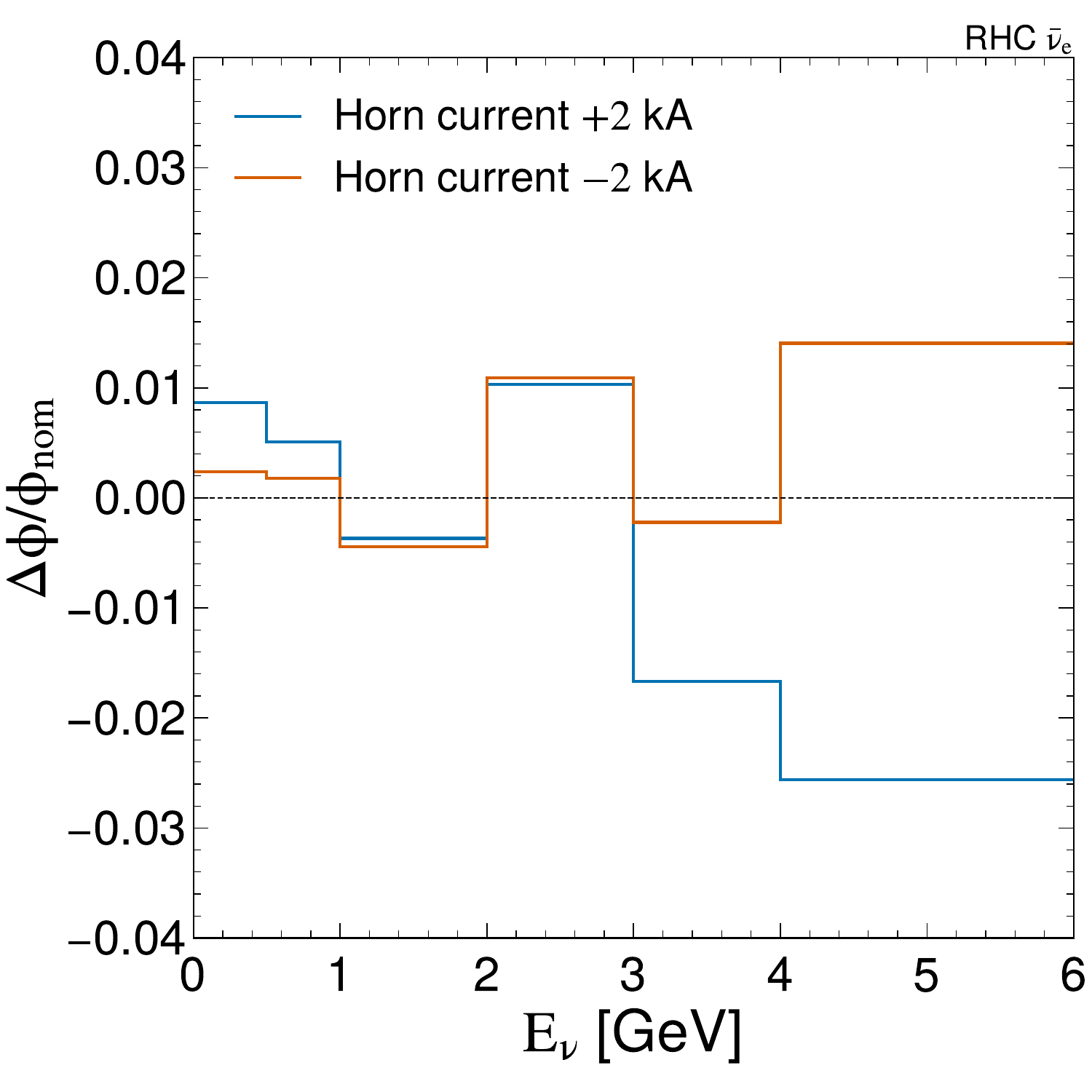}
    \includegraphics[width=0.25\textwidth]{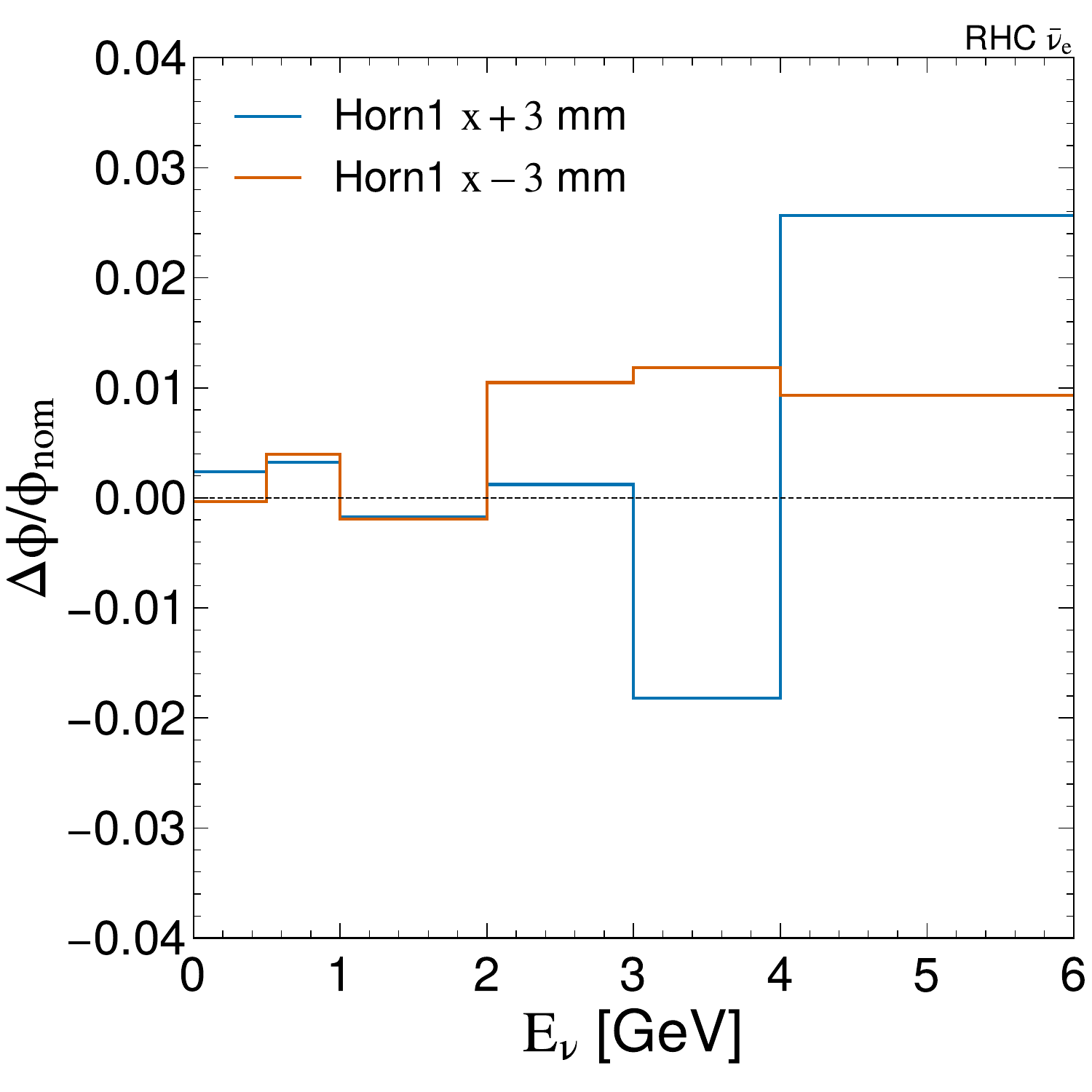}
    \includegraphics[width=0.25\textwidth]{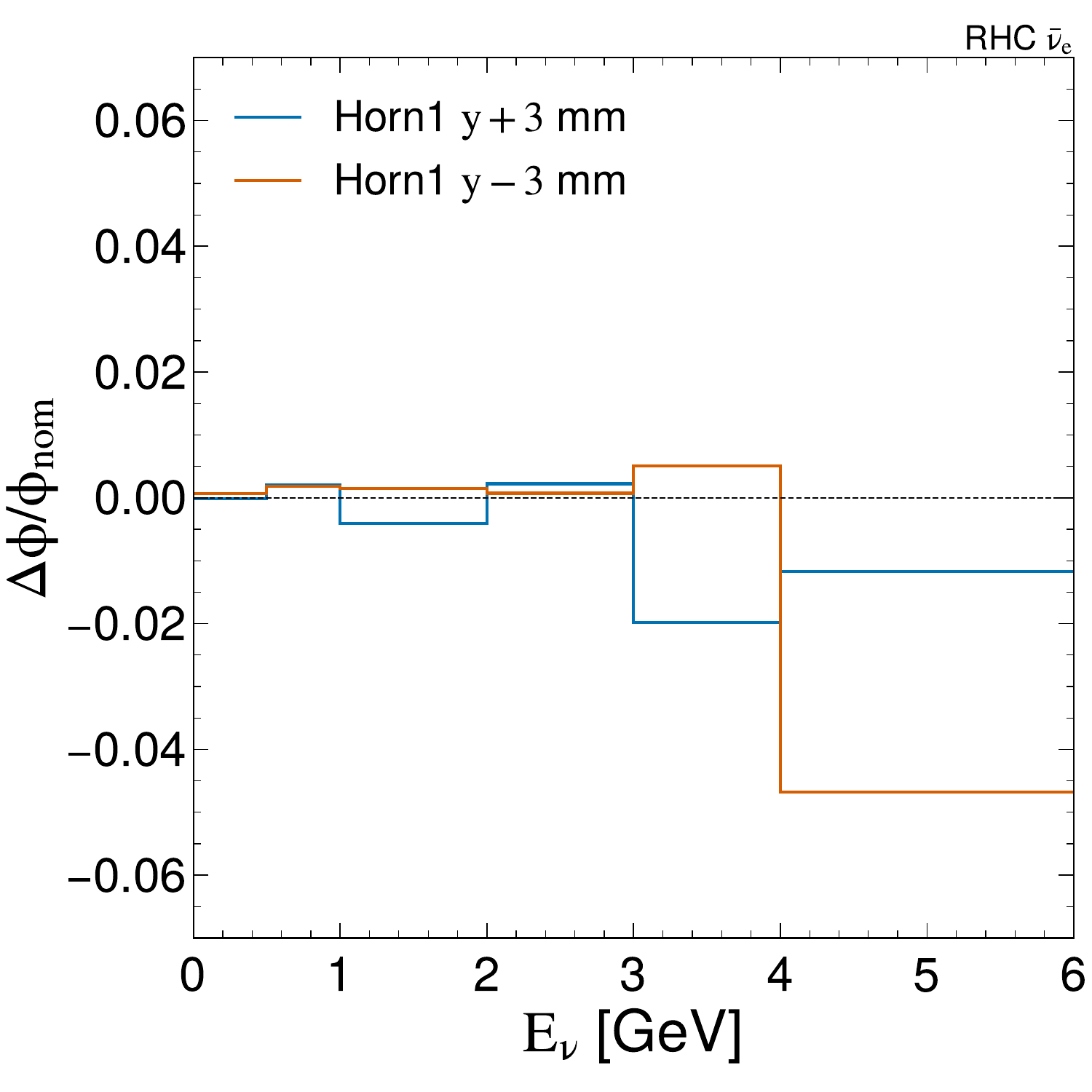}\\
    \includegraphics[width=0.25\textwidth]{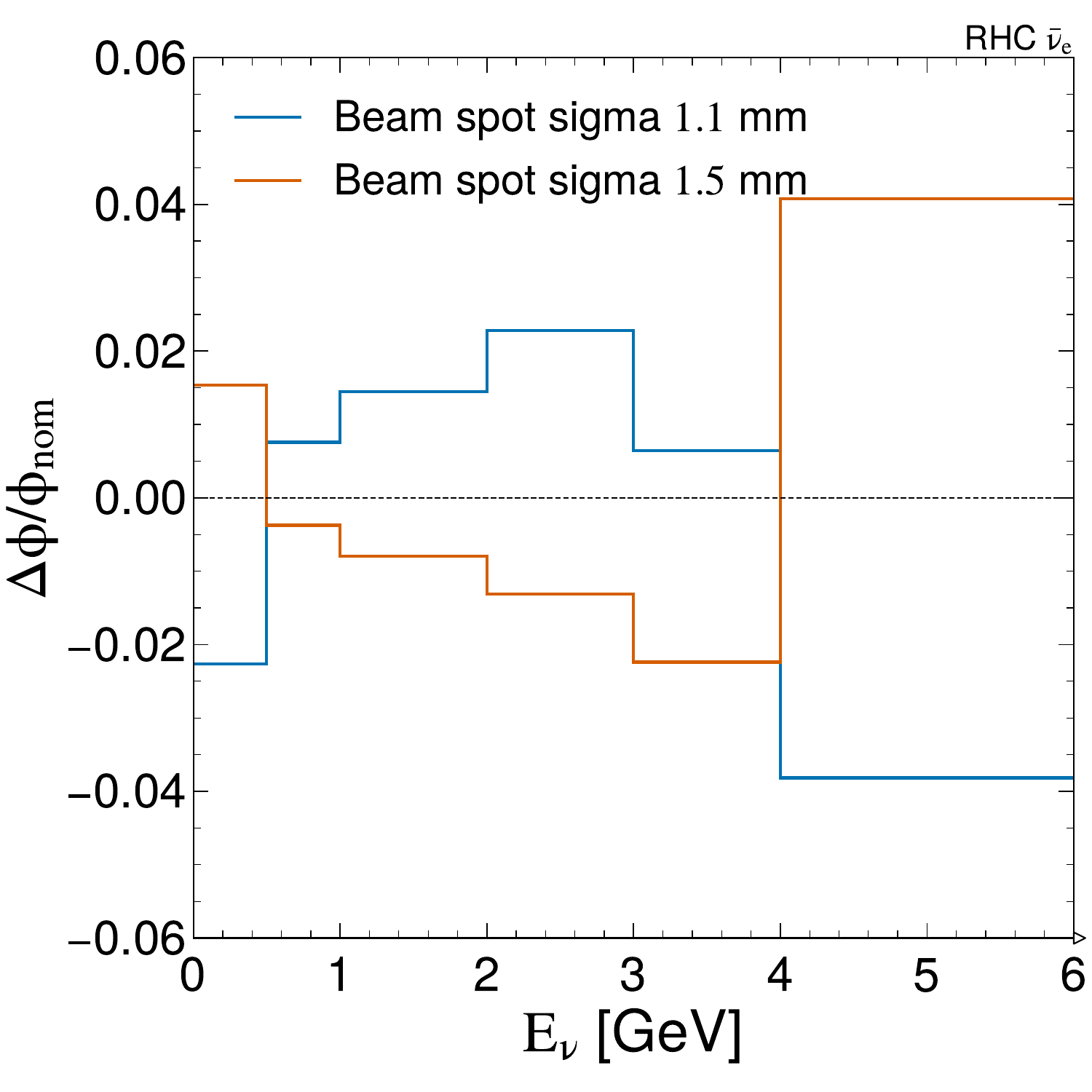}
    \includegraphics[width=0.25\textwidth]{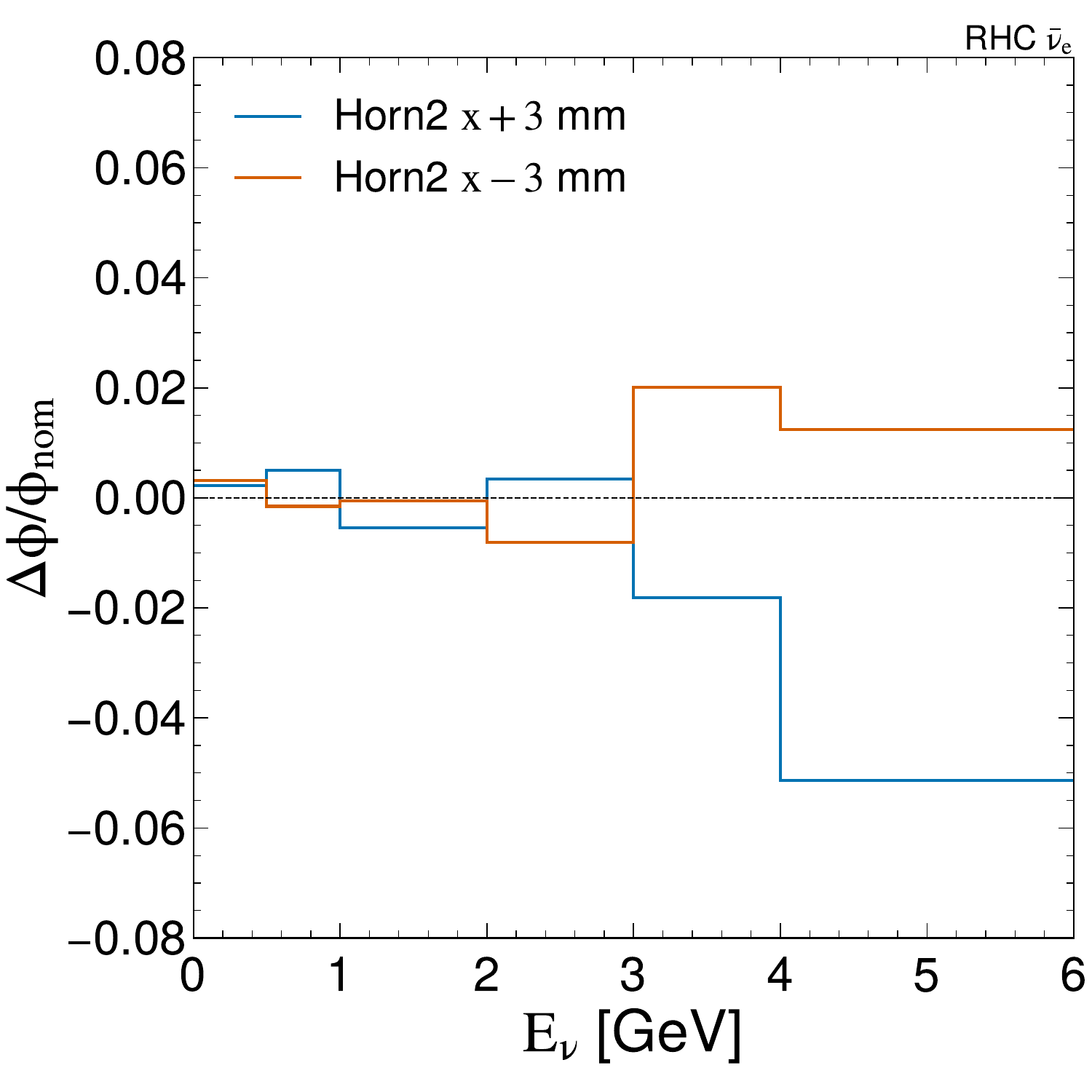}
    \includegraphics[width=0.25\textwidth]{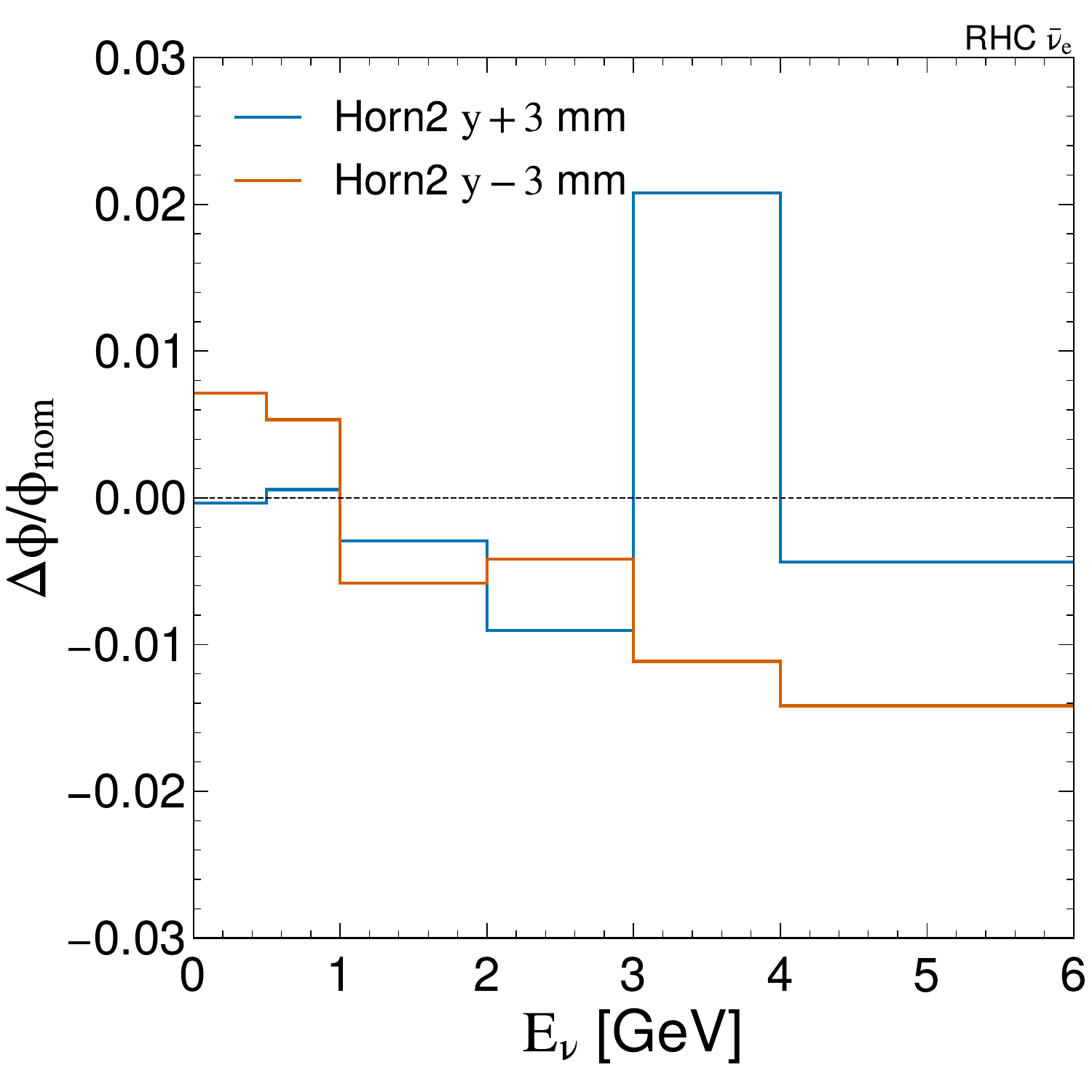}\\
    \includegraphics[width=0.25\textwidth]{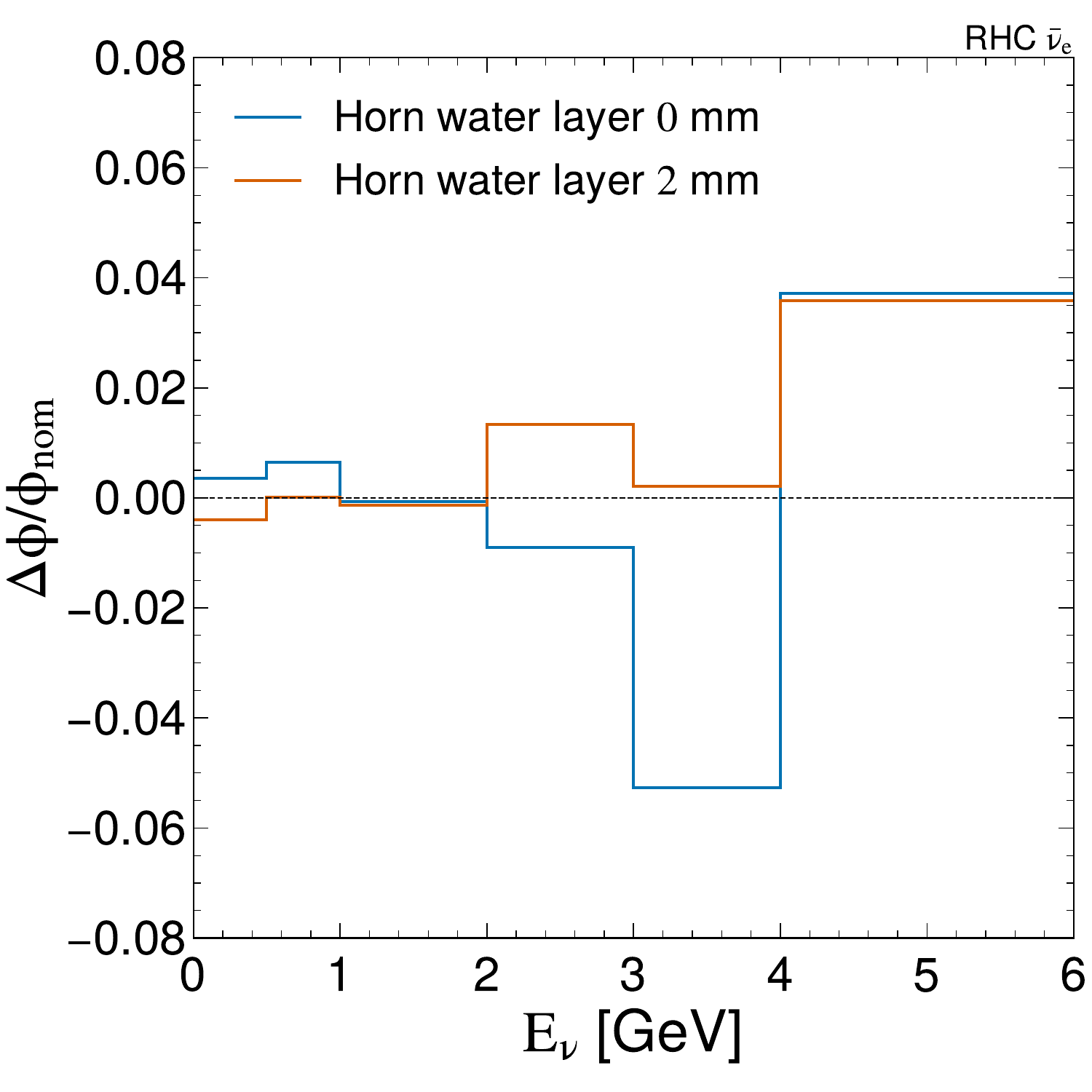}
    \includegraphics[width=0.25\textwidth]{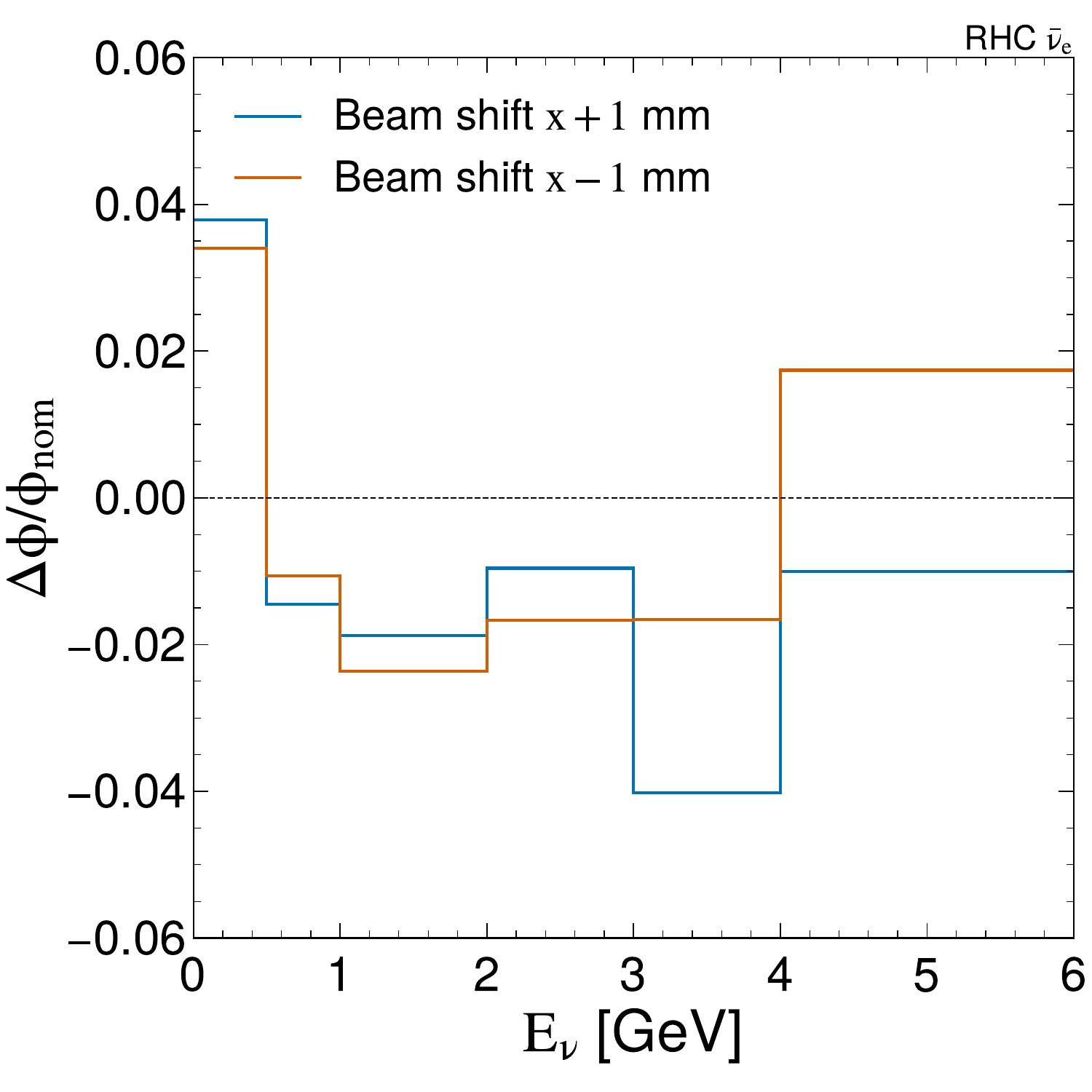}
    \includegraphics[width=0.25\textwidth]{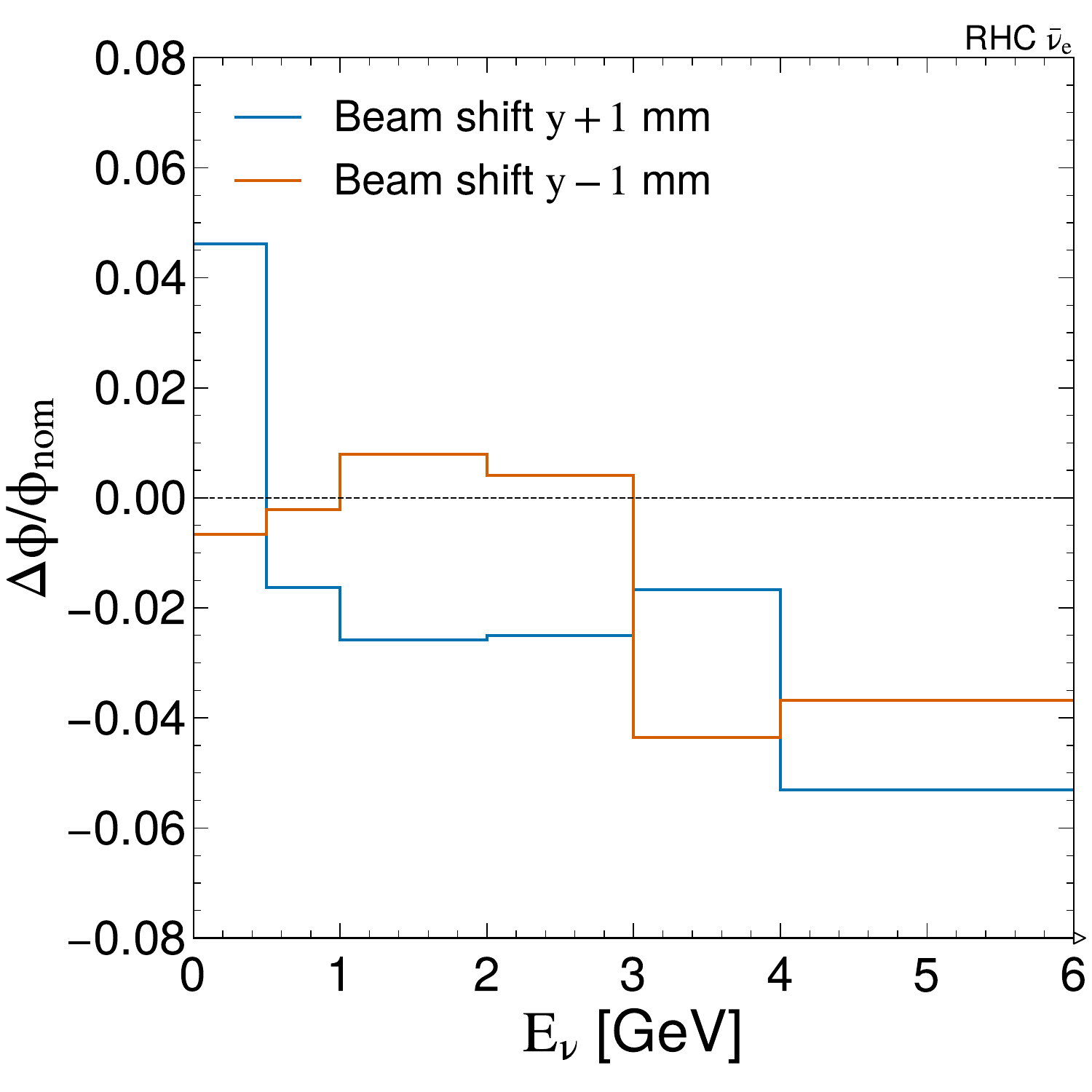}\\
    \includegraphics[width=0.25\textwidth]{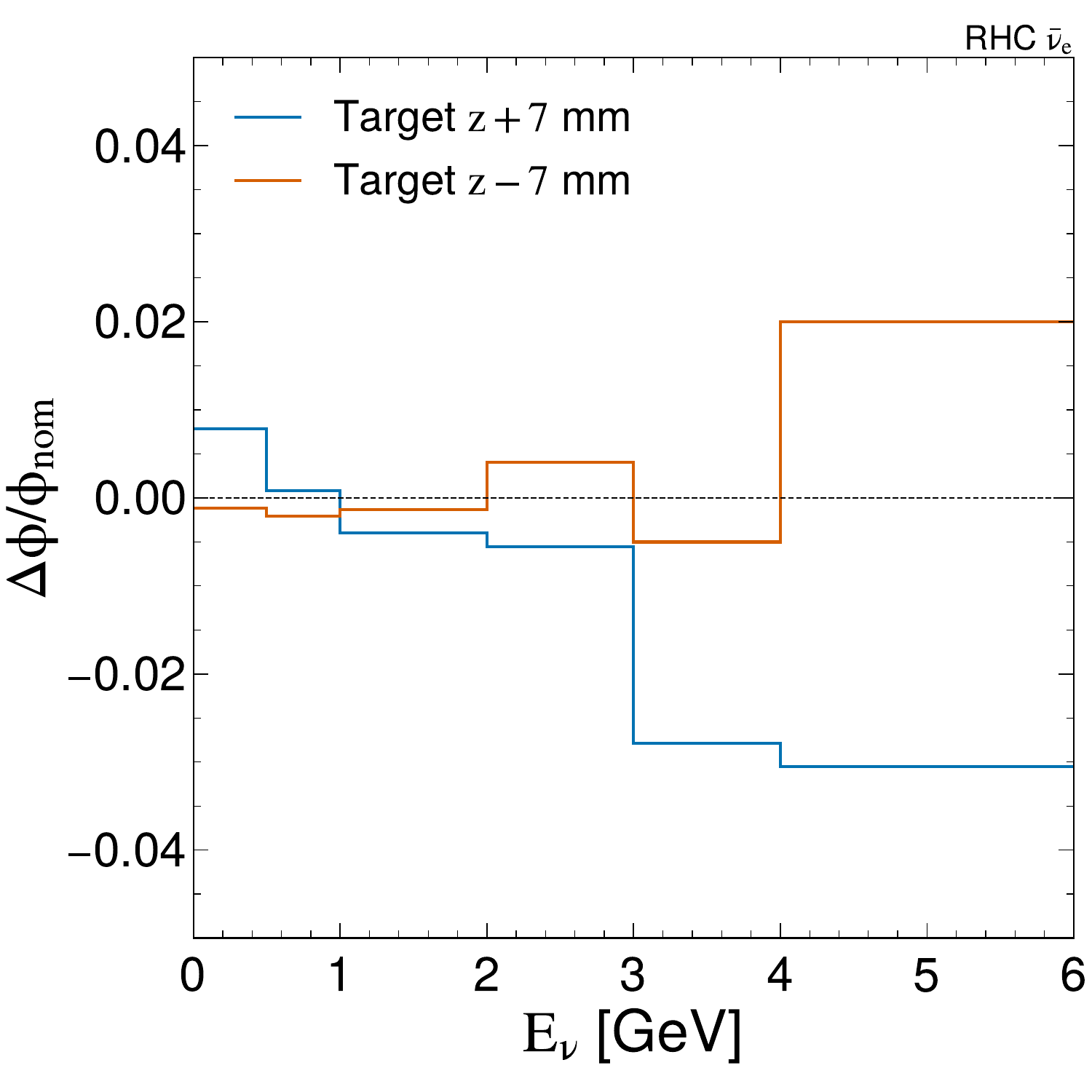}
    \caption[Beam Focusing Systematic Shifts (RHC, \nueb)]{Beam focusing systematic shifts in the fractional scale (RHC, \nueb).}
\end{figure}

%% file: beam_frac_uncertainties.tex
\clearpage
\section{Forward Horn Current}
\begin{figure}[!ht]
    \centering
    \includegraphics[width=0.48\textwidth]{UPDATED_fhc_numu_beam_fractional_uncertainties.pdf}
    \includegraphics[width=0.48\textwidth]{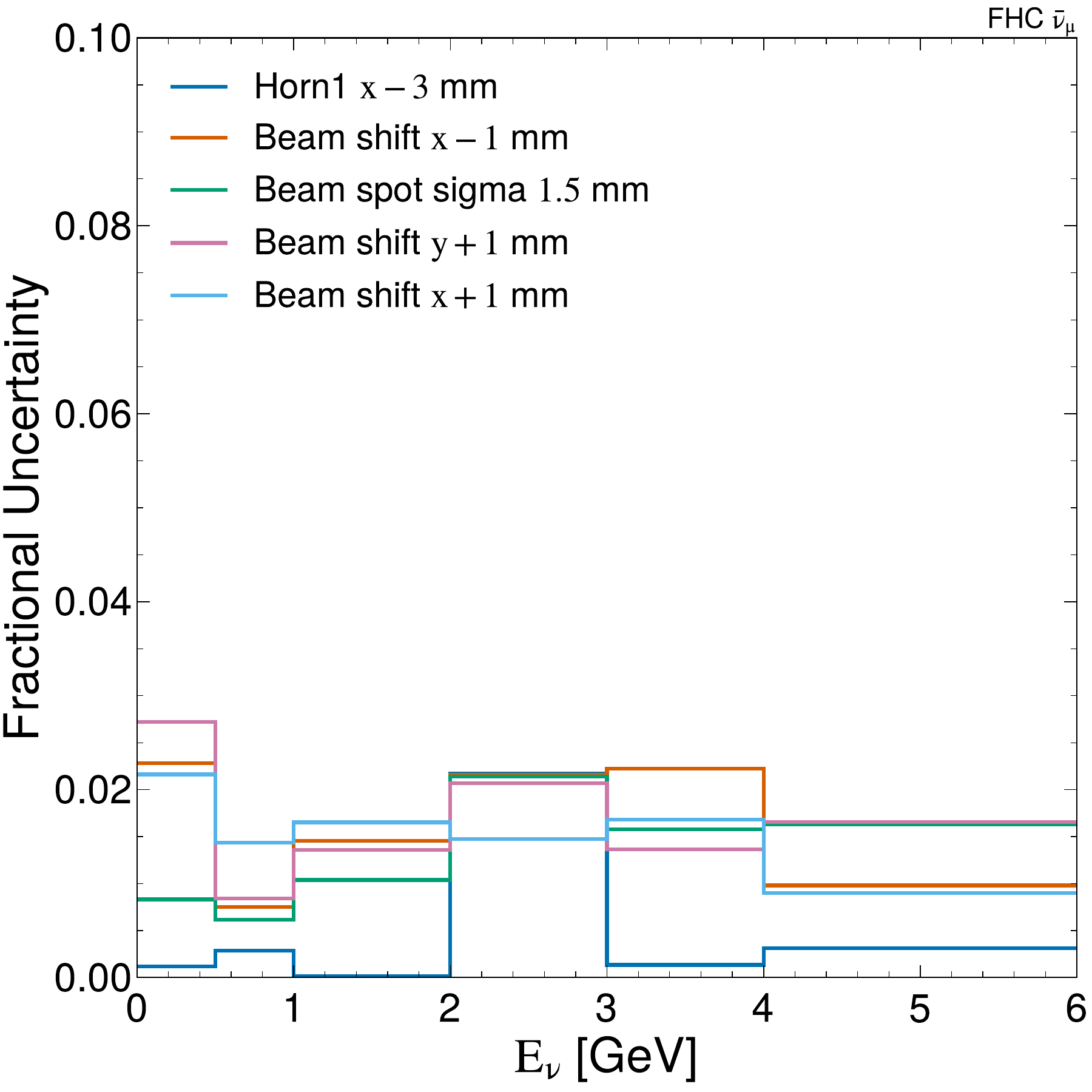}
    \includegraphics[width=0.48\textwidth]{UPDATED_fhc_nue_beam_fractional_uncertainties.pdf}
    \includegraphics[width=0.48\textwidth]{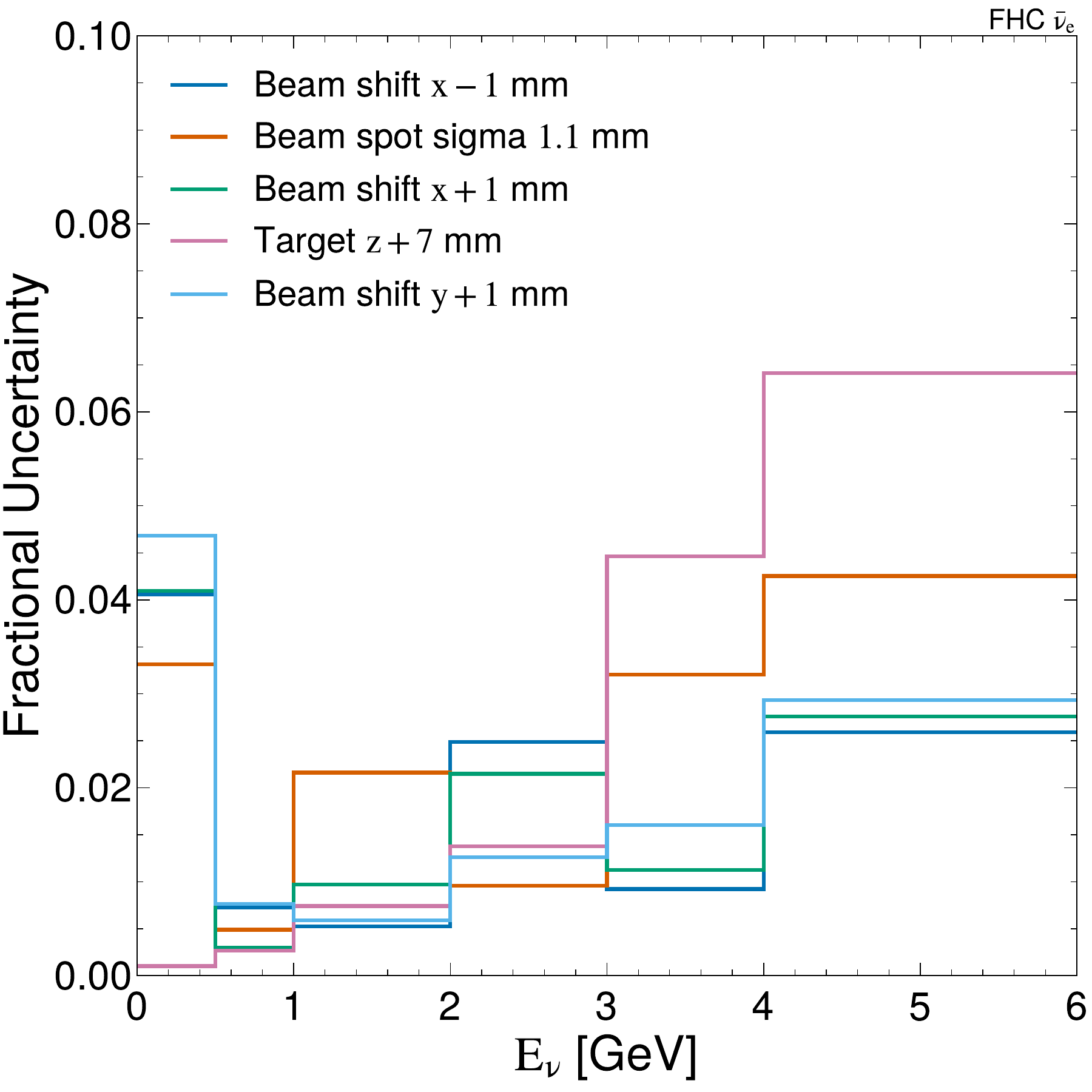}
    \caption[Beamline Focusing Systematic Uncertainties (FHC)]{Fractional uncertainties for the beam focusing systematics in the FHC mode.}
\end{figure}
\clearpage
\section{Reverse Horn Current}
\begin{figure}[!ht]
    \centering
    \includegraphics[width=0.48\textwidth]{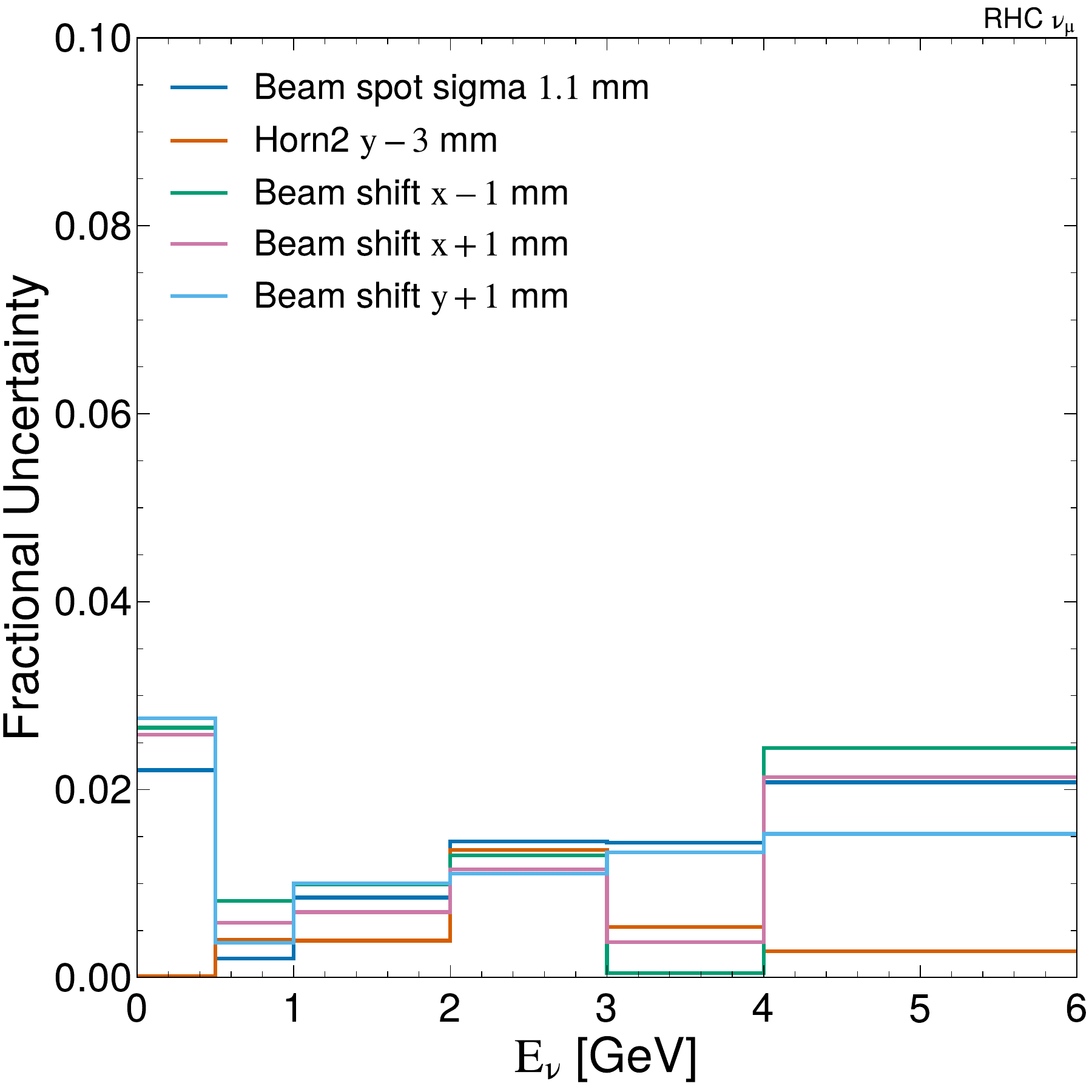}
    \includegraphics[width=0.48\textwidth]{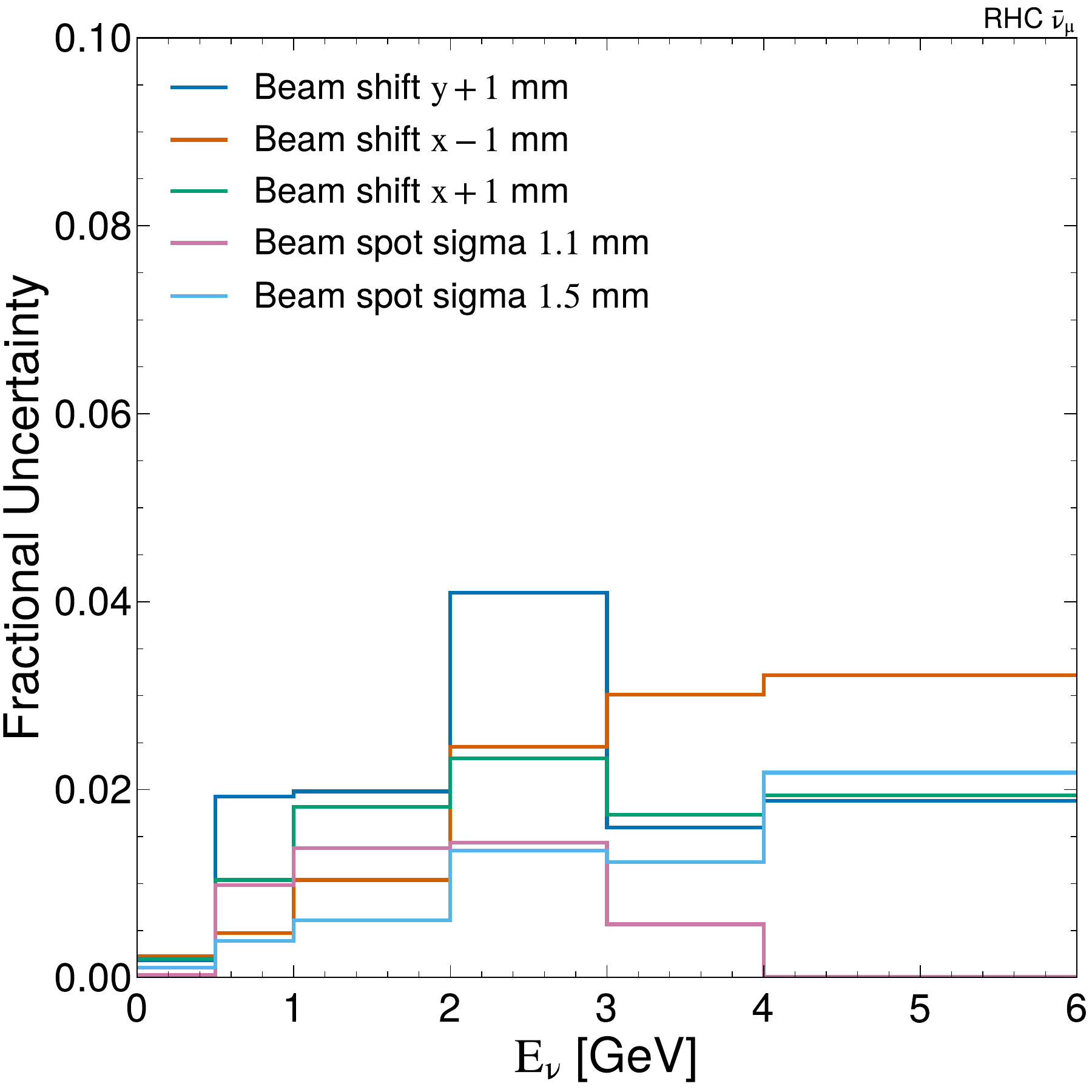}
    \includegraphics[width=0.48\textwidth]{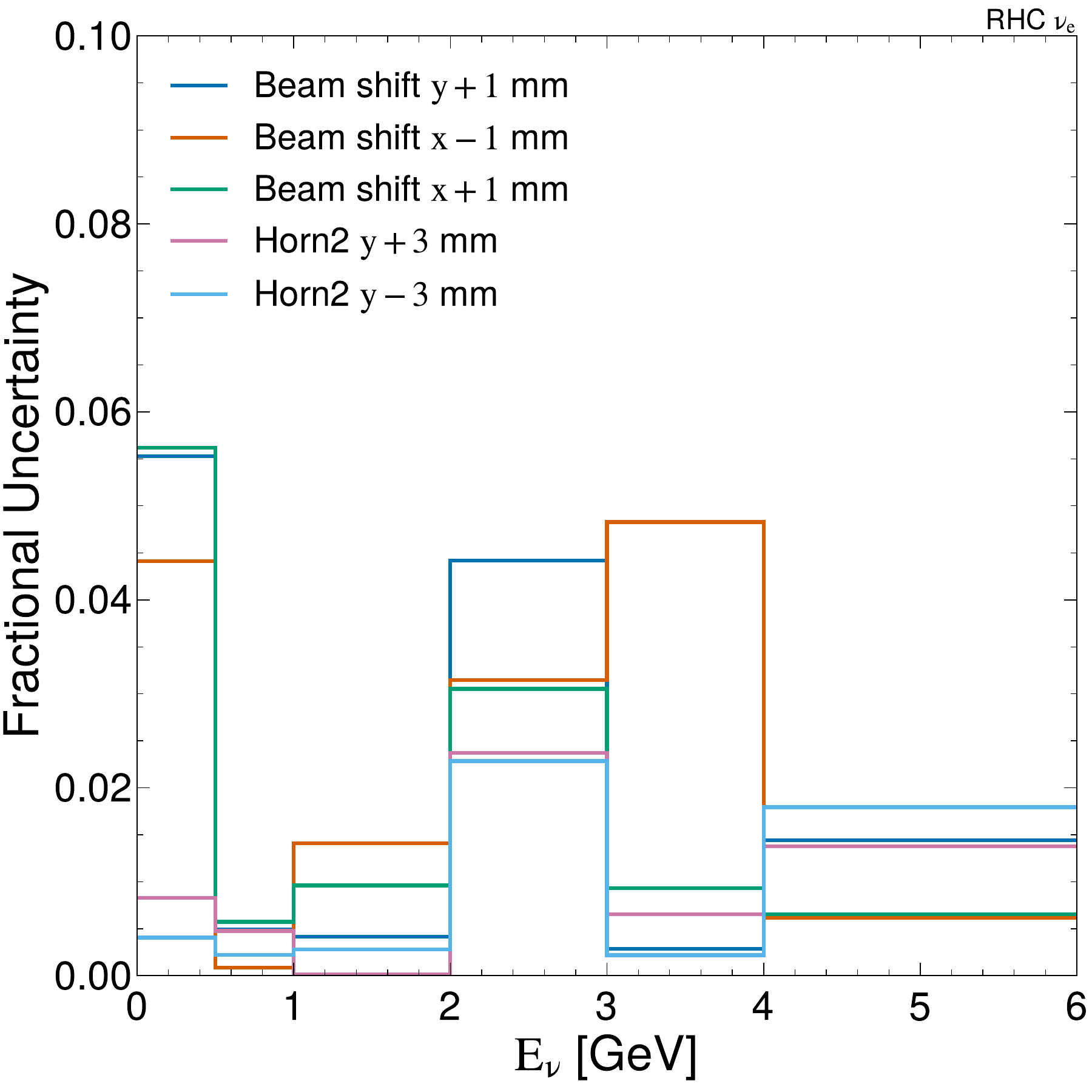}
    \includegraphics[width=0.48\textwidth]{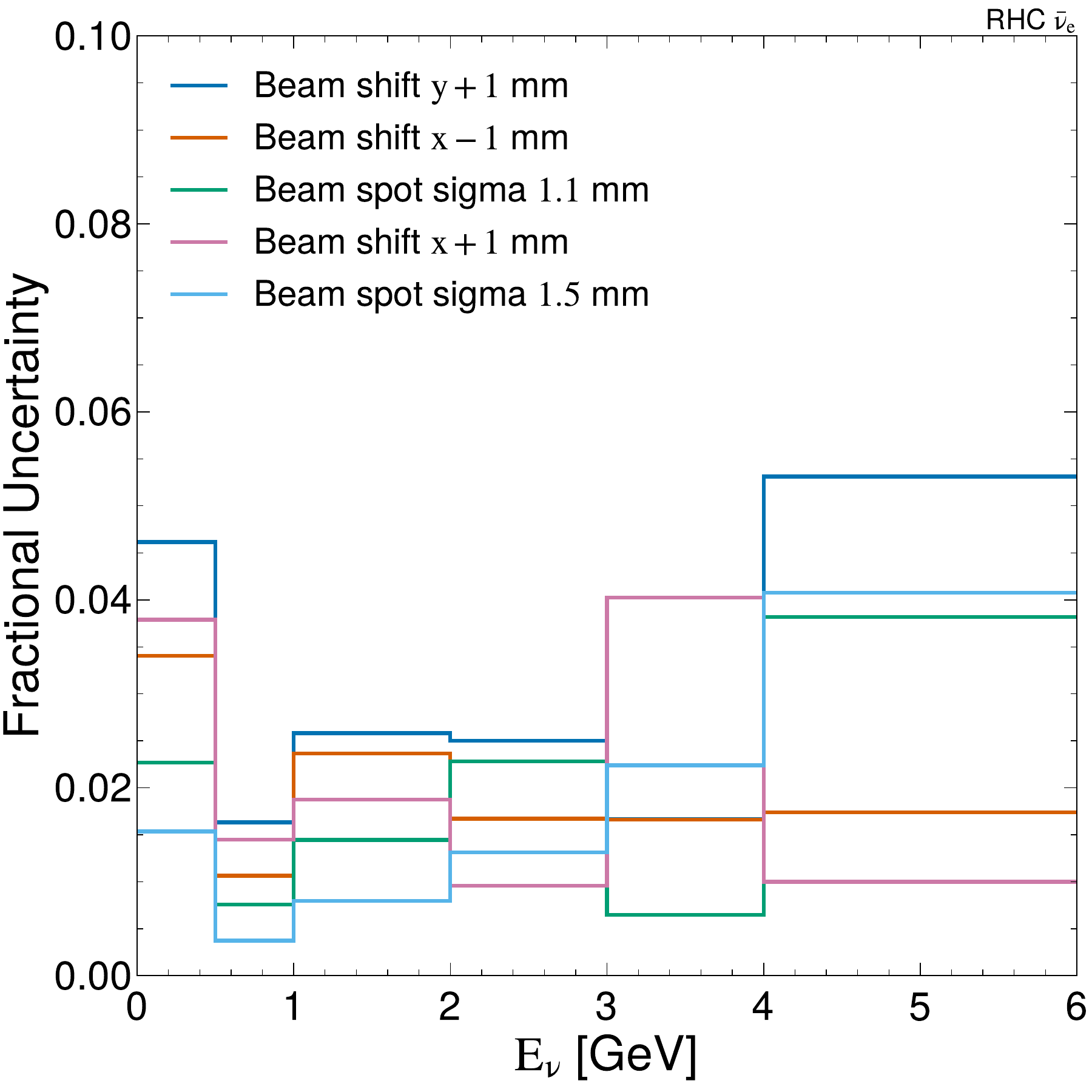}
    \caption[Beamline Focusing Systematic Uncertainties (RHC)]{Fractional uncertainties for the beam focusing systematics in the RHC mode.}
\end{figure}

%% file: beam_correlation_matrices.tex
\clearpage
\section{Covariance Matrices}
\begin{figure}[!ht]
    \centering
    \begin{subfigure}[]{0.27\textwidth}
\includegraphics[width=\textwidth]{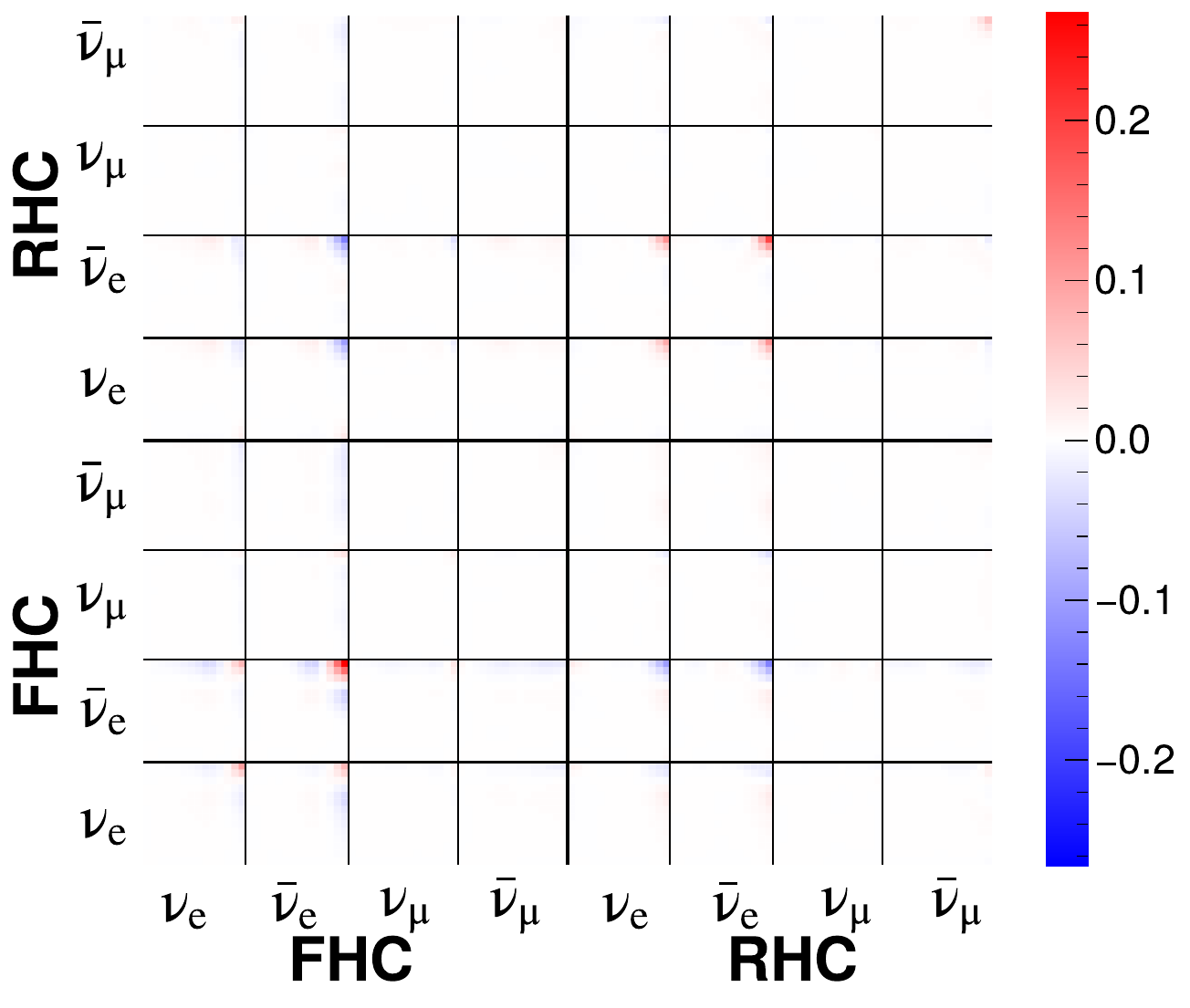}
        \caption{Total}
    \end{subfigure}
    \begin{subfigure}[]{0.27\textwidth}
\includegraphics[width=\textwidth]{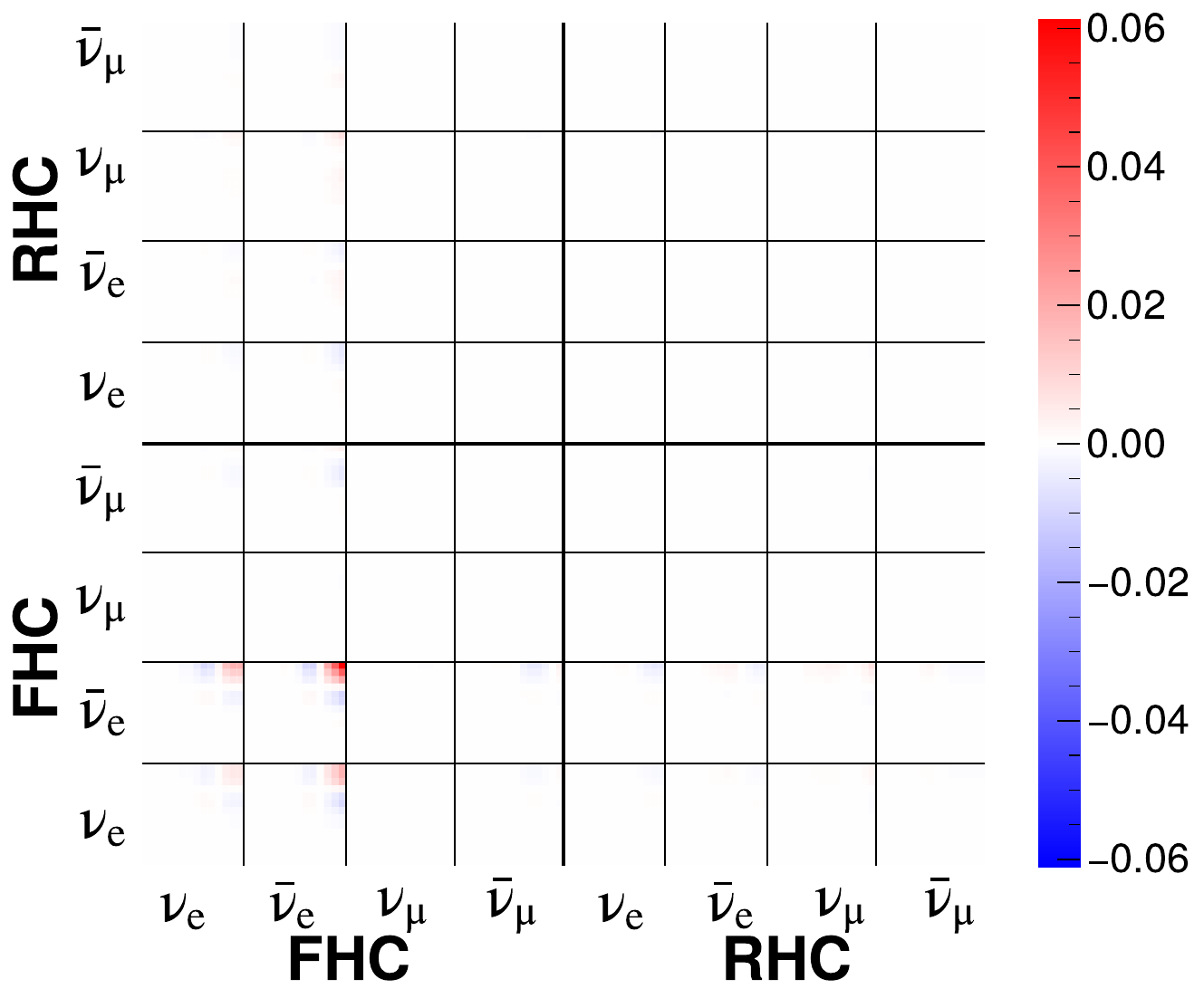}
        \caption{Beam Divergence}
    \end{subfigure}
    \begin{subfigure}[]{0.27\textwidth}
\includegraphics[width=\textwidth]{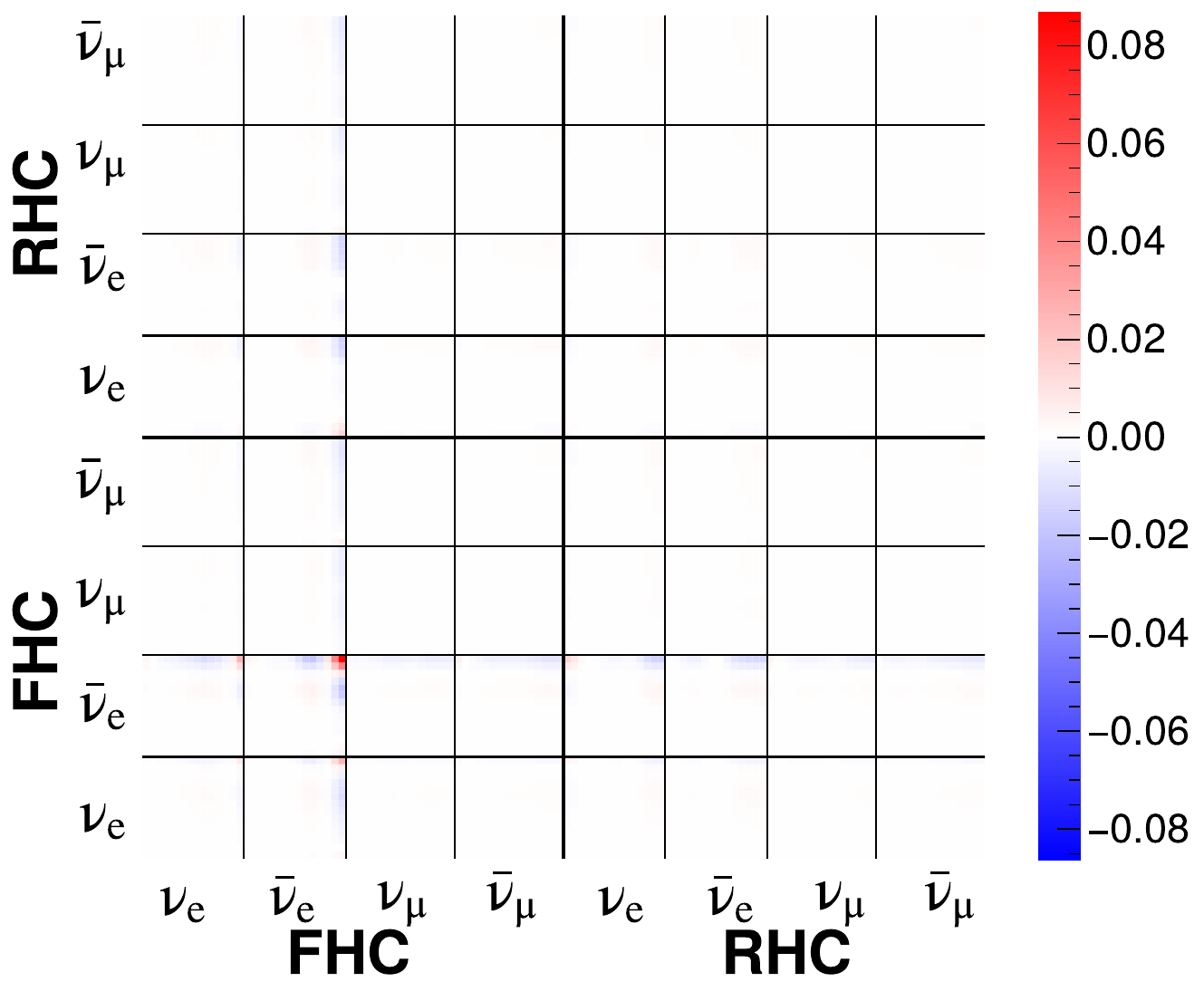}
        \caption{Beam Shift x}
    \end{subfigure}
    \begin{subfigure}[]{0.27\textwidth}
\includegraphics[width=\textwidth]{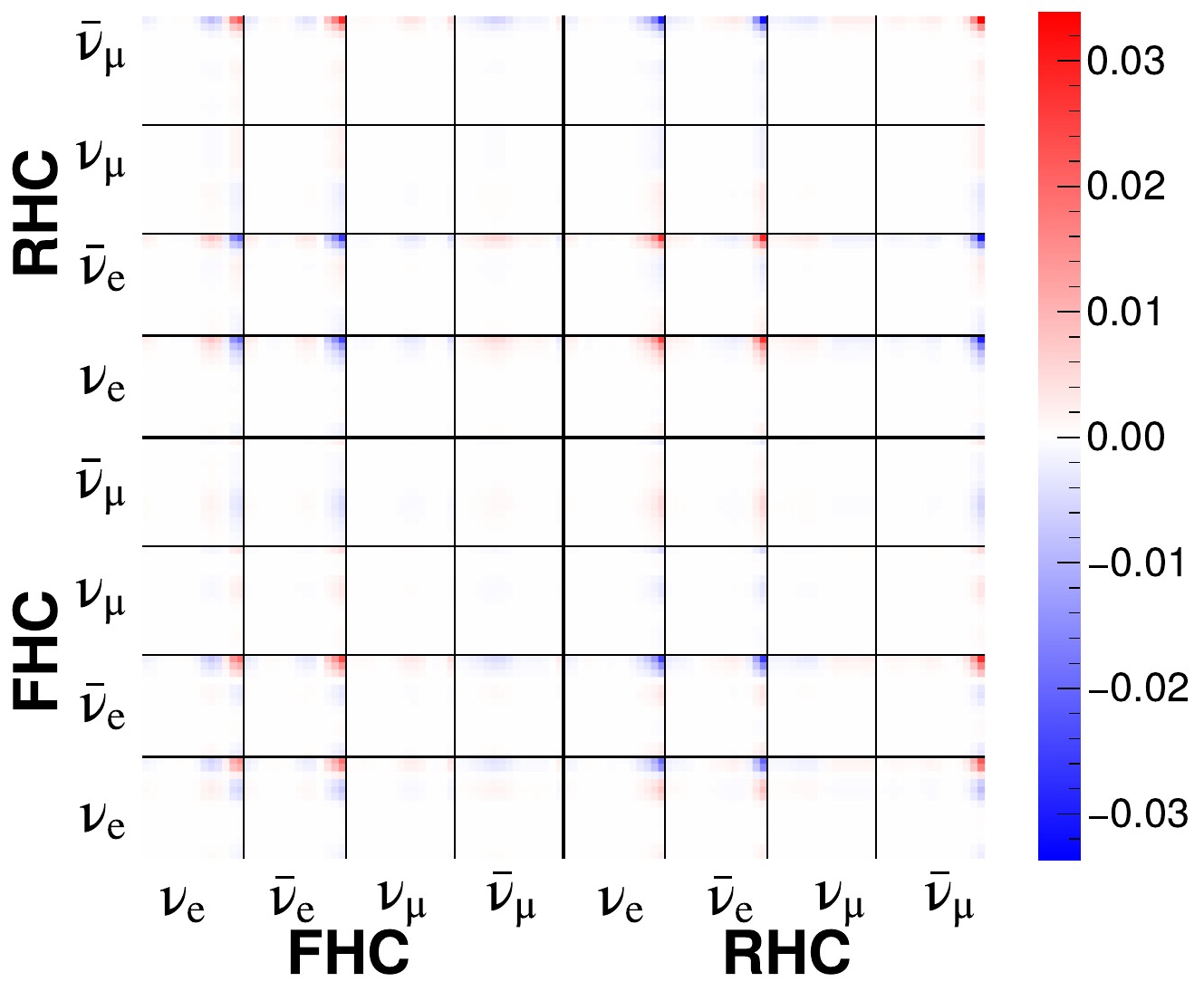}
        \caption{Beam Shift y \ensuremath{-1\sigma}}
    \end{subfigure}
    \begin{subfigure}[]{0.27\textwidth}
\includegraphics[width=\textwidth]{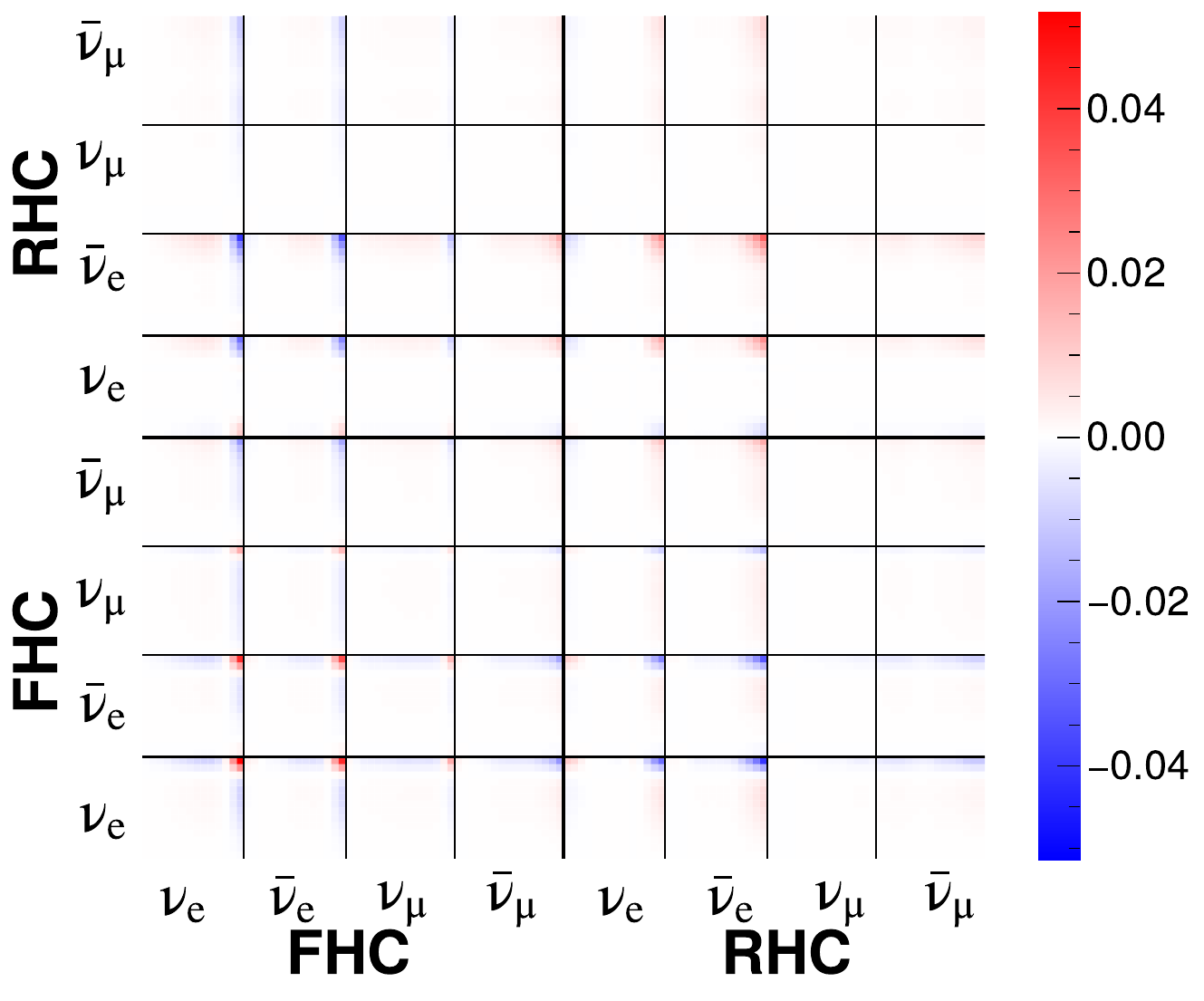}
        \caption{Beam Shift y \ensuremath{+1\sigma}}
    \end{subfigure}
    \begin{subfigure}[]{0.27\textwidth}
\includegraphics[width=\textwidth]{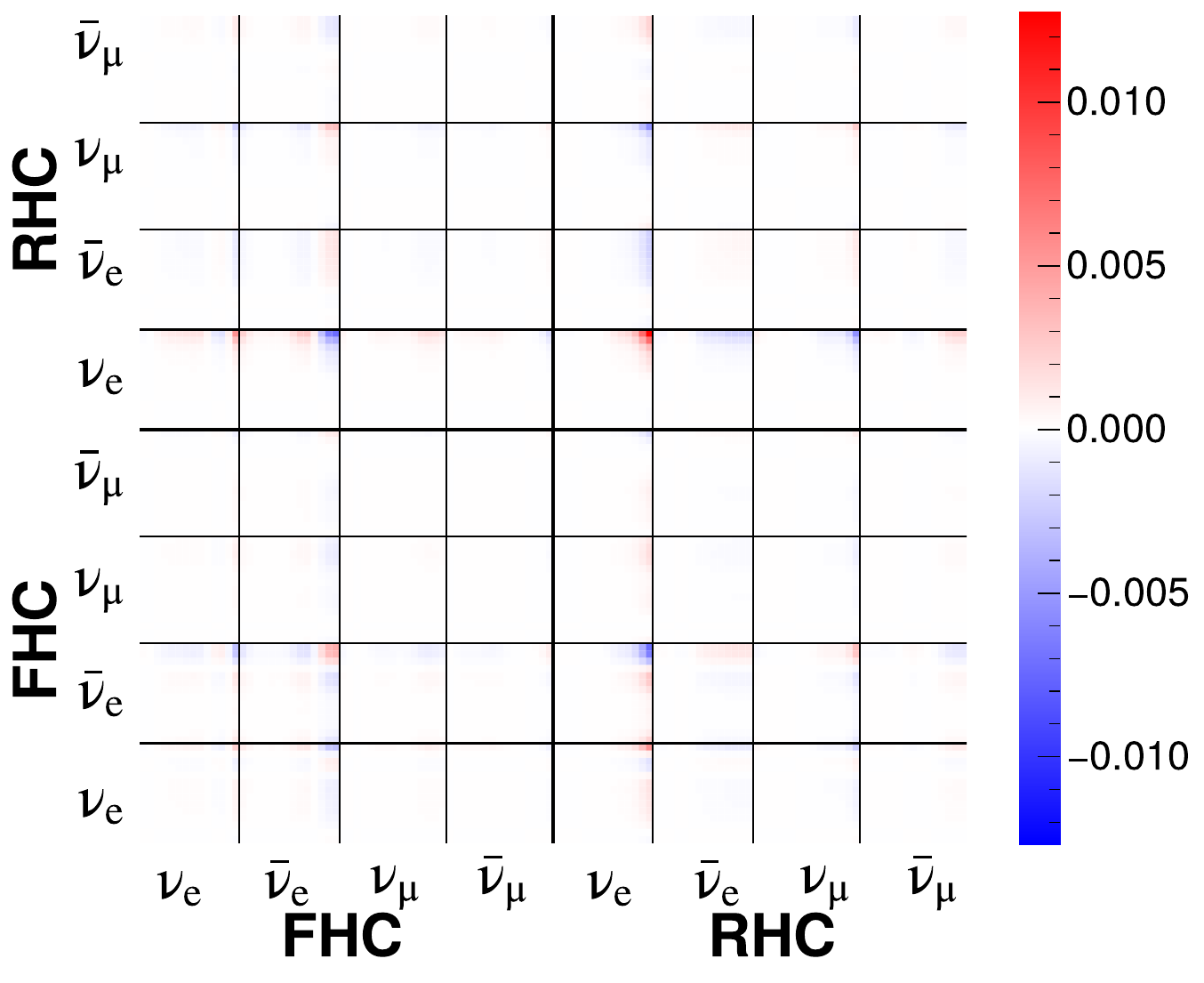}
        \caption{Beam Spot Size}
    \end{subfigure}
    \begin{subfigure}[]{0.27\textwidth}
\includegraphics[width=\textwidth]{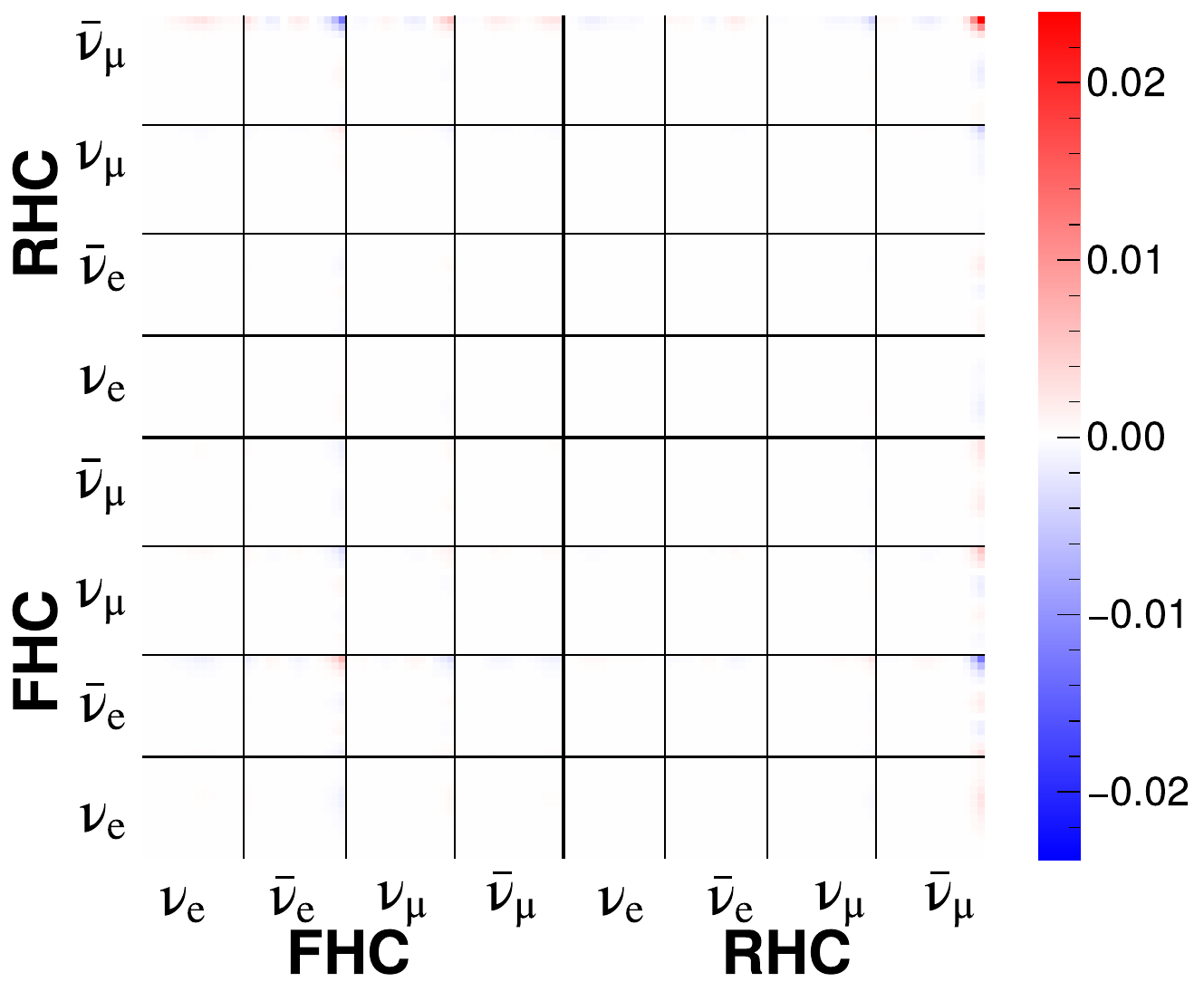}
        \caption{Horn Current \ensuremath{+1\sigma}}
    \end{subfigure}
    \begin{subfigure}[]{0.27\textwidth}
\includegraphics[width=\textwidth]{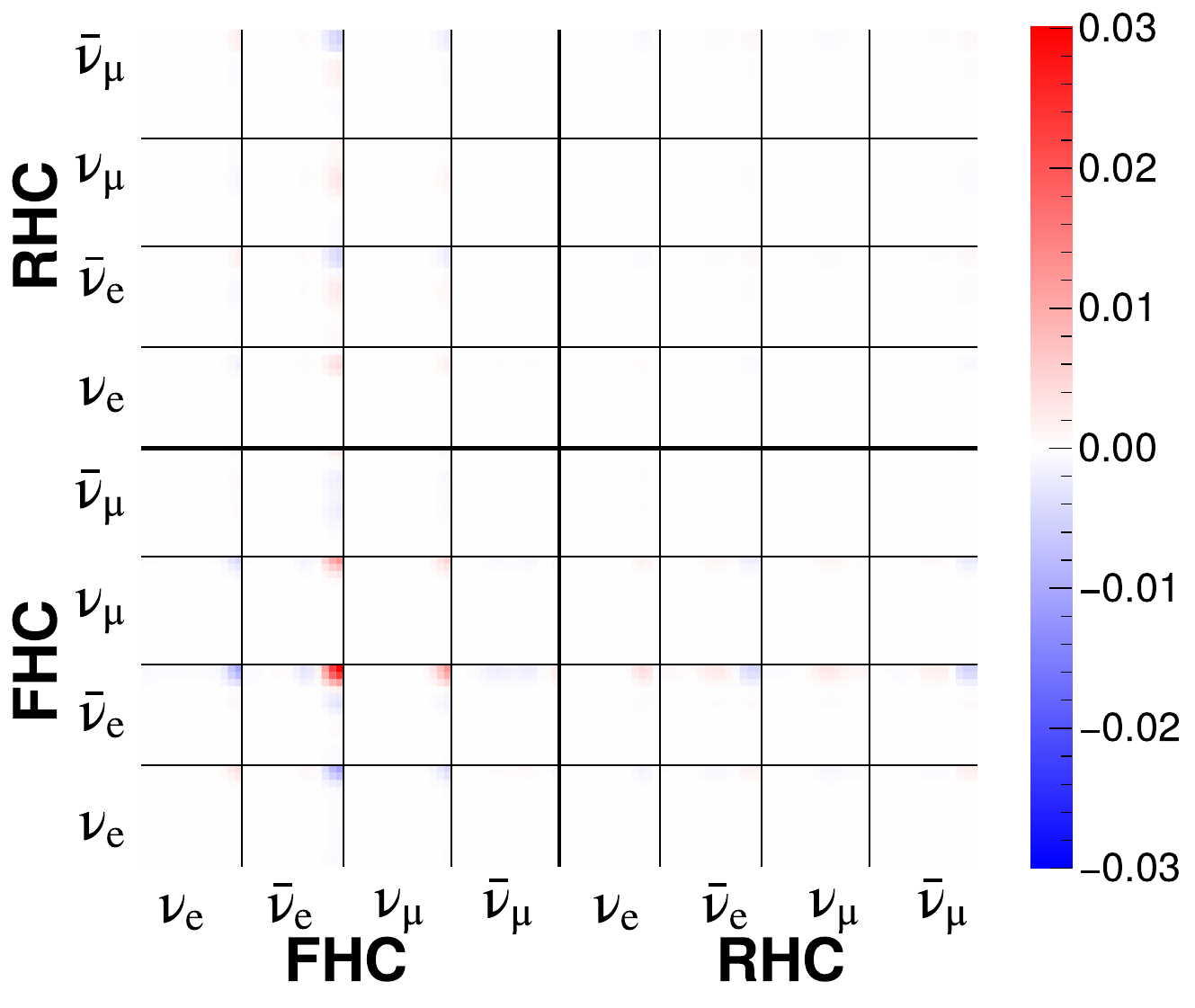}
        \caption{Horn 1 x-Position }
    \end{subfigure}
    \begin{subfigure}[]{0.27\textwidth}
\includegraphics[width=\textwidth]{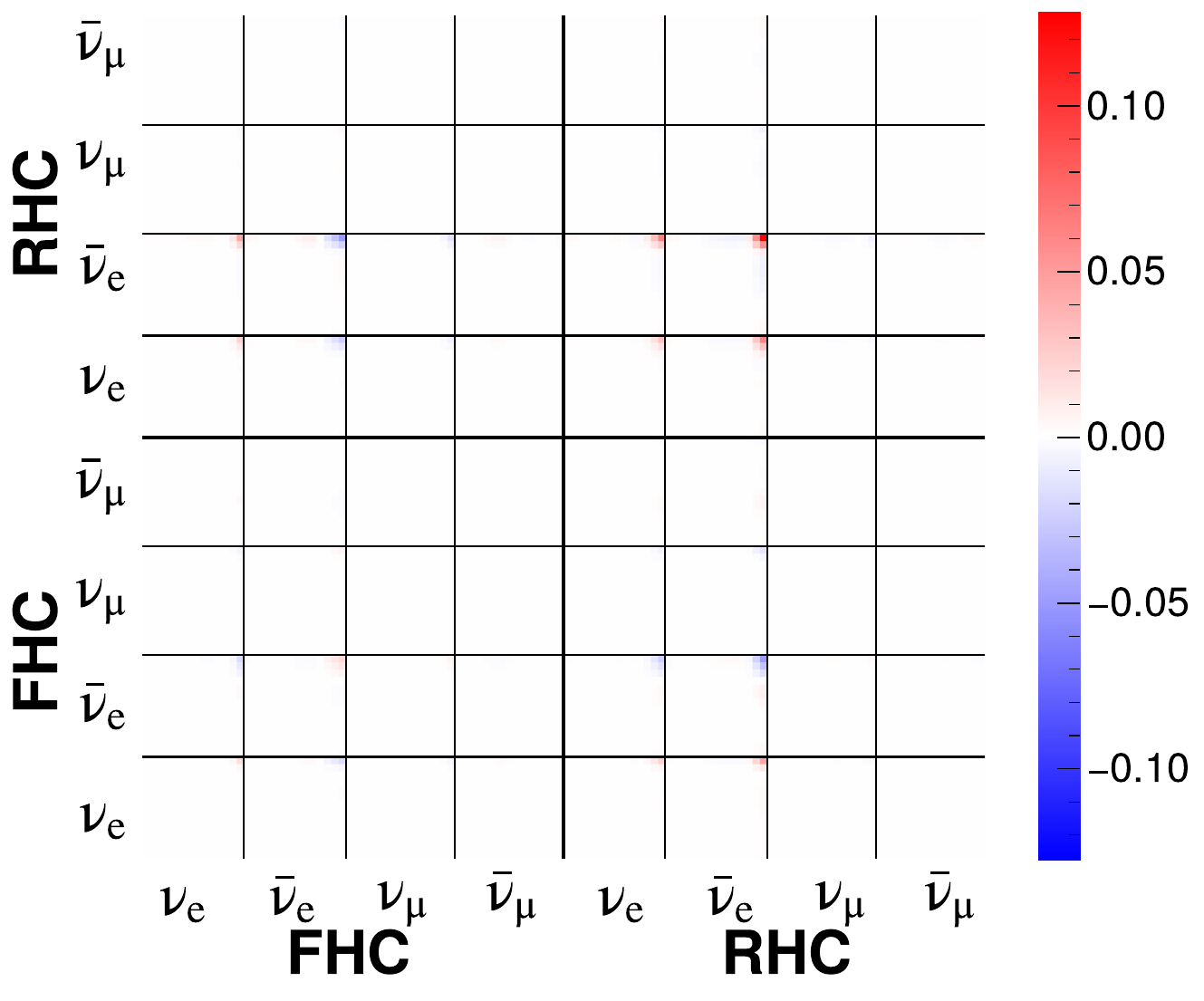}
        \caption{Horn 1 y-Position }
    \end{subfigure}
    \begin{subfigure}[]{0.27\textwidth}
\includegraphics[width=\textwidth]{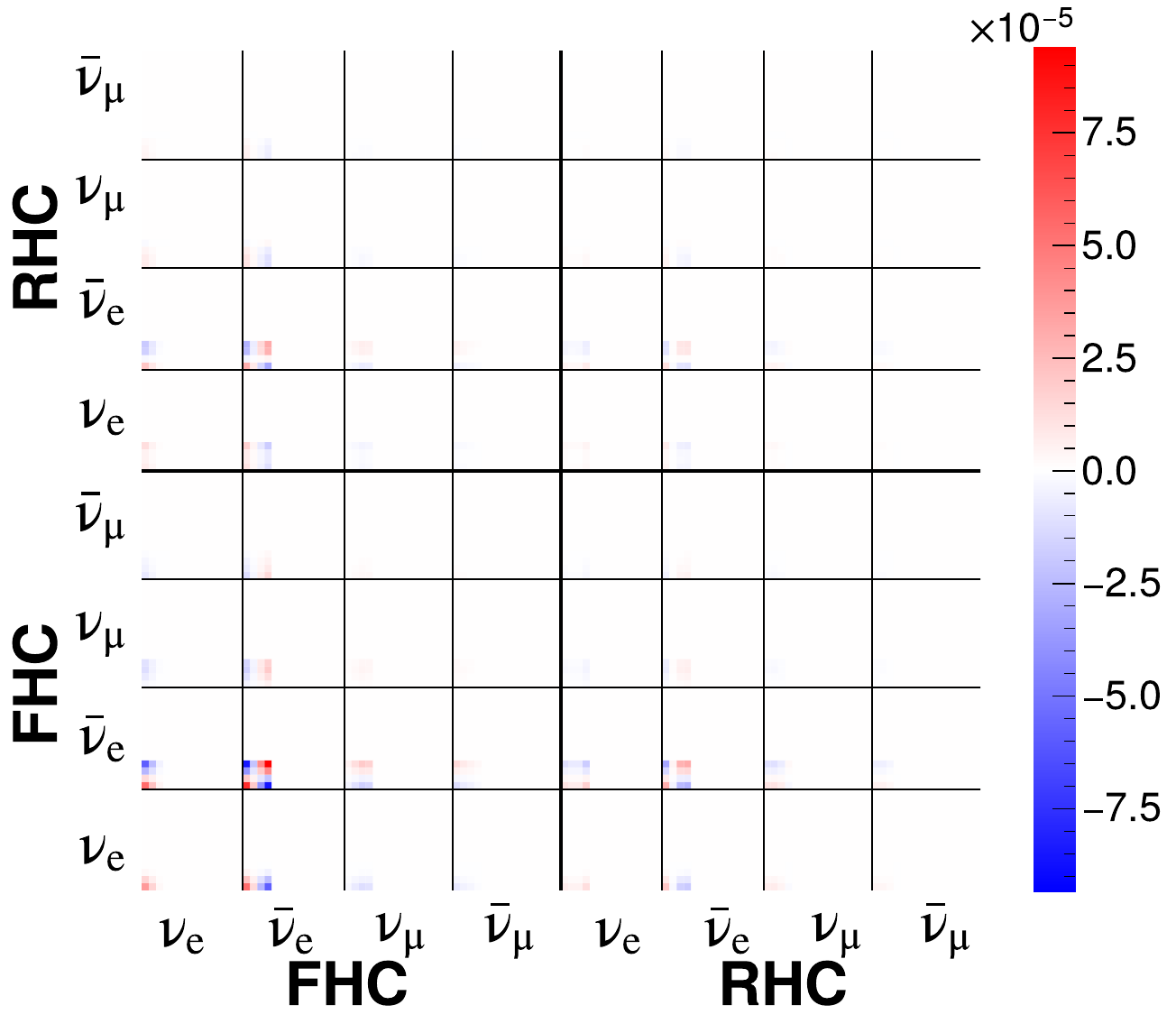}
        \caption{Horn Water Layer }
    \end{subfigure}
        \caption[Individual Beam Focusing Covariance Matrices]{All beam focusing systematic covariance matrices.}
\end{figure}

\clearpage
\section{Correlation Matrices}
\begin{figure}[!ht]
    \centering
    \begin{subfigure}[]{0.27\textwidth}
\includegraphics[width=\textwidth]{hcorr_beam_total_correlation_matrix.pdf}
        \caption{Total}
    \end{subfigure}
    \begin{subfigure}[]{0.27\textwidth}
\includegraphics[width=\textwidth]{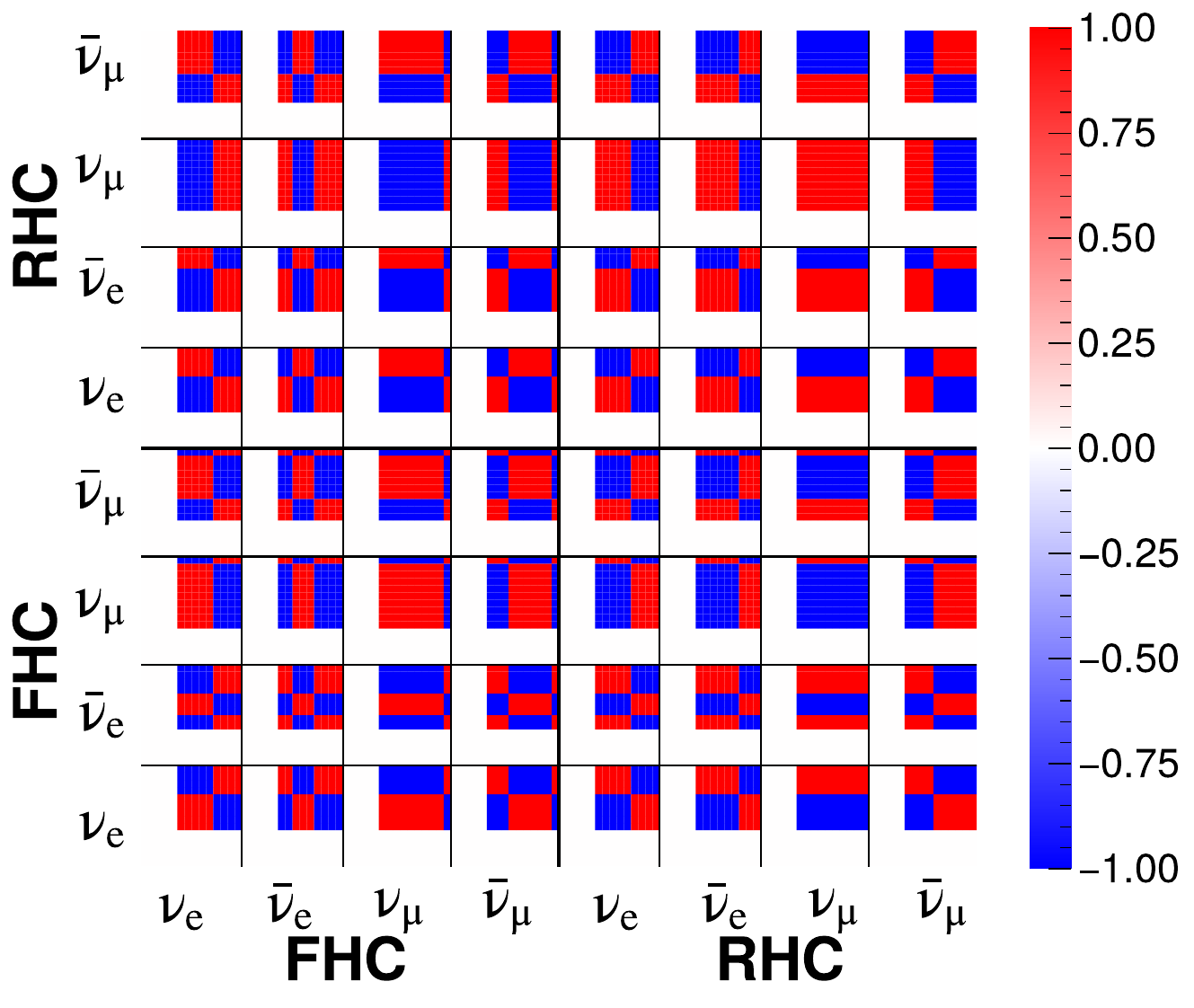}
        \caption{Beam Divergence}
    \end{subfigure}
    \begin{subfigure}[]{0.27\textwidth}
\includegraphics[width=\textwidth]{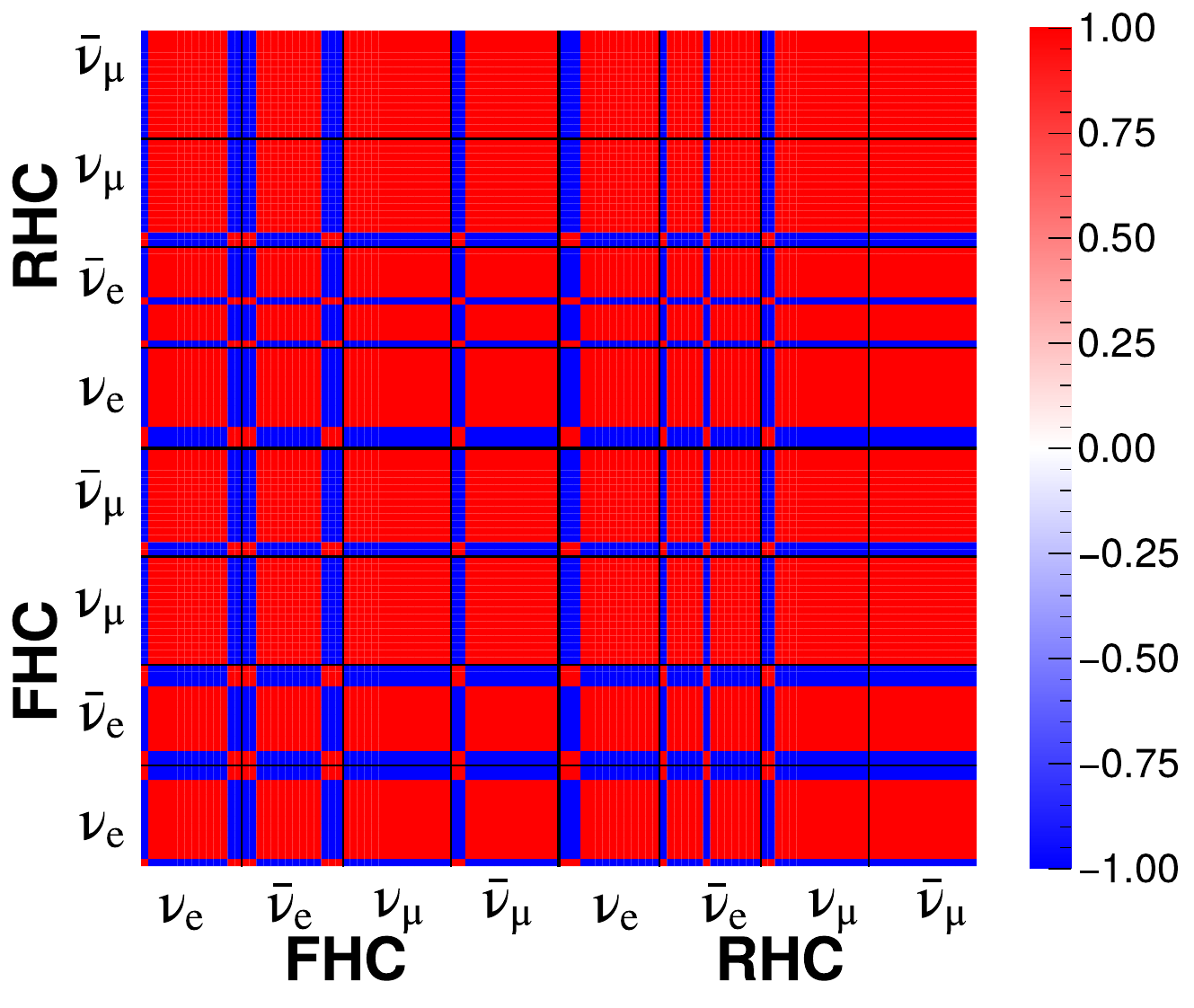}
        \caption{Beam Shift x}
    \end{subfigure}
    \begin{subfigure}[]{0.27\textwidth}
\includegraphics[width=\textwidth]{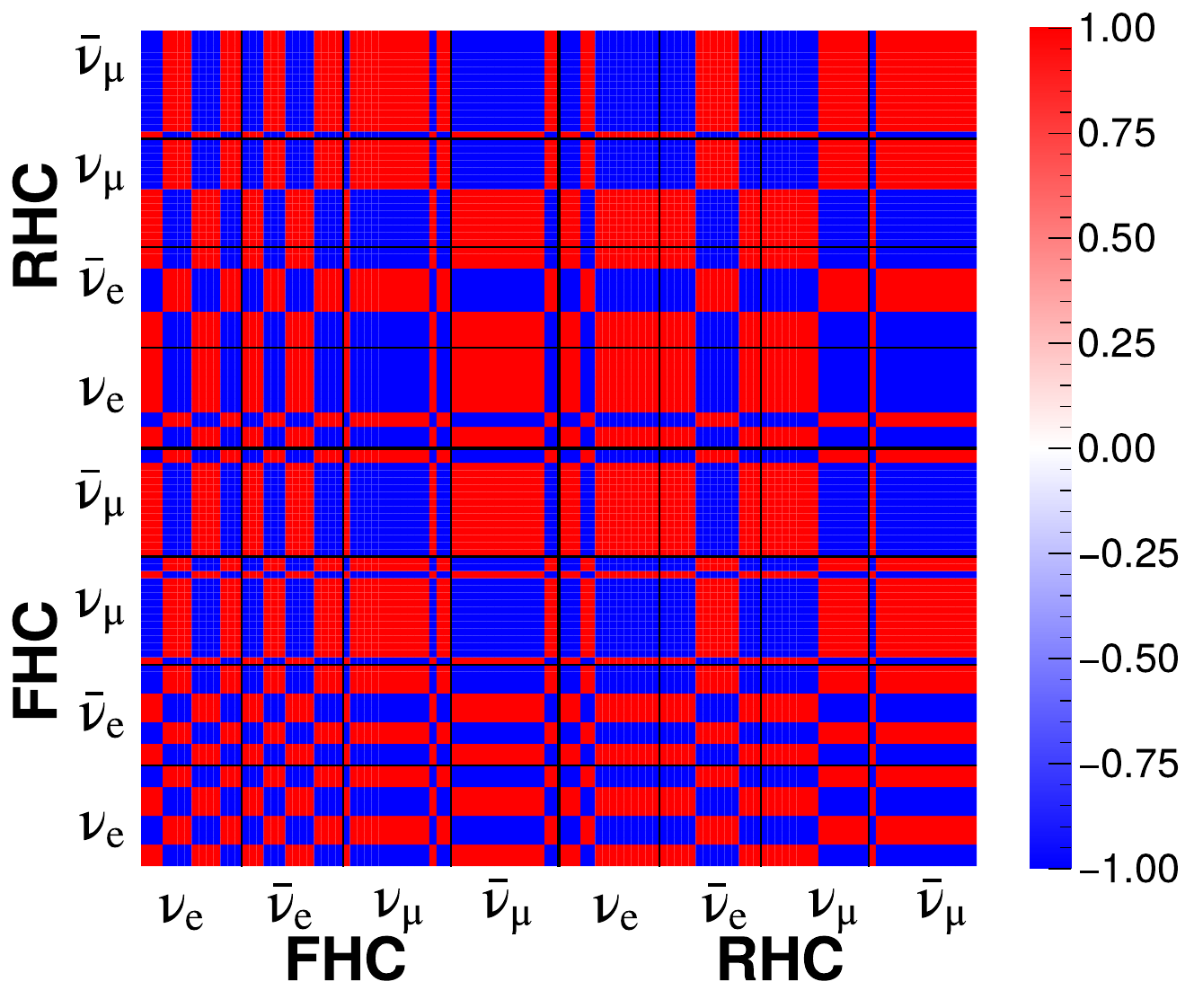}
        \caption{Beam Shift y \ensuremath{-1\sigma}}
    \end{subfigure}
    \begin{subfigure}[]{0.27\textwidth}
\includegraphics[width=\textwidth]{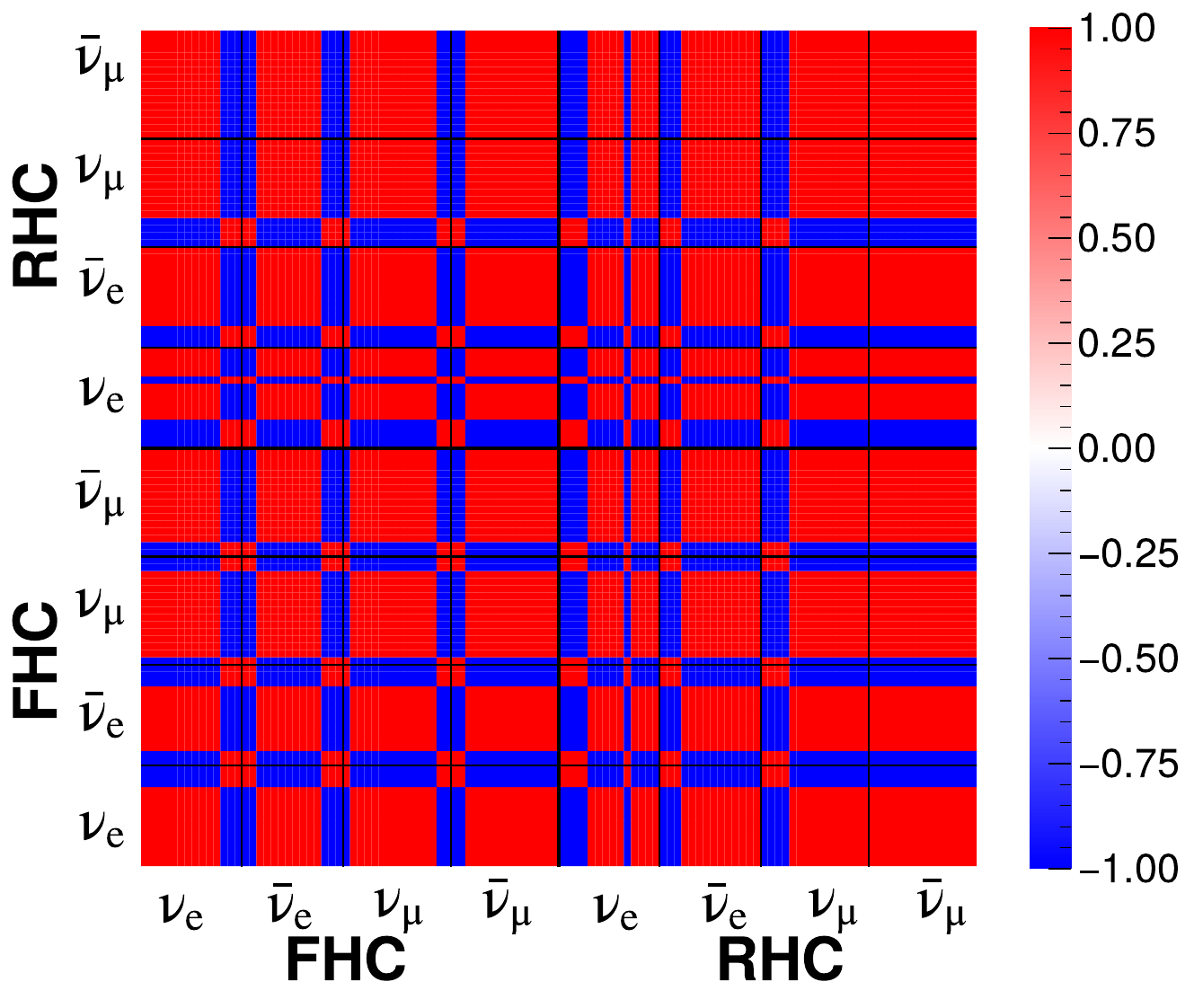}
        \caption{Beam Shift y \ensuremath{+1\sigma}}
    \end{subfigure}
    \begin{subfigure}[]{0.27\textwidth}
\includegraphics[width=\textwidth]{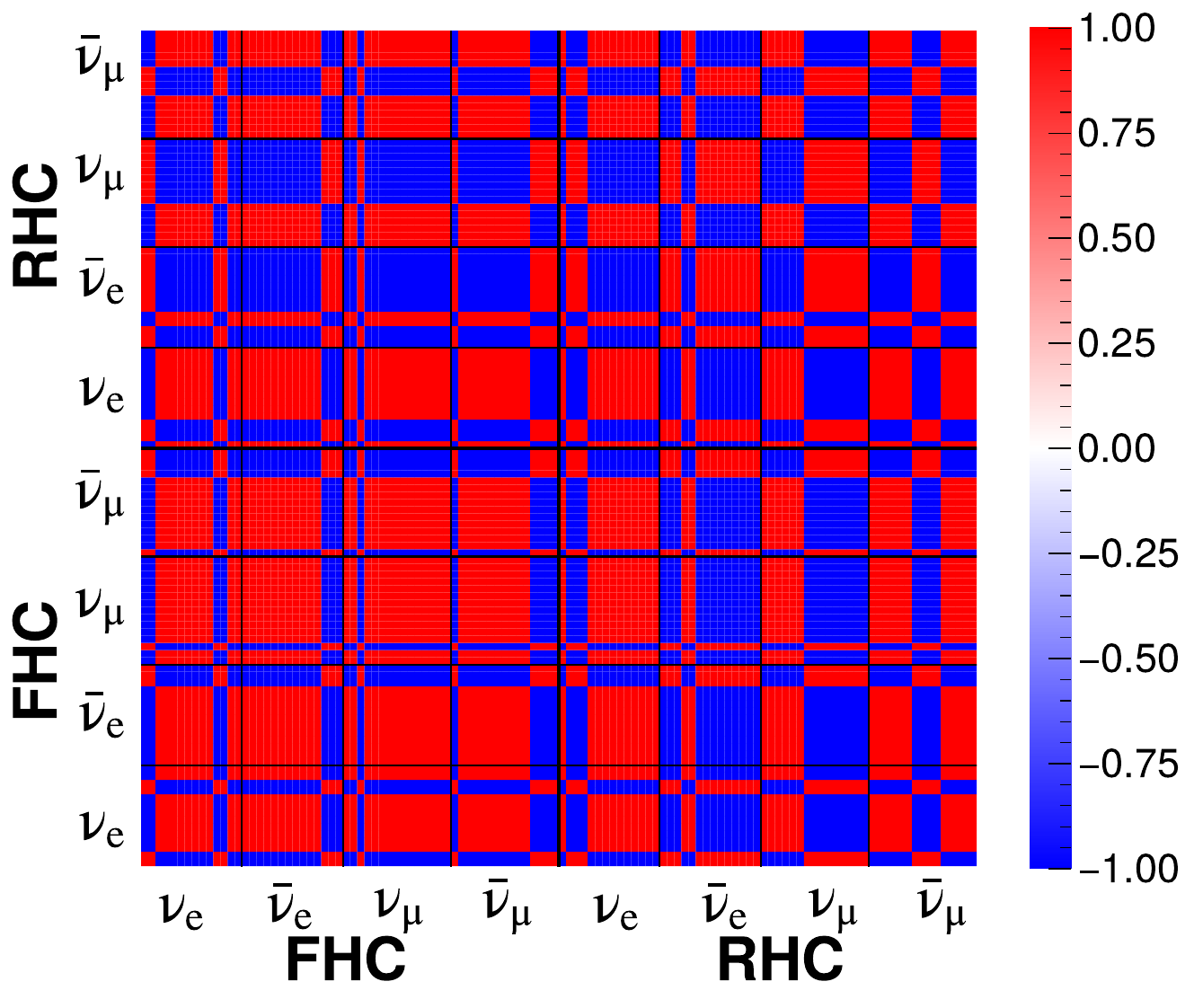}
        \caption{Beam Spot Size}
    \end{subfigure}
    \begin{subfigure}[]{0.27\textwidth}
\includegraphics[width=\textwidth]{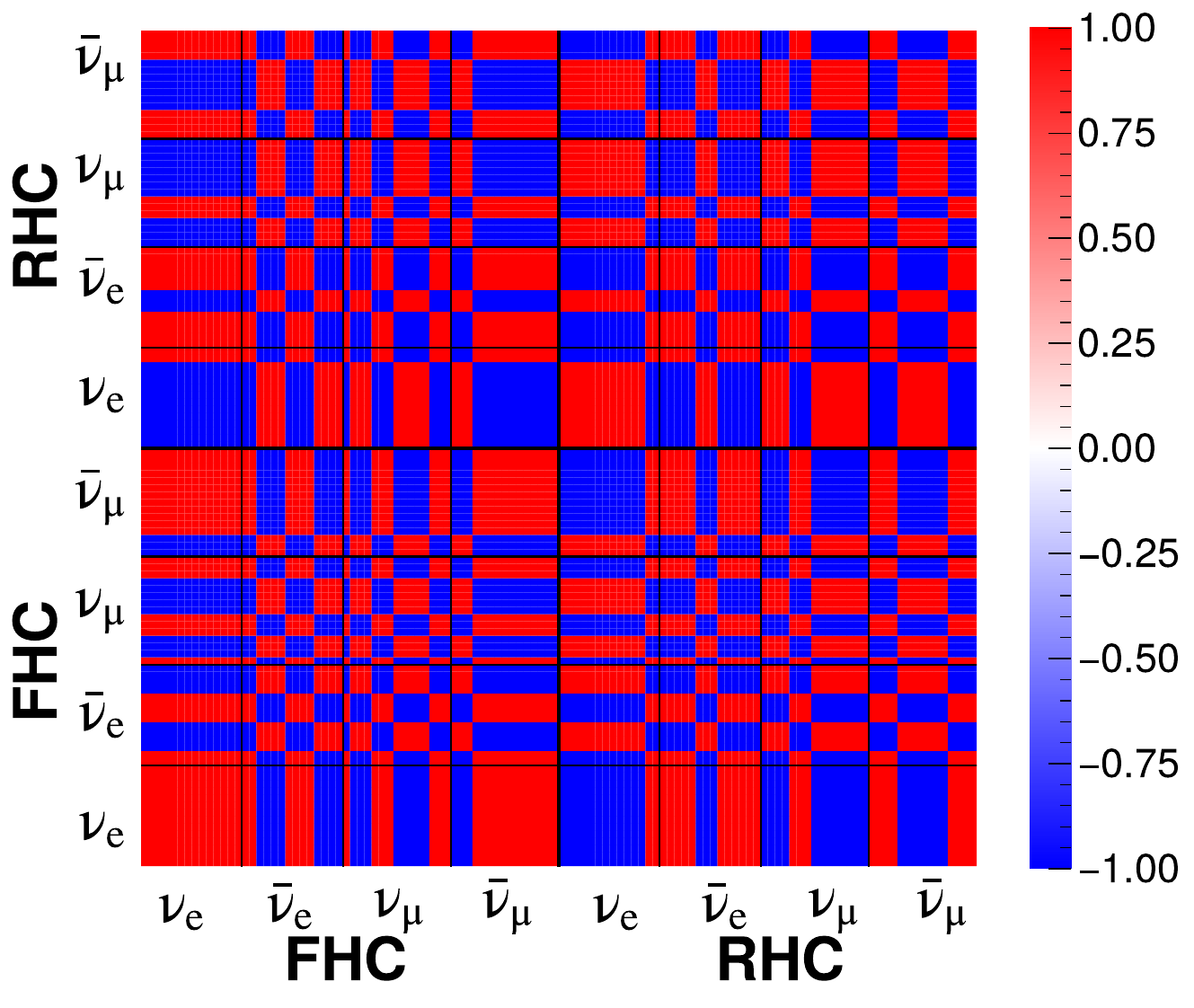}
        \caption{Horn Current \ensuremath{+1\sigma}}
    \end{subfigure}
    \begin{subfigure}[]{0.27\textwidth}
\includegraphics[width=\textwidth]{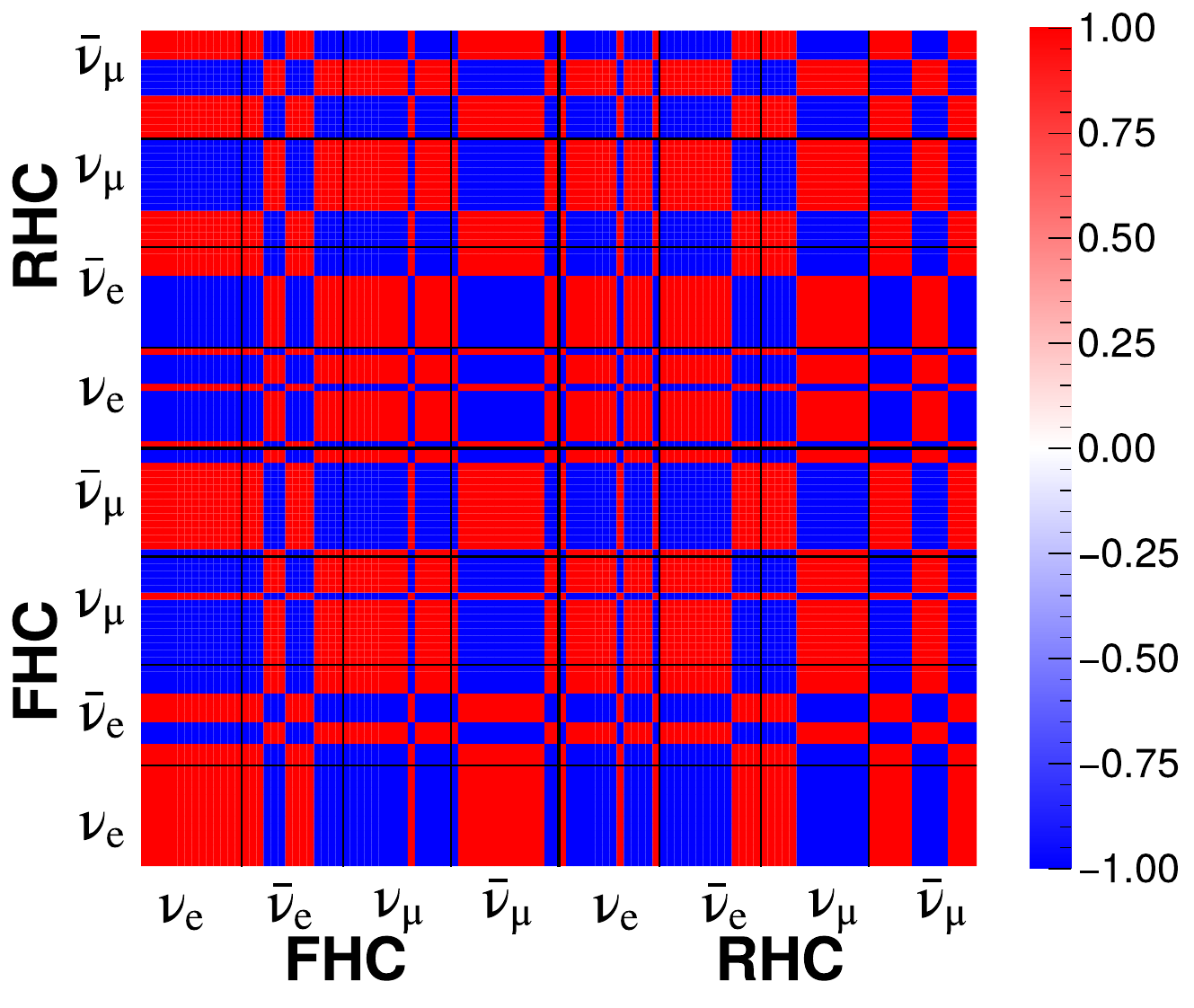}
        \caption{Horn 1 x-Position }
    \end{subfigure}
    \begin{subfigure}[]{0.27\textwidth}
\includegraphics[width=\textwidth]{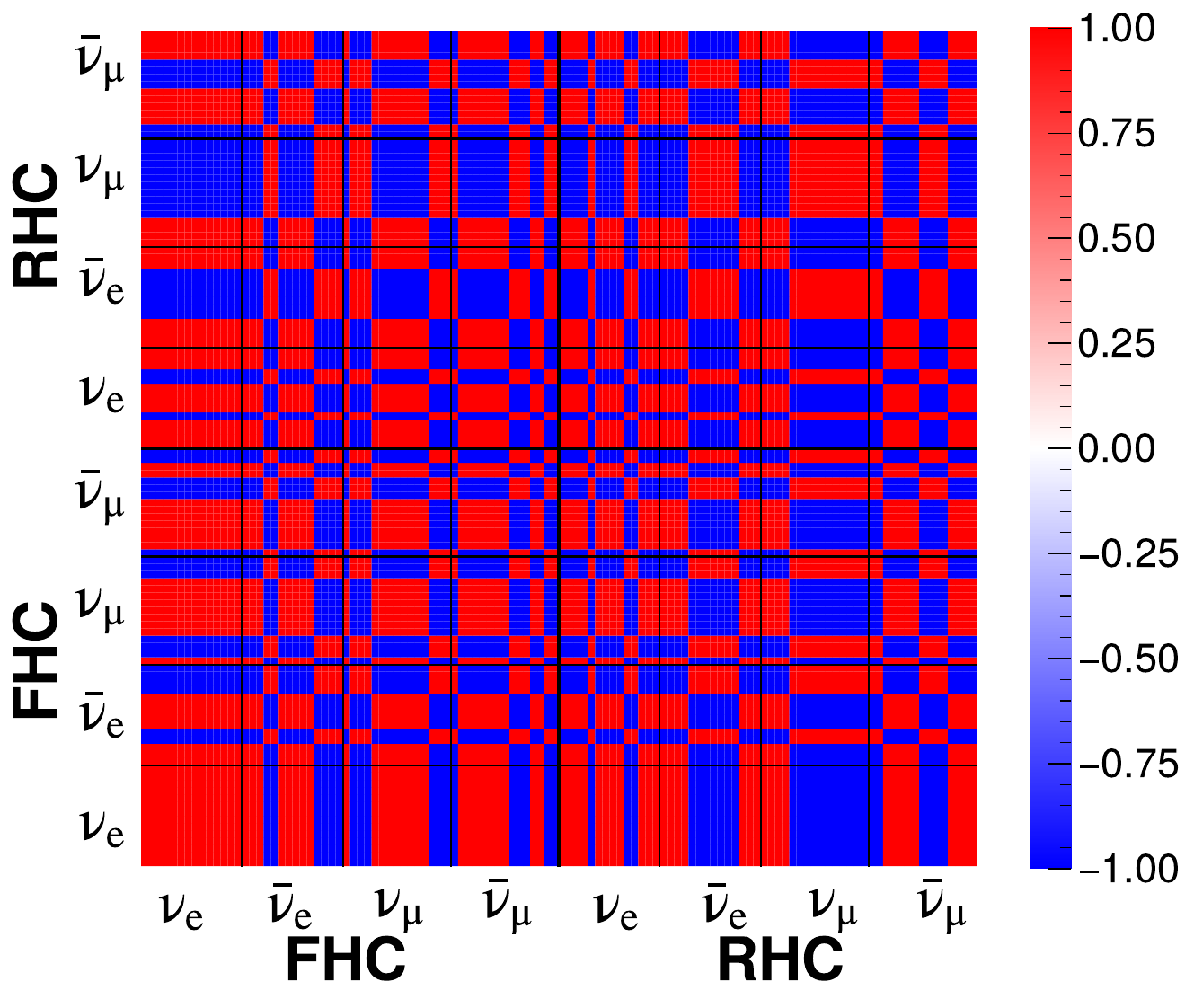}
        \caption{Horn 1 y-Position }
    \end{subfigure}
    \begin{subfigure}[]{0.27\textwidth}
\includegraphics[width=\textwidth]{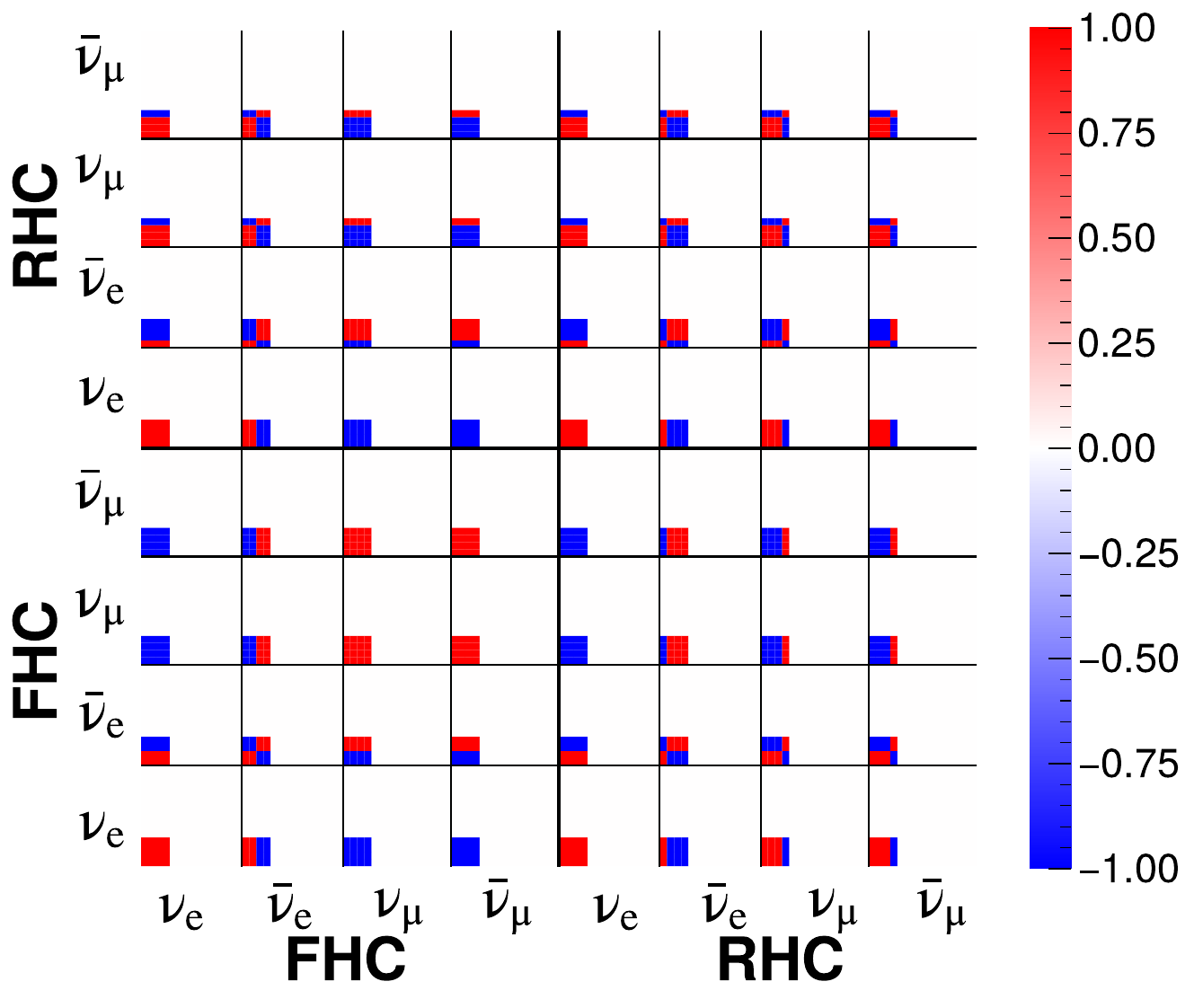}
        \caption{Horn Water Layer }
    \end{subfigure}
    \caption[Individual Beam Focusing Correlation Matrices]{All beam focusing systematic correlation matrices.}
\end{figure}

%% file: 1megawatt_upgrade_differences.tex
\clearpage
\begin{figure}[!ht]
    \centering
    \includegraphics[width=0.48\textwidth]{fhc_numu_numi_geometry_comparison.pdf}
    \includegraphics[width=0.48\textwidth]{fhc_nue_numi_geometry_comparison.pdf}
    \includegraphics[width=0.48\textwidth]{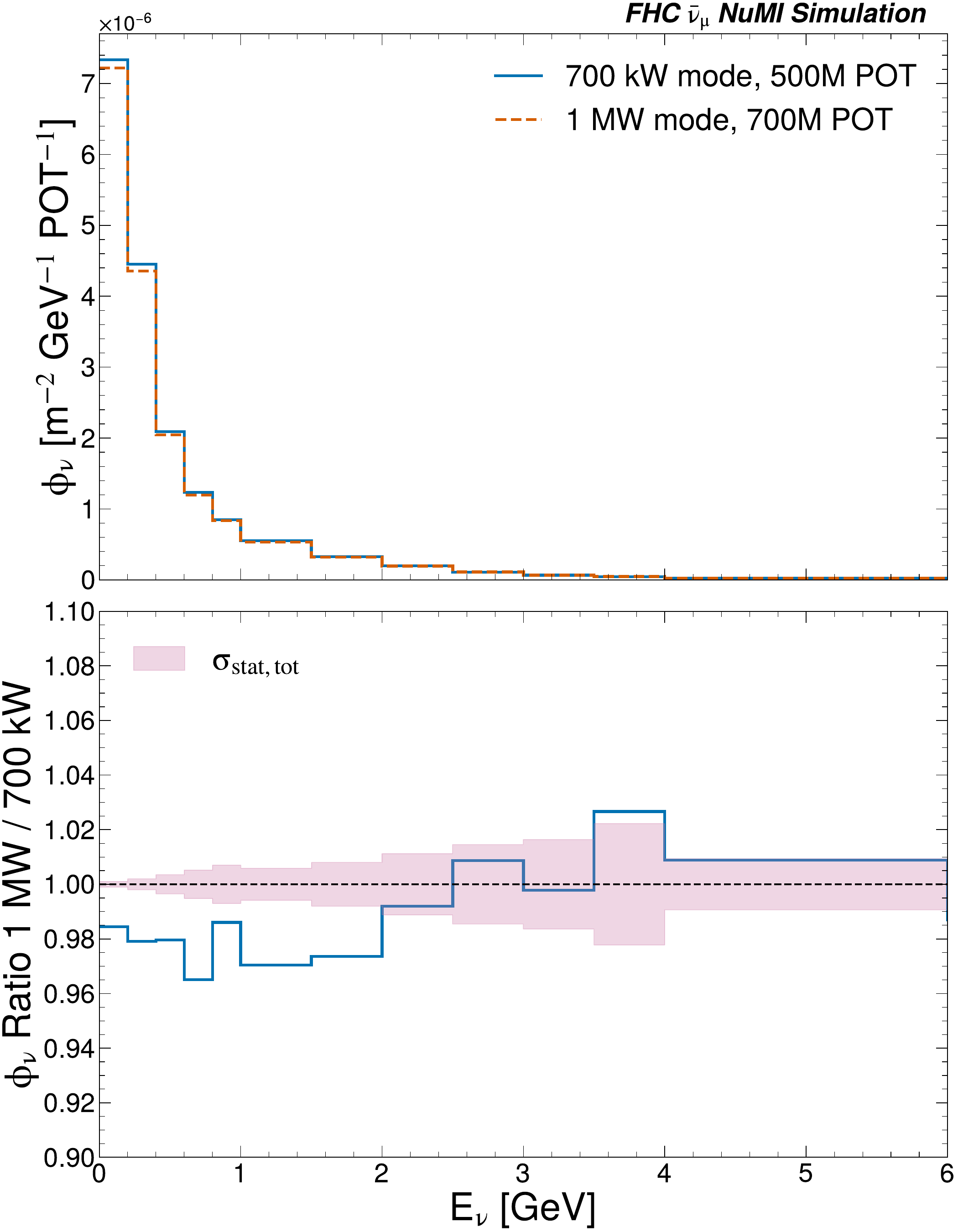}
    \includegraphics[width=0.48\textwidth]{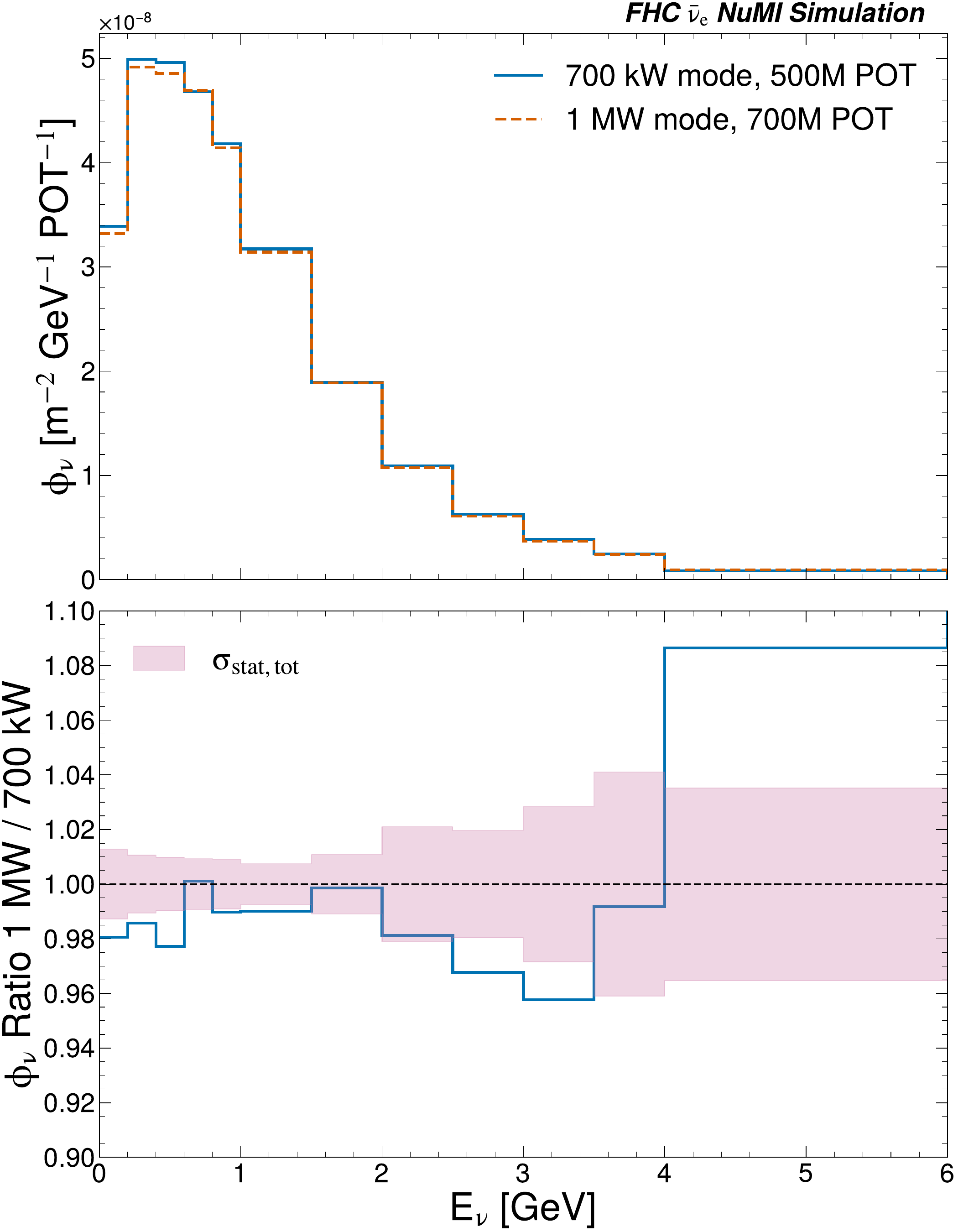}
    \caption[Comparison of the \SI{700}{kW} and \SI{1}{MW} NuMI Beam Geometries (FHC).]{Comparison of the \SI{700}{kW} and \SI{1}{MW} NuMI beam geometries (FHC).}
\end{figure}
\begin{figure}[!ht]
    \centering
    \includegraphics[width=0.48\textwidth]{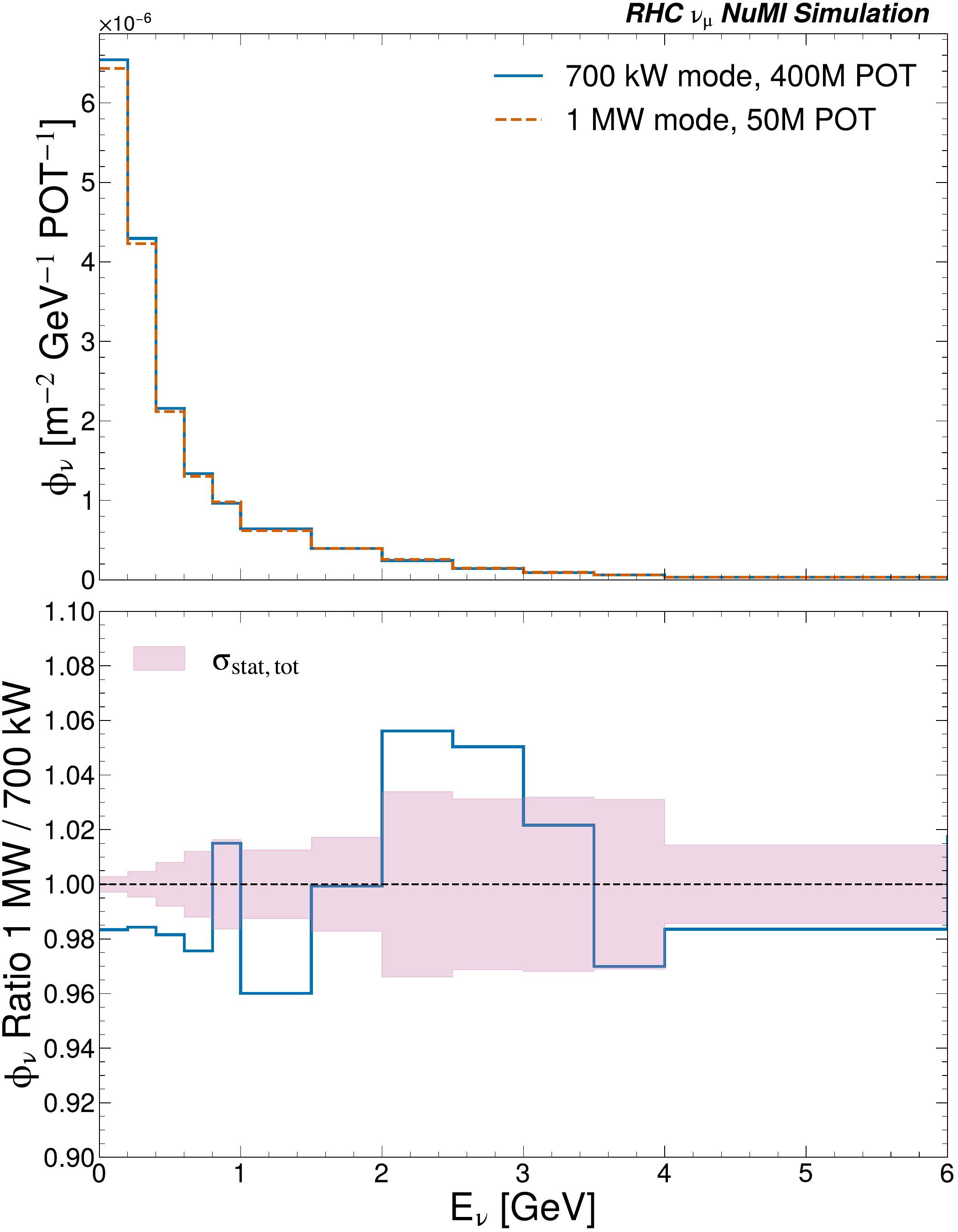}
    \includegraphics[width=0.48\textwidth]{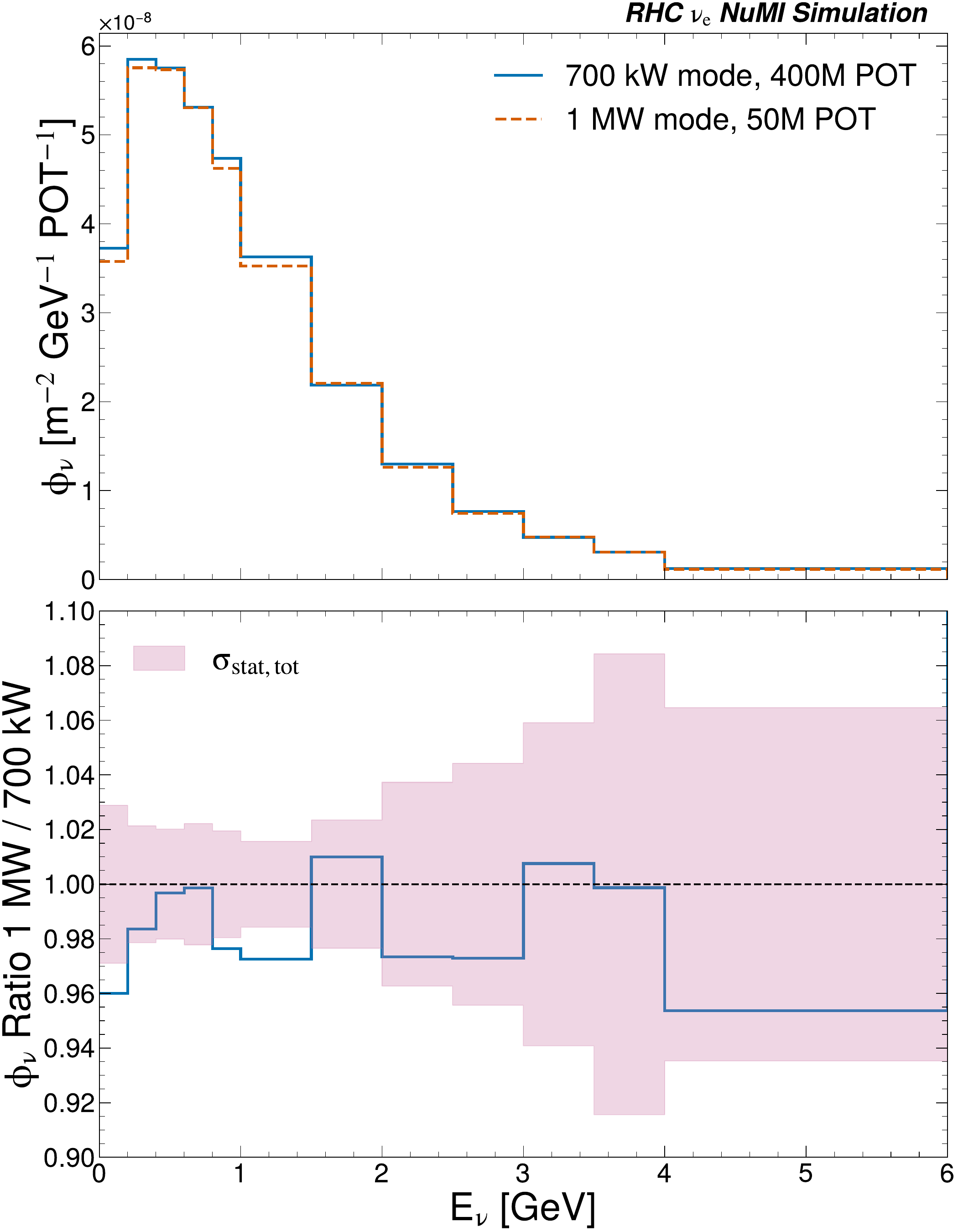}
    \includegraphics[width=0.48\textwidth]{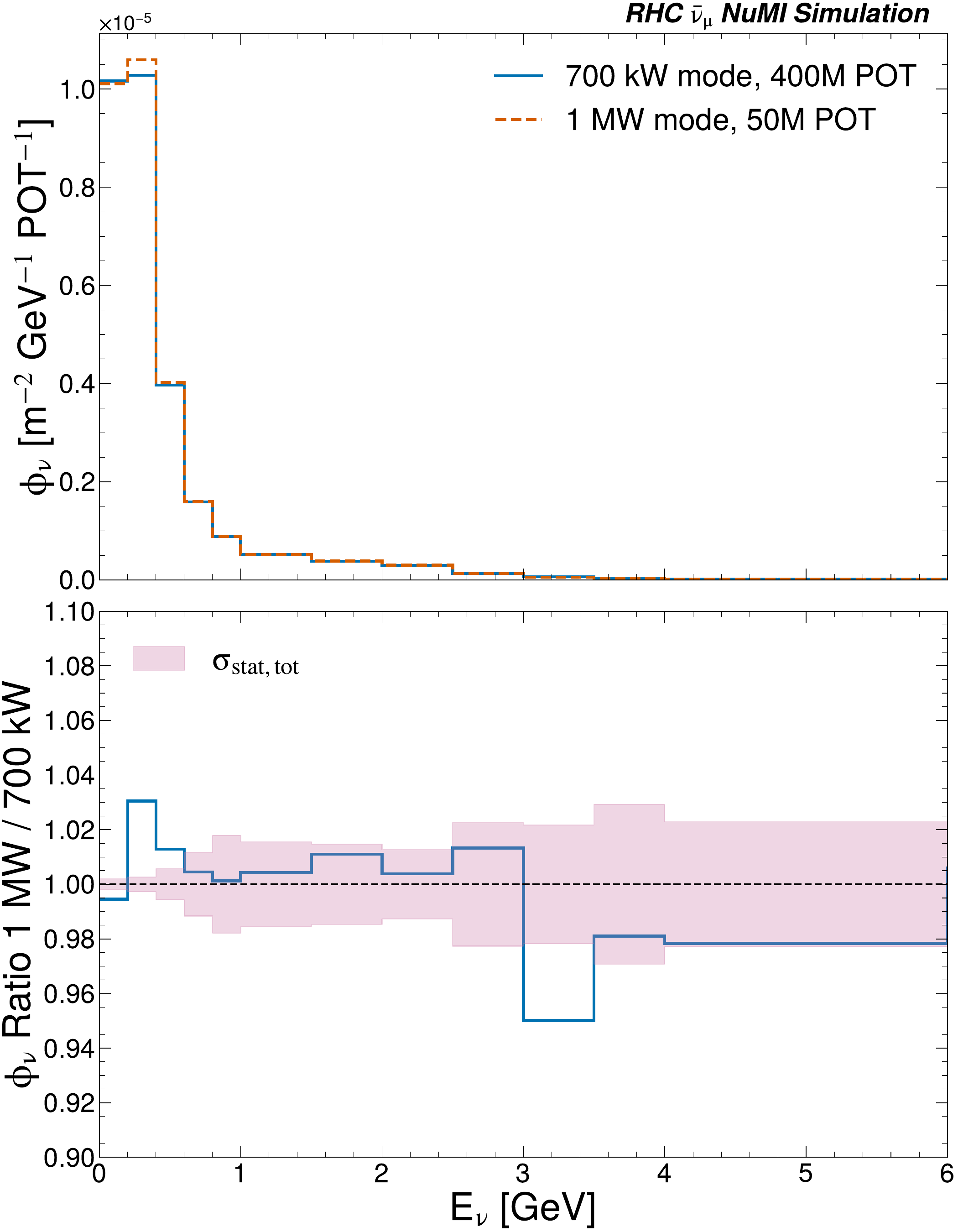}
    \includegraphics[width=0.48\textwidth]{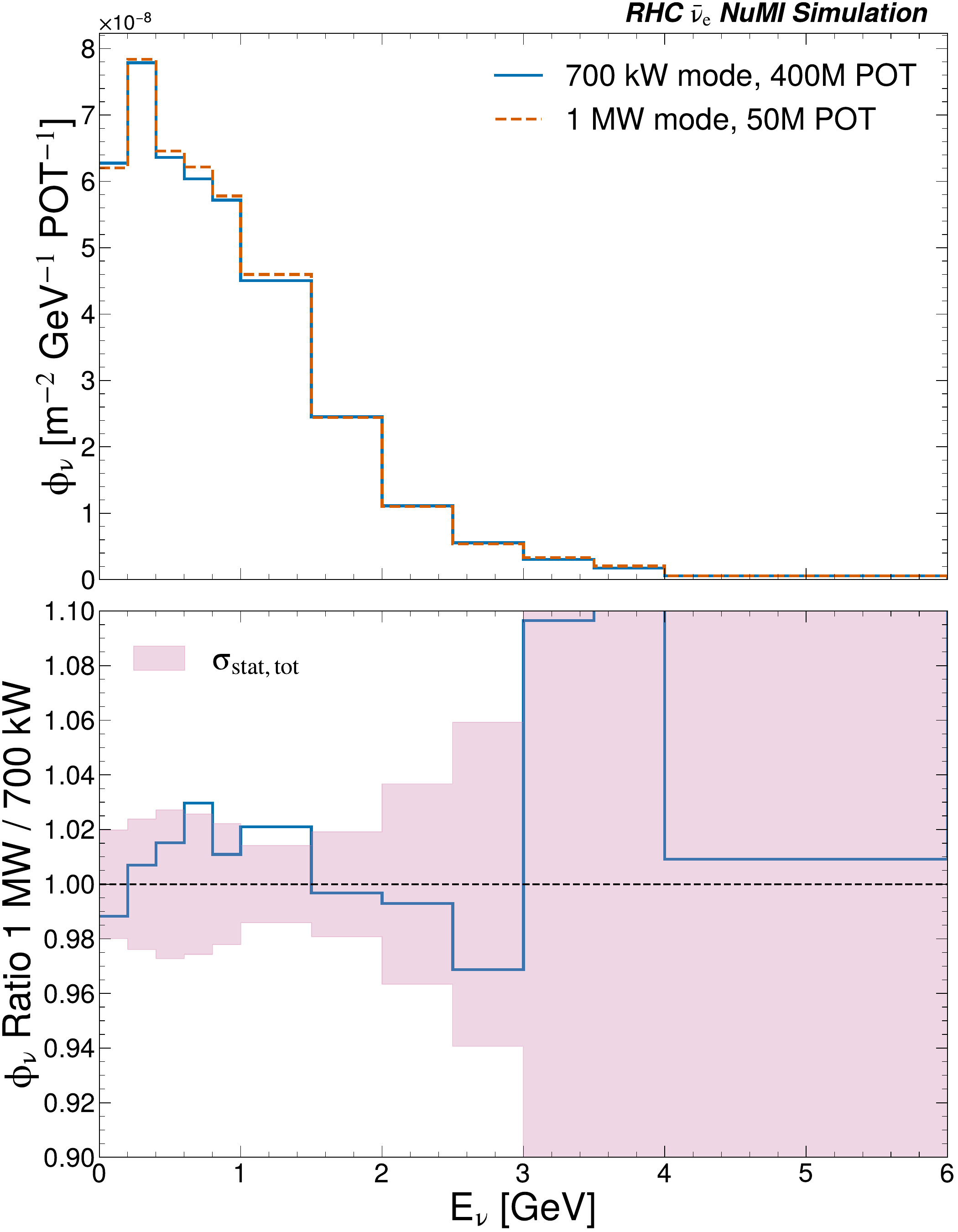}
    \caption{Comparison of the \SI{700}{kW} and \SI{1}{MW} NuMI Beam Geometries (RHC).}
\end{figure}

%% file: total_uncertainty.tex
\clearpage
The uncertainties presented in this appendix were produced using the simulation based on \acrshort{geant}-9.2p03.
These include uncertainty from hadron production modeling, NuMI beam focusing, additional uncertainty to account for differences between the \SI{700}{\kilo\watt} and \SI{1}{\mega\watt} beam geometries, and the statistical uncertainties present in the samples.
At the time of writing this thesis, this table has not yet been updated to reflect the uncertainties from the \acrshort{geant}-10.4 simulation.
\section{Forward Horn Current}
\begin{longtable}{lrrrrrrr}
\caption[Total Uncertainties (FHC)]{Total uncertainties for the \acrshort{fhc} beam mode.}
\label{tab:total_uncertainty_fhc}
\endfirsthead
\endhead
\toprule\toprule
Interval [GeV] & \nue & \nueb & $\nue + \nueb$ & \numu & \numub & $\numu + \numub$ & $\frac{\nue + \nueb}{\numu + \numub}$ \\
\midrule
$[0.0, 0.2]$   & 0.175 & 0.105 & 0.121 & 0.167 & 0.142 & 0.134 & 0.070 \\
$[0.0, 0.4]$   & 0.151 & 0.098 & 0.112 & 0.152 & 0.123 & 0.114 & 0.080 \\
$[0.0, 0.6]$   & 0.128 & 0.088 & 0.098 & 0.145 & 0.118 & 0.109 & 0.079 \\
$[0.0, 0.8]$   & 0.110 & 0.080 & 0.087 & 0.140 & 0.117 & 0.106 & 0.077 \\
$[0.0, 1.0]$   & 0.100 & 0.074 & 0.079 & 0.137 & 0.115 & 0.104 & 0.075 \\
$[0.0, 1.5]$   & 0.085 & 0.066 & 0.067 & 0.129 & 0.113 & 0.100 & 0.074 \\
$[0.0, 2.0]$   & 0.079 & 0.063 & 0.063 & 0.125 & 0.110 & 0.097 & 0.074 \\
$[0.0, 2.5]$   & 0.076 & 0.062 & 0.061 & 0.121 & 0.108 & 0.095 & 0.073 \\
$[0.0, 3.0]$   & 0.075 & 0.061 & 0.060 & 0.120 & 0.107 & 0.094 & 0.072 \\
$[0.0, 3.5]$   & 0.073 & 0.061 & 0.060 & 0.118 & 0.106 & 0.094 & 0.072 \\
$[0.0, 4.0]$   & 0.073 & 0.061 & 0.059 & 0.118 & 0.106 & 0.093 & 0.071 \\
$[0.0, 6.0]$   & 0.072 & 0.061 & 0.059 & 0.116 & 0.106 & 0.092 & 0.071 \\
$[0.0, 8.0]$   & 0.072 & 0.061 & 0.058 & 0.115 & 0.106 & 0.092 & 0.070 \\
$[0.0, 12.0]$  & 0.072 & 0.061 & 0.058 & 0.115 & 0.106 & 0.092 & 0.070 \\
$[0.0, 20.0]$  & 0.072 & 0.061 & 0.058 & 0.115 & 0.106 & 0.092 & 0.070 \\
\midrule
$[0.2, 20.0]$  & 0.072 & 0.061 & 0.058 & 0.115 & 0.106 & 0.092 & 0.070 \\
$[0.4, 20.0]$  & 0.066 & 0.061 & 0.055 & 0.094 & 0.092 & 0.076 & 0.067 \\
$[0.6, 20.0]$  & 0.061 & 0.060 & 0.051 & 0.077 & 0.077 & 0.061 & 0.053 \\
$[0.8, 20.0]$  & 0.059 & 0.061 & 0.050 & 0.068 & 0.067 & 0.053 & 0.043 \\
$[1.0, 20.0]$  & 0.058 & 0.061 & 0.049 & 0.062 & 0.063 & 0.049 & 0.038 \\
$[1.5, 20.0]$  & 0.058 & 0.061 & 0.048 & 0.060 & 0.062 & 0.047 & 0.035 \\
$[2.0, 20.0]$  & 0.058 & 0.064 & 0.049 & 0.059 & 0.065 & 0.046 & 0.035 \\
$[2.5, 20.0]$  & 0.060 & 0.075 & 0.050 & 0.064 & 0.070 & 0.049 & 0.036 \\
$[3.0, 20.0]$  & 0.066 & 0.108 & 0.058 & 0.064 & 0.076 & 0.049 & 0.042 \\
$[3.5, 20.0]$  & 0.080 & 0.143 & 0.072 & 0.063 & 0.082 & 0.050 & 0.060 \\
$[4.0, 20.0]$  & 0.103 & 0.156 & 0.091 & 0.064 & 0.087 & 0.052 & 0.079 \\
$[6.0, 20.0]$  & 0.129 & 0.165 & 0.117 & 0.065 & 0.093 & 0.054 & 0.104 \\
$[8.0, 20.0]$  & 0.268 & 0.350 & 0.272 & 0.094 & 0.143 & 0.079 & 0.291 \\
$[12.0, 20.0]$ & 0.462 & 0.743 & 0.503 & 0.205 & 0.261 & 0.178 & 0.600 \\
\bottomrule\bottomrule
\end{longtable}

\clearpage
\section{Reverse Horn Current}
\begin{longtable}{lrrrrrrr}
\caption[Total Uncertainties (RHC)]{Total uncertainties for the \acrshort{rhc} beam mode.}
\label{tab:total_uncertainty_rhc}
\endfirsthead
\endhead
\toprule\toprule
Interval [GeV] & \nue & \nueb & $\nue + \nueb$ & \numu & \numub & $\numu + \numub$ & $\frac{\nue + \nueb}{\numu + \numub}$ \\
\midrule
$[0.0, 0.2]$   & 0.110 & 0.148 & 0.115 & 0.146 & 0.165 & 0.136 & 0.057 \\
$[0.0, 0.4]$   & 0.102 & 0.131 & 0.107 & 0.130 & 0.148 & 0.118 & 0.071 \\
$[0.0, 0.6]$   & 0.091 & 0.115 & 0.096 & 0.126 & 0.141 & 0.112 & 0.074 \\
$[0.0, 0.8]$   & 0.082 & 0.102 & 0.085 & 0.124 & 0.136 & 0.110 & 0.074 \\
$[0.0, 1.0]$   & 0.076 & 0.093 & 0.078 & 0.122 & 0.133 & 0.108 & 0.075 \\
$[0.0, 1.5]$   & 0.068 & 0.081 & 0.068 & 0.119 & 0.127 & 0.104 & 0.075 \\
$[0.0, 2.0]$   & 0.066 & 0.077 & 0.065 & 0.115 & 0.123 & 0.100 & 0.075 \\
$[0.0, 2.5]$   & 0.065 & 0.076 & 0.065 & 0.112 & 0.121 & 0.098 & 0.074 \\
$[0.0, 3.0]$   & 0.065 & 0.076 & 0.064 & 0.111 & 0.119 & 0.097 & 0.073 \\
$[0.0, 3.5]$   & 0.065 & 0.076 & 0.064 & 0.110 & 0.118 & 0.096 & 0.073 \\
$[0.0, 4.0]$   & 0.065 & 0.075 & 0.064 & 0.110 & 0.118 & 0.096 & 0.073 \\
$[0.0, 6.0]$   & 0.065 & 0.073 & 0.063 & 0.110 & 0.117 & 0.095 & 0.072 \\
$[0.0, 8.0]$   & 0.065 & 0.072 & 0.063 & 0.109 & 0.116 & 0.095 & 0.072 \\
$[0.0, 12.0]$  & 0.065 & 0.072 & 0.063 & 0.109 & 0.116 & 0.095 & 0.072 \\
$[0.0, 20.0]$  & 0.065 & 0.072 & 0.063 & 0.109 & 0.116 & 0.095 & 0.072 \\
\midrule
$[0.2, 20.0]$  & 0.065 & 0.072 & 0.063 & 0.109 & 0.116 & 0.095 & 0.072 \\
$[0.4, 20.0]$  & 0.063 & 0.070 & 0.060 & 0.096 & 0.096 & 0.080 & 0.068 \\
$[0.6, 20.0]$  & 0.062 & 0.067 & 0.058 & 0.080 & 0.083 & 0.065 & 0.052 \\
$[0.8, 20.0]$  & 0.062 & 0.065 & 0.057 & 0.070 & 0.077 & 0.057 & 0.042 \\
$[1.0, 20.0]$  & 0.064 & 0.065 & 0.058 & 0.066 & 0.075 & 0.054 & 0.036 \\
$[1.5, 20.0]$  & 0.065 & 0.065 & 0.059 & 0.064 & 0.073 & 0.052 & 0.033 \\
$[2.0, 20.0]$  & 0.069 & 0.070 & 0.063 & 0.065 & 0.070 & 0.050 & 0.036 \\
$[2.5, 20.0]$  & 0.076 & 0.074 & 0.066 & 0.067 & 0.070 & 0.051 & 0.043 \\
$[3.0, 20.0]$  & 0.078 & 0.070 & 0.064 & 0.071 & 0.072 & 0.052 & 0.044 \\
$[3.5, 20.0]$  & 0.071 & 0.092 & 0.062 & 0.073 & 0.076 & 0.055 & 0.057 \\
$[4.0, 20.0]$  & 0.115 & 0.153 & 0.089 & 0.077 & 0.081 & 0.058 & 0.103 \\
$[6.0, 20.0]$  & 0.175 & 0.212 & 0.132 & 0.081 & 0.086 & 0.061 & 0.155 \\
$[8.0, 20.0]$  & 0.325 & 0.391 & 0.298 & 0.126 & 0.104 & 0.081 & 0.333 \\
$[12.0, 20.0]$ & 0.508 & 0.645 & 0.501 & 0.283 & 0.165 & 0.137 & 0.540 \\
\bottomrule\bottomrule
\end{longtable}

%% file: parent_decay_momenta.tex
\clearpage
\section{Forward Horn Current}
\begin{figure}[!ht]
\centering
\begin{subfigure}{0.23\textwidth}
    \includegraphics[width=\textwidth]{parent_momentum_new_geant_fhc_numu_pipm.pdf}
    \caption{$\pi^+ \to \numu$}
\end{subfigure}
\begin{subfigure}{0.23\textwidth}
    \includegraphics[width=\textwidth]{parent_momentum_new_geant_fhc_numu_Kpm.pdf}
    \caption{$K^+ \to \numu$}
\end{subfigure}
\begin{subfigure}{0.23\textwidth}
    \includegraphics[width=\textwidth]{parent_momentum_new_geant_fhc_numu_K0l.pdf}
    \caption{$K^0_L \to \numu$}
\end{subfigure}
\begin{subfigure}{0.23\textwidth}
    \includegraphics[width=\textwidth]{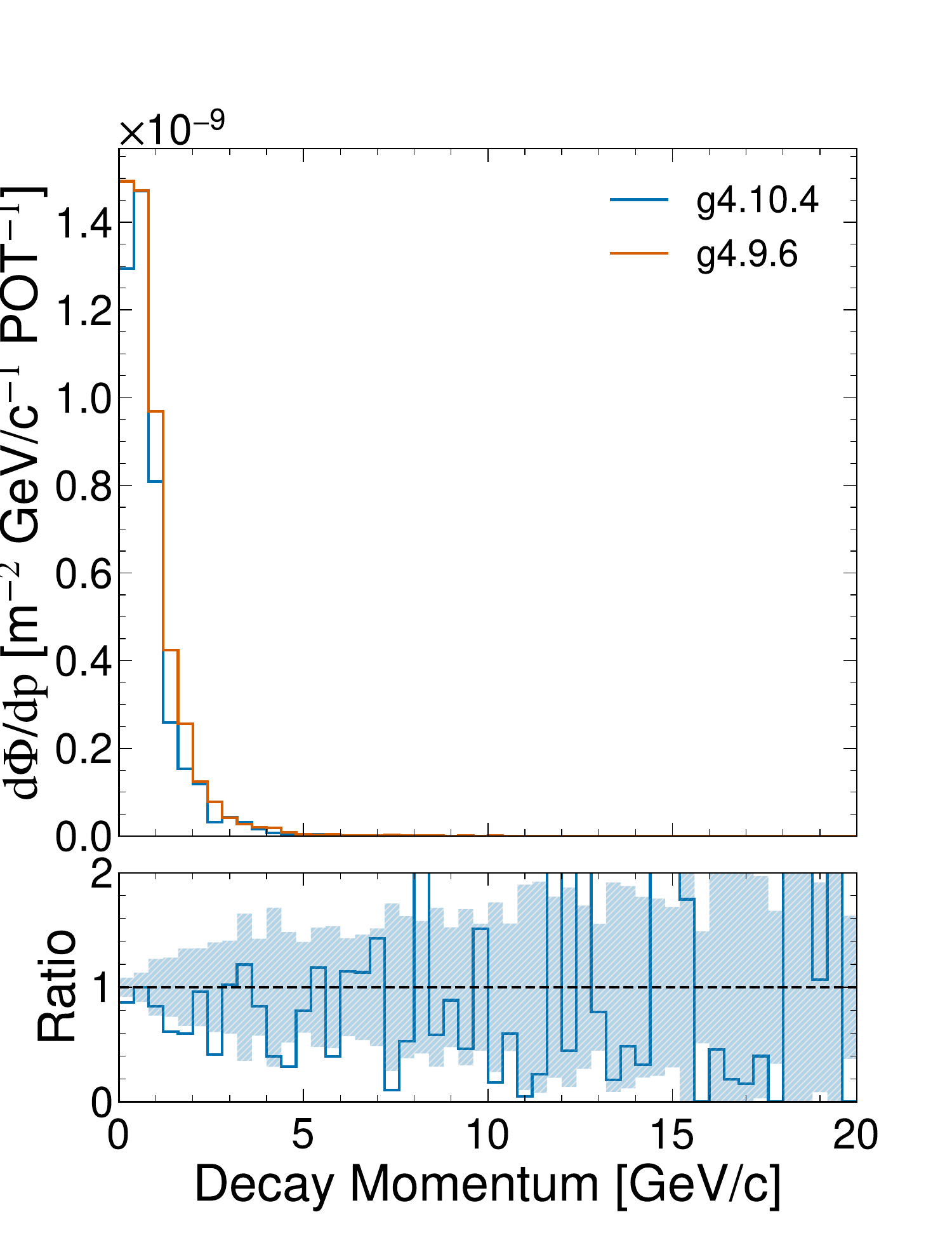}
    \caption{$\mu^- \to \numu$}
\end{subfigure}

\begin{subfigure}{0.23\textwidth}
    \includegraphics[width=\textwidth]{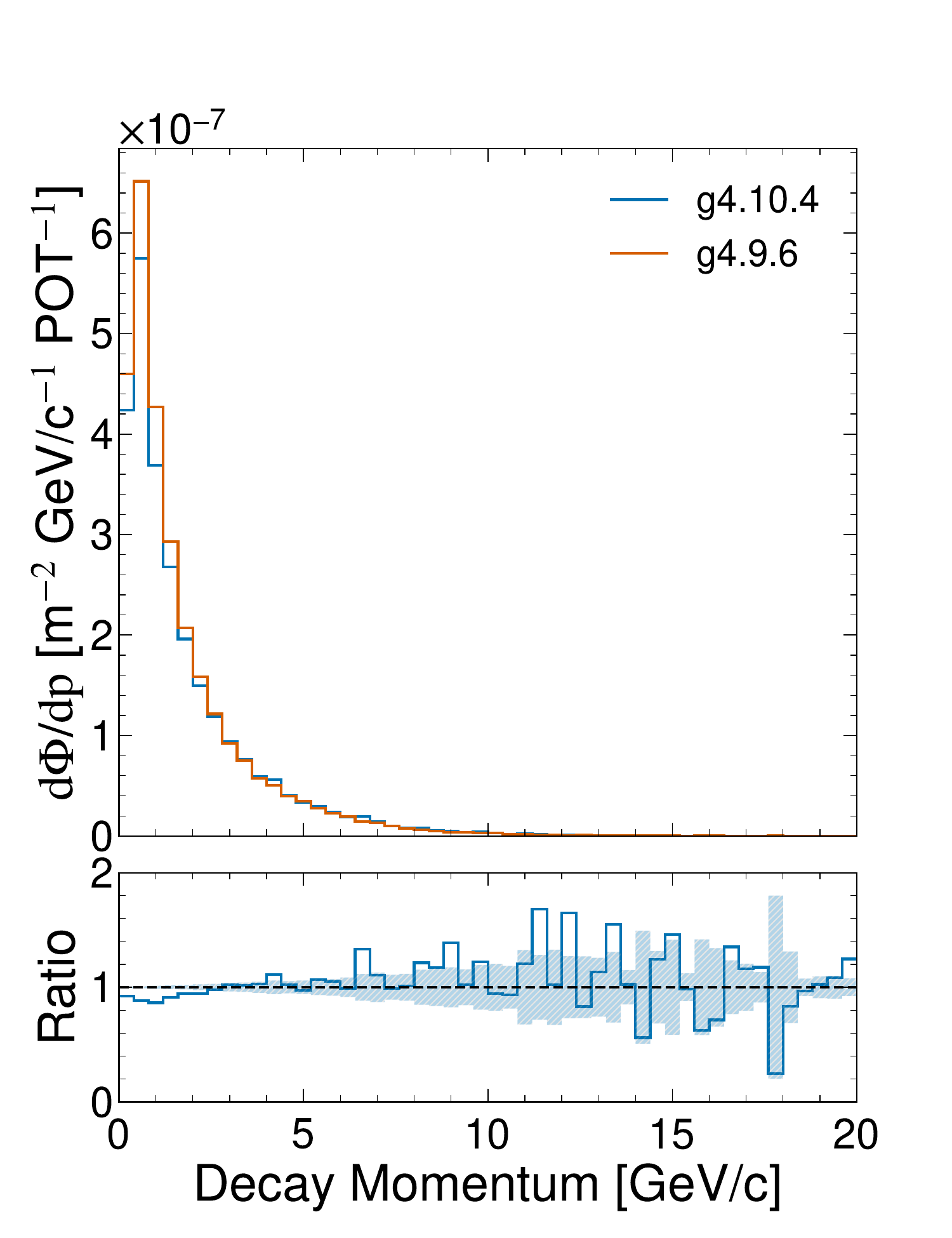}
    \caption{$\pi^- \to \numub$}
\end{subfigure}
\begin{subfigure}{0.23\textwidth}
    \includegraphics[width=\textwidth]{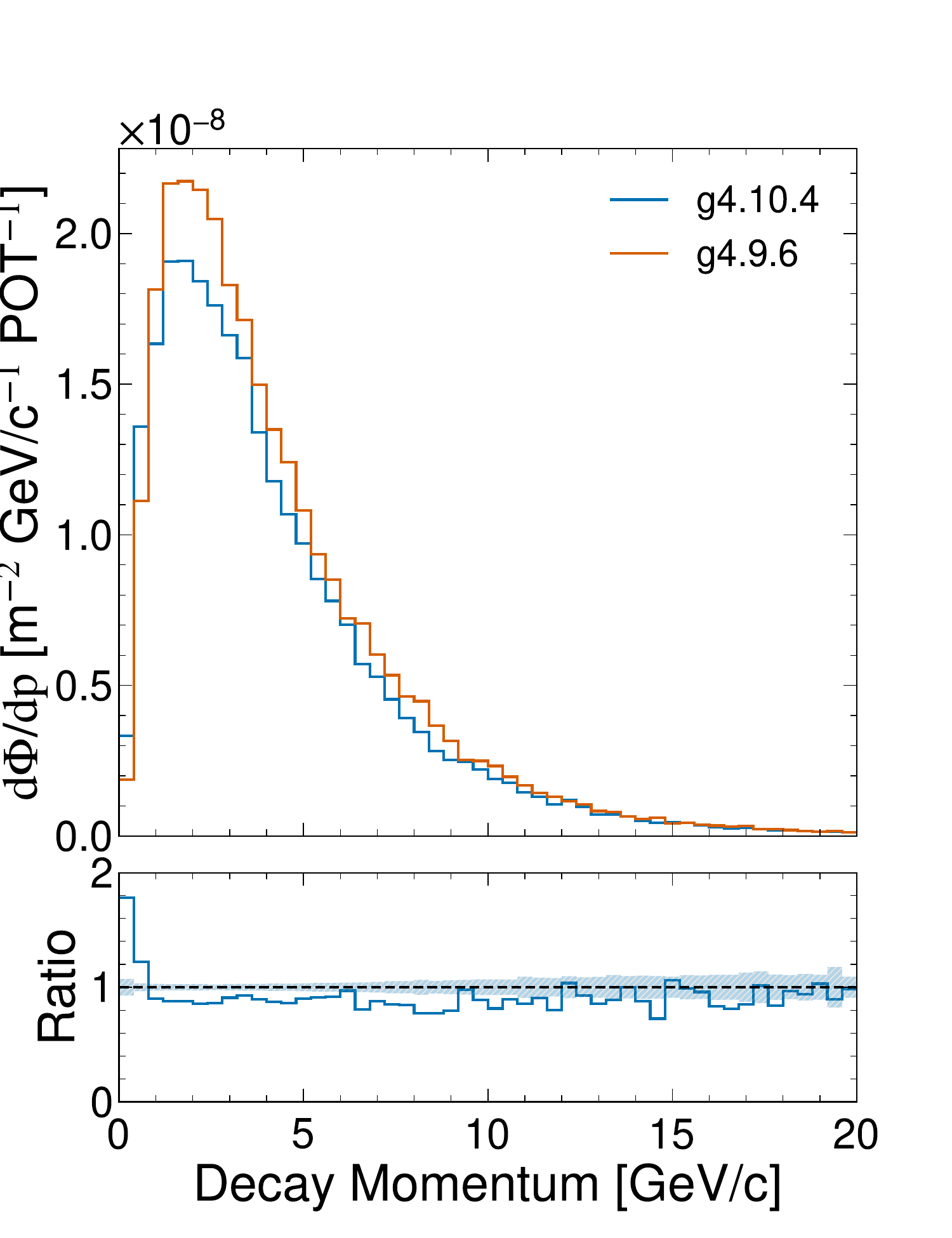}
    \caption{$K^- \to \numub$}
\end{subfigure}
\begin{subfigure}{0.23\textwidth}
    \includegraphics[width=\textwidth]{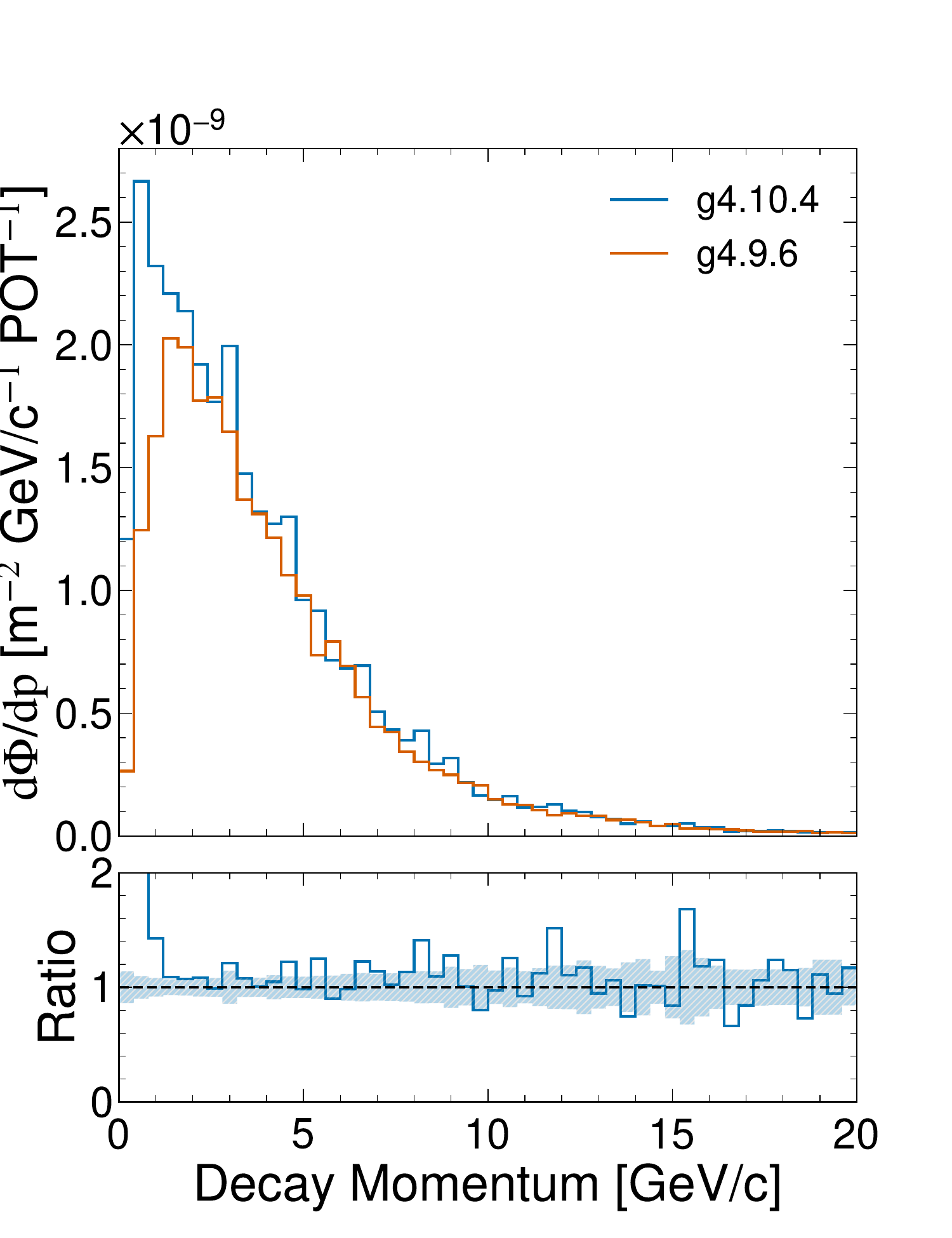}
    \caption{$K^0_L \to \numub$}
\end{subfigure}
\begin{subfigure}{0.23\textwidth}
    \includegraphics[width=\textwidth]{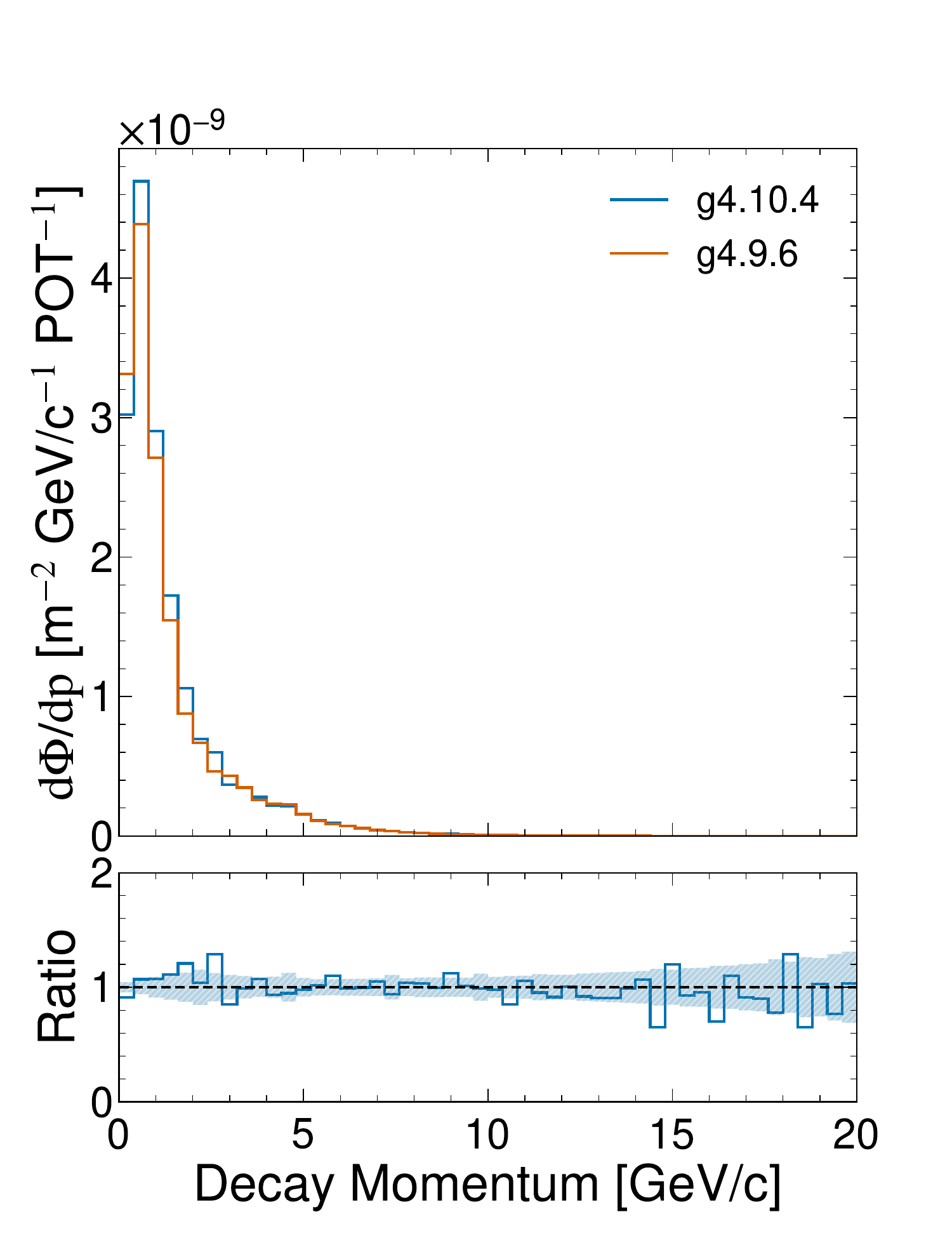}
    \caption{$\mu^+ \to \numub$}
\end{subfigure}
    \caption[Parent Decay Momenta (FHC, $\numu + \numub$)]{Parent decay momentum distributions for forward horn current muon neutrino and antineutrino modes.}
\end{figure}
\begin{figure}[!ht]
\centering

\begin{subfigure}{0.23\textwidth}
    \includegraphics[width=\textwidth]{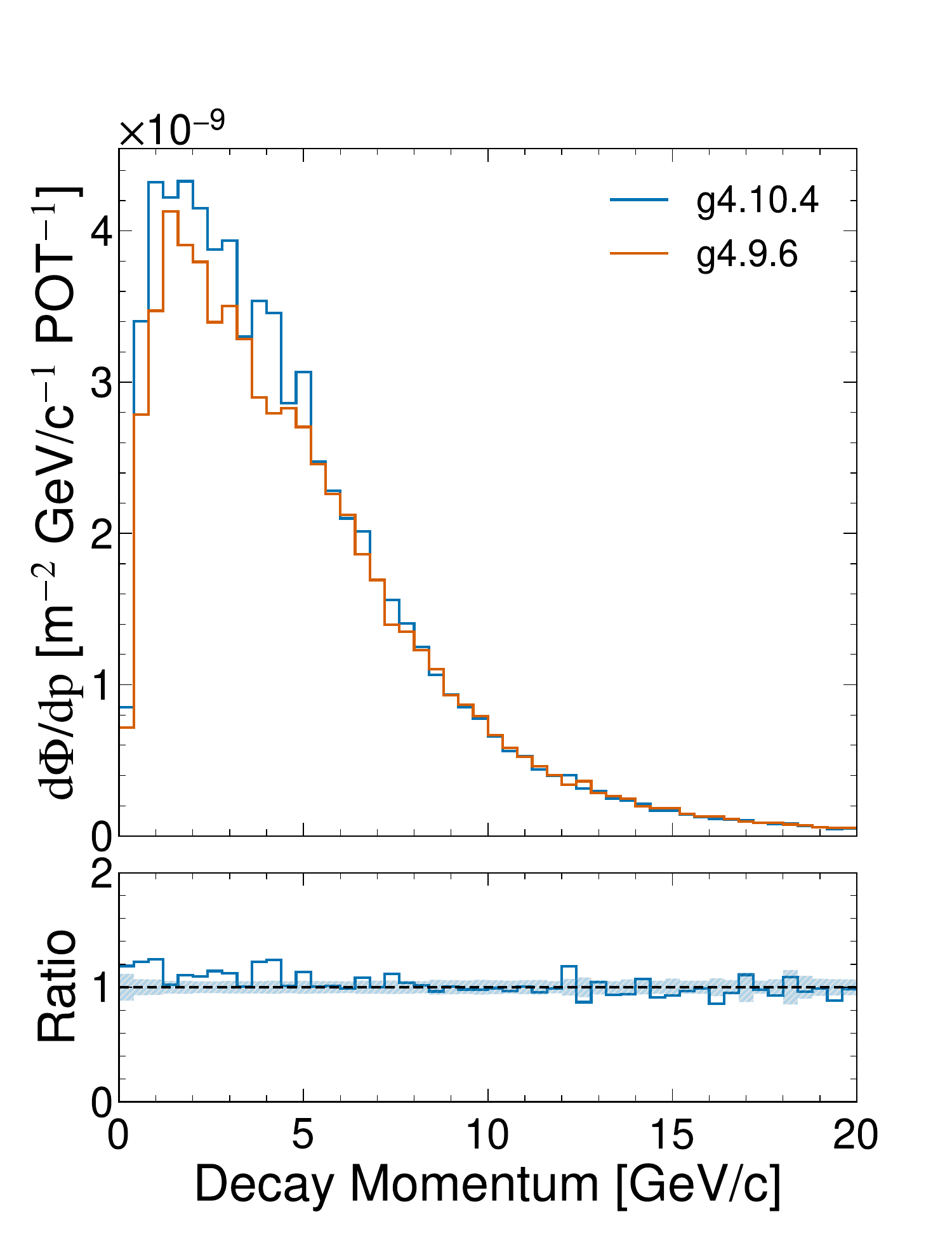}
    \caption{$K^+ \to \nue$}
\end{subfigure}
\begin{subfigure}{0.23\textwidth}
    \includegraphics[width=\textwidth]{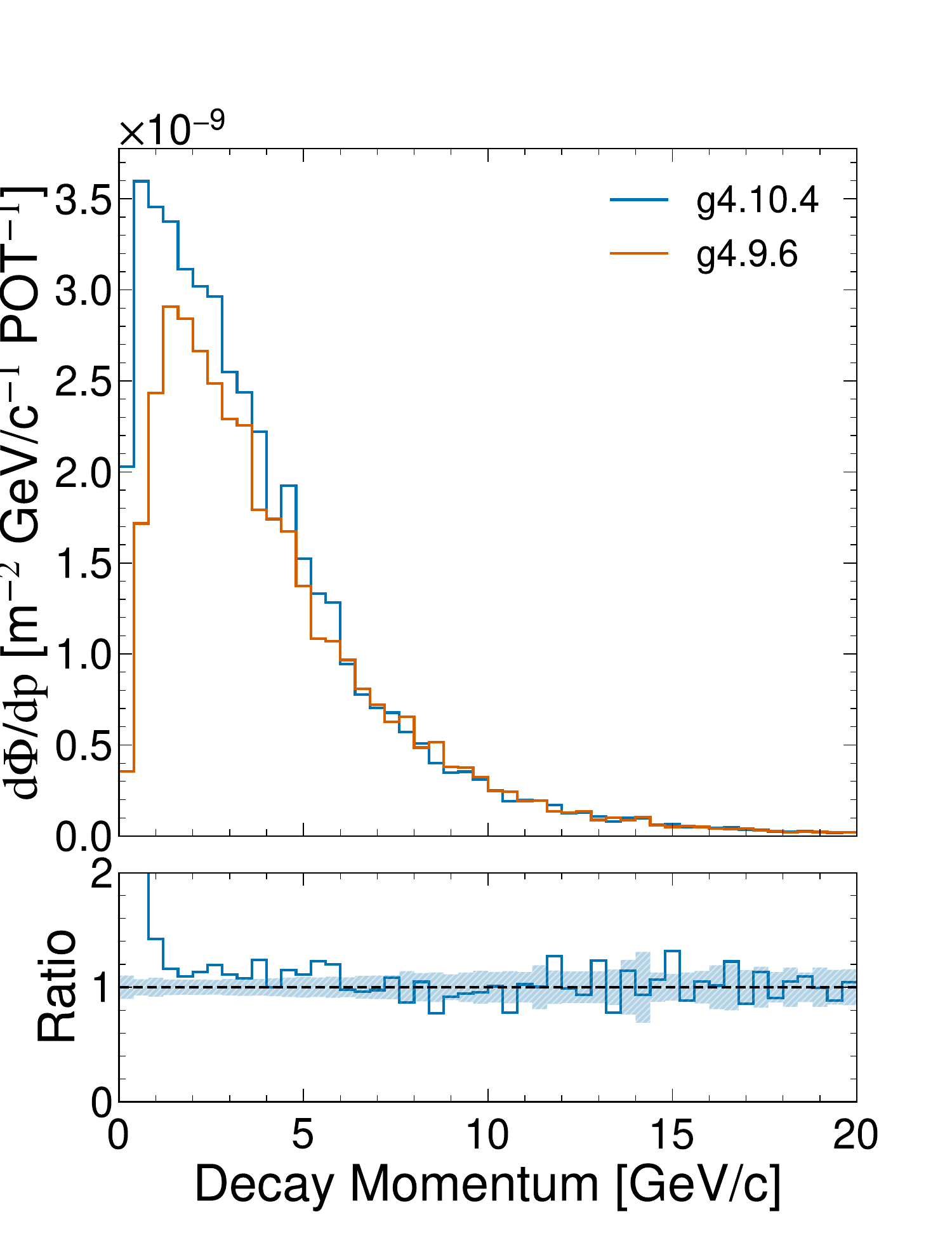}
    \caption{$K^0_L \to \nue$}
\end{subfigure}
\begin{subfigure}{0.23\textwidth}
    \includegraphics[width=\textwidth]{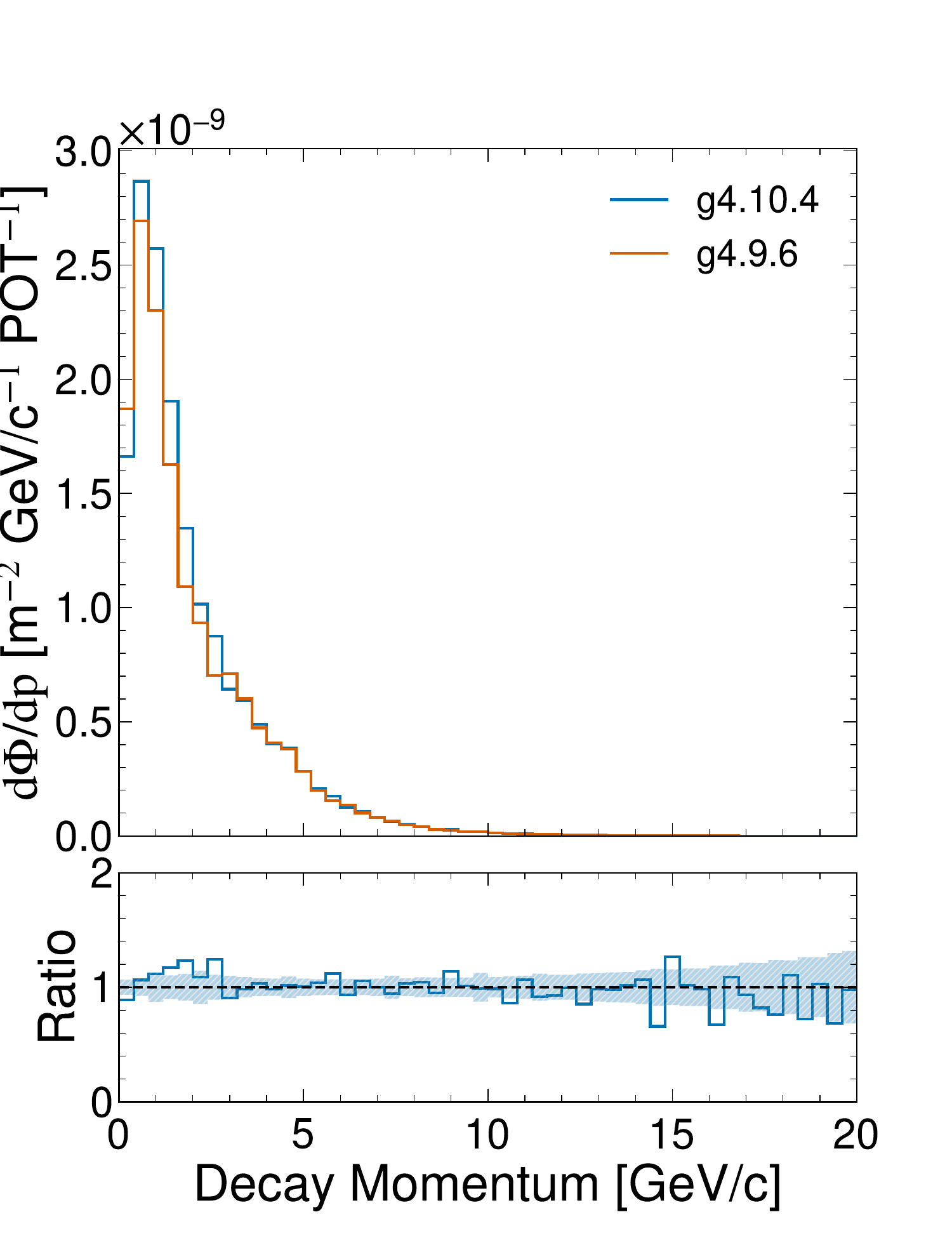}
    \caption{$\mu^+ \to \nue$}
\end{subfigure}

\begin{subfigure}{0.23\textwidth}
    \includegraphics[width=\textwidth]{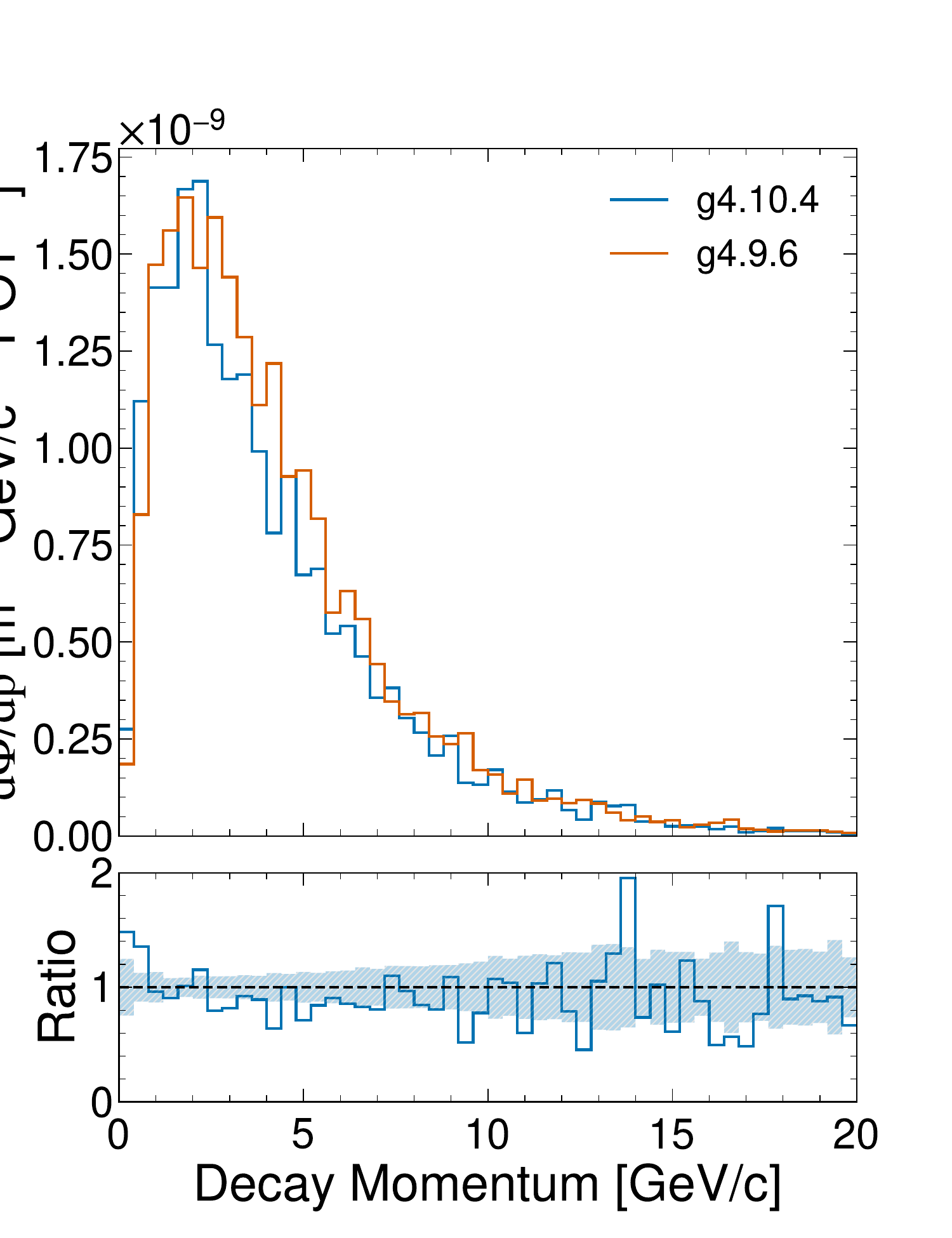}
    \caption{$K^- \to \nueb$}
\end{subfigure}
\begin{subfigure}{0.23\textwidth}
    \includegraphics[width=\textwidth]{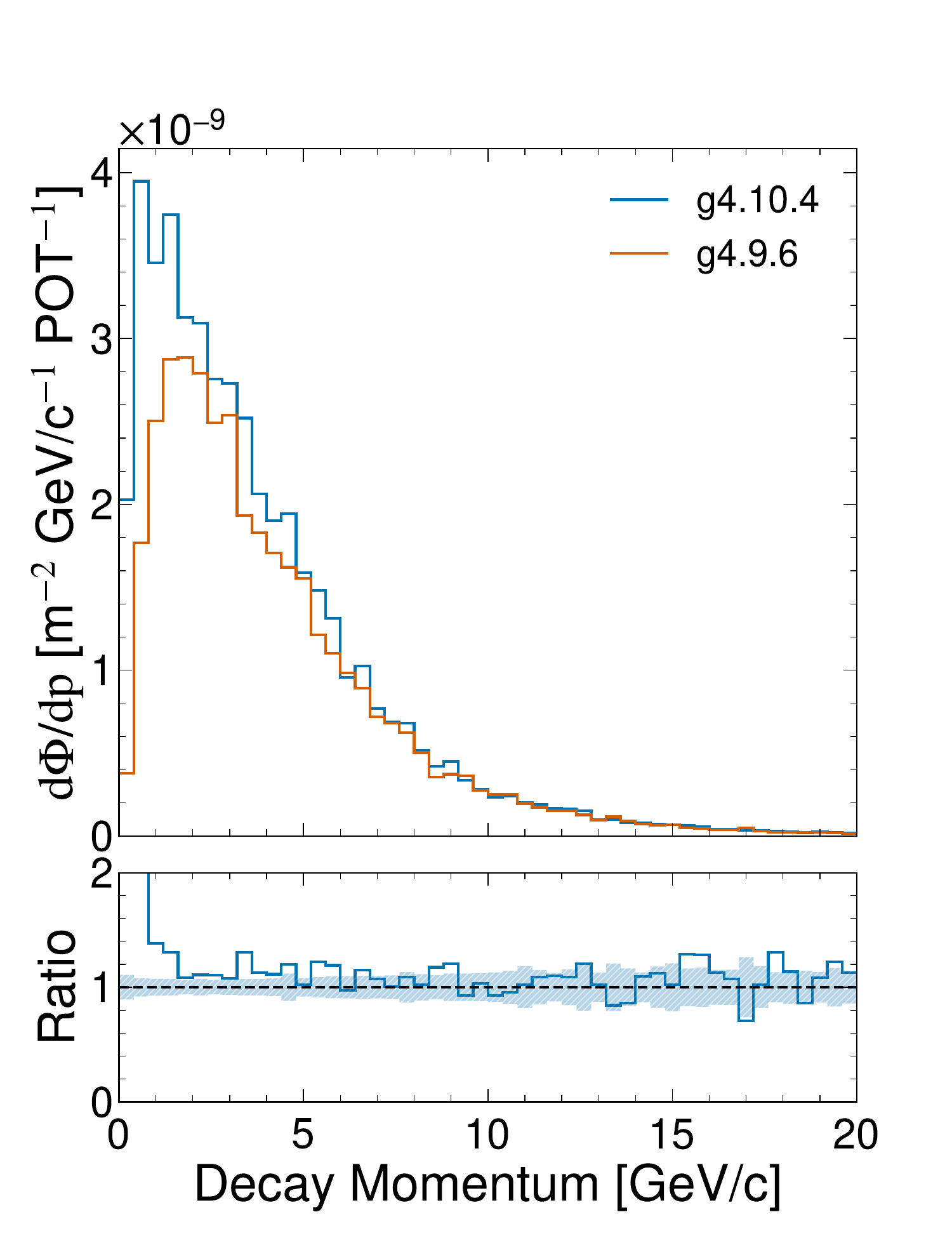}
    \caption{$K^0_L \to \nueb$}
\end{subfigure}
\begin{subfigure}{0.23\textwidth}
    \includegraphics[width=\textwidth]{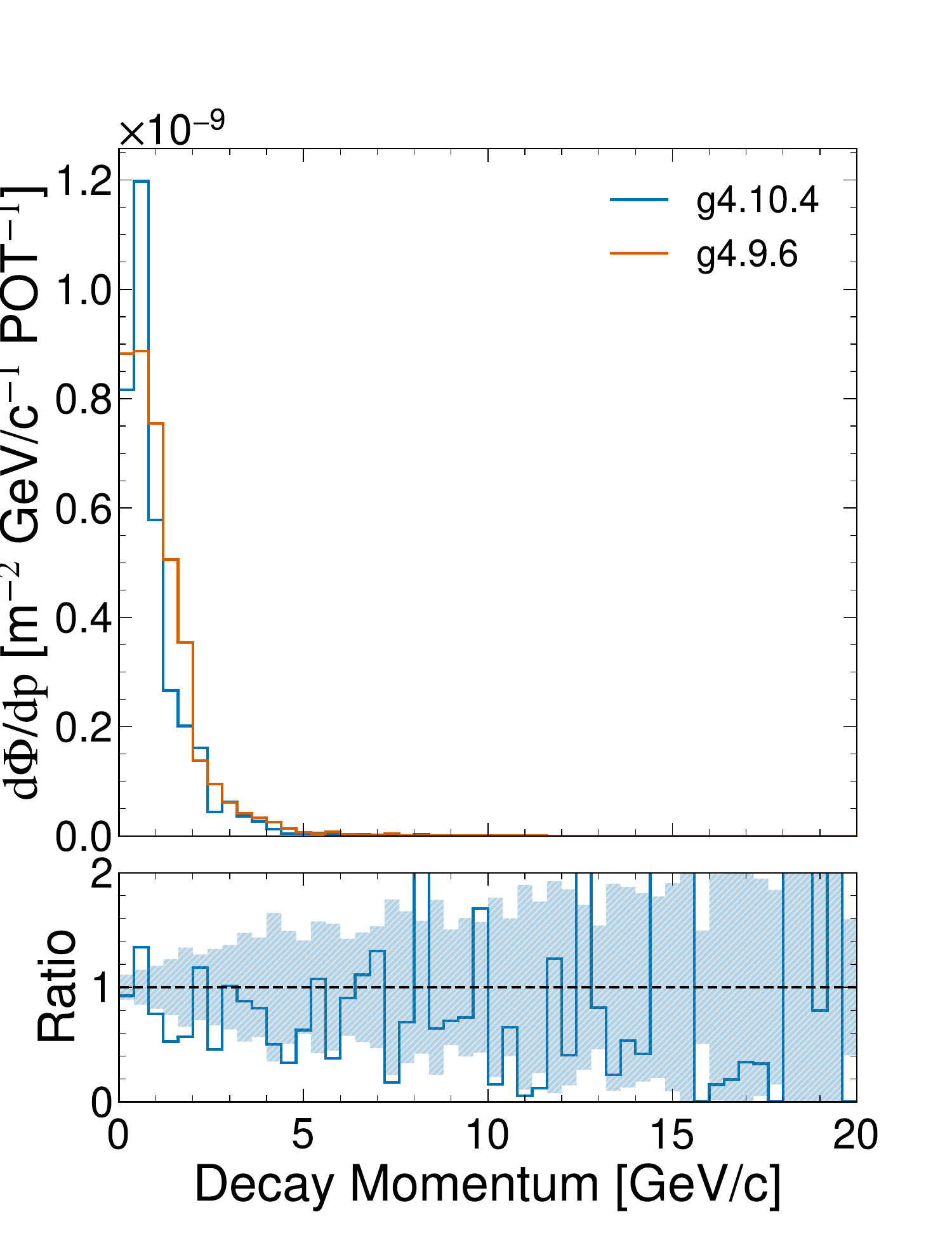}
    \caption{$\mu^- \to \nueb$}
\end{subfigure}
    \caption[Parent Decay Momenta (FHC, $\nue + \nueb$)]{Parent decay momentum distributions for forward horn current electron neutrino and antineutrino modes.}
\end{figure}
\clearpage

\section{Reverse Horn Current}
\begin{figure}[!ht]
\centering
\begin{subfigure}{0.23\textwidth}
    \includegraphics[width=\textwidth]{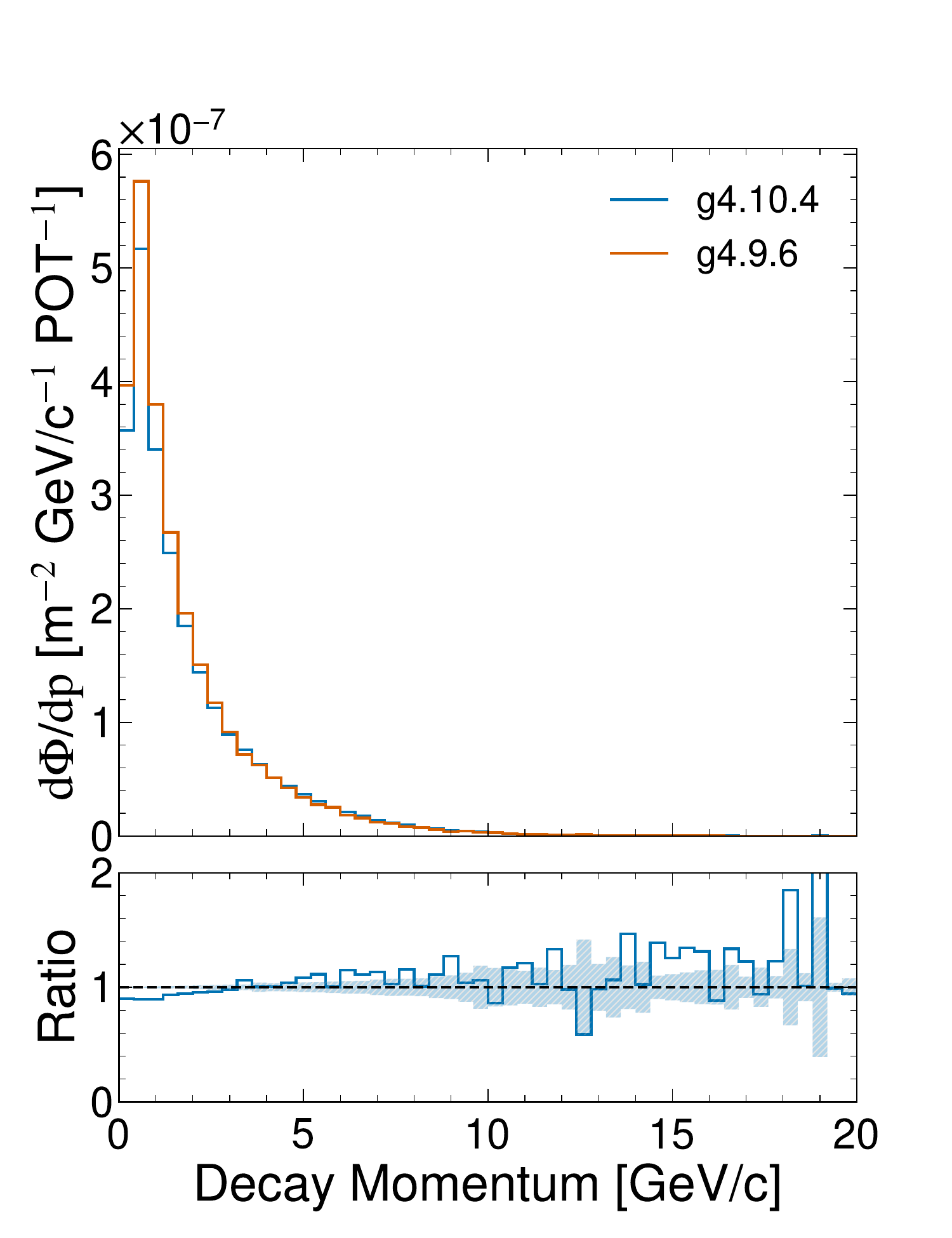}
    \caption{$\pi^+ \to \numu$}
\end{subfigure}
\begin{subfigure}{0.23\textwidth}
    \includegraphics[width=\textwidth]{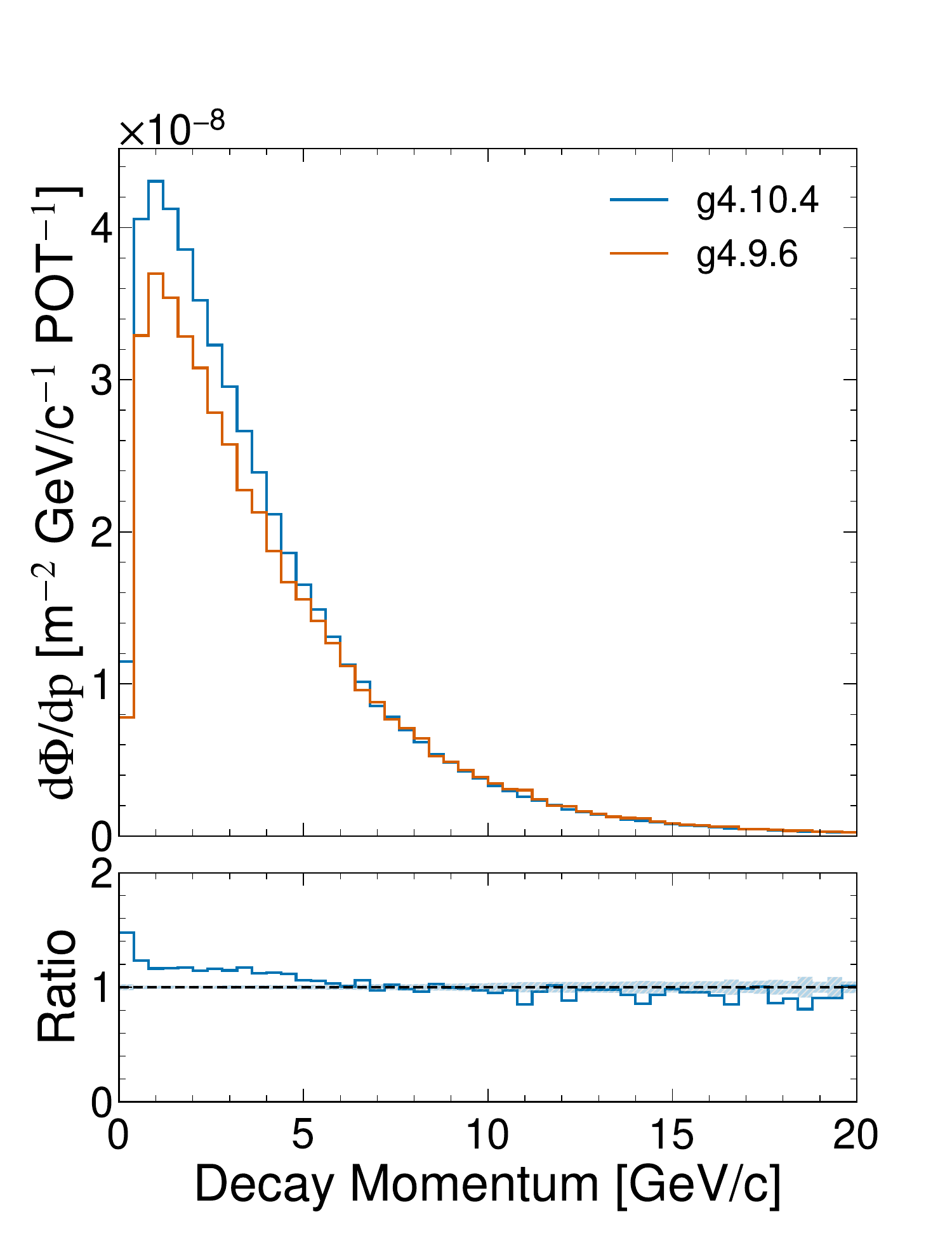}
    \caption{$K^+ \to \numu$}
\end{subfigure}
\begin{subfigure}{0.23\textwidth}
    \includegraphics[width=\textwidth]{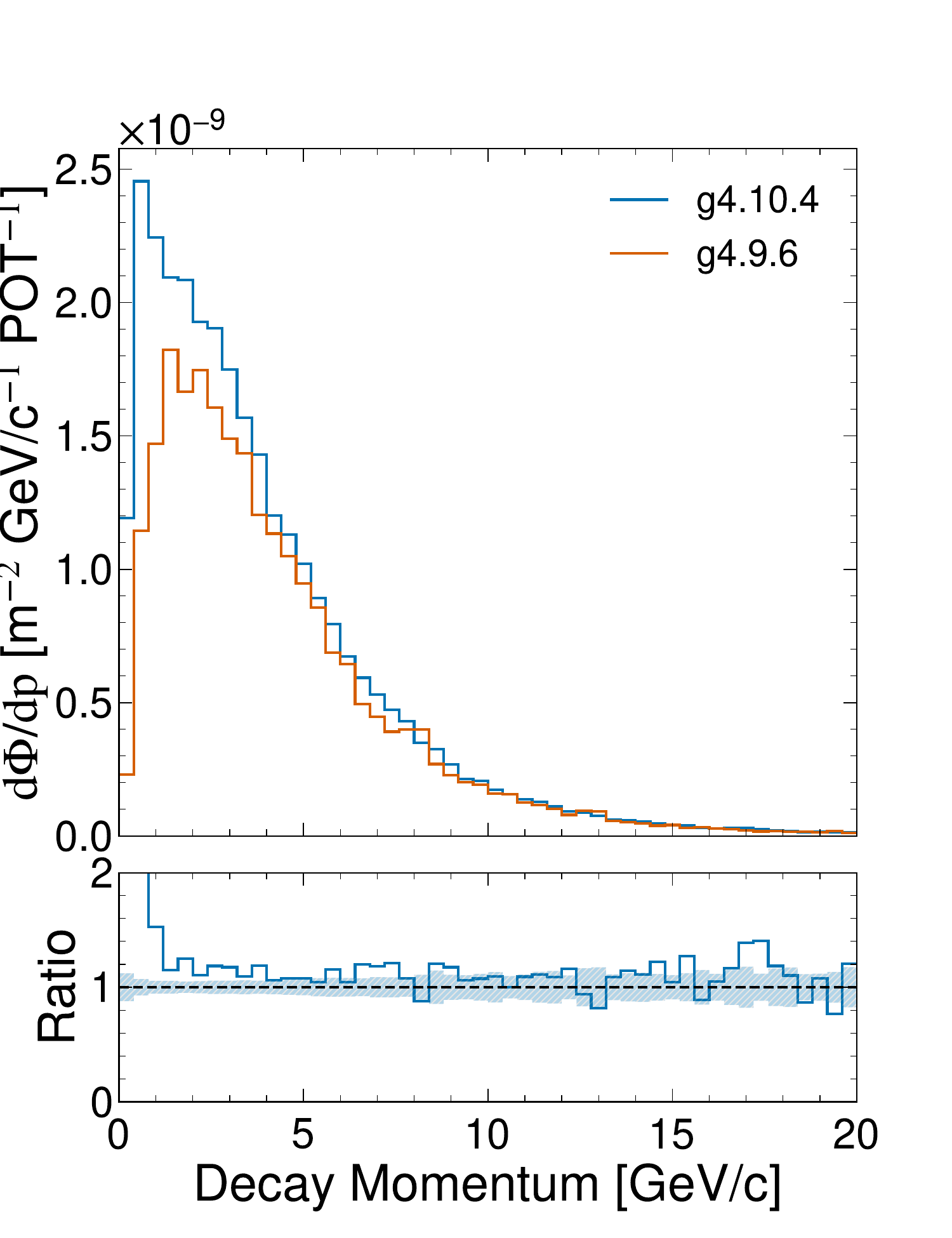}
    \caption{$K^0_L \to \numu$}
\end{subfigure}
\begin{subfigure}{0.23\textwidth}
    \includegraphics[width=\textwidth]{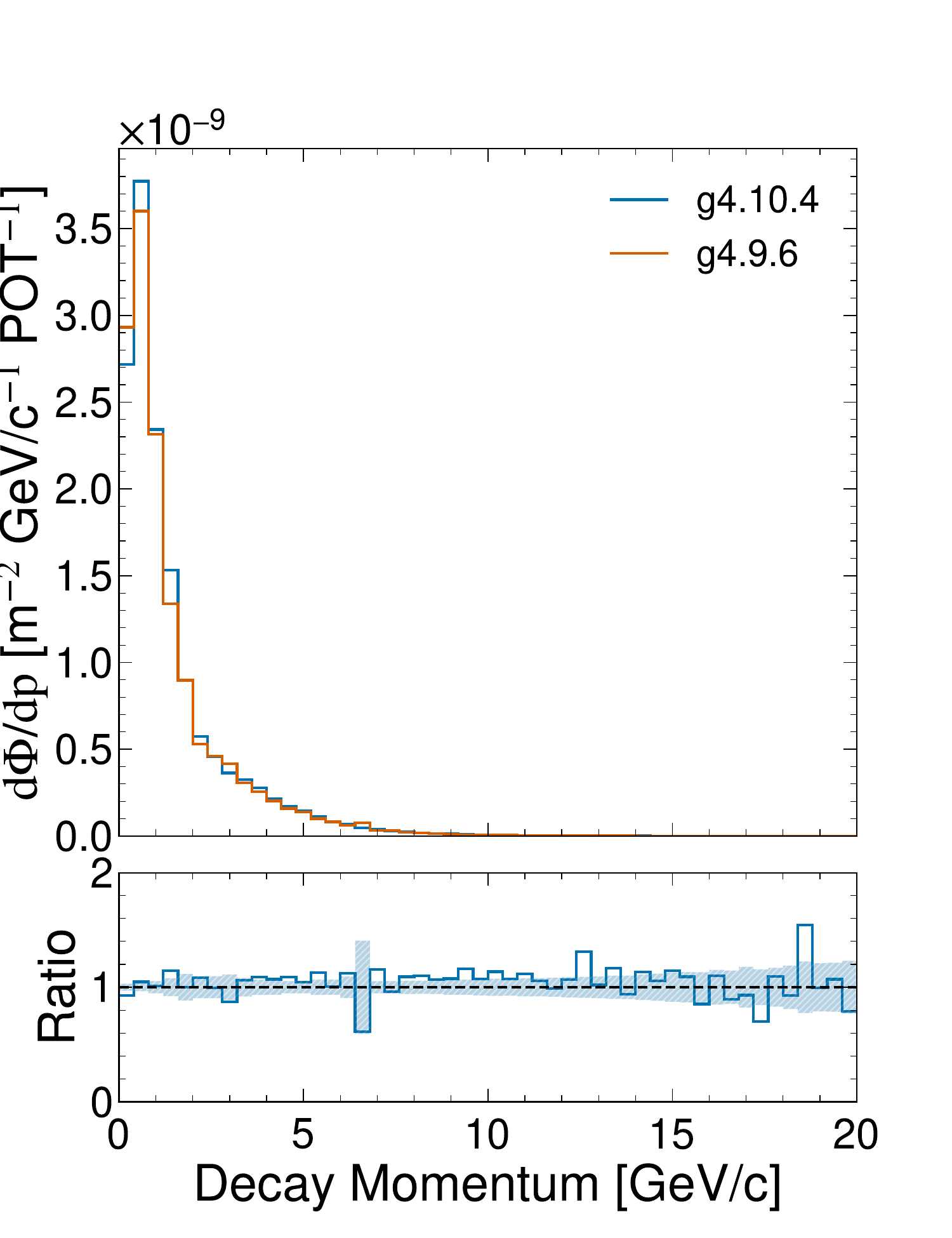}
    \caption{$\mu^- \to \numu$}
\end{subfigure}

\begin{subfigure}{0.23\textwidth}
    \includegraphics[width=\textwidth]{parent_momentum_new_geant_rhc_numubar_pipm.pdf}
    \caption{$\pi^- \to \numub$}
\end{subfigure}
\begin{subfigure}{0.23\textwidth}
    \includegraphics[width=\textwidth]{parent_momentum_new_geant_rhc_numubar_Kpm.pdf}
    \caption{$K^- \to \numub$}
\end{subfigure}
\begin{subfigure}{0.23\textwidth}
    \includegraphics[width=\textwidth]{parent_momentum_new_geant_rhc_numubar_K0l.pdf}
    \caption{$K^0_L \to \numub$}
\end{subfigure}
\begin{subfigure}{0.23\textwidth}
    \includegraphics[width=\textwidth]{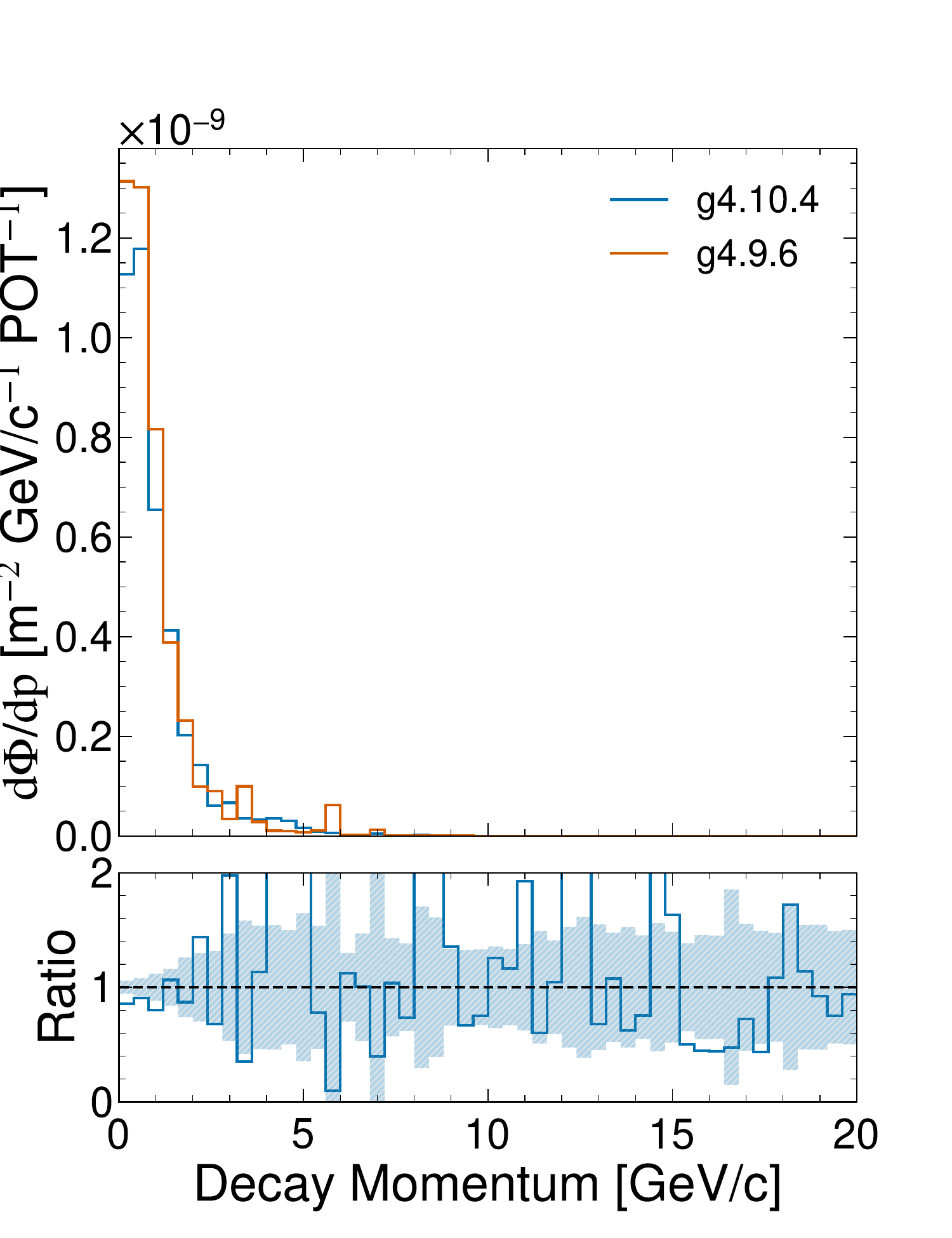}
    \caption{$\mu^+ \to \numub$}
\end{subfigure}
    \caption[Parent Decay Momenta (RHC, $\numu + \numub$)]{Parent decay momentum distributions for reverse horn current muon neutrino and antineutrino modes.}
\end{figure}
\begin{figure}[!ht]
\centering
\begin{subfigure}{0.23\textwidth}
    \includegraphics[width=\textwidth]{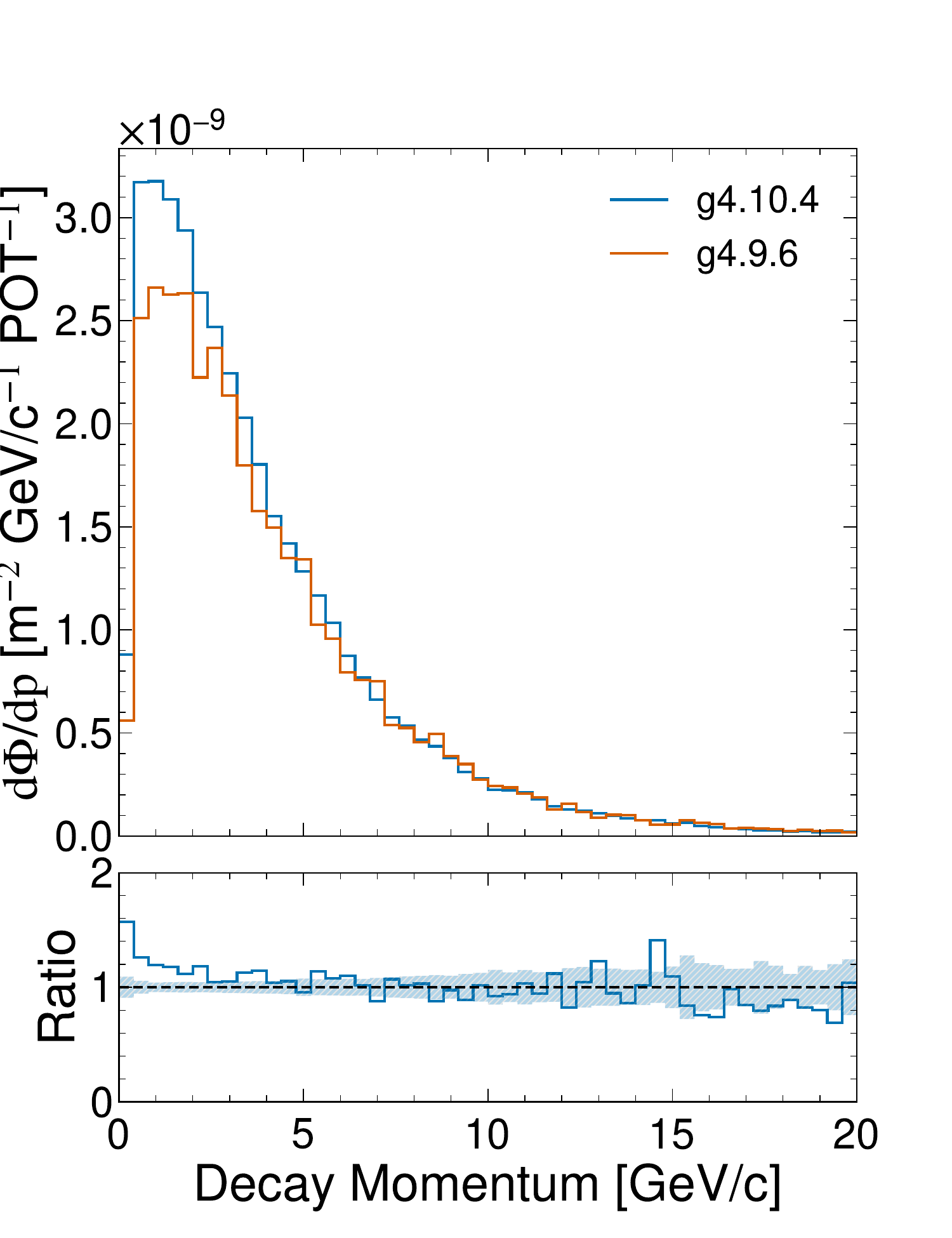}
    \caption{$K^+ \to \nue$}
\end{subfigure}
\begin{subfigure}{0.23\textwidth}
    \includegraphics[width=\textwidth]{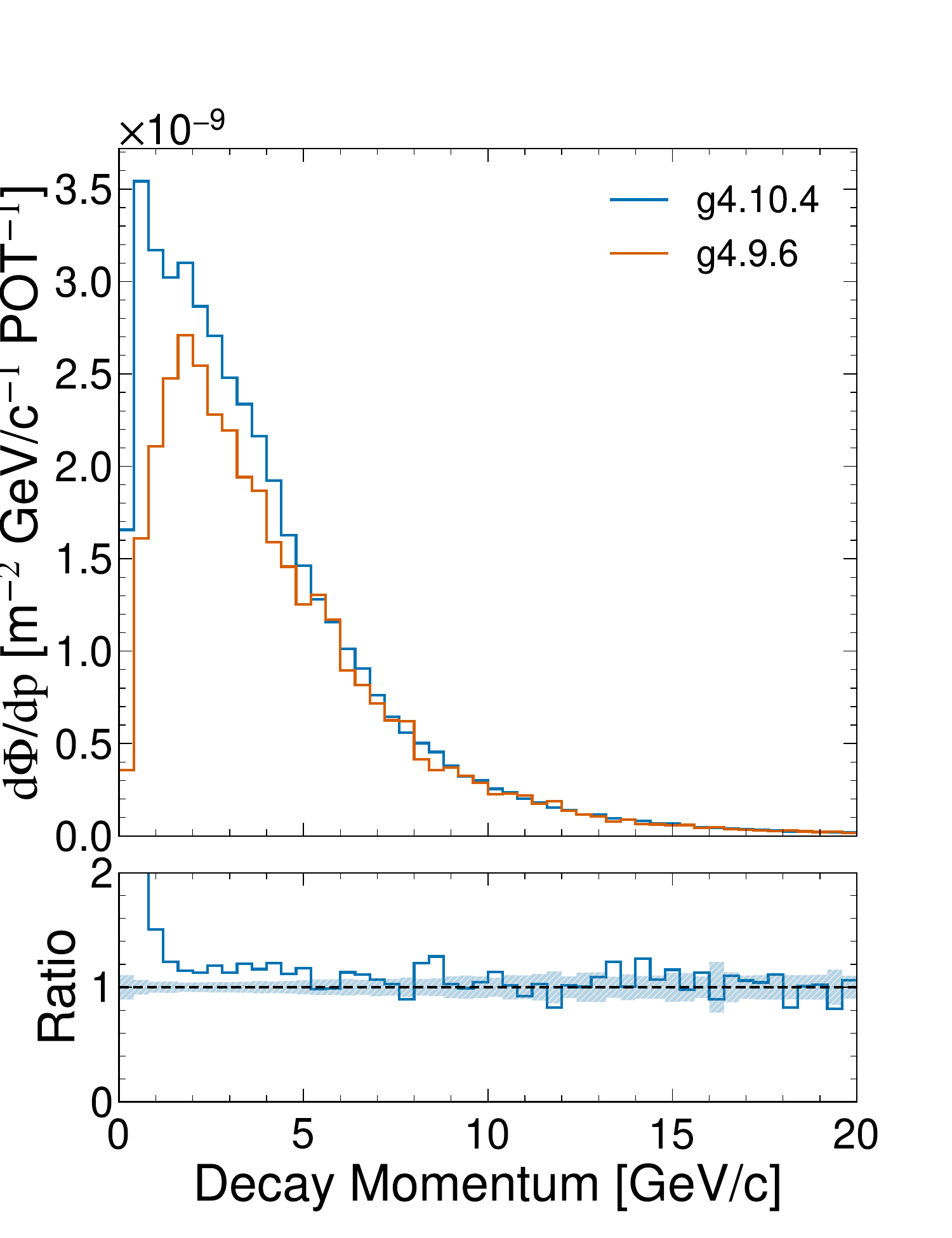}
    \caption{$K^0_L \to \nue$}
\end{subfigure}
\begin{subfigure}{0.23\textwidth}
    \includegraphics[width=\textwidth]{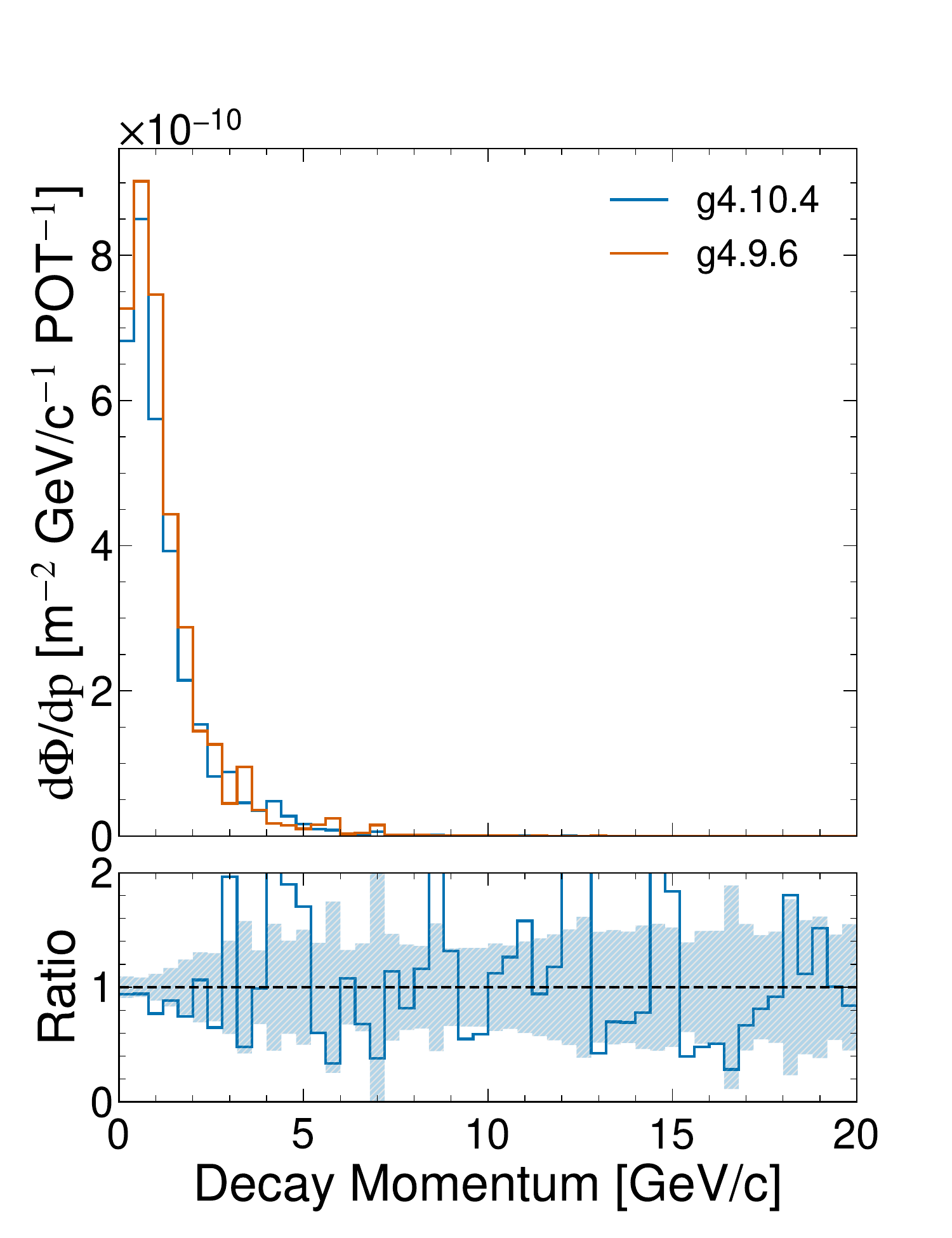}
    \caption{$\mu^+ \to \nue$}
\end{subfigure}

\begin{subfigure}{0.23\textwidth}
    \includegraphics[width=\textwidth]{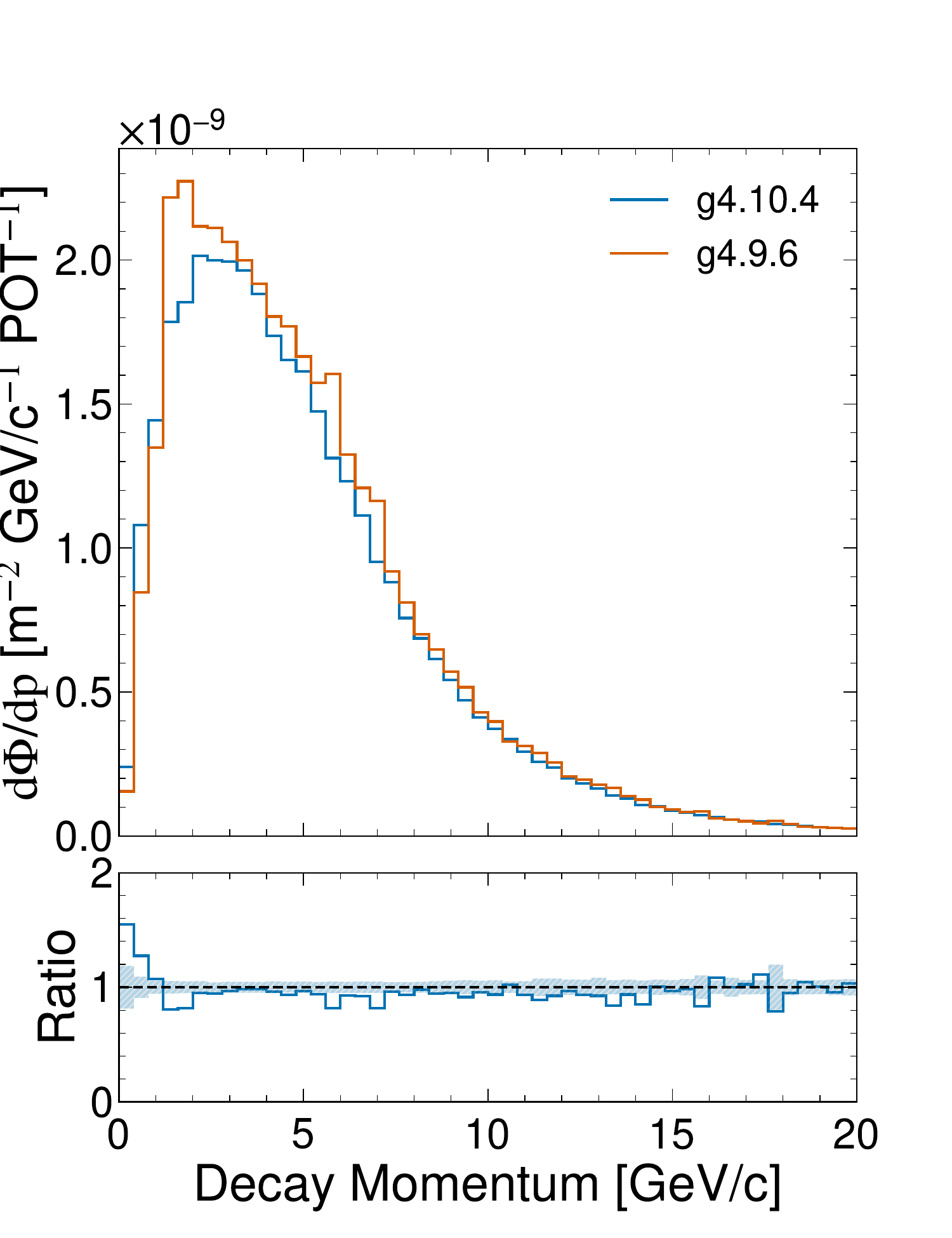}
    \caption{$K^- \to \nueb$}
\end{subfigure}
\begin{subfigure}{0.23\textwidth}
    \includegraphics[width=\textwidth]{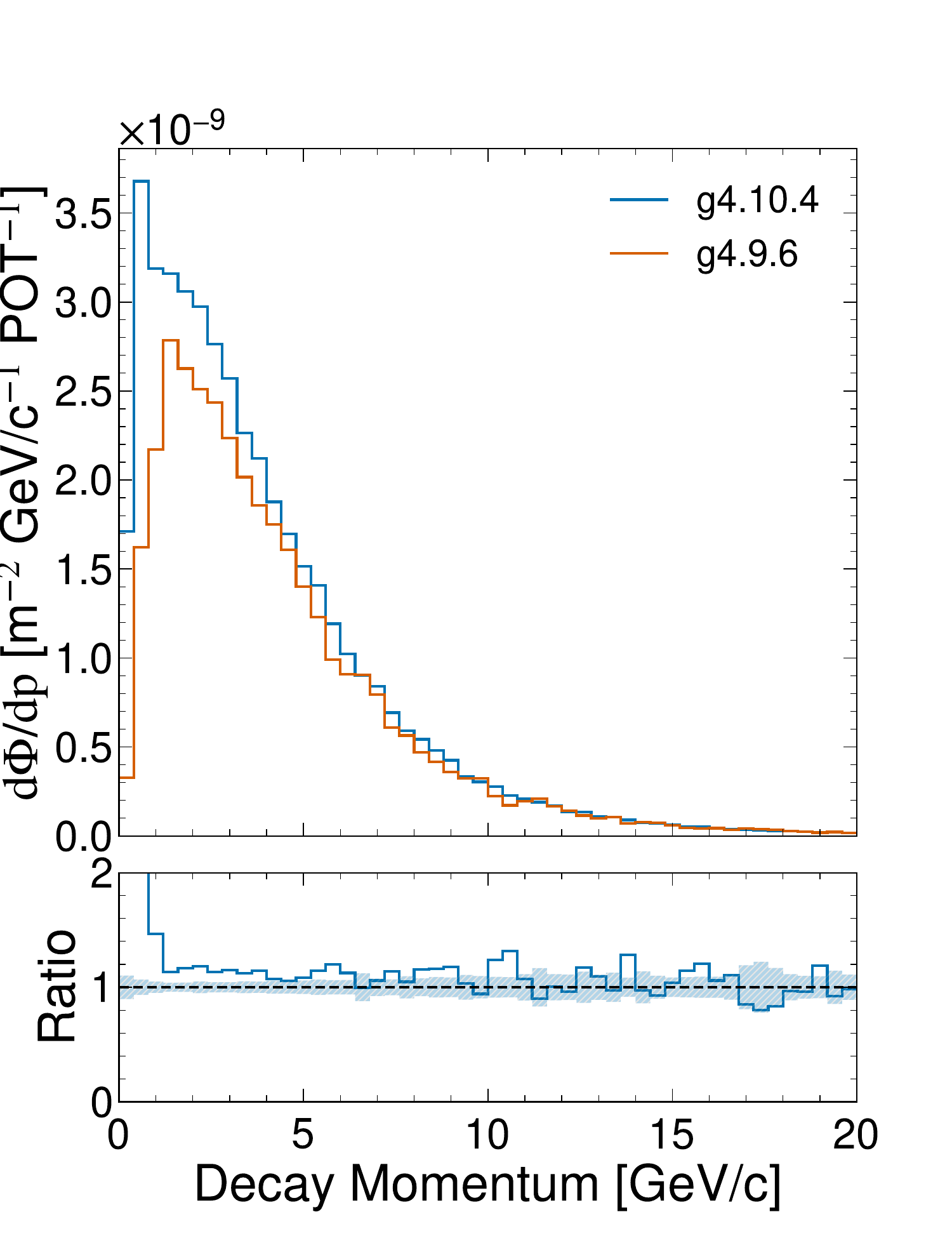}
    \caption{$K^0_L \to \nueb$}
\end{subfigure}
\begin{subfigure}{0.23\textwidth}
    \includegraphics[width=\textwidth]{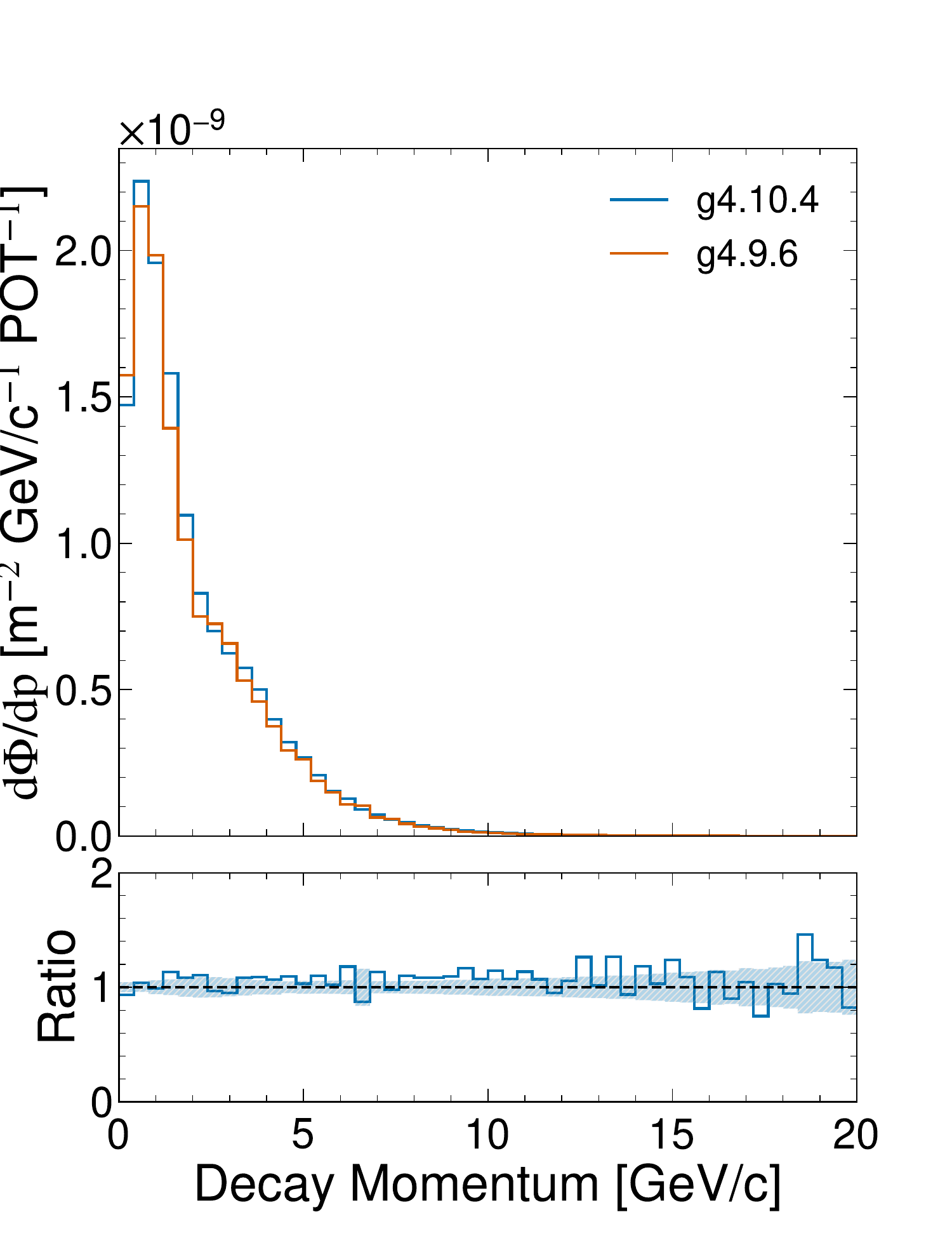}
    \caption{$\mu^- \to \nueb$}
\end{subfigure}
    \caption[Parent Decay Momenta (RHC, $\nue + \nueb$)]{Parent decay momentum distributions for reverse horn current electron neutrino and antineutrino modes.}
\end{figure}

%% file: parent_decay_angles.tex
\clearpage
\section{Forward Horn Current}
\begin{figure}[!ht]
    \centering
    \begin{subfigure}{0.23\textwidth}
        \includegraphics[width=\textwidth]{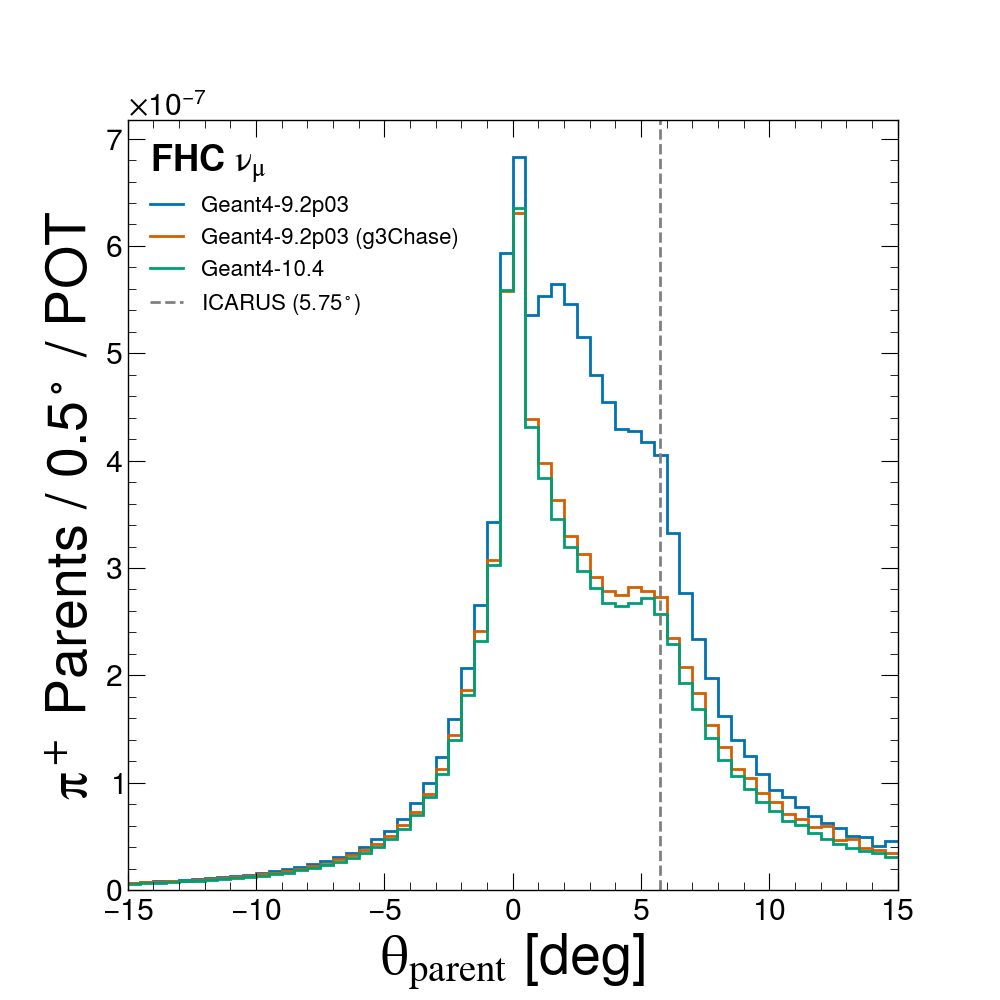}
        \caption{$\pi^+ \to \numu$}
    \end{subfigure}
    \begin{subfigure}{0.23\textwidth}
        \includegraphics[width=\textwidth]{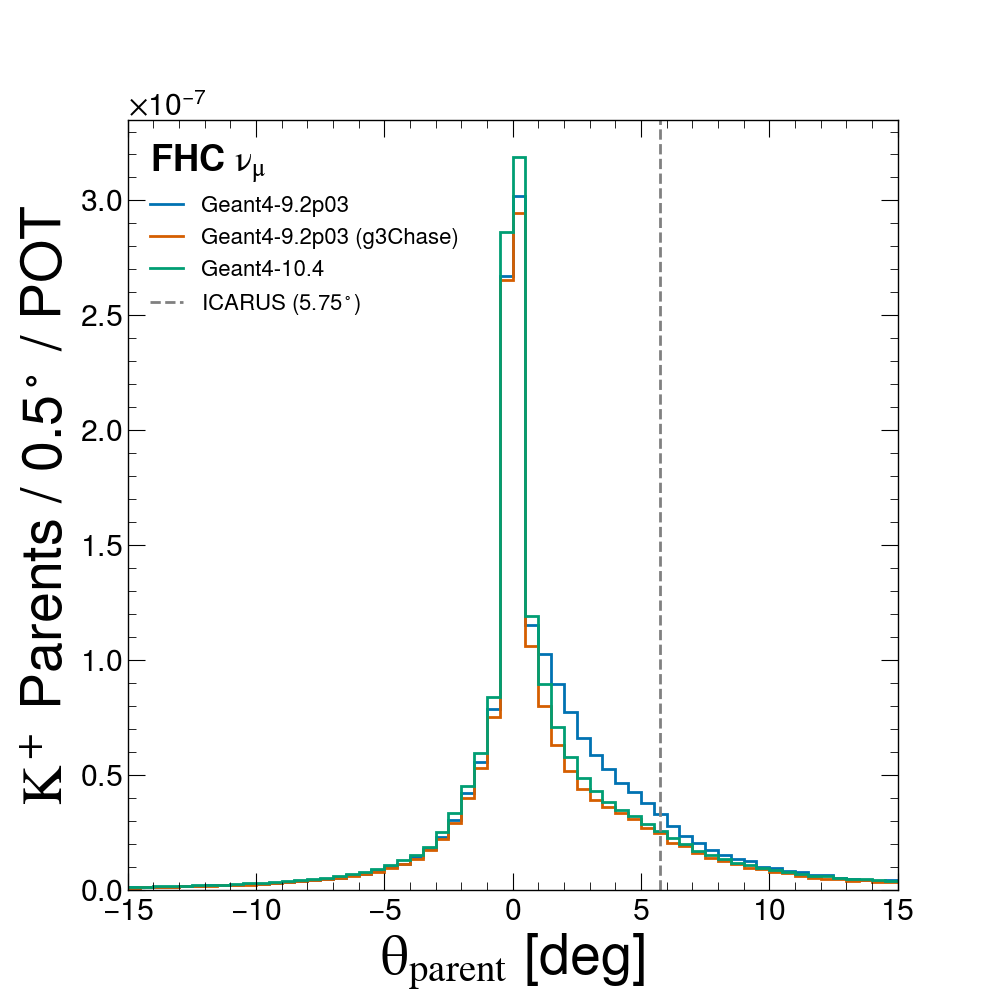}
        \caption{$K^+ \to \numu$}
    \end{subfigure}
    \begin{subfigure}{0.23\textwidth}
        \includegraphics[width=\textwidth]{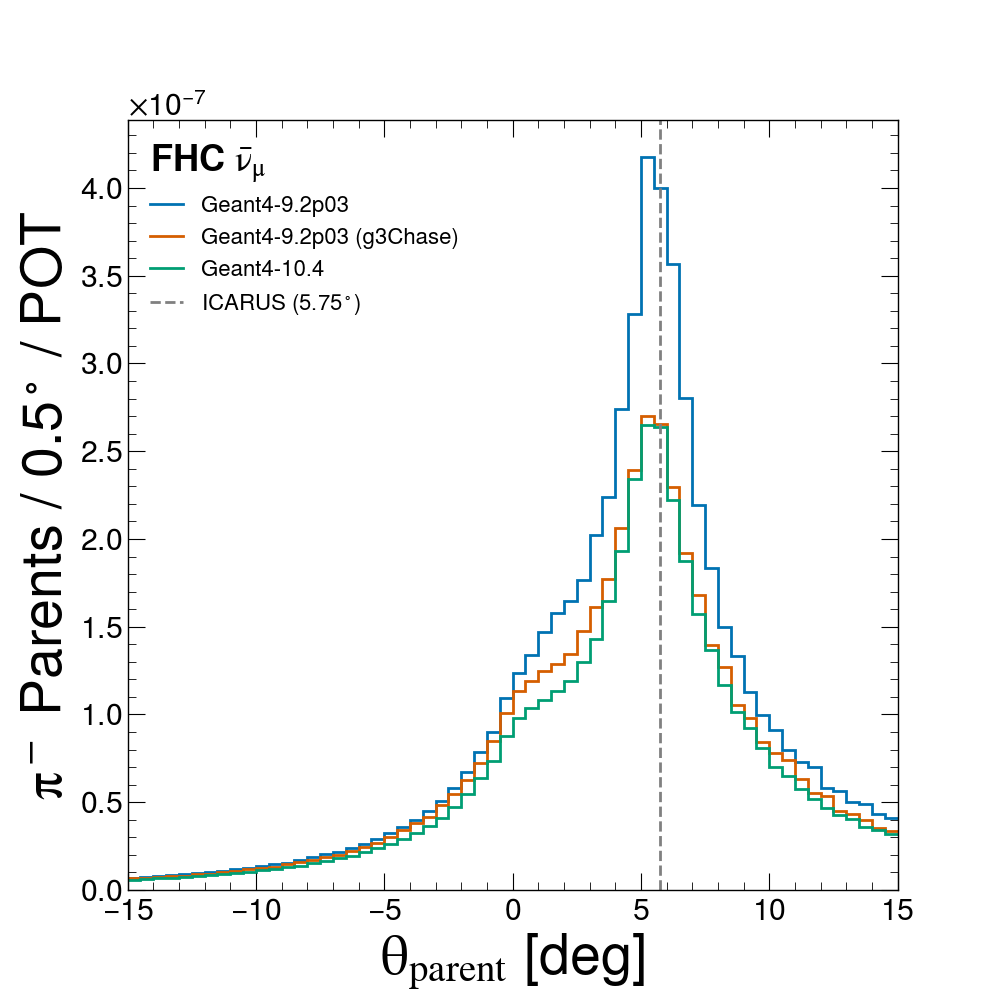}
        \caption{$\pi^- \to \numub$}
    \end{subfigure}
    \begin{subfigure}{0.23\textwidth}
        \includegraphics[width=\textwidth]{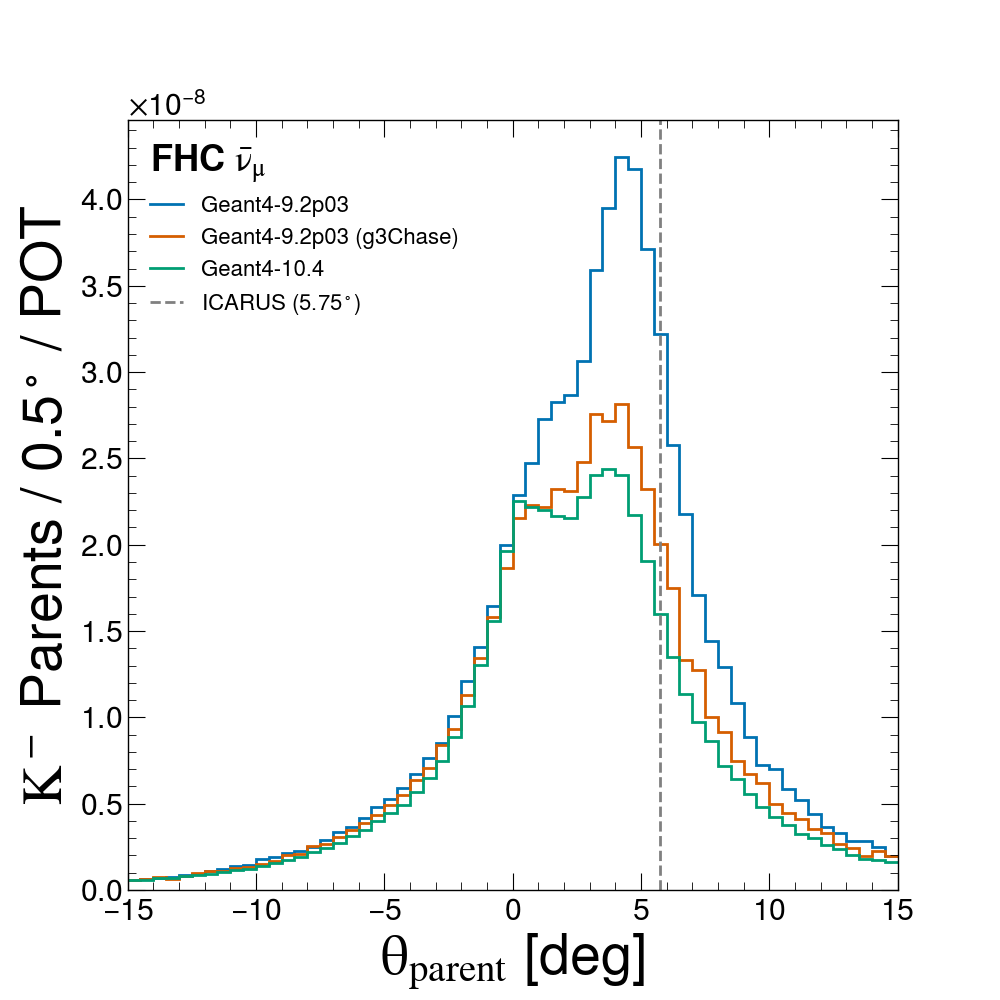}
        \caption{$K^- \to \numub$}
    \end{subfigure}
    \begin{subfigure}{0.23\textwidth}
        \includegraphics[width=\textwidth]{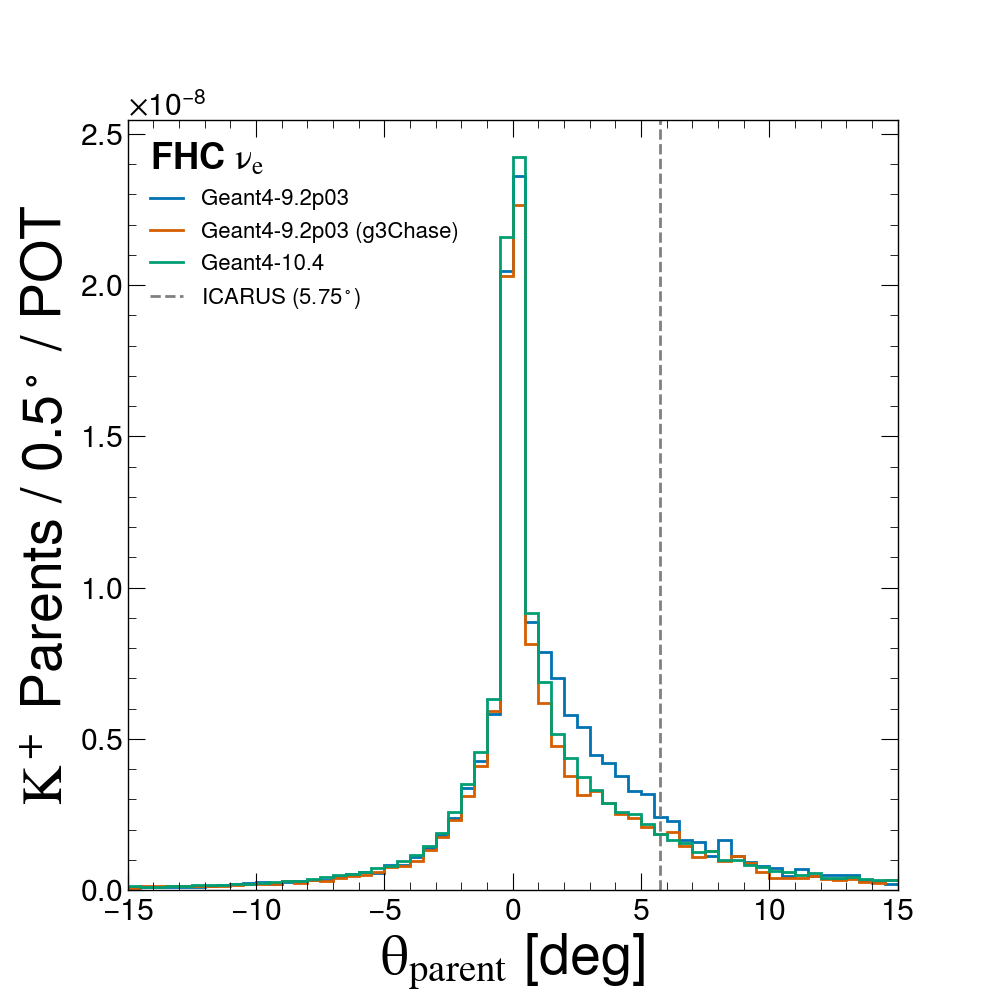}
        \caption{$K^+ \to \nue$}
    \end{subfigure}
    \begin{subfigure}{0.23\textwidth}
        \includegraphics[width=\textwidth]{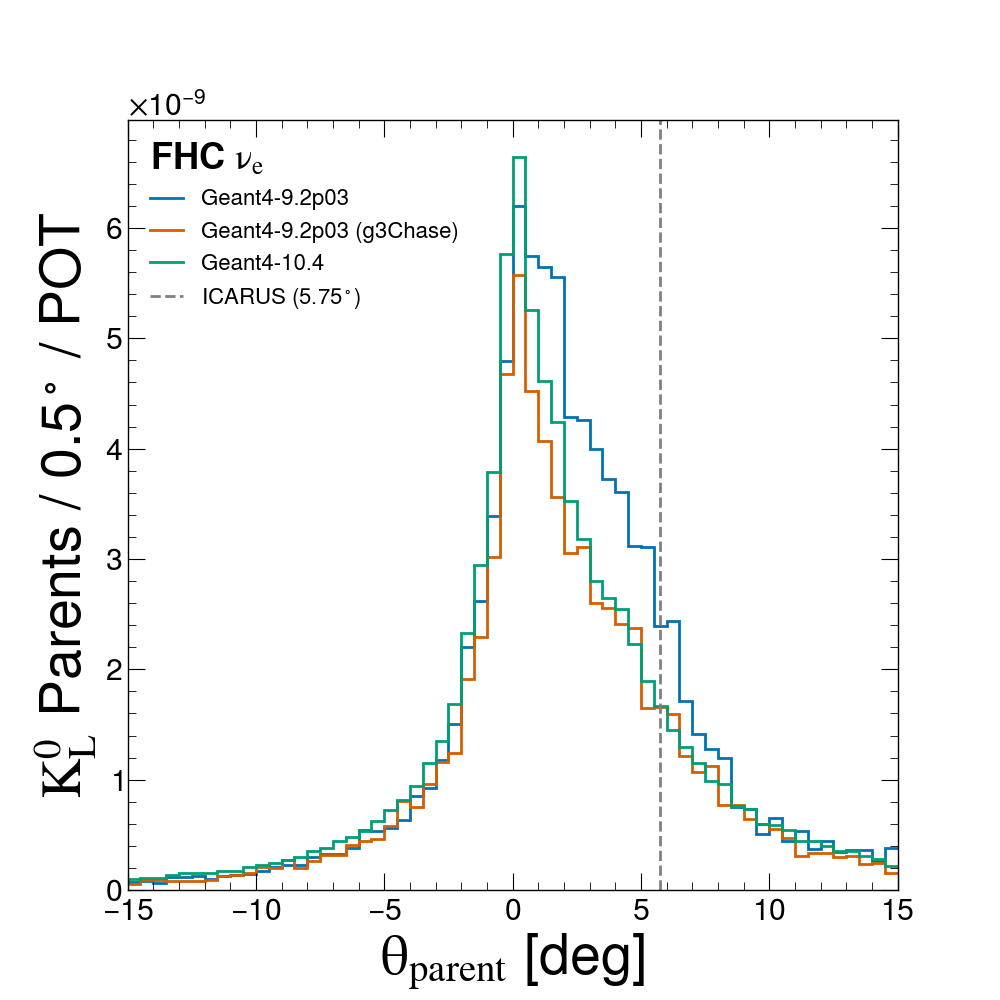}
        \caption{$K^0_L \to \nue$}
    \end{subfigure}
    \begin{subfigure}{0.23\textwidth}
        \includegraphics[width=\textwidth]{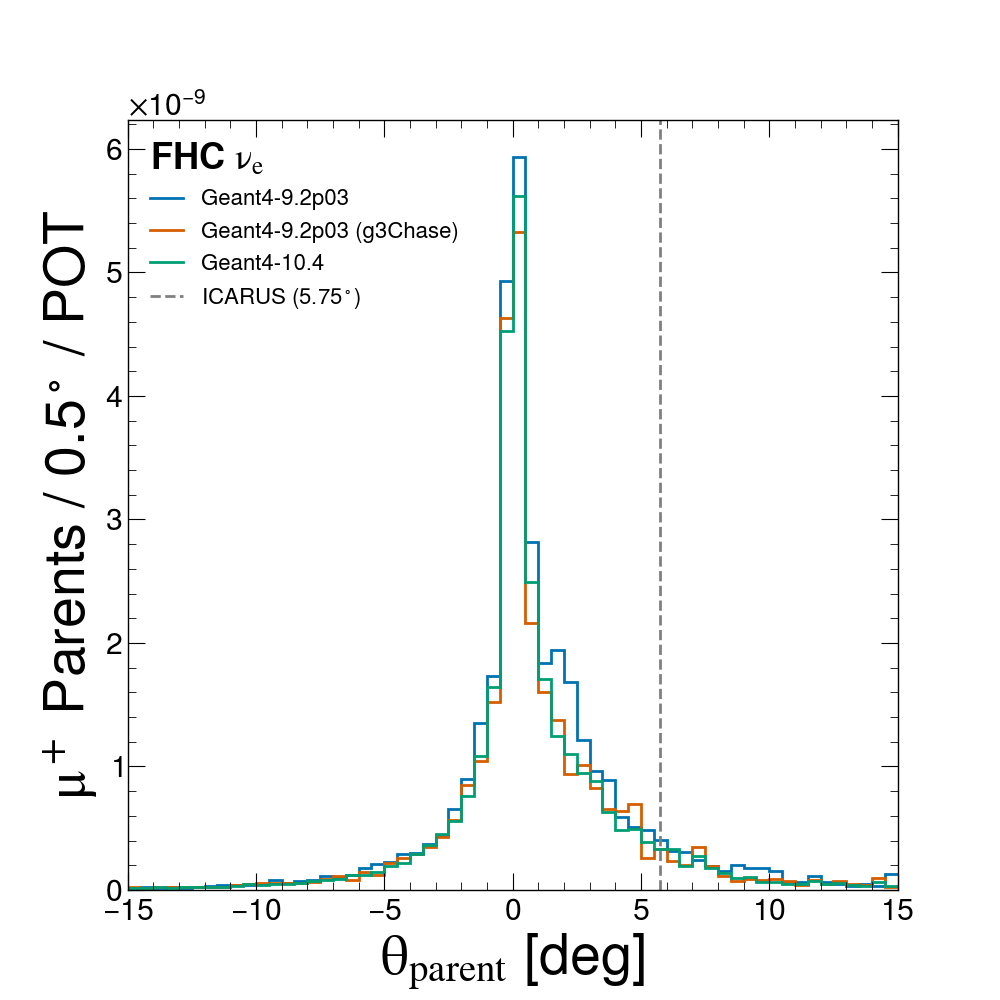}
        \caption{$\mu^+ \to \nue$}
    \end{subfigure}
    \begin{subfigure}{0.23\textwidth}
        \includegraphics[width=\textwidth]{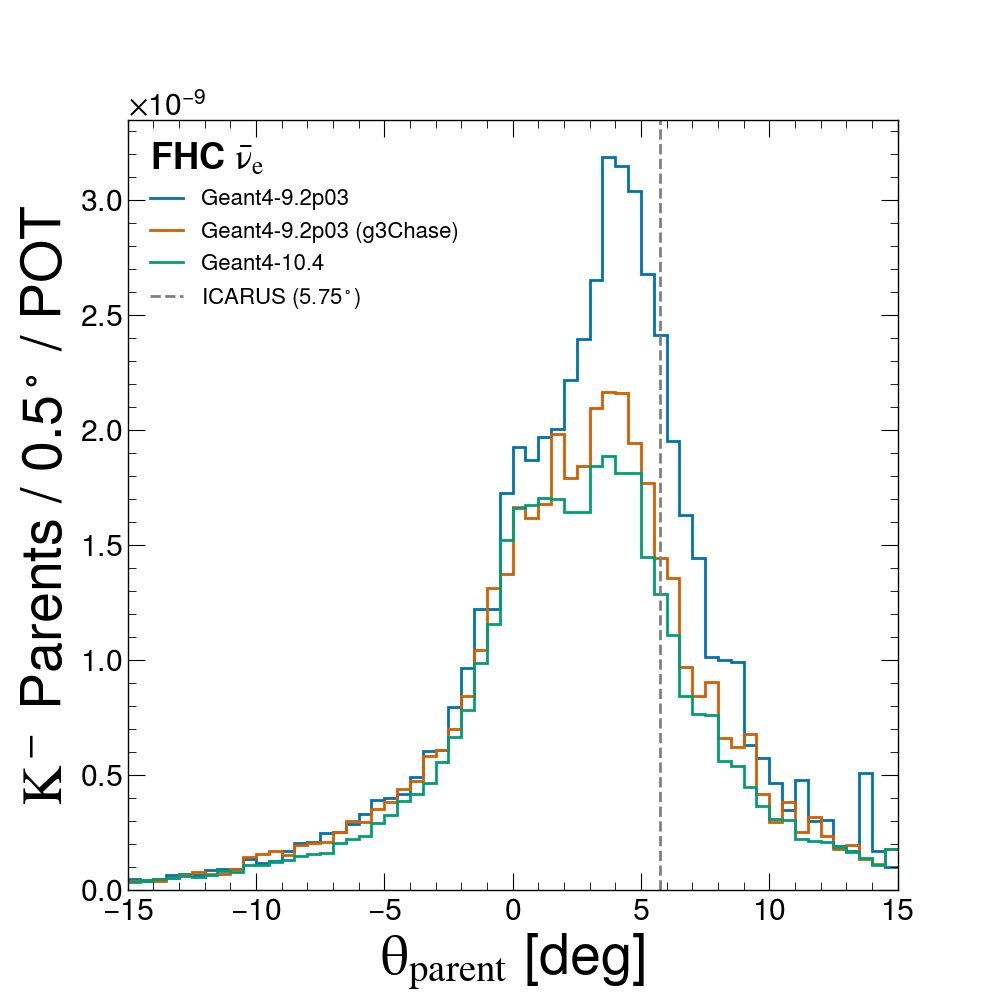}
        \caption{$K^- \to \nueb$}
    \end{subfigure}
    \begin{subfigure}{0.23\textwidth}
        \includegraphics[width=\textwidth]{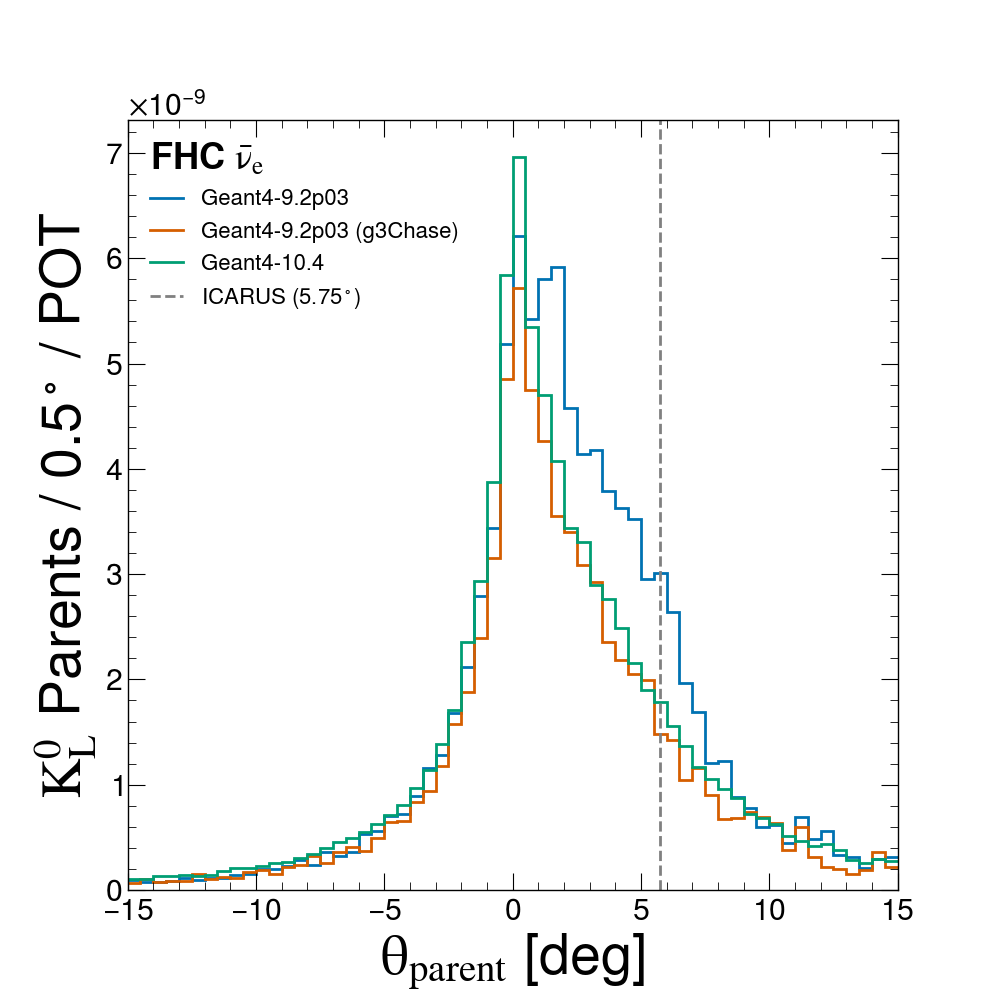}
        \caption{$K^0_L \to \nueb$}
    \end{subfigure}
    \begin{subfigure}{0.23\textwidth}
        \includegraphics[width=\textwidth]{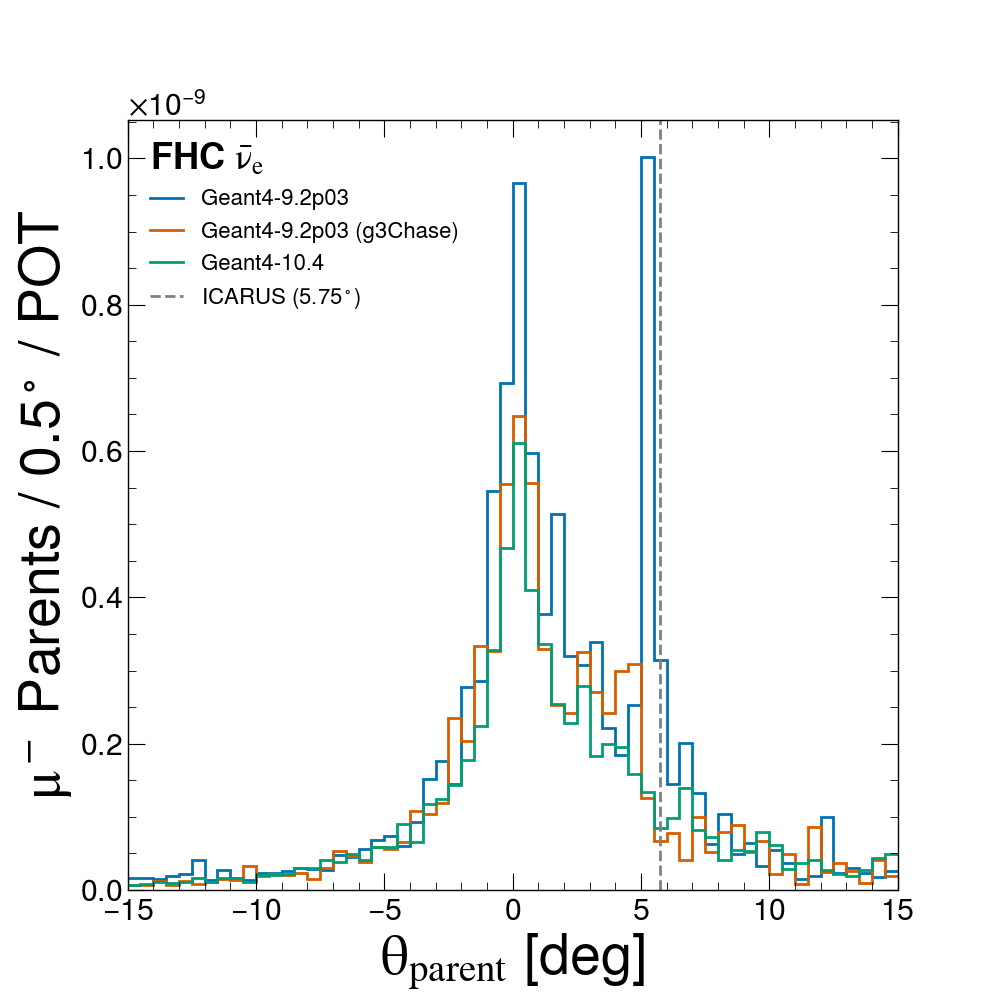}
        \caption{$\mu^- \to \nueb$}
    \end{subfigure}
    \caption[Parent Decay Angle (FHC)]{Parent decay angle distributions in the forward horn current mode.}
\end{figure}

\clearpage
\section{Reverse Horn Current}
\begin{figure}[!ht]
    \centering
    \begin{subfigure}{0.23\textwidth}
        \includegraphics[width=\textwidth]{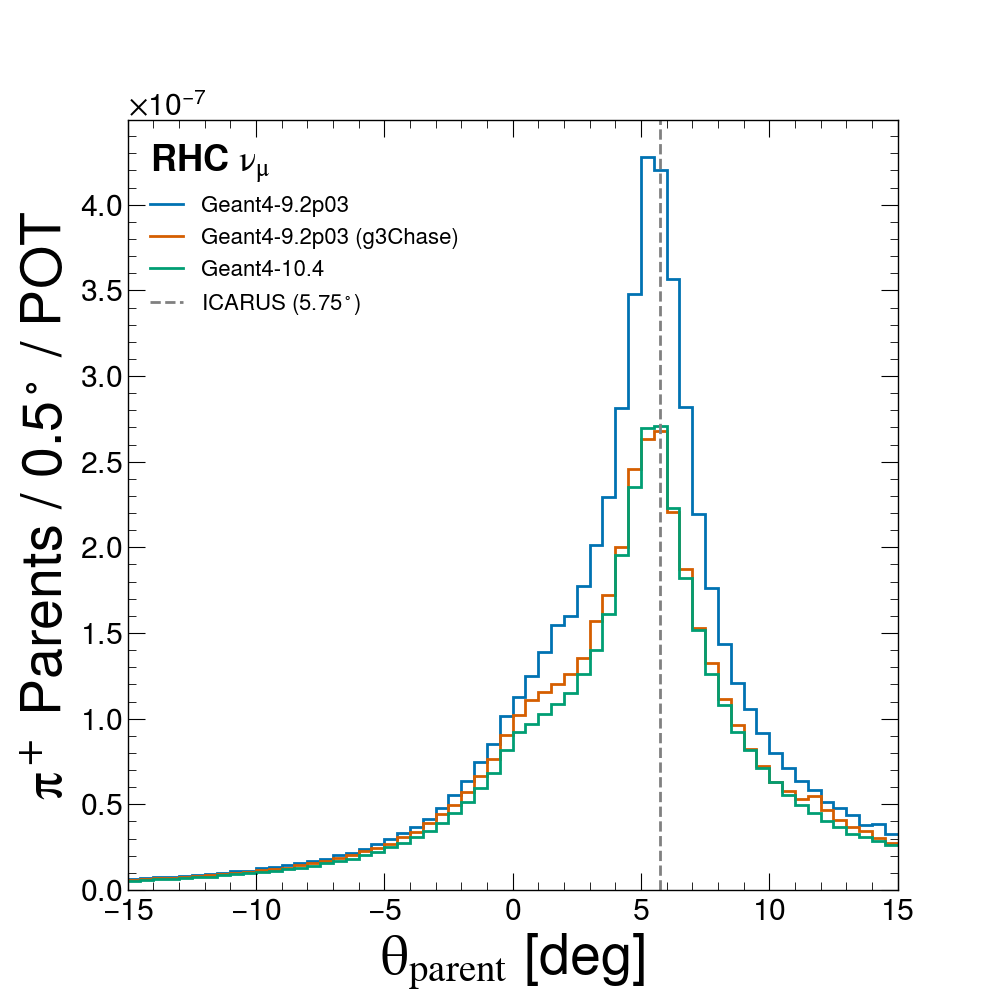}
        \caption{$\pi^+ \to \numu$}
    \end{subfigure}
    \begin{subfigure}{0.23\textwidth}
        \includegraphics[width=\textwidth]{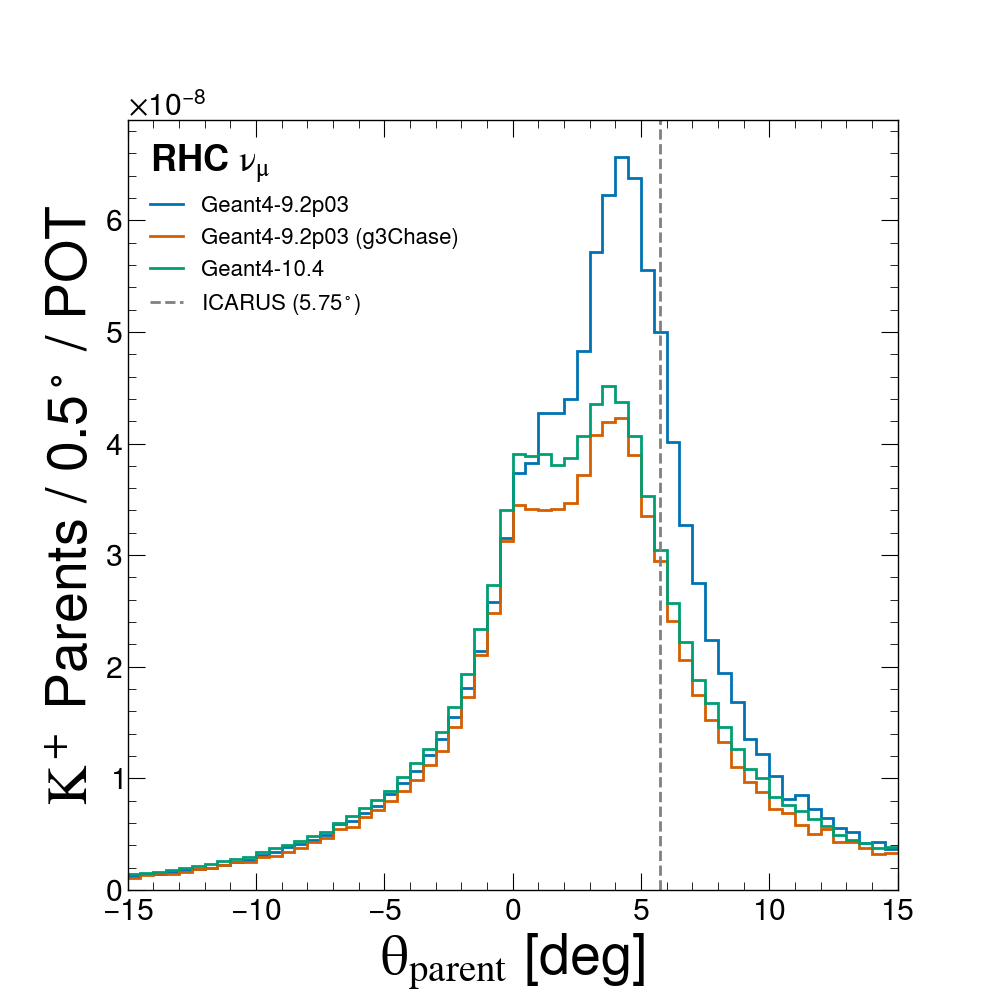}
        \caption{$K^+ \to \numu$}
    \end{subfigure}
    \begin{subfigure}{0.23\textwidth}
        \includegraphics[width=\textwidth]{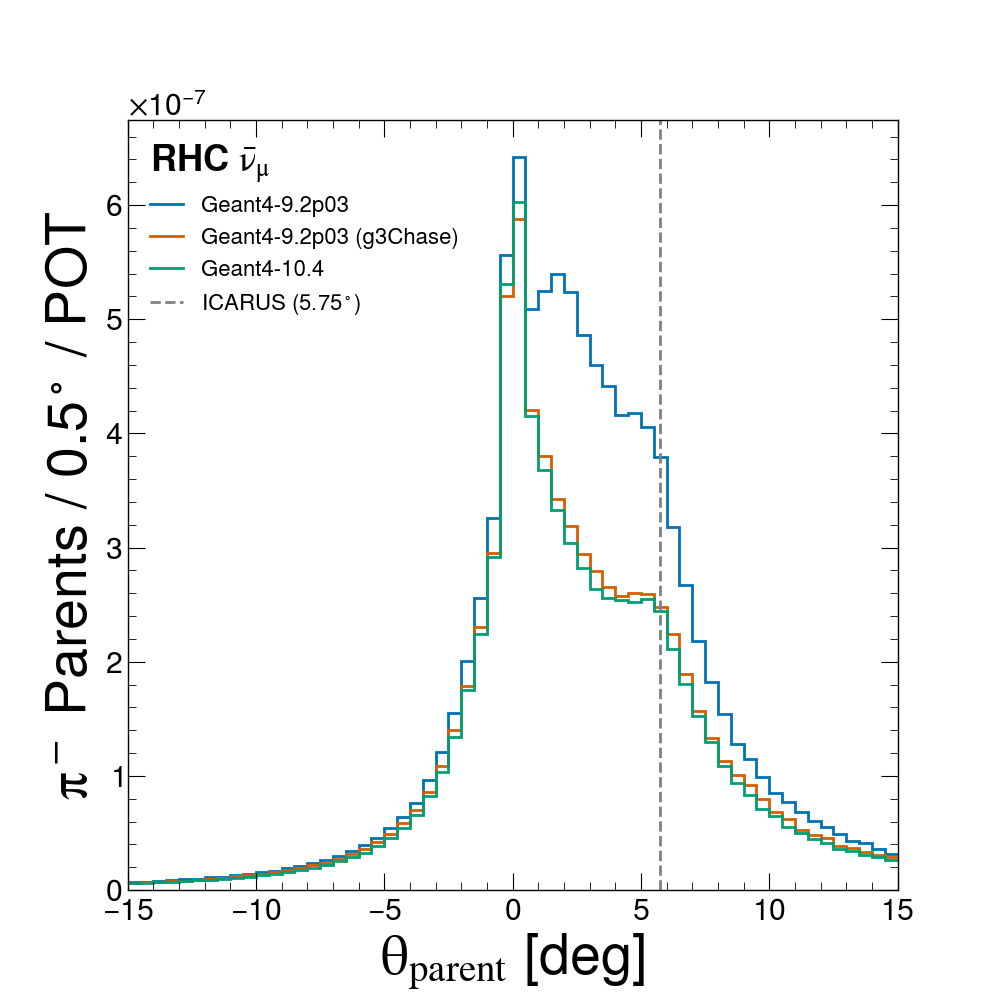}
        \caption{$\pi^- \to \numub$}
    \end{subfigure}
    \begin{subfigure}{0.23\textwidth}
        \includegraphics[width=\textwidth]{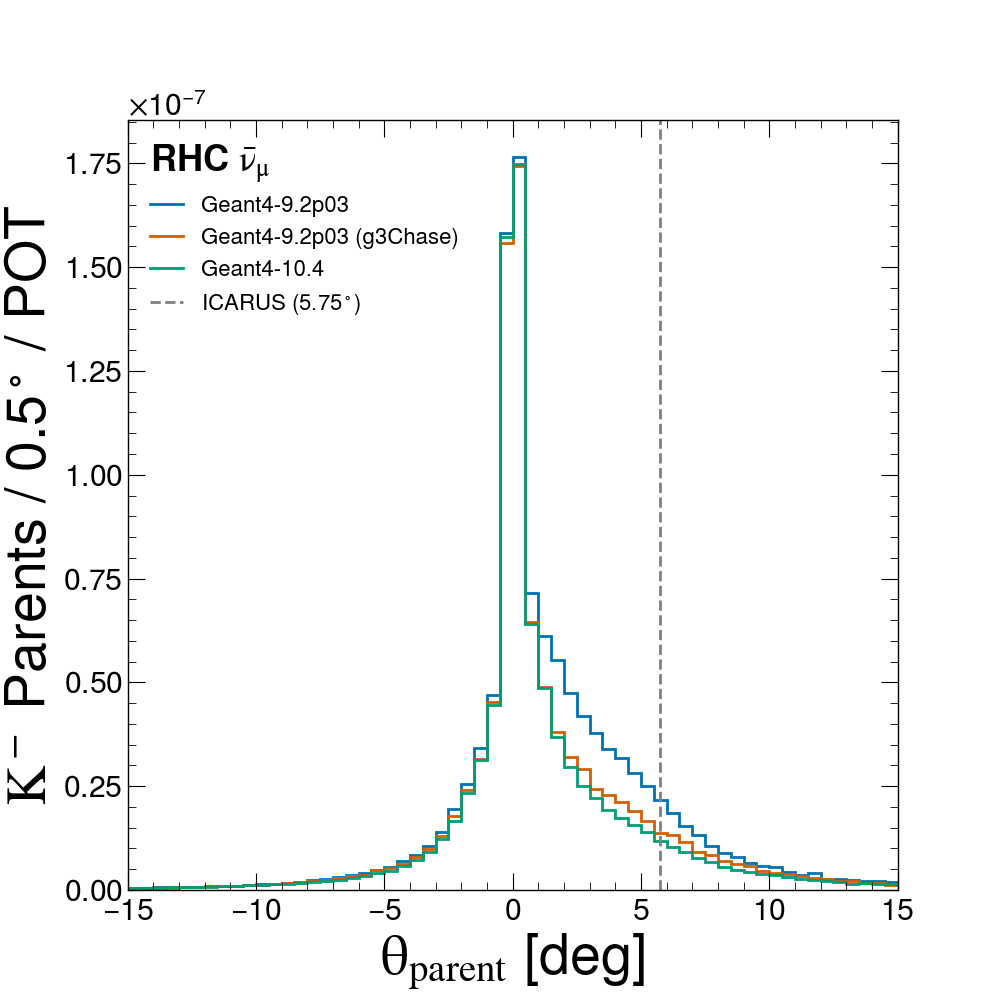}
        \caption{$K^- \to \numub$}
    \end{subfigure}
    \begin{subfigure}{0.23\textwidth}
        \includegraphics[width=\textwidth]{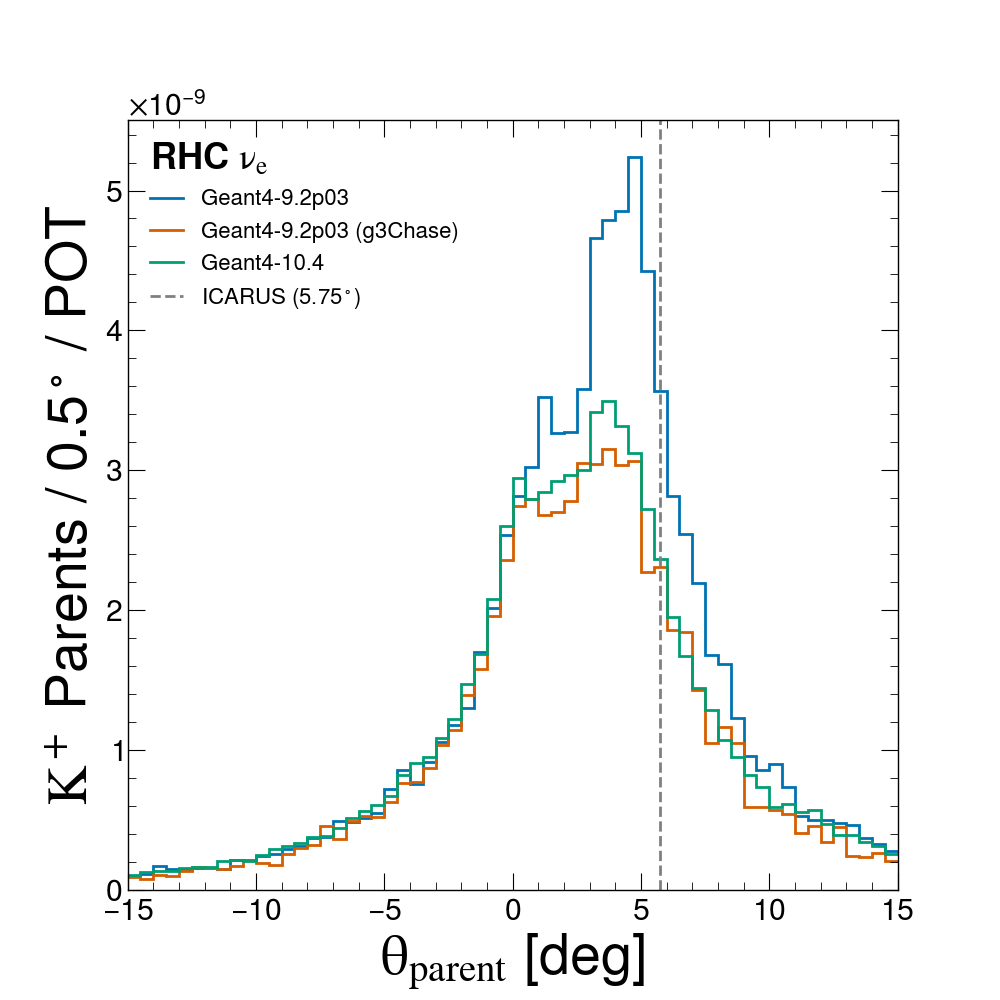}
        \caption{$K^+ \to \nue$}
    \end{subfigure}
    \begin{subfigure}{0.23\textwidth}
        \includegraphics[width=\textwidth]{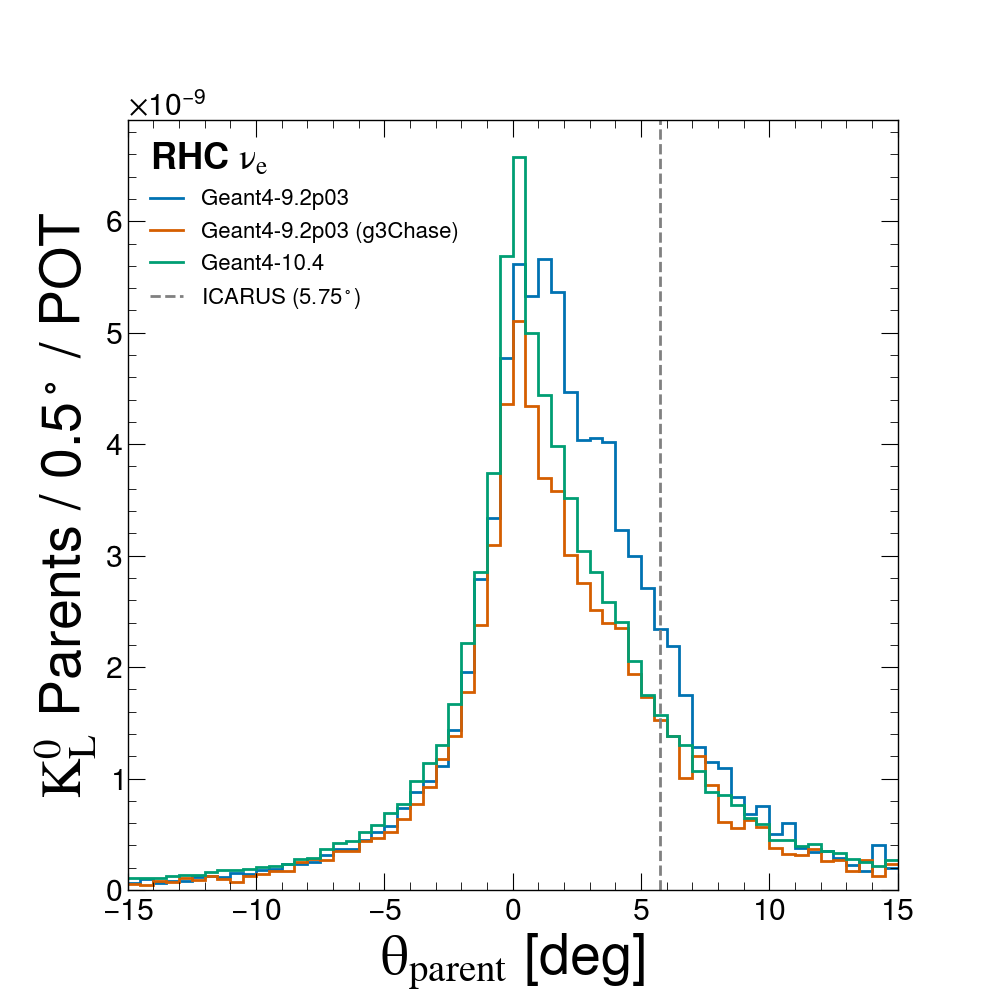}
        \caption{$K^0_L \to \nue$}
    \end{subfigure}
    \begin{subfigure}{0.23\textwidth}
        \includegraphics[width=\textwidth]{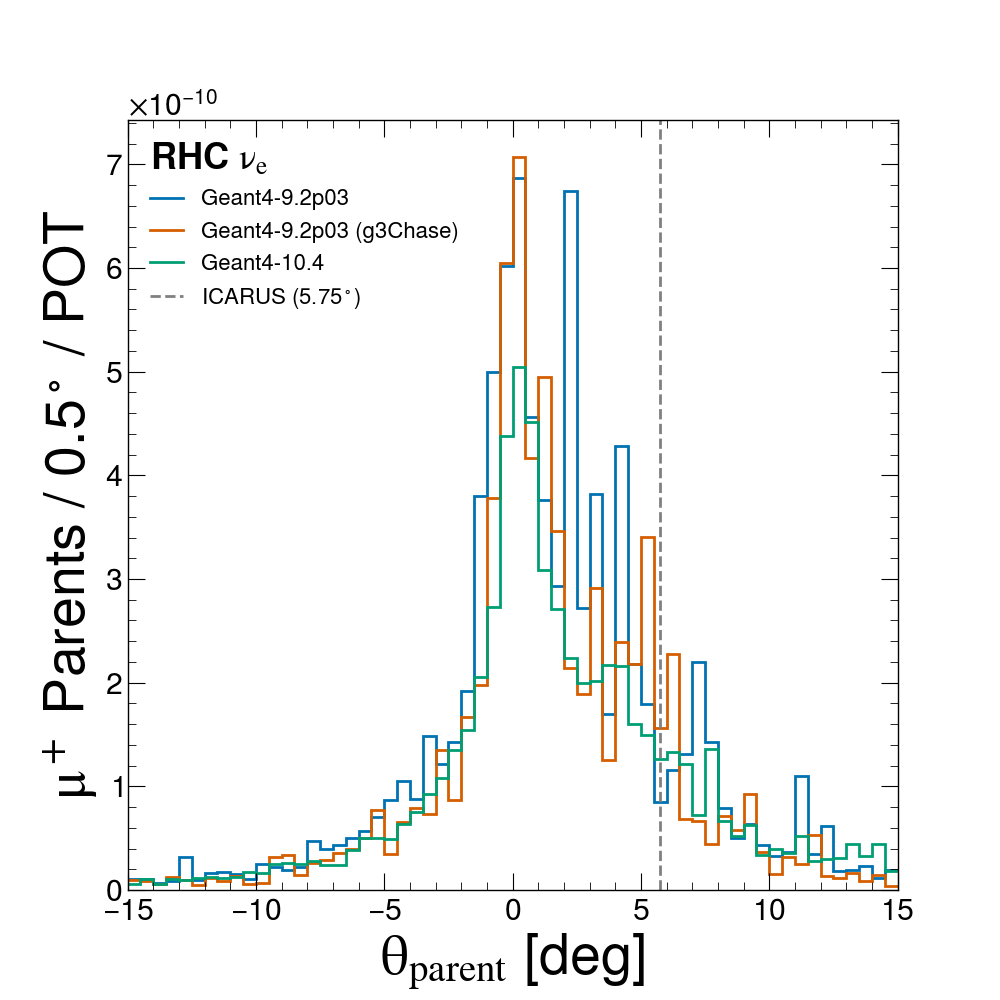}
        \caption{$\mu^+ \to \nue$}
    \end{subfigure}
    \begin{subfigure}{0.23\textwidth}
        \includegraphics[width=\textwidth]{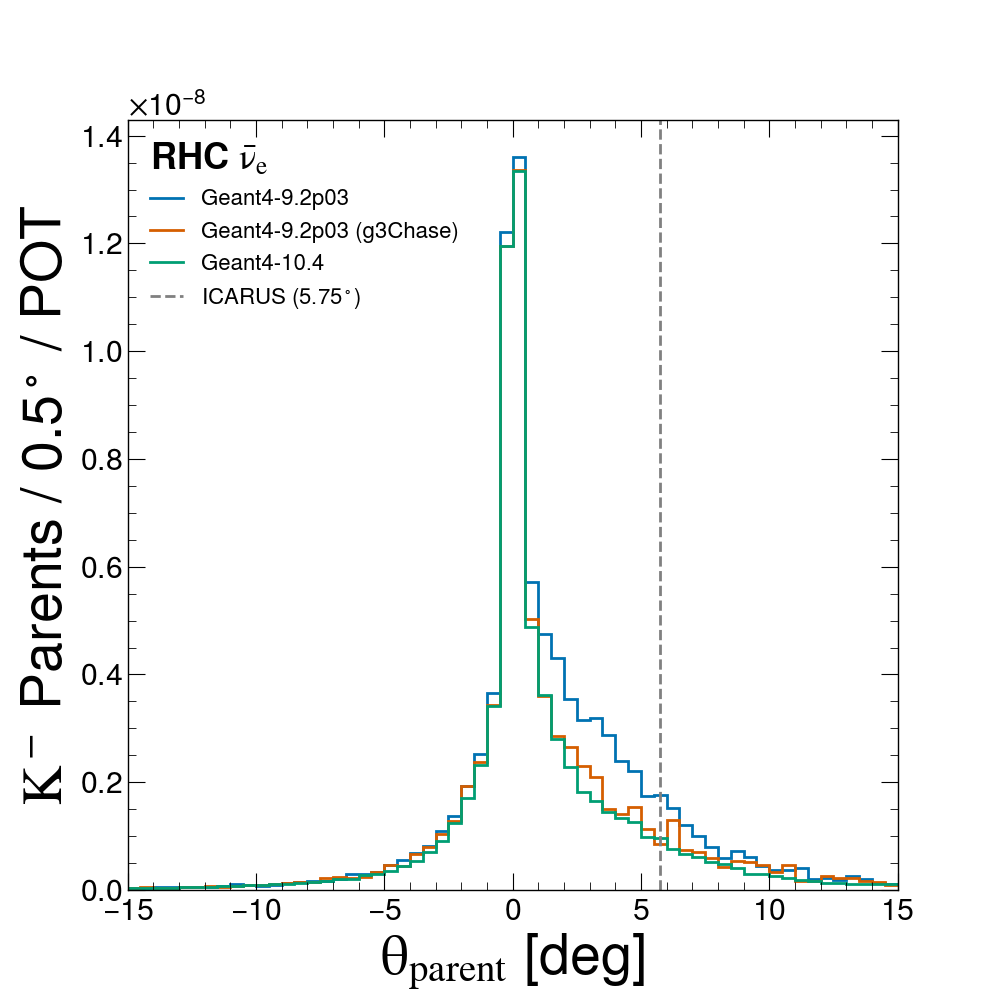}
        \caption{$K^- \to \nueb$}
    \end{subfigure}
    \begin{subfigure}{0.23\textwidth}
        \includegraphics[width=\textwidth]{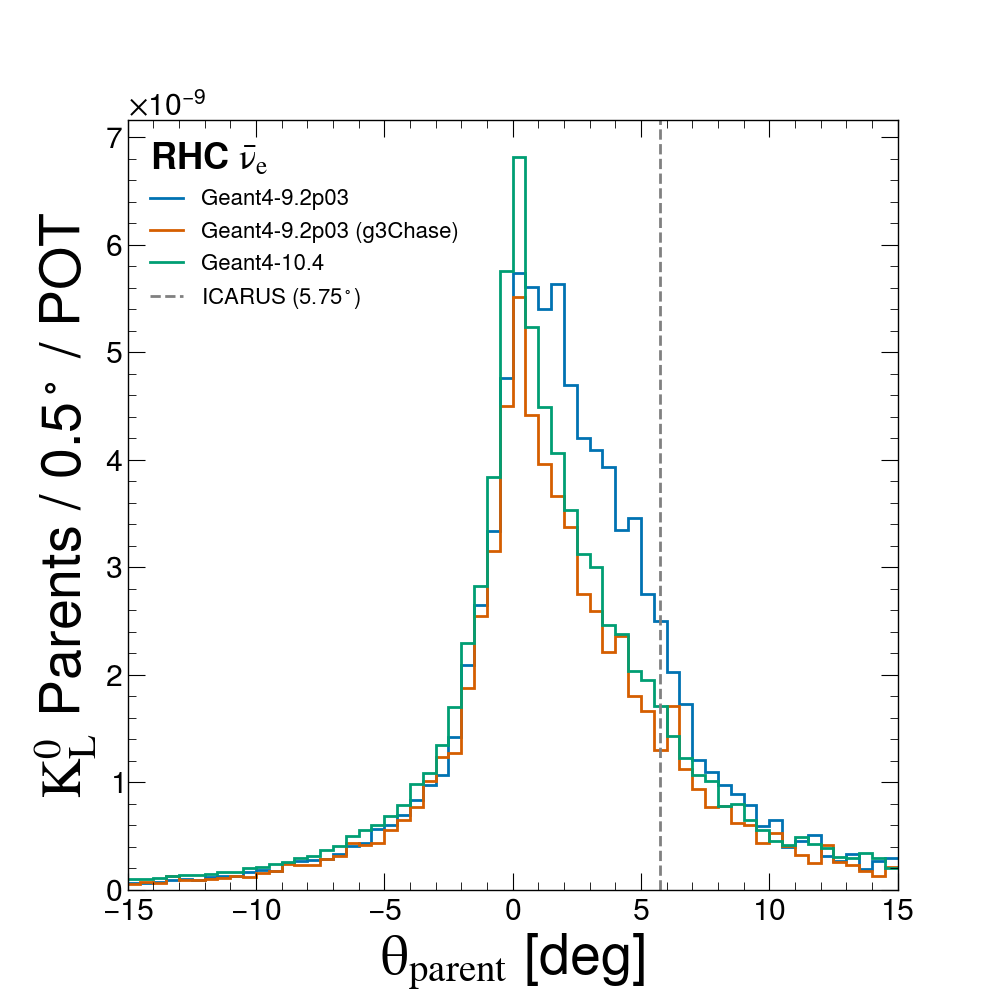}
        \caption{$K^0_L \to \nueb$}
    \end{subfigure}
    \begin{subfigure}{0.23\textwidth}
        \includegraphics[width=\textwidth]{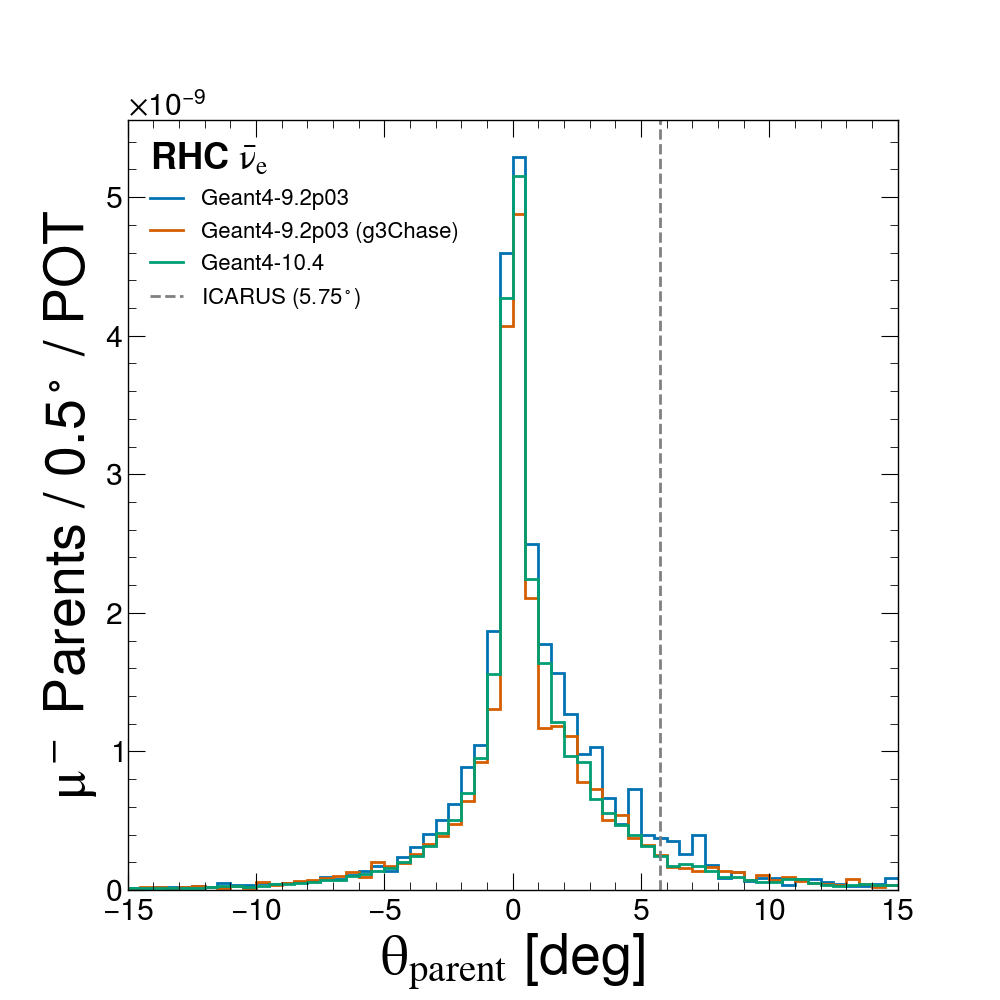}
        \caption{$\mu^- \to \nueb$}
    \end{subfigure}
    \caption[Parent Decay Angle (RHC)]{Parent decay angle distributions in the reverse horn current mode.}
\end{figure}

%% file: energy_vs_angles.tex
\clearpage
\section{Forward Horn Current}
\begin{figure}[!ht]
\centering
\begin{subfigure}{0.23\textwidth}
    \includegraphics[width=\textwidth]{g4Update_numu_piplus_angle.png}
    \caption{$\pi^+ \to \numu$}
\end{subfigure}
\begin{subfigure}{0.23\textwidth}
    \includegraphics[width=\textwidth]{g4Update_numu_kplus_angle.png}
    \caption{$K^+ \to \numu$}
\end{subfigure}

\begin{subfigure}{0.23\textwidth}
    \includegraphics[width=\textwidth]{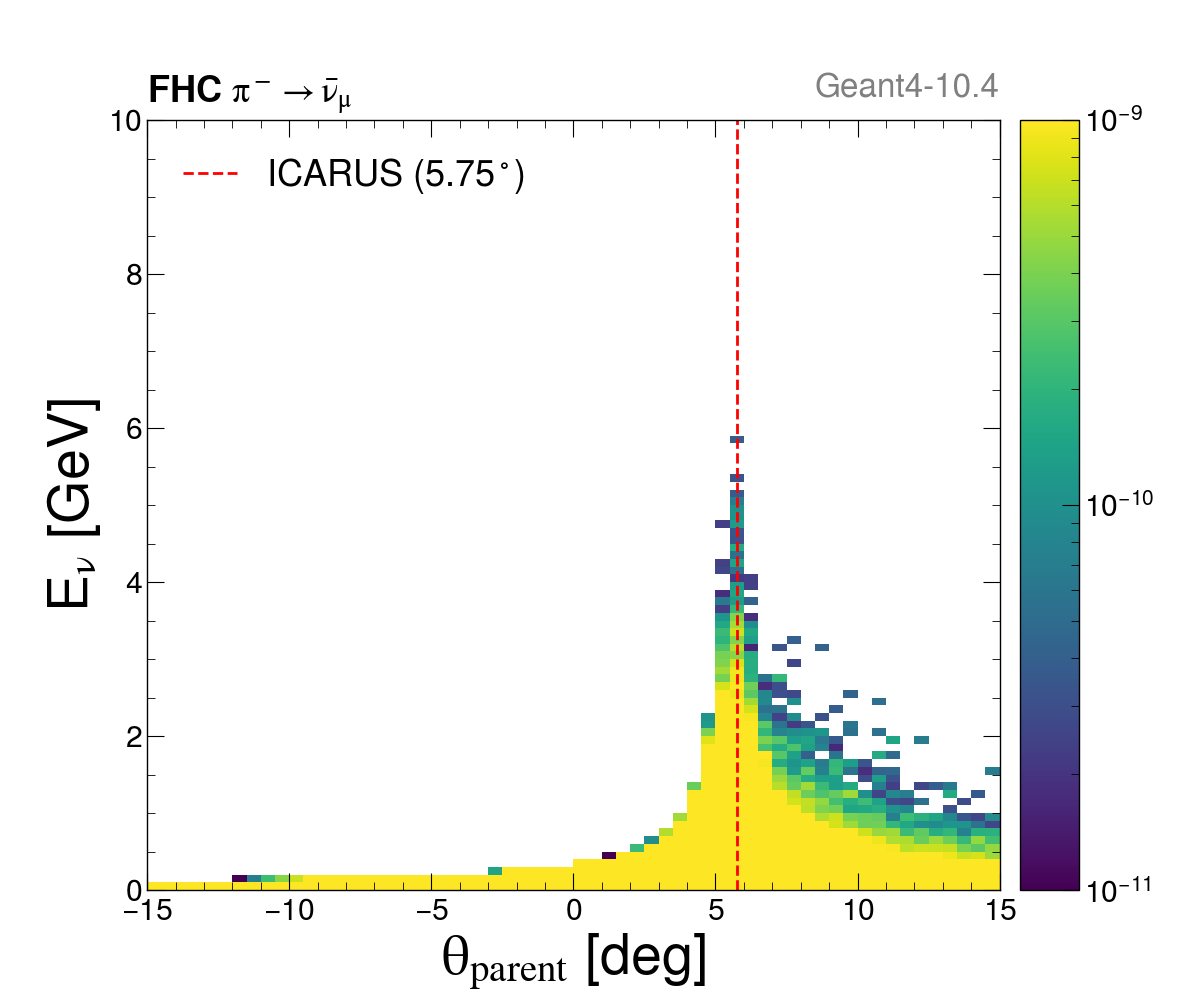}
    \caption{$\pi^- \to \numub$}
\end{subfigure}
\begin{subfigure}{0.23\textwidth}
    \includegraphics[width=\textwidth]{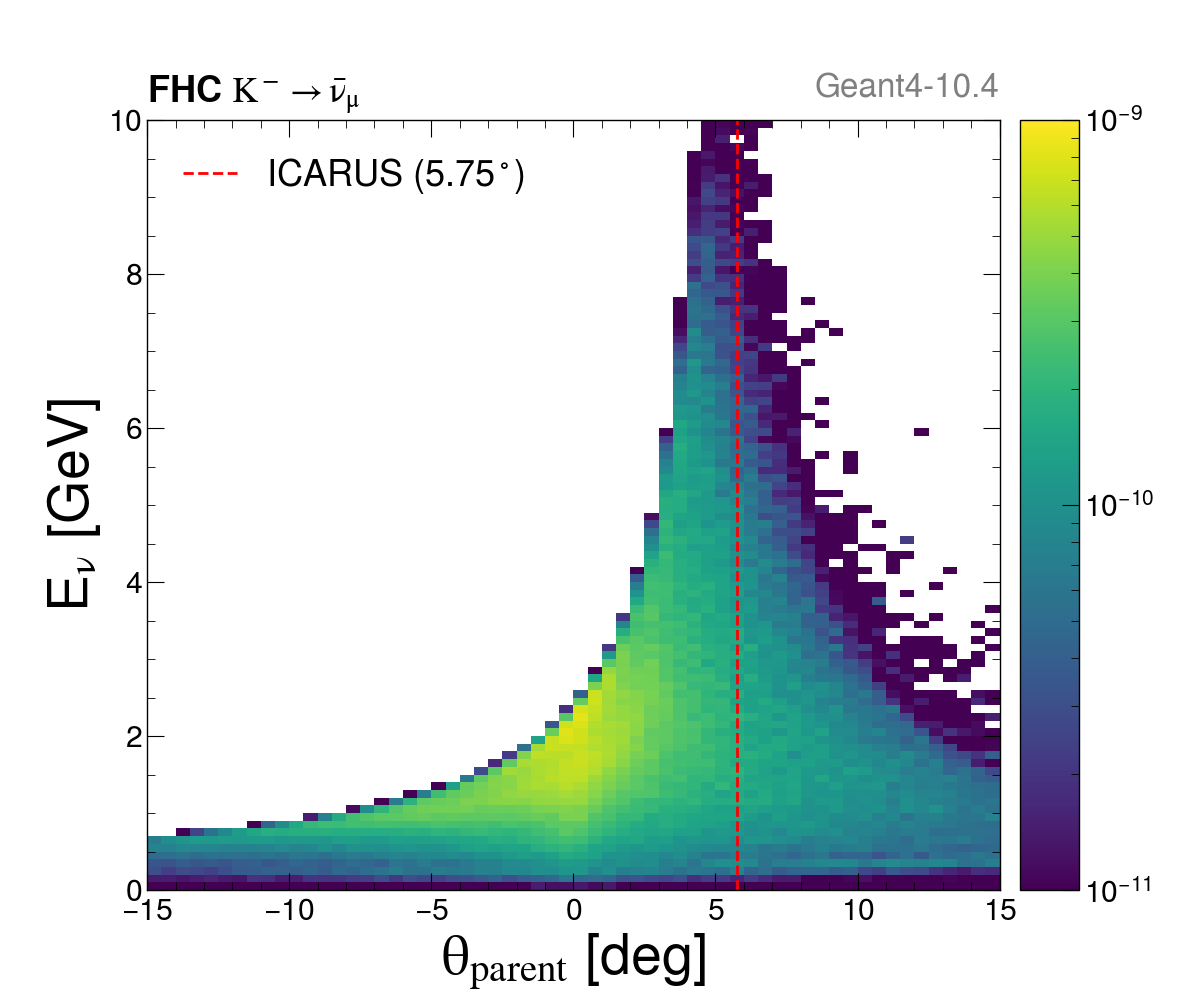}
    \caption{$K^- \to \numub$}
\end{subfigure}

\begin{subfigure}{0.23\textwidth}
    \includegraphics[width=\textwidth]{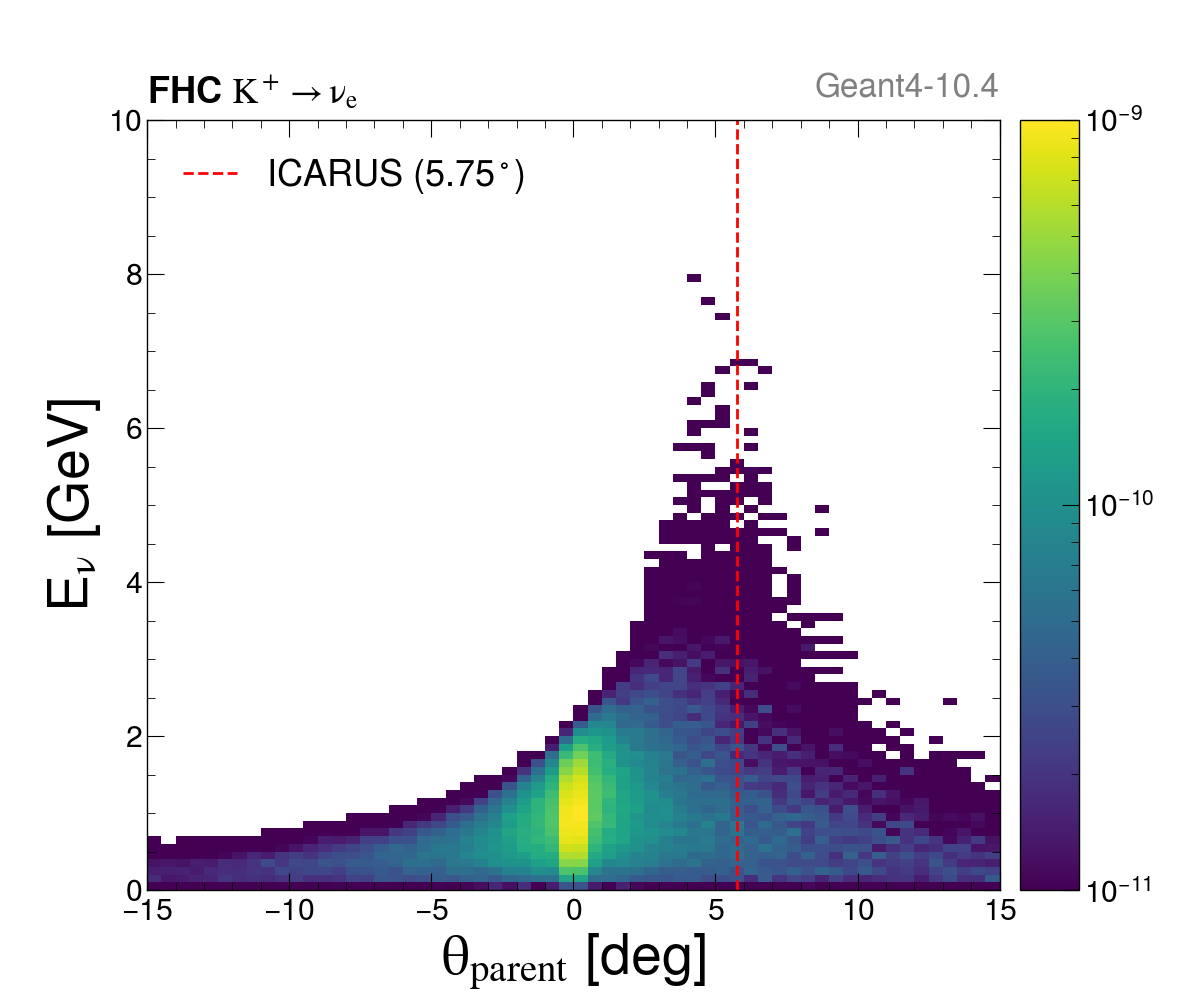}
    \caption{$K^+ \to \nue$}
\end{subfigure}
\begin{subfigure}{0.23\textwidth}
    \includegraphics[width=\textwidth]{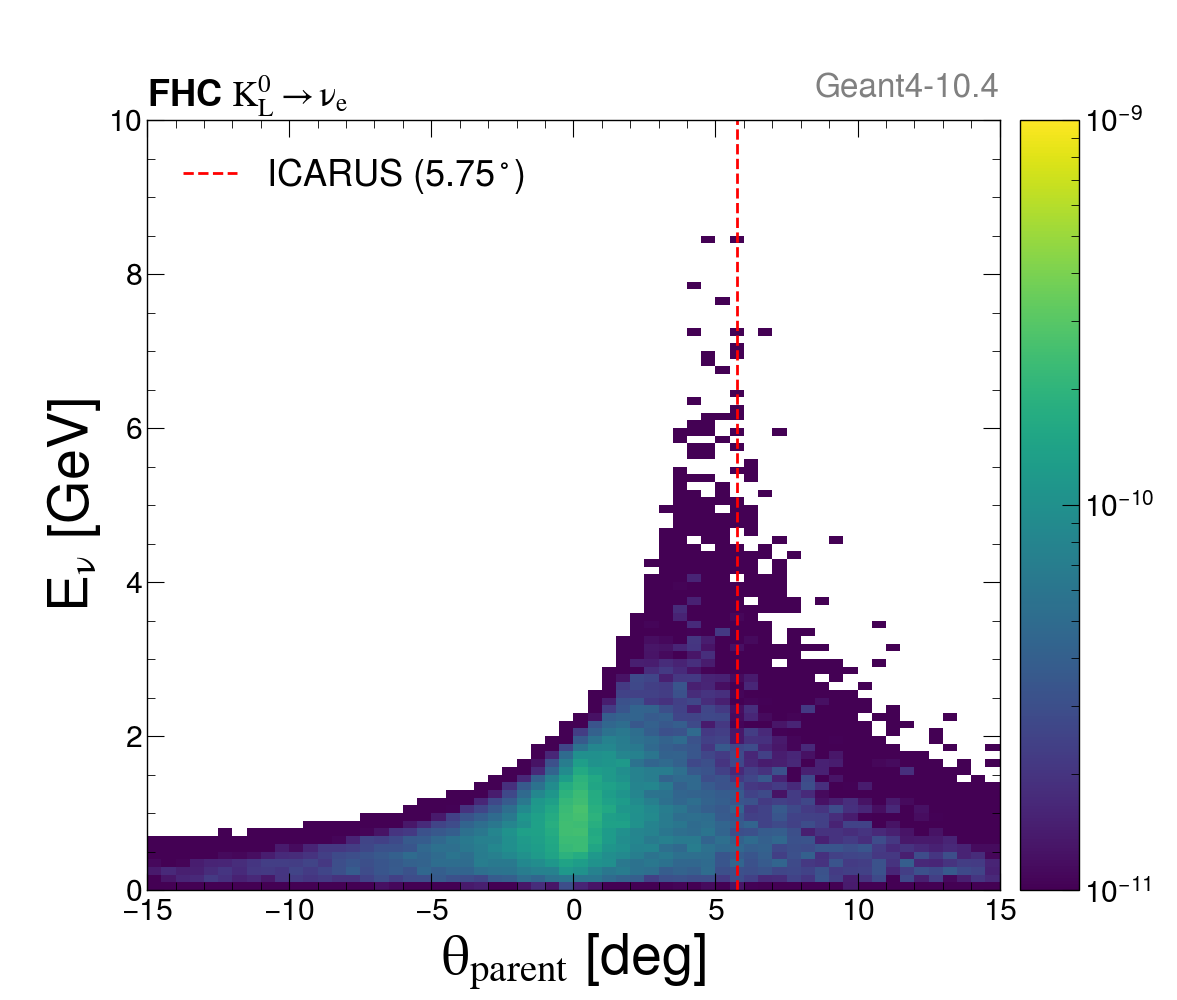}
    \caption{$K^0_L \to \nue$}
\end{subfigure}
\begin{subfigure}{0.23\textwidth}
    \includegraphics[width=\textwidth]{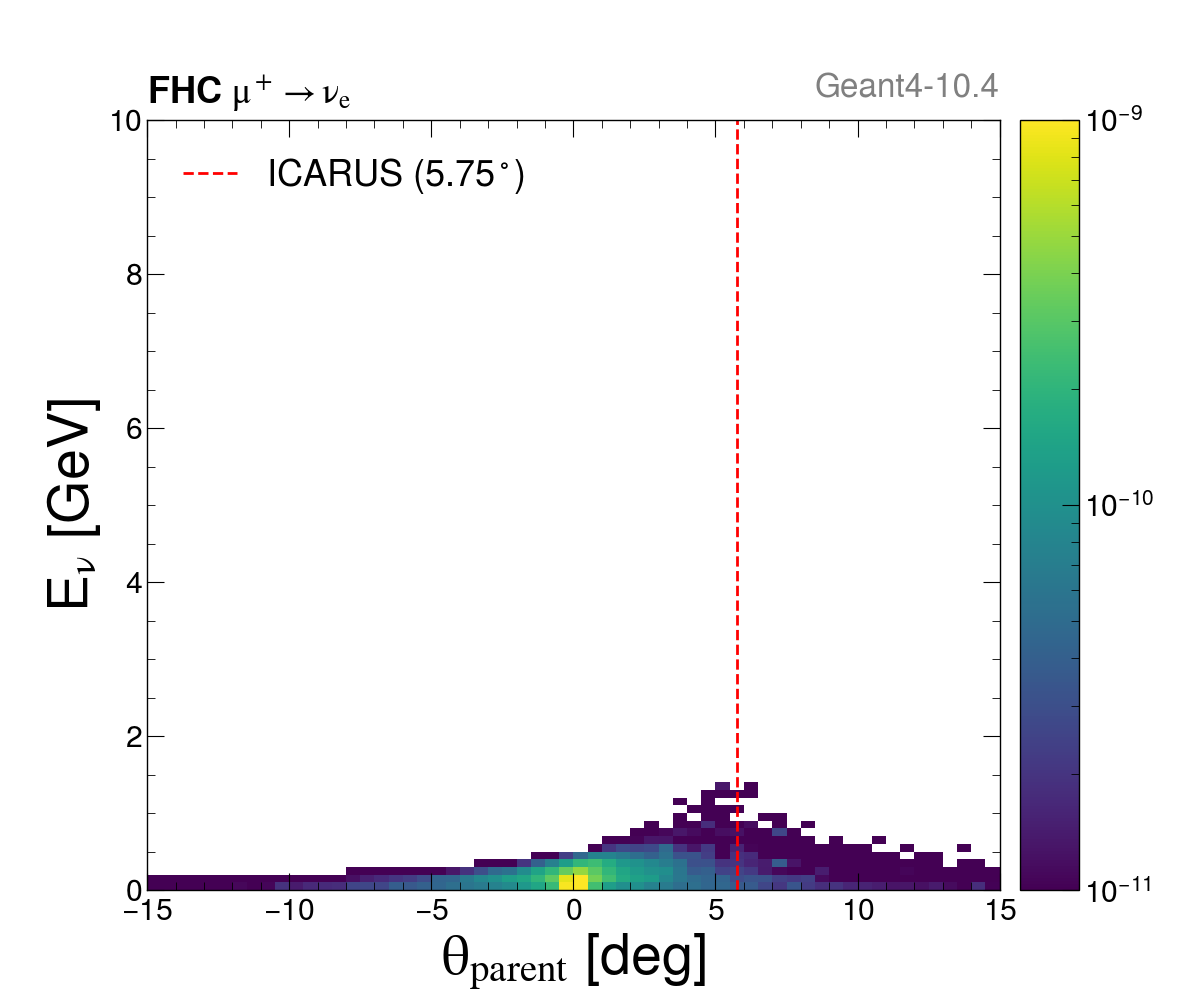}
    \caption{$\mu^+ \to \nue$}
\end{subfigure}

\begin{subfigure}{0.23\textwidth}
        \includegraphics[width=\textwidth]{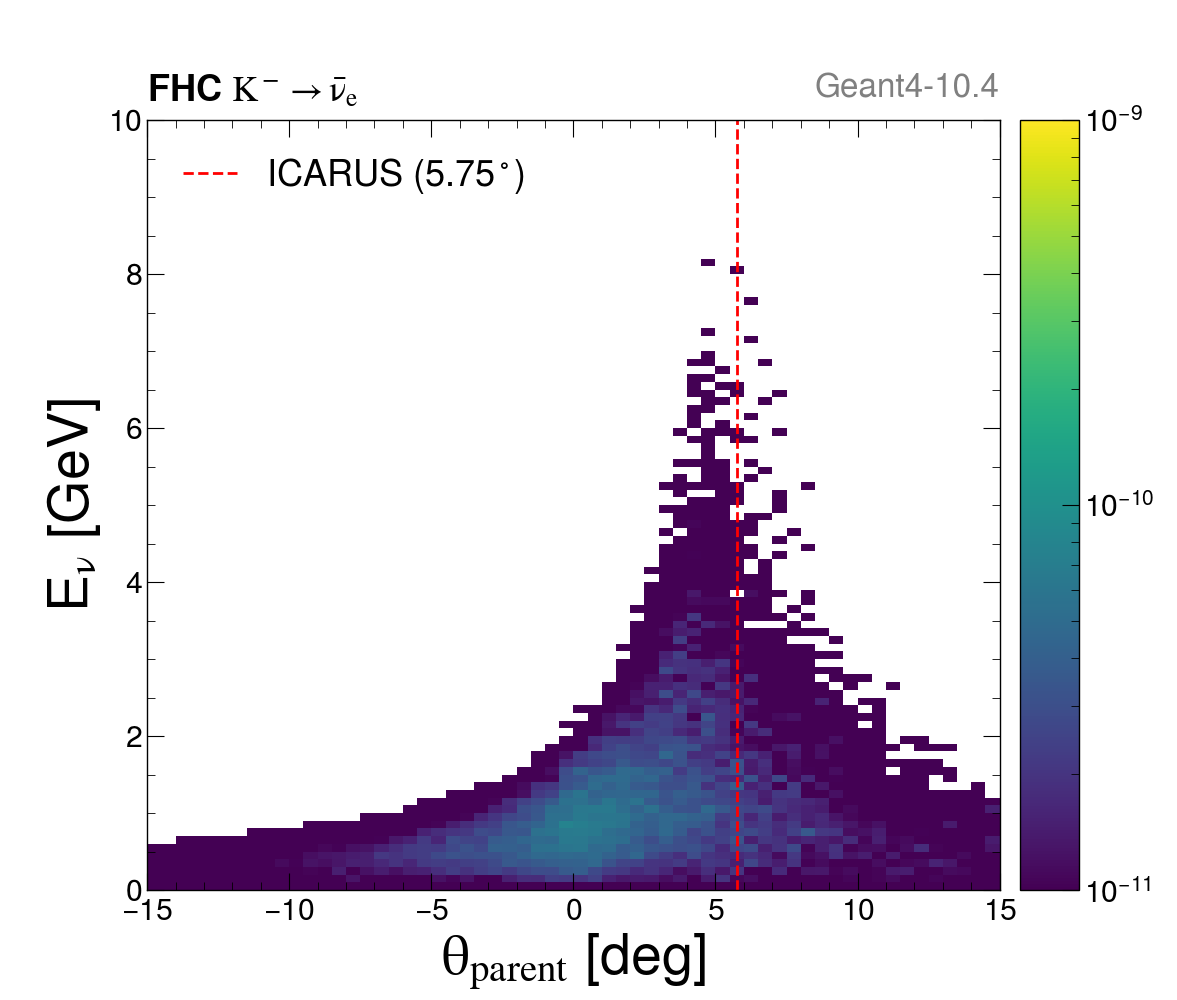}
            \caption{$K^- \to \nueb$}
\end{subfigure}
\begin{subfigure}{0.23\textwidth}
        \includegraphics[width=\textwidth]{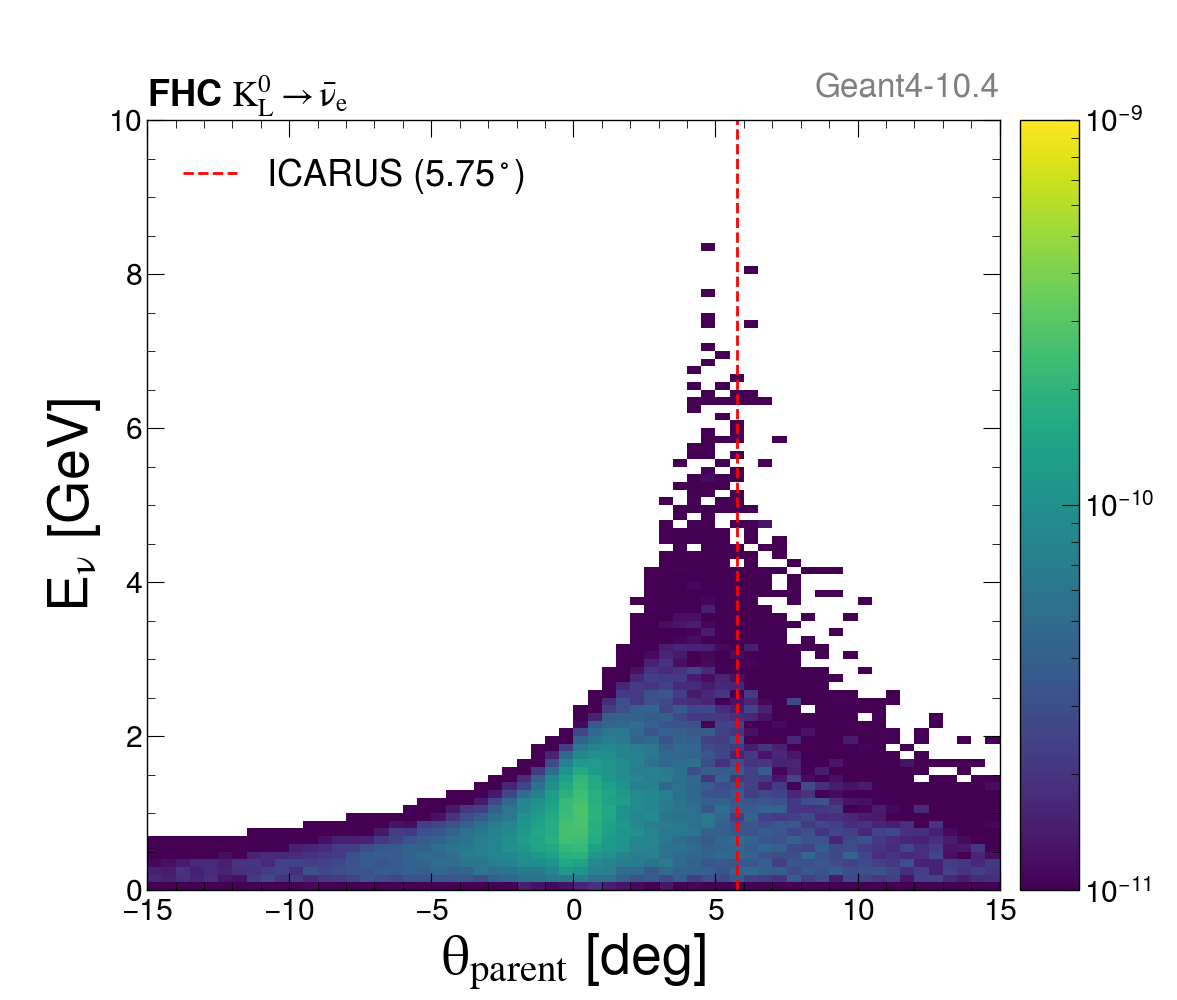}
            \caption{$K^0_L \to \nueb$}
\end{subfigure}
\begin{subfigure}{0.23\textwidth}
        \includegraphics[width=\textwidth]{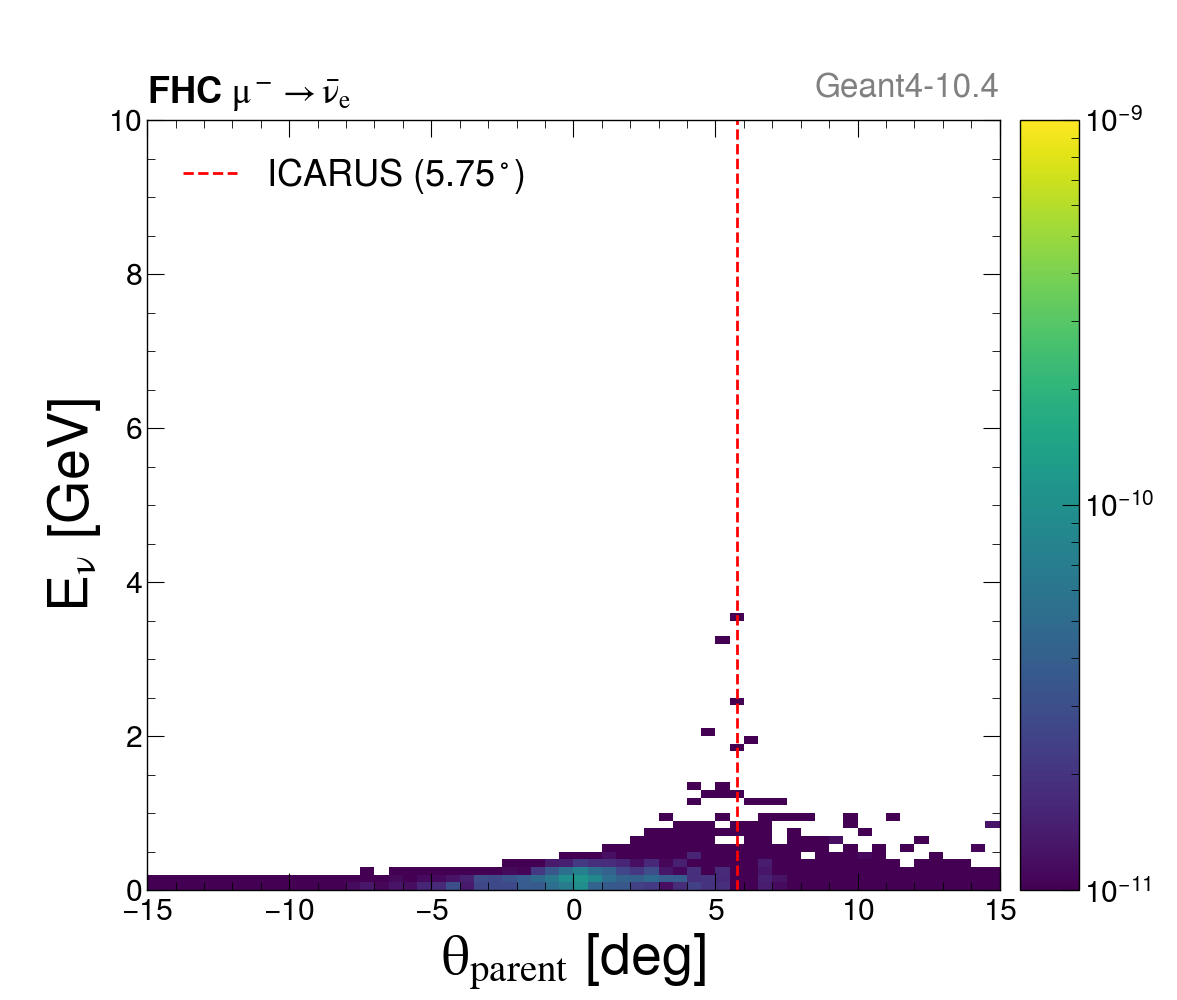}
            \caption{$\mu^- \to \nueb$}
\end{subfigure}
\caption[Parent Decay Angle vs. Neutrino Energy (FHC)]{Parent decay angle vs. neutrino energy distributions in the forward horn current mode.}
\end{figure}
\clearpage

\section{Reverse Horn Current}
\begin{figure}[!ht]
\centering
\begin{subfigure}{0.23\textwidth}
    \includegraphics[width=\textwidth]{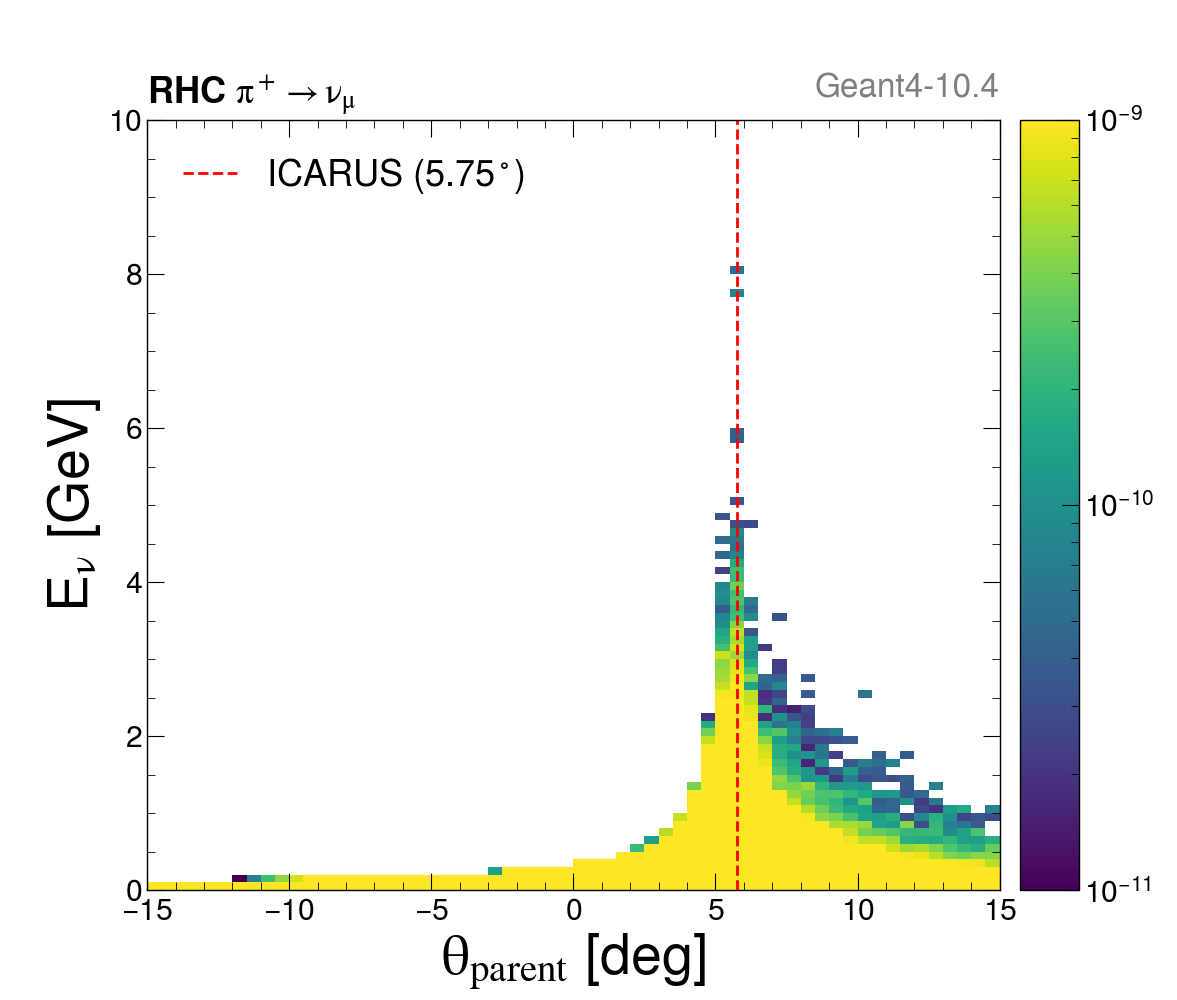}
    \caption{$\pi^+ \to \numu$}
\end{subfigure}
\begin{subfigure}{0.23\textwidth}
    \includegraphics[width=\textwidth]{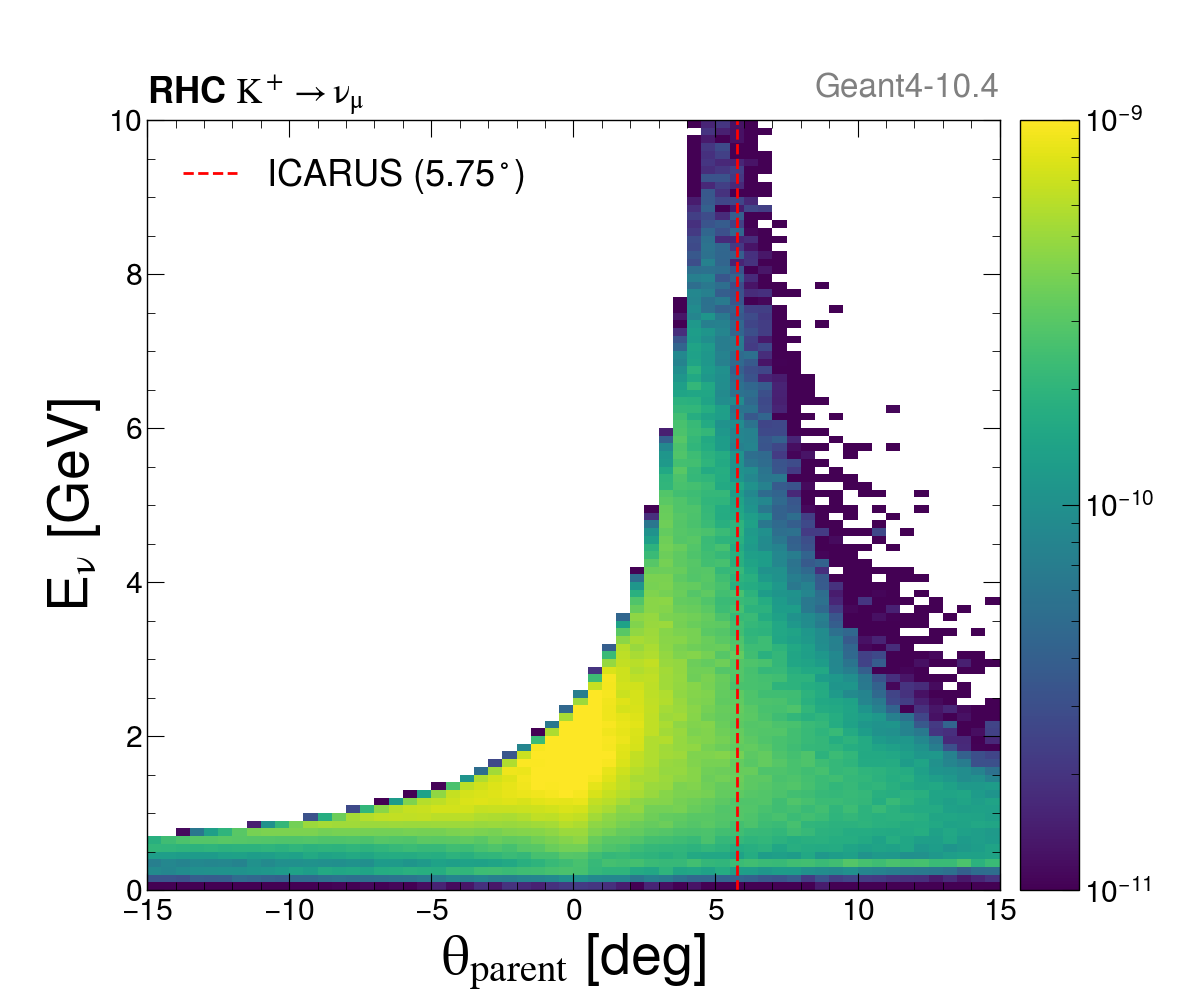}
    \caption{$K^+ \to \numu$}
\end{subfigure}

\begin{subfigure}{0.23\textwidth}
    \includegraphics[width=\textwidth]{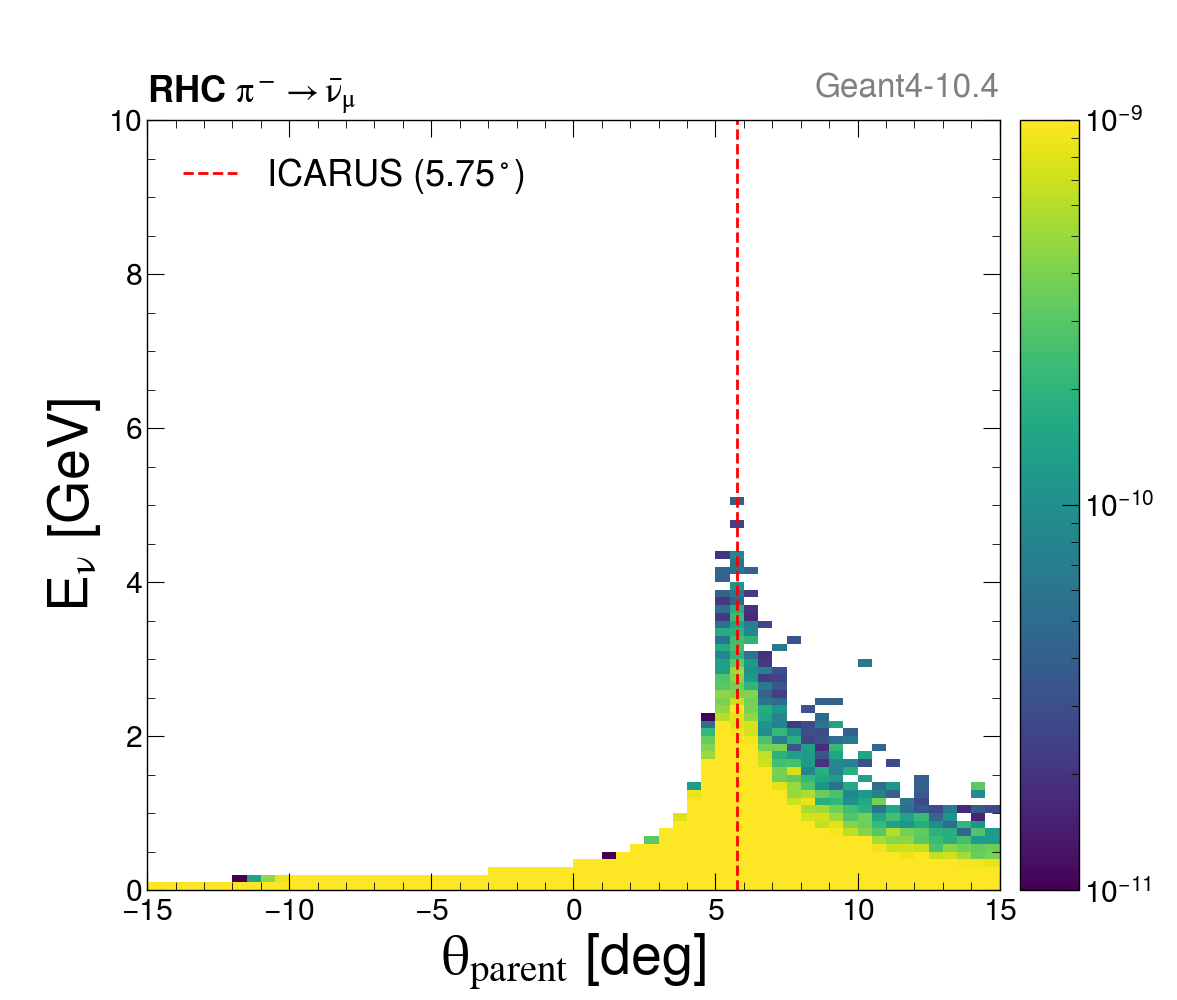}
    \caption{$\pi^- \to \numub$}
\end{subfigure}
\begin{subfigure}{0.23\textwidth}
    \includegraphics[width=\textwidth]{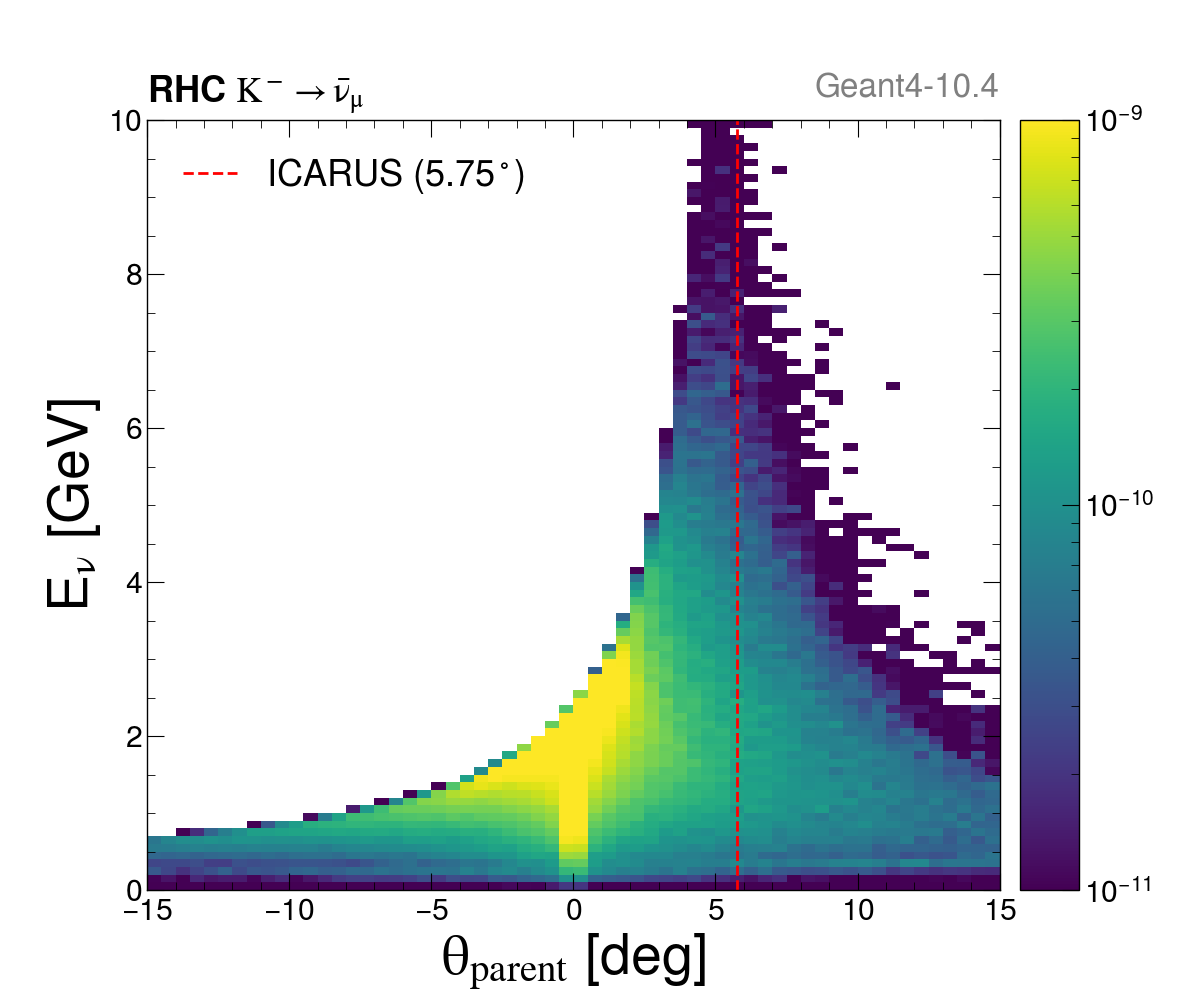}
    \caption{$K^- \to \numub$}
\end{subfigure}

\begin{subfigure}{0.23\textwidth}
    \includegraphics[width=\textwidth]{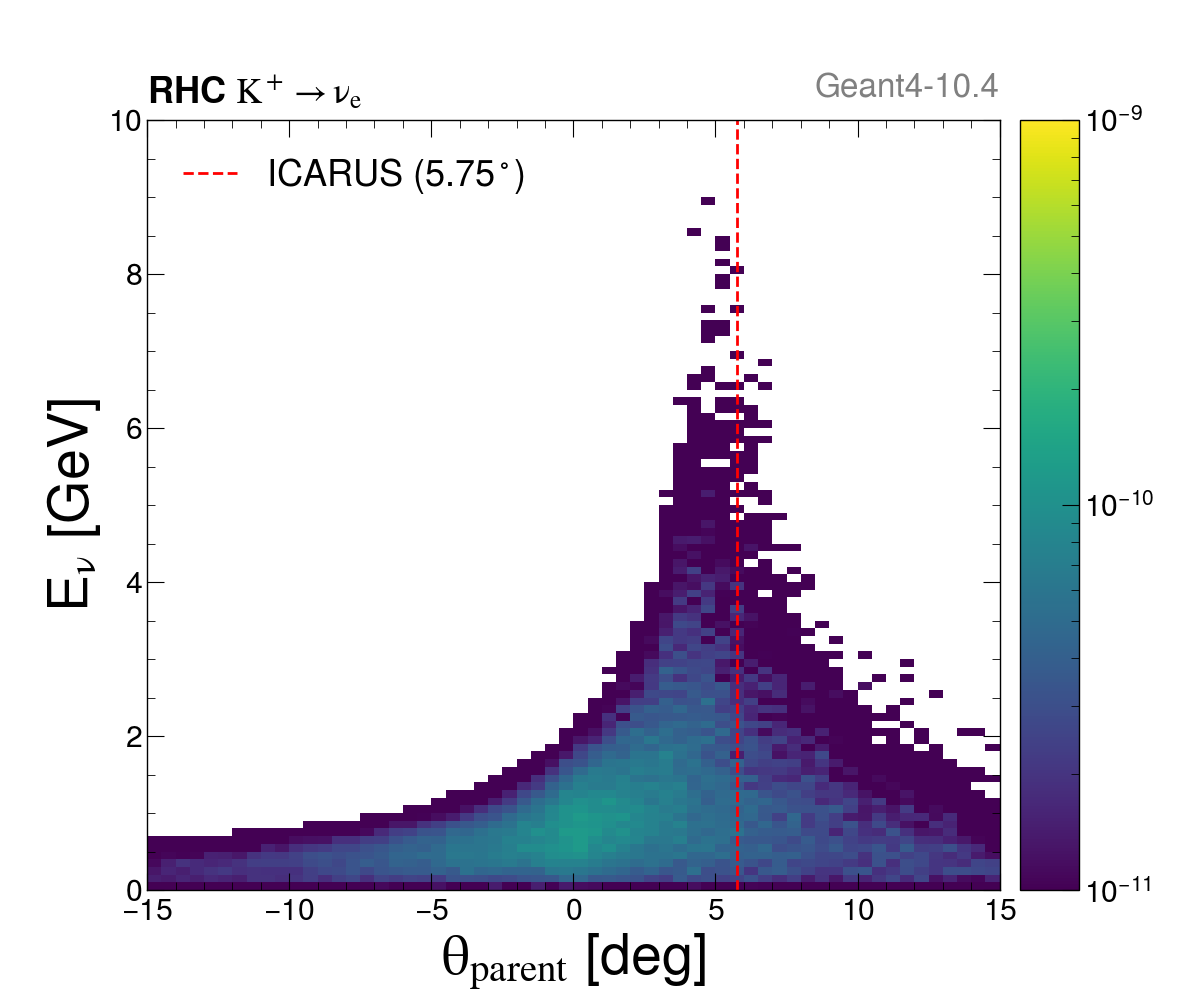}
    \caption{$K^+ \to \nue$}
\end{subfigure}
\begin{subfigure}{0.23\textwidth}
    \includegraphics[width=\textwidth]{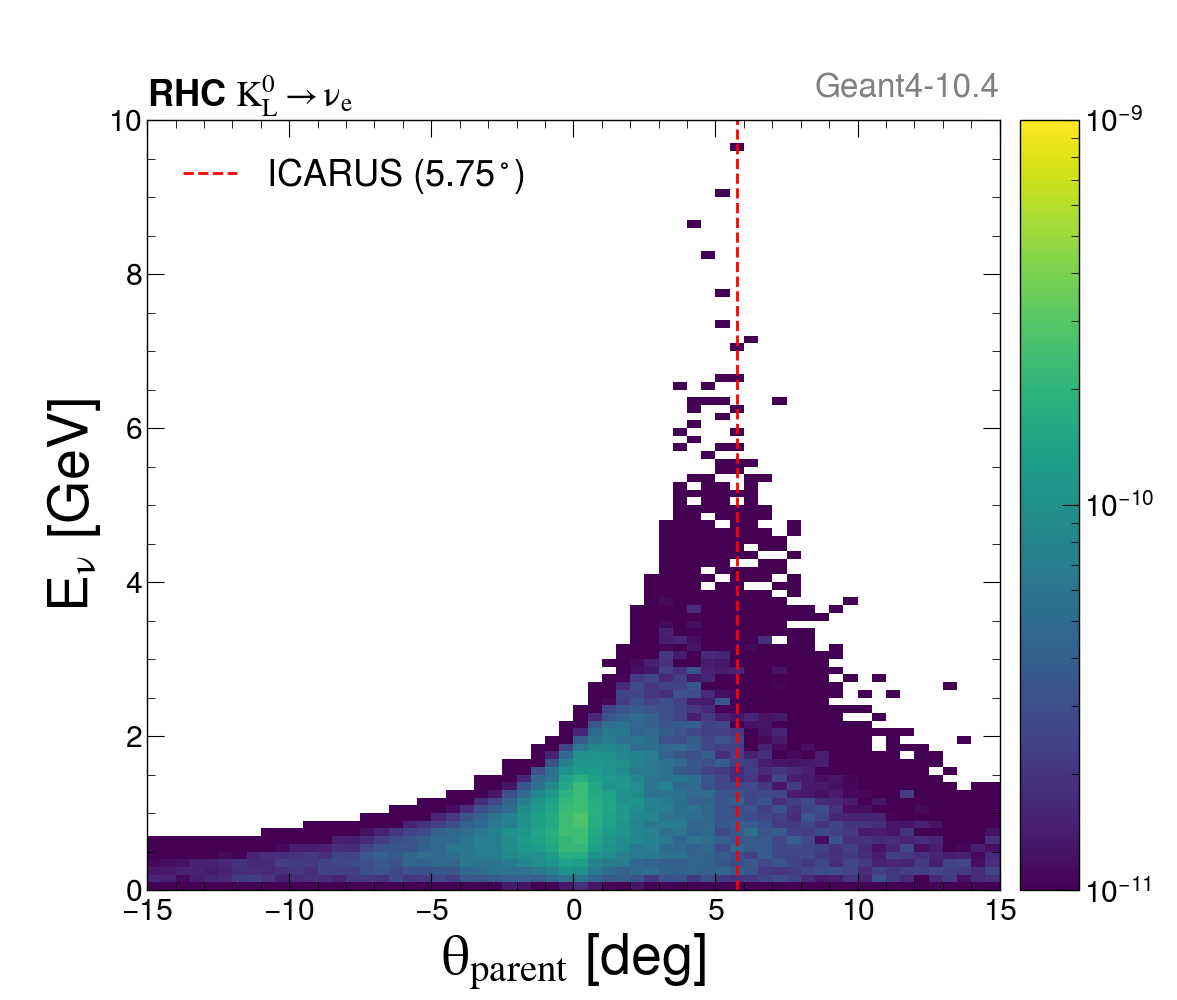}
    \caption{$K^0_L \to \nue$}
\end{subfigure}
\begin{subfigure}{0.23\textwidth}
    \includegraphics[width=\textwidth]{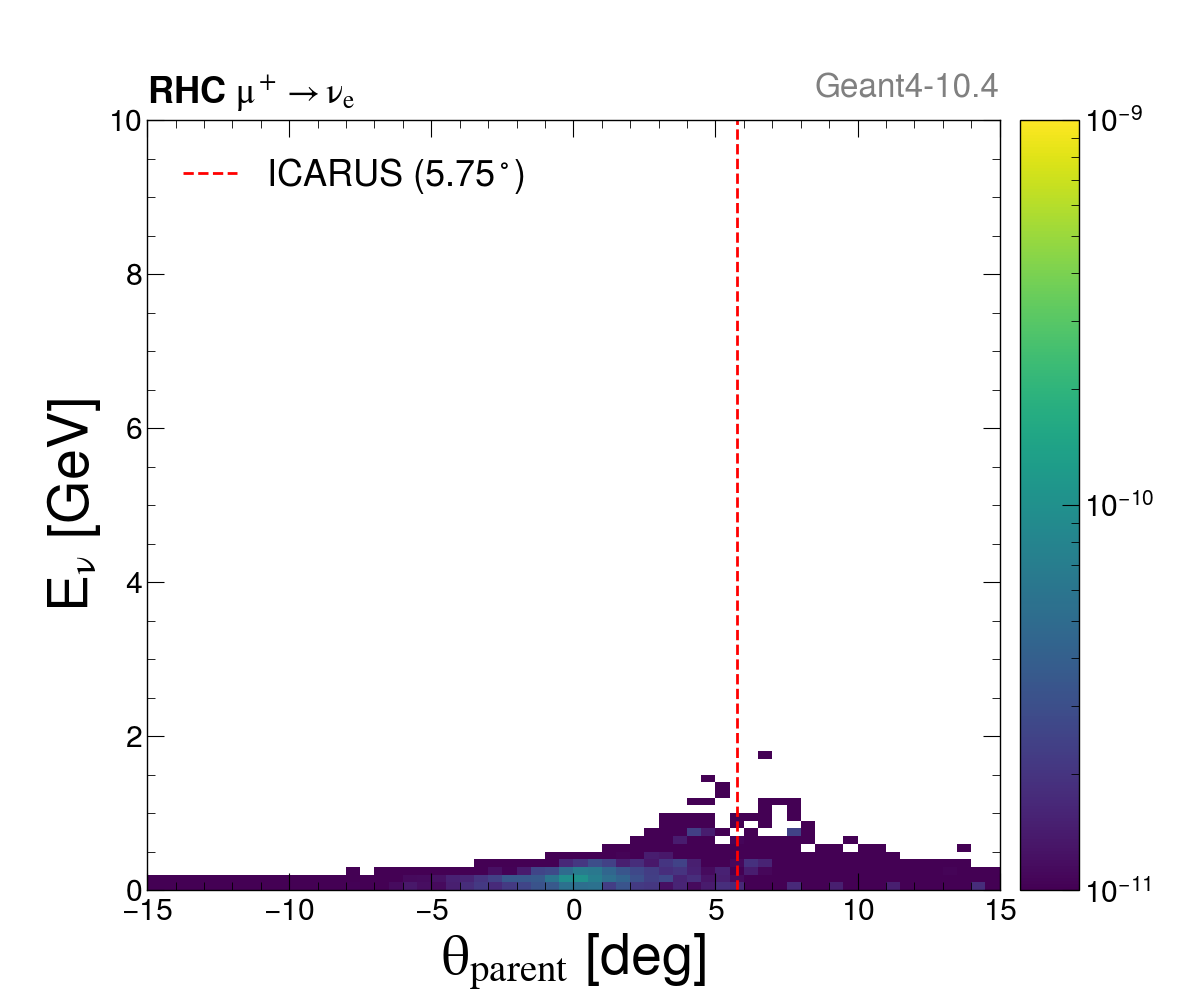}
    \caption{$\mu^+ \to \nue$}
\end{subfigure}

\begin{subfigure}{0.23\textwidth}
        \includegraphics[width=\textwidth]{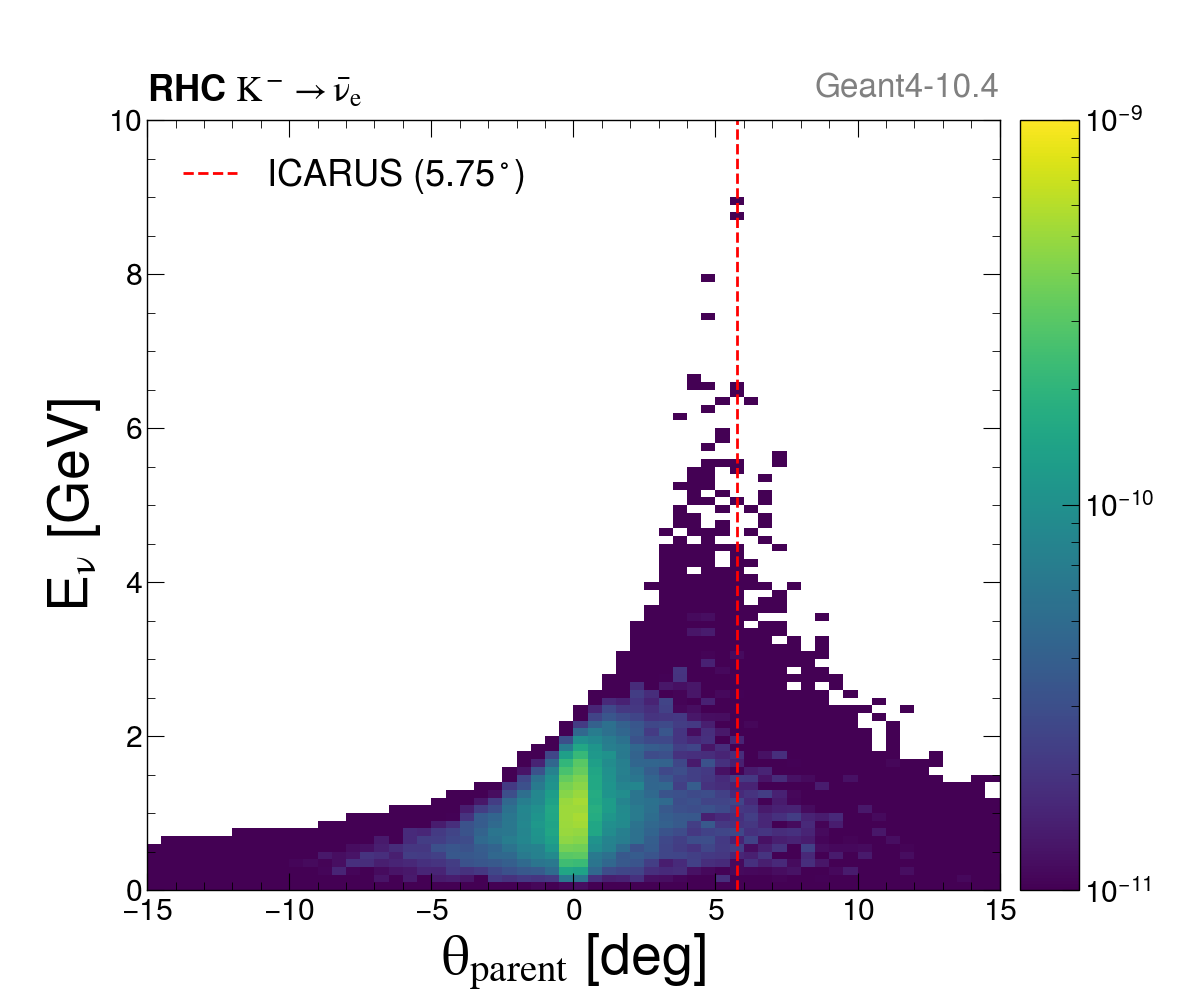}
            \caption{$K^- \to \nueb$}
\end{subfigure}
\begin{subfigure}{0.23\textwidth}
        \includegraphics[width=\textwidth]{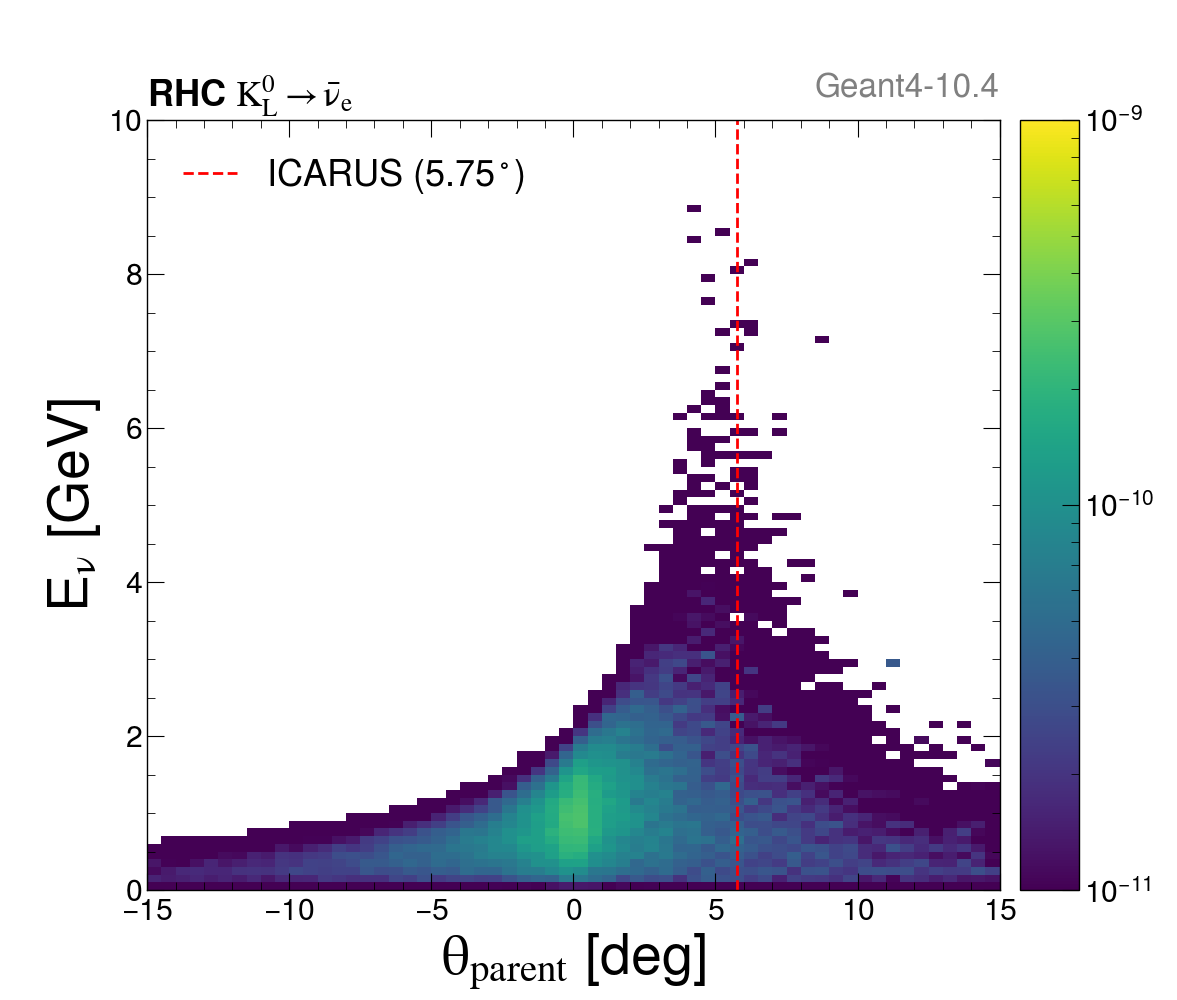}
            \caption{$K^0_L \to \nueb$}
\end{subfigure}
\begin{subfigure}{0.23\textwidth}
        \includegraphics[width=\textwidth]{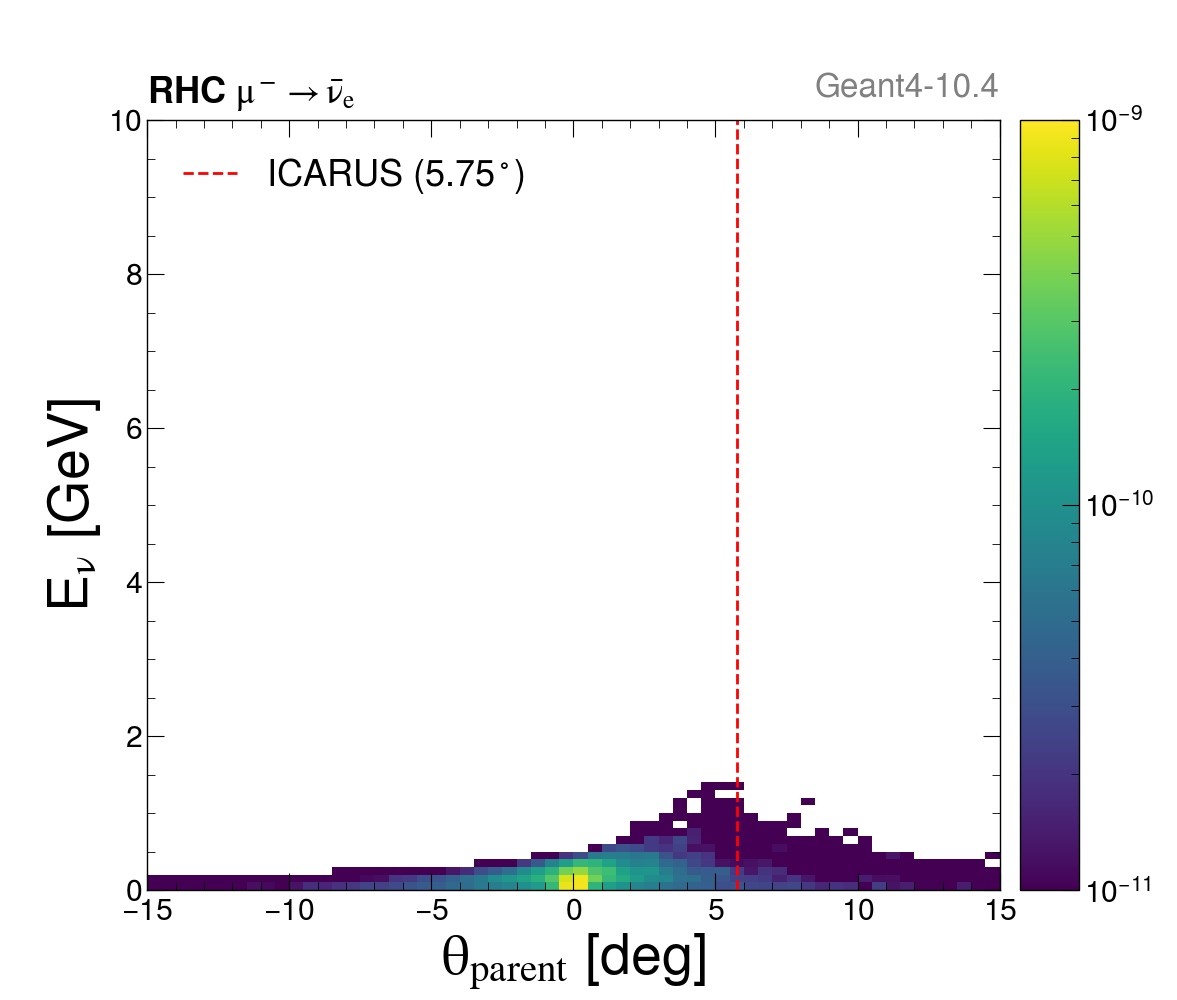}
            \caption{$\mu^- \to \nueb$}
\end{subfigure}
    \caption[Parent Decay Angle vs. Neutrino Energy (RHC)]{Parent decay angle vs. neutrino energy distributions in the reverse horn current mode.}
\end{figure}

%% file: flux_file_readme.tex
\textbf{GitHub Repository} \url{https://github.com/woodtp/flux-tool}\\

\noindent This package reads neutrino flux universes produced by Package to Predict the Flux (PPFX), and extracts a neutrino flux prediction with corresponding uncertainties.
All analysis products are output to a .root file specified in a config.toml. The package will also produce figures as pdf, png, and a .tex, for the majority of the products stored in the ROOT file.
\section{Prerequisites}

Before you begin, make sure you have the following prerequisites installed:

\begin{itemize}
	\item Python 3.11 or later: Visit the official Python website at \url{https://www.python.org/downloads} to download and install the latest version of Python.
	\item ROOT 6.28 or later: **Flux-Tool** requires ROOT/PyROOT version 6.28 or later. You can obtain ROOT from the official ROOT website at \url{https://root.cern/install}.
\end{itemize}

\section{Installation}

**Flux-Tool** is available for installation from PyPI, the Python Package Index. Follow the steps below to install the project from the terminal:

\begin{enumerate} 
	\item Create a virtual environment (optional but recommended):
	      \begin{verbatim} $ python -m venv myenv \end{verbatim}
	\item Activate the virtual environment:
	      \begin{verbatim}
    $ source myenv/bin/activate
\end{verbatim}
	\item Install Flux-Tool using pip:
	      \begin{verbatim}
    $ pip install flux-tool
\end{verbatim}
\end{enumerate}
\section{Usage}
\begin{verbatim}
$ flux_tool -h
usage: flux_uncertainties [-h] [-c CONFIG] [-p PRODUCTS_FILE] [-v] [-z]

This package coerces PPFX output into a neutrino flux prediction
with uncertainties, and stores various spectra related to the flux,
e.g., fractional uncertainties, covariance matrices, etc.

options:
  -h, --help            show this help message and exit
  -c CONFIG, --config CONFIG
                        specify the path to a toml configuration file
  -p PRODUCTS_FILE, --plots-only PRODUCTS_FILE
                        Specify path to an existing ROOT file for which
                        to produce plots
  -v, --verbose
  -z, --enable-compression
                        Enable compression of the output plots directory
\end{verbatim}
Alternatively, this package can be imported directly into an existing python script:

\begin{verbatim}
import flux_tool
\end{verbatim}

\subsection{Example config.toml}
\begin{verbatim}
# flux_tool configuration file

output_file_name = "out.root"
sources = "/path/to/directory/containing/input/histograms"

[Binning]
# Histogram bin edges for each neutrino flavor.
# Accepts:
#    1. an integer number of bins (between 0 and 20 GeV)
#    2. An array of bin edges (NOTE: they can be variable bin widths,
#       but must be monotonically increasing)
#    3. If unspecified, then fixed bin widths of 100 MeV is applied along
#       the [0, 20] GeV interval.
nue = 200
nuebar = [
  0.0,
  0.2,
  0.4,
  0.6,
  0.8,
  1.0,
  1.5,
  2.0,
  2.5,
  3.0,
  3.5,
  4.0,
  6.0,
  8.0,
  12.0,
]
numu = []
numubar = [
  0.0,
  0.2,
  0.4,
  0.6,
  0.8,
  1.0,
  1.5,
  2.0,
  2.5,
  3.0,
  3.5,
  4.0,
  6.0,
  8.0,
  12.0,
  20.0
]

  [PPFX]
# enable/disable specific PPFX reweight categories from
# appearing in the fractional uncertainty directory
# true = included, false = excluded
[PPFX.enabled]
attenuation = true
mesinc = true
mesinc_parent_K0 = true
mesinc_parent_Km = true
mesinc_parent_Kp = true
mesinc_parent_pim = true
mesinc_parent_pip = true
mesinc_daughter_K0 = true
mesinc_daughter_Km = true
mesinc_daughter_Kp = true
mesinc_daughter_pim = true
mesinc_daughter_pip = true
mippnumi = false
nua = true
pCfwd = false
pCk = true
pCpi = true
pCnu = true
pCQEL = false
others = true
thintarget = false

[Plotting]
draw_label = true # whether or not to draw the experiment label,
                  # e.g., ICARUS Preliminary
experiment = "ICARUS"
stage = "Preliminary"
neutrino_energy_range = [0.0, 6.0] # horizontal axis limits in [GeV]

[Plotting.enabled]
# Enable/disable specific plots from the visualization output
uncorrected_flux = true
flux_prediction = true
flux_prediction_parent_spectra = true
flux_prediction_parent_spectra_stacked = true
ppfx_universes = true
hadron_uncertainties = true
hadron_uncertainties_meson = true
hadron_uncertainties_meson_only = true
pca_scree_plot = true
pca_mesinc_overlay = true
pca_top_components = true
pca_variances = true
pca_components = true
hadron_covariance_matrices = true
hadron_correlation_matrices = true
beam_uncertainties = true
beam_covariance_matrices = true
beam_correlation_matrices = true
beam_systematic_shifts = true
\end{verbatim}

\section{Contents of the Output ROOT File}
\begin{itemize}
	\item beam\_samples
	      If provided to flux\_tool, copies of the systematically altered neutrino flux samples, including the nominal, are stored here.

	\item beam\_systematic\_shifts
	      Fractional shifts from the nominal, calculated for each flux sample in beam\_samples.
	\item covariance\_matrices
	      Contains all covariance and correlation matrices, organized into two subdirectories: one for hadron effects and another for beam effects (if applicable). Each covariance matrix is stored in 2 forms:

	      \begin{enumerate}
		      \item \verb+TH2D+ (prefixed hcov\_ or hcorr\_)
		      \item \verb+TMatrixD+ (prefixed cov\_ or corr\_)
	      \end{enumerate}

	      Covariance matrices with the \_abs suffix are in absolute units of the flux, whereas those without the suffix are normalized the PPFX universe mean, in the case of hadron systematics, or to the nominal beam run, in the case of the beam line systematics.
	      Each bin is labeled according to the combination of horn polarity, neutrino flavor, and energy bin number, e.g., fhc-nue-1.
	\item flux\_prediction
	      This directory holds a set of TH1D for each neutrino mode. The flux value is
	      extracted as the PPFX mean, while the uncertainties incorporate statistical,
	      hadron systematic, and beam line systematic (if applicable) uncertainties.
	\item fractional\_uncertainties
	      This directory contains two subdirectories, beam and hadron, containing the fractional contributions to the flux uncertainty for each effect.
	\item pca
	      This directory houses the outputs of the Principal Component Analysis of the hadron covariance matrix.
	      \begin{itemize}
		      \item \verb+eigenvectors/hevec_*+ Unit eigenvectors
		      \item \verb+principal_components/hpc_*+ principal components scaled by the square root of the corresponding eigenvalue and transposed into bins of neutrino energy
		      \item \verb+hcov_pca+ reconstructed hadron covariance matrix used for validation purpose.
		      \item \verb+heigenvals+ Each bin of this histogram (TH2D) holds the eigenvalues extracted from the PCA
		      \item \verb+heigenvals_frac+ same as the previous, but each eigenvalue is divided by
		            the sum of all eigenvalues such that each eigenvalue is represented as its contribution to the total variance.
	      \end{itemize}
	\item ppfx\_corrected\_flux
	      Directory containing the PPFX-corrected neutrino spectra. These histograms
	      are produced by calculating the means and sigmas of the flux distributions across
	      the 100 universes contained in ppfx\_output.
	\item ppfx\_flux\_weights
	      Directory containing TH1D for each horn-neutrino flavor combination, the bins of which contain weights that can be used to apply the PPFX  flux correction.
	\item ppfx\_output
	      Contains the original output received from PPFX, organized into two subdirectories corresponding to Forward Horn Current (FHC) and Reverse Horn Current (RHC). Each contains a nom subdirectory which holds the nominal (uncorrected) neutrino flux vs. energy spectrum, hnom\_nu*, in addition to the PPFX central value, hcv\_nu. Spectra broken down by parent hadron can be found under the parent subdirectory. The remaining subdirectories hold the universes for each hadron production
	      systematic:
	\item statistical\_uncertainties
	      Directory containing statistical uncertainties for every horn-neutrino flavor combination. Histograms with the suffix \_abs are in absolute units of the flux, and those without the suffix are in the fractional scale. The two matrices, hstatistical\_uncertainty\_matrix and statistical\_uncertainty\_matrix, are diagonal TH2D and TMatrixD, respectively, organizing the statistical uncertainties into a useful form to be added with covariance matrices.
	\item corr\_total
	      TMatrixD correlation matrix incorporating all sources of uncertainty
	\item cov\_total\_abs
	      TMatrixD covariance matrix in units of the flux, incorporating all sources of uncertainty
	\item hcorr\_total
	      TH2D correlation matrix incorporating all sources of uncertainty
	\item hcov\_total\_abs
	      TH2D covariance matrix in units of the flux, incorporating all sources of uncertainty
	\item matrix\_axis
	      TAxis with the binning and labels of all matrix axes
	\item xaxis\_variable\_bins
	      TAxis containing the binning applied to all spectra w.r.t. $E_\nu$ in GeV.
\end{itemize}